\documentclass[a4paper,12pt]{book}
\usepackage[english]{babel}
\usepackage[dvips]{graphicx}
\usepackage{fancyhdr}
\usepackage{latexsym}
\usepackage{amsfonts}
\usepackage{amssymb}
\usepackage{amsthm}
\usepackage{amsmath}
\usepackage{color}
\usepackage{colordvi}
\usepackage{axodraw}
\usepackage{array}

\usepackage{hyperref}

\fancypagestyle{plain}{ \fancyhead{}
 }

\fancypagestyle{index}{
\fancyhead[LE,RO]{\bfseries\thepage}
\fancyhead[LO]{\bfseries Contents}
\fancyhead[RE]{\bfseries Contents}

\addtolength{\headheight}{0.5pt}}

\fancypagestyle{introduction}{
\fancyhead[LE,RO]{\bfseries\thepage}
\fancyhead[LO]{\bfseries Introduction}
\fancyhead[RE]{\bfseries Introduction}

\addtolength{\headheight}{0.5pt}}

\fancypagestyle{references}{
\fancyhead[LE,RO]{\bfseries\thepage}
\fancyhead[LO]{\bfseries References}
\fancyhead[RE]{\bfseries References}

\addtolength{\headheight}{0.5pt}}

\fancypagestyle{appendixa}{
\fancyhead[LE,RO]{\bfseries\thepage}
\fancyhead[LO]{\bfseries Appendix A:  Structure functions in tau decays}
\fancyhead[RE]{\bfseries Appendix A:  Structure functions in tau decays}

\addtolength{\headheight}{0.5pt}}

\fancypagestyle{appendixb}{
\fancyhead[LE,RO]{\bfseries\thepage}
\fancyhead[LO]{\bfseries Appendix B: $\frac{\mathrm{d}\Gamma}{\mathrm{d}s_{ij}}$ in three-meson tau decays}
\fancyhead[RE]{\bfseries Appendix B: $\frac{\mathrm{d}\Gamma}{\mathrm{d}s_{ij}}$ in three-meson tau decays}

\addtolength{\headheight}{0.5pt}}

\fancypagestyle{appendixc}{
\fancyhead[LE,RO]{\bfseries\thepage}
\fancyhead[LO]{\bfseries Appendix C: Off-shell width of Vector resonances}
\fancyhead[RE]{\bfseries Appendix C: Off-shell width of Vector resonances}

\addtolength{\headheight}{0.5pt}}

\fancypagestyle{appendixd}{
\fancyhead[LE,RO]{\bfseries\thepage}
\fancyhead[LO]{\bfseries Appendix D: $\omega\rightarrow \pi^+\pi^-\pi^0$ within $R\chi T$}
\fancyhead[RE]{\bfseries Appendix D: $\omega\rightarrow \pi^+\pi^-\pi^0$ within $R\chi T$}

\addtolength{\headheight}{0.5pt}}

\fancypagestyle{appendixe}{
\fancyhead[LE,RO]{\bfseries\thepage}
\fancyhead[LO]{\bfseries Appendix E: Isospin relations between $\tau^-$ and $e^+e^-$ decay channels}
\fancyhead[RE]{\bfseries Appendix E: Isospin relations between $\tau^-$ and $e^+e^-$ decay channels}

\addtolength{\headheight}{0.5pt}}

\fancypagestyle{appendixf}{
\fancyhead[LE,RO]{\bfseries\thepage}
\fancyhead[LO]{\bfseries Appendix F:  Antisymmetric tensor formalism for meson resonances}
\fancyhead[RE]{\bfseries Appendix F:  Antisymmetric tensor formalism for meson resonances}

\addtolength{\headheight}{0.5pt}}

\fancypagestyle{appendixg}{
\fancyhead[LE,RO]{\bfseries\thepage}
\fancyhead[LO]{\bfseries Appendix G:  Successes of the large-$\N$ limit of $QCD$}
\fancyhead[RE]{\bfseries Appendix G:  Successes of the large-$\N$ limit of $QCD$}

\addtolength{\headheight}{0.5pt}}

\fancypagestyle{appendixh}{
\fancyhead[LE,RO]{\bfseries\thepage}
\fancyhead[LO]{\bfseries Appendix H:  Comparing theory to data}
\fancyhead[RE]{\bfseries Appendix H:  Comparing theory to data}

\addtolength{\headheight}{0.5pt}}

\fancypagestyle{conclusions}{
\fancyhead[LE,RO]{\bfseries\thepage}
\fancyhead[LO]{\bfseries Conclusions}
\fancyhead[RE]{\bfseries Conclusions}

\addtolength{\headheight}{0.5pt}}

\fancypagestyle{references}{
\fancyhead[LE,RO]{\bfseries\thepage}
\fancyhead[LO]{\bfseries Bibliography}
\fancyhead[RE]{\bfseries Bibliography}

\addtolength{\headheight}{0.5pt}}

\fancypagestyle{acknowledgements}{
\fancyhead[LE,RO]{\bfseries\thepage}
\fancyhead[LO]{\bfseries Acknowledgments}
\fancyhead[RE]{\bfseries Acknowledgments}

\addtolength{\headheight}{0.5pt}}

\title{Hadronic and radiative decays of the tau lepton: $\t^-\,\rightarrow\,(\pi\pi\pi)^-\nu_\t$, $\t^-\,\rightarrow\,(KK\pi)^-\nu_\t$, 
$\t^-\,\rightarrow\,\eta^{(\prime)}\pi^-\pi^0\nu_\t$, $\t^-\,\rightarrow\,(\pi/K)^- (\gamma)\nu_\t$}

\author{Pablo Roig Garc\'{e}s}

\begin{document}

\def\onebox{{\vbox{\hbox{$\sqr\thinspace$}}}}
\def\twobox{{\vbox{\hbox{$\sqr\sqr\thinspace$}}}}
\def\threebox{{\vbox{\hbox{$\sqr\sqr\sqr\thinspace$}}}}
\def\N{N_C}
\def\Tr{{\rm Tr}}
\def\CPT{\chi PT}
\def\RCPT{R\chi T}
\def\RCT{R\chi T}
\def\t{\tau}
\def\g{\gamma}
\def\p{\pi}
\def\mapright#1#2{\smash{
     \mathop{-\!\!\!-\!\!\!\rightarrow}\limits^{#1}_{#2}}}
     
\providecommand{\openone}{\leavevmode\hbox{\small1\kern-3.8pt\normalsize1}}

\newcommand{\cO}{{\cal O}}
\newcommand{\ket}{\,\rangle}
\newcommand{\bra}{\langle \,}
\newcommand{\strich}[1]{#1  \! \! \! \slash}
\newcommand{\gnr}{\Gamma_{\tau^- \to \nu_\tau P^-}}

\thispagestyle{empty}
\begin{center}
 \begin{huge}
  DEPARTAMENT DE F\'{I}SICA TE\`{O}RICA\\
\vspace{3.0cm}
  \textbf{Desintegraciones hadr\'{o}nicas y radiativas del lept\'on tau:
\vspace{0.1cm}
 $\t^-\,\rightarrow\,(P P P)^-\nu_\t$, $\t^-\,\rightarrow\,P^- \gamma \nu_\t$}\\
\vspace{3.0cm}
 PABLO ROIG GARC\'ES\\
\vspace{3.0cm}
 UNIVERSITAT DE VAL\`ENCIA\\
\vspace{0.1cm}
 Servei de Publicacions\\
\vspace{0.2cm}
 2010
 \end{huge}
\end{center}

\newpage
\thispagestyle{empty}
\begin{large}
\textbf{Aquesta Tesi Doctoral va ser presentada a Val\`{e}ncia el dia 15 de
novembre de 2010 davant un tribunal format per:\\
\\
\\
\\
- Dr. Arcadi Santamar\'{\i}a Luna\\
- Dr. Matthias Jamin\\
- Dr. Antonio Vairo\\
- Dr. Juan Jos\'{e} Sanz Cillero\\
- Dr. Ignasi Rosell Escrib\`{a}\\
\\
\\
\\
\\
Va ser dirigida per:\\
Dr. Jorge Portol\'es Ib\`a\~{n}ez\\
\\
\\
\\
\\
\\
\\
\\
$\textcopyright$Copyright: Servei de Publicacions\\
Pablo Roig Garc\'es\\
\\
\\
\\
\\
\\
\\
---------------------------------------------------------------------------------
\\
\\
Dip\`osit legal: V-3364-2011\\
I.S.B.N.: 978-84-370-7994-3\\
Edita: Universitat de Val\`encia\\
Servei de Publicacions\\
C/ Arts Gr\`afiques, 13 baix\\
46010 Val\`encia\\
Spain\\
Tel\`efon:(0034)963864115
}
\end{large}

\newpage
\thispagestyle{empty}

\begin{center}
\includegraphics[scale=0.125]{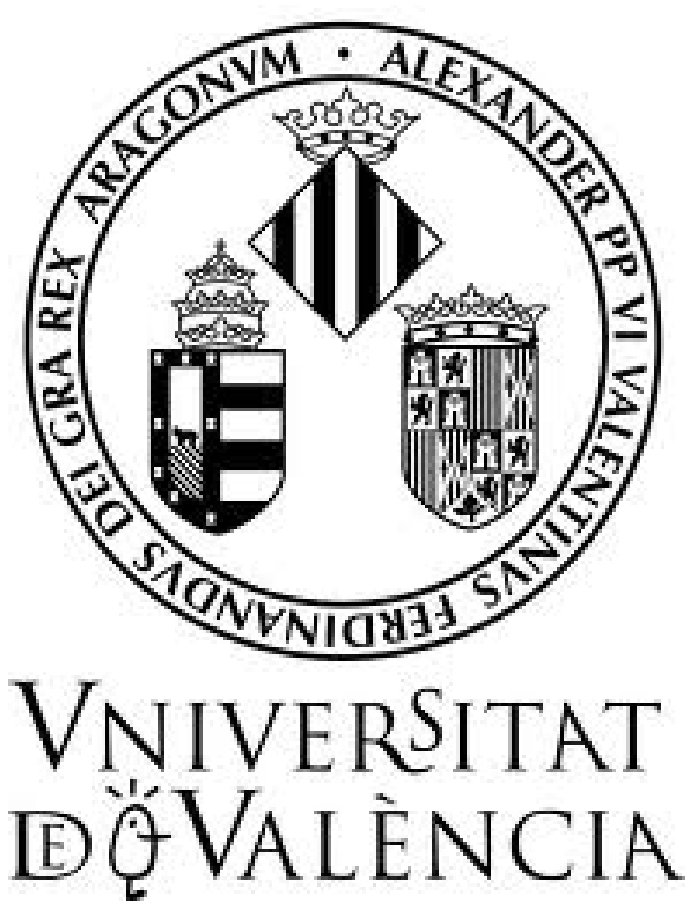}\\
\vspace{3cm}
\end{center}
\begin{huge}
\begin{center}
\textbf{Hadronic and radiative decays of the tau lepton:\\
\vspace{0.1cm}
$\t^-\,\rightarrow\,(\pi\pi\pi)^-\nu_\t$, $\t^-\,\rightarrow\,(KK\pi)^-\nu_\t$, 
$\t^-\,\rightarrow\,\eta^{(\prime)}\pi^-\pi^0\nu_\t$, $\t^-\,\rightarrow\,(\pi/K)^- (\gamma)\nu_\t$}
\end{center}
\end{huge}
\vspace{4 cm}

\begin{center}
\begin{large}
\textbf{
Pablo Roig Garc\'es}
\vspace{0.2cm}

IFIC, Departament de F\'{\i}sica Te\`{o}rica
\vspace{0.2cm}

PhD Thesis
\vspace{0.2cm}

PhD Advisor: Jorge Portol\'es Ib\'a\~{n}ez
\vspace{0.2cm}

Val\`encia, November, the 15th, 2010
\end{large}
\end{center}
1

\newpage
\thispagestyle{empty}
\phantom{hola!}
\vspace{3cm}
\hspace*{0.5cm}JORGE PORTOL\'ES IB\'A\~{N}EZ, Cient\'ifico Titular del Instituto de F\'{\i}sica Corpuscular de Val\`encia,\\
\\
\\
\\
\hspace*{2.0cm}CERTIFICA\\
\\
\\
\hspace*{2.0cm}Que la presente memoria ``Desintegraciones hadr\'{o}nicas y radiativas del\\
\hspace*{2.0cm} lept\'on tau'' ha sido realizada bajo su direcci\'on en el Departament de F\'{\i}-\\
\hspace*{2.0cm} sica Te\`orica de la Universitat de Val\`encia, por PABLO ROIG GARC\'ES\\
\hspace*{2.0cm} y constituye su Tesis para optar al grado de Doctor en F\'{\i}sica.\\
\\
\\
\hspace*{1.0cm}Y para que as\'i conste, en cumplimiento de la legislaci\'on vigente, presenta en el Departament de F\'{\i}sica Te\`orica 
de la Universitat de Val\`encia la referida Tesis Doctoral, y firma el presente certificado.\\
\\
\\
\\
\\
\\
\\
\\
\hspace*{8.0cm}Val\`encia, a 4 d'Octubre de 2010.\\
\\
\\
\\
\\
\\
\\
\\
\\
\\
\hspace*{10.0cm}Jorge Portol\'es Ib\'a\~{n}ez\\
2


\newpage
\thispagestyle{empty}
\phantom{hola!}
\vspace{3cm}
\begin{flushright}
\texttt{A mi madre, por todo y por siempre;\\ y a Maru, por 'aguantarme' durante\\ la redacci\'on de esta Tesis.}
\end{flushright}
\vspace{16.5cm}3

\frontmatter
\tableofcontents
\thispagestyle{index}
\include{contents}
\mainmatter
\chapter{Introduction}
\label{introduction}
\section{General introduction to the problem}  \label{Intro_introe}
\hspace*{0.5cm}This PhD Thesis studies those decays of the $\tau$ lepton including mesons among the final state particles,
that are called, for this reason, hadronic or semileptonic decays. Contrary to what happens to the other leptons (electron,
 $e^-$ and muon, $\mu^-$) its mass ($M_\tau\sim1.8$ GeV) is large enough to allow this kind of decays including light mesons 
(pions -$\pi$-, kaons -$K$- and eta particles -$\eta,\,\eta'$-). These processes include, in addition, the corresponding tau neutrino,
 $\nu_\t$, and may include (in the so-called radiative processes) multiple photons, $\gamma$. Although it is kinematically
 possible to produce other light mesons whose masses are lighter than $M_\tau$, the characteristic lifetime of these particles
 (resonances) is way too short to allow their detection. Notwithstanding, as we will see, the effect of their exchange is
 important in order to understand these decays. Although $M_\tau\sim2M_p$, where $M_p$ is the proton mass, since it is a bit smaller actually 
($M_\tau=1$.$777$ GeV and $2M_p=1$.$876$ GeV) and due to the conservation of baryon number (that would require producing them in
 baryon-antibaryon pairs, the lighter being proton-antiproton) $\t$ decays into baryons are forbidden.\\
\hspace*{0.5cm}Purely leptonic decays of the $\tau$ are processes mediated by the electroweak interaction, that today is adequately described in
 the framework of the Standard Model of Particle Physics ($SM$) \cite{SMEW}. In hadronic decays of the tau, the strong interaction acts, additionally. Although 
it is common lore that Quantum Chromodynamics, $QCD$ is the theory \cite{QCD} that describes it, we are not yet able to solve the problem at hand using 
only the $QCD$ Lagrangian, as we will explain.\\
\hspace*{0.5cm}Both difficulties mentioned above -the fact that the resonances are not asymptotic states, so that it is not possible to detect them,
 and our current inability to solve $QCD$ to find a solution to the problem- are telling us how interesting can the semileptonic tau decays be: 
 they provide a clean environment where to study the strong interaction at low and intermediate energies, because the electroweak part of the process
 is clean and under theoretical control; on the other side, given the range of energies the hadronic system can span 
 (essentially from the threshold for pion production, $\pi\sim0$.$14$ GeV, to $M_\tau$), the lightest resonances can be exchanged on-mass-shell (or simply 
on-shell, in what follows), in such a way that their effects are sizable and one can thus study their properties.\\
\hspace*{0.5cm}The strong interaction was discovered in the classical Rutherford's experiment, proving the existence of the atomic nucleus: it was a force of 
amazing strength and very short range. Soon after, this force started to be studied in scattering experiments between hadrons -as the particles experiencing 
strong interactions are called- (they were believed to be elementary then) instead of atomic nuclei. The large number of hadrons that was discovered in the '50s
 and '60s with the advent of the first particle accelerators suggested that these particles were not fundamental, analogously to what happened with the chemical
 elements, in the end constituted by protons and neutrons tighten in their nuclei. In order to complete the analogy, the systematics attached to the hadronic properties
 (in this context intimately related to the quantum numbers and specifically to the approximate flavour symmetries) helped to understand their substructure
 and Gell-Mann comprehended that hadrons had to be constituted by the so-called quarks, possessing an additional quantum number -christened as color-
 that solved all problems of adequacy of the observed spin and the quantum statistics obeyed by the particles.\\
\hspace*{0.5cm}However, whereas it is possible to unbind protons or neutrons from the nucleus -either through natural radioactivity
 or artificially- it has been impossible so far to free any constituent quark from the hadron to which it belongs. This property is known as
 confinement and, though there are theoretical reasons pointing to this phenomenon there is not yet any explanation of it. This fact
 imposes a difficulty when understanding the processes mediated by the strong interaction: while they are produced between quarks and gluons
 (the intermediary $QCD$ bosons, that also autointeract), our detectors are recording only (in this context) hadronic events, since 
quarks, antiquarks and gluons cluster immediately after they are produced into objects with vanishing total color charge: they hadronize.
 Contrary to the remaining known forces (electromagnetic, weak and gravitational), in $QCD$ the force does not diminish as the distance increases, 
but just the opposite -even though its range is very short-.\\
\hspace*{0.5cm}Although we will dwell into this later on, let it be enough for the moment saying that even though $QCD$ is the theory of strong interactions,
 we can not handle it in its non-perturbative regime and, therefore, computationally, its Lagrangian only allows to tackle analytically 
 inclusive processes at high energies ($E>2$ GeV typically) where a treatment in terms of quarks, antiquarks and gluons is
 meaningful. At the lower energies where our problem occurs we must search an alternative way. As it uses to happen in Physics, 
an appropriate choice of the degrees of freedom simplifies (or even allows) the solution. In our study of several exclusive tau decay channels 
it is evident that these will be the lighter mesons and resonances. The most convenient and rigorous way to do this is the use of Effective Field Theories 
($EFTs$), that preserve the symmetries of the fundamental theory and are written in terms of the relevant degrees of freedom in a given energy range.
 As $\tau$ decays typically happen in an energy scale densely populated by resonances, it would not be enough to employ Chiral Perturbation Theory ($\CPT$) 
\cite{Weinberg:1978kz, Gasser:1983yg, Gasser:1984gg} that includes only the lightest pseudoescalar  mesons ($\pi$, K, $\eta$). Instead, it will be necessary
 to incorporate the resonances as active degrees of freedom to extend the theory to higher energies: Resonance Chiral Theory ($R\chi T$) \cite{Ecker:1988te, 
Ecker:1989yg} is the tool allowing these developments.\\
\hspace*{0.5cm}In the remainder of the Introduction we will elaborate in greater detail on the relevant aspects of tau Physics that are not treated in 
later chapters and on $QCD$ and our limitations when implementing its solutions in Hadronic Physics at low and intermediate energies. After that,
 we will summarize some of the capital ideas underlying our theoretical framework concerning $EFTs$, Chiral Perturbation Theory,
 the large number of colours limit of $QCD$ and Resonance Chiral Theory itself. We will finish by enumerating the different chapters of this Thesis.\\
\section{$\t$ Physics}\label{Intro_TauPhysicse}
\begin{figure}[h]
\centering
\includegraphics[scale=0.8]{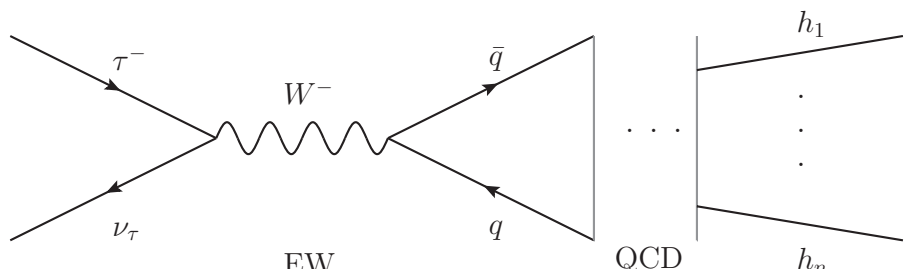}
\caption{Feynman diagram for the leading contribution to a generic hadronic decay of the $\tau$ lepton.}
\label{deshadtaugral}
\end{figure}

\hspace*{0.5cm}We begin with a short introduction to tau physics that will allow us to contextualize adequately this work: The $\tau$ lepton is
 a member of the third generation which decays into particles belonging to the first two and including light flavours, in addition to its neutrino, $\nu_\tau$.
 This is the reason why tau physics could give us
 useful hints in order to understand why there are (at least three) lepton and quark copies that only differ by their masses \cite{Amsler:2008zzb} (Table 1.1):
\\
\begin{table}[h!] \label{Table_Masses_SMe}
\begin{center}
\renewcommand{\arraystretch}{1.2}
\begin{tabular}{|c|c|c|c|c|}
\hline
Generation & quarks & $m_q$ & leptons & $m_l$ \\
\hline
1 & $d$ & $4$.$8^{+0.7}_{-0.3}$ MeV & $e^-$ & $\sim0$.$5110$ MeV \\
  & $u$ & $2$.$3^{+0.7}_{-0.5}$ MeV & $\nu_e$ & $< 2$ eV\\
2 & $s$ & $95\pm5$ MeV & $\mu^-$ & $\sim105$.$7$ MeV \\
  & $c$ & $1$.$275\pm0.025$ GeV & $\nu_\mu$ & $< 0$.$19$ MeV\\
3 & $b$ & $4$.$18\pm0.03$ GeV & $\tau^-$ & $\sim1$.$777$ GeV\\
  & $t$ & $173.5\pm0.6\pm0.8$ GeV & $\nu_\tau$ & $<18$.$2$ MeV\\
\hline
\end{tabular}
\caption{\small{Matter content of the $SM$ (updated from the PDG 2012 \cite{Beringer:1900zz}). Quark masses correspond to the $\overline{MS}$ scheme with renormalization scale $\mu=m_q$
 for heavy quarks ($c$, $b$) and $\mu=2$ GeV for light quarks. For the latter the given value of $m_q$ is an estimate for the so-called current quark mass.
 In the case of the $t$ quark it comes from direct Tevatron measurements. 
 Neutrino masses are the so-called ``effective masses``: $m_{\nu_k}^{2\,eff}=\sum_i |U_{ki}|^2 m_{\nu_i}^2$. However, cosmological data
 allows to set a much lower bound \cite{Komatsu:2010fb} $\sum_k(m_{\nu_k})<0$.$58$ eV.}}
\end{center}
\end{table}
\\
\hspace*{0.5cm}It seems reasonable that the heavier fermions will be more sensitive to the generation of fermion masses. Being the top quark
 too heavy to hadronize before decaying, $b$ quark and $\tau$ lepton physics seems promising in this respect.
 Although the value of $M_\tau$ does not allow for decays into charmed mesons (containing a $c$ quark), the $\t$ is the only lepton massive enough
 to decay into hadrons and thus relate somehow the (light) quark and lepton sectors.\\
\hspace*{0.5cm}The group led by M.~L.~Perl \cite{Perl:1975bf} discovered the $\tau$ in 1975. This constituted the first experimental evidence in favor of
 the existence of third generation particles and, consequently, the first indication that it was possible to accommodate $CP$ violation within the $SM$
 \cite{Kobayashi:1973fv}, that had already been observed \cite{Christenson:1964fg} in the neutral kaon system, through the
 Cabibbo, Kobayashi and Maskawa \cite{Kobayashi:1973fv, Cabibbo:1963yz} mixing matrix. Anyhow, this explanation is not enough to understand the enormous
 observed abundance of matter over antimatter in our Universe \cite{Komatsu:2010fb,Spergel:2003cb,Spergel:2006hy,Kowalski:2008ez,Komatsu:2008hk}, which represents one of the
 indications for the existence of some type of new physics beyond the $SM$~\footnote{The recent experimental results concerning
 the ratio of matter and antimatter galactic and extragalactic fluxes \cite{:2008zzr, Adriani:2008zq, Adriani:2008zr, Abdo:2009zk, Aharonian:2009ah, Adriani:2010rc, Adriani:2011xv, Adriani:2011cu} do not have 
-for the moment- a universally accepted explanation.}. $\t$ quantum numbers were established \cite{Perl:1977se} almost simultaneously to the discovery
 of the next particle of its generation, the $b$ quark \cite{Herb:1977ek}. The heaviest particle known so far, the top quark, was not detected until
 1995 \cite{Abe:1995hr, Abachi:1995iq}.\\
\hspace*{0.5cm}In the following, I will summarize briefly the Physics one can learn from $\tau$ decays \cite{Pich:2009zz, Pich:2008ni, Pich:2000rj, Barish:1987nj, Stahl:2000aq, Davier:2005xq}.\\
\hspace*{0.5cm}First of all, these decays allow to verify the universality of electroweak currents, both neutral and charged, with an amazing precision. Within the $SM$, the easiest 
leptonic $\tau$ decays are described by the following partial width \cite{Marciano:1988vm, Braaten:1990ef, Erler:2002mv}:
\begin{equation} \label{des_lep_taue}
\Gamma(\tau^-\,\to\,\nu_\tau\,l^-\,\overline{\nu}_l)\,=\,\frac{G_F^2\,M_\tau^5}{192\,\pi^3}\,f\left(\frac{m_l^2}{M_\tau^2}\right)\,r_{EW},
\end{equation}
where $f(x)=1-8x+8x^3-x^4-12x^2\mathrm{log}x$ and $r_{EW}$ includes the electroweak radiative corrections not incorporated in
 the Fermi constant, $G_F$, and the non-local structure of the $W$ propagator, $r_{EW}\sim0.9960$. Therefore, the ratio
\begin{equation}
\frac{\mathcal{B}(\tau\,\to\,e\nu_\tau\bar{\nu}_e)}{\mathcal{B}(\tau\,\to\,\mu\nu_\tau\bar{\nu}_\mu)}\,=\,\frac{f\left(\frac{m_e^2}{M_\tau^2}\right)}{f\left(\frac{m_\mu^2}{M_\tau^2}\right)}
\end{equation}
is fixed and allows to verify the universality of the charged electroweak currents. The resulting restriction is impressively strong \cite{PichTau12}:
\begin{equation}
 \frac{\mathcal{B}(\tau\,\to\,\mu\nu_\tau\bar{\nu}_\mu)}{\mathcal{B}(\tau\,\to\,e\nu_\tau\bar{\nu}_e)}\,=0.972559(5)\,,
\end{equation}
which is roughly three orders of magnitude more precise than the experimental ratio: $0.9761\pm0.0028$. The associated errors come from the current indetermination of $\sim0$.$16$ MeV
 in $M_\tau$. This uncertainty is much larger \cite{Abe:2006vf, Anashin:2007zz} in the ratio among the branching ratio, $BR$, to leptons and the tau lifetime: $\tau_\tau/\mathcal{B}(\tau\,\to\,e\nu_\tau\bar{\nu}_e)=(1632.9\pm0.6)\cdot10^{-15} s$
 -that is fixed using the value for the Fermi constant extracted from the $\mu$ decay- (it appeared to the fifth power in Eq. (\ref{des_lep_taue})). This uncertainty has already been  
 improved substantially by $BES-III$ and $KEDR$ \cite{Beringer:1900zz}. $KEDR$ foresees to reach a total error of $0$.$15$ MeV and $BES-III$ of only $0$.$10$ MeV that could further reduce the error in 
$\tau_\tau/\mathcal{B}(\tau\,\to\,e\nu_\tau\bar{\nu}_e)$. We will not discuss here another interesting issue as it is the determination of the $\tau$ mass close to its production threshold \cite{Asner:2008nq}.\\
\hspace*{0.5cm}In the $SM$, all leptons with the same electric charge have identical couplings to the $Z$ boson. This has been verified at $LEP$ and $SLC$
 for years. The precision reached for the neutral current axial-vector couplings ($a_i$) is even better than for the vectors ($v_i$) \cite{Pich:1999uk}:
\begin{equation}
\frac{a_\mu}{a_e}\,=\,1\mathrm{.}0001(14), \quad\frac{v_\mu}{v_e}\,=\,0\mathrm{.}981(82);\quad
\frac{a_\tau}{a_e}\,=\,1\mathrm{.}0019(15),\quad \frac{v_\tau}{v_e}\,=\,0\mathrm{.}964(32).
\end{equation}
\hspace*{0.5cm}From $\tau^-\,\to\,\nu_\tau\,l^-\,\overline{\nu}_l$ it was also obtained an upper limit for the standard and non-standard couplings 
\cite{Michel:1949qe, Bouchiat:1957zz, Kinoshita:1957zz, Kinoshita:1957zza, Fetscher:1986uj, Pich:1995vj}
 between left/right-handed ($L/R$) fermion currents: $RR$, $LR$ , $RL$, $LL$; scalars, vectors and tensors ($S$, $V$, $T$).
 It is remarkable that the current limits do not allow to conclude that the transition is of the predicted type: $V-A$ \cite{Feynman:1958ty}.
 Both these bounds and those on the '\textit{Michel parameters}'~\cite{Michel:1949qe}
 \footnote{These parameters describe the phase-space distribution in the $\mu$ and $\t$ leptonic decays.}
 will improve sizably in $BES-III$ \cite{Asner:2008nq, Shamov:2008zz}.\\
\hspace*{0.5cm}In the minimal $SM$ with massless neutrinos, there is an additive lepton number that is conserved separately for each generation
 (the so-called lepton flavour number). Notwithstanding, and after the evidence of the neutrino oscillation $\nu_\mu \to \nu_e$
announced by $LSND$ \cite{Athanassopoulos:1996jb, Aguilar:2001ty}, the confirmation through the oscillation measurements published by $SNO$ \cite{Ahmad:2001an, Ahmad:2002jz}
 and Super-Kamiokande \cite{Hirata:1992ku, Fukuda:1998mi} discards the former hypothesis of the minimal $SM$. There is therefore no doubt that this generation quantum number
conservation is violated in processes involving neutrinos \footnote{See the dedicated review of PDG2010 \cite{Nakamura:2010zzi}, for more details.}. Though the most natural 
thing would be that this also happens in processes with $e,\mu,\tau$, all current data
 are consistent with that conservation law in this lepton subsector. Despite the limits on lepton flavour violation coming from $\t$ decays are improving continuously
($br\leq10^{-7}$) \cite{Lees:2010ez, :2009tk, Aubert:2009ys, Aubert:2007kx, Aubert:2007pw, Aubert:2006cz, Aubert:2005wa, Aubert:2005tp, Aubert:2003pc, Miyazaki:2011xe, Miyazaki:2010qb, :2009wc,
Nishio:2008zx, Miyazaki:2008mw, Miyazaki:2007zw, Miyazaki:2007jp, Hayasaka:2007vc, Yusa:2006qq, Enari:2005gc, Enari:2004ax, Abe:2003sx}
, they are still far than the existing limits in $\mu$ decays with analogous violation \cite{Pich:2008ni}.\\
\hspace*{0.5cm}$FERMILAB$ in Chicago became a fundamental tool in the discovery of the other members of the third generation. There, the $b$ quark was discovered in 1977, 
the top quark in 1994-5 (with the help of the valuable LEP prophecy, as it will be explained) and, finally, the $DONUT$ experiment \cite{Kodama:2000mp} succeeded in 
the discovery of the remaining particle, the tau neutrino, in 2000. In fact, the best direct limits on the neutrino mass, $m_{\nu_\tau}$, come from hadronic $\t$ decays \cite{Barate:1997zg}:
 $\t^-\,\to\,\nu_\tau\,X^-$, where $X^-=(\pi\,\pi\,\pi)^-, (2K\,\pi)^-, (5\pi)^-$. Although -as it was already written- they are clearly superseded 
by the bounds coming from cosmological observations, our study might be able to improve the direct limits.\\
\hspace*{0.5cm}Besides, semileptonic $\t$ decays are the ideal benchmark to study strong interaction effects in very clean conditions, since the electroweak 
part of the decay is controlled theoretically -to much more precision than the hadronic uncertainties even taking the $LO$ contribution- and it does not get
 polluted by the hadronization that the process involves. These decays probe the hadronic matrix element of the left-handed charged current between the $QCD$
 vacuum and the final state hadrons, as it is represented in Figure \ref{deshadtaugral}. We will devote more attention to this topic, central in the Thesis,
 throughout this work.\\
\hspace*{0.5cm}A direct test of $QCD$ \cite{Pich:2008ge} can be made through the ratio
\begin{equation} \label{R_tau}
R_\t\,\equiv\,\frac{\Gamma(\t^-\,\to\,\nu_\tau\,\mathrm{hadrons}\,(\gamma))}{\Gamma(\t^-\,\to\,\nu_\tau\,e^-\,\nu_e\,(\gamma))}\,=\,R_{\t,V}\,+\,R_{\t,A}\,+\,R_{\t,S}\,,
\end{equation}
that splits the contributions from vector ($V$) and axial-vector ($A$) currents corresponding to an even/odd number of final-state pions
 from those with an odd number of kaons ($S$ is short for strangeness changing processes). Given channels (like $KK\pi$) can not be
 associated \textit{a priori} to $V$ or $A$ current. In this case, it is specially important to consider a study as ours in order to know the relative contributions
 of each one to the partial width of that channel.\\
\hspace*{0.5cm}The theoretical prediction requires the appropriate two-point correlation functions of left-handed ($LH$) quark currents:
$\, L^{\mu}_{ij} = \bar{\psi}_j \gamma^{\mu} (1-\gamma_5) \psi_i \;,\,\, $ ($i,j=u,d,s$):
\begin{equation} \label{eq:pi_ve}
\Pi^{\mu \nu}_{ij}(q) \equiv 
 i \int d^4x \, e^{iqx}\;
\langle 0|T(L^{\mu}_{ij}(x) L^{\nu}_{ij}(0)^\dagger)|0\rangle
=
  \left( -g^{\mu\nu} q^2 + q^{\mu} q^{\nu}\right) \, \Pi_{ij}^{(1)}(q^2)
  \,  +\,   q^{\mu} q^{\nu} \, \Pi_{ij}^{(0)}(q^2) \, .
\end{equation}
Using the property of analyticity, $R_\tau$ can be written as a contour integral in the complex $s$-plane, where the circuit is followed counterclockwise
 around a circle of radius $|s|=M_\tau^2$ centered at the origin:
\begin{equation} \label{eq:circlee}
 R_\tau \, =\, 6 \pi i \oint_{|s|=M_\tau^2} {ds \over M_\tau^2} \,
 \left(1 - {s \over M_\tau^2}\right)^2\,
\left[ \left(1 + 2 {s \over M_\tau^2}\right) \Pi^{(0+1)}(s)
         - 2 {s \over M_\tau^2} \Pi^{(0)}(s) \right] \, ,
\end{equation}
where \ $\Pi^{(J)}(s) \equiv |V_{ud}|^2 \, \Pi^{(J)}_{ud}(s) + |V_{us}|^2 \, \Pi^{(J)}_{us}(s)$.
In (\ref{eq:circlee}) one needs to know the correlators for complex $s\sim M_\tau^2$ only, that is larger than the scale associated to non-perturbative effects.
 Using the Operator Product Expansion, $OPE$ \cite{Wilson:1969zs, Wilson:1972ee, Zimmermann:1972tv}, to evaluate the contour integral
, $R_\tau$ can be written as an expansion in powers of $1/M_\tau^2$.\\
%
\hspace*{0.5cm}Therefore, the theoretical prediction of $R_{\t,V+A}$ can be written as follows \cite{Braaten:1991qm}:
\begin{equation}
R_{\t,V+A}\,=\,N_C\,|V_{ud}|^2\,S_{EW}\,(1\,+\,\delta'_{EW}\,+\,\delta_P\,+\,\delta_{NP})\,,
\end{equation}
where $N_C\,=\,3$ is the number of colours. $S_{EW}$ and $\delta'_{EW}$ contain the known electroweak corrections to leading and subleading orders
 in the logarithmic approximation. One can show and check that the non-perturbative corrections are small \cite{Braaten:1991qm}.
 The leading correction ($\sim20\%$) comes from the purely perturbative $QCD$ correction, $\delta_P$, that is known up to $\mathcal{O}\left(\alpha_S^4\right)$ \cite{Braaten:1991qm, Baikov:2008jh}
 and includes a resummation of the most relevant effects at higher orders \cite{Marciano:1988vm, Braaten:1990ef, Erler:2002mv,Baikov:2008jh,Le Diberder:1992te, Pich:1994zx, Pivovarov:1991rh}.
 The final result \cite{Davier:2005xq,Barate:1998uf,Ackerstaff:1998yj,Davier:2008sk} turns out to be extremely sensitive to the $\alpha_s(M_\t^2)$ value,
 allowing to determine very precisely the $QCD$ coupling that - in the $\overline{MS}$ scheme- is \cite{Pich:2010xb} \footnote{See in these references all works
 employed in this determination.}:
\begin{equation} \label{alphas_mtaue}
\alpha_s(M_\t^2)\,=\,0\mathrm{.}334\pm0\mathrm{.}014\,.
\end{equation}
\hspace*{0.5cm}When one uses the Renormalization Group Equations, $RGE$, to run this value up to the $Z$ scale \cite{Rodrigo:1997zd}
 one gets:
\begin{equation}
\alpha_s(M_Z^2)\,=\,0\mathrm{.}1204\pm0\mathrm{.}0016\,,
\end{equation}
while the value obtained in hadronic $Z$-boson decays \cite{Amsler:2008zzb} is \footnote{The complete QCD corrections to hadronic $Z$ decays at Order $\alpha_s^4$ in the limit of 
the heavy top quark mass have been computed recently \cite{Baikov:2012er}. These yield $\alpha_s(M_Z^2)\,=\,0$.$1190\pm0.0026$, completely dominated by the experimental error, 
and in even better agreement with the value extracted from $\tau$ decay data.}
\begin{equation} \label{alphas_MZe}
\alpha_s(M_Z^2)\,=\,0\mathrm{.}1176\pm0\mathrm{.}0020\,.
\end{equation}
Therefore, there is agreement between the extraction from hadronic tau decays and the direct measurement at the $Z$ peak, even with better precision in the 
first case. This provides a beautiful test of the change of the $QCD$ coupling with the scale, that is, a significant experimental test of \textit{asymptotic 
freedom}.\\
\hspace*{0.5cm}A comment with respect to the reliability of the errors in the result (\ref{alphas_mtaue}) is in order. This study assumes that the quark-hadron
 duality \cite{Poggio:1975af} is realized. Its violations \cite{Shifman:2000jv} and the error induced in the analysis that rely on it -like the determination
 of $\alpha_S$- is being an active area of research \cite{Cata:2005zj, Tesina_Martin, Cata:2008ye, Cata:2008ru, GonzalezAlonso:2010rn, Jamin:2011vd, Boito:2011qt, Boito:2012cr, Beneke:2012vb}
 recently, after being ignored systematically for years. Although the last works point to an understimation of the error induced,
 this issue is not clear yet and one would need on one hand more realistic theoretical models to estimate these violations 
(essentially only one is used that is based on resonances and less frequently, another one that is instanton based; both developed by Shifman \cite{Shifman:2000jv}), 
and on the other hand more quality experimental data to quantify precisely this effect.\\
\hspace*{0.5cm}The separate measurement of the decay widths of $|\Delta S|=0$ and $|\Delta S|=1$ processes gives us the opportunity to be
 sensitive to the effect of $SU(3)$ flavour symmetry breaking induced by the strange quark mass. Specifically, this happens through the differences
\begin{equation}
 \delta R_\t^{kl}=\frac{R_{\t,V+A}^{kl}}{|V_{ud}|^2}-\frac{R_{\t,S}^{kl}}{|V_{us}|^2}\approx24\frac{m_s^2(M_\t^2)}{M_\t^2}\Delta_{kl}(\alpha_S)-48\pi^2\frac{\delta O_4}{M_\t^4}Q_{kl}(\alpha_S)\,,
\end{equation}
where the spectral moments $R_\t^{kl}$ were introduced:
\begin{equation}
 R_\t^{kl}=\int_0^{M_\t^2}\mathrm{d}s\left( 1-\frac{s}{M_\t^2}\right)^k\left( \frac{s}{M_\t^2} \right)^l\frac{dR_\t}{ds}\,.
\end{equation}
\hspace*{0.5cm}The perturbative corrections $\Delta_{kl}(\alpha_S)$ and $Q_{kl}(\alpha_S)$ are known to $\mathcal{O}(\alpha_S^3)$ and $\mathcal{O}(\alpha_S^2)$,
 respectively \cite{Pich:1998yn, Pich:1999hc, Baikov:2004tk} and $\delta O_4$, proportional to the $SU(3)$ breaking- since it is to the difference of the $s$
 and $u$ quark condensates- is well determined \cite{Gamiz:2004ar}. Although in the future, with exceptional quality data, it would be possible to determine
 both $m_s(M_\t)$ and $|V_{us}|$ simultaneously analizing the whole set of spectral moments, in the most recent determination \cite{Jamin:2006tj} one fixes:
\begin{equation}
m_s(2\,\mathrm{GeV})\,=\,94\pm6\,\mathrm{MeV}\,,
\end{equation}
-obtained with the latest lattice determinations and using $QCD$ sum rules- in such a way that the most sensitive moment to $|V_{us}|$, with $kl=00$, 
allows to extract \cite{Gamiz:2007qs}:
\begin{equation}\label{V_us}
|V_{us}|=0\mathrm{.}2208\pm0\mathrm{.}0033_{\mathrm{exp}}\pm0\mathrm{.}0009_{\mathrm{th}}\,=\,0\mathrm{.}2208\pm0\mathrm{.}0034\,,
\end{equation}
that can be competitive with the standard extraction of $|V_{us}|$ from $K_{e3}$ decays \cite{Sciascia:2008fr, Antonelli:2010yf} and with the new proposals made for this 
determination. Moreover, the error associated to this extraction of $|V_{us}|$ can be reduced in the future since it is dominated by the experimental
 uncertainty that would be reduced notably in forthcoming years thanks to $B$-factory data \cite{O'Leary:2010af}. This is of great interest, since the value in Eq.(\ref{V_us}) 
is only marginally consistent with the PDG 2012 \cite{Beringer:1900zz} average: $|V_{us}|=0.2252\pm0.0009$. As suggested, another improvement that one could make would 
consist in fitting simultaneously $|V_{us}|$ and $m_s$ to a set of moments of the invariant mass distribution in hadronic tau decays, that would provide even more
 precision in the extraction of both parameters.\\
\hspace*{0.5cm}Up to now, all experimental results on the $\t$ are consistent with the $SM$. However, the analysis of $B$-factories data from $BaBar$ and $Belle$
 -and future experiments in the latter- or facilities dedicated to $\t-c$ production, as $BES$ are promising in order to obtain more and more stringent verifications
 of the $SM$ and explore the physics beyond it \footnote{In the Chapter \ref{Hadrondecays} we explain the importance of hadronic $\t$ decays in order to find out 
what the scalar sector of the $SM$ is and eventually to determine the mass of a light Higgs.}.\\
\section{$QCD$: The theory of strong interactions} \label{Intro_QCD:stronginteractiontheorye}
\begin{figure}[h]
\centering
\includegraphics[scale=0.7]{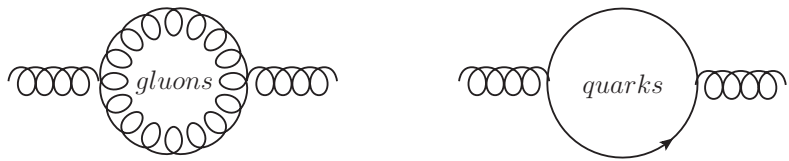}
\caption{Feynman diagrams for the one-loop contribution to the $\beta_{QCD}$ function.}
\label{diagram 1_loop_beta_function}
\end{figure}

\hspace*{0.5cm}Next we will introduce briefly $QCD$ in order to explain why it is not possible to solve the problems we are dealing with analytically and completely.\\
\hspace*{0.5cm}Deep inelastic scattering experiments at $SLAC$ \cite{Bloom:1969kc, Breidenbach:1969kd} allowed to conclude that the protons were not punctual
 particles. Instead, they have substructure made up of particles with fractional electric charge (\textit{quarks}). Although these quarks had been predicted
 theoretically while trying to find a scheme to clasify the large amount of mesons observed during the '60s -first \cite{Ne'eman:1961cd, GellMann:1961ky}-
 and when trying to understand how to apply the quantum statistics to all of them, and especially to the spin $3/2$ baryons -later on \cite{GellMann:1964nj}-,
 it was not clear that their existence went beyond a mathematical concept and all subsequent experiments failed in their attempt to isolate them as free particles.
 The two main features of strong interaction had manifested: asymptotic freedom at high energies and confinement of quarks in hadrons at low energies.\\
\hspace*{0.5cm}Different formal studies of non-abelian gauge theories \cite{Yang:1954ek, 'tHooft:1985ir} showed that the different $UV$ and $IR$ behaviours
 of this theory could be explained in terms of a non-conmuting algebra. Later on, the evidence that the baryon $\Delta^{++}$ existed led to the conclusion that
 there had to exist an additional quantum number -called colour- through the conservation of the Spin-Statistics theorem, and motivated the effort of the
 Scientific community that finally brought as a result the simultaneous explanation of all abovementioned phenomena through the picture presented by Fritzsch,
 Gell-Mann and Leutwyler \cite{QCD}, who identified $SU(3)$ - 3 being the number of different colours a quark can have- as the gauge group, basis for the
 construction of $QCD$. The theory remains thus invariant under local transformations of the $SU(3)$ colour group. There are a number of theoretical and experimental
 evidences supporting this picture \cite{Pich:1999yz}.\\
\hspace*{0.5cm}The local non-abelian gauge symmetry $SU(\N)$ for $n_f$ (number of flavours) quark matter fields \footnote{The gauge group fixes the bosons 
content -mediators of the theory- but it does not for the matter fields: its representation and number of copies is something that must be inferred from 
experimental data.} determines the $QCD$ Lagrangian, directly incorporating the interaction of these with the gluon gauge fields and the selfinteractions
 among these gluons. The $QCD$ Lagrangian is:
\begin{eqnarray} \label{fullLQCDe}
& & \mathcal{L}_{QCD} = \overline{q} \left(iD\hspace*{-0.26cm}/ \;-\mathcal{M}\right)q\,-\,\frac{1}{4}G^a_{\mu\nu}G^{\mu\nu}_a\,+\,\mathcal{L}_{\mathcal{GF+FP}}\,, \nonumber \\
& & D_\mu = \partial_\mu \, -\, ig_s G^a_\mu\frac{\lambda_a}{2}\,,\nonumber \\
& & G^a_{\mu\nu} = \partial_\mu G^a_\nu \,-\, \partial_\nu G^a_\mu \,+\, g_s f^{abc}G^b_\mu G^c_\nu,
\end{eqnarray}
where $a\,=\,1,\,\dots,\,8$, $G^a_\mu$ are the gluon fields and $g_{s}$ is the strong coupling constant. The quark field $q$ is a column vector
 with $n_f$ components in flavour space, $\mathcal{M}$ represents the quark mass matrix in flavour space and it is given by 
$\mathcal{M}\,=\,\mathrm{diag} \left(m_1,\, \dots,\,m_{n_f} \right)$, where $m_i$ are the different quark masses: $m_u,\,m_d,\,m_c,\,m_s,\,m_t,\,m_b$
 for $n_f\,=\,6$ within the $SM$, parameters whose value is not restricted by the gauge symmetry. As we see, the latter forbides a non-vanishing mass for the
 gluons and their couplings are universal, irrespective of flavour. The matrices $\frac{\lambda^a}{2}$ are the $SU(3)$ generators in the fundamental representation, 
 normalized as $\mathrm{Tr}\left(\frac{\lambda^a}{2}\,\frac{\lambda^b}{2}\right)\,=\,\frac{1}{2}\delta_{ab}$ and $f^{abc}$ are the structure constants of the
 gauge group, $SU(\N)$. Finally, the Fadeev-Popov term \cite{Faddeev:1967fc}, $\mathcal{L}_{\mathcal{GF+FP}}$, includes the anti-hermitian Lagrangian introducing
 the ghost fields and the gauge-fixing term:
\begin{eqnarray}
& & \mathcal{L}_{\mathcal{GF}}\,=\,-\frac{1}{2\xi}\,(\partial^\mu G_\mu^a)(\partial_\nu G^\nu_a)\,,\nonumber\\
& & \mathcal{L}_{\mathcal{FP}}\,=\,-\partial_\mu \overline{\phi}_aD^\mu\phi^a\,,\,\,\,D^\mu\phi^a\,\equiv\,\partial^\mu\phi^a\,-\,g_sf^{abc}\phi_b\,G^\mu_c\,,
\end{eqnarray}
where $\xi$ is the gauge parameter, and $\overline{\phi}^a$ (a = 1, $\dots$ , $N_C^2$-1) is a set of scalar, hermitian, massless and anticonmuting fields. 
The covariant derivative, $D^\mu\phi^a$, contains the needed coupling between ghost and gluon fields and $\mathcal{L}_{\mathcal{FP}}$ is antihermitian, as
it must, in order to introduce a explicit unitarity violation cancelling the non-physical probabilities corresponding to the longitudinal polarizations of gluons 
and restore the fundamental property of unitarity to the physical observables that are finally obtained. A very pedagogical explanation of this can be found in Ref. \cite{Pich:1999yz}. 
We do not discuss here the so-called $\theta$ term \cite{Wilczek:1977pj}, invariant under $SU(\N)$ and $CP$ violating if there is not any massless quark (which is not the case, see Table \ref{Table_Masses_SMe}). The 
most precise experiments do not indicate any $CP$ violation in strong interaction processes. This allegged violation would manifest, for instance, in a non-vanishing neutron
 dipole electric moment.  The experimental bound \cite{Amsler:2008zzb} is nine orders of magnitude smaller than a natural theoretical value.\\
\hspace*{0.5cm}The running of the coupling constant with the energy is behind the property of asymptotic freedom and seems to point to confinement as a natural 
consequence. The coupling $g_s$ that appears in the $QCD$ Lagrangian (\ref{fullLQCDe}) receives quantum corrections \cite{Pich:1999yz} that, at one loop, are 
given by the diagrams in Fig. \ref{diagram 1_loop_beta_function}.\\
\\
\hspace*{0.5cm}The $\beta_{QCD}$ function is defined through the use of the Renormalization Group Equations, $RGE$ \footnote{$RGE$ \cite{Stueckelberg:1953dz, GellMann:1954fq, Kadanoff:1966wm, Callan:1970yg, Symanzik:1970rt, Wilson:1970ag}
 are derived from the requirement that an observable can not depend on the arbitrary chosen renormalization scale and that the physics must be scale invariant.
 The last property implies that the Green functions have a well-defined behaviour under rescaling of the momenta appearing in them. This allows to relate 
the values of the renormalized quantities at different energies and also to calculate the anomalous dimensions, that modify the evolution with energy derived on 
dimensional grounds because of quantum effects.}). At one loop they are given by \cite{Gross:1973id, Gross:1973ju, Gross:1974cs, Politzer:1973fx, Politzer:1974fr}:
\begin{equation} \label{runninge} 
\beta_{QCD}\,=\, \mu \frac{\partial g_s}{\partial \mu} \,=\,
-\left(11-\frac{2n_f}{3}\right)\frac{g_s^3}{16\pi^2} \,,
\end{equation}
so that -at this order in the expansion-, is negative for $n_f\leq16$. $RGE$ imply that the renormalized coupling varies with the energy, so-called \textit{running coupling}.
 Being $\beta_{QCD}\leq0$, this will result in the decreasing of the renormalized coupling, $g_s^R$, when increasing the energy; that is, in asymptotic freedom.\\
\hspace*{0.5cm}We have implicitly assumed that the one-loop computation gives a reliable approach. When speaking about asymptotic behaviour this is the case,
 just remember how the values in Eqs. (\ref{alphas_mtaue}) and (\ref{alphas_MZe}) decreased with the energy taking into account higher-order loop effects. Indeed, the computations at 
(N)NNLO \cite{Larin:1993tp, vanRitbergen:1997va} support this reasoning.\\
\hspace*{0.5cm}Integrating the equation (\ref{runninge}) we find:
\begin{equation}
\alpha_s(q^2) \,=\, \frac{12\pi}{(33-2n_f)\log(q^2/\Lambda_{QCD}^2)} \,,
\end{equation}
where $\alpha_s \,\equiv\, g_s/4\pi$ has been defined. This equation depicts how the strong (renormalized) coupling running depends just on the $QCD$-scale,
 $\Lambda_{QCD}$, defined in terms of the renormalized coupling value at some renormalization scale, $\mu$, and $\mu$ itself by: 
\begin{equation}
\log(\Lambda_{QCD}^2)\,=\,\log \mu^2 \,-\, \frac{12\pi}{\alpha_s(\mu^2)(33-2n_f)}\,.
\end{equation}
\hspace*{0.5cm}$RGE$ and the experimental tests of them seem -Eqs. (\ref{alphas_mtaue}), (\ref{alphas_MZe})- to be in agreement with a very strong color 
interaction at low energies that can cause confinement. There is an easy intuitive picture of this phenomenon: when splitting two electric charges, the strength 
of the mutual interaction decreases (is screened) by the creation of dipoles between them. This effect corresponds to the term with $-2 n_f$ in Eq. (\ref{runninge}).
 In the case of colour charges, the different behaviour comes from the term including the $33$ in that equation. Gluon selfinteractions cause anti-screening 
and finally, it is not possible to keep on separating the quark-antiquark pair since, at some point, it becomes energetically favoured to create a new pair. In order to complete 
the intuitive analogy, one could compare this with magnets. When breaking one, there will always appear new ones, with oposed poles. It is impossible to isolate
 the magnetic monopole as it is isolating a colour charge.\\
\hspace*{0.5cm}More technically, the confining phase is defined in terms of the behaviour of the action of the Wilson loop \cite{Wilson:1974sk},
 that corresponds to the path followed by a quark-antiquark pair in four dimensions between its creation and annihilation points. In a non-confining theory,
 the action of this loop would be proportional to its perimeter. However, in a confining theory, the loop action would grow with the area. Since the perimeter of
 two open lines is equal to its sum, while the area goes to infinity, in a non-confining theory it would be possible to split the pair; while in the confining
 it would not be so. Although Wilson loops were introduced in order to have a non-perturbative formulation of $QCD$ and solve confinement, this has not been 
possible so far. Its influence -like many ideas that emerged trying to understand $QCD$- has been great, since it lead Polyakov \cite{Polyakov:1981rd}
to formulate string theories in a modern way.\\
\hspace*{0.5cm}Despite of what has been said, there could be an experimental way to come close to confinement. So far we have always considered field theories 
at finite temperature and density. At the begining of the Universe both were so high that chiral symmetry would be broken then and quarks and gluons would 
not have the time to hadronize because of their incessant interactions. This framework is being investigated in heavy ion experiments to try shed some light 
on the problem of confinement.\\
\hspace*{0.5cm}In summary, quantum corrections make that the strength of the interaction changes with the energy. In the case of $QCD$, it is very strong at 
low energies, so we will not be able to make a perturbative expansion in powers of the coupling constant and make useful computations in this way, because 
they will not converge since $\alpha_S\sim1$. Additionally, and due to confinement, one should find a way to relate the fundamental theory
 with quark, antiquark and gluon degrees of freedom with the mesons produced in hadronic tau decays. In the next sections and chapters we will see 
that the solution to both problems comes together: when one finds the appropriate degrees of freedom, we will understand how to build a reliable and 
useful computation.\\
\section{Quantum Effective Field Theories}\label{Intro_QEFTse}
\hspace*{0.5cm}The history of Physics is a history of the understanding of more and more numerous and diverse phenomena. In many cases, the comprehension of the new 
does not invalidate the description of the already-known, that is obtained as a particular limiting case of the new theories, whose range of applicability is larger. Sometimes, 
the old theory can be regarded as en effective theory of the new one in a determined subset of its application.\\
\hspace*{0.5cm}Some examples can illustrate this: at the beginning of the XIX century a correct description of electrostatics was already achieved. Diverse 
experiments due to \O{}rsted, Amp\`ere, Ohm and Faraday -among others- increased the number of phenomena to describe simultaneously including electrodynamics 
and magnetism with time-dependent fluxes. The whole set could be explained coherently through Maxwell equations. In them, the wave nature of light was described, 
showing it as an electromagnetic wave propagating at a given speed, $c$, that was a universal constant of the theory. In the limit $c\to\infty$, one loses 
the Maxwell's displacement current. As a consequence, the old theory (Amp\`ere's law) could be seen as a limiting case of the new one (Maxwell equations) 
when the appropriate parameter ($1/c$) was considered to be small. Amp\`ere's law can be considered as the first order in the expansion in $1/c$ of the so-called 
generalized Amp\`ere's law that could be obtained from the Maxwell equations. It is thus an $EFT$ of the former. In static phenomena a similar treatment, based 
on the complete Maxwell equations is unnecessary and Coulomb or Amp\`ere's laws are enough, obviously.\\
\hspace*{0.5cm}Newtonian mechanics is valid for a large number of situations in our everyday life. Notwithstanding this is not the case in the world of the 
 infinitely small or enormously fast. Quantum Mechanics generalizes it in the first case and Especial Relativity does it in the second. One of the fundamental 
hypothesis of the quantum theory is that the action is quantized in integer multiples of the reduced Planck's constant ($\hbar$), which allows to explain the emission 
blackbody spectra, for instance. The value of this constant in $IS$ units is so small that it becomes macroscopically irrelevant. For this reason it makes sense
 that the limit $\hbar\to 0$ of the quantum theory will bring us back to the classical theory that is this way an EFT of the former. Consequently, nobody would resort to
Quantum Mechanics to solve a macroscopic problem unless it is to illustrate an introductory lesson to the topic.\\
\hspace*{0.5cm}It can also be seen that classical Maxwell's electrodynamics is an $EFT$ of Quantum Electrodynamics, $QED$, appropriate in the limit $\hbar\to 0$. 
The theory that we have seen before as fundamental, it is from this point of view an $EFT$ of the next more fundamental theory. Again it is not necessary to solve 
the equation of motion of a macroscopic charged body in presence of an $EM$ field using the quantum theory. From the practical point of view, $EFTs$ are more useful 
than the fundamental ones in their subsectors of applicability.\\
\hspace*{0.5cm}But for the case that we work for a theory \textit{of everything} our theory will always be effective, and it will be better this way since one avoids 
complicating the problem without any need and the choice of variables is suitable to its description.  We still need to justify 
that this effective theory would be a quantum field theory, ($QFT$).\\
\hspace*{0.5cm}The most common method of studying $QFTs$ is based in the use of perturbation theory in powers of the corresponding coupling constant, that 
must be small for every term in the expansion to be smaller than the previous one so that we can cut our expansion at a given order -because the perturbative 
series is not exactly summable- and expect that the outcome approximates well the true result. Such an expansion does not make sense in our case of hadronic 
$\t$ decays, because of the value $\alpha_S\sim \mathcal{O}(1)$. Then, one has to find an alternative way to proceed.\\
\hspace*{0.5cm}In any case, it is convenient not to abandon $QFTs$, since their formalism warranties that the observables will fulfill all requirements 
of a relativistic quantum theory (as it must be the theory describing the Physics of our elementary particles): microcausality (if two space-time points are 
separated spatially, whatsoever operators defined in them satisfy trivial commutation or anticommutation rules -depending on their statistics-), unitarity 
(the sum of the probabilities of all possible events is unity), analyticity (the functions of the fields must be complex-differentiable in the vicinity of every point of its domain), 
Poincar\'{e} invariance (the symmetry group of Relativity), spin-statistics connexion theorem (Fermi-Dirac statistics for half-integer spin particles and Bose-Einstein statistics 
for particles with integer spin) and cluster decomposition (ensuring the locality of the theory, since sufficiently far away regions behave independently).\\
\hspace*{0.5cm}Although we have seen that the techniques of $QFTs$ are highly desirable we must admit that they are not enough on their own, because if one incorporates 
these very general principles into the theory one would need a lot of experimental information to characterize a theory and therefore make predictions. 
As we have seen before it is convenient to use $EFTs$. Therefore, it will be natural and adequate to employ quantum $EFTs$ in our problem.\\
\hspace*{0.5cm}In order to formulate them we need to identify the relevant degrees of freedom and the expansion parameter. Both things will happen generally at the same 
time, as we will see. There will be a typical scale, $\Lambda$, separating active and passive degrees of freedom. Particles with $m\ll\Lambda$ will be kept in the 
action while the heavy fields with $M\gg\Lambda$ will be functionally integrated out. We will consider the interactions among the lightest states that will be 
organized in a power series in $1/\Lambda$.
Since $m/\Lambda\ll1$ the effect of every consecutive term will be less than the previous one and we will be able to cut the expansion at a given order. Besides, 
we will be able to control the error introduced estimating the contribution of the first omitted term from the expansion parameter and the known terms.\\
\hspace*{0.5cm}We will close the section stating Weinberg's definition of $EFTs$ \cite{Weinberg:1978kz}: if -for a given set of degrees of freedom- we apply perturbation 
theory with the most general Lagrangian consistent with the assumed symmetries we will obtain the most general $S$ matrix elements -and therefore the observables 
that are constructed from them- that are consistent with analyticity, perturbative unitarity, cluster decomposition and the assumed symmetries.\\
\hspace*{0.5cm}We note that with respect to the most general formulation introduced before we are adding here the compromise with a choice of degrees of freedom
 and the assumption of the symmetries of the underlying theory. This approach will be reviewed later on, because it may be desirable to make a more elaborated  
approach including dynamical content of the underlying theory.\\
\section{Chiral Perturbation Theory}\label{Intro_ChPTe}
\hspace*{0.5cm}We have highlighted the concept of symmetry. Symmetries have always been the key to understand physical phenomena. On one hand they are expressed 
with the greatest mathematical rigour, on the other end they allow -in some cases- approximations, that are at the core of almost any realistic computation.\\
\hspace*{0.5cm}Which is the symmetry that we can employ to build our effective theory? The answer is neither easy nor immediate. One could think in some property 
directly related to the gauge group of the theory, with the property of colour. Due to hadronization, the possible structures with vanishing total colour charge 
are immediately fixed by the product of representations in group theory, since we know the representations of the gauge group (adjoint)
 and we have fixed that of matter (triplet and antitriplet for quarks and antiquarks, respectively). We can check that the mesons fulfill this condition, but we 
do not obtain any useful hint in order to develop our $EFT$. Indeed, assuming confinement, we observe that letting $N_C$ free is the only remaining possibility that 
we will consider next.\\
\hspace*{0.5cm}It will not be then a local gauge symmetry the one allowing us to built the $EFT$. Let us see which global symmetries the strong interaction has.
 We think first that in this Thesis we study processes that produce the lightest mesons: pions, kaons and eta particles. It is intuitive that the heavier quarks will not 
be active. Therefore, we consider the $QCD$ Lagrangian for light flavours: $u,\,d,\,s$, $n_f=3$ in (\ref{fullLQCDe}). If we neglect in first approximation the 
masses of these quarks $m_u\sim m_d\sim m_s \sim 0$, the $QCD$ Lagrangian is invariant under separate transformations of the $RH$ and $LH$ components of the quark 
fields, global transformations of the group $G\equiv SU(n_f)_L\otimes SU(n_f)_R$, the so-called chiral symmetry group.\\
\hspace*{0.5cm}Local symmetries determine the interaction -as in (\ref{fullLQCDe})-. There are two possibilities for globals symmetries: If both the Lagrangian 
and the vacuum are invariant under the group of transformations $G$ then the symmetry is manifest in the particles spectrum. However, even if the Lagrangian 
is invariant under transformations belonging to $G$, the vacuum is not, then the spectrum will reflect the symmetries of a certain subgroup $H$ of $G$, where 
both the Lagrangian and the vacuum will be symmetric under transformations of $H$, but only the Lagrangian will be invariant under all the group $G$. One speaks 
in this case of spontaneous symmetry breakdown of the symmetry $G\to H$. We also know that we will have as many massless scalar \footnote{Scalar stands for a spin-zero 
particle in this context. No reference is made to its intrinsic parity here.} particles (Goldstone bosons \cite{Goldstone:1961eq}) 
as broken generators. That is, the number of Goldstone bosons equals the difference between the number of generators in $G$ and $H$.\\
\hspace*{0.5cm}If we restore to phenomenology we observe that the lightest mesons can be classified in multiplets ($n_f=3$) of equal spin ($J$) and intrinsic 
parity ($P$), 
which corresponds to the representations of the group $SU(3)_V$. We also see that multiplets with opposite parity do not share mass: the vector multiplet ($J^P=1^-$)
 is lighter than that of axial-vectors ($1^+$); and that of pseudoscalar mesons ($0^-$) is much lighter than the scalars ($0^+$) or than the spin $1$ particles
 \footnote{We will call resonances all light-flavoured particles heavier than those belonging to the lightest multiplet $0^-$.}. In Chapter \ref{EFT} it is explained how these 
observations lead to the pattern of spontaneous breaking of the symmetry, which is $SU(3)_L\otimes SU(3)_R\to SU(3)_V$. There are $n_f^2-1=8$ broken generators, 
that would be the number of Goldstone bosons that we should observe. In fact, since the masses of the light quarks are small compared to the typical 
hadronization parameter, $\Lambda_{\chi SB}\sim 1$GeV, but not zero, we have in addition to the spontaneous breaking of the symmetry an explicit breaking of it 
 because $m_q\neq0,\, m_q=m_u,\,m_d,\,m_s$. That is why we observe $8$ particles with small but nonvanishing masses that we call pseudo-Goldstone bosons, $pGbs$, 
(for his origin in the spontaneous breaking of the symmetry and his mass in its explicit breaking). These are the pions, kaons and eta particles detected in 
our semileptonic tau decays: $\pi^\pm,\pi^0,\eta,K^\pm,K^0,\bar{K^0}$.\\
\hspace*{0.5cm}Now that we have the symmetry and our choice of degrees of freedom we have to worry about building the $EFT$ Lagrangian that contains them 
conveniently. Weinberg's theorem ensures that having done that, the perturbative treatment of it will lead to the most general $S$-matrix elements in a consistent
 way. The formalism that allows to build effective Lagrangians for symmetry groups that have been broken spontaneously is due to Callan, Coleman, Wess and Zumino \cite{Coleman:1969sm, Callan:1969sn}.
Its application to low-energy $QCD$ will allow us to write an $EFT$ describing the interaction among pseudo-Goldstone bosons. Moreover, since there is an 
energy gap between these particles and the next heavier ones, the effect of these heavier modes will be small and will allow to build an $EFT$ containing only 
these modes, Chiral Perturbation Theory, $\chi PT$ \cite{Gasser:1983yg, Gasser:1984gg}.\\
\hspace*{0.5cm} This theory has a natural expansion parameter in the ratio between masses or momenta of the pseudo-Goldstone bosons over the scale
 $\Lambda_{\chi SB}$, that will be much less than unity. All the initial problems are thus solved: $\chi PT$ is an $EFT$ built upon symmetries of $QCD$ in a 
specific subset of it (light flavours in low-energy processes where the only products are $pGbs$ and chiral symmetry is a good approximation) and with a expansion 
parameter that permits to do perturbation theory. However, since $M_\t\sim1.8$ GeV, the resonances could be active degrees of freedom, so that we will have to enlarge 
 $\chi PT$ to higher energies and include new degrees of freedom. Unfortunately, in this case it will be more complicated to proceed through the previous steps
to build the theory, as we will see.\\
\section{$QCD$ in the limit of a large number of coulours} \label{Intro_largeNclimite}
\hspace*{0.5cm}When we incorporate heavier particles the counting is broken, since the masses and momenta of these new degrees of freedom are of the same order or larger 
than $\Lambda_{\chi SB}$, in such a way that its ratio is no longer a good expansion parameter of the theory. We have another difficulty: there is no 
longer a large and well-defined energy gap separating the particles that are active degrees of freedom of the theory from those who will be integrated out 
because they are not. We will see that a solution to both problems can arrive by considering the large number of colours limit of $QCD$ \cite{'tHooft:1973jz, 'tHooft:1974hx, Witten:1979kh}.
 Anyhow, we should point out that as opposed to the low-energy sector with $\chi PT$, it is not known how to build an $EFT$ dual to $QCD$ in the intermediate 
energy range. The limit $N_C\to\infty$ is a tool that will allow to understand which are the dominant contributions and which are not important
 -among all allowed by symmetries- in our Lagrangian.\\
\hspace*{0.5cm}'t Hooft suggested considering $QCD$ in the limit when the number of colours of the gauge group goes to infinity \cite{'tHooft:1973jz}. 
His motivation was achieving a simpler theory that still kept some resemblance with the original one and from which one could derive qualitative properties 
-hopefully also quantitative- of the underlying one. In this limit $QCD$ is exactly soluble in two dimensions \cite{'tHooft:1974hx}, but not in four. 
 Still, if we assume that the theory is confining, a number of experimental features of $QCD$ can be derived, which suggests that this construction 
is a good approximation to nature. Among them we will highlight for the moment that:\\
- In the $N_C\to\infty$ limit mesons are free, stable (they do not decay) and do not interact among themselves. Meson masses have a smooth limit and there are 
infinite particles: a tower of excitations per each set of quantum numbers.\\
- At first order in the expansion in $1/N_C$, meson dynamics is described by tree level diagrams obtained with an effective local Lagrangian whose degrees of 
freedom are mesons, as it was discussed in the Weinberg's view of $\chi PT$.\\
\hspace*{0.5cm}At this point one can observe that there is a certain internal contradiction between the construction of $EFTs$ \`a la Weinberg and the expansion 
in $1/N_C$ for $QCD$ that should be solved in some way: on the one hand Weinberg's view is to define the particle content and the symmetries and then to build 
the most general Lagrangian consistent with the assumed symmetries and it guarantees that we will obtain the most general results through a perturbative approach. 
The problem is that the introduction of the resonances invalidates the former expansion parameter, that was successful for $\CPT$.\\
\hspace*{0.5cm}On the other side, the large number of colours limit of $QCD$ can help us to organize an expansion in  $1/N_C$, but it contradicts the ideas 
in the previous paragraph since one of its conclusions at lowest order is that we can not fix a priori the particle content of the $EFT$, for the consistency 
of the expansion we have to have infinite copies of every type of resonance.\\
\hspace*{0.5cm}Because of that we have two possibilities:\\
- Either we forget the requirement for the Weinberg's formulation of making a suitable choice of degrees of freedom for the energy range we are considering 
and we include the spectra demanded by the limit $N_C\to\infty$.\\
- Or we include the phenomenological spectra and depart from the $1/N_C$ counting.\\
\hspace*{0.5cm}One could think that incorporating subleading effects in $1/N_C$ we may be able to get the measured spectra. This idea can not become a reality 
for the moment because of the nature of the $1/N_C$ expansion in $QCD$. It is true that a given order in $\alpha_S$ there is a definite number of diagrams, 
and that they can be computed and their effects resummed, but this is not at all the case in $1/N_C$: at every order there are infinite diagrams, and nobody has 
been able to think of a mechanism able to study this question. In the framework of $EFTs$ based on this expansion there are studies investigating the $NLO$ 
in $1/N_C$.\\
\hspace*{0.5cm}Additionally, one can recall that the Weinberg's approach does not include any type of dynamical information on the underlying theory: this is the price 
to pay for its generality. In our case we will see that a theory with pseudo-Goldstone degrees of freedom and resonances, that respects the symmetries of low-energy 
$QCD$, and therefore reproduces $\chi PT$ at low momenta, based in the limit $N_C\to\infty$, is not compatible with the known asymptotic behaviour of $QCD$ 
at high energies. Since we want our theory to work up to some $E\sim2$ GeV and at these energies perturbative $QCD$ is already reliable, this should not happen. 
Then, the theory we need will require dynamical information from $QCD$ -this will allow it lo link the chiral and perturbative regimes in the sector of 
light-flavoured mesons- and, either renounce to the choice of the physical final states as degrees of freedom or to model the expansion in $1/N_C$. 
This is discussed in the next section.\\
\section{Resonance Chiral Theory}\label{Intro_RChTe}
\hspace*{0.5cm}Resonance Chiral Theory, $R\chi T$ \cite{Ecker:1988te, Ecker:1989yg}, includes the pseudo-Goldstone bosons and the resonances as 
active degrees of freedom of the theory and requires general properties of $QFTs$ and the invariance under $C$ and $P$ $QCD$ has. Their fundamental features 
are sketched in the following.\\
\hspace*{0.5cm}The low-energy limit of $R\chi T$ must be $\chi PT$. This property has been used to predict systematically the $LECs$ of $\chi PT$
 in terms of masses and couplings of the resonances when integrating these ones out of the action, at the chiral orders $\mathcal{O} \left(p^4\right)$ \cite{Ecker:1988te} 
and $\mathcal{O}\left(p^6\right)$ in the even \cite{Cirigliano:2006hb} and odd-intrinsic parity sectors \cite{Kampf:2011ty}, with $N_C\to\infty$ and requiring the $QCD$ 
high-energy behaviour.\\
\hspace*{0.5cm}The $\chi PT$ Lagrangian includes the octet of pseudo-Goldstone bosons. When extending $\CPT$, $R \chi T$ incorporates the resonances as
active degrees of freedom that are included in nonets, since octet and singlet of a $SU(N_C=3)$ group merge into a nonet for $N_C\to\infty$. The $\chi PT$ 
Lagrangian is built using the approximate chiral symmetry of massless $QCD$. After that, the spontaneous and explicit symmetry breaking is incorporated in 
exactly the same way as it happens in $QCD$. The nonets of resonances are added requiring the general $QFT$ properties and invariance under $C$ and $P$ and
the structure of the operators is determined by chiral symmetry. At first order in the expansion in $1/N_C$ the terms with more than a trace and the loops are 
suppressed. The first property permits to postpone some terms allowed by the symmetries of the Lagrangian and the second one its use at tree level, as it was already 
explained.\\
\hspace*{0.5cm}We remark that the theory determined by symmetries does not share yet some of the known properties of $QCD$ at high energies. Therefore, 
one must match the theory with asymptotic $QCD$ at the level of Green functions and/or form factors. The application of these properties determines a series 
of relations between the couplings of the theory that allows it to be predictive with less experimental information than otherwise. In this Thesis we obtain 
relations of this type on the form factors in two different type of processes that we will confront to those found in two- and three-point Green functions 
where the same couplings appear \footnote{There are not computations within $R\chi T$ of four-point Green functions, whose short-distance relations we could 
confront to ours.}. The nice $UV$ behaviour forbids terms with a lot of derivatives, which helps us to limit the number of operators 
in the Lagrangian, since the counting that worked in $\chi PT$ is now broken. The situation is not that easy, as we will comment later on, because consistency 
conditions may require the introduction of operators with more derivatives and some non-trivial relations among their couplings. Generally speaking, we do not 
include terms with a lot of derivatives because this would require fine-tuned relations to ensure the required cancellations needed at large momenta. In 
many cases the phenomenological success is, at the end of the day, the support of our approach.\\
\hspace*{0.5cm}There is an inconsistency between Weinberg's approach and the strict limit $N_C\to\infty$, but to our knowledge, there is no known way of 
implementing the infinite tower of resonances in a model independent way. Then, it seems reasonable to start studying easy processes with the minimum number 
of degrees of freedom involved. As we get more and more control in this approximation (or the data get more and more precise) we will be able to include more 
states if needed. This approach is practical in order to estimate the different coefficients of the theory and it also respects the goal of a good physical 
description, trying to do it in terms of the least number of variables.\\
\hspace*{0.5cm}Finally, our phenomenological study can not avoid introducing some properties that are higher orders in the $1/N_C$ expansion. In the energy range 
where the taus decay, resonances reach their on-shell condition and do indeed resonate due to their width, lower than its mass but typically non-negligible. Widths are a subleading effect. 
We will include them consistently within $R\chi T$, as we will see.\\
\section{Organization of the Thesis}\label{Intro_Organizatione}
\hspace*{0.5cm}As it has been said, our study adopts the approach of $EFTs$. For this reason we introduce its basics in Chapter \ref{EFT}. Three are the cornerstones 
of our work on the theory side: on the one hand ensuring the right limit at low energies, ruled by $\CPT$. On the other, the large number of colours ($N_C$) limit 
of $QCD$ applied to $EFT$ with hadronic degrees of freedom, in our case $R\chi T$. And finally, to warrant a behaviour at high energies in agreement with $QCD$ 
for the different form factors. The first and second question are considered in Chapters \ref{EFT} and \ref{RChT}, respectively, whereas the third one is introduced 
in Section \ref{LargeN_MatchingRChTQCD} and discussed in any particular application of the theory considered in later chapters, that are preceded by a brief 
summary of the theoretical studies undertaken and an overview on the essentials of exclusive hadronic tau decays (Chapter \ref{Hadrondecays}). The applications 
that we consider are: hadronic decays into three pions (Chapter \ref{3pi}) and with two kaons and a pion (Chapter \ref{KKpi}). We also include the three-meson decays including
 $\eta$ particles (Chapter \ref{eta}) and the radiative decays of the tau with a single meson $\t\to P^-\, \g\,\nu_\t$, where $P=\p,\,K$, in Chapter \ref{Pgamma}. 
With all of them we will improve exceptionally the control on the parameters of the resonance Lagrangian participating in the considered processes, both in the 
vector and in the axial-vector current and, therefore we will know better how to describe, in a theoretically sound way based on $EFTs$ and the symmetries of $QCD$, 
these $\tau$ decays. We will be able to take advantage of all these findings in the future, applying them to more complex processes. The thesis ends with the 
general conclusions on the work done.\\
\chapter{Effective Field Theories: Chiral Perturbation Theory} \label{EFT}
\section{Introduction} \label{EFT_intro}
\hspace*{0.5cm}Effective Field Theories are built upon two seemingly contradictory deep roots: the idea of symmetry and the usefulness of making justified 
approximations. That is because symmetry is linked to the mathematical structure behind and appears to be fundamental, while an approximation implicitly seems
to assume some deviations from Nature. We will clarify in which context -that of $EFTs$- both concepts join together in a rigorous approximation.\\
\hspace*{0.5cm}It is not true that if one had the exact solution to the complete theory, no one would use the $EFT$ instead. The $EFT$ is more convenient in 
its domain of applicability because it uses the right variables and exploits the hierarchy of physical scales in the problem. We will be more specific about this point later 
on.\\
\hspace*{0.5cm}Although the precise formulation of $EFTs$ has been reached in the last thirty years, the two main ideas named above are, in a sense, living within
 Physics for long. It is common lore that the choice of variables can make the problem easier. If not exactly realized in Nature, symmetries are sometimes given 
at a quite approximate level and allow for a parameterization of the problem that exploits that and renders the computation doable (it is advisable to use 
cylindrical coordinates to solve the Laplace Equation in a tube-shaped cavity, for instance). At the same time, either analytically or numerically, one can work 
corrections to this exact solution by including the symmetry breaking as it happens in reality, or faithfully modeling it. We will see quite generally how these 
ideas of symmetry and approximation apply.\\
\hspace*{0.5cm}One of the longstanding motivations in Physics has been that of pursuing the so-called \textit{theory of everything}. This theory would be 
of little practical use because the energy scales involved would be orders of magnitude higher than the ones we can probe experimentally (unless extreme phenomena 
near black holes are observed). That would be the situation for any theory that unifies Gravity with other forces, since its characteristic energy scale, the 
so-called Planck Mass ($M_{Pl}=(8\pi G)^{-1/2}\sim 10^{18}$ GeV), is outside our reach. Thus, this unification would be almost useless but from the point of view of the 
likely mathematical beauty. An $EFT$ description using just the active degrees of freedom in any specific setting is in order.\\
\hspace*{0.5cm}On the other hand, if we leave gravity aside, the situation changes: If there was exact unification of the $SM$ gauge groups \cite{Georgi:1974sy}
 -with maybe some others \cite{Senjanovic:1975rk}, adding extra particles \cite{Dimopoulos:1981zb}, using small additional dimensions to reduce $E\sim M_{Pl}$ 
\cite{Kaluza:1921tu, Klein:1926tv, Randall:1999ee}, etc.- at some energy scale ever accessible to experiments, then we would reaffirm our understanding on $SM$, gain more insight
 on some kind of Physics beyond it ($BSM$) discovered by that time and get a number of predictions testable in experiments.\\
\hspace*{0.5cm}Let us use the very well-known example of the hydrogen atom to explain how the relevant degrees of freedom arise naturally. A first description 
of the system is achieved by using the Schr\"odinger equation for the electron bounded to a proton by Coulomb's law. The only properties that count at this stage 
are the electron mass and charge (or, equivalently, the fine structure constant, $\alpha\equiv e^2/(4\pi)\sim1/137$). It does not essentially matter that the proton 
mass is not infinite, because it is much heavier than the electron one. The spin $1/2$ of the electron does not affect yet either. If one counts the mass scales 
that appear in the problem, one sees they are $m_e$ and $M_P$, being $m_e/M_P\sim5\cdot10^{-4}$. Any effect of $m_e/M_P\neq0$ will be a $\sim10^{-3}$ correction,
at most. It can be seen that the leading interaction involving spin (between the electron spin and the electrons' orbital angular momentum) are also suppressed
 with respect to the leading Coulombic interaction.
 The general feature we may extract is that the characteristic energy scale of the problem ($\Lambda$) is set by the electron mass and the strength of the 
interaction: $\Lambda\sim m_e \alpha$, the typical momentum (or inverse of length scale, the familiar Bohr radius) of the system. Therefore, the relevant 
degrees of freedom will correspond to particles with energies much lower than this one ($m,E\ll\Lambda$): ultrasoft photons with energy of the order of 
$m_e \alpha^2$, that sets the scale of energy splittings between levels, the Rydberg (or inverse of characteristic times). On the other hand, particles with 
much higher energies ($M\gg\Lambda$) will influence tinily the spectrum and thus can be integrated out from the action. This will be the case of the proton or 
soft and hard photons. But also of the $W$-boson, what justifies that electroweak corrections to this $QED$ bound state are marginal.\\
\hspace*{0.5cm}One has then the possibility of constructing the most general Lagrangian consistent with $QED$ symmetries including interactions among the lightest
 states and one will be able to organize them efficiently as an expansion in powers of $E/\Lambda$. We have found through this example the general rules for 
building $EFTs$: identifying the relevant energy scale of the problem, integrating out the heavier modes and building the most general Lagrangian consistent
 with symmetries involving the light modes: a tower of interactions that one will conveniently organize in powers of $E/\Lambda$. The procedure rests on Weinberg's
Theorem, that will be discussed in Section \ref{EFT_Validity}.\\
\hspace*{0.5cm}The fact that the heavier states can be integrated out (as explained in Section \ref{EFT_IntHM}) does not mean that they do not leave any mark
 in the low-energy Physics. The effect of these states on the $EFT$ is double: on the one side they pose symmetry requirements on the $EFT$ \footnote{For 
instance, in the non-relativistic $EFTs$ the relativistic invariance of the fundamental theory implies relations between the $LECs$ in the $EFT$ that are valid 
to all orders in the coupling constant.}, on the other hand they correct the values of the constants 
specifying the dynamics of the low-Energy theory ($LECs$), that are different in the full and in the effective theory in case they enter in both 
\footnote{This does not happen when the degrees of freedom are different in the fundamental and effective theory.}, see Section \ref{EFT_EffectsHM}.\\
\hspace*{0.5cm}If the underlying theory is weakly coupled at the scale $\Lambda$ one is able to compute explicitly the values of the $LECs$. Otherwise, one 
must rely on lattice evaluations or fix them phenomenologically as discussed in Section \ref{EFT_Weakly_and_Strongly}. In the first case, the same coupling constant will serve as an expansion parameter to apply 
the perturbative techniques, while in the second one it may be difficult to find such a parameter.\\
\hspace*{0.5cm}One important feature of $EFTs$ is that there is an infinite number of interacting terms in the $EFT$, which makes the theory 
non-renormalizable in the classical sense. However, this is not a problem once we understand that $EFTs$ add to the general characteristics of renormalizable 
theories the need of having a rule in order to estimate the size of these non-renormalizable terms. This will allow us to stop the expansion once we 
reach some desired maximum error associated to our computation. We will classify the terms in the Lagrangian according to some counting scheme that makes
 explicit the organization of all allowed interaction terms in powers of the expansion parameter. Then, at a given order, one will have infinites that will 
be renormalized by redefining the $LECs$ appearing at the next order in the expansion. If one regularizes using Dimensional Regularization (that preserves all
 symmetries of the theory in the renormalization procedure), renormalizability is assured since every infinite will be the coefficient of an operator respecting
 the symmetries and therefore already present in the effective Lagrangian at a higher order, determining an order-by-order renormalization. At the practical 
level, $EFTs$ are as renormalizable as those classically called that way. For a given asked accuracy we have to take into account terms in the expansion up 
to some order, that includes a finite number of terms, as in any renormalizable theory.\\
\hspace*{0.5cm}A general remark before closing this introductory section: It seems that $EFTs$ are the tool that solves everything and that
 is not true. The Hydrogen atom is a system with a well-defined hierarchy of scales and a non-relativistic nature. Then, the ratios $E/\Lambda$ and $v/c$ are 
two small magnitudes that work extremely well in the setting described above. Moreover, most of Particle Physics systems fulfill the non-trivial characteristic 
that there is a small number of quantities playing a role in the problem with some scaling among them. This allows (and suggests) an $EFT$ approach. Even in the
 cases where the scaling is not so well defined or the expansion parameter is not that small it is an advisable tool. However, in Chemistry (or even more 
in Biology) there appear an enormous amount of unrelated energy scales of comparable magnitude. Then even the extremely simple approach applied to the Hydrogen 
atom as first step will lead in an $EFT$ study to too cumbersome expressions to make any sense out of them. We are lucky that these techniques can be applied
 in Physics.\\
\hspace*{0.5cm}In the remainder of the chapter we will be more precise on the ideas sketched above. We have made extensive use of Refs.~\cite{Pich:1998xt, Manohar:1996cq, Dobado:1997jx, Georgi:1994qn, Manohar:1983md, Kaplan:1995uv, Kaplan:2005es, Georgi:1991ch}
in order to prepare this part.\\

\section{Validity of $EFTs$: Weinberg's Theorem} \label{EFT_Validity}
\hspace*{0.5cm}We will precise now the formulation of $EFTs$ \`a la Weinberg. It can be stated as a theorem \cite{Weinberg:1978kz}:\\
\hspace*{0.5cm}\textit{For a given set of asymptotic states, perturbation theory with the most general Lagrangian containing all terms allowed by the assumed symmetries
 will yield the most general $S$ matrix elements consistent with analyticity, perturbative unitarity, cluster decomposition and the assumed symmetries}.\\
\hspace*{0.5cm}$EFTs$ describe the physics at low energies, this defined with respect to some energy scale, $\Lambda$, characteristic of higher energy processes.
 Heavier states with $M \gtrsim \Lambda$ are integrated out from the action and the relevant degrees of freedom are those with masses $m<<\Lambda$. There is a 
well defined ordering in powers of $E/\Lambda$ for the infinite interactions among the light states one gets.\\
\hspace*{0.5cm}The view on renormalizability of $QFTs$ has changed through the years. For much time, it was claimed that for a $QFT$ to be renormalizable one
 needed that the Lagrangian contained only terms with dimension less or equal than that of the
 space-time, $D$. If operators of any dimension were allowed, one would need an infinite number of counterterms to absorb all the infinites and consequently an infinite number of unknown parameters
 condemning the theory to have no predictive power. Since the $EFT$ has an infinite number of terms ($\mathcal{L}_{eff}=\mathcal{L}_{\leq D}+\mathcal{L}_{D+1}+...$,
where only the first one is renormalizable in the classical sense) the conclusion seems devastating.\\
\hspace*{0.5cm}We have explained at the end of the previous section that the infinite number of terms in the $EFT$ will not cause any problem with respect to
the renormalization of the theory. For a definite number of powers in the ($E/\Lambda$) expansion, the symmetries of the $EFT$ allow only a finite number of
 operators in the Lagrangian. Consequently, there will be a finite number of counterterms that renormalize the theory at this order.\\
\hspace*{0.5cm}In addition to the problem of predictivity loss, one could think that as the energy of the process increases this 
tower of classically non-renormalizable interactions will give rise to a wild violation of unitarity at high energies. To clarify that this is not the case,
 we will classify the Quantum Field Theories according to their sensitivity to high energy \cite{Ecker:1994gg}:
\begin{enumerate}
\item Asymptotically free theories: Nothing in them signals a limiting energy beyond which they can no longer be employed.
\item Ultraviolet unstable theories: The theories themselves report about a limit energy range of applicability. This statement will be illustrated with
several examples in section \ref{EFT_Example}.
\end{enumerate}
\hspace*{0.5cm}$EFTs$ belong to the second group. The main difference with respect to the first ones comes from the appearance of new $LECs$ at every order 
in the perturbative expansion -that is not simply an expansion in the number of loops, as we will see-. It makes no sense to go further and further in this 
expansion indefinitely. Every new order in the series in ($E/\Lambda$) is intended to achieve a more detailed description. The accuracy reached with every
 new term goes as $\epsilon\lesssim\left( \frac{E}{\Lambda}\right)^{D_i^{max}-4} $, where $D_i^{max}$ is the highest dimension of all operators included 
\footnote{There will be quantum corrections to this estimate, that most of the time will be irrelevant. Anyway, the moral 
is that the error is essentially under control, as one would ask any theory for.}. Once
 we demand a limited precision, we know at which order in the expansion we shall stop. Moreover, if we desire to enlarge the applicability of the $EFT$ to higher energy physics
 the way out is not to include operators of higher and higher dimension, because as soon as $E\sim M_1$ -being $M_1$ the mass of the lightest initially integrated
 out particle- the Weinberg's theorem tells that the right procedure is to include it in the Lagrangian as an active field. This happens usually, and there is
 a formal way to deal with this successive incorporation of particles, that we will describe in the next section.\\
\hspace*{0.5cm}In $QCD$, confinement forbids quarks and gluons to be asymptotic states. Weinberg's theorem guarantees that writing out the most general Lagrangian
 in terms of hadrons -that can be thought as active degrees of freedom in a given subset of energies- consistent with the needed symmetries and respecting 
all the other stated conditions will bring us the most general observables consistent with the assumed symmetries and general properties of $QFTs$.\\

\section{Integrating out the heavy modes} \label{EFT_IntHM}
 \hspace*{0.5cm}We will explain here more formally the integration of heavy modes from the action that has been anticipated in the previous sections. We will use the path 
integral formalism and assume that the theory at high energies is known. The effective action $\Gamma_{\mathrm{eff}}$, will be written only in terms of
 the light modes and encodes all the information at low energies, where it yields the same $S$ matrix elements than the fundamental theory by construction.
 $S_{\mathrm{eff}}$ reads
\begin{equation}
e^{i \, S_{\mathrm{eff}}[\Phi_l]} \, = \, \int \, [\mathrm{d} \Phi_h] \,\, e^{iS[\Phi_l,\Phi_h]} \, , \label{heavymodes}
\end{equation}
where $\Phi_l$ and $\Phi_h$ refer to the light and heavy fields respectively and $S[\Phi_l,\Phi_h]$ is the action of the underlying theory where both modes 
are dynamical. The effective Lagrangian gets defined through
\begin{equation}
S_{\mathrm{eff}}[\Phi_l] \, = \, \int \mathrm{d}^4x \,\, \mathcal{L}_{\mathrm{eff}}[\Phi_l] \, .
\end{equation}
The effective action $S_{\mathrm{eff}}[\Phi_l]$ can be computed using the saddle point technique. The heavy field $\Phi_h$ is expanded around some field
 configuration $\overline{\Phi}_h$ as follows ($\Delta \Phi_h(x) \equiv \Phi_h(x)-\overline{\Phi}_h$)
\begin{eqnarray}\label{expansion_th}
S[\Phi_l,\Phi_h] &=& S[\Phi_l,\overline{\Phi}_h] \, + \, \int \mathrm{d}^4x \left. \frac{\delta S}{\delta \Phi_h (x)}
\right|_{\Phi_h=\overline{\Phi}_h} \Delta \Phi_h (x) \nonumber
\\&&  + \frac{1}{2} \int \mathrm{d}^4x\,\mathrm{d}^4y \left.\frac{\delta^2S}{\delta \Phi_h(x) \delta \Phi_h(y)}
\right|_{\Phi_h=\overline{\Phi}_h} \Delta \Phi_h(x) \Delta \Phi_h(y) \,+ \, \dots \,.
\end{eqnarray}
$\overline{\Phi}_h$ is chosen so that the second term in the $RHS$ of Eq.~(\ref{expansion_th}) vanishes to allow a (formal) Gaussian integration
\begin{equation}
\left.\frac{\delta S[\Phi_l,\Phi_h]}{\delta \Phi_h (x)}\right|_{\Phi_h=\overline{\Phi}_h} \, = \, 0 \, .
\end{equation}
With this choice Eq.~(\ref{heavymodes}) is
\begin{equation} \label{tree_loop}
e^{i \, S_{\mathrm{eff}}[\Phi_l]} \, = \, e^{i \, S[\Phi_l,\overline{\Phi}_h]} \, \int [\mathrm{d} \Phi_h] \,
e^{i \int \mathrm{d}^4x \, \mathrm{d}^4y \left\{ \frac{1}{2} \Delta \Phi_h(x) \, A(x,y) \, \Delta \Phi_h(y)+\dots \right\}}\,,
\end{equation}
where
\begin{equation}
A\left(x,y\right) \, \equiv \, \left.
\frac{\delta^2S}{\delta\Phi_h(x)\delta\Phi_h(y)}\right|_{\Phi_h=\overline{\Phi}_h}.
\end{equation}
We see from Eq.~(\ref{tree_loop}) that the first term corresponds to a tree level integration of the heavy field $\Phi_h$. The power counting of the $EFT$ will 
determine how the expansion is realized, as we will see in the case of $\chi PT$ in Chapter 2.\\
\hspace*{0.5cm}Two comments are in order: The procedure is iterative; one can have a pair of $\{EFT$-fundamental theory$\}$ valid up to some energy scale $\Lambda$.
 Then, for $E>\Lambda$, some other mode may become active and the fundamental theory in the previous step will become the $EFT$ in the next one. The other remark
concerns the actual use of that integration. The general procedure outlined above can involve complicated or unfeasible calculations. This may happen because
 the degrees of freedom are different in both theories and there is no unambiguous way of relating both, or because a perturbative treatment is not applicable.
 In these situations one may restore to phenomenology or lattice evaluations to do the computations. In any case, it is always possible to obtain some information
 on $e^{i \, S_{\mathrm{eff}}[\Phi_l]}$ from symmetry constraints stemming from the fundamental theory.\\

\section{Effect of heavy modes on low-energy Physics} \label{EFT_EffectsHM}
\hspace*{0.5cm}We have stated in Section \ref{EFT_intro} that although the heavy modes are non-dynamical at low energies they have an impact in the $LECs$ of the 
theory when integrating them out following the method in Sect.~\ref{EFT_IntHM} in going from the the fundamental theory to the effective one. We will 
see this general property at work in the case of $\chi PT$. There is again a theorem expressing this notion precisely, due to Appelquist and Carazzone \cite{Appelquist:1974tg}:\\
\hspace*{0.5cm}\textit{For a given renormalizable theory whose particles belong to different energy scales, that does not suffer Spontaneous Symmetry Breaking and does not 
have chiral fermions; the only effects of the heavy particles of characteristic mass $M$ in the physics of the light particles of masses around $m$ at low 
energies either appear suppressed by inverse powers of $M$ or through renormalization}.\\
\hspace*{0.5cm}The question is immediate: Does it apply in general? Does it to $QCD$?\\
\hspace*{0.5cm}$QCD$ is renormalizable \cite{Taylor:1971ff, 'tHooft:1972fi, Slavnov:1972fg, Becchi:1975nq} and we have discussed in the introduction that there are
six quarks with masses spanning four orders of magnitude. This suggests that one could very likely have hadrons belonging to different energy scales, and also
 dynamical gluons with very different energy-momentum. In principle, the first condition of the theorem does not seem impossible to meet, although one should
revise this assumption in any particular scenario. Moreover there is no spontaneous symmetry breaking associated to the $QCD$ vacuum. Finally its fermions are not chiral: left-handed and right-handed fermions do not couple 
differently to the gauge color group. All conditions of the Appelquist-Carazzone theorem are \textit{a priori} fulfilled. 

\section{Example} \label{EFT_Example}
\hspace*{0.5cm}We will introduce an example of $EFTs$ in order to help us illustrate some general characteristics of
 $EFTs$ that have already been discussed and will be of use in the next sections. We will see how:
\begin{itemize}
\item An $EFT$ indicates its border of applicability by itself.
\item $EFTs$ help to understand the physics involved giving us some hints that point to the more fundamental theory to be scrutinized in future experiments.
\end{itemize}

\hspace*{0.5cm}Inspired by the coupling of the electromagnetic current to photons, Fermi \cite{Fermi:1934sk} proposed as the basis of a theory of weak interactions 
a local current $\times$ current interaction among fermions that we can write -just for the lightest species- now as:
\begin{equation} \label{L-Fermi}
\mathcal{L} \, = \, -\frac{4G_F}{\sqrt{2}} \,\left[  V_{us}V^*_{ud} \, \left(
\overline{u} \gamma^\mu P_L s \right) \, \left( \overline{d} \gamma_\mu P_L u \right) \,+\, \left(
\overline{e} \gamma^\mu P_L \nu_e \right) \, \left( \overline{\nu}_\mu \gamma_\mu P_L \mu \right)\right]  \,
+ \, \cO \left(\frac{1}{M_W^4}\right) \,,
\end{equation}
where $P_L$ is the projector over left-handed states.\\
\hspace*{0.5cm}
Eq.~(\ref{L-Fermi}) and dimensional analysis imply that
$\sigma\left(\nu_\mu\,e^-\,\to\,\mu^-\,\nu_e\right)$ must diverge in the ultraviolet as $G_F^2\,s$, which signals again a more general theory, the $SM$.\\
\hspace*{0.5cm}Is this the end of the story? Coming back 
to the hierarchy problem
, see Table \ref{Table_Masses_SMe}
, there is a dimension five operator (thus, suppressed at low energies by the high-energy scale) that respects all gauge symmetries of the $SM$ and that
 will give mass to Majorana neutrinos \cite{Weinberg:1979sa}:
\begin{equation}
-\frac{1}{2\Lambda}\left( \overline{\widetilde{\ell}}_L\,\varphi\right)\,F\,\left( \widetilde{\varphi}^\dagger\,\ell_L \right)\,+\,\mathrm{h.c.} 
\end{equation}
through diagonalization of the mass-term \footnote{Left-handed leptons are collected in the $\ell$s, $\varphi$s include the Higgs field and $F$ is, in general
, non-diagonal in flavour space.}. Thus, $M^{\mathrm{Majorana}}\,\equiv\,\frac{v^2}{\Lambda}\,F$, where $v^2$ is related to the electroweak scale and $\Lambda$
 to the new Physics scale. The important lesson we learn from this particular example is that regarding the $SM$ as an $EFT$ we can go on learning about a more 
fundamental theory. To mention a recent example, let us note that these ideas have also been applied under the hypothesis of minimal flavor violation 
\cite{D'Ambrosio:2002ex}.\\

\hspace*{0.5cm}The same physical predictions in the full and effective theories should be expected around the heavy-threshold region. 
Thus, both descriptions are related through a so-called matching condition: the two theories (with and without the heavy field) should give the same $S$ 
matrix elements for processes involving light particles.\\
\hspace*{0.5cm}Until the matching conditions have not been taken into account, one is not dealing with the effective field theory, 
that is, the matching procedure is a fundamental step to develop effective approaches. We will illustrate how this works with 
 the Fermi theory, see Eq.~(\ref{L-Fermi}).\\

\hspace*{0.5cm}When the $W$-boson momentum is small compared to its mass, its propagator can be Taylor-expanded to give:
\begin{equation} \label{Taylor}
\frac{1}{p^2 \, - \, M_W^2} \, = \, - \frac{1}{M_W^2} \, \left( 1 \, + \, \frac{p^2}{M_W^2} \,+ 
\, \frac{p^4}{M_W^4} \, + \, \dots \right) \, ,
\end{equation}
and the lowest order in this expansion can be applied to the $SM$ tree-level result for $u\,s\,\rightarrow\,d\,u$ in the unitary gauge:
\begin{equation} \label{Fermi-ampli}
\mathcal{A} \, = \, \left( \frac{i \, g}{\sqrt{2}} \right)^2 \, V_{us}V^*_{ud} \, \left(
\overline{u} \gamma^\mu P_L s \right) \, \left( \overline{d} \gamma^\nu P_L u \right) \,
\left( \frac{-i \, g_{\mu\nu}}{p^2-M_W^2} \right) \, ,
\end{equation} to give:
 \begin{equation}
\mathcal{A} \, = \, \left( \frac{i}{M_W^2} \right) \, 
\left( \frac{i \, g}{\sqrt{2}} \right)^2 \, V_{us}V^*_{ud} \, \left(
\overline{u} \gamma^\mu P_L s \right) \, \left( \overline{d} \gamma_\mu P_L u \right) \,
+ \, \cO \left(\frac{1}{M_W^4}\right) \, .
\end{equation}
\hspace*{0.5cm}Comparing this to the amplitude obtained with Eq.~(\ref{L-Fermi}) we can match both theories up to $\cO \left(\frac{1}{M_W^4}\right)$ corrections by relating the corresponding coupling constants through the $W$ mass:
\begin{equation} 
\frac{G_F}{\sqrt{2}} \, = \, \frac{g^2}{8M_W^2} \,.
\end{equation}
\hspace*{0.5cm}Going further in the expansion of Eq.~(\ref{Taylor}) will require to include additional higher-dimension operators in Eq.~(\ref{L-Fermi}).\\
\\
\begin{figure}[h!]
\centering
\includegraphics[scale=0.8]{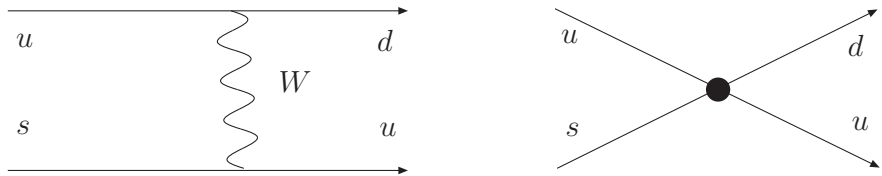}
\caption{Feynman diagrams for the flavour changing charged current process $u s\to d u$ at lowest order: within the fundamental electroweak theory through 
$W$-exchange (left) and using Fermi's $EFT$ (right). The effective local vertex is represented by the thick dot.}
\label{diagram Fermi&EW}
\end{figure}
\\
\hspace*{0.5cm}In this particular example the tree level matching is trivial because the $|\Delta S|\,=\,1$ processes are allowed without loops. For 
$|\Delta S|\,=\,2$, the $LO$ non-trivial contribution comes from the loop box diagram.\\
\hspace*{0.5cm}In the Fermi theory it has been possible to compute the $LECs$ in the $EFT$ from the fundamental theory. In our case, when studying low and 
intermediate energy $QCD$, it will not be possible to derive from $QCD$ the couplings of $R\chi T$, the matching conditions in this case will apply when 
demanding asymptotic $QCD$ behaviour to the Green functions and form factors obtained within $R\chi T$. That will impose some restrictions on the effective 
couplings of the theory, as we will see.\\

\section{Weakly and strongly coupled theories} \label{EFT_Weakly_and_Strongly}
\hspace*{0.5cm}
A weakly coupled 
theory is one in which perturbation theory applies in a given energy range, whereas it is strongly coupled in some energy interval if the couplings
 are there comparable to (or even greater than) unity and any ordering for the perturbation series makes no sense. One might try to understand why asymptotic
 states differ from interacting states for these strongly coupled theories as a consequence of this property. In practice, and apart from very few realistic
 exceptions (like, for instance, Bethe-Salpeter equation \cite{Salpeter:1951sz, Lepage:1977gd, Barbieri:1978mf} for bound states in $QFT$) we are only able 
to get rid of physical problems in $QFT$ by using perturbative methods: Either belonging to the original theory or to an $EFT$ (in fact, the naming non-perturbative 
methods has been generalized for this last case); so, it is clear that our inability for dealing with such kind of mathematical problems makes $EFTs$ even more necessary.\\

\hspace*{0.5cm}Finally, another consequence for weakly/strongly coupled theories is that in them the na\"if dimensional analysis is not/is modified by
 anomalous dimensions \footnote{The anomalous dimensions are quantum corrections to the classical operator dimensions. Their importance at different energy 
scales can be evaluated by using the so-called Renormalization Group Equations.}. Anomalous dimensions (together with a non-abelian gauge group) can explain
 how a theory can share ultraviolet freedom and infrared confinement, as it is the case for $QCD$. Ref.~\cite{Coleman:1974bu} gives an easy and illustrative 
example of this. We recall its main features in the following: Let us consider the two-dimensional Thirring model \cite{Thirring:1958in} for a fermionic field
 whose Lagrangian is
\begin{equation}
\mathcal{L} \, = \, \overline{\psi} \left(i \partial \! \! \! / \;  -  m \right) \psi \, - \,
\frac{1}{2}g\left(\overline{\psi}\gamma^\mu\psi \right)^2 \, .
\end{equation}
It can be shown that it is dual to the sine-Gordon model for a fundamental scalar field, with the Lagrangian:
\begin{equation}
\mathcal{L} \, = \, \frac{1}{2} \partial_\mu \phi \partial^\mu \phi \, + \, \frac{\alpha}{\beta^2} 
\cos \beta \phi \, ,
\end{equation}
where the couplings $g$ and $\beta$ are related by means of:
\begin{equation}
\frac{\beta^2}{4\pi} \, = \, \frac{1}{1 \, + \, g/\pi} \,,
\end{equation}
that indeed shows us that sine-Gordon strongly coupled model with $\beta^2 \, \approx \, 4\pi$ can be studied with the weakly coupled Thirring model 
($g \, \approx \, 0$). Then, we have two -equally valid and quite different- alternatives for describing the same theory (this happens because of the big 
anomalous dimensions that appear within strongly coupled theories at some scale. They can change drastically the behaviour of the different operators entering
 the $EFT$ when considering them at different energies). At a given energy scale, we can choose between a strongly coupled theory involving bosons that has 
big anomalous dimensions and a weakly coupled theory whose degrees of freedom are fermions with little anomalous dimensions. From the purely formal point of 
view, there is no reason to prefer one alternative to the other one, but in order to compute it is clear that the second option -that has a smooth perturbative
 behaviour- is more comfortable.\\
\hspace*{0.5cm}This example emphasizes again the importance of making a right selection of degrees of freedom. A theory that can be really involved at some 
energies (sine-Gordon model, that is strongly coupled) can be studied by means of another one, which is easier (Thirring model).\\
\hspace*{0.5cm}The parallelism with $QCD$ is tempting. The strongly coupled low-energy $QCD$ can be treated by means of a weakly coupled theory written in terms
 of bosons, and this will be much easier than if we would have tried to solve it using quarks and gluons as the relevant fields.\\

\section{Precise low-energy Physics as a probe for New Physics} \label{EFT_PreciselowEforNP}
\hspace*{0.5cm}Under the conditions of the Appelquist-Carazzone Theorem one sees that the effect of integrating out the heavy particles is to modify the values
 of the $LECs$ and impose symmetry restrictions. We enumerated how all conditions of the theorem (but for maybe the energy gap between particles in some regions
 of the spectrum) were accomplished for the theory of strong interactions. However one can have $EFTs$ applicable to sectors of the rest of the $SM$ that may 
not fulfill the theorem and bring valuable information about the Physics at higher energies by analyzing with precision the low-energy experiments because in 
this case the effect of heavy modes in low-energy Physics will not be that mild. This is indeed the case for the electroweak ($EW$) sector of the $SM$, where
 spontaneous symmetry breaking affects the $EW$ vacuum and fermions are chiral in the sense defined above.\\
\hspace*{0.5cm}Thus, we understand that when extremely precise $LEP$ data were analyzed and compared to theoretical computations including many quantum corrections, the $Z$
 width \cite{Veltman:1980fk}, and its decay into a $b$-$\overline{b}$ pair \cite{Bernabeu:1987me, Bernabeu:1990ws}, were shown to be so sensitive to the 
yet-undiscovered top quark, through $m_t$, that it indicated where to find it at $FERMILAB$, as we told in the introduction. This is a clear and historically
 interesting example of how immensely precise low-energy experiments can give us clues about where physics beyond our model waits hidden.\\
\hspace*{0.5cm}We will add two related examples of current interest: an electroweak precision test (measurement of the weak mixing angle) and the anomalous magnetic
 moment of the muon. We will be very schematic here just in order to highlight the point we wish to make, for a detailed analysis one can consult recent reviews 
on the topic \cite{Prades:2009tw, Jegerlehner:2009ry}, or Ref. \cite{Actis:2009gg}.\\
\hspace*{0.5cm}Precision tests of the $SM$ are promising places to look for physics $BSM$. An accurate measurement must be supplemented by very precise input
 parameters and higher order radiative corrections. At first sight it is striking that the measurement with finest precision is the main source of uncertainty in the
 end. That highest precision number is that of the fine structure constant, $\alpha$, determined from the measurement of the anomalous magnetic moment of the
 electron \cite{Hanneke:2008tm}, with amazing accuracy: $g_e/2 = 1$.$001 159 652 180 73(28) \Rightarrow \alpha^{-1} = 137$.$035 999 084(51)$, relying on perturbative
 $QED$ as summarized in Ref.\cite{Gabrielse:2006gg}. However, physics at higher energies is not described by this $\alpha$ measured at zero momentum transfer
 but for the one incorporating the quantum running. The shift of the fine structure constant from the Thompson limit to high energies involves necessarily
 a low-energy region in which non-perturbative hadronic effects spoil that astonishing precision. In particular, the effective fine structure constant at
the $Z$ pole plays an important role in $EW$ precision tests, like the weak mixing angle, $\theta_W$, related to $\alpha$, the Fermi constant, $G_F$,
 and $M_Z$ through \cite{Sirlin:1980nh, Sirlin:1989uf, Marciano:1980pb}:
\begin{equation}\label{Sirlin Relation}
 \mathrm{sin}^2\theta_W  \mathrm{cos}^2 \theta_W=\frac{\pi\alpha}{\sqrt{2}G_F M_Z^2(1-\Delta r)}\,,
\end{equation}
where $\Delta r$ incorporates the universal correction $\Delta \alpha (M_Z)$, the quadratic dependence on $m_t$ and all remaining quantum effects. In the $SM$,
 $\Delta r$ depends on various physical parameters including the mass of the Higgs Boson, $M_H$, still unknown \footnote{After the defence of this Thesis, both the ATLAS and 
CMS Collaborations have reported \cite{:2012gk, :2012gu} the observation of a Higgs-like particle with mass $\sim 125$ GeV.}. This way, the measurements of $\mathrm{sin}^2\theta_W$ 
can help to put indirect bounds on $M_H$ \cite{Schael:2005am, :2005ema, Passera:2008jk}. The error on $\Delta \alpha (M_Z)$ dominates the theoretical prediction.
 Here and in the case of the muon magnetic moment anomaly the source of the uncertainty is similar, and it depends on $R(s)$, defined as follows:
\begin{equation} \label{R(s)}
 R(s)=\frac{\sigma\left( e^+e^-\to \mathrm{had}(\gamma)\right)}{\sigma\left( e^+e^-\to \mu^+\mu^-(\gamma)\right)}\,.
\end{equation}
\hspace*{0.5cm}Specifically, the hadronic contribution $\Delta \alpha_{\mathrm{had}}^{(5)}(M_Z)$ of the five quarks lighter than the $Z$ boson can be related to
 Eq.~(\ref{R(s)}) via \cite{Cabibbo:1961sz}:
\begin{equation}
 \Delta \alpha_{\mathrm{had}}^{(5)}(M_Z)=-\left( \frac{\alpha M_Z^2}{3\pi}\right) \Re e \int_{m_\pi^2}^{\infty}\mathrm{d}s \frac{R(s)}{s(s-M_Z^2-i\epsilon)}\,, 
\end{equation}
where
\begin{equation} \label{R(s)_II}
 R(s)=\frac{\sigma_{\mathrm{had}}^0(s)}{4\pi\alpha^2/(3s)}\,,
\end{equation}
and $\sigma_{\mathrm{had}}^0(s)$ is the total cross section for the annihilation into any hadron with vacuum polarization and initial state $QED$ corrections
 subtracted off. As discussed extensively in the introduction, we are lacking a way of using the $QCD$ Lagrangian that allows to compute Eq.~(\ref{R(s)_II}) with 
enough accuracy to discriminate if there is new physics associated to the measurement of $\mathrm{sin}^2\theta_W$. The strategy is to use experimental data on 
the $e^+e^-$ annihilation into any hadronic state from threshold up to some energy ($\sim 2$ GeV) where we can already rely on perturbative $QCD$ supplemented
 by a motivated description of the lineshape of the many resonances appearing as sharp peaks in the hadronic cross section. Therefore, the theoretical prediction of
 $R(s)$ -and of the observables that depend on it- includes experimental information.\\
\hspace*{0.5cm}The current accuracy of this dispersion integral is at the level of $1\%$ and it is dominated by the measurements in the region below a few $GeV$
 \cite{Eidelman:1995ny,Davier:1997kw, Burkhardt:2001xp, Jegerlehner:2001wq, Jegerlehner:2003ip, Jegerlehner:2003qp, Jegerlehner:2003rx, Jegerlehner:2006ju, Jegerlehner:2008rs, Jegerlehner:2011ti,
 Hagiwara:2002ma, Hagiwara:2003da, Hagiwara:2006jt, Hagiwara:2011af, Prades:2009qp, Davier:2009zi, Davier:2009ag, Davier:2010nc, Benayoun:2009fz, Benayoun:2011mm, Benayoun:2012wc}.\\
\hspace*{0.5cm}As in the case of the fine-structure constant at the $Z$ scale we have just considered, the theoretical (in the sense commented above) prediction
 of the anomalous magnetic moment of the muon is dominated by the error on the hadronic vacuum polarization effects at non-perturbative energies. Using analyticity 
and unitarity it was shown \cite{Gourdin:1969dm} that it could be computed from the dispersion integral:
\begin{equation} \label{a_mu had LO}
 a_\mu^{had,LO}=\frac{1}{4\pi^3}\int_{4 m_\pi^2}^\infty \mathrm{d}s K(s) \sigma^0(s)=\frac{\alpha^2}{3\pi^2}\int_{4 m_\pi^2}^\infty \frac{\mathrm{d}s}{s}K(s)R(s)\,,
\end{equation}
where the kernel $K(s)\sim1/s$ is further enhancing the low-energy contributions. In particular, the dominant part is given by the two-pion vector form-factor ($F_V^{\pi\pi}(s)$):
\begin{equation} \label{a_mu had pipi}
 a_\mu^{had,\pi\pi}=\left( \frac{\alpha m_\mu}{6\pi}\right)^2 \int_{4 m_\pi^2}^\Lambda \frac{\mathrm{d}s}{s}\sigma_\pi^3|F_V^{\pi\pi}(s)|^2 K(s)\,,
\end{equation}
where $\sigma_\pi$ is defined immediately above Eq.~(\ref{B22}) and $\Lambda$ is the scale up to which we consider experimental data instead of the 
theoretical perturbative $QCD$ prediction.\\
\hspace*{0.5cm}A precise recent measurement of $\sigma(e^+e^-\to\pi^+\pi^-(\gamma))$ by the BaBar Collaboration \cite{Lees:2012cj} using the initial-state radiation method gives 
(taking into account the updated contribution of the other modes in ref.~\cite{Davier:2010nc})
$a_\mu^{had,LO}= (698$.$6\pm4$.$0)\cdot10^{-10}$, with similar results obtained by other groups.
 The current prediction is \cite{Davier:2010nc} $a_\mu^{SM}=116591865(49)\cdot10^{-11}$, to be compared to the experimental average
$a_\mu^{exp}=116592080(63)\cdot10^{-11}$, that yields for the difference $exp-SM$ $\Delta a_\mu = 215(80)\cdot 10^{-11}$, $2$.$7$ standard deviations. This is on the limit to point to new physics \cite{Passera:2010ev}. 
However, given the facts that:
\begin{itemize}
 \item There have been discrepancies \cite{Bini:2008zzh, Bondar:2009zz} in the shape of $R(s)$ (not that much in the integrated value) in the 
region of interest between the different experiments $KLOE$, $CMD2$, $SND$ and $BaBar$ that seemed to hint to underestimated systematical errors in the unfolding
 of the data. An increase of precision in the experimental measurement and a revision of the estimated uncertainties in the treatment of radiative corrections
 could help settle this issue.
 \item One should have an independent way of extracting $a_\mu^{had,LO}$ using $R_\tau$ -Eq.~(\ref{R_tau})- after an isospin rotation. This way one would have a theory prediction 
using $\tau$ data instead of $e^+e^-$ data. The results obtained with $\tau$ data seem to be slightly closer to the $SM$ ones ($2$.$3$ $\sigma$ away, \cite{Davier:2010nc}).
 \item A common treatment of radiative correction (including maybe more than one Monte Carlo generator) by the different collaborations would be desirable.
\end{itemize}
One should conclude that it is still early to claim for this $BSM$ physics.\\
\hspace*{0.5cm}The purpose of this section has been to show a few selected physical observables that allow for precision measurements at low energies that were sensitive to new 
physics at higher energies. The $SM$ would be the $EFT$ of the one describing all phenomena at this higher enery scale. In all cases this was possible because the 
$EW$ sector of the $SM$ did not fulfill the conditions of the Appelquist-Carazzone theorem, and thus the effect of heavier modes was not only in modifying the 
values of the $LECs$ and imposing additional symmetry properties. It is also instructive to see how in the cases of both $\mathrm{sin}\theta_W$ and $a_\mu$ this 
probe of $BSM$ physics is polluted by the hadronic uncertainty imposed by low-energy $QCD$, some aspects of which we are discussing through this Thesis.\\

\section{Summary of EFTs} \label{EFT_Summary}
\hspace*{0.5cm}We conclude the first block of the chapter with a recapitulation on the main features of EFTs. They are the following ones:
\begin{itemize}
\item Dynamics at low energies does not depend on details of physics at high energies.
\item One includes in the action only the relevant degrees of freedom according to the physics scale considered and to the particle masses. If there are large
 energy gaps, we can decouple the different energy scales, that is:
\begin{eqnarray*}
0\,\leftarrow\,m\,\ll\,E\,\ll\,M\rightarrow\,\infty\,.
\end{eqnarray*}
There is a well-defined perturbative way of incorporating finite corrections induced by these scales.
\item Exchanges mediated by heavy particles have been replaced by a set of local (non-renormalizable) operators involving only the light modes.
\item The $EFT$-Lagrangian is a sum of operators, $\mathcal{L} \, = \, \sum_i \, c_i O_i$, whose coefficients scale as $c_i\,\rightarrow\,E^{d_i-\gamma_i}\,c_i$.
 Here, $d_i$ comes from dimensional analysis and $\gamma_i$ is the anomalous dimension. Provided we have chosen the right degrees of freedom, anomalous dimensions
 are small and the leading behaviour at low energies is given by the lowest dimension operators. Then, going further in the expansion we improve our accuracy:
 to include all corrections up to order $1/E^p$, one should include all operators with dimension $\leq\,d_i-\gamma_i\,+\,p$, i.e., all terms with coefficients
 of dimension $\geq\,-p$. The number of operators to be considered at each order is finite.
\item Although $EFTs$ are not renormalizable in the classical sense -they are ultraviolet unstable-, they are order-by-order renormalizable for a given asked accuracy.
\item $EFTs$ have the same infrared behaviour than the underlying theory. On the contrary, $EFTs$ do not possess the same ultraviolet behaviour than the fundamental
 one, so we need to perform a matching procedure to ensure that they are equivalent at a given intermediate (matching) scale.
\item Whenever we respect symmetry principles for building the $EFT$, we will get the right theory written in terms of the variables we have chosen (\textit{Weinberg's theorem}).
\item Under some conditions, Sect.~\ref{EFT_EffectsHM}, the only remnants of the high-energy dynamics are in the $LECs$ and in the symmetries of the $EFT$ (\textit{Decoupling theorem}).
\end{itemize}

\section{Introduction to Chiral Perturbation Theory} \label{EFT_ChPT_intro}
\begin{figure}[h!] 
\centering
\includegraphics[scale=1]{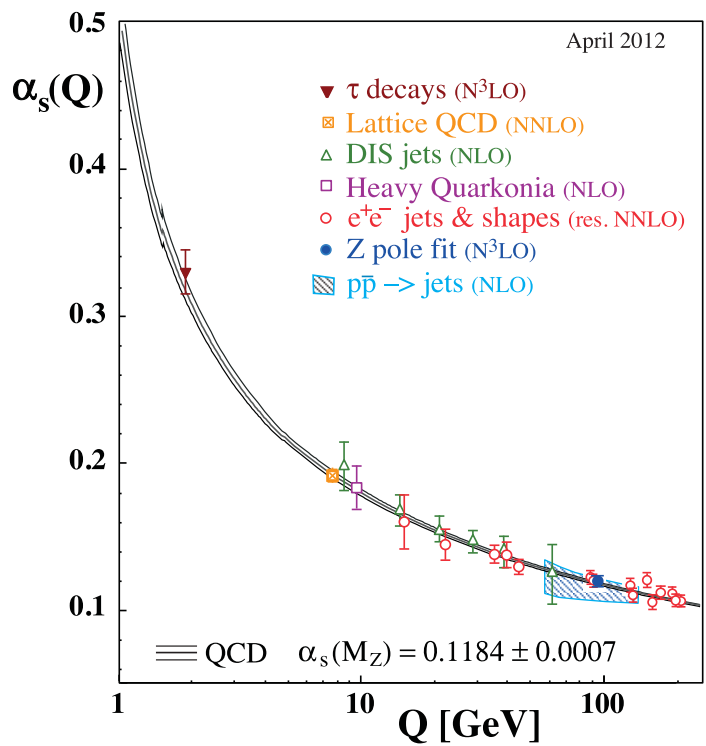} 
\caption{Summary of the values of $\alpha_S(\mu)$ at the values of $\mu$ where they are measured. The lines show the central values and
 the $\pm1\sigma$ limits of the $PDG$ \cite{Beringer:1900zz} average (see in Fig.\ref{Fig_alpha_S_input} the inputs used for this determination). 
The figure clearly shows the decrease in $\alpha_S$ when increasing $\mu$.}
\label{Fig_alpha_S_running}
\end{figure}

\hspace*{0.5cm}In this section we will introduce a paradigm of $EFTs$, $\chi PT$. We will need it to build the $R\chi T$ Lagrangian. The remainder of the chapter
will be devoted to it.\\
\hspace*{0.5cm}We have seen in the Introduction, Eq.~(\ref{runninge}), how the strong coupling evolves to smaller values with increasing energy
 and the converse in the other end of energies: it increases its value as the energy gets smaller and smaller. A look to Figure \ref{Fig_alpha_S_running}
could be instructive.\\
\\
\begin{figure}[h!] 
\centering
\includegraphics[scale=1]{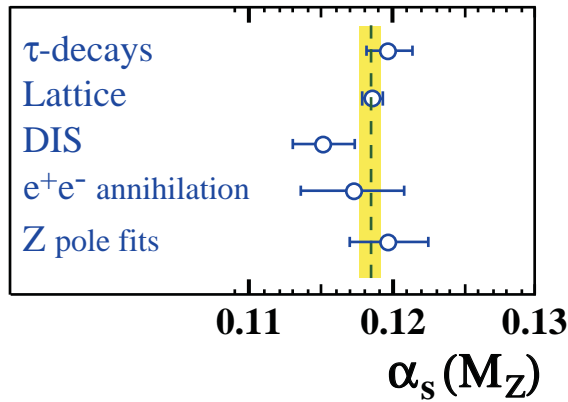} 
\caption{Summary of the values of $\alpha_S(M_Z^2)$ used as input for the world average value \cite{Beringer:1900zz}, $\alpha_S(M_Z^2)=0.1184(7)$.}
\label{Fig_alpha_S_input}
\end{figure}
\\
\hspace*{0.5cm}One sees that at $\mu\sim2$ GeV the value of $\alpha_S(\mu)$ is not yet that big to prevent a meaningful perturbative expansion
 in terms of it. Following the $RGE$ to extrapolate to lower values of $\mu$ it is found that at a typical hadronic scale $\mu\sim M_\rho, M_P$
 one can have $\alpha_S(\mu)\geq 0$.$5$ that jeopardizes that approach. Since hadronic decays of the $\t$ span the range $0$.$14-1$.$78$ GeV, one
 should find a way out that starts from $QCD$. One rigorous alternative is to simulate on the lattice the $QCD$ action. We will not report about
 this option here. Although it has proved to be very successful in many non-perturbative strong-interaction problems, the
 processes we study here have not been addressed by the lattice community yet. One can also construct models that keep this or that 
feature of $QCD$, but we do not find this alternative satisfactory. Finally, one can build an $EFT$ of $QCD$ for this subset of energies 
using as variables the active degrees of freedom as we will describe next.\\
\hspace*{0.5cm}In order to discuss the global symmetries of the $QCD$ Lagrangian, Eq.~(\ref{fullLQCDe}), we will restrict ourselves to the 
so-called light sector of $QCD$, with $n_f$ light flavours. In our case, $n_f = 3$: $u,\,d,\,s$, that are much lighter than the so-called heavy 
quarks $c,\,b,\,t$. The characteristic hadronic scale lies in between both regimes. Therefore, integrating out these three heavy quarks we go from
 \footnote{The effective strong coupling in the two theories are related by a matching condition, introduced in Sect.~\ref{EFT_Example}. This is discussed
 in Ref.~\cite{Rosell:2007kc}, for instance.} $QCD^{n_f=6}$ to $QCD^{n_f=3}$. The Lagrangian of $QCD^{n_f=3}$ in the limit of massless light quarks 
(so-called chiral limit) is:
\begin{equation} \label{QCD-chiral}
\mathcal{L}_{QCD}^0 \,=\, i \,\overline{q}_L D\! \! \! \! / \; q_L \,+\,i \,\overline{q}_R D\! \! \! \! / \; q_R
\,-\,\frac{1}{4}G^a_{\mu\nu}G^{\mu\nu}_a
\,,
\end{equation}
where the upper-index zero reminds us the limit taken, and we have employed the usual notation for the left(right)-handed spinors $q_L(q_R)$ and projectors:
 $q_L \equiv P_Lq$ ($q_R \equiv P_Rq$).\\
\hspace*{0.5cm}The Lagrangian, Eq.~(\ref{QCD-chiral}), is invariant under global transformations belonging to $SU(n_f)_L\otimes SU(n_f)_R\otimes U(1)_V\otimes U(1)_A$.
 $U(1)_V$ is trivially realized in the meson sector but it gives rise to baryon number conservation, $U(1)_A$ gets broken by anomalies and explains why the 
$\eta'$ is heavier than the $\eta$ or the kaons. Finally, it remains  $SU(n_f)_L\otimes SU(n_f)_R$, the so-called chiral group of transformations in flavour
 space, that acts on the chiral projections of the quark fields in the following way:
\begin{eqnarray} \label{chiraltransfqs}
q_L &\longrightarrow& q_L \,' \,=\,g_Lq_L\,, \nonumber \\
q_R &\longrightarrow& q_R \,' \,=\,g_Rq_R\,,
\end{eqnarray}
where $g_{L,R} \in SU(n_f)_{L,R}$.\\
\hspace*{0.5cm}This chiral symmetry, which should be approximately valid in the light quark sector ($u$, $d$, $s$), is however not seen in the
 meson spectrum (Table \ref{Spectrummesons}):\\
\begin{table}[h]  
\begin{center}
\renewcommand{\arraystretch}{1.2}
\begin{tabular}{|c|c|c|c|c|c|}
\hline
$J^P$&Particle&m (MeV)&$J^P$&Particle&m (MeV)\\
\hline
$0^-$ & $\pi^0$ & $\sim135$.$0$ & $0^+$ & a$_0$ & $\sim985$\\
      & $\pi^\pm$ & $\sim139$.$6$ &     & a$_0^\pm$ & $\sim985$\\
      & $\eta,\eta'$ & $\sim547$.$9$, $(957$.$8)$ & &$f_0$ & $\sim980$\\
      & $K^\pm$ & $\sim493$.$7$ & &$K^{*\pm}$& $\sim800$\\
      & $K^0, \overline{K}^0$ & $\sim497$.$7$ & & $K^{*0}, \overline{K}^{*0}$ & $\sim800$\\
\hline
$1^-$ & $\rho^0$ & $\sim775$.$8$ & $1^+$ & a$_1^0$ & $\sim1230$\\
      & $\rho^\pm$ & $\sim775$.$5$ & & a$_1^\pm$ & $\sim1230$\\
      & $\omega,\phi$ & $\sim782$.$7$, $(1019$.$5)$& & $h_1,f_1$ & $\sim1170$, $(1281$.$8)$\\
      & $K^{*\pm}$ & $\sim892$.$0$ & &$K_1^{*\pm}$& $\sim1273$\\
      & $K^{*0}, \overline{K}^{*0}$ & $\sim891$.$7$ & & $K_1^{*0}, \overline{K}_1^{*0}$ & $\sim1273$\\
\hline
\end{tabular}
\caption{\small{Spectrum of the lightest mesons \cite{Amsler:2008zzb}. The $f_0(500)$ \cite{Beringer:1900zz} or $\sigma$ \cite{Colangelo:2001df, Pelaez:2003dy, Caprini:2005zr}
is not included in the Table.}}
\label{Spectrummesons} 
\end{center}
\end{table}
\\
\hspace*{0.5cm}The conclusions we draw are the following ones:
\begin{itemize}
\item Mesons are nicely classified into $SU(3)_V$ representations.
\item Reversing intrinsic parity changes drastically the spectrum (just compare the masses of the octet of $J^P=0^-$ versus that with $J^P=0^+$, or $J^P=1^-$
 vs. $J^P=1^+$). Thus, a transformation involving $\gamma_5$ is not a symmetry of the spectrum. Chiral transformations seem not to be a symmetry of low-energy
 $QCD$.
\item The octet of $J^P=0^-$ stands out being its members much lighter than the ones in other octets.\\
\end{itemize}
\hspace*{0.5cm}Although chiral symmetry changes the parity of a given multiplet, the previous spectrum is not contradictory with it. We shall
 recall that there exist two ways of realizing a global symmetry: In the Wigner-Weyl way all the symmetries of the Lagrangian are shared by the vacuum of the
 theory and then they are manifest in the spectrum. In the Nambu-Goldstone way, there is a so-called spontaneous breakdown of the symmetry that makes 
the observed spectrum compatible with the underlying approximate symmetry.\\
\hspace*{0.5cm}There are two fundamental theorems concerning spontaneous symmetry breaking: Goldstone theorem \cite{Goldstone:1961eq, Goldstone:1962es}, which is devoted to 
global continuous symmetries and Higgs-Kibble theorem \cite{Englert:1964et, Higgs:1964ia, Higgs:1964pj, Higgs:1966ev, Kibble:1967sv}, that worries about local gauge symmetries.\\
\hspace*{0.5cm}Chiral symmetry is a global symmetry, then Goldstone theorem is the one applied here. We can state it in the following way: 
\textit{Given a global continuous symmetry of the Lagrangian: either the vacuum shares the symmetry of the Hamiltonian; or there appear spin zero massless particles
 as a display of Spontaneous Symmetry Breaking. In the last case, for every spontaneously broken generator, the theory must contain a massless particle, the 
so-called Goldstone boson}.\\
\hspace*{0.5cm}Vafa and Witten showed \cite{Vafa:1983tf} that the ground state of the theory must be invariant under vector transformations, so that Spontaneous
 Symmetry Breaking cannot affect the vector part of the chiral subgroup ($V^a_\mu \equiv R^a_\mu \,+\,L^a_\mu$), but the axial one 
($A^a_\mu \equiv R^a_\mu \,-\,L^a_\mu$).\\
\hspace*{0.5cm}Let us consider \cite{Pich:1995bw, Scherer:2002tk} a Noether charge $Q$, and assume the existence of an operator $\cO$ that satisfies
\begin{equation} \label{SCSB}
\bra 0|[Q,\cO]|0 \ket \, \ne \, 0 \, ;
\end{equation}
the only possibility for this to be valid is that $Q|0 \ket\,\ne \, 0$. Goldstone theorem states there exists a massless state $| G \ket$ such that
\begin{equation} \label{SCSB2} 
\bra 0 | J^0 | G \ket \bra G | \cO | 0 \ket \,\ne 0 \,. 
\end{equation}
\hspace*{0.5cm}It is important to notice that the quantum numbers of the Goldstone boson are dictated by those of $J^0$ and $\cO$. The quantity in the left-hand
 side of Eq.~(\ref{SCSB}) is called the order parameter of $SSB$.\\
\hspace*{0.5cm}Considering that $U(1)_A$ is affected by anomalies, only $SU(n_f)_A$ can be concerned with the Goldstone theorem. Then, for $n_f=3$ and for the 
lightest quark flavours ($u$, $d$, $s$) we end up with eight broken axial generators of the chiral group and, correspondingly, eight pseudoscalar Goldstone 
states $| G^a\ket$, which can be identified with the eight lightest hadrons (three $\pi$s, four $K$s and the $\eta$, see Table \ref{Spectrummesons}), their 
(relatively) small masses being generated by the explicit breaking of chiral symmetry induced by the quark mass matrix entering the $QCD$ Lagrangian. The 
corresponding operators, $\cO^a$, must be pseudoscalars. The simplest possibility is  $\cO^a=\overline{q}\gamma_5 \lambda^a q$, which satisfies
\begin{equation}
\bra 0|[Q^a_A,\overline{q}\gamma_5 \lambda^b q]|0 \ket \, = \,-\frac{1}{2}\,\bra 0|\overline{q}\left\lbrace \lambda^a,\lambda^b \right\rbrace q|0 \ket \,= -\frac{2}{3}\, \delta^{ab}\, 
\bra 0 | \overline{q}q | 0 \ket \,. 
\end{equation}
\hspace*{0.5cm}The quark condensates
\begin{equation} \label{quarkcondensate} 
\bra 0 | \overline{u}u | 0 \ket \,=\,
\bra 0 | \overline{d}d | 0 \ket \,=\,
\bra 0 | \overline{s}s | 0 \ket \,\ne \,0
\end{equation}
are then the natural order parameters of Spontaneous Chiral Symmetry Breaking ($S\chi SB$).\\
\section{Different representations for the Goldstone fields} \label{EFT_Different_representations_pGs}
\hspace*{0.5cm}Based on the previous reasoning, our basic assumption is the pattern of $S\chi SB$:
\begin{equation}
G\,\equiv\,SU(3)_L\otimes\,SU(3)_R\,\,\begin{scriptsize}\overrightarrow{S\chi SB}\end{scriptsize}\,\,H\,\equiv\,SU(3)_V.
\end{equation}
\hspace*{0.5cm}Since there is a mass gap between the lightest multiplet of pseudoscalar particles and the rest of the spectrum, we can easily apply the Weinberg's 
approach and formulate an $EFT$ dealing only with these modes.\\
\hspace*{0.5cm}The general formalism for $EFT$-Lagrangians with $SSB$ was worked out by Callan, Coleman, Wess and Zumino ($CCWZ$) \cite{Coleman:1969sm, Callan:1969sn}.
 A very clear explanation can be found in Ref.~\cite{Manohar:1996cq}.\\
\hspace*{0.5cm}Consider a theory in which a global symmetry group $G$ is spontaneously broken down to one of its subgroups, $H$. The vacuum manifold is the 
coset space $G$/$H$.\\ 
\hspace*{0.5cm}The set of coordinates we choose has to be able to describe the local orientation of the vacuum for small fluctuations around the standard 
vacuum configuration. Let $\Xi(x)\in G$ be the rotation matrix that transforms the standard vacuum configuration to the local field one. Due to the invariance
 of the vacuum under $H$ transformations, $\Xi$ happens to be not unique; namely, $\Xi(x)\,h(x)$ -where $h$ $\in$ $H$- gives the same configuration. In the 
present case, $\Xi(x)\in$ $O(N)$ and we can parameterize any vector $\phi$ by means of the suitable $\Xi$ matrix:
\begin{equation}
\phi(x) \,=\, \Xi(x) \left( \begin{array}{c} 0 \\ 0 \\ . \\ . \\ . \\ 0 \\  v \end{array} \right) \,.
\end{equation}
The same configuration $\phi(x)$ can also be described by $\Xi(x)\,h(x)$. In our example, $h(x)$ is a matrix of the form:
\begin{equation}
h(x)\,=\,\left( \begin{array}{cc} h'(x) & 0 \\ 0 & 1 \end{array} \right) \,,
\end{equation}
with $h'(x)$ is an arbitrary $O(N-1)$ matrix, since:
\begin{equation}
\left( \begin{array}{cc} h'(x) & 0 \\ 0 & 1 \end{array} \right)\,\left( \begin{array}{c} 0 \\ 0 \\ . \\ . \\ . \\ 0 \\  v \end{array} \right)\,=\,\left( \begin{array}{c} 0 \\ 0 \\ . \\ . \\ . \\ 0 \\  v \end{array} \right)\,.
\end{equation}
\hspace*{0.5cm}The $CCWZ$ prescription is to pick a set of broken generators $X$, and choose
\begin{equation} \label{CCWZ}
\Xi (x)\,=\, e^{i\,X\cdot \pi(x)}\, ,
\end{equation}
where $\pi(x)$ describes the Goldstone modes.\\
\hspace*{0.5cm}Under a global transformation $g$, the matrix $\Xi(x)$ changes to a new matrix $g$ $\Xi(x)$ because $\phi(x)\,\rightarrow\,g\phi(x)$, that it is
 not in the standard form of Eq.~(\ref{CCWZ}), but can be written as
\begin{equation}\label{CCWZ2a}
g \, \Xi\,=\, \Xi '\, h \, ,
\end{equation}
that is usually turned into
\begin{equation} \label{CCWZ2b}
\Xi(x) \, \longrightarrow g \, \Xi(x) \, h^{-1}(g,\,\Xi(x))\, .
\end{equation}
\hspace*{0.5cm}$CCWZ$ formalism is characterized by equations (\ref{CCWZ}) and (\ref{CCWZ2b}) for the pseudo-Goldstone boson ($pG$) fields and their transformation
 law. The transformation $h$ appearing there is non-trivial because the Goldstone boson manifold is curved. Any other choice gives the same results as $CCWZ$ formalism for all observables,
 such as the $S$ matrix, but does not give the same off-shell Green functions.\\
\hspace*{0.5cm}The $CCWZ$ prescription in Eq.~(\ref{CCWZ}) says nothing about which set of broken generators we would better choose. Depending on our choice, we will have a 
different base. There are two that have become standard in order to write the $QCD$ chiral Lagrangian, the so-called $\xi$-basis and the $\Sigma$-basis \cite{Manohar:1996cq}.
 Each of them brings us a different equivalent parameterization (commonly called $U$ and $u$, respectively).\\
\hspace*{0.5cm}There are many simplifications that occur for $QCD$ because the coset space $G/H$ is isomorphic to a Lie group.\\
\hspace*{0.5cm}Let $X^a\,=\,T^a_L\,+\,T^a_R$ be our choice of broken generators.\\
\hspace*{0.5cm}An element $g$ $\in$ $G$ can be written as:
\begin{equation} \label{g-xi}
g\,=\,\left( \begin{array}{cc} L & 0 \\ 0 & R \end{array} \right)\,,
\end{equation}
where $L(R)\,\in\,SU(3)_{L(R)}$. The unbroken transformations are of the form (\ref{g-xi}),  with $L\,=\,R\,=\,U$,
\begin{equation} 
g\,=\,\left( \begin{array}{cc} U & 0 \\ 0 & U \end{array} \right)\,.
\end{equation}
\hspace*{0.5cm}Now, using the $CCWZ$ recipe, Eq.~(\ref{CCWZ}):
\begin{equation}
\Xi(x)\,=\,e^{\,i\,X\cdot \pi(x)}\,=\,exp\left( \begin{array}{cc} i\,T\cdot \pi & 0 \\ 0 & -i\,T\cdot \pi \end{array}
\right)\,=\,\left( \begin{array}{cc} \xi(x) & 0 \\ 0 & \xi^\dagger(x) \end{array} \right) \, ,
\end{equation}
where
\begin{equation}\label{xi}
\xi\,=\,e^{\,i\,X\cdot \pi} \,
\end{equation}
stands for the upper block of $\Xi(x)$. The transformation rule Eq.~(\ref{CCWZ2b}) gives
\begin{equation}
\left( \begin{array}{cc} \xi (x) & 0 \\ 0 & \xi^\dagger(x) \end{array} \right) \, \rightarrow \,
\left( \begin{array}{cc} L & 0 \\ 0 & R \end{array} \right)
\left( \begin{array}{cc} \xi(x) & 0 \\ 0 & \xi^{\dagger}(x) \end{array} \right)
\left( \begin{array}{cc} U^{-1}(x) & 0 \\ 0 & U^{-1}(x) \end{array} \right) \, ,
\end{equation}
and, consequently, the transformation rule for $\xi$,
\begin{equation} \label{transfxi}
\xi(x)\,\longrightarrow\, L \, \xi(x)\, U^{-1}(x) \,=\, U(x) \, \xi(x)\, R^{\dagger} \,,
\end{equation}
which defines $U$ in terms of $L$ ($R$) and $\xi$. Instead, if we choose $X^a\,=\,T^a_L$ as the basis for broken generators, we will have $U\,=\,R$, and
\begin{equation} \label{transfsigma}
\Sigma(x)\,\longrightarrow\, L \, \Sigma(x)\, R^{\dagger}\,.
\end{equation}
\hspace*{0.5cm}Finally, comparing Eqs.~(\ref{transfxi}) and (\ref{transfsigma}), one concludes that $\Sigma$ and $\xi$ are related by
\begin{equation} \label{equivxisigma}
\Sigma(x)\,=\,\xi^2(x)\,.
\end{equation}
\hspace*{0.5cm}In the context of $\CPT$, everybody writes $U(x)$ instead of $\Sigma(x)$ and $u(x)$ substitutes $\xi(x)$. It is also more common to employ
 $\Phi(x)$ for the coordinates of the Goldstone fields. We will follow this notation from now on.\\
\hspace*{0.5cm}The Goldstone boson nature restricts these fields to be angular variables, thus dimensionless. It is convenient to work with boson fields of 
mass dimension one, which motivates the standard choice:
\begin{equation} \label{uU} 
u\,=\,e^{i\,T\cdot\,\Phi/F},\,\,\,\,U\,=\,u^2\,  
\end{equation}
where $F\,\sim\,92$.$4$ MeV is the pion decay constant.
\begin{equation} \label{PhiGoldstones} 
\Phi(x)\,=\,\sqrt{2}\,T_a \Phi^a(x)\,=\frac{1}{\sqrt{2}} \sum_{a=1}^8  \lambda_a \Phi^a\,=\,\left( \begin{array} {ccc}
\frac{1}{\sqrt{2}}\pi^{0}+\frac{1}{\sqrt{6}}\eta_8 & \pi^{+} & K^{+} \\
\pi^{-} & -\frac{1}{\sqrt{2}}\pi^{0}+\frac{1}{\sqrt{6}}\eta_8 & K^{0} \\
K^{-} & \overline{K}^{0} & -\frac{2}{\sqrt{6}}\eta_8 \end{array} \right) \,,
\end{equation}
where the Gell-Mann matrices in flavour space, $\lambda_a$, -which are the fundamental representation of $SU(3)$- have been introduced with the same normalization
 as for $SU(N_C)$ generators of $QCD$.\\
\hspace*{0.5cm}Notice that $U(\Phi)$ transforms linearly under the chiral group, but the induced transformation on the $pG$ fields is highly non-linear.\\
\hspace*{0.5cm}There is abundant good literature available on this topic and its specific application to $\chi PT$ \cite{Ecker:1994gg, Pich:1995bw, Scherer:2002tk, Leutwyler:1993iq, Bijnens:1993xi, Donoghue:1992dd}.\\
\section{Lowest order Lagrangian. Method of external currents}\label{EFT_externalcurrents}
\hspace*{0.5cm}In order to obtain an $EFT$ realization of $QCD$ at low energies for the light quark sector, we should write the most general Lagrangian involving
 the matrix $U(\Phi)$ (or $u(\Phi)$), which respects chiral symmetry. The Lagrangian can be organized in terms of increasing powers of momentum or,  equivalently,
 of derivatives (the subindex $2n$ refers to that):
\begin{equation} \label{expansion}
\mathcal{L}_{\CPT}\,=\,\sum_{n=1}^\infty \mathcal{L}_{2n} \,,
\end{equation}
being the dominant behaviour at low energies given by the terms with the least number of derivatives. Unitarity of $U$ obliges two derivatives to be present 
for having a non-trivial interaction. At lowest order, the effective chiral Lagrangian in the $U$-formalism is uniquely given by the term:
\begin{equation} \label{U-p2}
\mathcal{L}_2 \,=\, \frac{F^2}{4} \bra \partial_\mu U^{\dagger} \partial^\mu U \ket \, .
\end{equation}
where $\bra A \ket$ is short for trace in flavour space of the matrix $A$.
\\
\hspace*{0.5cm}Expanding the matrix that exponentiates the $pG$ fields in a power series in $\Phi$, we get the Goldstone kinetic terms plus a tower of interactions
 increasing in the number of pseudoscalars. It is a capital fact that all interactions among the Goldstones can be predicted in terms of a single coupling, $F$.
 The non-linearity of the $EFT$-Lagrangian relates the amplitudes of processes involving a different number of $pGs$, allowing for absolute predictions in 
terms of $F$. This sector was thoroughly studied by Weinberg \cite{Weinberg:1978kz, Weinberg:1966kf, Weinberg:1966fm, Weinberg:1968de}.\\
\hspace*{0.5cm}But the lightest mesons do not interact solely due to elastic scattering among themselves. In addition to the strong interaction, they also experience 
electromagnetic and (semileptonic) electroweak interactions and this has to be taken into account. In order to compute the associated Green functions, 
we will follow the procedure employed by Gasser and Leutwyler, who developed $\CPT$ consistently to one loop (\cite{Gasser:1983yg, Gasser:1984gg, Gasser:1984ux}).
 We extend the chiral invariant $QCD$ massless Lagrangian, Eq.~(\ref{QCD-chiral}), by coupling the quarks to external Hermitian matrix fields $v_\mu,\,a_\mu,\,s,\,p$ 
\footnote{We do not include the tensor source \cite{Cata:2007ns}.}:
\begin{equation}
\mathcal{L}_{QCD}\,=\,\mathcal{L}_{QCD}^{0} \,+\,
\overline{q}\gamma^\mu\left(v_\mu+\gamma_5a_\mu\right)q\,-\,\overline{q}\left(s-i\gamma_5p\right)q \, .
\end{equation}
\hspace*{0.5cm}External photons and $W$ boson fields are among the gauge fields and (pseudo)sca\-lar fields provide a very convenient way of incorporating explicit
 $\chi$SB through the quark masses (see Eq.~(\ref{Mq}), where $\mathcal{M}$ is defined.):
\begin{eqnarray} \label{breakingexternalsources}
 r_\mu  \, & \rightarrow\, &  r_\mu + e \mathcal{Q} A_\mu\,,\nonumber\\
 \ell_\mu \, & \rightarrow\, & \ell_\mu + e \mathcal{Q} A_\mu \,+\,\frac{2\,e}{\sqrt{2}\mathrm{sin}\theta_W}\left(W^\dagger T_+\,+\,h.c.\right)\,,\nonumber\\
 s \, & \rightarrow\,  & s \,+ \,\mathcal{M}\,,
\end{eqnarray}
being
\begin{equation}  \label{Q T+}
 \mathcal{Q} = \left( \begin{array}{ccc} \frac{2}{3} & 0 & 0 \\
                       0 & -\frac{1}{3} & 0 \\
                       0 & 0 &  -\frac{1}{3} 
                      \end{array} \right) \;\;,\;\; T_+ =\left( \begin{array}{ccc} 0 & V_{ud} & V_{us} \\
                       0 & 0 & 0 \\
                       0 & 0 & 0 
                      \end{array} \right)\,.
\end{equation}
\hspace*{0.5cm}Inclusion of external fields promotes the global chiral symmetry to a local one:
\begin{eqnarray} \label{chiraltransfsources}
q \, & \rightarrow & \,g_R\,q_R\,+\,g_L\,q_L\,,\nonumber \\
s\,+\,ip \, & \rightarrow & \, g_R\,(s\,+\,ip)\,g_L^\dagger\, ,\nonumber \\
\ell_\mu \, &  \rightarrow  & \, g_L\,\ell_\mu\, g_L^{\dagger}\,+\,ig_L\,\partial_\mu \,g_L^{\dagger}\, ,\nonumber \\ 
r_\mu  \, &  \rightarrow & \, g_R \,r_\mu\, g_R^{\dagger}\,+\,ig_R\,\partial_\mu\, g_R^{\dagger}\, .
\end{eqnarray}
where we have introduced the definitions $r_\mu \equiv v_\mu+a_\mu$ and $\ell_\mu \equiv v_\mu - a_\mu$; and requires the introduction of a covariant derivative,
 $D_\mu U$, and associated non-Abelian field-strength tensors, $F_{L,R}^{\mu\nu}$:
\begin{eqnarray} \label{F_L,R}
D_\mu U\,=\,\partial_\mu U\,-\,ir_\mu U\,+\,iU\ell_\mu\,\,,\,\,D_\mu U\,\rightarrow\,g_R \,D_\mu\, U\, g_L^{-1},\nonumber \\
F_x^{\mu\nu}\,=\,\partial^\mu\,x^\nu\,-\,\partial^\nu\,x^\mu\,-\,i\left[ x^\mu,\,x^\nu\right]\,\,,\,\,x\,=\,r,\,\ell. 
\end{eqnarray}
\hspace*{0.5cm}The transformations of the external sources under the discrete symmetries $P//C$ are as follows:
\begin{eqnarray} \label{discretetransfsources}
 s\,+\,ip \, & & \rightarrow \, s\,-\,ip \,//\,(s\,-\,ip)^\top ,\nonumber \\
\ell_\mu \, & & \rightarrow \,r^\mu  \,//\, -r_\mu^\top ,\nonumber \\
r_\mu \, & & \rightarrow \,\ell^\mu \,//\, -\ell_\mu^\top \nonumber \,.
\end{eqnarray}
\hspace*{0.5cm}The power of the external field technique is exhibited when computing chiral Noether currents. Green functions are obtained as functional 
derivatives of the generating functional, $Z\left[ v^\mu,\,a^\mu,\,s,\,p\right]$, defined via the path-integral formula
\begin{equation}
\mathrm{exp}\left\lbrace iZ\right\rbrace\,=\,\int\mathcal{D}q\,\mathcal{D}\overline{q}\, \mathcal{D}G_\mu\,\mathrm{exp}\left\lbrace i\int\mathrm{d}^4x\,\mathcal{L}_{\mathcal{QCD}}\right\rbrace\,=\,\int\mathcal{D}U \mathrm{exp}\,\left\lbrace i\int\mathrm{d}^4x\,\mathcal{L}_{\mathrm{eff}}\right\rbrace\,.
\end{equation}
\hspace*{0.5cm}At lowest order in momenta, $Z$ reduces to the classical action, $S_2\,=\,\int\mathrm{d}^4x\,\mathcal{L}_2$, and the currents can be trivially computed by
 taking suitable derivatives with respect to the external fields. In particular, one realizes the physical meaning of the pion decay constant, $F$, defined as
\begin{equation} \label{F} 
\bra0|(J^\mu_A)^{12}|\pi^+(p)\ket\,\equiv\,i\sqrt{2}F\,p^\mu\,.
\end{equation}
\\      
\hspace*{0.5cm}The locally chiral invariant Lagrangian of lowest order describing the strong, electromagnetic and semileptonic weak interactions of mesons 
was given by Gasser and Leutwyler \cite{Gasser:1983yg, Gasser:1984gg}:
\begin{equation} \label{L-Op2-U}
\mathcal{L}_2\,=\,\frac{F^2}{4}\,\bra D_\mu U D^\mu U^{\dagger}\,+\,\chi U^{\dagger}\,+\,\chi^{\dagger} U\ket\,,\,\,\,\chi\,=\,2B(s\,+\,ip).
\end{equation}
\hspace*{0.5cm}The two $LECs$ that characterize completely the $\cO(p^2)$-chiral Lagrangian are related to the pion decay constant and to the quark condensate in the chiral limit:
\begin{eqnarray} \label{F,B} 
F_\pi & = & F\left[ 1\,+\,\cO(m_q)\right]\,=\,92\mathrm{.}4\, \mathrm{MeV}\,,\nonumber\\
\bra 0 | \overline{u}u | 0 \ket & = & -F^2\,B\left[ 1\,+\,\cO(m_q)\right].  
\end{eqnarray}
\hspace*{0.5cm}A consistent chiral counting must be developed to organize the infinite allowed terms in the Lagrangian. Depending on the actual relation of 
these two $LECs$ one could have different $EFTs$ for low-energy $QCD$. This illustrates the fact that the Weinberg's approach to $EFTs$ does only rely on 
symmetries but does not have dynamical content incorporated. This issue is studied in the next section.\\
\section{Weinberg's power counting rule}\label{EFT_Weinbergs_counting_rule}
\hspace*{0.5cm}Chiral Lagrangians were originally organized in a derivative expansion based on the following chiral counting rules (see Table \ref{StandardChiralCounting}).\\
\begin{table}[h]
\begin{center}
\renewcommand{\arraystretch}{1.2}
\begin{tabular}{|c|c|}
\hline
Operator&$\cO$\\
\hline
$U$ & $p^0$\\
$D_\mu U$, $v_\mu$, $a_\mu$  & $p$\\
$F_{L,R}^{\mu\nu}$ & $p^2$ \\
$s$, $p$ & $p^2$\\
\hline
\end{tabular}
\caption{\small{Chiral counting in Standard $\CPT$.}}
\label{StandardChiralCounting}
\end{center}
\end{table}
\hspace*{0.5cm}Using Eqs.~(\ref{uU}), (\ref{PhiGoldstones}) and setting the external scalar field equal to the quark mass matrix -that is, explicitly breaking chiral symmetry in the same way it happens in $QCD$-,
\begin{equation} \label{Mq} 
s\,=\,\mathcal{M}_q\,=\,\left( \begin{array}{ccc} m_u & 0 & 0 \\ 0 & m_d & 0 \\ 0 & 0 & m_s \end{array} \right) \, ,
\end{equation}
one can straightforwardly read off from Eq.~(\ref{L-Op2-U}) the pseudoscalar meson masses to leading order in $m_q$:
\begin{eqnarray} \label{mesonquarkmasses}
m_{\pi^\pm}^2 & = & 2\widehat{m}B\,,\nonumber\\
m_{\pi^0}^2 & = & 2\widehat{m}B\,-\,\varepsilon\,+\,\cO(\varepsilon^2) ,\nonumber\\
m_{K^\pm}^2 & = & (m_u\,+\,m_s)B\,,\nonumber\\
m_{\overline{K}^0,\,K^0}^2 & = & (m_d\,+\,m_s)B\,,\nonumber\\
m_{\eta_8}^2 & = & \frac{2}{3}(\widehat{m}\,+\,2m_s)B\,+\,\varepsilon\,+\,\cO(\varepsilon^2),
\end{eqnarray}
where \cite{Pich:1995bw}
\begin{eqnarray}
\widehat{m}\,=\,\frac{1}{2}\,(m_u\,+\,m_d)\,,\,\,\,\varepsilon\,=\,\frac{B}{4}\,\frac{(m_u-m_d)^2}{(m_s-\widehat{m})}\,.
\end{eqnarray}
\hspace*{0.5cm}With the quark condensate assumed to be non-vanishing in the chiral limit ($B\neq0$), these relations explain the chiral counting rule in Table \ref{StandardChiralCounting}.\\
\hspace*{0.5cm}Up to this point, $\CPT$ is a very elegant way of understanding the phenomenological successes obtained in the pre-$QCD$ era. The well-known relations already got with
 current algebra techniques are recovered using Eqs.~(\ref{L-Op2-U}) and (\ref{Mq}):
\begin{eqnarray}
F^2\,m_\pi^2 & = & -2\widehat{m}\,\bra 0 | \overline{q}\,q | 0 \ket\,\,\,\,\,\,\cite{GellMann:1968rz}\\
B & = & \frac{m_\pi^2}{2\widehat{m}}\,=\,\frac{m_{K^+}^2}{m_s\,+\,m_u}\,=\,\frac{m_{K^0}^2}{m_s\,+\,m_d}\,\,\,\,\,\,\cite{GellMann:1968rz}\,,\,\cite{Weinberg:1977ma},\\
3m_{\eta_8}^2 & = & 4m_{K}^2\,-\,m_{\pi}^2\,\,\,\,\,\,\cite{GellMann:1957zzb}\,,\,\cite{Okubo:1961jc},
\end{eqnarray}
but the real power of $\CPT$ -as an $EFT$- lies in the fact that it gives a perfectly defined way of taking into account the next orders in the chiral expansion and the quantum corrections.\\
\hspace*{0.5cm}$\CPT$ is based on a two-fold expansion: as a low-energy effective theory, it is an expansion in small momenta. On the other hand, it is also an expansion in the quark masses,
 $m_q$, around the chiral limit. In full generality, the Lagrangian is:
\begin{equation}
\mathcal{L}_{\CPT}\,=\,\sum_{i,j}\mathcal{L}_{ij}\,,\quad\mathcal{L}_{ij}\,=\,\cO(p^i\,m_q^j)\,.
\end{equation}
\hspace*{0.5cm}The two expansions become related by Eq.~(\ref{mesonquarkmasses}). If the quark condensate is non-vanishing in the chiral limit, meson masses squared
 start out linear in $m_q$. Assuming the linear terms to give the dominant behaviour there, we end up with the standard chiral counting with $m_q\,\sim\,\cO(p^2)$
 and
\begin{equation}
\mathcal{L}_{eff}\,=\,\sum_d \mathcal{L}_d\,,\quad\mathcal{L}_d\,=\,\sum_{i\,+\,2j\,=\,d}\mathcal{L}_{ij}\,.
\end{equation}
\hspace*{0.5cm}In short, Eq.~(\ref{L-Op2-U}) has two $LECs$: $F$ and $B$. $\CPT$ assumes $B$ to be big compared to $F$ and organizes the chiral counting according to that assumption.
 Generalized $\chi PT$ \cite{Fuchs:1991cq, Stern:1993rg, Knecht:1993eq} was developed as a scheme adapted for much smaller values of $B$, a picture that is not supported
 by lattice evaluations of the condensate \cite{Giusti:1998wy, Hernandez:1999cu, Chiu:2003iw, Becirevic:2004qv, Gimenez:2005nt, McNeile:2005pd, DeGrand:2006uy, :2009fh}
 or estimations based on sum rules \cite{Spiridonov:1988md, Chetyrkin:1994qu, Chetyrkin:1996xa, Dosch:1997wb, Jamin:2002ev, Dominguez:2007hc, Bordes:2010wy}.
 Another issue that is still unsolved concerns the possible instabilities due to vacuum fluctuations of sea $q-\bar{q}$ pairs, as the number $n_f$ of light fermions increases 
\cite{DescotesGenon:1999uh, DescotesGenon:1999zj, DescotesGenon:2000di, DescotesGenon:2001tn, DescotesGenon:2002yv, DescotesGenon:2003cg, DescotesGenon:2007ta, Bernard:2010ex, Bernard:2012fw, Bernard:2012ci}.\\
\hspace*{0.5cm}In order to build higher orders in the chiral expansion it is better to use the chiral tensor formalism that uses $u$ instead of $U$ as the exponential
 non-linear realization of the $pGs$. This is so because the building blocks in the $U$-formalism ($U$, $F_{L,R}^{\mu,\nu}$, $\chi$) do not transform in the
 same way under the chiral group. While this is not an issue for dealing with the lowest order Lagrangians, in can make very difficult to determine the minimal set 
of independent operators at higher chiral orders.\\
\hspace*{0.5cm}The Lagrangian must be chiral symmetric, hermitian, and Lorentz, parity ($P$) and charge conjugation ($C$) invariant. In the $u$-formulation 
one uses traces of chiral tensors either transforming as
\begin{equation}\label{goodtransformation}
X\,\rightarrow \, h(g,\,\Phi) \, X \,h(g,\,\Phi)^\dagger \,,
\end{equation}
or being chiral invariant.\\
\hspace*{0.5cm}With this purpose, we define the chiral tensors:
\begin{eqnarray} \label{L-Op2-u}
u_\mu &=& i\left\{ u^\dagger \left( \partial_\mu-ir_\mu \right) u - u \left( \partial_\mu-i l_\mu
\right)u^\dagger \right\} \, ,\nonumber \\
\chi_{\pm} &=& u^\dagger\, \chi\, u^\dagger \pm u\, \chi^\dagger\, u \, ,
\end{eqnarray}
The lowest order chiral Lagrangian that can be written respecting also all other symmetries is then
\begin{equation} \label{p2-u}
\mathcal{L}_2 \,=\, \frac{F^2}{4} \bra u_\mu\, u^\mu\, + \,\chi_+ \ket \, .
\end{equation}
\hspace*{0.5cm}Explicit $\chi SB$ is incorporated through $\chi_+$, where $\chi$ is given by Eq.~(\ref{L-Op2-U}). Then, in the isospin limit,
\begin{equation}
\chi\,=\,2\,B\,s\,=\,\left( \begin{array}{ccc} m_\pi^2 & 0 & 0 \\ 0 & m_\pi^2 & 0 \\ 0 & 0 &
2m_K^2-m_\pi^2 \end{array} \right) \, .
\end{equation}
\hspace*{0.5cm}The content of Eqs.~(\ref{L-Op2-U}) and (\ref{p2-u}) is exactly the same.\\

\hspace*{0.5cm}Now we will explain why the expansion in an $EFT$ may not be simply an expansion in the number of loops. This is indeed what happens in $\chi PT$.\\
\hspace*{0.5cm}Consider an arbitrary complex Feynman diagram involving just $pGs$. We recall the expansion in $\CPT$, Eq.~(\ref{expansion}). The $LO$ Lagrangian
 supplies $\cO(p^2)$ vertices, the $NLO$ one $\cO(p^4)$ couplings, and so on and so forth. Using $N_d$ to denote the number of vertices obtained employing 
the Lagrangian of order $\cO(p^d)$, remembering that $pG$ fields have mass dimension one and that any momentum running inside a loop is to be integrated over
 four dimensions -after proper renormalization-, one may conclude that for this generic diagram all powers of momentum will fulfill the relation:
\begin{equation} \label{powersmomentumFD}
D\,=\,4L\,-\,2B_I\,+\,\sum_{d}N_d\,d\,,
\end{equation}
being $L$ the number of loops and $B_I$ the number of internal boson lines, respectively.\\
\hspace*{0.5cm}Moreover, there is a topological relation for any connected Feynman diagram
\begin{equation}
L\,=\,B_I\,-\,\left( \sum_{d}N_d\,-\,1\right)
\end{equation}
that one can use to erase $B_I$ from Eq.~(\ref{powersmomentumFD}) to end up with $\mathrm{Weinberg's}$ power counting rule \cite{Weinberg:1978kz}:
\begin{equation} \label{Weinbergcounting}
D\,=\,2L\,+\,2\,+\,\sum_{d} N_d (d-2) \,.
\end{equation}
\hspace*{0.5cm}It is straightforward to read off from (\ref{Weinbergcounting}) the exact ordering of the chiral expansion:
\begin{itemize}
\item $D$ $=$ $2$ corresponds to $L$ $=$ $0$ and $N_2$ $=$ $1$, i.e., tree level contributions obtained using Eq.~(\ref{L-Op2-U}) -or (\ref{p2-u})-. This makes 
sense: we recover the predictions of old current algebra as the dominant very low-energy behaviour in $\CPT$.
\item $D$ $=$ $4$ is obtained either with $L = 0$ and $N_4$ $=$ $1$, or with $L$ $=$ $1$ and arbitrary insertions of $N_2$. Tree level contributions coming 
from $\mathcal{L}_4$ are to be balanced with one-loop diagrams formed with $\mathcal{L}_2$. And so on. This is in agreement with the order-by-order renormalization
 of $\chi PT$: the divergences generated at a given chiral order are renormalized by the appropriate counterterms appearing at the next order.
\end{itemize}
\hspace*{0.5cm}However, this is not the whole story. For $\CPT$ to be dual to $QCD$ at low energies it must satisfy classical symmetries slightly modified 
by quantum properties. Classical symmetries can be swapped away by anomalies, which are long-distance non-perturbative effects. And this is what happens with
 $U(1)_A$ for the massless $QCD$ Lagrangian, Eq.~(\ref{QCD-chiral}). To ensure the duality, every aspect of low-energy $QCD$ must be realized in the same way in
 $\CPT$ and, particularly, one must add a term that reproduces this anomaly, as it is discussed in the next section. Moreover, it is not possible to write a
 generating functional for massless $QCD$ that is simultaneously invariant under the subsets of $V$ and $A$ transformations of the chiral group. It is mandatory
 to include a term that mimics this behaviour whose degrees of freedom are $pGs$. This task was carried out by Wess, Zumino and Witten \cite{Wess:1971yu, Witten:1983tw}, 
who wrote such a kind of functional, $WZWf$. It happens to start contributing at $\cO(p^4)$. Therefore, three contributions shape $NLO$ chiral expansion that
 are, schematically: $L$ $=$ $0$ with $N_4$ $=$ $1$, $L$ $=$ $1$ $\forall$ $N_2$, and $WZWf$.\\
\section{NLO in the chiral expansion}\label{EFT_NLO_in_chiral_expansion}
\hspace*{0.5cm}The lowest order Lagrangian is $\mathcal{O}\left(p^2\right)$ (Eq.~(\ref{p2-u})) for even intrinsic parity and $\mathcal{O}\left(p^4\right)$ (the $WZWf$) in the odd-intrinsic parity sector.\\
\hspace*{0.5cm}There is a new operator that enters the $\cO(p^4)$ Lagrangian in the $u$-formalism:
\begin{eqnarray} \label{op-p4}
f^{\mu\nu}_\pm &=& u\,F_L^{\mu\nu}\,u^\dagger \, \pm \, u^\dagger\, F_R^{\mu\nu}\,u \, ,
\end{eqnarray}
that collects the left(right)-handed field-strength tensors presented in Eq.~(\ref{F_L,R}). The covariant derivative in this formalism reads:
\begin{equation} \label{covariantderivative}
\nabla_\mu \, X \,=\, \partial_\mu X \,+\, \left[ \Gamma_\mu,\,X \right],
\end{equation}
defined in terms of the chiral connection
\begin{equation}
\Gamma_\mu \,=\,\frac{1}{2} \left\{ u^\dagger \left( \partial_\mu\,-\,ir_\mu \right) u \,+ \,
u \left( \partial_\mu\,-\,i l_\mu
\right)u^\dagger \right\} \,,
\end{equation}
for the covariant derivative to be transformed in the same way as $X$ does, Eq.~(\ref{goodtransformation}).\\
\hspace*{0.5cm}It easy to check that $(\nabla_\mu X)^\dagger\,=\,\nabla_\mu(X^\dagger)$. The connection does not transform covariantly as $X$. It will be 
useful when introducing the chiral multiplets of resonances in $\RCT$, which transform in the same fashion. With $\Gamma_\mu$ one may also build the covariant
 tensor
\begin{equation}
\Gamma_{\mu\nu}\,=\,\partial_\mu \Gamma_\nu\,-\,\partial_\nu \Gamma_\mu\,+\,\left[ \Gamma_\mu,\Gamma_\nu \right].
\end{equation}
\hspace*{0.5cm}The other $\cO(p^2)$ operators transforming covariantly, Eq.~(\ref{goodtransformation}) are:
\begin{eqnarray} \label{hmunu}
u_\mu u_\nu\,,\nonumber\\
h_{\mu\nu}\,=\,\nabla_\mu u_\nu\,+\,\nabla_\nu u_\mu\,,
\end{eqnarray}
\hspace*{0.5cm}where it has been used that $u_\mu$ can be written as
\begin{eqnarray}
u_\mu\,=\,i\,u^\dagger D_\mu U u^\dagger\,=\,-i\,u D_\mu U^\dagger\,=\,u_\mu^\dagger,
\end{eqnarray}
and is traceless.\\
\hspace*{0.5cm}The relevant transformation properties of chiral tensors in this formalism are shown in Table \ref{chiraltransfu}. Taking these into account,
 we achieve the most general $\cO(p^4)$ chiral Lagrangian written in terms of them:
\begin{eqnarray} \label{L-Op4-u}
\mathcal{L}_4 &=& L_1 \bra u_\mu u^\mu \ket^2 \,+\, L_2 \bra u_\mu u^\nu \ket \bra u^\mu u_\nu \ket \,+\,
L_3 \bra u_\mu u^\mu u_\nu u^\nu \ket \,+\, L_4 \bra u_\mu u^\mu \ket \bra \chi_+ \ket \nonumber \\
& &  + \, L_5 \bra u_\mu u^\mu \chi_+ \ket \,+\, L_6 \bra \chi_+ \ket^2 \,+\, L_7 \bra \chi_-\ket^2
\,+\, L_8/2 \, \bra \chi_+^2 + \chi_-^2 \ket \nonumber \\ & & 
-\,i\, L_9 \bra f^{\mu\nu}_+ u_\mu u_\nu \ket \, 
+\, L_{10}/4 \, \bra f_{+ \mu\nu}f_+^{\mu\nu}-f_{-\mu\nu}f_-^{\mu\nu}\ket \nonumber \\
& & 
+\,i\,L_{11} \bra \chi_-(\nabla_\mu u^\mu + i/2\, \chi_-)\ket \,
-\,L_{12}\bra 
(\nabla_\mu u^\mu + i/2 \, \chi_-)^2 \ket \nonumber \\ & &  
+\,H_1/2\, \bra f_{+ \mu\nu}f_+^{\mu\nu}+f_{-\mu\nu}f_-^{\mu\nu}\ket \,+\,
H_2/4 \, \bra \chi_+^2-\chi_-^2 \ket \,,
\end{eqnarray}
where the terms whose coefficients are $L_{11}$ and $L_{12}$ do vanish considering the equations of motion ($EOM$) of $\cO(p^2)$. The $EOM$ for $\mathcal{L}_2$
 is:
\begin{equation} \label{EOM_p2}
\cO^{EOM}_2(u)\,=\,\nabla_\mu u^\mu\,-\frac{i}{2}\,\left( \chi_- -\frac{1}{n_f}\bra\chi_-\ket\right)\,=\,0;  
\end{equation}
$L_{11}$ and $L_{12}$ can now be skipped at $\cO(p^4)$ using that:
\begin{equation}
\mathcal{L}_4^{\mathrm{off-shell}}\,=\, L_{11}\bra\chi_- \cO^{EOM}_2(u)\ket\,-\,L_{12}\bra\cO^{EOM}_2(u) \cO^{EOM}_2(u)^{\dagger}\ket.
\end{equation}
\hspace*{0.5cm}The terms with coefficients $H_1$ and $H_2$ are contact terms relevant for the renormalization of $\CPT$.\\
\begin{table} [h]
\begin{center}
\renewcommand{\arraystretch}{1.2}
\begin{tabular}{|c|c|c|c|c|} 
\hline
Operador & $P$ & $C$ & h.c. & $\chi$ order \\
\hline
$u_\mu$ &  $-u^\mu$ & $u_\mu^T$ & $u_\mu$ & $p$\\
$\chi_\pm$ & $\pm \chi_\pm$ & $\chi_\pm^T$ & $\pm \chi_\pm$ & $p^2$\\
$f_{\mu\nu\,\pm}$ & $\pm f^{\mu\nu}_\pm$ & 
$\mp f_{\mu\nu\,\pm}^T$ & $f_{\mu\nu\, \pm}$ & $p^2$\\
\hline
$\Phi$& -$\Phi$ & $\Phi^T$ & $\Phi$ & 1\\
$u$ & $u^{\dagger}$ & $u^T$ & $u^{\dagger}$ & 1\\
$\Gamma^\mu$ & $\Gamma_\mu$ & -$\Gamma^{\mu\,T}$ & -$\Gamma^\mu$ & $p$\\
$u_{\mu\nu}$ &  $-u^{\mu\nu\,\dagger}$ & $u_{\mu\nu}^T$ & $u_{\mu\nu}^\dagger$ & $p^2$\\
$\nabla_\mu u_\nu$ & -$\nabla^\mu u^\nu$ & $(\nabla_\mu u_\nu)^T$ & $\nabla_\mu u_\nu$ & $p^2$\\
\hline
\end{tabular}
\caption{\small{Transformation properties under $C$, $P$ and hermitian conjugation of the chiral tensors and other useful structures in the $u$-formalism.
 $T$ means transposed.}}
\label{chiraltransfu}
\end{center}
\end{table}
\hspace*{0.5cm}The renormalization of $\CPT$ needed to work at $\cO(p^4)$ was accomplished in Refs. \cite{Gasser:1983yg, Gasser:1984gg}. The divergences that
 arise using $\mathcal{L}_2$ at one-loop are of order $\cO(p^4)$ and are renormalized with the $LECs$ of $\mathcal{L}_4$:
\begin{eqnarray}\label{gamma}
L_i&=&L_i^r(\mu)\,+\, \Gamma_i \frac{\mu^{D-4}}{32\pi^2} \left\{ \frac{2}{D-4}\,+\,C\right\}\,, \quad 
\nonumber \\
H_i&=&H_i^r(\mu)\,+\, \tilde{\Gamma}_i \frac{\mu^{D-4}}{32\pi^2} \left\{ \frac{2}{D-4}\,+\,C\right\}\,,
\end{eqnarray}
where $D$ is the space-time dimension and $C$ is a constant defining the renormalization scheme.\\
\hspace*{0.5cm}The renormalized couplings, $L_i^r(\mu)$ do depend on the arbitrary renormalization scale $\mu$. This dependence cannot survive in any physical
 observable. As it had to happen, it is canceled out with that coming from the loop in any physically meaningful quantity.\\
\\
\hspace*{0.5cm}As we said, the odd-intrinsic parity sector starts at $\cO\left(p^4\right)$. Its appearance is due to one of the anomalies affecting the chiral
 group $U(3)_L\otimes U(3)_R$. On the one hand there is the anomaly which makes that the classical symmetry $U(1)_A$ is lost at the quantum level. Apart from group 
color factor, it is identical to the case of the axial anomaly of $QED$, discovered perturbatively in one-loop computations \cite{Adler:1969gk, Bell:1969ts} 
by Adler, Bell and Jackiw. Later on, the proof that this result does not receive radiative corrections \cite{Adler:1969er} by Adler and Bardeen insinuated 
that anomalies could have a non-perturbative nature, as it was shown by Fujikawa, using the path-integral formalism \cite{Fujikawa:1979ay, Fujikawa:1980eg}.\\
\hspace*{0.5cm}On the other hand, there is an anomaly that affects the whole $U(3)_L\otimes U(3)_R\sim U(3)_V\otimes U(3)_A$ symmetry group that originates in the fact that it is not
 possible to preserve the simultaneous invariance of the generating functional under vector and axial-vector transformations. Wess and Zumino \cite{Wess:1971yu} were
 the first to obtain a functional generating this anomaly that affects chiral transformations written in terms of $pGs$. Operatively, it is more useful the
 one derived later by Witten \cite{Witten:1983tw}, that I will present here following the discussion in Ref.~\cite{Bijnens:1994qh}.\\
\hspace*{0.5cm}The fermionic determinant does not allow for a chiral invariant regularization. Given the transformations
\begin{equation}
g_R\,=\,1\,+\,i(\alpha(x)+\beta(x))\,,\,\,g_L\,=\,1\,+\,i(\alpha(x)-\beta(x)),
\end{equation}
the conventions in the definition of the fermionic determinant may be chosen to preserve the invariance of the generating functional, $Z$, either under $V$ 
transformations, or under the $A$ ones; but not both simultaneously. Choosing to preserve invariance under the transformations generated by the vector current, 
the change in $Z$ only involves the difference $\beta(x)$ between $g_R$ and $g_L$.
\begin{eqnarray}
\delta Z & = &-\int\mathrm{d}x\bra \beta(x) \Omega(x)\ket\,,\\
\Omega(x) & = &\frac{N_C}{16\pi^2}\varepsilon^{\alpha \beta \mu \nu}\left[ v_{\alpha \beta}\,v_{\mu \nu}\,+\,\frac{4}{3}\,D_\alpha\, a_\beta\, D_\mu a_\nu\,+\,\frac{2i}{3}\,\left\lbrace \ v_{\alpha \beta},\,a_\mu\, a_\nu \right\rbrace\right.\nonumber\\
& & \left. \qquad\qquad+\,\frac{8i}{3}\,a_\mu \,v_{\alpha\beta}\, a_\nu\,+\,\frac{4}{3}\,a_\alpha a_\beta \, a_\mu\, a_\nu \right]\,,\\
v_{\alpha\beta} & = & \partial_\alpha\, v_\beta\,-\,\partial_\beta v_\alpha\,-\,i\left[v_\alpha,\,v_\beta \right]\,,\\
D_\alpha\, a_\beta & = & \partial_\alpha\, a_\beta\,-\,i\left[v_\alpha,\,a_\beta \right]\,.
\end{eqnarray}
\hspace*{0.5cm}Notice that $\Omega$ only depends on the external fields, $v_\mu$ and $a_\mu$, and that the quark masses do not occur.\\
\hspace*{0.5cm}The explicit form for the functional $Z\left[ U, \ell, r\right]$ that reproduces the chiral anomaly given by Witten \cite{Witten:1983tw} is:
\begin{eqnarray} \label{Z_WZW} 
Z\left[ U,\, \ell,\, r\right]_{\mathrm{WZW}}& = & -\frac{i\,N_C}{240\pi^2} \int_{M^5}\mathrm{d}^5x\varepsilon^{ijklm}\bra \Sigma^L_i\,\Sigma^L_j\,\Sigma^L_k\,\Sigma^L_l\,\Sigma^L_m\ket\\
& & -\,\frac{iN_C}{48\pi^2}\,\int\mathrm{d}^4x\varepsilon_{\mu\nu\alpha\beta}\left( W(U,\, \ell,\, r)^{\mu\nu\alpha\beta}\,-\,W(\textbf{1},\, \ell,\, r)^{\mu\nu\alpha\beta}\right)\,,\nonumber\\
W(U,\, \ell,\, r)_{\mu\nu\alpha\beta} & = & \bra U \,\ell_\mu \,\ell_\nu \,\ell_\alpha\, U^\dagger\, r_\beta\,+\,\frac{1}{4}\,U\,\ell_\mu\, U^\dagger\, r_\nu \,U \,\ell_\alpha \,U^\dagger\, r_\beta\,+\,i\,U\,\partial_\mu\, \ell_\nu\, \ell_\alpha\, U^\dagger\, r_\beta\nonumber\\
& & +\,i\,\partial_\mu \,r_\nu\, U\, \ell_\alpha\, U^\dagger\, r_\beta\,-\,i\,\Sigma_\mu^L \,\ell_\nu\, U^\dagger \,r_\alpha \,U \,\ell_\beta\,+\,\Sigma_\mu^L \,U^\dagger\, \partial_\nu\, r_\alpha \,U\, \ell_\beta\nonumber\\
& &-\,\Sigma_\mu^L\, \Sigma_\nu^L \,U^\dagger \,r_\alpha\, U \,l_\beta\,+\,\Sigma_\mu^L \,l_\nu\, \partial_\alpha\, l_\beta\,+\,\Sigma_\mu^L\, \partial_\nu\, l_\alpha\, l_\beta\nonumber\\
& &-\,i \,\Sigma_\mu^L \,\ell_\nu\, \ell_\alpha \,\ell_\beta\,+\,\frac{1}{2}\,\Sigma_\mu^L\, \ell_\nu\, \Sigma_\alpha^L \,\ell_\beta\,-\,i\,\Sigma_\mu^L\, \Sigma_\nu^L\, \Sigma_\alpha^L\, \ell_\beta\ket\nonumber\\
& &-\,(L\,\longleftrightarrow\,R)\,\,,\,\,N_C\,=\,3\,,\nonumber\\
& & \,\Sigma_\mu^L\,=\,U^\dagger\, \partial_\mu\, U\,\,,\,\,\Sigma_\mu^R\,=\,U \,\partial_\mu\, U^\dagger\,, \label{W_WZW} 
\end{eqnarray}
where $(L\,\longleftrightarrow\,R)$ stands for the exchange
\begin{equation} \label{LRexchange}
U\,\longleftrightarrow\,U^\dagger\,\,,\quad\ell_\mu\,\longleftrightarrow\,r_\mu\,\,,\quad\Sigma_\mu^L\,\longleftrightarrow\,\Sigma_\mu^R\,.
\end{equation}
\hspace*{0.5cm}The first term in Eq.~(\ref{Z_WZW}) bears the mark of the anomaly: It is a local action in five dimensions that can not be written as a finite 
polynomial in $U$ and $D_\mu U$ in four dimensions. This term involves at least five pseudoscalar fields and will not contribute either to the hadronic three 
meson decays of the $\tau$ or to its radiative decays with one meson in the final state. Eqs.~(\ref{Z_WZW}) and (\ref{W_WZW}) contain all the anomalous 
contributions to electromagnetic and semileptonic weak meson decays: $\pi^0\to\gamma\gamma$, $\pi\to e\nu_e\gamma$, etc. For further discussions on this 
topic see also Ref.~\cite{Callan:1983nx}.\\
\section{NNLO overview and scale over which the chiral expansion is defined}\label{EFT_NNLO_chiral_expansion}
\hspace*{0.5cm}$\CPT$ is an expansion in powers of momentum over a typical hadronic scale that we can understand in two equivalent ways:
\begin{itemize}
\item $pGs$ stand out due to $S\chi SB$. This generates -through quantum effects-, the $S\chi SB$-scale, $\Lambda_\chi$, as a natural parameter over which the
 chiral expansion is defined.
\item Decoupling theorem told us that one of the effects of heavy integrated-out particles in the physics of light modes appears in inverse powers of these 
larger masses. Then, we expect typical masses of the lowest-lying resonances to provide this scale, as well.
\end{itemize}
\hspace*{0.5cm}I will start considering the appearance of that scale through loops. For this, let us consider scattering among $pGs$ at $\cO(p^4)$. Apart from the tree level
 contribution of $\mathcal{L}_4$, there will be that given by $\mathcal{L}_2$ at one-loop, whose amplitude is of order
\begin{equation}
I\,\sim \, \int \frac{\mathrm{d}^4 p}{(2\pi)^4}\frac{p^2}{F^2}\frac{p^2}{F^2}\frac{1}{p^4}\,,
\end{equation}
where $1/p^4$ comes from the internal boson propagators and each interacting vertex of $\mathcal{L}_2$ gives -after expanding the $LO$ Lagrangian up to 
terms with four powers of $\Phi$, that is, four $pGs$- a factor $(p/F)^2$ \footnote{All vertices will have some momentum $p$, but other external ones too.
 The $q$ below intends to represent these ones.}. One can estimate this integral as
\begin{equation}
I\,\sim\, \frac{q^4}{16\pi^2} \frac{1}{F^4} \log \mu \,,
\end{equation}
where $\mu$ is the renormalization scale. On the other hand, the tree level interaction given by $\mathcal{L}_4$ has the shape $L_i^r\,(q/F)^4$. We know the 
total amplitude is scale independent which implies that a shift in the scale $\mu$ has to be balanced by that of the tree level $\mathcal{L}_4$ 
contribution. The loop-related factor $1/16\,\pi^2$ must be also in $\mathcal{L}_4$.\\
\hspace*{0.5cm}We can write the $\mathcal{L}_{eff}$ as:
\begin{equation}
\mathcal{L}\,=\,\frac{F^2}{4}\left[ \widetilde{\mathcal{L}}_2 \,+\,
\frac{\widetilde{\mathcal{L}}_4}{\Lambda_\chi^2}\,+\,\frac{\widetilde{\mathcal{L}}_6}{\Lambda_\chi^4} \dots\right]\,, 
\end{equation}
where $1/\Lambda_\chi$ gives the expansion of the $EFT$-Lagrangian in powers of $q$/$\Lambda_\chi$. It is straightforward to check that this also happens as
 the chiral order in the expansion is increased. Taking into account the loop-related factor, we estimate
\begin{equation}
\Lambda_\chi \, \sim\, 4\pi F \, \sim\, 1.2\, \mathrm{GeV}\,.
\end{equation}
\hspace*{0.5cm}This dimensional analysis suggests that the $n-pGs$ vertex will receive a contribution from the $\cO(p^m)$ Lagrangian that will go as
\begin{equation}
F^2\Lambda_\chi^2 \left( \frac{\Phi}{F} \right)^n \left( \frac{\partial}{\Lambda_\chi} \right)^m \,,
\end{equation}
and, consequently, the $LECs$ in the Lagrangian will be of order
\begin{equation}\label{coefficient}
\frac{F^2}{\Lambda_\chi^{m-2}}  \,\sim\, \frac{F^{4-m}}{(4\pi)^{m-2}} \,.
\end{equation}
\hspace*{0.5cm}On should be aware that $\Lambda_\chi \sim 1$ GeV does not mean $\CPT$ can be applied up to this energy. The complementary point of view 
explained at the beginning helps to understand this. $M_\rho\sim0$.$8$ GeV sentences all $\CPT$ attempts to explain physics from this region on to failure 
unless one explicitly incorporates resonances as active degrees of freedom, as $\RCPT$ does.\\
\hspace*{0.5cm}$M_\rho\sim0$.$8$ GeV is to be regarded as a clear upper bound for the validity of $\CPT$ and corresponds to the typical size of the counterterm
 corrections. On the other hand, $4\pi F \, \simeq \, 1$.$2$ GeV is the scale associated to quantum effects 
through loop corrections.\\
\hspace*{0.5cm}There is still one more reason for using $\RCPT$ when we go over $0$.$5$ GeV, or so. The lowest order contributions in the chiral expansion 
lose importance continuously and we are forced to go further and further in the expansion to reach the same accuracy. Two $LECs$ specify the $\cO(p^2)$ Lagrangian, 
$10$ appear at $\cO(p^4)$ but $90$ are challenging us and the number of different experimental data we can collect to fix them at $\cO(p^6)$. Although not all of them 
enter a given process, the description becomes pretty much easier when resonances become active variables, as we will see. The $\cO\left(p^6\right)$ $\CPT$ 
Lagrangian was developed and the renormalization program was accomplished in Refs.~\cite{Fearing:1994ga, Bijnens:1999sh, Bijnens:1999hw, Bijnens:2001bb}.\\

\chapter{The Large $N_C$ limit and Resonance Chiral Theory}\label{RChT}
\section{Introduction} \label{LargeN_intro}
\hspace*{0.5cm}We finished the last Chapter by recalling some of the motivations for enlarging $\CPT$ and extend it to higher energies. The problem is that 
as soon as we try to do it the chiral counting gets broken because the momentum of the $pGs$ can become comparable to $\Lambda_\chi$ or $M_\rho$. Thus, there 
is no immediate parameter to build the expansion upon.\\
\hspace*{0.5cm}In many other instances in $QCD$, this difficulty does not occur. $QCD$ is perturbative at high energies, so the same strong coupling is a useful
 expansion parameter in that energy region. There can be a double expansion (in $\alpha_S$ and $1/M_Q$) because pole masses are good parameters for a quick convergence
 of the perturbative series when studying heavy quarks: Heavy Quark Effective Theory ($HQET$) -for just one heavy quark- \cite{Isgur:1989ed, Isgur:1989vq, Eichten:1989zv, Georgi:1990um, Grinstein:1990mj, Manohar:2000dt}, 
or (potential) Non-Relativistic $QCD$ ($(p)NRQCD$) -if both quarks are heavy- \cite{Caswell:1985ui, Lepage:1992tx, Bodwin:1994jh, Pineda:1997bj, Brambilla:1999qa, Luke:1999kz, Brambilla:1999xf, Brambilla:2004jw, Brambilla:2004wf, 
Brambilla:2010cs}.\\
\hspace*{0.5cm}In addition to the lack of a natural expansion parameter in the region where the light-flavoured resonances pop up, there is some thinking in $\CPT$, 
that might guide the strategy to follow. While in the $EFTs$ listed in the previous paragraph the expansion parameters are quantities appearing in the $QCD$
 Lagrangian, this is not the case for $\CPT$: the expansion involves the momenta and masses of the $pGs$ and $\Lambda_\chi$ (not the quarks and gluons and 
$\Lambda_{QCD}$).\\
\hspace*{0.5cm}The proposed expansion parameter, $1/N_C$, will indeed be useful to describe the physics of light-flavoured mesons. Moreover, it can be used to understand 
qualitatively some results in $\CPT$ in terms of a quantity that defines the gauge group of the strong interactions in $QCD$.\\
\section{1/$\N$ expansion for QCD} \label{LargeN_1/N QCD}
\begin{figure}[h!]
\centering
\includegraphics[scale=0.7]{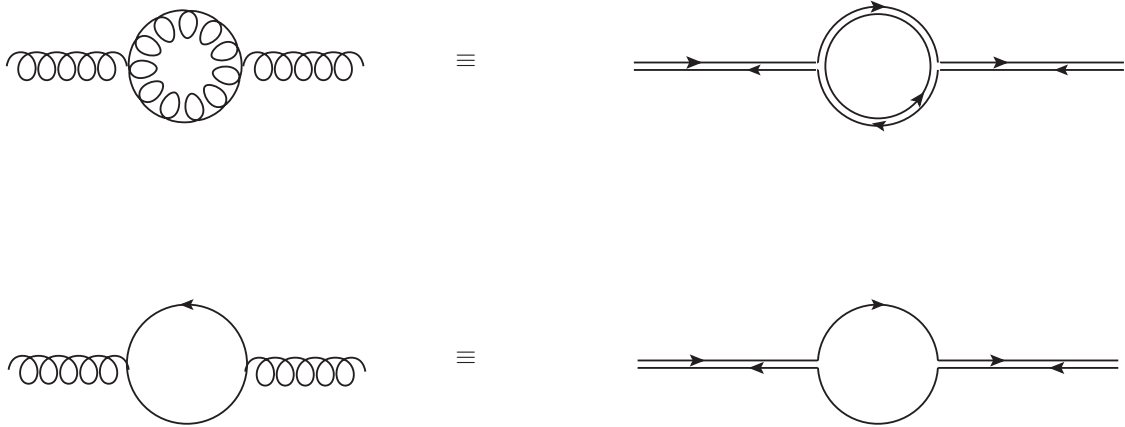}
\caption{Feynman diagrams for the $LO$ contribution to the gluon self-energy: using the usual notation (left), and with the double line one (right).}
\label{diagram gluon_selfenergy_doubleline}
\end{figure}

\begin{figure}[h!]
\centering
\includegraphics[scale=0.7]{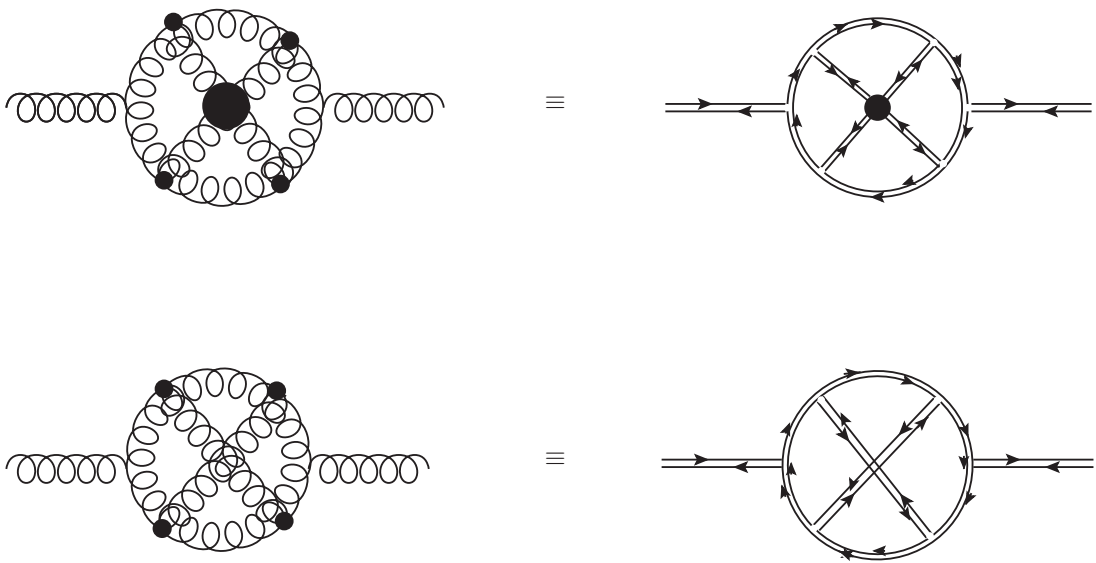}
\caption{Comparison between planar and non-planar diagrams for gluon self-energy. The four-gluon vertex is depicted by the thickest dot. Three gluons vertices are 
highlighted by a thick dot. Other superpositions of two gluon lines correspond to crossings and not to intersections.}
\label{diagram planar_and_nonpdiags}
\end{figure}

\begin{figure}[h!]
\centering
\includegraphics[scale=0.7]{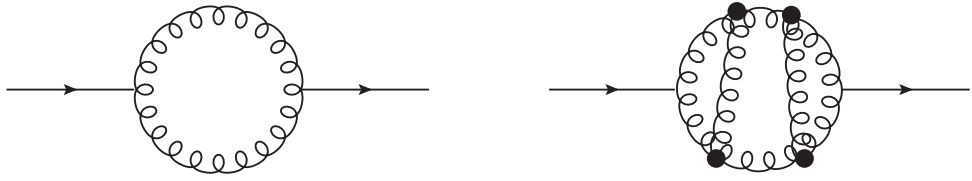}
\caption{Two $LO$ contributions in 1/$\N$ to gluon self-energy. Gluon vertices are highlighted by a thick dot.}
\label{diagram gluon_selfenergy}
\end{figure}

\hspace*{0.5cm}'t Hooft \cite{'tHooft:1973jz} had the seminal idea of generalizing $QCD$ from a theory of three colours to the case with $\N$ colours. Though a priori this 
can be regarded as an unnecessary artifact that will make things even more difficult, this is not -at all- the case, and $QCD$ gets simplified in the large-$\N$
 limit becoming even solvable in one spatial plus one time dimension \cite{'tHooft:1974hx}.\\
\hspace*{0.5cm}Later on, many papers appeared guided by 't Hooft's idea, Ref.~\cite{Witten:1979kh} is the capital one, but see also Refs.~\cite{Witten:1978bc, Di Vecchia:1980ve, Witten:1980sp}.
 This part of the chapter is mainly based on them and on Refs.~\cite{Coleman, Manohar:1998xv, Pich:2002xy, Kaiser:2000gs}.\\
\hspace*{0.5cm}As we will see, there are many phenomenological facts that find their only explanation on large-$\N$ arguments. This is, at the end of the day,
 the strongest support the 1/$\N$ expansion for $QCD$ has.\\
\hspace*{0.5cm}Recall the $QCD$ Lagrangian, (\ref{fullLQCDe}). From it, we can read off the Feynman rules obtained for all $QCD$ vertices and see how the coupling
 of a fermionic line to a gluon is $\cO(g_s)$, exactly as the three gluon vertex. The four gluon interaction is $\cO(g_s^2)$, each quark loop runs over
 three colours and each gluon loop over eight possibilities, corresponding to the number of generators of $SU(3)_C$. Eight is not so much larger than three, but
 in the large-$\N$ limit, the number of gluon states ($N_C^2-1$) is really huge compared to that of quarks ($N_C$). It is also reasonable to approximate 
$N_C^2-1$ by $N_C^2$, that is, to consider $U(N_C)$ instead of $SU(N_C)$. The first ingredient of the large-$\N$ limit of $QCD$ is to take into account that
 gluon states are more important than quark states. The second one comes from asking a finite behaviour in this limit for the quantum corrections and happens
 to modify the usual counting in powers of $g_s$ for the vertices reminded before.\\
\hspace*{0.5cm}Prior to that, it is useful to introduce in this context the double line notation for the gluon lines. This way we represent each gluon as a
 quark-antiquark pair, an approximation that becomes exact in the large-$\N$ limit. For the gluon selfenergy, we will have the diagrams displayed in Fig.~\ref{diagram gluon_selfenergy_doubleline}.
\\
\\
\hspace*{0.5cm}Within the new notation, each line represents a given colour propagating. After fixing external colour indices, there is no remaining freedom
 in the quark-loop contribution, but there is still an inner loop in the purely gluon contribution over which $\N$ colours can run.\\
\hspace*{0.5cm}If we now use the Feynman rules obtained from Eq. (\ref{fullLQCDe}), we perceive that the second diagram in Figure \ref{diagram gluon_selfenergy_doubleline}
behaves, in the large-$\N$ limit, as $g_s^2$, while the first one diverges: it goes as $g_s^2\,N_C$. As we want gluon self-energy (and the $\beta$-function)
 to be finite in the limit $N_C\rightarrow\infty$, we are led to redefine $g_s$ as $\tilde{g}_s\equiv\frac{g_s}{\sqrt{N_C}}$. This also modifies the Feynman rules. Now, gluon
 coupling to a fermionic current or three gluon vertex will be of $\cO\left(\frac{1}{\sqrt{N_C}}\right)$; and four gluon local interaction of $\cO\left(\frac{1}{N_C}\right)$.
 This way, the beta function reads:
\begin{equation} \label{betafunctionlargeN}
 \mu \frac{\mathrm{d}\tilde{g}_s}{\mathrm{d}\mu}\,=\,-\left(11-2\frac{n_f}{N_C}\right)\,\frac{\tilde{g}_s^2}{48\pi^2}\,,
\end{equation}
and we keep the hadronization scale $\Lambda_{QCD}$ independent of the number colours when it is taken to be large. One can see from Eq.(\ref{betafunctionlargeN})
 that quark loops are suppresed with respect to gluon loops in the large-$N_C$ limit.\\
\hspace*{0.5cm}Let us consider now the diagrams of Figure \ref{diagram planar_and_nonpdiags}.
\\
\\
\hspace*{0.5cm}The upper diagram in Figure \ref{diagram planar_and_nonpdiags} is planar -all superposition of lines correspond to intersections- , while the 
lower one is not -some of them are just crossings-. For the first one, the counting is $\left(\frac{1}{\sqrt{\N}}\right)^6\,\frac{1}{\N}\,\N^4\,=\,1$, of 
the same order than the purely gluon contribution in Figure \ref{diagram gluon_selfenergy_doubleline}.\\
\hspace*{0.5cm}For the lower one, we have $\left(\frac{1}{\sqrt{\N}}\right)^6\,\N\,=\,\frac{1}{\N^2}$. This is because the diagram is non-planar and the number
 of colour loops has decreased from four to one and the central vertex has disappeared.\\
\\
\hspace*{0.5cm}We have seen explicitly how, in the easiest examples, two selection rules arise:
\begin{itemize}
\item Non-planar diagrams are suppressed by the factor $\frac{1}{\N^2}$. 
\item Internal quark loops are suppressed by the factor $\frac{1}{\N}$.
\end{itemize}
\hspace*{0.5cm}Planar diagrams, with arbitrary exchanged gluons do dominate (and the obliged external quark loop to be a meson).\\
\hspace*{0.5cm}To see that all this is always true it is convenient to rescale fermion and gluon fields:
\begin{eqnarray} \label{rescaled fields}
\tilde{G}_\mu^a\,=\,\frac{\tilde{g}_S}{\sqrt{\N}}\,G_\mu^a\,,\\
\tilde{q}\,=\,\frac{1}{\sqrt{\N}}\,q\,,
\end{eqnarray}
whence
\begin{equation} \label{L_QCD_redefinedfields} 
\mathcal{L}_{QCD}\, =\, \N \left[ \overline{\tilde{q}} \left(i\tilde{D}\hspace*{-0.26cm}/ \,-\mathcal{M}\right)\tilde{q}\,-\,\frac{1}{4\tilde{g}_S^2}\,\tilde{G}^a_{\mu\nu}\tilde{G}^{\mu\nu}_a\right]\,,
\end{equation}
and all the counting is simply in powers of $\frac{1}{\N}$. It can be seen \cite{Manohar:1998xv} that the order in this expansion for any connected vacuum 
diagram is related to a topological invariant, the Euler-Poincar\'e characteristics, which allows to demonstrate the above properties in full generality 
independently of the number of exchanged gluons. The conclusion we draw is clear: In the large-$\N$ limit, Feynman diagrams are planar and without internal 
quark loops.\\
\hspace*{0.5cm}It is worth to stress that the $\frac{1}{\N}$-expansion for $QCD$ has to be regarded in a different way than usual expansions in perturbation theory.
 Expanding in powers of the coupling constant gives us Feynman diagrams and, at any given order, there are a finite number of them contributing to a process.
 Looking at the commented equivalence among diagrams belonging to Figure \ref{diagram gluon_selfenergy_doubleline}, which is clearly the $LO$ Feynman diagram,
 and to Figure \ref{diagram planar_and_nonpdiags}; a general fact is enlightened: In order to obtain a given order in $\frac{1}{\N}$, we need an infinite number
 of Feynman diagrams. The diagrams collected in Figure \ref{diagram gluon_selfenergy} are both of $LO$ in 1/$\N$ and one could think of much more complicated ones.\\
\\
\section{$N_C$ counting rules for correlation functions} \label{LargeN_countingrules}
\hspace*{0.5cm}Two related basic assumptions are made for studying meson dynamics in the large-$\N$ limit:
\begin{itemize}
\item $QCD$ remains to be a confining theory for $\N\to\infty$.
\item Therefore, the sum of the dominant planar diagrams is responsible of confinement in this limit.
\end{itemize}
\hspace*{0.5cm}We will consider quark and gluon composite operators whose quantum numbers are able to create a meson \footnote{The large-$\N$ limit is also useful
 to understand some properties of baryons. For a review on this topic, see Refs.~\cite{Manohar:1998xv, Dashen:1993jt, Dashen:1994qi, Jenkins:1998wy}.} (they must be color-singlet thus). The 
aim is to understand some salient features of meson phenomenology by looking to gauge invariant operators that cannot be splitted into separate gauge invariant
 pieces. For the quark operators, \footnote{Symmetries also allow mixed operators composed by quarks and gluons and glueballs made up of gluons only.} the 
suitable bilinears are named according to the same spin-parity assignments used for mesons: scalars ($\overline{q}\,q$), pseudoscalars ($\overline{q}\,\gamma_5\,q$),
 vectors ($\overline{q}\,\gamma^\mu\,q$), or axial-vectors ($\overline{q}\gamma^\mu\,\gamma_5 q$). We will represent them generically as $\tilde{\cO}_i(x)$. We will use again
 the method of external sources (or currents, $J_i(x)$) coupled to them. In order to be consistent with the counting for the rescaled fields introduced in Eq. (\ref{rescaled fields}),
 the right expression to add will be \cite{Manohar:1998xv} $N_C\,J_i(x)\,\tilde{\cO}_i(x)$ that will keep all the selection rules told before. Correlations functions
 are obtained functionally differentiating the generating functional $W(J)$ with respect to the sources:
 \begin{equation}
\bra \tilde{\cO}_1\,\tilde{\cO}_2\dots\tilde{\cO}_z\ket_C\,=\,\frac{1}{iN_C}\,\frac{\partial}{\partial J_1}\,\frac{1}{iN_C}\,\frac{\partial}{\partial J_2}\dots\,\frac{1}{iN_C}\,\frac{\partial}{\partial J_z}W(J)|_{J=0}\,,
\end{equation}
and each additional functional differentiation (\textit{i.e.} each source insertion) is weighted by a factor $\sim\frac{1}{N_C}$. It can be shown that $\cO(\N^2)$
 contributions stem from planar vacuum-like diagrams only with gluon lines. They can contribute to correlation functions of purely gluon operators. $n$-point
 Green function of purely gluon operators will be of $\cO(\N^{2-n})$. An $r$-meson vertex is of order $\N^{1-r/2}$. Quark bilinear operators start contributing
 at $\cO(\N)$ -which corresponds to a quark loop in the outermost border-, being $\cO(\N^{1-n})$ the corresponding $n$-point Green function.\\
\hspace*{0.5cm}Considering that with $\N=3$ the symmetric wave-function in colour space of a meson is written as
\begin{equation}
M\,=\,\frac{1}{\sqrt{3}}\,\sum_{i=1}^3\overline{q}_i\,q_i\,,
\end{equation}
in the large-$\N$ limit this will be
\begin{equation}
M\,=\,\frac{1}{\sqrt{N_C}}\,\sum_{i=1}^{\N}\overline{q}_i\,q_i\,,
\end{equation}
providing an amplitude for creating a meson that is -as it should be- independent of $\N$. This property applies also for glueballs.\\
\hspace*{0.5cm}For any arbitrary number of currents, the dominant contribution will be
\begin{equation} \label{bilinears}
\bra T\left(J_1 \dots J_n \right) \ket \, \sim \, \cO(N_C) \, ,
\end{equation}
given by diagrams with one external quark loop and arbitrary insertions of gluon lines that do not spoil the planarity of the diagram.\\
\begin{figure}[h!]
\centering
\includegraphics[scale=0.7]{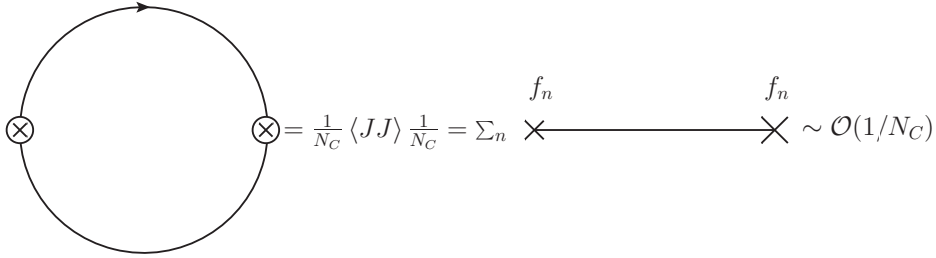}
\caption{Basic diagram for the $2$-point correlator and representation as a sum of tree-level diagrams with meson exchange.}
\label{diagram 2pointcorrelator_as_mesonexchange}
\end{figure}
\\
\hspace*{0.5cm}In Ref.~\cite{Manohar:1998xv}, it is shown that the action of $J(x)$ over the vacuum will create only one-meson states in the large-$\N$ limit. 
Taking this into account, the two-current correlator is of the form:
\begin{equation} \label{2point}
\bra J(k)\,J(-k) \ket \, =\, \sum_n \frac{f_n^2}{k^2-m_n^2} \, ,
\end{equation}
where the sum extends over infinite meson states of mass $m_n$ and decay constant $f_n$ defined through $f_n\,=\,\bra 0|J|n\ket$. This is a capital result
 from which we can derive most of meson phenomenology in this limit:
\begin{itemize}
\item Being $\bra J J \ket\sim\cO(\N)$, $f_n\,\sim\,\cO(\sqrt{\N})$ because $k^2$ has nothing to do with the $\N$-counting. Meson decay constants are $\cO(\sqrt{\N})$.
\item Moreover, the whole denominator must be $\cO(1)$, whence $m_n\sim\cO(1)$, too. Meson masses are said to have smooth large-$\N$ limit \footnote{This argument is used 
\cite{Pelaez:2006nj} to show that the $q\bar{q}$ component of the $f_0(500)$ or $\sigma$ meson is not predominant.}.
\item $\bra J(k)\,J(-k) \ket$ is known to have a logarithmic behaviour at large momentum \cite{Brodsky:1973kr, Brodsky:1974vy, Lepage:1979zb, Lepage:1980fj},
 $k$. Therefore, if we are to obtain these logarithms adding terms going as 1/$k^2$, we will need an infinite number of them. Despite seeming quite surprising
 at first, the conclusion is clear: There are infinite mesons in the large-$\N$ limit.
\item The poles of (\ref{2point}) are all in the real axis. Because the instability of a particle is translated into an imaginary part in its propagator 
that has to do with its decay width, we deduce that mesons are stable for $\N\rightarrow\infty$.
\end{itemize}
\hspace*{0.5cm}To sum up, the two-point correlator is reduced in the large-$\N$ limit to the addition of tree-level diagrams in which $J(-k)$ creates a 
meson with amplitude $f_n$ that propagates according to $\frac{1}{k^2-m_n^2}$ being annihilated by $J(k)$ with the same amplitude.\\
\hspace*{0.5cm}It is straightforward to generalize this result for an arbitrary number of currents.\\
\hspace*{0.5cm}It can be shown that, in the large-$\N$ limit \cite{Pich:2002xy}:
\begin{itemize}
\item $n$-point Green functions are given by sums of tree level diagrams obtained by using a phenomenological Lagrangian written in terms of freely propagating
 mesons that accounts for local effective interactions among $m\leq n$ of them.
\item Mesons do not interact, because both the $m$-meson vertex and the matrix element creating $m$ mesons from the vacuum are $\cO(\N^{1-m/2})$, suppressed
 in this limit.
\item The same can be applied to gluon states by considering gluon currents, $J_G\,=\,\bra G^{\mu\nu\,a}\,G_{\mu\nu\,a}\ket$. The $n$-point Green function
 is $\cO(\N^2)$ and a $g$-gluon operator vertex is $\cO(\N^{2-g})$; so gluon states are also free, stable and non-interacting in the strict limit.
\item The mixed correlator with $m$ quark-bilinears and $g$ gluon operators is $\cO(\N)$, but the local vertex among all them is  $\cO(\N^{1-g-m/2})$, that
 is suppressed, too. Gluon and meson states do decouple in the large-$\N$ limit, being their ensemble suppressed by 1/$\sqrt{\N}$.
\end{itemize}
\hspace*{0.5cm}The discussion for $3-$ and $4-$point correlators given by meson exchange is portrayed by Figure \ref{diagram 3and4pointcorrelatorasmesonexchange} where the counting in 1/$\N$ is given.\\
\begin{figure}[h!]
\centering
\includegraphics[scale=0.9]{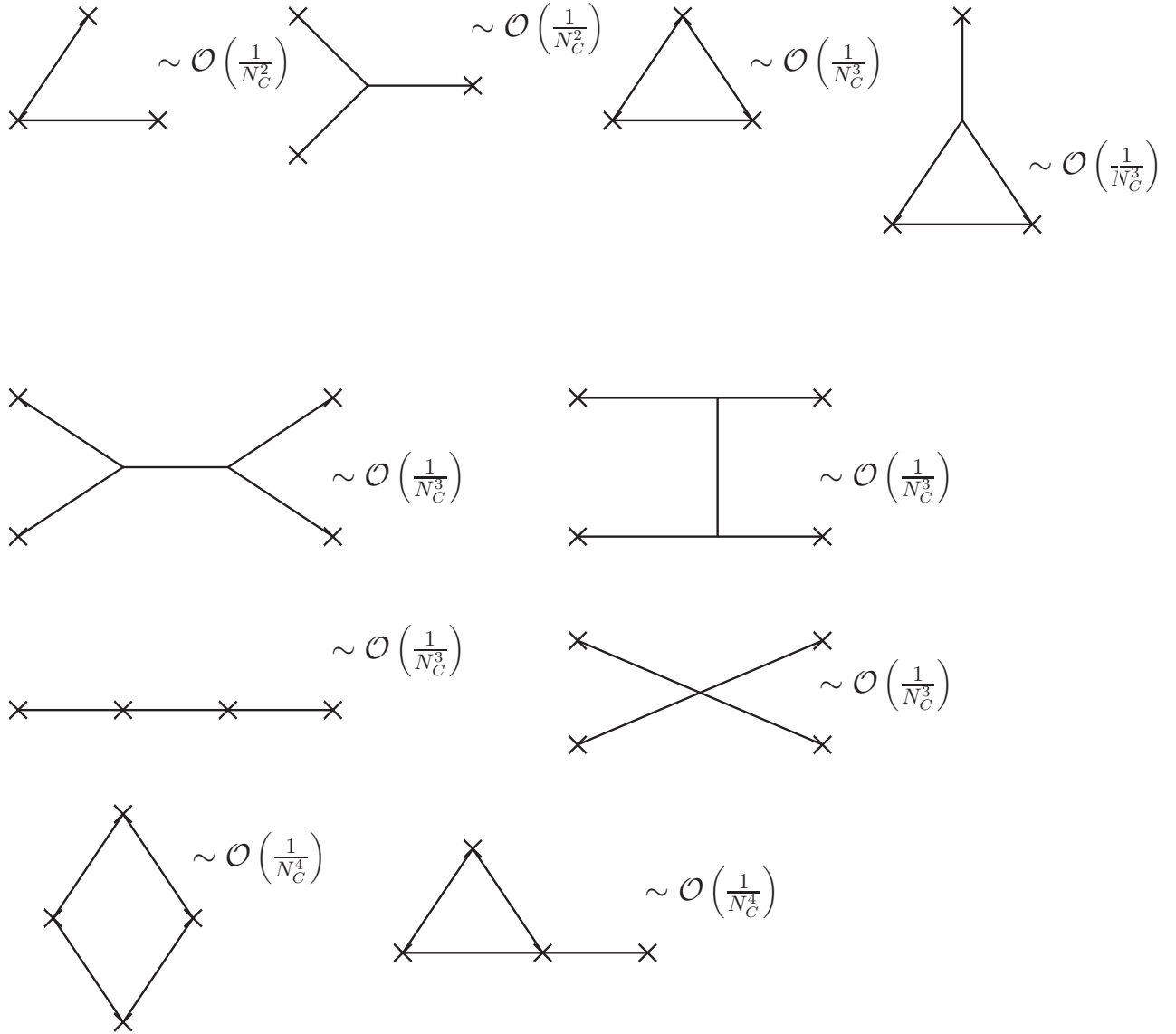}
\caption{$3$ and $4$-point correlators given by meson exchanges: Tree-level diagrams are dominant and every quark loop is suppressed by one power of $1/N_C$.
 The counting in $1/N_C$ is given for all of them taking into account that every source brings in a $1/N_C$ factor and that $r$-meson vertices introduce the
 factor $\N^{1-r/2}$.}
\label{diagram 3and4pointcorrelatorasmesonexchange}
\end{figure}
\\
\hspace*{0.5cm}
When considering the different Green functions, one wants to 
guarantee that all the poles are originated by the tree-level diagrams obtained from an $EFT$-Lagrangian. To show this, one needs to restore to unitarity 
and crossing symmetry. Crossing means that every pole appearing in a given channel will manifest in all others related to the previous one by crossing. 
Unitarity guarantees that every pole in a given diagram will reappear each time this particular topology occurs as a subdiagram in a 
higher-order Green function. All amplitudes are thus produced by tree level exchanges with vertices given in an $EFT$-Lagrangian.\\

\hspace*{0.5cm}Any model or theory based on the $1/N_C$ expansion will naturally raise the question of 
the error associated to that expansion, which can be naively estimated as $1/N_C\sim30\%$. It is natural to object to the 
convergence of a series in $1/N_C$ that ends up being $1/3$ in the real world. The associated error of the size of the expansion 
parameter would make our effort useless. Fortunately, this does not seem to be the case. Just to give a counterexample, a look at the results of 
Ref.~\cite{Guerrero:1997ku} will quickly suggest that the actual error can be much smaller, which is noteworthy, 
especially taking into account that in the quoted reference, 
an impressive agreement with data in $e^+e^-\to\pi^+\pi^-$ up to $s=1$ GeV$^2$ was obtained in terms of just one parameter, $M_\rho$.

Let us recall how this can be possible. First, the large-$N_C$ expansion of QCD is not an expansion in the usual 
perturbative sense. For instance, in QED, the perturbative 
expansion means, firstly, that the diagrams with less photon couplings to fermionic lines dominate. 
After that, when we compute diagrams both at tree level, and including loops, we realize 
that the expansion parameter is $\alpha=e^2/(4\pi)^2$ and the coefficients of the series are small compared to $\alpha$:
 every order we go further in the expansion  the error reduces  by $\sim 1/\alpha\simeq$ 137.
 In the large-$N_C$ expansion of QCD, we know first, which diagrams are leading-order (planar diagrams with gluon exchanges)
and which ones are suppressed. Unfortunately, 
the expansion has a fundamental subtlety that prevents one from  determining the expansion parameter after that: there are infinite 
diagrams at any given order in the expansion, so that one cannot perform a calculation at both LO and NLO and 
compare them to know what the expansion parameter is. It is known, however, that diagrams with internal quark loops are suppressed as $1/N_C$ and 
that non-planar diagrams are suppressed as $1/N_C^2$. Moreover, if the number of quark flavours, $n_f$, 
is not considered to be smaller than $N_C$ the former diagrams can 
even scale as $n_f/N_C$. These kind of contributions would be responsible for the mixing of $q\bar{q}$ 
and $q\bar{q}q\bar{q}$ states. The fact that this effect is not observed 
in Nature and the success of the quark model classification of mesons in $q\bar{q}$ multiplets suggests 
that the coefficients of these diagrams with internal quark loops are 
tiny, in such a way that the first non-negligible correction would come from the non-planar diagrams, 
suppressed as $1/N_C^2\sim10\%$. This reasoning may explain why 
the large-$N_C$ expansion is a such a good approximation for low and intermediate-energy $QCD$, given the 
phenomenological accomplishments of its applications in meson effective field theories \cite{Pich:2002xy} 
and the corroborated predictions given by the large-$N_C$ limit both for $\chi PT$ 
\cite{Ecker:1994gg, Pich:1995bw} and for $R \chi T$~\cite{Ecker:1988te} coupling constants. 
All these reasons seem to suggest that, quite generally, some factor comes to complement 1/$N_C$ for the value of the 
expansion parameter to be reduced and the relative accuracy to be increased.

This conclusion is supported by the investigation of some $\mathcal{O}(p^4)$ and $\mathcal{O}(p^6)$ couplings of $\chi PT$, 
which are done modelling the $NLO$ expansion in 
$1/N_C$ of $R\chi T$ \cite{Rosell:2004mn, Rosell:2006dt, Portoles:2006nr, Pich:2008jm, Rosell:2009yb, Pich:2010sm}. 
According to the size of the corrections, we judge that 15$\%$ can be a 
reasonable general estimate (see, however, our discussion in Appendix C of Ref.~\cite{Shekhovtsova:2012ra} on the possible variations 
on the predictions of the couplings obtained in the $N_C\to\infty$ 
limit). Noticeably, the actual expansion parameter can be computed for $R\chi T$ in the study of the vector form factor 
of the pion at $NLO$ in the $1/N_C$ expansion 
\cite{SanzCillero:2009pt}, yielding
\begin{equation}
 \alpha_V\,=\,\frac{n_f}{2}\frac{2G_V^2}{F^2}\frac{M_V^2}{96\pi F^2}\,,
\end{equation}
which, at lowest order, is the ratio of the vector width and mass, $\alpha_V\sim0.2$, agreeing with the previous discussion.

Moreover, we should emphasize that our approach goes beyond the $N_c\to\infty$ limit. We will see that the lowest order in the 1/$N_C$ expansion for the theory 
in terms of mesons is supplemented by the leading higher-order correction, namely by including the resonance (off-shell) widths for the wide states $\rho$, 
$K^\star$ and $a_1$ and constant widths for the narrow $\omega$ and $\phi$ states. This seems to point to smaller errors than those characteristic of the $LO$ contribution 
in the 1/$N_C$ expansion and may be able to explain, altogether, an eventual fine agreement with data.
\section{Resonance Chiral Theory} \label{LargeN_RChT}
\hspace*{0.5cm}Our methodology stands on the construction of an action, with the relevant degrees of freedom (Weinberg's approach), led by the chiral symmetry
 and the known asymptotic behaviour of form factors and Green functions driven by large-$N_C$ $QCD$. The large-$N_C$ expansion of $SU(N_C)$ $QCD$ implies that,
 in the $N_C\to \infty$ limit, the study of Green functions of $QCD$ currents can be carried out through the tree level diagrams of a Lagrangian theory that 
includes an infinite spectrum of non-decaying states. Hence the study of the resonance energy region can be performed by constructing such a Lagrangian theory.
 The problem is that we do not know how to implement an infinite spectrum in a model independent way. However, it is well known from the phenomenology that the
 main role is always played by the lightest resonances. Accordingly it was suggested in Refs. \cite{Ecker:1988te, Ecker:1989yg} that one can construct a suitable
 Lagrangian involving the lightest nonets of resonances and the octet of Goldstone bosons states ($\pi$, K and $\eta$). This is indeed an appropriate tool
 to handle the hadronic decays of the tau lepton. The guiding principle in the construction of such a Lagrangian is chiral symmetry. When resonances are integrated
 out from the theory, i.e. one tries to describe the energy region below such states ($E \ll M_\rho$), the remaining setting is that of $\chi PT$, that was 
described in Chapter one and in the previous section in the context of the large-$N_C$ limit. Then, $\RCT$ is a link between the chiral and asymptotic regimes 
of $QCD$ and a very useful tool to understand intermediate-energy $QCD$ dynamics \cite{Portoles:2010yt}.\\
\hspace*{0.5cm}The path-integral formalism is adequate to explain what we are doing. Starting from the generating functional, $Z$, of $QCD$, one obtains the
 different Green functions taking the suitable functional derivatives of $Z$. Depending on up to which energy we consider the interesting physics scale to be,
 the heavier integrated-out degrees of freedom will be all meson resonances ($\CPT$), or just the charmed -and even heavier- mesons ($\RCPT$). That is,
\begin{eqnarray} \label{path-resonance}
e^{\,i\,Z}&=&\int \mathcal{D}q\,\mathcal{D}\overline{q}\,\mathcal{D}G_\mu \, e^{\,i\,\int \mathrm{d}^4x \,
\mathcal{L}_{QCD}}\nonumber \\
&=&\int
\mathcal{D}u\,\prod_{i=1}^\infty\mathcal{D}V_i\,\prod_{j=1}^\infty\mathcal{D}A_j\,
\prod_{k=1}^\infty\mathcal{D}S_k\,
\prod_{m=1}^\infty\mathcal{D}P_m\, e^{\,i\int
\mathrm{d}^4x \mathcal{L}_{R\chi T}(u,V_i,A_j,S_k,P_m)}\nonumber \\
&=&\int \mathcal{D}u \, e^{\,i\int \mathrm{d}^4x \mathcal{L}_{\chi
PT}(u)} \, ,
\end{eqnarray}
where $V$, $A$, $S$, $P$ designates the type of resonance: vector ($1^{--}$), axial-vector ($1^{++}$), scalar ($0^{++}$) and pseudoscalar ($0^{-+}$). Integrating the resonances
 out of the action reproduces $\CPT$-$pGs$ interaction simply modifying the $\CPT$-$LECs$. In Eq. (\ref{path-resonance}) an infinite number of resonances have been
 considered per each set of quantum numbers, as $\N\to\infty$ tells. There are two approaches to this: the single resonance approximation ($sra$) considers 
that the lowest-lying multiplets are able to collect the bulk of the dynamical information and these are the only resonant degrees of freedom kept active in 
the action. Since there is an infinite number of Green functions, it is obviously impossible to satisfy all  matching conditions with asymptotic $QCD$ with these
 few resonances. The minimal hadronic approximation ($mha$) generalizes the $sra$ by including into the action the minimal number of resonances that allows 
fulfilling the $QCD$ short-distance constraints in the considered amplitude.\\
\hspace*{0.5cm}$\RCPT$ includes explicitly the $\CPT$ at $\cO(p^2)$ but, instead of adding to it the next order in the chiral expansion (which is obtained upon integration of the resonances, as it is 
explained below), that piece is supplemented with a Lagrangian describing interactions among $pGs$ and resonances. The more convenient chiral tensor formalism in terms of $u(x)$ and all the other structures
 introduced in (\ref{op-p4}) for the external sources and the $pGs$ (\ref{uU}), (\ref{PhiGoldstones}), (\ref{L-Op2-u}) -whose transformation properties under
 $C$, $P$ and Hermitian conjugation were collected in Table \ref{chiraltransfu}- are employed to write:\\
\begin{equation} \label{def}
\mathcal{L}_{\RCPT}(u,V,A,S,P)\,=\,\mathcal{L}_{\,\CPT}^{(2)}(u)\,+\,\mathcal{L}_R(u,V,A,S,P)\, .
\end{equation}
\hspace*{0.5cm}For the resonance fields, the observed multiplets tell us that only octets and singlets in flavour space occur; so reffering to them as $R$
 and $R_1$, respectively, the non-linear realization of the chiral group $G$, will be given by \footnote{There are many possible ways to transform the resonance
 fields that lead to the same transformation under the vector group and it can be shown that they are all equivalent after a field redefinition. Since we are 
working in the $u$-basis the most convenient choice is the one in Eq. (\ref{tr}).}
\begin{equation} \label{tr}
R \, \rightarrow \, h(g,\,\Phi) \, R \, h(g,\,\Phi)^\dagger \, , \qquad R_1 \, \rightarrow \, R_1 \, .
\end{equation}
\hspace*{0.5cm}For the first vector nonet \footnote{In the $\N\to\infty$ limit octet and singlet converge to a nonet.}, we will have:
\begin{eqnarray} \label{vector-res}
V_{\mu\nu} & = & \frac{1}{\sqrt{2}}\,\sum_{a=0}^8 \lambda_a V^a_{\mu\nu}  \\
& = & \left( \begin{array} {ccc}
\frac{1}{\sqrt{2}}\rho^{0}+\frac{1}{\sqrt{6}}\omega_8+\frac{1}{\sqrt{3}}\omega_1 & \rho^{+} & K^{*\,+} \\
\rho^{-} & -\frac{1}{\sqrt{2}}\rho^{0}+\frac{1}{\sqrt{6}}\omega_8+\frac{1}{\sqrt{3}}\omega_1 & K^{*\,0} \\
K^{*\,-} & \overline{K}^{*\,0} & -\frac{2}{\sqrt{6}}\omega_8+\frac{1}{\sqrt{3}}\omega_1 \end{array} \right)_{\mu\nu} \,,\nonumber
\end{eqnarray}
where the antisymmetric tensor formulation for vector fields has been introduced instead of that due to Proca, that may be more familiar \footnote{The appendix E is 
specially devoted to this topic.}. With this description one is able to collect, upon integration of resonances, the bulk of the low-energy couplings at $\cO(p^4)$
 in $\chi PT$ without the inclusion of additional local terms \cite{Ecker:1989yg}. In fact it is necessary to use this representation if one does not include
 the $\mathcal{L}^{(4)}_{\CPT}$ in the Lagrangian theory. Analogous studies at higher chiral orders have been carried out at $\cO(p^6)$ in the even- \cite{Cirigliano:2006hb} and 
odd-intrinsic \cite{Kampf:2011ty} parity sectors. We will assume that no $\mathcal{L}^{(n)}_{\CPT}$ with $n = 8, 10, ...$ need to be included in the theory.\\
We will write explicitely the axial-vector octet since it is also of major importance in the processes considered in this Thesis:
\begin{eqnarray} \label{axial-res}
A_{\mu\nu} & = & \frac{1}{\sqrt{2}}\,\sum_{a=0}^8 \lambda_a A^a_{\mu\nu}  \\
& = & \left( \begin{array} {ccc}
\frac{1}{\sqrt{2}}\mathrm{a}_1^{0}+\frac{1}{\sqrt{6}}h_1+\frac{1}{\sqrt{3}}f_1 & \mathrm{a}_1^{+} & K_{1A}^{*\,+} \\
\mathrm{a}_1^{-} & -\frac{1}{\sqrt{2}}\mathrm{a}_1^{0}+\frac{1}{\sqrt{6}}h_1+\frac{1}{\sqrt{3}}f_1 & K_{1A}^{*\,0} \\
K_{1A}^{*\,-} & \overline{K}_{1A}^{*\,0} & -\frac{2}{\sqrt{6}}h_1+\frac{1}{\sqrt{3}}f_1 \end{array} \right)_{\mu\nu} \,.\nonumber
\end{eqnarray}
One can proceed analogously for $S$ and $P$-multiplets \cite{Ecker:1988te} and for the particles with $J^{PC}=1^{+-}$ \cite{Ecker:2007us}.\\
\hspace*{0.5cm}The formulation of a Lagrangian theory that includes both the octet of Goldstone mesons and $U(3)$ nonets of resonances is carried out through
 the construction of a phenomenological Lagrangian \cite{Coleman:1969sm, Callan:1969sn} where chiral symmetry determines the structure of the operators.
\hspace*{0.5cm}In order to construct the relevant Lagrangians, we need to introduce the covariant derivative (\ref{covariantderivative}), dictated by the local
 nature of the non-linear realization of $G$ in $R$ (\ref{tr}), in such a way that
\begin{equation}
\nabla_\mu R \rightarrow  h(g,\,\Phi) \, \nabla_\mu R \, h(g,\,\Phi)^\dagger \,.
\end{equation}
\hspace*{0.5cm}The transformation properties of resonance fields under $P$, $C$ and Hermitian conjugation that are needed to write the Effective Lagrangian are
 collected in Table \ref{TRCPT}. For other structures appearing in the later extensions of the original Lagrangian and their transformation properties, see Section
 \ref{LargeN_ExtensionsL}.\\
\begin{table} 
\begin{center}
\renewcommand{\arraystretch}{1.2}
\begin{tabular}{|c|c|c|c|} 
\hline
Operador & $P$ & $C$ & h.c. \\
\hline
$V_{\mu\nu}$ &  $V^{\mu\nu}$ & $-V_{\mu\nu}^T$ & $V_{\mu\nu}$ \\
$A_{\mu\nu}$ & $-A^{\mu\nu}$ & $A_{\mu\nu}^T$ & $A_{\mu\nu}$ \\
$S$ & $S$ & $S^T$ & $S$\\ 
$P$ & $-P$ & $P^T$& $P$\\
\hline
\end{tabular}
\caption{\small{Transformation properties of resonances under $P$, $C$ and Hermitian conjugation.}} \label{TRCPT}
\end{center}
\end{table}
\\
\hspace*{0.5cm}We have to build the most general Lagrangian involving all the pieces that respect the assumed symmetries. This means considering 
$\cO(p^2)$-$pG$ tensors together with one resonance field. If we think about $\pi\,\pi$-scattering again, this amounts to consider as the first correction 
to the tree level amplitude not only the one-loop diagrams obtained with arbitrary insertions of $\mathcal{L}_2$ vertices, but also meson-resonance exchange
 amid both pairs of pions.\\
\hspace*{0.5cm}For the kinetic terms, bilinears in the meson fields are considered including the covariant derivative \footnote{It reduces to the ordinary 
one in the case of singlets.} and incorporating by hand the corresponding mass of the octet or singlet. 
The pattern of preserved $SU(3)_V$ justifies the same mass for all members of the representation of the symmetry group and 
it cannot be determined from first principles, but fitted to the experiment. The discrepancy among resonance masses within the same multiplet has two sources: 
on the one hand it corresponds to $SU(3)_V$ breaking operators and on the other hand to $NLO$ corrections in $1/\N$.\\
\hspace*{0.5cm}The construction of the interaction terms involving resonance and Goldstone fields is driven
by chiral and discrete symmetries with a generic structure given by
\begin{equation} \label{genericstructure}
 \mathcal{O}_i\,\sim\,\left\langle R_1 R_2 ... R_j \chi^{(n)}(\phi)\right\rangle\,, 
\end{equation}
where $\chi^{(n)}(\phi)$ is a chiral tensor that includes only Goldstone and auxiliary fields. It transforms like $R$ in Eq. (\ref{tr}) and has chiral counting 
$n$ in the frame of $\CPT$. This counting is relevant in the setting of the theory because, though the resonance theory itself has no perturbative expansion,
 higher values of $n$ may originate violations of the proper asymptotic behaviour of form factors or Green functions. As a guide we will include at least 
those operators that, contributing to our processes, are leading when integrating out the resonances. In addition we do not include operators with higher-order
 chiral tensors, $\chi^{(n)}(\phi)$, that would violate the $QCD$ asymptotic behaviour unless their couplings are severely fine tuned to ensure the needed 
cancellations of large momenta.\\
\hspace*{0.5cm}Guided by these principles and considering only one resonance field, the Lagrangian that was obtained in Ref.~\cite{Ecker:1988te} is
\begin{equation} \label{tot}
\mathcal{L}_R \,=\, \sum_{R=V,\,A,\,S,\,P} \left\{ \mathcal{L}_{\,\mathrm{kin}}(R)\,+\,\mathcal{L}_2(R) \right\} \, ,
\end{equation}
where the kinetic term \footnote{This naming can be a little bit confusing because this term also includes interactions hidden in the covariant derivative 
part.} is
\begin{eqnarray} \label{Rkin}
\mathcal{L}_{\,\mathrm{kin}}(R) &=& 
-\frac{1}{2} \bra \nabla^\lambda R_{\lambda\mu} \nabla_\nu R^{\nu\mu}\,-\,
\frac{1}{2}M_R^2 R_{\mu\nu}R^{\mu\nu} \ket 
 \, , \quad R=V,A \quad , \nonumber \\
\mathcal{L}_{\,\mathrm{kin}}(R) &=& \frac{1}{2} \bra \nabla^\mu R\nabla_\mu R\,-\,
M_R^2 R^2 \ket 
 \, , \quad R=S,P \quad ,
\end{eqnarray}
and $M_R$ 
 stands for the 
nonet mass in the chiral limit. The purely interacting term, $\mathcal{L}_2(R)$, is given by
\begin{eqnarray}\label{R}
\mathcal{L}_2[V(1^{--})] &=& \frac{F_V}{2\sqrt{2}}\,\bra V_{\mu\nu} f^{\mu\nu}_+ \ket \,+\,
\frac{i\, G_V}{2\sqrt{2}} \bra V_{\mu\nu} [u^\mu,\,u^\nu] \ket \, , \nonumber \\
\mathcal{L}_2[A(1^{++})] &=& \frac{F_A}{2\sqrt{2}}\,\bra A_{\mu\nu} f^{\mu\nu}_- \ket\, , \nonumber \\
\mathcal{L}_2[S(0^{++})] &=& c_d \bra S\,u_\mu u^\mu\ket\,+\,c_m\bra
S\,\chi_+\ket\,+\,\tilde{c}_d\,S_1\bra u_\mu u^\mu \ket \,+\, \tilde{c}_m\,S_1\bra\chi_+\ket \, ,
\nonumber \\
\mathcal{L}_2[P(0^{-+})] &=& i\,d_m \bra P \chi_- \ket\,+\,i\tilde{d}_m\,P_1\bra\chi_-\ket \, ,
\end{eqnarray}
where all couplings are real and it has been considered only the octet for $V$ and $A$, because  $\bra f^{\mu\nu}_\pm\ket\,=\,\bra[u^\mu,\,u^\nu]\ket\,=\,0$ 
forbides couplings for $V$ and $A$ singlets at this chiral order.
\\
\hspace*{0.5cm}
 We also assume exact $SU(3)$ symmetry in the construction of the interacting terms,
 i.e. at level of couplings. Deviations from exact symmetry in hadronic tau decays have been considered in \cite{Moussallam:2007qc}. However we do not include
 $SU(3)$ breaking couplings because we are neither considering next-to-leading corrections in the $1/N_C$ expansion. These corrections have already been considered
within $\RCT$ in Refs.~\cite{Rosell:2007kc, Rosell:2004mn, Rosell:2006dt, Portoles:2006nr, Pich:2008jm, Rosell:2009yb, Pich:2010sm, SanzCillero:2009pt, Rosell:2005ai, SanzCillero:2007ib, SanzCillero:2009ap}.\\
\section{Matching $\RCPT$ with QCD asymptotic behaviour} \label{LargeN_MatchingRChTQCD}
\hspace*{0.5cm}The long distance features of $QCD$ \cite{Shifman:1978bx, Shifman:1978by} have to be inherited by $\RCPT$ as made precise by the matching 
conditions. At high energies, $\RCPT$ must match the $OPE$, and this will impose some relations among its couplings. These relations will depend upon the 
Lagrangian we choose. Due to historical reasons we will explain here the results obtained with the kinetic pieces and the interactions terms in Eq.~(\ref{R}) 
restricting our attention to those that can be relevant in the processes we examine, namely the spin-one resonances. 
 The number of relations that we obtain will influence decisively in the predictability of the theory. Working in the $sra$, we find \cite{Pich:2002xy}
\begin{itemize}
\item Vector Form Factor.\\
\hspace*{0.5cm}At $LO$ in $1/\N$, the form factor of the pion is given within $\RCT$ by
\begin{equation}
\mathcal{F}(q^2)\,=\,1\,+\,\frac{F_V\,G_V}{F^2}\frac{q^2}{M_V^2-q^2} \, .
\end{equation}
and $QCD$ short-distance behaviour \cite{Floratos:1978jb} dictates $\Im$m $\Pi_V(q^2)\,\rightarrow\,$const. as $q^2\rightarrow\infty$, which results \footnote{The imaginary
 part of the Vector-Vector correlator is given by the Vector form factor. Although the former has contributions from all possible $n$-meson form factors, it is sensible to
require that every one vanishes asymptotically in order to fulfill that constraint. See the discussions below eqs.(\ref{constr}) and (\ref{FNdR}), respectively.} in a relation for the resonance couplings
\begin{equation} \label{rest1}
F_V\,G_V \,=\,F^2 \,.
\end{equation}
\item Axial Form Factor.\\
\hspace*{0.5cm}We consider the axial form factor
$G_A(t)$ governing the matrix element 
$\langle \gamma | A_{\mu} | \pi \rangle$ \cite{Gasser:1983yg}.
Extracting $G_A(t)$ from the $\left\langle VAP\right\rangle$ 
Green function by setting the pion massless, one finds
\begin{equation}
\label{ganew} 
G_A(t) = \frac{F^2}{M_V^2} \; \frac{b_1+b_3 t}{M_A^2-t} ~,
\end{equation} 
where $b_1$ and $b_3$ are unknowns. Demanding that the form factor $G_A(t)$ vanishes for large $t$ 
\cite{Ecker:1989yg,Lepage:1979zb,Lepage:1980fj,Brodsky:1981rp}, we
obtain
\begin{equation}
\label{b3} 
b_3=0~.
\end{equation}
Using the $\RCT$ Lagrangian, Eq. (\ref{R}), under the hypothesis of single resonance exchange, one finds \cite{Ecker:1989yg, Cirigliano:2004ue}
\begin{equation}
\label{gaold}
G_A(t) = \frac{2F_V G_V - F_V^2}{M_V^2} + \frac{F_A^2}{M_A^2-t} \; \;.
\end{equation}
Requiring $G_A(t)$ to vanish for $t \to \infty$ implies the relation
$F_V = 2 G_V$, one version of the so-called KSFR relation 
\cite{Kawarabayashi:1966kd,Riazuddin:1966sw}. The inclusion of 
bilinear resonance couplings modifies the form factor as given
in Eq.~(\ref{ganew}) \cite{Cirigliano:2004ue} with $b_1=M_A^2 - M_V^2, ~b_3=0$, and it induces
a correction to the KSFR relation:
\begin{eqnarray} \label{rest2}
\displaystyle\frac{2 F_V G_V - F_V^2}{2 F^2} = 1 -
\displaystyle\frac{F_V^2}{2 F^2} = \displaystyle\frac{M_A^2 - 2 M_V^2}
{2(M_A^2 - M_V^2)}  \; \; .
\end{eqnarray}
\item Weinberg's sum rules.\\
\hspace*{0.5cm}The two-point function of a vector correlator between left-handed and right-handed quarks defines the mixed correlator
\begin{equation}
\Pi_{LR}(q^2)\,=\,\frac{F^2}{q^2}\,+\,\frac{F_V^2}{M_V^2-q^2}\,-\,\frac{F_A^2}{M_A^2-q^2} \,.
\end{equation}
Gluon interactions safeguard chirality, so $\Pi_{LR}$ must fulfill a non-subtracted dispersion relation. Moreover, it must vanish in the chiral limit faster
 than $1/(q^2)^2$ as $q^2\rightarrow\infty$. This implies \cite{Weinberg:1967kj} the relation for the couplings:
\begin{equation}\label{rest3}
F_V^2-F_A^2\,=\,F^2\,\, ,\qquad M_V^2\,F_V^2\,-\,M_A^2\,F_A^2\,=\,0\,.
\end{equation}
\end{itemize}
\hspace*{0.5cm}Considering the above restrictions (\ref{rest1}), (\ref{rest2})
and (\ref{rest3})
, we are able to write all 
decay constants in terms of $F$ 
 and the resonance masses:
\begin{eqnarray} \label{matching}
F_V^2\,=\,F^2\frac{M_A^2}{M_A^2-M_V^2}\,,\quad\;F_A^2\,=\,F^2\frac{M_V^2}{M_A^2-M_V^2}\,,\quad\;G_V^2\,=\,F^2\,\left( 1\,-\,\frac{M_V^2}{M_A^2}\right)
. \nonumber \\
\end{eqnarray}
\hspace*{0.5cm}Finally, we will see that applying the $QCD$-ruled short-distance behaviour to the decays $\t^-\to P^- \gamma\nu_\t$ computed using  $\RCT$ 
\cite{GuoRoig} allows to relate $V$ and $A$ masses
\begin{eqnarray}\label{masses}
2\,M_A^2\,=\,3\,M_V^2\,,
\end{eqnarray}
a result that reproduces the one obtained in Ref.~\cite{Dumm:2009kj} for the form-factor $\mathcal{F}_{\pi\gamma^*\gamma}$.\\
\hspace*{0.5cm}These relations guarantee the matching among $QCD$ and its $EFT$, $\RCPT$, for the considered Green functions. Let us, however, make a caveat about 
phenomenology and $QCD$ at this point.\\
\hspace*{0.5cm}There are infinite Green functions both in $QCD$ and also in its $EFTs$. In perturbative $QCD$, all of them are described in terms of a single
 coupling, $\alpha_s$. In non-perturbative $QCD$ this is clearly not the case. Then, the situation changes and, unless we bear this in mind, we can arrive 
-or seem to arrive, to be precise-, to inconsistencies.\\
\hspace*{0.5cm}There is no difference between considering one set or another of Green functions in high-energy $QCD$. The situation is opposite in its low 
and intermediate-energy regime. For instance, we have seen that for a set consisting of vector, 
and axial-vector form factors of $pGs$ and $LR$ 
 two-point correlators the relations (\ref{rest1}), (\ref{rest2}) and (\ref{rest3}) 
 ensure $QCD$ asymptotic behaviour for the $EFT$ working in the $sra$.\\
\hspace*{0.5cm}But, once we go further and study three-point correlators and form factors involving three particles in the final state it is likely 
that either the previous relations get modified \textit{in some cases}, or -what it is preferable- we admit that the $sra$ is a valid approach if we do not 
intend to describe all $QCD$ Green functions at the same time \footnote{The validity of these assumptions within large-$\N$ $QCD$ is studied in Refs. 
\cite{Peris:1998nj, Knecht:1999gb, Peris:2000tw, deRafael:2002tj}.}. Otherwise, we are forced to incorporate a second multiplet of resonances in the 
(axial-)vector case.\\
\hspace*{0.5cm}These discrepancies among $QCD$-asymptotic restrictions for the parameters entering $\mathcal{L}_R$ has already been found and discussed \cite{Cirigliano:2004ue}.
 However, the understanding of this issue is evolving as more works are concluded. Our position towards this problem will get defined in later chapters 
concerning the practical applications of the theory. We will see that we will arrive to consistent relations for the radiative decays of the tau with one meson 
and for the three meson decays. However, it is very likely that they will not coincide with the relations one could find studying four-point Green functions or four meson form factors. 
In any case, the study of these is a too involved task that we do not consider for the time being.\\
\section{Extensions of the original Lagrangian} \label{LargeN_ExtensionsL}
\subsection{Even-intrinsic parity sector}
\hspace*{0.5cm}We recall the purpose of the original paper were $\RCPT$ was borned \cite{Ecker:1988te}, it was to build a sound theory including resonances within
 the chiral framework that respected all principles and symmetries governing light-flavoured $QCD$ and that was able to reproduce the $\cO(p^4)$ even-intrinsic
 parity chiral Lagrangian upon integration of the resonances.\\
\hspace*{0.5cm}In order to construct (\ref{tot}), (\ref{Rkin}), (\ref{R}), $\cO(p^2)$-$pG$ tensors together with one resonance field were enough to accomplish that
 purpose.\\
\hspace*{0.5cm}One may wonder why we intend to extend the original Lagrangian in $\RCT$, while for $\CPT$ the decision consists in going one order further 
in the chiral expansion. The nature of $pGs$ is completely different to that of resonances. Whereas the first ones transform non-linearly under the vector 
subgroup, the second ones do it linearly. This results in a huge, fundamental difference. Processes involving different number of $pGs$ are related. For 
instance, all $2n$-$pGs$ $\to$ $2n$-$pGs$ scattering processes are connected at a given order. As the easiest example, all of them are written in terms 
of the pion decay constant at $LO$; but the divergence structure -before renormalizing- of say $12$ $\pi$ $\to$ $12$ $\pi$ is given by that of the 
simplest process $2$ $\pi$ $\to$ $2$ $\pi$, as well.\\
\hspace*{0.5cm}On the contrary, resonances are not free excitations (even in absence of $\chi SB$); so that any time we want to consider physics involving 
one more multiplet of resonances, we have to extend our Lagrangian to include it relying again on the same symmetry principles that guided the construction of the already
 existing pieces.\\
\hspace*{0.5cm}The analysis of $\t^-\,\to\,(\pi\,\pi\,\pi)^-\,\nu_\tau$ within $\RCT$ \cite{GomezDumm:2003ku} could not ignore the relevance of 
the axial-vector $a_1$-resonance exchange within this decay. 
The contribution given by the chain $a_1\,\to\,\rho\,\pi\,\to\,\pi\,\pi\,\pi$ driven by vector 
exchange was accounted for by going one step beyond the work in Ref.~\cite{Ecker:1988te} including bilinear terms in the resonance fields that lead to a coupling
 $a_1\,\rho\,\pi$, hence only the generalization including one pseudoscalar was considered in the quoted paper.\\
\hspace*{0.5cm}The most general Lagrangian respecting all the assumed symmetries and including one $\cO(p^2)$ chiral tensor, one vector and one axial-vector
 resonance fields can be written \cite{GomezDumm:2003ku}
\begin{equation} \label{LVAPa} 
\mathcal{L}_2^{VAP}\,=\,\sum_{i=1}^{5}\lambda_i\mathcal{O}^i_{VAP}\,,
\end{equation}
where $\lambda_i$ are new unknown real adimensional couplings, and the operators $\mathcal{O}^i_{VAP}$ constitute the complete set of operators for building 
vertices with only one pseudoscalar~\footnote{For a larger number of $pGs$, additional operators may emerge.}, which are given by
\begin{eqnarray}\label{LVAPb}
& & \mathcal{O}_{VAP}^1 \, = \, \bra\left[V^{\mu\nu} ,\,A_{\mu\nu}\right] \chi_-\ket\,,\nonumber\\
& & \mathcal{O}_{VAP}^2\,=\,i\bra\left[V^{\mu\nu} ,\,A_{\nu\alpha}\right]h_\mu^\alpha \ket\,,\nonumber\\
& & \mathcal{O}_{VAP}^3\,=\,i\bra\left[ \nabla^\mu V_{\mu\nu},\,A^{\nu\alpha}\,\right] u_\alpha\ket\,,\nonumber\\
& & \mathcal{O}_{VAP}^4\,=\,i\bra\left[ \nabla^\alpha V_{\mu\nu},\,A_\alpha^{\;\;\nu}\right] u^\mu\ket\,,\nonumber\\
& & \mathcal{O}_{VAP}^5\,=\,i\bra\left[ \nabla^\alpha V_{\mu\nu},\,A^{\mu\nu}\right] u_\alpha\ket\,,
\end{eqnarray}
where it has been used $h^{\mu\nu}$ defined in Eq. (\ref{hmunu}).
 As we are only interested in tree level diagrams, the $\cO(p^2)$ $\CPT$ $EOM$, (\ref{EOM_p2}), 
has been used in $\mathcal{L}_2^{VAP}$ in order to eliminate one of the possible operators.\\
\hspace*{0.5cm}Explicit computation of the Feynman diagrams involved in this process -and in all applications studied in this Thesis- show that all the contributions
 coming from $\mathcal{L}_2^{VAP}$ can be written in terms of only three combinations of their couplings
\begin{eqnarray} \label{lambdas0,',''}
& & \lambda_0\,=\,-\frac{1}{\sqrt{2}}\left[ 4\lambda_1\,+\,\lambda_2\,+\,\frac{\lambda_4}{2}\,+\,\lambda_5 \right]\,,\nonumber\\
& & \lambda'\,=\,\frac{1}{\sqrt{2}}\left[ \lambda_2\,-\,\lambda_3\,+\,\frac{\lambda_4}{2}\,+\,\lambda_5\right] \,,\nonumber\\
& & \lambda''\,=\,\frac{1}{\sqrt{2}}\left[ \lambda_2\,-\,\frac{\lambda_4}{2}\,-\,\lambda_5\right] \,.
\end{eqnarray}
\subsection{Odd-intrinsic parity sector}
\hspace*{0.5cm}Now we turn to the $WZW$ anomalous term, we recall that it is the $LO$ contribution in the odd-intrinsic parity sector -$\cO(p^4)$ in the chiral 
counting-. In Ref. \cite{RuizFemenia:2003hm}, that is the fundamental reference for this section, a suitable Lagrangian was built and odd-intrinsic parity processes were examined
 within $\RCT$. It is a common practice to assume that upon integration of the resonances in this Lagrangian one could saturate the values of the $\cO(p^6)$ $LECs$
 of $\CPT$ in the anomalous sector, analogously as it happens in the even-intrinsic parity sector \footnote{This was checked only recently \cite{Kampf:2011ty}.}.\\
\hspace*{0.5cm}The quoted reference has an added interest, because it gave rise to a set of works studying the behaviour of 3-point Green functions in $\RCT$ \cite{Cirigliano:2006hb, Cirigliano:2004ue, Cirigliano:2005xn}.\\
\hspace*{0.5cm}Several authors \cite{Pallante:1992qe, Moussallam:1994xp, Moussallam:1997xx, Knecht:2001xc, Bijnens:2003rc} started to analyze systematically a set
 of $QCD$ three-point functions that were free of perturbative contributions from $QCD$ at short distances \footnote{They vanish in absence of $S\chi SB$ for
 massless quarks.}, a fact that made more reliable a smooth matching of the $OPE$ result went down to low energies and the $EFT$ description by a theory 
including resonances.\\
\hspace*{0.5cm}It was shown in Ref.~\cite{Knecht:2001xc} that while
 the ansatz derived from the lowest meson dominance approach to the large-$\N$ limit of $QCD$ incorporates by construction the right short-distance behaviour
 ruled by $QCD$, the same Green functions as calculated with a resonance Lagrangian, in the vector field representation, are incompatible with the $OPE$ outcome.
The authors pointed out that these discrepancies cannot be repaired just by introducing the chiral Lagrangian of $\cO(p^6)$ \footnote{
When one considers the pion form factor calculated within the Resonance Theory both in the vector and in the antisymmetric tensor formalisms \cite{Ecker:1988te},
 compatilibity with high-energy $QCD$ constraints is found in the latter case without introducing $\mathcal{L}_{\CPT}^{(4),\,\mathrm{even}}$. In the former 
case, the asymptotic behaviour is not good but upon introducing the $\mathcal{L}_{\CPT}^{(4),\,\mathrm{even}}$ the required falloff is recovered. This possibility
 of including the Lagrangian at the next order in the chiral expansion does not yet yield the proper ultraviolet behaviour in the odd-intrinsic parity sector
 when working with the vector field formalism.}, \footnote{We are not referring to the chiral counting in the framework of $\RCT$, where this is known to be
 lost. We recall the remark in the paragraph including Eq. (\ref{genericstructure}).}. New terms including resonance fields and higher-order derivatives are
 needed in this case in the vector-field representation, but the general procedure remains unknown.\\
\hspace*{0.5cm}This can be a serious drawback for any $EFT$ involving resonances as active fields. Ref.~\cite{RuizFemenia:2003hm} studies one class of Green
 functions analyzed in Ref.~\cite{Knecht:2001xc} in the odd-intrinsic parity sector with antisymmetric tensor formalism for the resonances. This required the
 introduction of an odd-intrinsic parity Lagrangian in the formulation of Ref.~\cite{Ecker:1988te} containing all allowed interactions between two vector objects
 (either currents or resonances) and one pseudoscalar meson. I will introduce this extension of the original Lagrangian of $\RCPT$ in the following.\\
\hspace*{0.5cm}In principle, taking into account Weinberg's power counting rule and resonance exchange among vertices with pseudoscalar legs; at $\cO(p^4)$
 in the even-intrinsic parity sector one needs to treat on the same footing $\mathcal{L}_2$ at one loop, $\mathcal{L}_4$ at tree-level and $\mathcal{L}_2(R)$.
 In Ref.~\cite{Ecker:1989yg} it was shown that at this order in the chiral counting, the Effective Lagrangian 
$\mathcal{L}_{R\chi T}\,\equiv\,\mathcal{L}_2\,+\,\mathcal{L}_R$ is enough to satisfy the high-energy $QCD$ constraints.\\
\hspace*{0.5cm}Analogously, for the odd-intrinsic parity sector, three different sources might be considered:
\begin{itemize}
\item The $WZW$ action \cite{Wess:1971yu, Witten:1983tw}, which is $\cO(p^4)$ and fulfills the chiral anomaly,
\item Chiral invariant $\varepsilon_{\mu\nu\rho\sigma}$ terms involving vector mesons which, upon integration, will start to contribute at $\cO(p^6)$ in the
 antisymmetric tensor formalism, and
\item The relevant operators in the $\cO(p^6)$ $\CPT$ Lagrangian.
\end{itemize}
\hspace*{0.5cm}The odd-intrinsic parity Lagrangians with resonances have already been studied in order to consider the equivalence for reproducing the one-loop
 divergencies of the $WZW$ action among different representations for the resonance fields \cite{Pallante:1992qe}. This procedure has also been thought to 
estimate the couplings appearing the $\cO(p^6)$ chiral Lagrangian \cite{Prades:1993ys}.\\
\hspace*{0.5cm}Within the antisymmetric tensor formalism, all the needed building blocks have already been introduced. Chiral invariance of the generating 
functional, together with Lorentz, Parity and Charge conjugation invariance and Hermiticity of the Lagrangian determine an independent set of operators for 
$VVP$ and $VJP$ Green functions to be~\footnote{The convention for the Levi-Civita density is $\varepsilon_{0123}=+1$ and $J$ is short for external vector 
current.}~\footnote{One more operator arises for the singlet $VVP$ and $VJP$ Green functions, respectively, which affect netural current processes, see Ref.\cite{Chen:2012vw}.}\\
$\bullet$ $VJP$ terms:
\begin{eqnarray} \label{VJPops}
& & \mathcal{O}_{VJP}^1\,=\,\varepsilon_{\mu\nu\rho\sigma}\,\bra \left\lbrace V^{\mu\nu},\,f_+^{\rho\alpha}\right\rbrace \nabla_\alpha u^\sigma\ket\,,\nonumber\\
& & \mathcal{O}_{VJP}^2\,=\,\varepsilon_{\mu\nu\rho\sigma}\,\bra \left\lbrace V^{\mu\alpha} ,\,f_+^{\rho\sigma}\right\rbrace \nabla_\alpha u^\nu\ket\,,\nonumber\\
& & \mathcal{O}_{VJP}^3\,=\,i\,\varepsilon_{\mu\nu\rho\sigma}\,\bra \left\lbrace V^{\mu\nu},\,f_+^{\rho\sigma}\right\rbrace \chi_-\ket\,,\nonumber\\
& & \mathcal{O}_{VJP}^4\,=\,i\,\varepsilon_{\mu\nu\rho\sigma}\,\bra V^{\mu\nu}\left[ f_-^{\rho\sigma},\,\chi_+\right] \ket\,,\nonumber\\
& & \mathcal{O}_{VJP}^5\,=\,\varepsilon_{\mu\nu\rho\sigma}\,\bra \left\lbrace \nabla_\alpha V^{\mu\nu},\,f_+^{\rho\alpha}\right\rbrace u^\sigma\ket\,,\nonumber\\
& & \mathcal{O}_{VJP}^6\,=\,\varepsilon_{\mu\nu\rho\sigma}\,\bra \left\lbrace \nabla_\alpha V^{\mu\alpha},\,f_+^{\rho\sigma}\right\rbrace u^\nu \ket\,,\nonumber\\
& & \mathcal{O}_{VJP}^7\,=\,\varepsilon_{\mu\nu\rho\sigma}\,\bra \left\lbrace\nabla^\sigma V^{\mu\nu},\,f_+^{\rho\alpha} \right\rbrace u_\alpha\ket\,.
\end{eqnarray}
\\
$\bullet$ $VVP$ terms:
\begin{eqnarray} \label{VVPops}
& & \mathcal{O}_{VVP}^1\,=\,\varepsilon_{\mu\nu\rho\sigma}\,\bra \left\lbrace V^{\mu\nu},\,V^{\rho\alpha}\right\rbrace \nabla_\alpha u^\sigma\ket\,,\nonumber\\
& & \mathcal{O}_{VVP}^2\,=\,i\,\varepsilon_{\mu\nu\rho\sigma}\,\bra \left\lbrace V^{\mu\nu},\,V^{\rho\sigma}\right\rbrace \chi_-\ket\,,\nonumber\\
& & \mathcal{O}_{VVP}^3\,=\,\varepsilon_{\mu\nu\rho\sigma}\,\bra \left\lbrace \nabla_\alpha V^{\mu\nu},\,V^{\rho\alpha}\right\rbrace u^\sigma\ket\,,\nonumber\\
& & \mathcal{O}_{VVP}^4\,=\,\varepsilon_{\mu\nu\rho\sigma}\,\bra \left\lbrace \nabla^\sigma V^{\mu\nu},\,V^{\rho\alpha}\right\rbrace u_\alpha\ket\,.
\end{eqnarray}
\hspace*{0.5cm}The Schouten identity,
\begin{equation}
g_{\rho\sigma}\varepsilon_{\alpha\beta\mu\nu}\,+\,g_{\rho\alpha}\varepsilon_{\beta\mu\nu\sigma}\,+\,g_{\rho\beta}\varepsilon_{\mu\nu\sigma\alpha}\,+\,g_{\rho\mu}\varepsilon_{\nu\sigma\alpha\beta}\,+\,g_{\rho\nu}\varepsilon_{\sigma\alpha\beta\mu}\,=\,0\,,
\end{equation}
has been employed to reduce the number of independent operators.\\
\hspace*{0.5cm}The authors of Ref. \cite{Pallante:1992qe} also built $VVP$ operators in the antisymmetric tensor formalism but applying the $LO$ $EOM$ to 
reduce the number of operators to three. This is a valid procedure provided one is only interested in on-shell degrees of freedom; but particles inside 
Green functions are not on their mass-shell. The resonance Lagrangian for the odd-intrinsic parity sector will thus be defined as
\begin{eqnarray} \label{LVodd} 
\mathcal{L}_V^{odd} & = & \mathcal{L}_{VJP}\,+\,\mathcal{L}_{VVP}\,,\nonumber\\
\mathcal{L}_{VJP}\,=\,\sum_{a=1}^7\,\frac{c_a}{M_V}\,\mathcal{O}_{VJP}^a &, &\,\,\,\,\mathcal{L}_{VVP}\,=\,\sum_{a=1}^4d_a\,\mathcal{O}_{VVP}^a\,;
\end{eqnarray}
where the octet mass, $M_V$, has been introduced in $\mathcal{L}_{VJP}$, in order to define dimensionless $c_a$ couplings. The set defined above is a complete basis for constructing vertices with only one-pseudoscalar; for a larger number of pseudoscalars 
additional operators may emerge.\\
\hspace*{0.5cm}As discussed, the $\cO(p^6)$ $\CPT$ Lagrangian in the odd-intrinsic parity sector has to be considered, as well. Two operators may contribute at LO in 1/$\N$ to the $\bra VVP\ket$ Green function:
\begin{equation} \label{Loddp6} 
\mathcal{L}_{\CPT}^{(6),\,\mathrm{odd}}\,=\,i\,\varepsilon_{\mu\nu\alpha\beta}\,\left\lbrace t_1\bra \chi_- f_+^{\mu\nu}f_+^{\alpha\beta}\ket\,-\,i\,t_2\bra \nabla_\lambda f_+^{\lambda\mu}\left\lbrace  f_+^{\alpha\beta},\,u^\nu \right\rbrace \ket \right\rbrace\,,
\end{equation}
where the $t_i$ $LECs$ are not fixed by symmetry requirements. The operators in Eq. (\ref{Loddp6}) belong both to the $EFT$ where resonances are still active 
fields and to that one where they have been integrated out. Hence in the latter case, we can split the couplings as $t_i\,=\,t_i^R\,+\,\hat{t}_i$, where $t_i^R$
 is generated by the integration of the resonances and $\hat{t}_i$ stands for the surviving $\cO(p^6)$ $\CPT$ contribution when the resonances are still 
active. 
In analogy with the even-intrinsic parity sector, the $\hat{t}_i$ are negligible compared to the resonance contributions, which means that the ${t}_i$ are generated 
completely through interaction of vectors. Accordingly, we should not include $\mathcal{L}_{\CPT}^{(6),\,\mathrm{odd}}$ in our study to avoid double counting of 
degrees of freedom. 
 Then, the relevant effective resonance 
theory will be given by:
\begin{equation}
Z_{\mathrm{\RCT}}[v,\,a,\,s,\,p]\,=\,Z_{WZW}\left[v,\,a\right]\,+\,Z_{V\chi}^{\mathrm{odd}}[v,\,a,\,s,\,p]\,,
\end{equation}
where $Z_{V\chi}^{\mathrm{odd}}$ is generated by $\mathcal{L}_\chi^2$ in (\ref{p2-u}), $\mathcal{L}_V$ in (\ref{tot}), (\ref{Rkin}), (\ref{R}) and 
$\mathcal{L}_V^{\mathrm{odd}}$ in (\ref{LVodd}).\\
\\
\hspace*{0.5cm}The $VVP$ Green function is
\begin{equation} \label{Pi_VVP}
\left( \Pi_{VVP}\right)^{(abc)}_{(\mu\nu)}(p,\,q)\,=\,\int \mathrm{d}^4x\int \mathrm{d}^4y\, e^{i(p\cdot x\,+\,q\cdot y)}\,\bra0|T\left[ V_\mu^a(x)\,V_\nu^b(y)\,P^c(0)\right] |0\ket\,.
\end{equation}
\hspace*{0.5cm}Provided a Green function is related to an order parameter of $QCD$, it vanishes in the chiral limit to all orders in perturbation theory,
 so that there is no term in the $OPE$ expansion that goes with the identity. This is specially nice regarding the matching with 
the $\RCT$ result, that will never include such a kind of term including the identity. This is the case for $\bra VVP\ket$ Green functions, and also for 
$\bra VPPP\ket$ Green functions like those related to $\t$ decays into three mesons but, for instance, it is no longer so in $\bra VVVV\ket$ Green functions
 (like, for example, light-by-light scattering processes). It is conventionally written that we can rely on matching $\RCT$ with the $OPE$ at such low energies
 as $2$ $GeV$ when there are order parameters involved. Otherwise, the matching becomes very cumbersome.\\
\hspace*{0.5cm}The $\bra VVP\ket$ Green function is built within $\RCPT$ in \cite{RuizFemenia:2003hm}. When the limit of two momenta becoming large at the 
same time is taken, one finds compatibility with the $QCD$ short-distance constraints, provided the following conditions among the 
$\mathcal{L}_V^{\mathrm{odd}}$ couplings hold
\begin{eqnarray} \label{OPErelations_for_c,d} 
& & 4c_3\,+\,c_1 \, = \, 0\,,\nonumber\\
& & c_1\,-\,c_2\,+\,c_5=0\,,\nonumber\\
& & c_5\,-\,c_6=\,\frac{N_C}{64\pi^2}\,\frac{M_V}{\sqrt{2}F_V}\,,\nonumber\\
& & d_1\,+\,8d_2=\,-\frac{N_C}{64\pi^2}\,\frac{M_V^2}{F_V^2}\,+\,\frac{F^2}{4F_V^2}\,,\nonumber\\
& & d_3\,=\,-\frac{N_C}{64\pi^2}\,\frac{M_V^2}{F_V^2}\,+\,\frac{F^2}{8F_V^2}\,.
\end{eqnarray}
 Being the couplings in the (odd-intrinsic parity) Effective Lagrangian independent of $pG$ masses the
 result turns out to be general.\\
\hspace*{0.5cm}Now it comes the crucial point. The obtained $\bra VVP\ket$ Green function reproduces the lowest meson dominance ansatz in \cite{Moussallam:1994xp}:
\begin{equation} \label{VVPansatz} 
\Pi_{VVP}(p^2,\,q^2,\,(p+q)^2)\,=\,-\frac{\bra\overline{\psi}\psi\ket_0}{2}\cdot\frac{(p^2+q^2+r^2)-\frac{N_C}{4\pi^2}\frac{M_V^4}{F^2}}{(p^2-M_V^2)(q^2-M_V^2)r^2}\,.
\end{equation}
\hspace*{0.5cm}The previous ansatz (\ref{VVPansatz}) recovers the lowest meson dominance estimates for the $LECs$ derived in Ref.~\cite{Knecht:2001xc}. Their authors found 
impossible to reproduce them working with the vector representation for the resonances, not even paying the price of introducing local contributions from 
the $\cO(p^6)$ chiral Lagrangian. They suggested that the problem could be due to the Effective Lagrangian approach and unlikely to be cured by using other
 representations for the resonance fields. The work undertaken in Ref. \cite{RuizFemenia:2003hm} contradicts this assertion for the $\bra VVP\ket$ Green 
function in the odd-intrinsic parity sector.\\
\hspace*{0.5cm}The derived Lagrangian, Eq. (\ref{LVodd}), was tested through the computation of the decay width for the process $\omega\,\to\,\pi\,\gamma$ that
 was completely predicted thanks to the relations (\ref{OPErelations_for_c,d}). This calculation pop up the question about the validity of Vector meson 
dominance-assumption \cite{Sakurai, Kaiser:2005eu}. It was found that the direct vertex was larger than expected, even comparable to the $\rho$-mediated process. Anyway, these 
results agree with the large-$\N$ limit of $\RCPT$, in which both contributions are of the same order in the expansion. This feature was confirmed through 
the computation of other channels: In particular, $\omega\to3\pi$ showed that Vector meson dominance hypothesis was at variance with the experimental value 
for the decay width. This confirmed what was suggested before: that local $VPPP$ vertices in the odd-intrinsic parity sector are relevant.\\
\\
\hspace*{0.5cm}We present here the last extension \cite{Tesina, Roig:2007yp} of the original Lagrangian that was first applied to study the $K \overline{K}\pi$ decay
 modes of the $\tau$ lepton. We have found that the most general $VPPP$ Lagrangian in the odd-intrinsic parity sector is
\begin{equation} \label{LVPPP}
{\cal L}_{VPPP} = \sum_{i=1}^5 \, \frac{g_i}{M_V} \, {\cal O}^i_{VPPP} \, ,
\end{equation}
where, in the chiral limit and using the Schouten identity, three new operators arise
\begin{eqnarray} \label{VPPP_chirallimit} 
{\cal O}_{VPPP}^1 & = & i \,\varepsilon_{\mu\nu\alpha\beta} \, \left\langle
V^{\mu\nu} \, \left( \, h^{\alpha\gamma} u_{\gamma} u^{\beta} - u^{\beta} u_{\gamma}
h^{\alpha\gamma} \right) \right\rangle \, , \nonumber\\
{\cal O}_{VPPP}^2 & = & i\, \varepsilon_{\mu\nu\alpha\beta} \, \left\langle
V^{\mu\nu} \, \left( \, h^{\alpha\gamma} u^{\beta} u_{\gamma} - u_{\gamma} u^{\beta}
h^{\alpha\gamma} \, \right) \right\rangle \, , \nonumber\\
{\cal O}_{VPPP}^3 & = & i \,\varepsilon_{\mu\nu\alpha\beta} \, \left\langle
V^{\mu\nu} \, \left( \, u_{\gamma} h^{\alpha\gamma} u^{\beta}  -  u^{\beta}
h^{\alpha\gamma} u_{\gamma} \, \right) \right\rangle \, .
\end{eqnarray}
\hspace*{0.5cm}Apart from these ones, when the chiral limit is not taken, two new operators have to be taken into account
\begin{eqnarray} \label{VPPP_chiralcontributions} 
{\cal O}_{VPPP}^4 & = &  \varepsilon_{\mu\nu\alpha\beta} \, \left\langle
\left\lbrace  \,V^{\mu\nu} \,,\,  u^{\alpha}\, u^{\beta}\, \right\rbrace \,{\cal \chi}_{-} \right\rangle \, , \nonumber\\
{\cal O}_{VPPP}^5 & = &  \varepsilon_{\mu\nu\alpha\beta} \, \left\langle
 \,u^{\alpha}\,V^{\mu\nu} \, u^{\beta}\, {\cal \chi}_{-} \right\rangle \, .
\end{eqnarray}
\hspace*{0.5cm}From the previous distinction, we can guess that matching at high energies will give us some information on the three couplings that survive
 in the chiral limit, but for the others it is likely that only phenomenological information will shed light on them.\\
\hspace*{0.5cm}Notice that we do not include analogous pieces to Eqs. (\ref{VJPops}) and (\ref{LVPPP}) with an axial-vector resonance, that 
would contribute to the hadronization of the axial-vector current. This has been thoroughly studied in Ref.~\cite{GomezDumm:2003ku} (see also 
\cite{Dumm:2009va}, this picture is supported by the conclusions in Chapter \ref{3pi}) in the description of the $\tau \rightarrow \pi \pi 
\pi \nu_{\tau}$ process and it is shown that no $\langle A \chi^{(4)}(\varphi) \rangle$ operators are needed to describe its hadronization. 
Therefore those operators are not included in our minimal description of the relevant form factors appearing in later chapters.\\
\subsection{Concluding remarks}
\hspace*{0.5cm}There are other extensions of the Lagrangian which will not be used in this Thesis. For the even-intrinsic parity sector, the reader can find 
in Ref.~\cite{Cirigliano:2006hb} the minimal set of operators corresponding to the coupling of a resonance and an $\cO(p^4)$ chiral tensor, 
and trilinear resonance terms without any chiral tensor. For the odd-intrinsic parity part, in Ref.~\cite{Mateu:2007tr}
 there were written new pieces for the couplings among the two vector objects and a $pG$ or a vector source: $\mathcal{O}_{V_1V_2P}$ and $\mathcal{O}_{V_1V_2J}$.\\
\\
\hspace*{0.5cm}All the introduced extensions of the original Lagrangian will play a r\^ole in the three meson decays and one meson radiative decays of the $\t$
 examined in this Thesis. As we will see, the contribution of $VPPP$ vertices in the odd-intrinsic parity sector turns out to be fundamental for the decay 
$\omega\to3\pi$ (see Appendix D). Therefore, apart from its own interest, we can take advantage of it to get additional restrictions on the new couplings introduced throughout
 this section. One of the targets of our work is to gain more control over the new couplings of the resonance Lagrangian which we have just introduced and thus 
improve our quantitative understanding of intermediate-energy meson dynamics.\\
\chapter{Hadronic decays of the $\tau$ lepton}\label{Hadrondecays}
\section{Introduction} \label{Hadrondecays_Intro}
\hspace*{0.5cm}In this chapter we want to set the model independent description of the hadronic decays of the $\tau$ which we study. Using Lorentz invariance 
and general properties of $QCD$ one can decompose any amplitude participating in a given process in terms of a set of scalar quantities that only depend 
on kinematical invariants, the so-called form factors. As explained in Appendix A, this description is equivalent to that in terms of 
structure functions.\\
\hspace*{0.5cm}Moreover, we also desire to explain the three different approaches to describe the 
the involved
form factors and to illustrate
 others than ours. This will let appreciate better the improvements introduced by our study compared to previous approaches.\\
\hspace*{0.5cm}As we have discussed in previous chapters we still do not know a way to derive the form factors related to the hadronization of $QCD$ 
currents in the intermediate energy region. In view of this, three major approaches have been developed to tackle these problems \cite{Portoles:2004vr}:
\begin{itemize}
 \item The first approach is the one motivated by the discussion in the previous chapters and followed in this Thesis. It consists in exploiting the power 
of the $EFT$ framework \`a la Weinberg -Section \ref{EFT_Validity}- supplemented by some dynamical content of the problem at hand, namely the known short-distance 
behaviour of $QCD$ -Section \ref{LargeN_MatchingRChTQCD}- and the large-$\N$ limit of $QCD$ -Chapter \ref{RChT}-.
 \item The second approach is that of modeling phenomenological Lagrangians. Their actions are written in terms of hadron fields but employing ad-hoc assumptions
 whose link with $QCD$ is not clear and which are introduced in order to get a simpler theory. As an example, we have the suggested Hidden Symmetry or Gauge
 Symmetry Lagrangians mentioned briefly in appendix F.
 \item Finally, we have another -more comfortable though less satisfactory- way of facing the problem. It is to propose dynamically driven parametrizations.
 They provide an amplitude suggested by the assumed dynamics: resonance dominance, polology, etc. The expressions one obtains are much easier than those given
 by more based approaches, as we will see giving some examples later on. Numerical fit to data is quicker and the accordance between the theoretical expressions
 and the experiments is, often, remarkable. Notwithstanding, the connection between their ad-hoc parameters and $QCD$ is missing. According to our understanding, 
the point is not just to get an impressive fit to the experimental points but to understand the hadronization of the $QCD$ currents in these particular 
processes.\\
\end{itemize}
\subsection{Breit-Wigner approach}
\hspace*{0.5cm}As we want to compare our results to those obtained within the Breit-Wigner models, we describe their main features in this section.\\
\hspace*{0.5cm}Since long time ago, it is well-known that any hadronization process occurring in the resonance energy region will be dominated by the contribution of these 
resonance states. The application of this resonance dominance to hadronic tau decays has a long history \cite{Fischer:1979fh, kuhn:1982di, Pich:1987qq, Pich:1989pq}. 
A model based on these ideas that became very popular is due to K\"uhn and Santamar\'{\i}a \cite{Kuhn:1990ad}.\\
\hspace*{0.5cm}In any of these cases the parametrization is accomplished by combining Breit-Wigner factors ($BW_R(Q^2)$) according to the expected resonance
 dominance in each channel \footnote{Consequently, they do not depend only on $Q^2$, the total hadronic momenta, but also on other Lorentz invariants depending
 on the considered channel. The $Q^2$ in parenthesis intends to be a shorthand notation we will keep in the following.}, that is,
\begin{equation}
F(Q^2)\,=\,\mathcal{N}\,\,f\left(\alpha_i,\,\mathrm{BW}_{\mathrm{R_i}}(Q^2)\right)\,,
\end{equation}
where $\mathcal{N}$ is a normalization and the former expression is not linear, in general, in the Breit-Wigner terms. Data are analyzed by fitting the $\alpha_i$
 parameters and the masses and on-shell widths entering the Breit-Wigner factors. Two main models of parametrizations have been employed:\\
\\
a) K\"uhn-Santamar\'{\i}a Model (KS)\\
\hspace*{0.5cm}The $BW$ form factors are given by \cite{Fischer:1979fh, kuhn:1982di, Pich:1987qq, Pich:1989pq, Kuhn:1990ad}
\begin{equation}
\mathrm{BW}_{\mathrm{R_i}}^{\mathrm{KS}}(Q^2)\,=\,\frac{M_{\mathrm{R_i}}^2}{M_{\mathrm{R_i}}^2-Q^2-i\sqrt{Q^2}\Gamma_{\mathrm{R_i}}(Q^2)}\,,
\end{equation}
that vanishes in the high-$Q^2$ region, as demanded by short-distance $QCD$.\\
\\
b) Gounaris-Sakurai Model (GS)\\
\hspace*{0.5cm}It was originally developed to study the r\^ole of the $\rho(770)$ resonance in the vector form factor of the pion \cite{Gounaris:1968mw} still
 in the current algebra era. It has been applied over the years to other hadronic resonances \cite{Barate:1997hv, Anderson:1999ui, Athanas:1999xf} 
by the experimental collaborations. The $BW$ function now reads\\
\begin{equation}
\mathrm{BW}_{\mathrm{R_i}}^{\mathrm{GS}}(Q^2)\,=\,\frac{M_{\mathrm{R_i}}^2+f_{\mathrm{R_i}}(0)}{M_{\mathrm{R_i}}^2-Q^2+f_{\mathrm{R_i}}(Q^2)-i\sqrt{Q^2}\Gamma_{\mathrm{R_i}}(Q^2)}\,,
\end{equation}
where $f_{\mathrm{R_i}}(Q^2)$ encodes information of the off-shell behaviour of the considered resonance. For the particular case of the $\rho(770)$, it can
 be read off from Ref.~\cite{Gounaris:1968mw}.\\
\hspace*{0.5cm}In both models the normalization is fixed in order to reproduce the $\CPT$ $\cO(p^2)$ behaviour at $Q^2<<M_\rho^2$. The experimental groups use
 to believe that the difference among the predictions of these two models gives an estimate of the theoretical error \footnote{In fact, Ref.~\cite{Kuhn:1990ad}
already used it with this purpose.}. As we will see, the simplicity of these models is irrelevant if they fail to verify 
properties coming from $QCD$ itself, as it happens to be the case, both in the three meson modes and the radiative decays with one meson. It is true that we
 learn things about the resonance structure using these models to fit the data, but it is not -as argued many times- that the (occasionally) little discrepancy
 among themselves in the observables (values of masses, on-shell widths and branching ratios and shape of the spectra and Dalitz plots) can be regarded as a
 proof of the rightness of both models.\\
\subsection{Model independent description. General case}
\hspace*{0.5cm}Within the Standard Model \footnote{A nice and short introductory description can be found in Refs. \cite{Pich:1995ft, Pich:2007vu}.} the matrix amplitude for the exclusive hadronic decays of the $\t$, $\t^-\,\to\,H^-\nu_\t$,
 is generically given by
\begin{equation}
\mathcal{M}\,=\,\frac{G_F}{\sqrt{2}}V_{CKM}^{ij}\,\overline{u}_{\nu_\t}\,\gamma^\mu\,(1-\gamma_5)\,u_\t\,\mathcal{H}_\mu\,,
\end{equation}
where $G_F$ is the Fermi constant, $V_{CKM}^{ij}$ the corresponding element of the $CKM$ matrix, and
\begin{equation} \label{defhadcurr}
\mathcal{H}_\mu\,=\,\bra H|(V_\mu-A_\mu)\,e^{i\mathcal{L}_{\mathrm{QCD}}}|0\ket\,,
\end{equation}
is the matrix element of the left-handed current that has to be evaluated in the presence of the strong $QCD$ interactions.\\
\hspace*{0.5cm}Symmetries help us to decompose $\mathcal{H}_\mu$ in terms of the allowed Lorentz structures of implied momenta and a set of scalar functions of kinematical
 invariants, the hadronic form factors, $F_i$, of $QCD$ currents,\\
\begin{equation}
\mathcal{H}_\mu\,=\,\sum_i\underbrace{(\dots)^i_\mu}_{\mathrm{Lorentz\,\,structure}}\, F_i(Q^2,\dots)\,;
\end{equation}
that are universal in the sense of being independent on the initial state, describing therefore the hadronization of $QCD$ currents.\\
\hspace*{0.5cm}This decomposition is studied in detail in Appendix A, where it is also derived the equivalence among using form factors
 or structure functions to describe these hadronic decays. We just recall for the moment that the decay width for a given channel is obtained by 
integrating over the $n$ hadrons plus one neutrino differential phase space, the hadron and lepton tensors are defined from the following~\footnote{$\overline{\sum}_{s,s'}$ corresponds to the averaged sum over polarizations.}
\begin{equation}
\overline{\sum}_{s,s'}\,\mathcal{M}\mathcal{M}^\dagger\equiv\left( \frac{1}{2\times\frac{1}{2}+1}\right)^2\,\frac{G_F^2}{2}\,|V_{CKM}|^2\,\mathcal{H}_{\mu\nu}\,\mathcal{L}^{\mu\nu}\,,
\end{equation}
where $\mathcal{H}_\mu$ is the hadron current defined in Eq.~(\ref{defhadcurr}), $\mathcal{L}^{\mu\nu}$ carries information on the lepton sector and 
$\mathrm{d}\mathcal{PS}$ stands for the differential element for phase space integration:
\begin{equation}
 \mathcal{H}_{\mu\nu}\equiv\,\mathcal{H}_\mu \mathcal{H}_\nu\;,\;\;\mathcal{L}^{\mu\nu}\,=\,\sum_{s,s'} \overline{u}_\tau(\ell,s)\gamma^\mu(1-\gamma_5)u_{\nu_\tau}(\ell',s')\gamma^\nu(1-\gamma_5)u_\tau(\ell,s)\,.
\end{equation}
Then, one has
\begin{equation}
\mathrm{d}\Gamma\,=\,\frac{G_F^2}{4M_\t}\,|V_{CKM}^{ij}|^2\,\mathcal{L}_{\mu\nu}\,\mathcal{H}^\mu\,\mathcal{H}^{\nu\dagger}\,\mathrm{d}\mathcal{PS}\,,
\end{equation}
with
\begin{equation}
\mathcal{L}_{\mu\nu}\,\mathcal{H}^\mu\,\mathcal{H}^{\nu\dagger}\,=\,\sum_X \,L_X\,W_X\,,
\end{equation}
where $W_X$ are the structure functions defined in the hadron rest frame.\\
\hspace*{0.5cm}
The hadron structure functions, $W_X$, can be written in terms of the form factors and kinematical components and
 the study of spectral functions or angular distributions of data allow us to reconstruct them. Their number depends on the final state, being $4$ in the case 
of two mesons and $16$ for three. Either form factors or, equivalently, structure functions, are the target to achieve.\\
\section{One meson radiative decays of the $\t$} \label{Hadrondecays_One_meson_decays}
\subsection{Model independent description} \label{Hadrondecays_One_meson_decays_MI}
\begin{figure}[h!]
\centering
\includegraphics[scale=0.6]{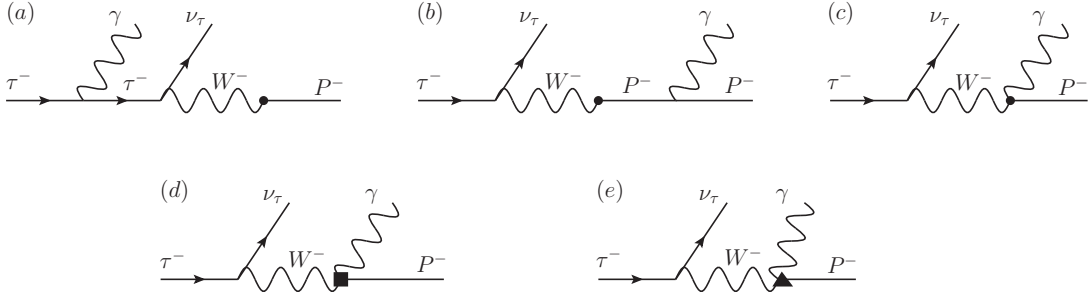}
\caption{Feynman diagrams for the different kinds of contributions to the radiative decays of the tau including one meson, as explained in the main text. The
 dot indicates the hadronization of the $QCD$ currents. The solid square represents the $SD$ contribution mediated by the axial-vector current and the solid
 triangle the $SD$ contribution via the vector current.}
\label{diagrams general decomposition amplitude in radiative decays tau with one meson}
\end{figure}

\hspace*{0.5cm}For the decay modes with lowest multiplicity, $\t\to P^-\nu:\t$, $P=\pi,K$, the relevant matrix elements are already known from the measured decays $\pi^-\to\mu^-\nu_\mu$ 
and $K^-\to\mu^-\nu_\mu$. The corresponding $\tau$ decay widths can then be predicted rather accurately. These predictions are in good agreement with the 
measured values, and provide a quite precise test of charged current universality.\\
\hspace*{0.5cm}When one considers the emission of a photon things change and they provide dynamical information \cite{Decker:1993ut} about the hadronic matrix 
elements of the $L_\mu= \left( V_\mu-A_\mu\right) $ current.\\
\hspace*{0.5cm}The process is $\tau^-(p_\t) \to \nu_\tau(q) P^-(p) \gamma(k)\,$. The kinematics of this decay is equivalent to that of the radiative pion decay
 \cite{Brown:1964zza}. We will use $t := (p_\t - q)^2 = (k + p)^2$. In complete analogy to the case of the radiative pion decay \cite{Bae67}, the matrix element
 for the decay $\tau^- \to P^- \gamma \nu_\t$ can be written as the sum of four contributions:
\begin{equation}
   \mathcal{M} \left[\tau^-(p_\t) \to \nu_\tau(q) P^-(p) \gamma(k)\right]
   = \mathcal{M}_{IB_\tau} + \mathcal{M}_{IB_P} + \mathcal{M}_{V} + \mathcal{M}_{A}\,,
\end{equation}
with \footnote{Notice that $i$ and minus factors differ with respect to Ref.~\cite{Decker:1993ut} ($DF$). Moreover, our form factors have dimension of inverse mass 
while theirs are dimensionless. In their work, this factor of $(\sqrt{2}m_\pi)^{-1}$ in the form factors is compensated by defining the sum over polarizations 
of the matrix element squared with an extra $2 m_\pi^2$ factor. This should be taken into account to compare formulae in both works using that 
$F_V(t)^{DF}\,=\,\sqrt{2}m_\pi F_V(t)^{Our}\,,\;\;F_A(t)^{DF}\,=\,2\sqrt{2}m_\pi F_A(t)^{Our}$.}
\begin{eqnarray} \label{Gral_IB}
  i \mathcal{M}_{IB_{\tau}} & = &  G_F\, V_{CKM}^{ij}\, e\, F_P\, p_\mu\,
\epsilon_\nu(k)\, L^{\mu \nu}\,,
   \nonumber \\
  i \mathcal{M}_{IB_{P}} & = & \, G_F\, V_{CKM}^{ij}\, e\, F_P \,\,\epsilon^\nu(k)\,
\left( \frac{2\,p_\nu \, (k\, +\, p)_\mu}
      {m_P^2\, -\, t} \,+\, g_{\mu\nu} \right)\, L^\mu\,, \nonumber \\
  i \mathcal{M}_V & = &  i \,G_F \,V_{CKM}^{ij}\, e\, F_V(t)\, \epsilon_{\mu \nu \rho \sigma}
     \epsilon^\nu(k)\,  k^\rho\, p^\sigma\,  L^\mu\,,  \nonumber \\
  i \mathcal{M}_{A} & = & G_F\, V_{CKM}^{ij}\, e\, F_A(t)\, \epsilon^\nu(k)\, \left[ (t\,-\,2\,m_P^2) \,g_{\mu\nu}\,-\,2\,p_\nu k_\mu \right] L^\mu\,,
\end{eqnarray}
where $e$ is the electric charge and $\epsilon_\nu$ is the polarization vector of the photon. $F_V(t)$ and $F_A(t)$ are the so called structure dependent form 
factors. Finally $L^\mu$ and $L^{\mu\nu}$ are lepton currents defined by
\begin{eqnarray}
   L^\mu & = & \bar{u}_{\nu_\t}(q) \gamma^\mu (1-\gamma_5) u_\t(p_\t)\,, \nonumber \\
   L^{\mu \nu} & = & \bar{u}_{\nu_\t}(q) \gamma^\mu (1-\gamma_5)
\frac{\strich{k} - \strich{p}_\t
      - M_\t}{(k - p_\t)^2 - M_\t^2} \gamma^\nu u_\t(p_\t)\,.
\end{eqnarray}
The notation introduced for the independent amplitudes describes the four kinds of contributions: $\mathcal{M}_{IB_{\tau}}$ is the bremsstrahlung off the tau, 
(Figure \ref{diagrams general decomposition amplitude in radiative decays tau with one meson}(a)), $\mathcal{M}_{IB_{P}}$ is the sum of the $pG$ bremsstrahlung
 (Figure \ref{diagrams general decomposition amplitude in radiative decays tau with one meson}(b)), and the seagull diagram 
(Figure \ref{diagrams general decomposition amplitude in radiative decays tau with one meson}(c)), $\mathcal{M}_{V}$ is the structure dependent vector (Figure 
\ref{diagrams general decomposition amplitude in radiative decays tau with one meson}(d)) and $\mathcal{M}_{A}$ the structure dependent axial-vector 
contribution (Figure \ref{diagrams general decomposition amplitude in radiative decays tau with one meson}(e)). Our ignorance of the exact mechanism of 
hadronization is parametrized in terms of the two form factors $F_A(t)$ and $F_V$(t). In fact, these form factors are the same functions of the momentum 
transfer $t$ as those in the radiative pion decay, the only difference being that $t$ now varies from $0$ up to $M_\tau^2$ rather than just up to $m_\pi^2$.\\
\hspace*{0.5cm}The two matrix elements $\mathcal{M}_{IB_{\tau}}$ and $\mathcal{M}_{IB_{P}}$ are not separately gauge invariant, but their sum, ie. the (total)
matrix element for internal bremsstrahlung $IB$
\begin{equation}
   \mathcal{M}_{IB} = \mathcal{M}_{IB_{\tau}} + \mathcal{M}_{IB_{P}}\,,
\end{equation}
is indeed gauge invariant, as $\mathcal{M}_V$ and $\mathcal{M}_A$ are. We also define the (total) structure dependent radiation $SD$ by
\begin{equation}
   \mathcal{M}_{SD} = \mathcal{M}_{V} + \mathcal{M}_{A}\,.
\end{equation}
\hspace*{0.5cm}The spinor structure can be rearranged to give
\begin{eqnarray}
  i \mathcal{M}_{IB} & = & G_F\, V_{ij}^{CKM}\, e\, F_P\, M_\t \bar{u}_{\nu_\t}(q) (1\,+\,\gamma_5) \left[\frac{p_\t \cdot \epsilon}{p_\t \cdot k} - \frac{p \cdot \epsilon}
{p \cdot k} - \frac{\strich{k}\strich{\epsilon}}{2 p_\t \cdot k} \right] u_\t(p_\t)\,,  \\
  i \mathcal{M}_{SD} & = & G_F\, V_{ij}^{CKM}\, e \left\{ i \epsilon_{\mu \nu \rho \sigma} L^\mu \epsilon^\nu k^\rho p^\sigma F_V(t)
      + \bar{u}_{\nu_\t}(q) (1\,+\,\gamma_5) \left[ (t\,-\,m_P^2) \strich{\epsilon} - 2 (\epsilon \cdot p) \strich{k} \right] u(p_\t) F_A(t) \right\} \,. \nonumber
\end{eqnarray}
The square of the matrix element is then given by
\begin{equation}
  \overline{| \mathcal{M} |^2} = \overline{| \mathcal{M}_{IB} |^2} + 2 \overline{\Re e (\mathcal{M}_{IB} \mathcal{M}_{SD}^\star)} + \overline{| \mathcal{M}_{SD} |^2}\,,
\end{equation}
where the bar denotes summing over the photon polarization and neutrino spin and averaging over the tau spin.\\
\hspace*{0.5cm}We follow Ref.~\cite{Decker:1993ut} and divide the decay rate as follows: the internal bremsstrahlung part $\Gamma_{IB}$ arising from 
$\overline{| \mathcal{M}_{IB} |^2}$, the structure dependent part $\Gamma_{SD}$ coming from $\overline{| \mathcal{M}_{SD} |^2}$, and the interference part 
$\Gamma_{INT}$ stemming from $2 \overline{\Re e (\mathcal{M}_{IB} \mathcal{M}_{SD}^\star)}$. Furthermore $\Gamma_{SD}$ is subdivided into the vector-vector 
($\Gamma_{VV}$), the axial-vector---axial-vector ($\Gamma_{AA}$) and the vector---axial-vector interference term $\Gamma_{VA}$. Similarly $\Gamma_{INT}$ gets 
splitted into the internal bremsstrahlung-vector interference $\Gamma_{IB-V}$ and the internal bremsstrahlung--axial-vector interference $\Gamma_{IB-A}$ parts.
 Thus, one has
\begin{eqnarray} \label{parts Gamma radiative decay tau one pG}
    \Gamma_{total} & = & \Gamma_{IB} + \Gamma_{SD} + \Gamma_{INT}\,, \nonumber \\
    \Gamma_{SD} & = & \Gamma_{VV} + \Gamma_{VA} + \Gamma_{AA}\,, \nonumber \\
    \Gamma_{INT} & = & \Gamma_{IB-V} + \Gamma_{IB-A}\,.
\end{eqnarray}
\hspace*{0.5cm}It is convenient to introduce the dimensionless variables $x$ and $y$:
\begin{equation}
 x := \frac{2 p_\tau \cdot k}{M_\t^2}\,,\quad \quad y := \frac{2 p_\tau \cdot p}{M_\t^2}\,.
\end{equation}
In the tau rest frame $x$ and $y$ are the energies $E_\gamma$ and $E_\pi$ of the photon and the pion, respectively, expressed in units of $M_\t/2$:
\begin{equation} \label{x,y_defs}
   E_\gamma = \frac{M_\t}{2} x \,,\quad \quad E_\pi    = \frac{M_\t}{2} y\,.
\end{equation}
 Eq. (\ref{x,y_defs}) sets the scale for the photons to be considered as ''hard'' or ''soft''. This means that the formulae for internal bremsstrahlung should be 
similar for radiative tau and pion decay, once they are expressed in terms of $x$ and $y$, as it is the case, albeit photons of comparable softness will 
have very different energies in both cases.\\
\hspace*{0.5cm}The kinematical boundaries for $x$ and $y$ are given by
\begin{eqnarray}
   0 \leq x \leq & 1 - r_P^2\,,\quad \quad 1 - x + \frac{r_P^2}{1 -x} \leq y \leq 1 + r_P^2\,,
\end{eqnarray}
where
\begin{equation}
 r_P^2:= \left( \frac{m_P}{M_\t} \right)^2 \sim ^{0\mathrm{.}006}_{0\mathrm{.}077} \ll 1\,,
\end{equation}
where the upper figure corresponds to $P=\pi$ and the lower one to $P=K$. It is also useful to note that
\begin{equation}
   p \cdot k = \frac{M_\t^2}{2} (x + y - 1 - r_P^2)\,,\quad\quad   t := (p_\t - q)^2 = (k + p)^2 = M_\t^2 (x + y - 1)\,.
\end{equation}
The differential decay rate is given by \cite{PartKin}
\begin{equation}
  d\Gamma(\tau^- \rightarrow \nu_\tau P^- \gamma) =
  \frac{1}{512 \pi^5 E_{\tau}} \delta^{(4)} (k + p + q - p_\t)
  \overline{| \mathcal{M} |^2} \frac{d^3 \vec{k} d^3 \vec{p} d^3 \vec{q}}{E_{\gamma} E_{\pi} E_{\nu}}\,,
\end{equation}
where the bar over the matrix element denotes summing over the photon polarization and neutrino spin and averaging over the tau spin. Choice of the tau rest 
frame, integration over the neutrino momentum, $\vec{p}$, and the remaining angles and introduction of $x$ and $y$ yields
\begin{equation}
   \frac{d^2 \Gamma}{dx\, dy} = \frac{m_\t}{256 \pi^3} \overline{| \mathcal{M} |^2}\,.
\end{equation}
The integration over $y$ gives the photon spectrum
\begin{equation}
    \frac{\mathrm{d}\Gamma}{\mathrm{d}x} = \int_{1 - x + \frac{r_P^2}{1-x}}^{1 + r_P^2} \mathrm{d}y \, \frac{\mathrm{d}^2 \Gamma}{\mathrm{d}x\, \mathrm{d}y}\,.
\end{equation}
Because of the infrared divergence of the internal bremsstrahlung a low-energy cut must be introduced for the photon energy, by requiring $x \geq x_0$ one
obtaines the integrated decay rate
\begin{equation}
   \Gamma (x_0) = \Gamma (E_0) = \int_{x_{0}}^{1 - r_P^{2}} \mathrm{d}x \frac{\mathrm{d} \Gamma}{\mathrm{d}x}\,,
\end{equation}
that does depend on the photon energy cut-off ($E_0 = \frac{M_\t}{2} x_0$). Instead of $x$ and $y$ one can use $x$ and $z$, where $z$ is the scaled momentum
 transfer squared:
\begin{equation}
   z = \frac{t}{M_\t^2} = x + y - 1\,,
\end{equation}
whose kinematical boundaries are
\begin{equation}
   z - r_P^2 \leq x  \leq 1 - \frac{r_P^2}{z}\,,\quad \quad r_P^2 \leq z \leq 1\,.
\end{equation}
Integration of $\frac{\mathrm{d}^2 \Gamma}{\mathrm{d}x\, \mathrm{d}y}$ over $x$ yields the spectrum in $z$, i. e. the spectrum in the invariant mass of
 the $P$-photon system:
\begin{equation}
   \frac{\mathrm{d}\Gamma}{dz} (z) = \frac{\mathrm{d}\Gamma}{\mathrm{d}z} \left(\sqrt{t}\right) = \int_{z - r_P^2}^{1 - r_P^2/z} \mathrm{d}x\frac{\mathrm{d}^2 \Gamma}{\mathrm{d}x\, \mathrm{d}y} (x, y=z-x+1)\,.
\end{equation}
The integrated rate for events with $t \geq t_0$ is then given by
\begin{equation}
   \Gamma(z_0) = \Gamma\left(\sqrt{t_0}\right) = \int_{z_0}^{1} \mathrm{d}z \frac{\mathrm{d}\Gamma}{\mathrm{d}z} (z)\,.
\end{equation}
We note that $z_0$ is both an infrared and a collinear cut-off.\\
\hspace*{0.5cm}In terms of the quantities defined in Eq.~(\ref{parts Gamma radiative decay tau one pG}) the differential decay rate is
\begin{eqnarray}
   \frac{\mathrm{d}^2\Gamma_{IB}}{\mathrm{d}x \, \mathrm{d}y} & = & \frac{\alpha}{2 \pi} f_{IB}\left(x,y,r_P^2\right) \frac{\gnr}{\left(1-r_P^2\right)^2}\,, \\
   \frac{\mathrm{d}^2\Gamma_{SD}}{\mathrm{d}x \, \mathrm{d}y} & = & \frac{\alpha}{8 \pi} \frac{M_\t^4}{F_P^2} \left[ |F_V(t)|^2 f_{VV}\left(x,y,r_P^2\right) +
 4 \Re e(F_V(t) F_A^\star(t))f_{VA}\left(x,y,r_P^2\right) +\right.\nonumber\\
& & \left. 4 |F_A(t)|^2 f_{AA}(x,y,r_P^2) \right] \frac{\gnr}{(1-r_P^2)^2}\,, \nonumber \\
   \frac{\mathrm{d}^2\Gamma_{INT}}{\mathrm{d}x \, \mathrm{d}y} & = &\frac{\alpha}{2 \pi}\frac{M_\t^2}{F_P}\left[ f_{IB-V}\left(x,y,r_P^2\right) \Re e(F_V(t))
 + 2 f_{IB-A}\left(x,y,r_P^2\right) \Re e(F_A(t)) \right] \frac{\gnr}{\left(1-r_P^2\right)^2}\,,\nonumber
\end{eqnarray}
where
\begin{eqnarray}
   f_{IB} \left(x,y,r_P^2\right) & = & \frac{[r_P^4 (x + 2) - 2 r_P^2 (x + y) + (x + y - 1)\left(2 - 3x + x^2 + xy\right)]\left(r_P^2 - y + 1\right)}{\left(r_P^2 - x - y +1\right)^2 x^2} \,,\nonumber \\
   f_{VV} \left(x,y,r_P^2\right) & = & - [r_P^4 (x + y) + 2 r_P^2 (1 - y) (x + y) + (x + y - 1)\left(-x + x^2 - y + y^2\right)]\,,\nonumber \\
   f_{AA} \left(x,y,r_P^2\right) & = & f_{VV}\left(x,y,r_P^2\right)\,,\nonumber \\
   f_{VA}\left(x,y,r_P^2\right) & = & - [r_P^2 (x + y) + (1 - x - y)(y-x)] \left(r_P^2 - x - y + 1\right) \,, \nonumber \\
   f_{IB-V}\left(x,y,r_P^2\right) & = & - \frac{\left(r_P^2 - x - y + 1\right)\left(r_P^2 - y + 1\right)}{x} \,,\nonumber \\
   f_{IB-A}\left(x,y,r_P^2\right) & = & - \frac{[r_P^4 - 2 r_P^2(x + y) + (1 - x + y) (x + y - 1)]\left(r_P^2 - y + 1\right)}{\left(r_P^2 - x - y + 1\right) x}\,.
\end{eqnarray}
In the approximation $r_P^2 \approx 0$ (vanishing pion mass) these formulae simplify to
\begin{eqnarray}
   \label{eqnib2}
   f_{IB} (x, y, 0) & = &  \frac{[ 1+(1-x)^2 -x(1-y)](1-y)}{(x + y - 1) x^2}\,,\nonumber \\
   f_{VV} (x,y,0) & = & \left(x - x^2 + y - y^2\right)(x + y - 1)\,,\nonumber \\
   f_{VA} (x,y,0) & = & (x + y - 1)^2 (x - y)\,,\nonumber \\
   f_{IB-V} (x,y,0) & = & \frac{(x + y - 1)(1 - y)}{x}\,,\nonumber \\
   f_{IB-A} (x,y,0) & = & \frac{(x - y - 1)(1 - y)}{x}\,.
\end{eqnarray}
We note that the radiative decay rate has been expressed in terms of the rate of the non-radiative decay ($\tau^- \to \nu_\tau P^-$):
\begin{equation} \label{definition Gamma non radiative}
   \gnr = \frac{G_F^2  |V_{CKM}^{ij}|^2 F_P^2}{8 \pi} M_\t^3 (1 - r_P^2)^2\,.
\end{equation}
\hspace*{0.5cm}We finish this section by presenting the analytical expressions for the invariant mass spectrum:
\begin{eqnarray}
   \frac{\mathrm{d}\Gamma_{IB}}{\mathrm{d}z} & = & \frac{\alpha}{2 \pi} \left[ r_P^4 (1 -z) + 2 r_P^2 \left(z - z^2\right)
   - 4z + 5 z^2 - z^3 + \right.\nonumber \\
   &  & \left. + \left(r_P^4 z + 2 r_P^2 z - 2z -2z^2 + z^3\right) \mathrm{ln} z \right] \frac{1}{z^2 - r_P^2 z} \frac{\gnr}{\left(1 - r_P^2\right)^2}\,,\nonumber \\
   \frac{\mathrm{d}\Gamma_{VV}}{\mathrm{d}z} & = & \frac{\alpha}{24 \pi} \frac{M_\t^4}{F_P^2} \frac{(z - 1)^2 \left(z - r_P^2\right)^3 (1 + 2z)}{z^2} |F_V(t)|^2 
\frac{\gnr}{(1 - r_P^2)^2} \,,\nonumber \\
   \frac{\mathrm{d}\Gamma_{VA}}{\mathrm{d}z} & = & 0 \,,\nonumber \\
   \frac{\mathrm{d}\Gamma_{AA}}{\mathrm{d}z} & = &  \frac{\alpha}{6 \pi} \frac{M_\t^4}{F_P^2} \frac{(z - 1)^2 \left(z - r_P^2\right)^3 (1 + 2z)}{z^2} |F_A(t)|^2
   \frac{\gnr}{\left(1 - r_P^2\right)^2}\,, \nonumber \\
   \frac{\mathrm{d}\Gamma_{IB-V}}{\mathrm{d}z} & = & \frac{\alpha}{2 \pi} \frac{M_\t^2}{F_P} \frac{(z-r_P^2)^2 (1 -z  + z \ln z)}{z} \Re e(F_V(t))
   \frac{\gnr}{\left(1 - r_P^2\right)^2}\,, \nonumber \\
   \frac{\mathrm{d}\Gamma_{IB-A}}{\mathrm{d}z} & = & -\frac{\alpha}{\pi} \frac{M_\t^2}{F_P} \left[ r_P^2 (1 - z) - 1 -z + 2 z^2 + \right. \nonumber \\
   & & \left. + \left(r_P^2 z - 2z - z^2\right) \ln z \right]\frac{z-r_P^2}{z} \Re e(F_A(t)) \frac{\gnr}{\left(1 - r_P^2\right)^2}\,.
\end{eqnarray}
The interference terms $IB-V$ and $IB-A$ are now finite in the limit $z \to r_P^2$, which proves that their infrared divergencies are integrable.\\
\hspace*{0.5cm}Although the above formulae have been noted in Ref.\cite{Decker:1993ut}, we independently calculate them
\footnote{Typoes in Refs.~\cite{Decker:1993ut, KimResnick, Geng:2003th} have been detected through our calculation. The minus sign 
difference in the definition of the $IB$ part has been taken into account.}
and explicitly give them here for completeness. Moreover we would like to point out that due to the fact that our definitions of the form-factors
 $F_V(t)$ and $F_A(t)$ differ from the ones given in Ref.~\cite{Decker:1993ut}, as we have mentioned before, there are some subtle
 differences in the above formulae between ours and theirs.\\
\subsection{Breit-Wigner models}\label{Hadrondecays_One_meson_decays_BW}
\hspace*{0.5cm}These processes were studied by Decker and Finkemeier in a series of papers \cite{Decker:1993ut, Decker:1993py, Decker:1994ea, Decker:1994dd}.
 Their parametrization respected the chiral limit ($t=0$) for the vector form factor, as given by the Wess-Zumino term. However, for the axial-vector form 
factor it was fixed to the value of $F_A(t=0)$ in the radiative decay of the pion. This way, not only the value at threshold but also the low-$t$-dependence
 of the amplitude deviates from the $QCD$ prediction, which is not satisfactory. Moreover, the off-shell widths employed for the vector resonances were just 
phase-space motivated, while the one for the axial-vector resonance a$_1$ employed the ad-hoc expression in the $KS$ model. High-energy $QCD$ behaviour of the
 form factors was properly implemented. Finally, the addendum to Ref.~\cite{Decker:1993ut} change the relative sign between the $IB$ and $SD$ contributions, 
and this has not been confirmed by any later independent study, so this is another motivation for our work.\\
\hspace*{0.5cm}Our study, included in the chapter \ref{Pgamma} intends to go beyond these approximations and provide a more based description of these decays. They 
are still undetected, a feature that makes them more interesting, as it is strange according to the estimations of the decay width of these processes. For reference, the 
decay modes reported by the $PDG$ \cite{Amsler:2008zzb} are reviewed in Table \ref{Listdecaystau}.\\
\begin{table}[h!]  
\begin{center}
\renewcommand{\arraystretch}{1.2}
\begin{tabular}{|c|c|c||c|c|c|}
\hline
$n$ & Decay mode & BR($\%$) & $n$ & Decay mode & BR($\%$)\\
\hline
$0$ & $e^-\overline{\nu}_e$ & $17$.$85(5)$ & & $\eta \overline{K}^0 \pi^-$ & $2$.$2(7)\cdot10^{-2}$\\
    & $\mu^-\overline{\nu}_\mu$ & $17$.$36(5)$ & & $\eta K^-\pi^0$ & $1$.$8(9)\cdot10^{-2}$\\
    & $ e^- e^- e^+ \overline{\nu}_e$ & $2$.$8(1$.$5)\cdot10^{-3}$ & & $\pi^- 2\eta$ & $<1$.$1\cdot10^{-2}$ $^a$\\
    & $ \mu^- e^- e^+ \overline{\nu}_\mu$ & $< 3$.$6 \cdot10^{-3}$ $^b$ & & $K^+ 2K^-$ & $1$.$58(18)\cdot10^{-3}$ $^c$\\
$1$ & $\pi^-$ & $10$.$91(7)$ & & $\eta' \pi^- \pi^0$& $<8$.$0\cdot10^{-3}$ $^b$\\
      & $K^-$ & $6$.$95(23)\cdot10^{-1}$ & $4$ & $\pi^+ 2\pi^-\pi^0$ & $4$.$59(7)$ \\
$2$ & $\pi^-\pi^0$ & $25$.$52(10)$ &  & $\pi^- 3\pi^0$ & $1$.$04(7)$\\
    & $\pi^-\overline{K}^0$ & $8$.$4(0$.$4)\cdot10^{-1}$ & & $K^-\pi^+\pi^-\pi^0$ & $1$.$35(14)\cdot10^{-1}$\\
    & $K^- \pi^0$ & $4$.$28(15)\cdot10^{-1}$ & & $K^- K^0 2\pi^0$ & $< 1$.$6\cdot10^{-2}$ $^a$\\
    & $K^- K^0$ & $1$.$58(16)\cdot10^{-1}$ &  & $K^- 3\pi^0$ & $4$.$7(2$.$1)\cdot10^{-2}$\\
    & $K^- \eta$ & $2$.$7(6)\cdot10^{-2}$&  & $\pi^-\pi^0 K^0 \overline{K}^0$ & $3$.$1(2$.$3)\cdot10^{-2}$\\
    & $\pi^- \eta$ & $<1$.$4\cdot10^{-2}$ $^a$ &  & $\pi^- 2\pi^0 \overline{K}^0$ & $2$.$6(2$.$4)\cdot10^{-2}$\\
    & $\eta' \pi^-$ & $<7$.$4\cdot10^{-3}$ $^b$ & &$\eta 2\pi^- \pi^+$ & $2$.$3(5)\cdot10^{-2}$ \\
$3$ & $2\pi^- \pi^+$ & $9$.$32(7)$ & & $\eta 2\pi^0 \pi^-$ & $1$.$5(5)\cdot10^{-2}$\\
    & $\pi^- 2\pi^0$ & $9$.$27(12)$ & & $2K^- K^+ \pi^0$ & $<4$.$8\cdot10^{-4}$ $^b$\\
    & $\pi^-\pi^+K^-$& $3$.$41(16)\cdot10^{-1}$ & & $2\eta \pi^-\pi^0$ & $<2$.$0\cdot10^{-2}$\\
    & $\pi^-\pi^0\overline{K}^0$ & $3$.$9(4)\cdot10^{-1}$ & $5$ & $2\pi^- \pi^+ 2\pi^0$ & $7$.$6(5)\cdot10^{-1}$\\
    & $\pi^-\pi^0\eta$ & $1$.$81(24)\cdot10^{-1}$ & & $\pi^- 4\pi^0$ & $7$.$6(5)\cdot10^{-1}$ \\
    & $\pi^-K^0\overline{K}^0$ & $1$.$7(4)\cdot10^{-1}$ & &$K^- 4\pi^0$ & $-$ \\
    & $K^-K^0\pi^0$ & $1$.$58(20)\cdot10^{-1}$ & $6$ & $2\pi^- \pi^+ 3\pi^0$ & $-$ \\
    & $\pi^-K^+K^-$ & $1$.$40(5)\cdot10^{-1}$ &  & $3\pi^- 2\pi^+ \pi^0$ & $-$ \\
    & $K^- 2\pi^0$ & $6$.$3(2$.$3)\cdot10^{-2}$ & & & \\
\hline                 
\end{tabular}
\caption{\small{Decays of the $\t$ according to the number of mesons, $n$, and the BR \cite{Amsler:2008zzb}. For the decay $\t^-\,\to\,\nu_\t\,X^-$,
 $X^-$ is displayed in the table. $^a$: with 95 $\%$ CL. $^b$: with 90 $\%$ CL. $^c$: However one should take into account the very recent measurement 
by the $Belle$ collaboration \cite{:2010tc} giving a BR of $(3.29\pm0.17^{+0.19}_{-0.20})\cdot 10^{-5}$. $-$: The PDG does not give a bound for these channels.}}
\label{Listdecaystau} 
\end{center}
\end{table}
\\
\section{Two meson decays of the $\t$} \label{Hadrondecays_Two_meson_decays}
\subsection{Model independent description} \label{Hadrondecays_Two_meson_decays_MI}
\hspace*{0.5cm}The vector form factor of the pion, $F_V^\pi(s)$ is defined through:
\begin{equation} \label{definition_VFF_Pion}
\Big<\pi^+(p')\pi^-(p)\Big|\frac{1}{\sqrt{2}}(\overline{u}\gamma^\mu u-\overline{d}\gamma^\mu d)\Big|0\Big>\,=\,(p'-p)^\mu F_V^\pi(s)\,,
\end{equation}
where $s\,=\,(p+p')^2$ ($s$ will be defined analogously throughout this section), and the participating current is the third component of the vector one of the $SU(3)$
 flavor symmetry of $QCD$. The matrix element of Eq.~(\ref{definition_VFF_Pion}) is related by chiral symmetry to the one appearing in $\tau$ decays \footnote{We do not 
discuss the decays including, in addition, a photon. See e.g. Ref.~\cite{Cirigliano:2002pv} for details.}
\begin{equation} \label{definition_VFF_Pion_Taudecays}
\bra\pi^-(p_{\pi^-})\pi^0(p_{\pi^0})|\overline{d}\gamma^\mu u|0\ket\,=\,\sqrt{2}(p_{\pi^-}-p_{\pi^0})^\mu F_V^\pi(s)\,.
\end{equation}
\hspace*{0.5cm}The associated vector and scalar form factors entering the decay $\tau^-\to K^-\pi^0\nu_\t$ are defined through:
\begin{equation}
\bra K^-(p_K)\pi^0(p_\pi)|\overline{s}\gamma^\mu u|0\ket\,=\,\frac{1}{\sqrt{2}}\left[ \left(p_K-p_\pi-\frac{\Delta_{K\pi}}{s}q \right)^\mu F_V^{K^-\pi^0}(s)+\frac{\Delta_{K\pi}}{s}q^\mu F_S^{K^-\pi^0}(s)\right]\,.
\end{equation}
\hspace*{0.5cm}The different kaon and pion masses imply the appearance of the scalar form factor, $F_S^{K^-\pi^0}(s)$, that accounts for the non-conserving 
vector current part of the decay, $\,q^\mu\,=\,(p_K\,+\,p_\pi)^\mu\,,s\,=\,q^2$ \footnote{Similarly, there is also a non-vanishing scalar form factor in the isospin limit in the $K/\pi^-$ $\eta^{(\prime)}$ matrix elements. Its 
discussion for these decay modes below is restricted, however, to the vector form factors.} and
\begin{equation} \label{DeltaKpi} 
\Delta_{K\pi}\,=\,m_K^2-m_\pi^2\,. 
\end{equation}
\hspace*{0.5cm}Chiral symmetry dictates then the structure of the process $\tau\to \bar{K}^0\pi^-\nu_\t$:
with the changes:
\begin{equation}\label{Changes between different charge states in F^{Kpi}}
 K^-\pi^0 \to \bar{K}^0\pi^-\,,\quad\overline{s}\gamma^\mu u\to \overline{u}\gamma^\mu s\,,\quad F_{V,\,S}^{K^-\pi^0}(s)\to \sqrt{2}\,F_{V,\,S}^{\bar{K}^0\pi^-}(s)\,.
\end{equation}
\hspace*{0.5cm}An equivalent description is given in terms of the pseudoscalar $\Big[$ $F_P^{K\pi}(s)= \left(F_S^{K\pi}(s)-F_V^{K\pi}(s)\right)\frac{\Delta_{K\pi}}{s}$ $\Big]$ and vector
 form factors.\\
\hspace*{0.5cm}The vector form factor into two kaons is probed through $\tau^-\to K^-K^0\nu_\tau$:
\begin{equation}
\bra K^0(p_0)K^-(p_-)|\overline{d}\gamma^\mu u|0\ket\,=\,F_V^K(s)(p_0-p_-)^\mu\,,
\end{equation}
where, as in the case of the pion form factor, the vector current is conserved in the isospin limit
.\\
\hspace*{0.5cm}The decay $\tau^-\to\eta\pi^-\nu_\tau$ can only be produced in the $SM$ as an isospin violating effect \cite{Pich:1987qq}, since it has opposite $G$-parity 
to the participating vector current. 
 The coupling to the vector current occurs via an $\eta-\pi^0$ mixing. The related matrix element will exhibit the structure \cite{Tisserant:1982fc, Neufeld:1994eg}
\begin{equation}
\bra \eta(p_\eta)\pi^-(p_\pi)|\overline{d}\gamma^\mu u|0\ket\,=\,\sqrt{\frac{2}{3}}\,F_V^{\pi\eta}(s)\frac{m_d-m_u}{m_d+m_u}\frac{m_\pi^2}{m_\eta^2-m_\pi^2}(p_\pi-p_\eta)^\mu\,.
\end{equation}
\hspace*{0.5cm}Finally, the vector form factor contribution to the $K^-\eta^{(\prime)}$ decay modes can be parametrized as follows
\begin{equation}
\bra K^-(p_K)\eta^{(\prime)}(p_\eta^{(\prime)})|\overline{s}\gamma^\mu u|0\ket\,=\,\sqrt{\frac{3}{2}}\,F_V^{K\eta^{(\prime)}}(s)(p_K-p_\eta^{(\prime)})^\mu\,.
\end{equation}

\hspace*{0.5cm}The differential decay rate for the process $\tau\rightarrow \nu_{\tau}h_1(p_1) h_2(p_2)$ is obtained from
\begin{equation}
d\Gamma(\tau\rightarrow \nu_{\tau}h_1 h_2)=\frac{1}{2M_\t} \frac{1}{2} \frac{G_F^{2}}{2}\,|V_{CKM}^{ij}|^2 \,\left\{L_{\mu\nu}H^{\mu\nu}\right\}\,\mathrm{d} \mathcal{PS}^{(3)}\,.
\label{decay into two mesons}
\end{equation}
\hspace*{0.5cm}In order to disentangle the angular dependence it is useful to introduce suitable linear combinations of density matrix elements of the hadronic
 system \footnote{The general procedure is studied in Ref.~\cite{Kuhn:1992nz}, where it is shown that the angular dependencies can be isolated by introducing
 $16$ combinations of defined symmetry.}
\begin{equation}
L_{\mu\nu}H^{\mu\nu}=2(M_\t^2-s)\,(\,\bar{L}_{B}W_{B}+\bar{L}_{SA}W_{SA}+\bar{L}_{SF}W_{SF}+\bar{L}_{SG}W_{SG}\,)\,,
\label{lh}
\end{equation}
We note that the most general decomposition of $L_{\mu\nu}H^{\mu\nu}$ (for two body decays) in terms of density matrix elements (or structure functions) 
$W_X$ of the hadronic system has two additional terms $\bar{L}_{A}W_{A}+\bar{L}_{E}W_{E}$ \cite{Kuhn:1992nz}. However, $W_A$ and $W_E$ vanish in the case of tau 
decays into two pseudoscalar mesons. Nonvanishing $W_A$ and $W_E$ arise for example in decay modes with a vector and a pseudoscalar \cite{Decker:1992jy}.\\
\hspace*{0.5cm}The hadron structure functions are related to the vector and scalar form factors as follows:
\begin{eqnarray} \label{W_i in two meson decay} 
W_B (s)   &=& 4 (\vec{p}_1)^2\,|F_V(s)|^2\,,\;\;\;W_{SA} (s)=s\,\frac{\Delta_{12}^2}{s^2}  |F_S(s)|^2\,,\\
W_{SF} (s) &=& 4\sqrt{s}|\vec{p}_1|  \, \Re e\left[F_V(s)F_S^*(s)\right] \,,\nonumber\\
W_{SG} (s) &=& -4\sqrt{s}|\vec{p}_1|  \, \Im m\left[F_V(s)F_S^*(s)\right]\,,\nonumber
\end{eqnarray}
where $|\vec{p}_1|=p_1^z$ is the momentum of $h_1$ in the rest frame of the hadronic system:
\begin{equation}
p_{1}^{z}=\frac{1}{2\sqrt{s}}\sqrt{\left[s-m_1^{2}-m_2^{2}\right]^2-4m_{1}^{2}m_{2}^{2}}\quad,\;\;\; E^2_1=(p_1^{z})^2+m^2_1\,. 
\end{equation}
The hadron structure functions $W_{B,\,SA,\,SF,\,SG}$ are linearly related to the density matrix elements of the hadronic system:
\begin{equation}
W_{B} \, = \, H^{33}\,,\;W_{SA} \, = \, H^{00}\,,\quad W_{SF} \, = \,  H^{03}+H^{30}\,,\;W_{SG} \, = \, -i(H^{03}-H^{30})\,.
\end{equation}
Finally, the differential decay rate $d\Gamma/ds$ yields
\begin{eqnarray} \label{eqndiff}
\frac{d\Gamma}{ds}&=&\frac{G_F^{2}|V_{CKM}^{ij}|^2}{2^8\pi^3}2N_{12}^2\frac{(M_\t^{2}-s)^{2}}{M_\t s^{3/2}}\,\,|{p}_{1}^z|\,\,\frac{2s+M_\t^2}{3M_\t^2}
 \left\{ W_B(s) \,+ \,\frac{3M_\t^2}{2s+M_\t^2} \,W_{SA}(s) \right\}\,,\nonumber\\
\end{eqnarray}
where $N_{12}$ stands for the coupling of the vector current to the mesons $1$ and $2$ normalized to the $\pi^-\pi^0$ case (i.e. $N_{K^-\pi^0}=1/2$).
\subsection{Theoretical descriptions of the form factors}\label{Hadrondecays_Two_meson_decays_Theo_desc}
\hspace*{0.5cm}There is a great amount of data available on $F_V^\pi(s)$, Eq.~(\ref{definition_VFF_Pion}), because it appears in the hadron matrix element entering
 the process $e^+e^-\to\pi^+\pi^-$ where there are many precise measurements 
\cite{Akhmetshin:2003zn, Aulchenko:2006na, Akhmetshin:2006bx, Achasov:2006vp, Aloisio:2004bu, :2008en, :2009fg, Ambrosino:2010bv} and, in the isospin limit, of the decay: $\t^-\to\pi^-\pi^0\nu_\t$,
 where the latest data was published by the $Belle$ Collaboration \cite{Fujikawa:2008ma}. $F_V^\pi(s)$ has been studied within $\CPT$ up to $\cO(p^6)$ 
\cite{Gasser:1984ux, Gasser:1990bv, Colangelo:1996hs, Bijnens:1998fm, Bijnens:2002hp}, so the very-low energy description is really accurate.\\
\hspace*{0.5cm}The energies going from $M_\rho$ to $\sim1$.$2$ GeV are dominated by the $\rho$ (770) so that this resonance can be characterized through the 
study of this form factor very well. Ref.~\cite{Guerrero:1997ku} attempted to improve the $\cO(p^4)$ $\CPT$ result \footnote{The leading $\cO(p^4)$ $\CPT$ terms are also 
reproduced \cite{Guerrero:1998hd}.}by matching it at higher energies with an 
Omn\`es solution \cite{Omnes:1958hv} of the dispersion relation satisfied by the vector form factor of the pion. This way, the description keeps its validity 
up to $1$ GeV, approximately. Some years later, the unitarization approach was used \cite{Oller:2000ug} to obtain a good description of data in this region,
 as well. The $KS$-model \cite{Kuhn:1990ad} also parametrized this decay. We will discuss this model in more detail along its description of the $3\pi$ channel in 
Section \ref{Hadrondecays_Three_meson_decays_Theo_desc}.\\
\hspace*{0.5cm}A model independent parametrization of this form factor built upon an Omn\`es solution for the dispersion relation has also been considered 
\cite{Pich:2001pj, Pich:2002ne, De Troconiz:2001wt, Hanhart:2012wi, DY}. Combining this procedure with $\RCT$ \cite{Pich:2001pj} improves the previous approach (it extends
 now to $\sim 1$.$3$ GeV) if one includes information on the $\rho'$ (1450) through $\pi\pi$ elastic phase-shift input in the Omn\`es solution~\footnote{$\rho''$ (1700) 
exchange has been included in an extension of these works~\cite{Roig:2011iv}, and was coded in TAUOLA \cite{Shekhovtsova:2012ra}.}.\\
\hspace*{0.5cm}It is clear that $\rho'$ (1450) and $\rho''$ (1700) will play the main r\^ole in the energy region up to $2$ GeV. However, the proposed parametrizations
 including both resonances only allow to quantify the relative strength of each one and the likely interference amid these resonances, the possible presence 
of a continuum component, etc. are completely lost with the most simple approaches. Within $\RCT$, the $\rho'$ (1450) was incorporated through a Dyson-Schwinger-like
resummation \cite{SanzCillero:2002bs}, and the ideas of the $\N\to\infty$ limit and vector meson dominance were used in Ref.~\cite{Bruch:2004py}, including 
a pattern of radial excitations expected from dual resonance models. They included the three lightest $\rho$ resonances plus a tower of infinite number of zero-width 
higher-excited states in the spirit of large-$\N$ \cite{Dominguez:2001zu}. Using the hidden gauge formalism, Ref.~\cite{Benayoun:2007cu} has emphasized the r\^ole of the 
$\rho-\omega-\phi$ mixing in this form factor (see also Refs.\cite{Benayoun:2009fz,O'Connell:1995wf,O'Connell:1995xx,Gardner:1997ie,Benayoun:2009im}). 
Using Pad\'e approximants Refs.~\cite{Masjuan:2008fv, Masjuan:2008xg} have studied all available space-like data on this form factor up to $Q^2=12$ GeV$^2$.\\
\hspace*{0.5cm}The last years have witnessed the discrepancy between $e^+e^-\to\pi^+\pi^-$ and $\t^-\to\pi^-\pi^0\nu_\t$ predictions
 \cite{Akhmetshin:2003zn, Aulchenko:2006na, Akhmetshin:2006bx, Achasov:2006vp, Aloisio:2004bu, :2008en, Fujikawa:2008ma} for $F_V^\pi(s)$. There 
have been some theoretical studies \cite{Davier:2009ag, Cirigliano:2002pv, Ghozzi:2003yn, FloresBaez:2006gf, FloresTlalpa:2006gs} of radiative corrections in the $\t$ decay 
mode, but the difference is not fully accounted for yet. Two-loop ($\cO(p^6)$) representations of low-energy pion form factors and $\pi-\pi$ scattering phases in the presence 
of isospin breaking have also been computed recently \cite{DescotesGenon:2012gv}.\\
\hspace*{0.5cm}There is also a large amount of good quality data on the $K\pi$ form factors. In addition to the $e^+e^-$ data from $E865$ \cite{Appel:1999yq},
 $CLEO$ data appeared on $\tau$ decays \cite{Pedlar:2005sj}, and two high-precision studies of the related $\tau$ decays were recently published by $BaBar$ 
\cite{Aubert:2007jh} -for the charge channel $\tau^{-} \to K^{-} \pi^0 \nu_{\tau}$- and $Belle$ \cite{:2007rf}-$\tau^- \to K_S \pi^- \nu_\tau$-. A comparison of 
the newest experiments with the Monte Carlo expectations for the $\pi^-\pi^0$ and $(K \pi)^-$ meson modes is presented in Ref.~\cite{Actis:2009gg}.\\
\hspace*{0.5cm}The form factor $F_V^{K\pi}(s)$ was computed at $\cO(p^4)$ in $\CPT$ in Ref.~\cite{Gasser:1984ux}. In Ref.~\cite{Bijnens:2002hp} the $\CPT$ 
analysis is performed within the three flavour framework at next-to-next-to-leading order. This provides a good description of the very-low energy spectrum. 
In order to extend it to higher energies, in Ref.~\cite{Lowe:2005xe} the Linear sigma model, a quark-triangle model and Vector meson dominance have been used. The 
comparison to data \cite{Appel:1999yq} favours the last one. Simple Breit-Wigner models supplemented with vector meson dominance have also been used \cite{Finkemeier:1995sr,
Finkemeier:1996dh}. They suffered the same problems explained in Section \ref{Hadrondecays_One_meson_decays_BW}.\\
\hspace*{0.5cm}Both the vector and the scalar form factor have been reviewed recently. In Ref.~\cite{Jamin:2006tk}, the distribution function for this decay has 
been obtained with the relevant vector and scalar form factors presented above computed within $\RCT$ and taking into account additional constraints from 
dispersion relations and short-distances. The dynamically generated $K^*_0$ (800) should be the resonance starring at the scalar form factor, whether $K^*$ 
(1410) will modify a bit the more prominent contribution of $K^*$ (892) to $F_V^{K\pi}(s)$, as the results confirm \cite{Jamin:2008qg} when confronting it 
to Ref.~\cite{:2007rf}. In Ref.~\cite{Jamin:2004re}, the knowledge of $\cO(p^6)$ chiral $LECs$ and of light quark masses has been improved studying 
$F_0^{K\pi}(s)$. These form factors have been studied using analyticity constraints and taking into account isospin violating corrections by Moussallam 
\cite{Moussallam:2007qc}. This strategy was also followed in Ref.~\cite{Boito:2008fq} but sticking to the exact $SU(2)$ limit. The scalar form factor 
has also been studied \cite{Bernard:2007tk, Bernard:2009ds} matching $\CPT$ to a dispersive representation. Lately, it has been realized that there is anticorrelation 
in the parameters describing the vector form factor in $F_V^{K\pi}(s)$ from tau decays and $K_{\ell3}$ decays \cite{Boito:2008fq, Boito:2010me}, which has 
allowed to reach smaller errors in the slope and curvature of this form factor.\\
Refs. \cite{TesisGuerrero, Guerrero:1998ei} analyzed the two-kaon vector form factor, $F_V^K(s)$, in much the same way as done for two pions \cite{Guerrero:1997ku}. By that time, 
the experimental data \cite{Barate:1997hv, Asner:1996ht} were not in enough agreement with each other to check the proposed expression. New finer results have 
been published since then \cite{Barate:1999hi, Abbiendi:1999pm}, so a dedicated study of these modes within $\RCT$ employing all present knowledge of $EFTs$, short-distance $QCD$, the large-$\N$
 expansion, analyticity and unitarity would be desirable specially in light of forthcoming data from $BaBar$ and $Belle$.\\
\hspace*{0.5cm}We turn to the $\t$ decays into $\eta$ modes: the $\pi^-\eta^{(')}$ channel has been observed recently for the first time \cite{EPS09}, while for the $K^-\eta$ meson system there 
are recent measurements already published \cite{Inami:2008ar, delAmoSanchez:2010pc} by both the $Belle$ and $BaBar$ Collaborations. The smaller $BR$ for the first one is 
consistent with the findings of Ref.~\cite{Pich:1987qq} summarized before. A first description of this decay was attempted at the beginning of the eighties \cite{Paver:1981ph} 
and revisited recently \cite{Paver:2010mz, Paver:2011md} and the main features were already established few years later \cite{Pich:1987qq, Zachos:1987td}. The $\CPT$ result 
at $\cO(p^4)$ \cite{Gasser:1984ux} was extended to higher invariant masses of the hadronic system in Ref.~\cite{Bramon:1987zb}. Even the isospin 
breaking corrections have been computed for this mode \cite{Neufeld:1994eg}. Again a study along the lines proposed in this Thesis would be interesting. The 
decay $\t^-\to K^-\eta\nu_\tau$ has been considered besides the $\CPT$ computation at $\cO(p^4)$ \cite{Gasser:1984ux} only in Ref.~\cite{Li:1996md}. The relevant strangeness conserving scalar form factors 
have been computed within a coupled channel unitary framework recently \cite{Guo:2011pa, Guo:2012ym, Guo:2012yt}.\\
\section{Three meson decays of the $\t$} \label{Hadrondecays_Three_meson_decays}
\subsection{Model independent description} \label{Hadrondecays_Three_meson_decays_MI}
\hspace*{0.5cm}The hadronic matrix element for the considered decays may be written as \cite{Kuhn:1992nz, Colangelo:1996zp}
\begin{eqnarray} \label{generaldecomposition_3mesons}
\bra (PPP)^-|(V-A)^\mu|0\ket & = & 
V_1^\mu\, F_1^A(Q^2,\,s_1,\,s_2)\,+\,V_2^\mu\, F_2^A(Q^2,\,s_1,\,s_2)\\
& + & Q^\mu\, F_3^A(Q^2,\,s_1,\,s_2)\,+\,i\,V_3^\mu\, F_4^V(Q^2,\,s_1,\,s_2)\,,\nonumber
\end{eqnarray}
where
\begin{eqnarray} \label{Set_of_independent_Vectors_3meson}
& & V_1^\mu\, = \, \left( g^{\mu\nu} - \frac{Q^{\mu}Q^{\nu}}{Q^2}\right) \, 
(p_1 - p_3)_{\nu} \,\,\, , \quad V_2^\mu\, = \, \left( g^{\mu\nu} - \frac{Q^{\mu}Q^{\nu}}{Q^2}\right) \, 
(p_2 - p_3)_{\nu} \, ,\nonumber\\
& & V_3^\mu\, = \,\varepsilon^{\mu\alpha\beta\gamma}\,p_{1\alpha}\,p_{2\beta}\,p_{3\gamma}\,\,\,,\quad Q^\mu\,=\,(p_1\,+\,p_2\,+\,p_3)^\mu\,,\quad s_i\,=\,(Q\,-\,p_i)^2,
\end{eqnarray}
the upper indices on the form factors stand for the participating current, either the axial-vector ($A$), or the vector one ($V$) and not for the quantum 
numbers carried by them; notice that $F_3^A(Q^2,\,s_1,\,s_2)$ is the pseudoscalar form factor that accounts for a $J^P\,=\,0^-$ transition. $F_1^A(Q^2,\,s_1,\,s_2)$
 and $F_2^A(Q^2,\,s_1,\,s_2)$ are the axial-vector form factors that carry $J^P\,=\,1^+$ degrees of freedom. Finally, $F_4^V(Q^2,\,s_1,\,s_2)$ is the vector 
form factor, that has $J^P\,=\,1^-$
.\\
\hspace*{0.5cm}There are other properties that one can derive in full generality. For instance, due to the chiral Ward identity relating axial-vector and 
pseudoscalar currents, conservation of the first one in the chiral limit imposes that $F_3^A(Q^2,\,s_1,\,s_2)$ must vanish with the square of a $pG$ mass and hence,
 its contribution may be suppressed. There are, of course, other constraints coming from $SU(2)$ or $SU(3)$ flavour symmetries for a given mode, like 
those stating that the form factors for the decays $\t^-\to\pi^0\pi^0\pi^-\nu_\t$ and $\t^-\to\pi^-\pi^-\pi^+\nu_\t$ are identical in the $SU(2)$ limit, as it 
happens for the form factors in the decays $\t^-\to K^0 \bar{K}^0\pi^-\nu_\t$ and $\t^-\to K^+ K^- \pi^-\nu_\t$. There are other symmetry requirements: in 
$\t^-\to(3\pi)^-\nu_\t$, Bose-Einstein symmetry implies that $F_2^A(Q^2,\,s_1,\,s_2)\,=\,F_1^A(Q^2,\,s_2,\,s_1)$ and $G$-parity forbids axial-vector current 
contributions to the decays of the $\t$ into $\eta\pi^-\pi^0$ and $\eta\eta\pi^-$\cite{Pich:1987qq}. This kind of relations will be discussed and used in the
 following chapters.\\
\hspace*{0.5cm}We consider a general three meson decay of the $\tau$: $\tau^-(\ell,s)\,\longrightarrow\,\nu_\tau(\ell',s')\,+\,h_1(p_1,z_1)\,+\,h_2(p_2,z_2)\,
+\,h_3(p_3,z_3)$. The polarizations ($s,s',z_1,z_2,z_3$) will play no role in the following since we will assume the tau to be unpolarized. Then, the 
differential phase space for a generic channel is given by
\begin{equation}
\mathrm{d}\Gamma(\tau^-\,\longrightarrow\,(3h)^- \nu_\tau)\,=\,\frac{1}{2M_\tau}\,\frac{G_F^2}{2}|V_{ij}^{CKM}|^2\mathcal{L}_{\mu\nu}H^{\mu\nu} \mathrm{d}\mathcal{PS}^{(4)}\,,
\end{equation}
where the phase space-integration involves the three mesons and one neutrino in the final state:
\begin{equation}
\mathrm{d}\mathcal{PS}^{(4)}\,=\,(2\pi)^4\delta^4(\ell-\ell'-p_1-p_2-p_3)\frac{\mathrm{d}^3\overrightarrow{\ell'}}{(2\pi)^3 2E_\nu}\frac{\mathrm{d}^3\overrightarrow{p_1}}{(2\pi)^3 2E_1}\frac{\mathrm{d}^3\overrightarrow{p_2}}{(2\pi)^3 2E_2}\frac{\mathrm{d}^3\overrightarrow{p_3}}{(2\pi)^3 2E_3}\,.
\end{equation}
In order to obtain the differential decay width as a function of $Q^2$ (the so-called spectral function), the integration over $\int\mathrm{d}\mathcal{PS}^{(4)}$ 
is carried out in two steps:
\begin{equation}
\mathrm{d}\Gamma\,=\,\prod_{i=1}^{3}\int\frac{\mathrm{d}^3\overrightarrow{\ell'}}{(2\pi)^3 2E_\nu}\frac{\mathrm{d}^3\overrightarrow{p_i}}{(2\pi)^3 2E_i}(2\pi)^4\delta\left(\ell-\ell'-\sum_{i=1}^3p_i\right)\overline{\sum}|\mathcal{M}|^2\,,
\end{equation}
provided we use the K\"allen's trick to split the integrations by introducing a Dirac delta as follows:
\begin{eqnarray}
\frac{\mathrm{d}\Gamma_{\tau^-\rightarrow (h_1h_2h_3)^-\nu_\tau}}{\mathrm{d}Q^2} & = & \int\frac{\mathrm{d}^3\overrightarrow{p_\nu}}{(2\pi)^3 2E_\nu}\delta\left(Q^2-(p_\tau-p_\nu)^2\right)\times\nonumber\\
& & \int\prod_{i=1}^{3}\frac{\mathrm{d}^3\overrightarrow{p_i}}{(2\pi)^3 2E_i}(2\pi)^4\delta\left(Q-\sum_{i=1}^3p_i\right)\overline{\sum}|\mathcal{M}|^2\,.
\end{eqnarray}
We have:
\begin{eqnarray} \label{before}
& & \int\frac{\mathrm{d}^3\overrightarrow{p_\nu}}{(2\pi)^3 2E_\nu}\delta\left(Q^2-(p_\tau-p_\nu)^2\right)\,=\,\int\frac{\mathrm{d}^3\overrightarrow{p_\nu}}{(2\pi)^3 2E_\nu}\delta\left(Q^2-M_\tau^2-m_\nu^2+2p_\tau p_\nu\right)\,=\nonumber\\
& & =\,\frac{1}{(2\pi)^2}\int \frac{\mathrm{d}|\overrightarrow{p_\nu}||\overrightarrow{p_\nu}|^2}{E_\nu}\delta\left(Q^2-M_\tau^2-m_\nu^2+2M_\tau E_\nu\right)\,=\nonumber\\
& & =\,\frac{1}{(2\pi)^2}\int\frac{\mathrm{d}|\overrightarrow{p_\nu}||\overrightarrow{p_\nu}|^2}{E_\nu}\frac{\delta\left(|\overrightarrow{p_\nu}|-|\tilde{\overrightarrow{p_\nu}}|\right)}{2M_\tau\frac{|\overrightarrow{p_\nu}|}{E_\nu}}\,=\nonumber\\
& & =\,\frac{1}{(2\pi)^2}\frac{1}{(2M_\tau)^2}|\tilde{\overrightarrow{p_\nu}}|\,=\,
\frac{1}{(2\pi)^2}\frac{\lambda^{1/2}(Q^2,M_\tau^2,m_\nu^2)}{4M_\tau^2}\,.
\end{eqnarray}
where it has been taken into account that decay widths are defined in the rest frame of the decaying particle, and $|\tilde{\overrightarrow{p_\nu}}|\,=\,\frac{\lambda^{1/2}(Q^2,M_\tau^2,m_\nu^2)}{2M_\tau}$.\\
The integration over $\int\mathrm{d}\Pi_3$ is left involving the momenta $p_i$. It yields \cite{PartKin} (notice that a factor $(2\pi)^{-9}$ is not included 
in the definition of $\mathrm{d}\Pi_3$ immediately below):
\begin{equation} \label{d3Pi}
\int\mathrm{d}\Pi_3\,:=\,\int\mathrm{d}s\,\mathrm{d}t\,\delta\left(s-(Q-p_3)^2\right)\,\delta\left(t-(Q-p_2)^2\right)\frac{\mathrm{d}\mathcal{PS}^{(3)}}{(2\pi)^4}\,=\,\frac{\pi^2}{4Q^2}\int\mathrm{d}s\,\mathrm{d}t\,,
\end{equation}
where $s\equiv(p_1+p_2)^2\equiv s_{12}$, $t\equiv(p_1+p_3)^2\equiv s_{13}$, and $u\equiv(p_2+p_3)^2\equiv s_{23}=Q^2-s-t+m_1^2+m_2^2+m_3^2$. Using Eq.~(\ref{d3Pi}), one has:
\begin{eqnarray} \label{fulldGammadQ2}
& & \frac{\mathrm{d}\Gamma_{\tau^-\rightarrow (h_1h_2h_3)^-\nu_\tau}}{\mathrm{d}Q^2}\,=\,\frac{G_F^2|V_{ij}^{CKM}|^2}{128(2\pi)^5 M\tau}\frac{\lambda^{1/2}(Q^2,M_\tau^2,m_\nu^2)}{M_\tau^2}\frac{1}{Q^2}\times\nonumber\\
& & \left\lbrace \omega_{SA}(Q^2,s,t)\left(\Sigma_{\tau\nu}-\frac{\Delta_{\tau\nu}^2}{Q^2}\right)-\frac{1}{3}\frac{\overline{\omega}(Q^2,M_\tau^2,m_\nu^2)}{Q^2}\left(\omega_A(Q^2,s,t)+\omega_B(Q^2,s,t)\right) \right\rbrace \,,\nonumber\\
\end{eqnarray}
where the integrated structure functions have been defined as follows:
\begin{equation} \label{integrated structure functions}
\omega_{SA,\,A,\,B}(Q^2)\,=\,\int\mathrm{d}s\,\mathrm{d}t\,W_{SA,\,A,\,B}(Q^2,s,t)\,.
\end{equation}
The other definitions employed include the so-called weak matrix element:\\
$\overline{\omega}(Q^2,M_\tau^2,m_\nu^2)\equiv(M_\tau^2-Q^2)(M_\tau^2+2Q^2)-m_\nu^2(2M_\tau^2-Q^2-m_\nu^2)$, and 
$\Sigma_{\tau\nu}\equiv M_\tau^2+m_\nu^2\,,\,\Delta_{\tau\nu}\equiv M_\tau^2-m_\nu^2$. $W_{SA,A,B}(Q^2,s,t)$ correspond to the structure functions in 
Ref.~\cite{Kuhn:1992nz}. In terms of the form factors and set of independent vectors in Eqs.~(\ref{generaldecomposition_3mesons}), (\ref{Set_of_independent_Vectors_3meson})
 they are
\begin{eqnarray}\label{structure functions three meson case}
 W_{SA} & = & \left[Q^\mu\, F_3^A(Q^2,s_1,s_2)\right]\left[Q_\mu\, F_3^A(Q^2,s_1,s_2)\right]^*=Q^2\,|F_3^A(Q^2,s_1,s_2)|^2\,,\nonumber\\
W_A & = & -\left[V_1^\mu\,F_1^A(Q^2,s_1,s_2)+V_2^\mu\,F_2^A(Q^2,s_1,s_2)\right]\times\nonumber\\
& & \left[V_{1\,\mu}\,F_1^A(Q^2,s_1,s_2)+V_{2\,\mu}\,F_2^A(Q^2,s_1,s_2)\right]\,,\nonumber\\
W_B & = & \left[V_{3\,\mu}\,F_4^V(Q^2,s_1,s_2)\right]\left[V_3^\mu\,F_4^V(Q^2,s_1,s_2)\right]^*\,.
\end{eqnarray}
In the excellent limit of vanishing neutrino mass, the $Q^2$-spectrum is simply given by:
\begin{equation}\label{fulldGammadQ2vanishingnumass}
 \frac{d \,\Gamma}{d \,Q^2} \, = \, \frac{G_F^2 \, |V_{ij}^{CKM}|^2}{128 \, (2 \pi)^5 \, M_{\tau}} \,
\left( \frac{M_{\tau}^2}{Q^2}-1 \right)^2 \, \int \, ds \, dt \, 
\left[ W_{SA} + \frac{1}{3} \left(  1 + 2 \frac{Q^2}{M_{\tau}^2} \right) \left( W_A + W_B \right) 
\right] \, .
\end{equation}
The limits of integration when obtaining the full width are the following ones:
\begin{equation} \label{integrationlimits}
 \int_{Q^2_{min}}^{Q^{2\,max}}\mathrm{d}Q^2\int_{s_{min}}^{s^{max}}\mathrm{d}s\int_{t_{min}}^{t^{max}}\mathrm{d}t\,,
\end{equation}
\begin{equation}
t_{min}^{max} (Q^2,s)\, = \, \frac{1}{4s}\,\left\lbrace  \left( Q^2+m_1^2-m_2^2-m_{3}^{2}\right) ^{2}-\left[ \lambda^{1/2}\left( Q^2,s, m_{3}^2\right) \mp  \lambda^{1/2}\left(m_1^{2},m_2^{2},s\right) \right] ^{2} \right\rbrace\,,
\end{equation}
\begin{eqnarray}
 s_{min} \, = \, (m_1+m_2)^2\,\,,\quad s^{max} \, = \,(\sqrt{Q^2}-m_3)^2\,\,\,;\nonumber\\
 Q^2_{min} \, = \,(m_1+m_2+m_3)^2\,\,\,,\quad  Q^{2\,max} \, = \,(M_\tau-m_\nu)^2\,.
\end{eqnarray}
\subsection{Recent experimental data}\label{Hadrondecays_Three_meson_decays_Exp_data}
\hspace*{0.5cm}The $BaBar$ \cite{Aubert:2007mh} and $Belle$ \cite{:2010tc} collaborations have recently reported the measurement of the branching 
fractions of various particle combinations (any combination of pions and kaons) in the decay to three charged hadrons. The mass spectra have not been analysed 
yet. Previous studies of the mass spectra were done by the $CLEO$ group \cite{Asner:1999kj}, and the $ALEPH$ \cite{Barate:1998uf}, $DELPHI$ \cite{Abreu:1998cn}
and $OPAL$ \cite{Ackerstaff:1997dv} collaborations on the $3\pi$ mode. $CLEO$ studied with detail also the $KK\pi$ modes \cite{Liu:2002mn, Coan:2004ep, Duboscq:2004zb}.
 The $3K$ modes have been observed by $Babar$ \cite{Aubert:2007mh} and $Belle$ \cite{Inami:2006vd}. Recently, the $Belle$ Collaboration performed a detailed 
study of various decays with the $\eta$ meson in the final state \cite{Inami:2008ar}.\\
\subsection{Theoretical description of the form factors}\label{Hadrondecays_Three_meson_decays_Theo_desc}
\hspace*{0.5cm}Even before the discovery of the tau lepton, its mesonic decays and the relation between these ones and the hadronic states produced in $e^+e^-$ 
annihilation were studied \cite{Tsai:1971vv, Thacker:1971hy}. The late seventies witnessed the pioneering work of Fischer, Wess and Wagner \cite{Fischer:1979fh},
 that employed the method of phenomenological Lagrangians to derive relations between different $n$-pion modes in terms of the pion decay constant, $F$. The odd-intrinsic 
parity sector was also studied along these lines \cite{Kramer:1984cy}. Ref.~\cite{Gilman:1984ry} used
 isospin invariance and measurements on $e^+e^-$ annihilation to relate several channels, with important results for the decays involving $\eta$ mesons in Ref.~\cite{Gilman:1987my}. 
Ref.~\cite{Berger:1987rj} attempted a Lagrangian description of the 
$3\pi$ decays including resonances and, explicitly, the a$_1$-$\pi$-$\rho$ vertex \footnote{The a$_1$ dominance in heavy lepton decays was proposed in 1971 \cite{Tsai:1971vv},
 4 years before the tau lepton was actually discovered.}. However, the model was not consistent with $\CPT$ at $\cO(p^4)$ and, moreover 
made the mistake of not including energy-dependent widths for these spin-one resonances. The model by Isgur, Morningstar and Reader \cite{Isgur:1988vm} 
violated chiral symmetry. Braaten \textit{et. al.} \cite{Braaten:1987jh, Braaten:1989zn} used an $EFT$ approach, based on $U(3)_L\otimes U(3)_R$ for the vector
 resonances and respecting chiral symmetry up to $\cO(p^2)$ for the $pGs$. However, it left aside the axial-vector mesons that happened to dominate these decays
 and assumed hidden local symmetry. Ref.~\cite{Feindt:1990ev} used the isobar model that violates $3$-particle unitarity. In addition, the model did not 
respect chiral symmetry constraints. The works by B.~A.~Li \cite{ Li:1996md, Li:1995aw, Li:1995tv, Li:1996ks, Li:1997tj} covered in a unified framework the 
most interesting decay channels. His theory was based on chiral symmetry for the $pGs$ and the resonances were incorporated following $U(3)_L\otimes U(3)_R$.
 The author introduced an unjustified energy-dependent $\rho$-$\pi$-$\pi$ vertex. Studies using old current algebra techniques were also undertaken years ago
 \cite{Biswas:1980dz, Beldjoudi:1994hm}.\\
\hspace*{0.5cm}The KS model \cite{Kuhn:1990ad} was a step forward, because in the zero-momentum limit it recovered the $\CPT$ results and, additionally, it  
provided more realistic off-shell widths for the involved resonances accounting both for vector and axial-vector states. However, as it was shown later, it was 
inconsistent with chiral symmetry \footnote{One could think that it is not that important to fulfill the $NLO$ results 
in $\CPT$ while one is attempting a description in the $GeV$ region. This is not true. The spectra are very sensitive to the normalization and low-energy 
dependence of the form factors which are transported to the rest of the spectrum, as it is illustrated in Fig.~\ref{fig:lowQ23pi}.} at $\cO(p^4)$ 
\cite{GomezDumm:2003ku, Portoles:2000sr} \footnote{The $\CPT$ result at $\cO(p^4)$ \cite{Colangelo:1996hs} 
checked that $\CPT$ could only describe this decay in a tiny window of phase space. This low-energy part motivated also the study \cite{Girlanda:1999fu} aimed 
to find hints on the mechanism of dynamical chiral symmetry breaking.}. Moreover, the proposed widths are only phase-space motivated. Although they worked 
quite well, there is some lack of dynamics in them, a feature that should not be satisfactory. The KS model was a major achievement in $LEP$ times to understand $\tau$
 data. Nowadays, the much more precise data samples and the finer understanding of the Lagrangian approach to the intermediate-energy meson dynamics demands
 the latter to be applied by the experimental community. We will report about its application to the three-meson decays modes of the tau in Chapters 6, 7 and 8. 
The recent study in Ref.~\cite{Vojik:2010ua} includes also the sigma contribution to this channel and reports that the a$_1'$ effect is needed to improve 
significantly the agreement with the data. However, it violates chiral symmetry already at $LO$ since it does not include the diagrams with pions in $\CPT$ at $\cO(p^2)$.\\
\hspace*{0.5cm}Other three meson channels were studied following the KS model \cite{Finkemeier:1995sr, Decker:1992kj, Decker:1992rj, Decker:1993ay}. In addition 
to the comments we made to the original work, several other issues enter in these cases \cite{Roig:2008ev, Roig:2008xt, Roig:2008xta, Roig:2009ff}: Some of the intermediate exchanged 
resonances in a given channel that are allowed by quantum numbers are not included in the model and, moreover, the treatment of spin-one resonances is inconsistent:
 the $\rho$ (1450) has noticeable different mass and width in the axial-vector and vector current form factors and there are two multiplets of vector resonances 
in the axial-vector current while three in the vector current, a very unnatural phenomenon. It seems difficult to explain why the $\rho(1700)$ happens to be so 
important in the vector form factor, given its high mass. The KS model and its generalizations were implemented in the renowned $TAUOLA$ library for tau decays 
\cite{Jadach:1990mv, Jadach:1990mz, Jezabek:1991qp, Jadach:1993hs}. Later on, this parametrization of the hadronic matrix elements was complemented by others 
based on experimental data by the $CLEO$ \footnote{The $CLEO$ parametrization was private, reserved for the use of the collaboration until published in 
\cite{Golonka:2003xt} .} and Novosibirsk groups
 \cite{Bondar:2002mw}. Currently, we are improving some of the proposed currents \cite{Shekhovtsova:2012ra} using the results in this Thesis.\\
\hspace*{0.5cm}We end the section by quoting early studies that constructed the form factors using the chiral symmetry results at low energies and experimental 
information to extend it to the GeV-scale, on the $KK\pi$ \cite{GomezCadenas:1990uj} and $\eta\pi\pi$ modes~\cite{Kramer:1984pm, Kramer:1986tm, Kramer:1988ud}.\\
\section{Decays of the $\t$ including more mesons} \label{Hadrondecays_Many_meson_decays}
\subsection{Model independent description} \label{Hadrondecays_Many_meson_decays_MI}
\hspace*{0.5cm}A model independent description of the four pion tau decays (and of its electroproduction) can be found in Ref.~\cite{Ecker:2002cw}, where the structure dependent 
form factors were included based on $\omega$, $a_1$ and double $\rho$ exchange and by performing a resummation of the pion form factor. Ref.~\cite{Decker:1994af} builds the 
amplitude for the $4\pi$ decay assuming some decay chains and that the vertex functions are given by their on-shell structure and are transverse. This work was 
 consequence of a previous study of $\tau\to\omega\pi^- \nu_\tau$ \cite{Decker:1992jy}. In Refs.~\cite{Kuhn:1998rh, Czyz:2000wh} isospin symmetry is used in order to 
determine that all decay channels $\t^-\to(4\pi)^-\nu_\t$ can be parametrized in terms of form factors that depend just on one quantity once the symmetries 
associated to relabeling the different $4\pi$ momenta have been used. Moreover, the form factors appearing in $e^+e^-\to (4\pi)^0$ can be obtained with the 
same single function. Isospin symmetry was systematically used for the first time in multimeson tau decays in Refs.~\cite{Rouge:1995cc, Rouge:1997kj}, where
 the meson channels $K\bar{K}+n\pi$, $n\pi$, $(2n+1)\pi$, $2n\pi$, and $K+n\pi$ were examined.\\
\hspace*{0.5cm}The decays with five mesons have been addressed in Ref.~\cite{Kuhn:2006nw}, but the proposed hadronic matrix elements rely on the assumed substructure 
of the process.\\
\subsection{Experimental data}
\label{Hadrondecays_Many_meson_decays_Exp_data}
\hspace*{0.5cm}Tau decays into four-meson modes have been measured very recently in the $B$-factories, and more data was collected by $ALEPH$ \cite{Buskulic:1996qs} and
 $CLEO$ \cite{Edwards:1999fj, Arms:2005qg}. $BaBar$ measured the $BR$ for the mode $(K \pi)^0 K^- \pi^0$ through $K^{*0}K^- \pi^0$ \cite{Aubert:2008nj} and 
$Belle$ for the mode $2\pi^-\pi^+\eta$~\cite{:2008uy}. The five charged meson modes have been measured by $BaBar$ \cite{Aubert:2005waa} with much larger statistics than $CLEO$ 
\cite{Gibaut:1994ik} achieved. Finally, the six-pion final state was studied also by the $CLEO$ Collaboration \cite{Anastassov:2000xu}.\\
\subsection{Theoretical description of the form factors}
\label{Hadrondecays_Many_meson_decays_Theo_desc}
\hspace*{0.5cm}Besides the references quoted in Sect.~\ref{Hadrondecays_Many_meson_decays_MI}, G.~L\'opez Castro \textit{et. al.} have investigated the 
$\tau^-\to(\omega/\phi) P^-\nu_\tau$ decays ($P=\pi,\,K$) within a VMD model \cite{LopezCastro:1996xh, FloresTlalpa:2007bt}.
\hspace*{0.5cm}In Ref.~\cite{Sobie:1999ke} comparisons of five- and six-pion $\tau$ decay data with the isospin relations indicates that the final states in
 these decays tend to involve an $\omega$ resonance. If the decay dynamics of the seven-pion $\tau$ decays are similar to the five- and six-pion decays, 
then isospin relations could explain a low branching ratio limit on the $\tau\to4\pi^-3\pi^+ \nu_\tau$ decay.\\
\section{Hadronic $\t$ decays in Higgs physics at the LHC}\label{Tau_at_LHC}
\begin{figure}[h!]
\centering
\includegraphics[scale=0.8]{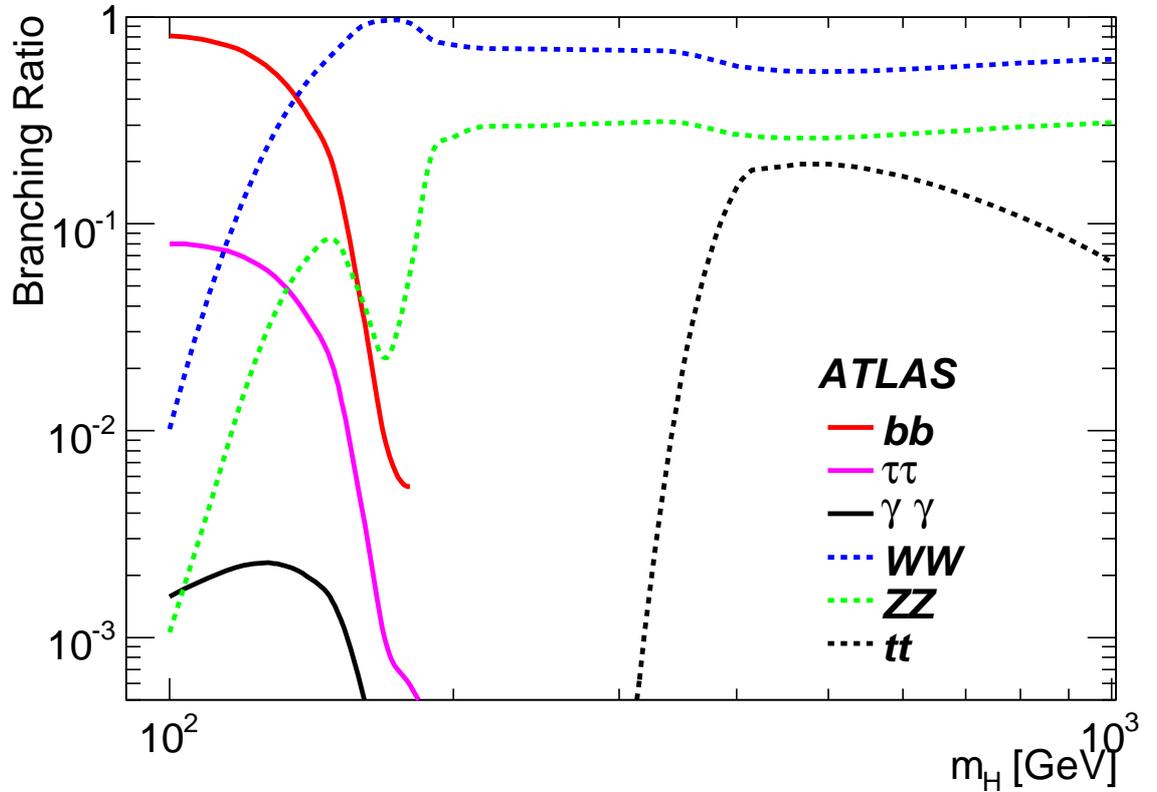}
\caption{\small{Branching ratio for the relevant decay modes of the $SM$ Higgs boson as a function of its mass.}}
\label{Fig:Higgsdecays}
\end{figure}

\begin{figure}[h!]
\centering
\includegraphics[scale=0.8]{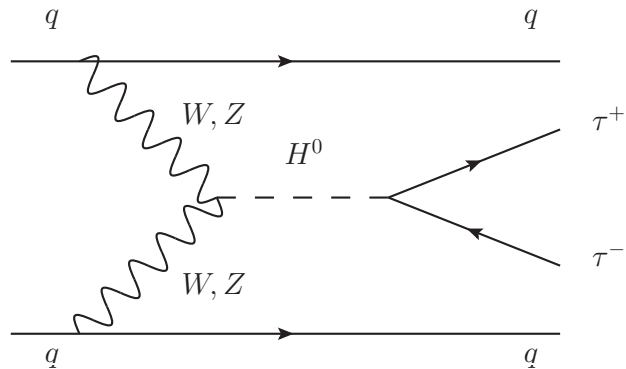}
\caption{\small{Feynman diagram for the lowest order Higgs production via $VBF$ and subsequent decay $H^0\to\t^+ \t^-$.}}
\label{Fig:VBF}
\end{figure}

\begin{figure}[h!]
\centering
\includegraphics[scale=0.8]{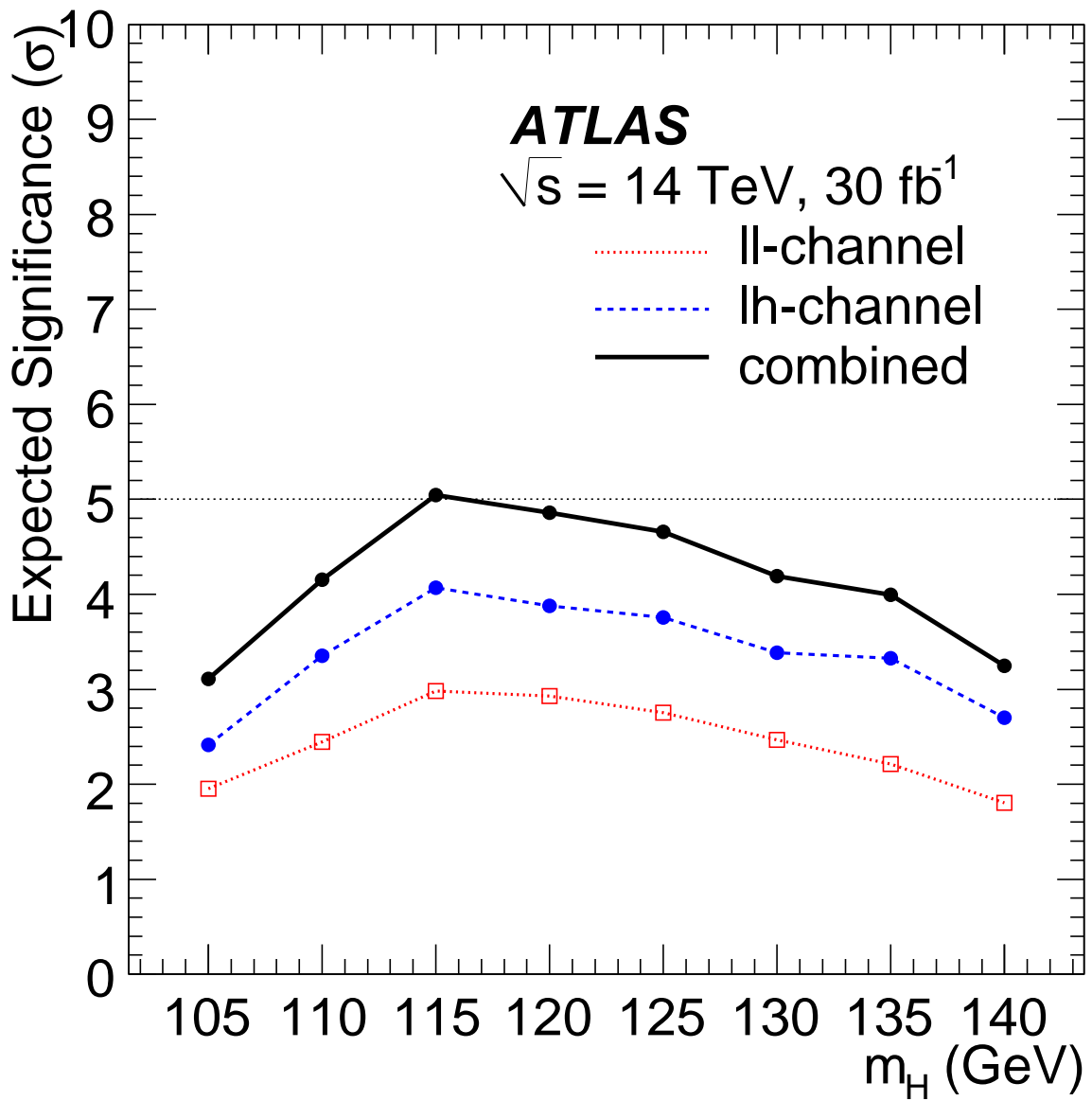}
\caption{\small{Expected signal significance for several masses based on fitting the $m_{\t\t}$ spectrum in $H^0\to\t^+ \t^-$ with $30 fb^{-1}$ of data (one 
year of data taking). From Ref.~\cite{Aad:2008zzm}. In the TAU10 Conference (13-19.09.2010), R.Goncalo reported on behalf of the $ATLAS$ Coll. that the 
$hh$ mode was at an advanced stage for being incorporated in these plots soon. However, figures were not available yet.}}
\label{Fig:Htautau}
\end{figure}

\begin{figure}[h!]
\centering
\includegraphics[scale=0.8]{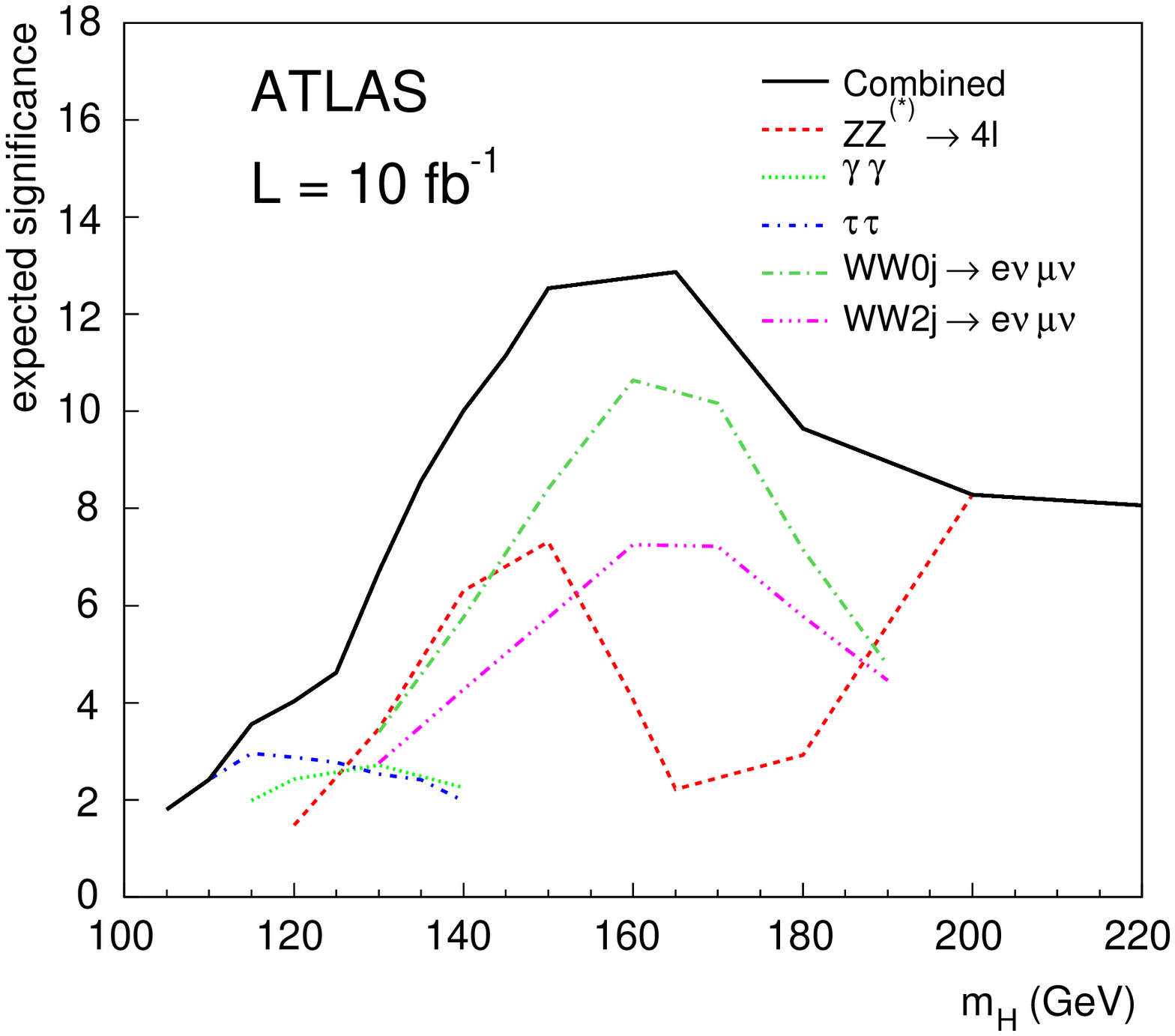}
\caption{\small{The median discovery significance for the $SM$ Higs boson for the various channels as well as the combination for the integrated luminosity 
of $10 fb^{-1}$ for the lower mass range. From Ref.~\cite{Aad:2008zzm}.}}
\label{Fig:Hdiscoveryall}
\end{figure}

\hspace*{0.5cm}In this section we will report briefly on the importance of mastering hadronic decays of the tau in Higgs physics at the $LHC$ \footnote{See the two notes 
added afted this Thesis was defended at the end of this section.}. One of the main goals of the $ATLAS$ \cite{Aad:2008zzm} and $CMS$ \cite{Ball:2007zza} experiments at $LHC$ 
is the search for the Higgs boson and the source of electroweak symmetry breaking. Both detectors are capable of doing that for any possible range of masses: $114$.$4$ GeV (direct exclusion limit at $95\%$ confidence level 
obtained by $LEP$ \cite{Barate:2003sz}) to $1$ TeV \footnote{One should also note 
the exclusion bounds set by the Tevatron experiments, $CDF$ and $D0$. In the combined analysis, \cite{Aaltonen:2011gs} with up to 7.1 $fb^{-1}$ of data analyzed at $CDF$, and up to 
8.2 $fb^{-1}$ at $D0$, the 95$\%$ confidence level upper limits on Higgs boson production is a factor of 0.54 times the SM cross section for a Higgs boson mass of 
165 GeV. The region $158<m_H<173$ GeV is excluded at 95$\%$ confidence level.} -there are many reasons to believe that the TeV scale is an upper limit for the Higgs mass, 
see for instance \cite{Quigg:2009vq} and references therein-. In the low-mass region ($m_H<130$ GeV) the decays of the Higgs boson into two photons or into 
two taus are the most promising for discovery \footnote{The decays into $WW^\star\to4\ell$ and $ZZ^\star\to2\ell2\nu_\ell$ can also bring valuable information in this region, though.}. Irrespective of the value of $m_H$ its decays 
into taus will be important to measure its couplings, spin and $CP$ properties \cite{Plehn:2001nj, Duhrssen:2004cv, Ruwiedel:2007zz}. The 
production and decay of the $\tau$ leptons are well separated in space and time providing potential for unbiased measurements of the polarization, spin 
correlations, and the parity of the resonances decaying into $\t$ leptons \cite{Czyczula:2012ny, Banerjee:2012ez}. The excellent knowledge of $\t$ decay modes from low-energy experiments indeed 
makes this an ideal signature for observations of new physics. In the context of the Minimal Supersymmetric Standard Model ($MSSM$),
 the branching ratio of a $H\to \gamma\gamma$ decay is generally suppressed which makes the search for the decay $H\to \t\t$ very important \footnote{The decays $H\to \gamma\gamma$ and $H\to \t\t$ 
have also been searched for at the Tevatron experiments: $CDF$ \cite{Aaltonen:2011rj} for the diphoton decay and $DO$ \cite{Abazov:2011jh} and $CDF$ \cite{Aaltonen:2012jh} for the $\t\t$ decay.}. This section is 
mainly based in Refs.~\cite{Aad:2008zzm, Ball:2007zza, Aad:2009wy, RichterWas:2009wx}.\\
\hspace*{0.5cm}In Figure \ref{Fig:Higgsdecays} the branching fractions of the $SM$ Higgs boson are shown as a function of $m_H$. Immediately after they 
are opened, the $WW$ and $ZZ$ decay modes dominate over all others ($t\bar{t}$ can barely reach $20\%$, as we can see from the curve appearing for $m_H\gtrsim 300$ GeV). All other fermionic modes are only relevant for 
the Higgs boson masses below $2 (M_W-\Gamma_W)$. These modes show peculiar structures with a peak corresponding to both weak bosons being on-shell. We see also a valley in 
the $WW$ channel corresponding to the peak in that of $ZZ$, since one is plotting the branching ratio. The decay $H\to b\bar{b}$ is dominating below $140$ GeV. However, even though it was included as a possible 
channel that could help the Higgs discovery in the low-mass case up to 2005 \cite{D'Hondt:2005im}, a re-evaluation of the $QCD$ backgrounds swapped it away 
in later reports \cite{Aad:2008zzm, Ball:2007zza}. Thus, the decays $H\to \t\t$ (with $br\sim8\%$ for $120<m_H<140$ GeV) and $H\to \gamma\gamma$ ($br\sim2
\cdot10^{-3}$) would be the way to discover the Higgs boson in the low-mass case. In particular, the design of the detectors in the $ATLAS$ 
and $CMS$ experiments makes --always attaching to this low-mass case-- the $H\to\t\t$ decay the most promising signature in the former and the $H\to\gamma
\gamma$ in the latter.\\
\\
\hspace*{0.5cm}Although the reconstruction of $\t$ leptons is usually understood as a reconstruction of the hadronic decay modes, since it would be difficult 
to distinguish lepton modes from the primary electrons and muons, a dedicated effort has been devoted to electron and muon vetoing to reduce their background, 
so that all possible decays of a $\t^+\t^-$ pair: hadron-hadron ($hh$), lepton-lepton ($\ell\ell$) or mixed ($\ell h$) can be detected. The following generic 
nomenclature for $\t$ decays is used from the detection point of view: single-prong means that exactly one charged meson (most frequently a $\pi$) is 
detected in the reconstructed decay, while three-prong means that there are three charged mesons detected. It is understood that one can generally tell a 
charged lepton from a charged meson and the small fraction ($0$.$1\%$) of five-prong decays is usually too hard to detect in a jet environment. The transverse 
momentum range of interest at $LHC$ spans from below $10$ GeV to $500$ GeV which makes necessary that at least two detection strategies are developed, as we 
will comment later on. As one can read from Table \ref{Listdecaystau}, $\t$ leptons decay hadronically in $64$.$8\%$ of the cases, while in $17$.$8\%$ ($17$.$4\%$) 
of the cases they decay to an electron (muon). It is interesting to note that the $\t\to\pi^\pm \nu_\tau$ decay represents only the $22$.$4\%$ of single-prong 
hadronic decays, so the detection of $\pi^0$ particles in $\t\to n\pi^0 \pi^\pm \nu_\tau$ is fundamental. For the three-prong $\t$ decays, the $\t\to
3\pi^\pm \nu_\tau$ decay contributes $61$.$6\%$ and the  $\t\to n\pi^0 3\pi^\pm \nu_\tau$ only about one third. Although the decays containing only pions 
dominate, there is a small percentage of  decays containing $K^\pm$ that can be identified as for states with $\pi^\pm$ from the detector point of view. A 
small fraction of states with $K_S^0$ cannot be easily classified as one- or three- prong decays since the $K_S^0$ decays significantly both to two 
charged and two neutral pions. In any study performed so far other multi-prong hadronic modes have been neglected.\\
\hspace*{0.5cm}Three-prong decays of the $\t$ (essentially $\t\to n\pi^0 3\pi^\pm \nu_\tau$) have the additional interest of allowing for the reconstruction 
of the $\t$ decay vertex. This is possible because $c \t_\t\sim87\mu m$ that one can separate with the inner silicon detector tracking system. The transverse 
impact parameter of the $3 \pi^\pm$ can be used to distinguish them from objects originated at the production vertex. As we stated before, this allows for a 
full treatment of spin effects. This has been done within the framework of the $ATLAS$ Monte Carlo simulation and events were generated using $PYTHIA$ 6.4
\cite{Sjostrand:2006za} interfaced with $TAUOLA$ \cite{Davidson:2008ma, Davidson:2010rw, Davidson:2010ew} \footnote{See also Sect. \ref{KKpi_Intro}.}. Full spin correlations in production and decay of $\t$ leptons 
were implemented. The associated spin properties in gauge boson, Higgs boson or $SUSY$ cascade decays carry information on the polarization of the decaying 
resonance: $\t$ leptons from $W\to\t\nu_\t$ and $H^\pm\to\t\nu_\t$ will be completely longitudinally polarized, with $P_\t=+1$ and $P_\t=-1$, respectively. 
As a result, the charged to total visible energy distributions for one-prong decays will be different in these cases, permitting their differentiation 
unambiguously. At the $LHC$ this effect can be used to suppress the background from the former and enhance observability of the latter \cite{Assamagan:2002in}. 
The $\t$ polarization could also be used to discriminate between $MSSM$ versus extra dimension scenarios \cite{Assamagan:2001jw}. On the contrary, $\t$ leptons 
from neutral Higgs boson decays are effectively not polarized and those coming from $Z$ decays obey a complicated function of the center-of-mass energy of 
the system and the angle of the decay products \cite{Pierzchala:2001gc}. In the cleaner environment of a lepton collider, like the $ILC$, building variables 
sensitive to the longitudinal and transverse spin correlations may lead to a $CP$ measurement of the Higgs boson \cite{Bower:2002zx, Desch:2003rw}.\\
\hspace*{0.5cm}Two complementary algorithms for $\t$-identification and reconstruction have been studied:
\begin{itemize}
 \item A track-based algorithm~\cite{trackbased}, which relies on tracks reconstructed in the inner detector and adopts an energy-flow approach based only on 
tracks and the energy in the hadronic calorimeter. It starts from seeds built from few (low multiplicity) high-quality tracks collimated around the leading 
one. It has been optimized for visible transverse energies in the range $10-80$ GeV, that corresponds to $\t$-decays from $W\to\t\nu_\t$ and $Z\to \t\t$ 
processes.
 \item A calorimeter-based algorithm~\cite{calobased}, which relies on clusters reconstructed in the hadronic and electromagnetic calorimeters and builds the 
identification variables based on information from the tracker and the calorimeters. It has been optimized for visible transverse energies above $30$ GeV, 
which corresponds to hadronic $\t$-decays from heavy Higgs-boson production and decay.
\end{itemize}
\hspace*{0.5cm}Whereas the track-based algorithm has been tuned to preserve similar performance for single- and three-prong decays, the calorimeter-based 
algorithm has been tuned to provide the best possible rejection at medium-to-high energies and it is therefore more performant for single-prong decays than 
the track-based algorithm. Depending on the specific process and scenario under study, the trigger requirements are different, a complete description can be
 found in Ref.~\cite{Aad:2008zzm}.\\
\hspace*{0.5cm}The Higgs-boson can be produced via four different mechanisms at hadron colliders. Although the largest production cross-section for $m_H \lesssim 1$ 
TeV is always that of gluon fusion, $gg\to H$, which is mediated at lowest order by a $t$-loop, the cleanest signal in the $H\to \t\t$ channel is due to the 
so-called Vector Boson Fusion ($VBF$) channel \cite{Rainwater:1998kj, Asai:2004ws}, that is represented in Figure~\ref{Fig:VBF}.\\
\\
\hspace*{0.5cm}The expected performance in the $ATLAS$ experiment will be adequate to extract $\t$ signals in early $LHC$ data from $W\to\t\nu_\t$ and 
$Z\to \t\t$ decays. These signals are important to establish and calibrate the $\t$ identification performance with early data. The study of dijet events 
from $QCD$ processes will allow a determination of $\t$ fake rates. It is expected that such rates can be measured with a statistical precision at the percent 
level or better already with data corresponding to an integrated luminosity of $100$ $pb^{-1}$ \footnote{We give some numbers to make easier this and subsequent 
figures: although the design luminosity of the $LHC$ is $10^{34} cm^{-2} s^{-1}$, it is still a bit optimistic to count on $10^{32} cm^{-2} s^{-1}$ for the 
first year of operation. In this case, one could expect $\sim30 fb^{-1}$ at the end of the first year. In fact the nominal luminosity is $66$.$2 fb^{-1}/$year, 
so $100\, pb^{-1}\,=\,0$.$1 fb^{-1}$ would be achieved very early because the instantaneous luminosity for the very first measurements was expected to be at the 
level of $10^{31} cm^{-2} s^{-1}$.}. In any case, despite the advances in theoretical tools and extraordinarily detailed simulation of the $ATLAS$ detector, 
it is preferable to estimate backgrounds from data rather than relying entirely on Monte Carlo simulations. All estimates can thus get sizable corrections 
as a result.\\
\hspace*{0.5cm}Once this first stage is completed, one would be ready to search for the $qqH\to qq\t\t$ decays via a Higgs boson produced in association 
with two jets. This analysis requires excellent performance from every $ATLAS$ detector subsystem; $\t$ decays imply the presence of electrons, muons, 
pions and a few kaons, and missing transverse momentum, while the $VBF$ process introduces jets that tend to be quite forward in the detector. Due to the 
small rate of signal production and large backgrounds, particle identification must be excellent and optimized specifically for this channel. Furthermore, 
triggering relies on the lowest energy lepton triggers or exceptionally challenging tau trigger signatures. The $ATLAS$ collaboration has estimated the 
sensitivity based on $\ell\ell$ and $\ell h$ modes \cite{Aad:2008zzm}. The $hh$ channel has also been investigated and gives similar results for signal and 
non-$QCD$ backgrounds as the other channels. However, due to the challenge of predicting the $QCD$ background the estimated sensitivity for this mode was 
not reported.\\
\hspace*{0.5cm}The signal events are produced with significant transverse momenta, so the $\t$ from the decay are boosted which causes their decay products 
being almost collinear in the lab frame. The di-tau invariant mass can be therefore reconstructed in the collinear approximation \footnote{i.e., one assumes 
that the $\t$ direction is given by their visible decay products: leptons or hadrons.}. The mass resolution is $\sim 10$ GeV, leading to a $\sim3$.$5\%$ precision 
on the mass measurement with $30 fb^{-1}$ of data (one year of data taking). In the more recent analysis particular emphasis is put on data-driven background 
estimation strategies. Expected signal significance for several masses based on fitting the $m_{\t\t}$ spectrum is shown in Figure \ref{Fig:Htautau}. The 
results obtained neglecting pileup effects indicate that a $\sim 5\sigma$ significance can be achieved for the Higgs boson mass in the range of special 
interest: $115-125$ GeV after collecting $30 fb^{-1}$ of data and combining the $\ell\ell$ and $\ell h$ channels. The effects induced by the event pile-up have 
not been fully addressed yet. As it is intuitive, the hadronic decay gives more constraints, since there is only one neutrino that escapes detection, while two 
are unobserved in the lepton case. Unfortunately the $QCD$ background prevents the usage of the $hh$ mode for the moment, that could further improve the 
discovery potential in these decays. One can check in Figure~\ref{Fig:Hdiscoveryall} -note that this corresponds to one third of the luminosity taken as reference 
previously- that this is the gold-plated mode in the $ATLAS$ experiment for $m_H\lesssim 135$ GeV \footnote{In the $CMS$ experiment the $H\to\gamma\gamma$ 
decay mode has a significance of one sigma more than $H\to\t \t$ in the energy range of interest \cite{Ball:2007zza}. Remarkably, the two photon mode at 
$CMS$ and the two tau mode at $ATLAS$ have a similar significance at the discovery level in the mass range $115$ GeV$\leftrightarrow 120$ GeV with $30 fb^{-1}$ 
of data.}.\\
\\
\\
\hspace*{0.5cm}We will not cover in detail the relevance of hadronic tau decays in Higgs searches in the context of the $MSSM$. We will 
just recall the most prominent features. The topic is studied in depth in Refs.~\cite{Aad:2008zzm, Ball:2007zza, Aad:2009wy, RichterWas:2009wx, 
Abdullin:1998pm, Carena:2002qg, Schumacher:2004da}. The $LHC$ has a large potential in the investigation of the $MSSM$ Higgs sector. The Higgs 
couplings in the $MSSM$ are different to those in the $SM$. In particular, for large Higgs masses ($m_H>160$ GeV) its decays into weak gauge 
bosons are either suppressed or absent in the case of the pseudoscalar Higgs, $A$. On the other hand, the coupling to third generation 
fermions is strongly enhanced for large regions of the parameter space which makes the decays into $\t$ leptons even more interesting. The 
search for light neutral Higgs boson is based on the same channels as for the $SM$ case, with more relevance of $H\to\t\t$ for larger masses 
in some subsets of the parameter space, due to enhanced couplings. In addition to this, $A\to\t\t$ is also relevant for large values of 
tan$\beta$ \footnote{The ratio of the values of the two Higgs condensates.}. In both decay channels, the $\ell h$ detection mode would 
provide again the highest sensitivity. A final promising decay channel is $H^\pm\to\t^\pm\nu_\t$, that would unambiguously proof the existence 
of physics beyond the $SM$. For a high $SUSY$ mass scale this charged Higgs boson could be the first signal of new physics (and indication 
for $SUSY$) discovered. The first results of the dedicated searches have already been presented by the $CMS$ \cite{Chatrchyan:2011nx} and $ATLAS$ \cite{Aad:2011rv} 
Collaborations setting stringent bounds in the $MSSM$ parameter space.\\
\\
\hspace*{0.5cm}Note added (I): After this PhD Thesis was defended, the $LHC$ experiments have published promising data pointing to the existence of a light Higgs. Especifically, the $ATLAS$ Collaboration 
\cite{:2012si} excludes $m_H$ in the ranges $[112.9-115.5]$ GeV, $[131-238]$ GeV and $[251-466]$ GeV at the $95 \%$ confidence level, while the range $[124-519]$ GeV is expected 
to be excluded in the absence of a signal. An excess of events is observed around $m_H \sim 126$ GeV with a local significance of $3.5$ standard deviations. The local 
significance of $H \to \gamma \gamma$, the most sensitive channel in this mass range, is 2.8 sigma. The global probability for the background to produce such a fluctuation 
anywhere in the explored Higgs boson mass range $[110-600]$ GeV is estimated to be $\sim1.4\%$. The $CMS$ Collaboration \cite{Chatrchyan:2012tx} excludes the standard model 
Higgs boson in the mass range $[127-600]$ GeV at 95$\%$ confidence level, and in the mass range $[129-525]$ GeV at 99$\%$ confidence level. An excess of events above the expected 
SM background is observed at the low end of the explored mass range making the observed limits weaker than expected in the absence of a signal. The largest excess, with a local 
significance of $3.1$ sigma, is observed for a Higgs boson mass hypothesis of $124$ GeV. The global significance of observing an excess with a local significance greater than $3.1$ 
sigma anywhere in the search range $[110-600]$ ($[110-145]$) GeV is estimated to be $1.5$ sigma ($2.1$ sigma). Respectively, the searches made by these two experiments in the diphoton 
Higgs decay channel are reported in Refs.~\cite{:2012sk, Chatrchyan:2012tw}. In the first one, the largest excess with respect to the background-only hypothesis is observed at $126.5$ GeV, 
with a local significance of $2.8$ standard deviations (the second one gives the value quoted above, $124$ GeV, with a significance of $3.1$ sigma). Finally, the search for the neutral 
Higgs bosons decaying to tau pairs has already been reported by $CMS$ \cite{Collaboration:2012vp} yielding important restrictions in the $MSSM$ parameter space, but not so much in the 
$SM$ Higgs boson searches. There is agreement that more data is required to draw definitive conclusions. However, it is very promising that forthcoming data taken during 2012 can help to 
finally unveil the nature of the electroweak symmetry breaking in the SM.\\
\\
\hspace*{0.5cm}Note added (II): A Higgs-like boson has been discovered by both the ATLAS \cite{:2012gk} and CMS \cite{:2012gu} experiments at CERN. Now, one of the most interesting questions is 
to clarify if its couplings correspond to the predicted particle or not. In this respect, the two significant differences with the predictions that were observed (higher rate in the di-photon channel and 
lower rate in the di-tau decay) seem to be shrinking according to the results presented at the Hadron Collider Physics Symposium 2012.

\chapter{$\tau^-\to (\pi\pi\pi)^- \nu_\tau$ decays}\label{3pi}
\section{Introduction}\label{3pi_Intro}
\hspace*{0.5cm}In this chapter we will discuss the hadronic form factors and related observables appearing in $\tau^-\to (\pi\pi\pi)^- \nu_\tau$
 decays. These processes are a very clean scenario to learn about the axial-vector current, because the vector current contribution is 
forbidden by $G$-parity in the isospin limit. Moreover, the starring r\^ole of the lightest vector and axial-vector resonances will allow to study in detail the 
properties of the latter, since the first one is extremely well known from $e^+e^-\to\pi^+\pi^-$ and $\tau^-\to\pi^-\pi^0\nu_\tau$ decays. At the same 
time, this will be a stringent test of the joint consistency of the proposed width for a given definition of mass \cite{Boito:2008fq}.\\
\hspace*{0.5cm}The $\tau \rightarrow \pi \pi \pi \nu_{\tau}$ decay is thus driven by the hadronization of the axial-vector current. Within 
the resonance chiral theory, and considering the large-$N_C$ expansion, this process has been studied in Ref.~\cite{GomezDumm:2003ku}. In the
 light of later developments we revise here \cite{Dumm:2009va} this previous work by including a new off-shell width for the a$_1(1260)$ 
resonance that provides a good description of the $\tau \rightarrow \pi \pi \pi \nu_{\tau}$ spectrum and branching ratio. We also consider 
the r\^ole of the $\rho(1450)$ resonance in these observables. Thus we bring in an overall description of the $\tau \rightarrow \pi \pi \pi 
\nu_{\tau}$ process in excellent agreement with our present experimental knowledge.\\
\hspace*{0.5cm}The significant amount of experimental data on $\tau$ decays, in particular, $\tau\rightarrow \pi\pi\pi\nu_{\tau}$ branching 
ratios and spectra~\cite{Barate:1998uf}, encourages an effort to carry out a theoretical analysis within a model-independent framework capable
 to provide information on the hadronization of the involved $QCD$ currents. A step in this direction has been done in Ref.~\cite{GomezDumm:2003ku},
where the $\tau \rightarrow \pi \pi \pi \nu_{\tau}$ decays have been analyzed within the resonance chiral theory (R$\chi$T)~\cite{Ecker:1988te,Ecker:1989yg}.
 As explained in detail in earlier chapters, this procedure amounts to build an effective Lagrangian in which resonance states are treated as
active degrees of freedom. Though the analysis in Ref.~\cite{GomezDumm:2003ku} allows to reproduce the experimental data on $\tau \rightarrow
 \pi \pi \pi \nu_{\tau}$ by fitting a few free parameters in this effective Lagrangian, it soon would be seen that the results of this fit 
are not compatible with theoretical expectations from short-distance $QCD$ constraints \cite{Cirigliano:2004ue}. We believe that the inconsistency
 can be attributed to the usage of an ansatz for the off-shell width of the a$_1(1260)$ resonance, which was introduced ad-hoc in Ref.~\cite{GomezDumm:2003ku}.
 The aim of our work was to reanalyse $\tau\to\pi\pi\pi\nu_\tau$ processes within the same general scheme, now considering the energy-dependent
 width of the a$_1(1260)$ state within a proper R$\chi$T framework. The last issue, that is one of the major developments of our work is considered in 
detail in Section \ref{3pi_a1width}.\\
\hspace*{0.5cm}Although this chapter is based in Ref.\cite{Dumm:2009va}, the material covered in Sects. \ref{3pi_LowE}, \ref{3pi_dGdsij} 
and \ref{3pi_SF} is presented in this Thesis for the first time.\\
\section{The axial-vector current in $\tau^-\to (\pi\pi\pi)^- \nu_\tau$ decays} \label{3pi_FF}
\hspace*{0.5cm}Our effective Lagrangian will include the pieces given in Eqs. (\ref{p2-u}), (\ref{R}) and (\ref{LVAPb}) \footnote{Notice that
 we only consider the effect of spin-one resonances. Given the vector character of the $SM$ couplings of the hadronic matrix elements in $\tau$
 decays, form factors for these processes are ruled by vector and axial-vector resonances. Notwithstanding those form factors are given,
in the $\tau \rightarrow PPP \nu_{\tau}$ decays, by a four-point Green function where other quantum numbers might play a role, namely scalar 
and pseudoscalar resonances \cite{Jamin:2006tj, Jamin:2000wn, Buettiker:2003pp, DescotesGenon:2006uk}. Among these, in the three pion tau decay
 modes, the lightest state -that one could expect to give the dominant contribution- is the $\sigma$ or $f_0(500)$. As we assume the $N_C 
\rightarrow \infty$ limit, the nonet of scalars corresponding to the $f_0(500)$ is not considered. This multiplet is generated by rescattering
 of the ligthest pseudoscalars and then subleading in the $1/N_C$ expansion \cite{Cirigliano:2003yq} (See, however, Ref.~\cite{Nieves:2011gb}).}. 
These decays are worked out considering exact isospin symmetry, so the corresponding hadronic matrix elements will be
\begin{equation} \label{eq:tmu1}
T_{\pm \mu}(p_1,p_2,p_3) \,  =  \,
 \langle  \pi_1(p_1)\pi_2(p_2)\pi^{\pm}(p_3)  |\, A_\mu\,
 e^{i {\cal L}_{QCD}} | 0  \rangle \, .
\end{equation}
Outgoing states $\pi_{1,2}$ correspond here to $\pi^-$ and $\pi^0$ for upper and lower signs in $T_{\pm\mu}$, respectively. The hadronic tensor 
is written in terms of three form factors following Eq. (\ref{generaldecomposition_3mesons}), with $F_4^V(Q^2,\,s_1,\,s_2)\,=\,0$ because we have 
no vector current contribution. Since the contribution of $F_3^A(Q^2,\,s_1,\,s_2)$ -carrying pseudoscalar degrees of freedom- to the spectral
 function of $\tau\to\pi\pi\pi\nu_\tau$ goes like $m_{\pi}^4/Q^4$ and, accordingly, it is very much suppressed with respect to those coming from 
$F_1^A(Q^2,\,s_1,\,s_2)$ and $F_2^A(Q^2,\,s_1,\,s_2)$, we will not consider it in the following.\\
\begin{figure}[t]
\begin{center}
\vspace*{0.9cm}
\includegraphics[scale=0.7,angle=0]{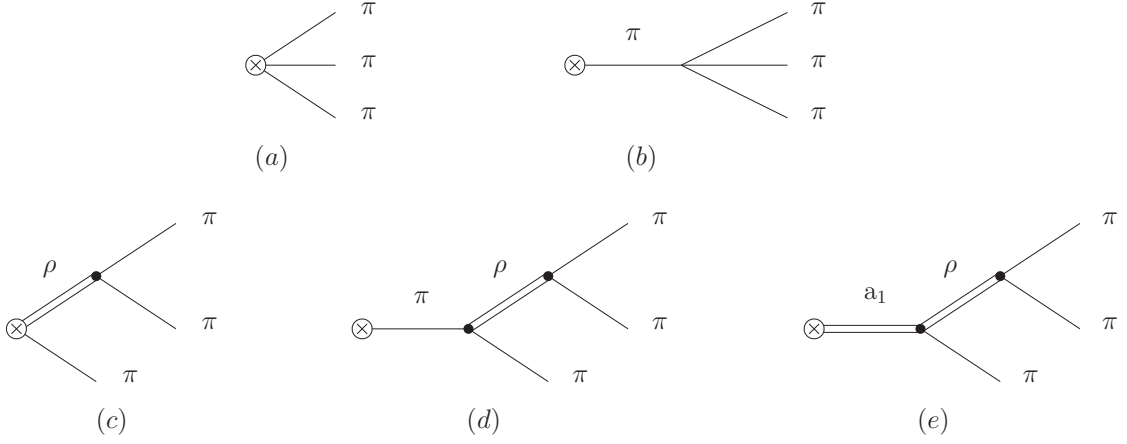}
\caption[]{\label{fig:tau3pidiags} \small{Diagrams contributing to the hadronic axial-vector form factors $F_i$~: (a) and (b) contribute to $F_1^{\chi}$,
(c) and (d) to $F_1^R$ and (e) to $F_1^{RR}$.}}
\end{center}
\end{figure}
\hspace*{0.5cm}The evaluation of the form factors $F_1$ and $F_2$ within the context of $R\chi T$ has been carried out in 
Ref.~\cite{GomezDumm:2003ku}. One has~:
\begin{equation}
 F_{\pm i} \ = \ \pm \left(
 F_i^{\chi} \, + \, F_i^{\mathrm{R}} \, + \, F_i^{\mathrm{RR}}\right)
\ ,\qquad i=1,2\ ,
\end{equation}
where the different contributions correspond to the diagrams in Figure~\ref{fig:tau3pidiags}. In terms of the Lorentz invariants $Q^2$, $s=(p_1+p_3)^2$,
 $t=(p_2+p_3)^2$ and $u=(p_1+p_2)^2$ (notice that $u = Q^2-s-t+3m_\pi^2$) these contributions are given by~\cite{GomezDumm:2003ku}
\begin{eqnarray}
\label{eq:t1r}
F_1^{\chi}(Q^2,s,t) & = & - \frac{2\sqrt{2}}{3 F} \nonumber \\
F_1^{\mathrm{R}}(Q^2,s,t) & = & \frac{\sqrt{2}\,F_V\,G_V}{3\,F^3} \left[
\, \frac{3\,s}{s-M_V^2} \, - \,
\left( \frac{2 G_V}{F_V} - 1 \right) \, \left(
\, \frac{2 Q^2-2s-u}{s-M_V^2} \, + \, \frac{u-s}{t-M_V^2} \,
\right)\right] \;\; \nonumber \\
F_1^{\mathrm{RR}}(Q^2,s,t) & = & \frac{4 \, F_A \, G_V}{3 \,F^3} \,
 \frac{Q^2}{Q^2-M_A^2} \, \bigg[- \, (\lambda' + \lambda'')
 \, \frac{3\,s}{s-M_V^2} \,  \\
 & & \qquad\qquad\qquad\qquad\ \ + \, \, H(Q^2,s) \, \frac{2 Q^2 + s -
u}{s-M_V^2} \, + \,  H(Q^2,t) \, \frac{u-s}{t-M_V^2}\bigg] \ ,\nonumber
\end{eqnarray}
where
\begin{equation} \label{eq:fq2}
H(Q^2,x)  =  - \,\lambda_0\, \frac{m_\pi^2}{Q^2} \, +  \,
\lambda'\, \frac{x}{Q^2} \, + \,  \lambda''  \; ,
\end{equation}
$\lambda_0$, $\lambda'$ and $\lambda''$ being linear combinations of the $\lambda_i$ couplings that can be read in
Eq.~(\ref{LVAPa}). Bose symmetry under the exchange of the two identical pions in the final state implies that the form factors 
$F_1$ and $F_2$ are related by $F_2(Q^2,s,t) = F_1(Q^2,t,s)$.\\
\hspace*{0.5cm}The resonance exchange approximately saturates the phenomenological values of the ${\cal O}(p^4)$ couplings in the standard
$\chi PT$ Lagrangian. This allows to relate both schemes in the low energy region, and provides a check of our results in the limit $Q^2\ll
M_V^2$. This check has been performed \cite{GomezDumm:2003ku}, verifying the agreement between our expression Eq.\ (\ref{eq:t1r}) ---two-resonance 
exchange terms do not contribute at this order--- and the result obtained within $\chi PT$ in Refs.\ \cite{Colangelo:1996hs, Decker:1993ay} 
coming from saturation by vector meson resonances of the ${\cal O}(p^4)$ couplings~:
\begin{equation}
\label{chpt}
\left. T_{\pm\mu}^{\chi PT}\right|_{1^+}\, = \,
\mp \,\frac{2\sqrt{2}}{3 F}
\left[ \left( 1 + \frac{3\,s}{2\,M_V^2} \right) V_{1\mu} +
\left( 1 + \frac{3\,t}{2\,M_V^2} \right) V_{2\mu} \right]\,
+ \mbox{ chiral loops }  \, + \, {\cal O}(p^6) \; \;.
\end{equation}
\hspace*{0.5cm}As an aside, it is worth to point out that this low--energy behaviour is not fulfilled by all phenomenological models proposed
 in the literature. In particular, in the widely used $KS$ model \cite{Kuhn:1990ad} the hadronic amplitude satisfies
\begin{equation}
T_{\pm\mu}^{(KS)}\;\;
\mapright{\; \; s,t\,\ll\, M_V^2 \; \; }
\, \mp \,\frac{2\sqrt{2}}{3 F}
\left[ \left( 1 + \frac{s}{M_V^2} \right) V_{1\mu} +
\left( 1 + \frac{t}{M_V^2} \right) V_{2\mu} \right] \;.
\label{ks}
\end{equation}
Thus, while the lowest order behaviour is correct (it was constructed to be so), it is seen that the $KS$ model fails to reproduce the 
$\chi PT$ result at the next--to--leading order. Accordingly this model is not consistent with the chiral symmetry of $QCD$.\\
\section{Short-distance constraints ruled by $QCD$} \label{3pi_Shortdistance}
\hspace*{0.5cm}Besides the pion decay constant $F$, the above results for the form factors $F_i$ depend on six combinations of the coupling 
constants in the Lagrangian ${\cal L}_{\rm R \chi T}$, namely $F_V$, $F_A$, $G_V$, $\lambda_0$, $\lambda'$ and $\lambda''$ and the masses 
$M_V$, $M_A$ of the vector and axial-vector nonets. All of them are in principle unknown parameters. However, it is clear that 
${\cal L}_{\rm R \chi T}$ does not represent an effective theory of $QCD$ for arbitrary values of its couplings. Though the determination of 
the effective parameters from the underlying theory is still an open problem, one can get information on the couplings by assuming that the 
resonance region ---even when one does not include the full phenomenological spectrum--- provides a bridge between the chiral and perturbative
regimes~\cite{Ecker:1989yg}. This is implemented by matching the high energy behaviour of Green functions (or related form factors) evaluated
within the resonance theory with asymptotic results obtained in perturbative $QCD$~\cite{Ecker:1989yg, Cirigliano:2004ue, RuizFemenia:2003hm, 
Cirigliano:2005xn, Moussallam:1997xx, Knecht:2001xc, Mateu:2007tr, Amoros:2001gf}. In the $N_C \rightarrow \infty$ limit, and within the 
approximation of only one nonet of vector and axial-vector resonances, the analysis of the two-point Green functions $\Pi_{V,A}(q^2)$ and the
 three-point Green function $VAP$ of $QCD$ currents with only one multiplet of vector and axial-vector resonances lead to the following 
constraints \cite{Pich:2002xy}~:\\
\begin{itemize}
\item[i)] By demanding that the two-pion vector form factor vanishes at high momentum transfer one obtains the condition $F_V \, G_V =
F^2$~\cite{Ecker:1989yg}.
\item[ii)] The first Weinberg sum rule~\cite{Weinberg:1967kj} leads to $F_V^2 - F_A^2 = F^2$, and the second Weinberg sum rule gives $F_V^2 \,
M_V^2 \, = \, F_A^2 \, M_A^2$~\cite{Ecker:1988te}.
\item[iii)] The analysis of the VAP Green function~\cite{Cirigliano:2004ue} gives for the coupling combinations $\lambda_0$, $\lambda'$ and
$\lambda''$ entering the form factors in Eq.~(\ref{eq:t1r}) the following results~:
\begin{eqnarray}
 \lambda' & = & \frac{F^2}{2 \, \sqrt{2} \, F_A \, G_V} \; = \;
\frac{M_A}{2 \, \sqrt{2} \, M_V} \,, \label{eq:lam1} \\[3.5mm]
\lambda'' & = & \frac{2 \, G_V \, - F_V}{2 \, \sqrt{2} \, F_A} \; = \;
\frac{M_A^2 - 2 M_V^2}{2 \, \sqrt{2} \, M_V \, M_A} \, , \label{eq:lam2} \\[3.5mm]
4 \, \lambda_0 & = & \lambda' + \lambda'' \; = \; \frac{M_A^2-M_V^2}{\sqrt{2} \, M_V \, M_A} \, , \label{eq:lam3}
\end{eqnarray}
where the second equalities in Eqs.~(\ref{eq:lam1}) and (\ref{eq:lam2}) are obtained using the above relations i) and ii).
\end{itemize}
As mentioned above, $M_V$ and $M_A$ stand for the masses of the vector and axial-vector resonance nonets, in the chiral and large-$N_C$ limits. 
A phenomenological analysis carried out in this limit~\cite{Mateu:2007tr} shows that $M_V$ is well approximated by the $\rho(770)$ mass, whereas 
for the axial-vector mass one gets $M_{{\rm a}_1}^{1/N_C} \equiv M_A = 998 (49)$~MeV (which differs appreciably from the presently accepted
 value of $M_{\mathrm{a}_1}$(1260)$=1230\pm40$ MeV).\\
\hspace*{0.5cm}In addition, one can require that the $J=1$ axial-vector spectral function in $\tau\to\pi\pi\pi\nu_\tau$ vanishes for large 
momentum transfer. We consider the axial two--point function $\Pi_A^{\mu\nu}(Q^2)$, which plays in $\tau\to\pi\pi\pi\nu_\tau$ processes the 
same role than the vector--vector current correlator does in the $\tau \rightarrow \pi \pi \nu_{\tau}$ decays, driven by the vector form factor.
 The goal will be to obtain $QCD$--ruled constraints on the new couplings of the resonance Lagrangian. As these couplings do not depend on 
the Goldstone masses we will work in the chiral limit but our results will apply for non--zero Goldstone masses too. In the chiral limit the 
$\Pi_A^{\mu\nu}(Q^2)$ correlator becomes transverse, hence we can write
\begin{equation}
\Pi_A^{\mu\nu}(Q^2)=(Q^\mu Q^\nu-g^{\mu\nu}Q^2)\,\Pi_A(Q^2)\,.
\end{equation}
As in the case of the pion and axial form factors, the function $\Pi_A(Q^2)$ is expected to satisfy an unsubtracted dispersion relation.
This implies a constraint for the $J=1$ spectral function Im$\Pi_A(Q^2)$ in the asymptotic region, namely\cite{Floratos:1978jb}
\begin{equation}
{\rm Im}\Pi_A (Q^2)\,\,
\mapright{\; \; \;  Q^2 \rightarrow \infty \; \; \; }{} \; \; 
\frac{N_C}{12\,\pi} \; \; .
\label{constr}
\end{equation}
Now, taking into account that each intermediate state carrying the appropriate quantum numbers yields a positive contribution to Im$\Pi_A (Q^2)$,
 we have
\begin{equation}
{\rm Im}\Pi_A (Q^2)\,\geq\,
-\frac{1}{3\,Q^2} \int \mathrm{d}\Phi\; \left(T^\mu|_{1^+}\right)
\, \left(T_\mu|_{1^+}\right)^\ast\,,
\end{equation}
d$\Phi$ being the differential phase space for the three--pion state. The constraint in Eq.\ (\ref{constr}) then implies
\begin{equation}
\lim_{Q^2\to\infty}\;
\int_0^{Q^2} ds\; \int_0^{Q^2-s} dt\; \frac{W_A}{(Q^2)^2} = 0 \,,
\label{cond}
\end{equation}
where $W_A$ is the structure function defined in Eq.~(\ref{structure functions three meson case}) \footnote{As expected from partial conservation
 of the axial-vector current ($PCAC$), the analogous relation is automatically fulfilled by $W_{SA}$.}. It can be seen that the condition in 
Eq.~(\ref{cond}) is not satisfied in general for arbitrary values of the coupling constants in the chiral interaction Lagrangian. In fact, it
 is found that this constraint leads to the relations in Eqs.~(\ref{eq:lam1}) and (\ref{eq:lam2}), showing the consistency of the procedure 
\footnote{These results and those in Section \ref{KKpi_QCDconstraints} have been obtained using the program $MATHEMATICA$ \cite{Mathematica}.}.\\
\hspace*{0.5cm}The above constraints allow in principle to fix all six free parameters entering the form factors $F_i$ in terms of the vector
 and axial-vector masses $M_V$, $M_A$. However the form factors in Eq.~(\ref{eq:t1r}) include zero--width $\rho$(770) and a$_1$(1260) propagator
 poles, which lead to divergent phase--space integrals in the calculation of $\tau\to\pi\pi\pi\nu_\tau$ decay widths. As stated above, in 
order to regularize the integrals one should take into account the inclusion of resonance widths, which means to go beyond the leading order 
in the $1/N_C$ expansion. To account for the inclusion of $NLO$ corrections we perform the substitutions~:
\begin{equation}
 \frac{1}{M_{R_j}^2-q^2} \; \; \longrightarrow \;\; \frac{1}{M_{j}^2-q^2- \, i \,
M_{j} \, \Gamma_{j}(q^2)} \; ,
\label{eq:subst}
\end{equation}
Here $R_j=V,A$, while the subindex $j=\rho,{\rm a}_1$ on the right hand side stands for the corresponding physical state.\\
\hspace*{0.5cm}The substitution in Eq.~(\ref{eq:subst}) implies the introduction of additional theoretical inputs, in particular, the behaviour
 of resonance widths off the mass shell. This issue is studied in detail in Appendix C 
and Section \ref{3pi_a1width}. In the 
following, we will compare it both to the popular width developed in the $KS$ model~\cite{Kuhn:1990ad} and to the proposal of the earlier study within
 $R\chi T$, where this off-shell width was added by hand.\\
\subsection{Expressions for the off-shell width of the a$_1$ resonance}\label{3pi_a1width}
\begin{figure}[h]
\begin{center} \label{fig_transverse_part_orrelators_axial}
\vspace*{0.9cm}
\includegraphics[scale=1.0,angle=0]{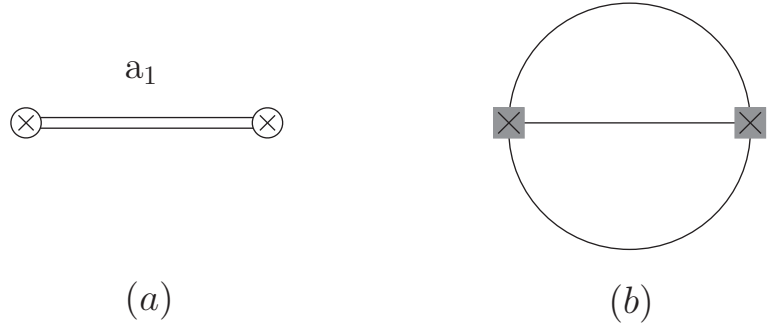}
\caption{ \small{Diagrams contributing to the transverse part of the correlator of axial-vector currents in 
Eq.~(\ref{eq:correa2}). Diagram (a) gives $\Pi_{(0)}^{{\rm a}_1}$ and diagram (b) provides $\Pi_{(1)}^{{\rm a}_1}$. The squared
axial-vector current insertion in (b) corresponds to the sum of the diagrams in Figure~\ref{fig:tau3pidiags}. The double line in (a) indicates the 
a$_1$ resonance intermediate state. Solid lines in (b) indicate any Goldstone bosons that carry the appropriate quantum numbers. }}
\end{center}
\end{figure}

\begin{figure}[!t]
\begin{center}
\vspace*{0.9cm}
\includegraphics[scale=0.62,angle=-90]{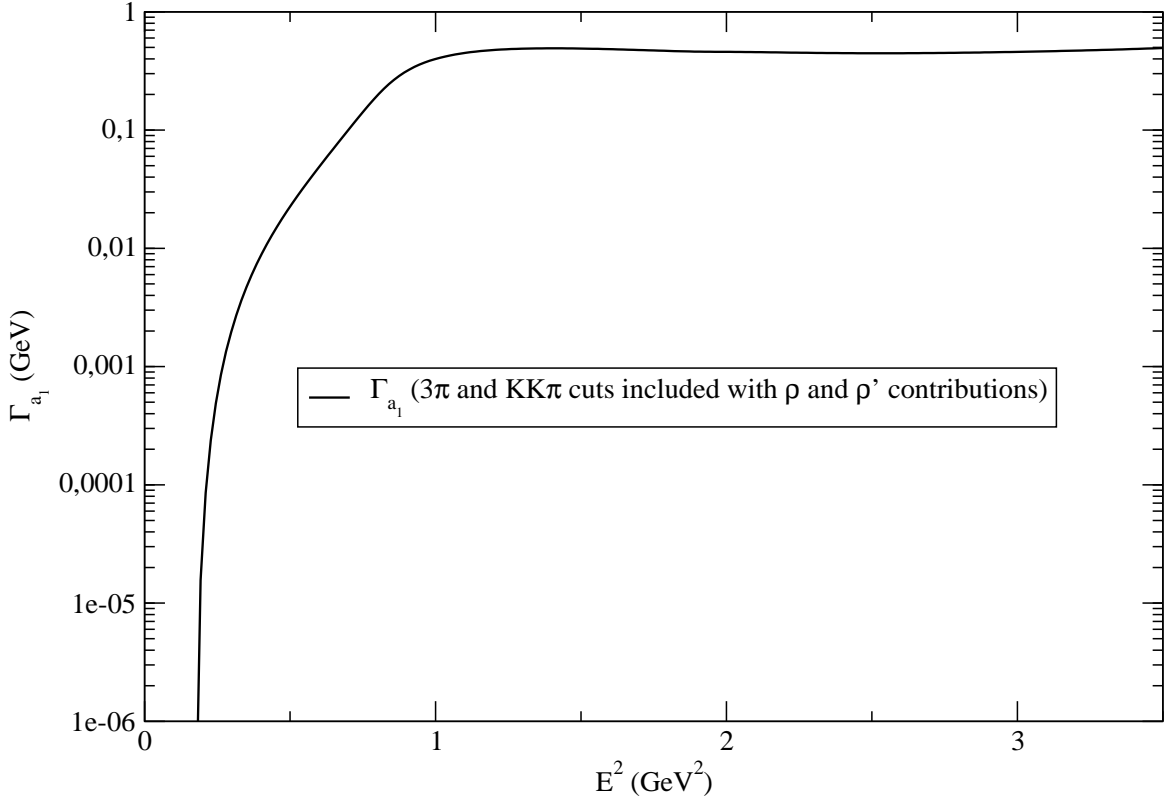}
\caption[]{\label{fig:Gamma_a1} \small{Plot of our expression for $\Gamma_{\mathrm{a}_1}(Q^2)$ using the values of the couplings discussed in Sect. 
\ref{3pi_Pheno}.}}
\end{center}
\end{figure}

\hspace*{0.5cm}The definition that is given in Appendix C 
for the spin-one resonance width -and applied for the vector case-
holds for axial-vector mesons as well, but it would amount to evaluate the axial-vector-axial-vector current correlator with absorptive 
cuts of three $pGs$ (two-loops diagrams) within $\RCT$. This motivated the chiral based off-shell behaviour proposal in Ref.~\cite{GomezDumm:2003ku},
 an oversimplified approach in which the a$_1$ width was written in terms of three parameters, namely the on-shell width
$\Gamma_{{\rm a}_1}(M_{{\rm a}_1}^2)$, the mass $M_{{\rm a}_1}$ and an exponent $\alpha$ ruling the asymptotic behaviour~:
\begin{equation} \label{old a1 width}
\Gamma_{\mathrm{a}_1}(q^2)\,=\,\Gamma_{\mathrm{a}_1}(M_A^2)\,\frac{\phi(q^2)}{\phi(M_A^2)}\,\left(\frac{M_A^2}{q^2}\right)^\alpha\,\theta\left(q^2\,-\,9m_\pi^2\right)\,
\end{equation}
where
\begin{eqnarray}
\phi(q^2) & = & q^2\,\int\,\mathrm{d}s\,\mathrm{d}t\left\lbrace V_1^2\,|\mathrm{BW}_\rho(s)|^2\,+\,V_2^2\,|\mathrm{BW}_\rho(t)|^2\right.\nonumber\\
& & \left.+2(V_1\cdot V_2)\,\Re e\left[ \mathrm{BW}_\rho(s)\,\mathrm{BW}_\rho(t)^*\right] \right\rbrace\,,
\end{eqnarray}
and
\begin{equation}
\mathrm{BW}_\rho(q^2)\,=\,\frac{M_V^2}{M_V^2\,-\,q^2\,-\,iM_V\,\Gamma_\rho(q^2)}\,
\end{equation}
is the usual Breit-Wigner function for the $\rho$ (770) meson resonance shape, the energy-dependent width $\Gamma_\rho(q^2)$ is given by 
Eq.~(\ref{rhowidth}), and the integral extends over the $3\pi$ phase space. The vectors $V_1$ and $V_2$ and the Mandelstam variables $s$ 
and $t$ entering the function $\phi(x\,=\,q^2,\,M_A^2)$ are defined following the general conventions given in Sect. \ref{Hadrondecays_Three_meson_decays_MI}.
 One can check them explicitly in Ref.~\cite{GomezDumm:2003ku}.\\
\hspace*{0.5cm}A fundamental result of this Thesis is the improvement in the description of the off-shell axial-vector widths. We will follow 
the paper \cite{Dumm:2009va} in our explanation.\\
We propose here a new parameterization of the a$_1(1260)$ width that is compatible with the R$\chi$T framework used throughout our analysis.
 As stated, to proceed as in the $\rho$ meson case, one faces the problem of dealing with a resummation of two-loop diagrams in the two-point
 correlator of axial-vector currents. However, it is still possible to obtain a definite result by considering the correlator up to the 
two-loop order only. The width can be defined in this way by calculating the imaginary part of the diagrams through the well-known 
Cutkosky rules.\\
\hspace*{0.5cm}Let us focus on the transversal component, $\Pi_T(Q^2)$, of the two-point Green function~:
\begin{eqnarray}\label{eq:correa1}
\Pi_{\mu\nu}^{33} & = & i \;\int\;{\rm d}^4x\; e^{iQ\cdot x}
\;\langle\, 0|\, T[A_\mu^3(x)\,A_\nu^3(0)]\, |0\, \rangle \nonumber \\
& = & (Q^2\,g_{\mu\nu}\,-\,Q_\mu Q_\nu)\;\Pi_{T}(Q^2) \, + \, Q_{\mu} Q_{\nu} \, \Pi_L(Q^2) \ ,
\end{eqnarray}
where $A_{\mu}^i = \overline{q} \gamma_{\mu} \gamma_5 \frac{\lambda^i}{2} q$. We will 
assume that the transversal contribution is dominated by the $\pi^0$ and the neutral component of the ${\rm a}_1(1260)$ triplet~: 
$\Pi_T(Q^2) \simeq \Pi^{\pi^0}(Q^2) +  \Pi^{{\rm a}_1}(Q^2)$.
Following an analogous procedure to the one in Ref.~\cite{GomezDumm:2000fz}, we write $\Pi^{{\rm a}_1}(Q^2)$ as the sum
\begin{equation} \label{eq:correa2}
\Pi^{{\rm a}_1}(Q^2) \ = \ \Pi^{{\rm a}_1}_{(0)}\; + \;\Pi^{{\rm a}_1}_{(1)}\; +
\;\Pi^{{\rm a}_1}_{(2)}\; + \;\dots \ \ ,
\end{equation}
where $\Pi^{{\rm a}_1}_{(0)}$ corresponds to the tree level amplitude, $\Pi^{{\rm a}_1}_{(1)}$ to a two-loop order contribution,
 $\Pi^{{\rm a}_1}_{(2)}$ to a four-loop order contribution, etc. The diagrams to be included are those which have an absorptive part in 
the $s$ channel. The first two terms are represented by diagrams (a) and (b) in Figure~\ref{fig_transverse_part_orrelators_axial}, 
respectively, where effective vertices denoted by a square correspond to the sum of the diagrams in Figure~\ref{fig:tau3pidiags}. Solid lines in the 
diagram (b) of Figure~\ref{fig_transverse_part_orrelators_axial} correspond to any set of light pseudoscalar mesons that carry the 
appropriate quantum numbers to be an intermediate state.\\
\hspace*{0.5cm}The first term of the expansion in Eq.~(\ref{eq:correa2}) arises from the coupling driven by $F_A$ in the effective 
Lagrangian~(\ref{R}). We find
\begin{equation}
\Pi^{{\rm a}_1}_{(0)} \ = \ -\,\frac{F_A^2}{M_{{\rm a}_1}^2-Q^2} \ .
\label{pi0}
\end{equation}
Thus, if the series in Eq.~(\ref{eq:correa2}) can be resummed one should get
\begin{equation}
\Pi^{{\rm a}_1}(Q^2) \ = \ -\,\frac{F_A^2}{M_{{\rm a}_1}^2-Q^2 + \Delta(Q^2)} \ ,
\label{pitot}
\end{equation}
and the energy dependent width of the a$_1(1260)$ resonance can be defined by
\begin{equation}
M_{{\rm a}_1}\,\Gamma_{{\rm a}_1}(Q^2)\ = \ - \, {\rm Im}\,\Delta(Q^2) \ .
\end{equation}
Now if we expand $\Pi^{{\rm a}_1}(Q^2)$ in powers of $\Delta$ and compare term by term with the expansion in Eq.~(\ref{eq:correa2}), from
 the second term we obtain
\begin{equation}
\Delta(Q^2) \ = \
-\,\frac{(M^2_{{\rm a}_1}-Q^2)}{\Pi^{{\rm a}_1}_{(0)}}\ \Pi^{{\rm a}_1}_{(1)} \ .
\end{equation}
The off-shell width of the a$_1(1260)$ resonance will be given then by
\begin{equation}
\Gamma_{{\rm a}_1}(Q^2)\ = \
\,\frac{(M^2_{{\rm a}_1}-Q^2)}{M_{{\rm a}_1}\,\Pi^{{\rm a}_1}_{(0)}}
\ {\rm Im}\,\Pi^{{\rm a}_1}_{(1)} \ .
\label{gamma_pi1}
\end{equation}
\hspace*{0.5cm}As stated, $\Pi^{{\rm a}_1}_{(1)}$ receives the contribution of various intermediate states. These contributions can be calculated within
 our theoretical R$\chi$T framework from the effective Lagrangian in Eqs.~(\ref{R}), (\ref{LVAPb}), (\ref{VJPops}), (\ref{VVPops}), (\ref{VPPP_chirallimit})
 and (\ref{VPPP_chiralcontributions}). In particular, for the intermediate $\pi^+\pi^-\pi^0$ state one has
\begin{eqnarray}
\Pi^{{\rm a}_1}_{(1)}(Q^2) & = & \frac{1}{6Q^2}\;\int\;
\frac{{\rm d}^4p_1}{(2\pi)^4}\;\frac{{\rm d}^4p_2}{(2\pi)^4}\
T^\mu_{1^+}\; T_{1^+\mu}^\ast\
\prod_{i=1}^3\; \frac{1}{p_i^2-m_\pi^2+i\epsilon} \ ,
\end{eqnarray}
where $p_3 = Q -p_1-p_2$, and $T_{1^+}$ is the $1^+$ piece of the hadronic tensor in Eq.~(\ref{eq:tmu1}),
\begin{equation}
T_{1^+}^{\mu} \; = \; V_1^\mu\,F_1 \, + \,  V_2^\mu\,F_2 \ .
\end{equation}
When extended to the complex plane, the function $\Pi^{{\rm a}_1}_{(1)}(z)$ has a cut in the real axis for $z\geq 9m_\pi^2$, where 
${\rm Im}\,\Pi^{{\rm a}_1}_{(1)}(z)$ shows a discontinuity. The value of this imaginary part on each side of the cut can be calculated 
according to the Cutkosky rules as~:
\begin{equation}
{\rm Im}\,\Pi^{{\rm a}_1}_{(1)}(Q^2\pm i\epsilon) = \mp\frac{i}{2}\;\frac{1}{6Q^2}
\;\int\; \frac{{\rm d}^4p_1}{(2\pi)^4}\;\frac{{\rm d}^4p_2}{(2\pi)^4}\
T^\mu_{1^+}\; T_{1^+\mu}^\ast\ \prod_{i=1}^3\; (-2i\pi)\; \theta(p_i^0)\;
\delta(p_i^2-m_\pi^2) \ ,
\end{equation}
with $p_3 = Q -p_1-p_2$ and $Q^2 > 9 m_{\pi}^2$. After integration of the delta functions one finds
\begin{equation}
{\rm Im}\,\Pi^{{\rm a}_1}_{(1)}(Q^2\pm i\epsilon) = \pm\;\frac{1}{192\,Q^4}
\;\frac{1}{(2\pi)^3}\;\int\;{\rm d}s\,{\rm d}t \
T^\mu_{1^+}\; T_{1^+\mu}^\ast \ ,
\end{equation}
where the integrals extend over a three-pion phase space with total momentum squared $Q^2$. Therefore, the contribution of the 
$\pi^+\pi^-\pi^0$ state to the a$_1(1260)$ width will be given by
\begin{equation} \label{eq:Gamma_a1_pi}
\Gamma_{{\rm a}_1}^{\pi}(Q^2) \ = \ \frac{-1}{192(2\pi)^3 F_A^2 M_{{\rm a}_1}}\,
\left( \frac{M_{{\rm a}_1}^2}{Q^2}-1 \right)^2 \; \int
\mathrm{d}s\, \mathrm{d}t\; T^{\mu}_{1^+}\; T^{\ast}_{1^+\mu} \; .
\end{equation}
In the same way one can proceed to calculate the contribution of the intermediate states $K^+K^-\pi^0$, $K^0\bar K^0\pi^0$, $K^-K^0\pi^+$
 and $K^+\bar K^0\pi^-$. The corresponding hadronic tensors $T^K_{1^+}$ can be obtained from Ref.~\cite{Dumm:2009kj} (see Sect. \ref{KKpi_FF}). Additionally one could
 consider the contribution of $\eta \pi \pi$ and $\eta\eta\pi$ intermediate states. However, the first one vanishes in the isospin limit 
because of $G$-parity (see Chapter \ref{eta}) and the second one is suppressed by a tiny upper bound for the branching ratio 
\cite{Amsler:2008zzb,Gilman:1987my} and they will not be taken into account.\\
\hspace*{0.5cm}In this way we have \footnote{It is important to stress that we do not intend to carry out the resummation of the
series in Eq.~(\ref{eq:correa2}). In fact, our expression in Eq.~(\ref{gamma_pi1}) would correspond to the result of the resummation if
this series happens to be geometric, which in principle is not guaranteed \cite{GomezDumm:2000fz}.}
\begin{equation} \label{eq:Gamma_a1_tot}
\Gamma_{{\rm a}_1}(Q^2) \ = \ \Gamma_{{\rm a}_1}^\pi(Q^2)\,  \theta(Q^2-9m_{\pi}^2) \
+ \ \Gamma_{{\rm a}_1}^K(Q^2) \, \theta(Q^2-(2 m_K+m_{\pi})^2)  \ ,
\end{equation}
where
\begin{equation} \label{eq:Gamma_a1_pimod}
\Gamma_{{\rm a}_1}^{\pi,K}(Q^2) \ = \ \frac{-S}{192(2\pi)^3 F_A^2 M_{{\rm a}_1}}\;
\left( \frac{M_{{\rm a}_1}^2}{Q^2}-1 \right)^2 \; \int \mathrm{d}s\, \mathrm{d}t\;
T^{\pi,K\mu}_{1^+}\; T^{\pi,K\ast}_{1^+\mu} \ .
\end{equation}
Here $\Gamma_{{\rm a}_1}^{\pi}(Q^2)$ recalls the three pion contributions and $\Gamma_{{\rm a}_1}^K(Q^2)$ collects the contributions of 
the $KK\pi$ channels. In Eq.~(\ref{eq:Gamma_a1_pimod}) the symmetry factor $S = 1/n!$ reminds the case with $n$ identical particles in the
 final state. It is also important to point out that, contrarily to the width proposed in Ref.~\cite{GomezDumm:2003ku} 
[$\Gamma_{{\rm a}_1}(Q^2)$, in Eq.~(\ref{old a1 width})], the on-shell width $\Gamma_{{\rm a}_1}(M_{{\rm a}_1}^2)$ is
now a prediction and not a free parameter.\\
\hspace*{0.5cm}With this off-shell width a very accurate description of the related observables will be given in later sections \footnote{Our computation, based on the optical 
theorem, does not give the corresponding real part of the loop function, which we have disregarded. Although this approximation might be supported numerically it 
induces a violation of analiticity. We have also neglected systematically three-body final-state interactions for the moment. They may be important in the available
phase space \cite{Anisovich:1996tx, Niecknig:2012sj}.}. In other formalisms, like that of the hidden local symmetry chiral models \cite{Achasov:2010ex} one needs to restore to a extremely unnatural off-shell width description that reaches 
the value of $10$ GeV for $Q^2\sim2$.$5$ GeV$^2$. In Figure~\ref{fig:Gamma_a1} our expression for $\Gamma_{\mathrm{a}_1}$ is plotted as a function of the 
invariant mass squared the hadronic system has.\\
\section{Phenomenology of the $\tau^-\to(\pi\pi\pi)^-\nu_\tau$ process}\label{3pi_Pheno}
\subsection{The contribution of the $\rho(1450)$} \label{3pi_rhoprime}
\begin{figure}[!t]
\begin{center}
\vspace*{0.9cm}
\includegraphics[scale=0.62,angle=-90]{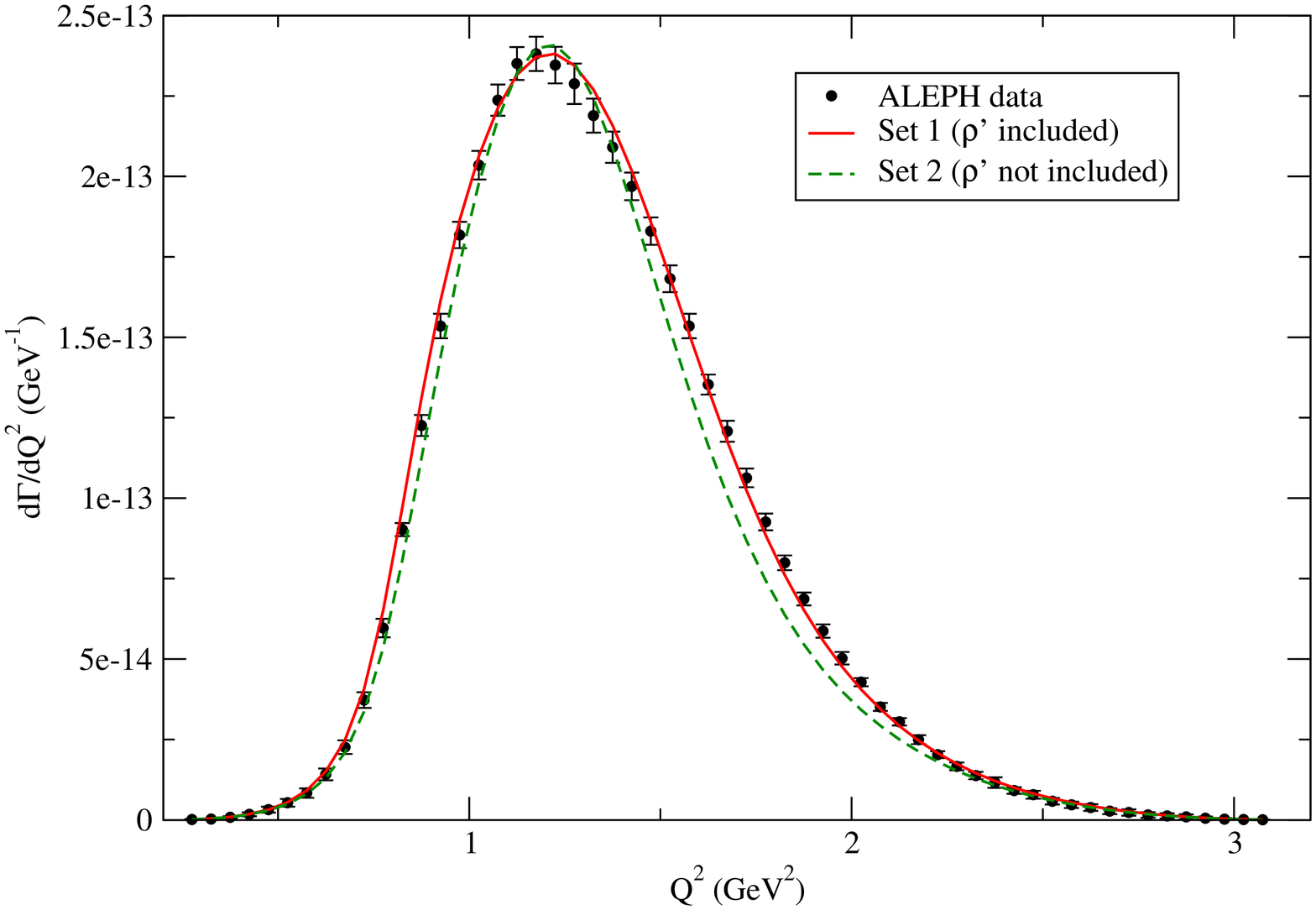}
\caption[]{\label{fig:spectralfunction3pi} \small{Comparison between the theoretical $M_{3\pi}^2$-spectra of the $\tau^- \rightarrow \pi^+ \pi^- \pi^-
\nu_{\tau}$  with $ALEPH$ data \cite{Barate:1998uf}. Set~1 corresponds to the values of the parameters~: $F_V = 0\mathrm{.}180 \, \mathrm{GeV}$, 
$F_A = 0\mathrm{.}149 \, \mathrm{GeV}$, $M_{\mathrm{a}_1} = 1\mathrm{.}120 \, \mathrm{GeV}$, $\beta_{\rho'} = -0\mathrm{.}25$, $M_A \simeq 0\mathrm{.}91 \, \mathrm{GeV}$. Set~2 
corresponds to the values of the parameters~: $F_V = 0\mathrm{.}206 \, \mathrm{GeV}$, $F_A = 0\mathrm{.}145 \, \mathrm{GeV}$, $M_{\mathrm{a}_1} = 1\mathrm{.}150 \, \mathrm{GeV}$,
 $\beta_{\rho'}=0$, i.e. without the inclusion of the $\rho'$. In the case of Set 2 the overall normalization of the spectrum has been 
corrected by a $15 \%$ to match the experimental data.}}
\end{center}
\end{figure}

\hspace*{0.5cm}It turns out that, though some flexibility is allowed around the predicted values for the parameters, the region between
$1$.$5-2$.$0$ GeV$^2$ of the three pion spectrum is still poorly described by the scheme we have proposed here. This is not surprising as
the $\rho(1450)$, acknowledgeably rather wide, arises in that energy region. We find that it is necessary to include, effectively, the role 
of a $\rho' \equiv \rho(1450)$, in order to recover good agreement with the experimental data. The $\rho'$ belongs to a second, heavier, 
multiplet of vector resonances that we have not considered in our procedure. Its inclusion would involve a complete new set of analogous 
operators to the ones already present in $\mathcal{L}_{R\chi T}$, Eqs.~(\ref{R}), (\ref{LVAPb}), with the corresponding new couplings. This
 is beyond the scope of our analysis. However we propose to proceed by performing the following substitution in the $\rho(770)$ propagator~\footnote{This replacement does 
not alter the short-distance vanishing of the form factor and it changes its chiral limit accordingly to the contribution expected from the second multiplet of vector resonances.}:
\begin{equation}
 \frac{1}{M_{\rho}^2-q^2-iM_{\rho} \Gamma_{\rho}(q^2)} \longrightarrow
\frac{1}{1+\beta_{\rho'}} \, \left[ \frac{1}{M_{\rho}^2-q^2-iM_{\rho} \Gamma_{\rho}(q^2)} \, + \,
\frac{\beta_{\rho'}}{M_{\rho'}^2-q^2-iM_{\rho'} \Gamma_{\rho'}(q^2)} \right] \, ,
\end{equation}
where as a first approximation the $\rho'$ width is given by the decay into two pions~:
\begin{eqnarray}
 \Gamma_{\rho'}(q^2) & = &  \Gamma_{\rho'}(M_{\rho'}^2) \, \frac{M_{\rho'}}{\sqrt{q^2}}\,
\left( \frac{p(q^2)}{p(M_{\rho'}^2)} \right)^3 \, \theta ( q^2 - 4 \,m_{\pi}^2 ) \, , \\ \nonumber
p(x) & = &  \frac{1}{2} \, \sqrt{x-4 m_{\pi}^2} \, .
\end{eqnarray}
For the numerics we use the values $M_{\rho'} = 1$.$465$ GeV and $\Gamma_{\rho'}(M_{\rho'}^2) = 400$ MeV as given in Ref.~\cite{Amsler:2008zzb}.
 We find that a good agreement with the spectrum, $d \Gamma/dQ^2$, measured by $ALEPH$ \cite{Barate:1998uf} is reached for the set of values~:
\begin{eqnarray} \label{eq:set1}
F_V \, = \, 0\mathrm{.}180 \, \mathrm{GeV} \; & \; , \;  & \;  F_A \, = \, 0\mathrm{.}149 \, \mathrm{GeV} \; \; \; \; \; \; \;  , \; \; \; \;
\beta_{\rho'} \, = -0\mathrm{.}25  \;,  \nonumber \\
M_V \, = \, 0\mathrm{.}775 \, \mathrm{GeV}  \; & \; , \;  & \;
M_{K^*} \, = \, 0\mathrm{.}8953 \, \mathrm{GeV}\;  \; \; , \; \; \; \;
M_{{\rm a}_1} \, = \, 1\mathrm{.}120 \, \mathrm{GeV}  \, ,
\end{eqnarray}
that we call Set~1. The corresponding width is $\Gamma (\tau \rightarrow \pi \pi \pi \nu_{\tau}) = 2\mathrm{.}09 \, \times \, 10^{-13}$ GeV, in
excellent agreement with the experimental figure $\Gamma (\tau \rightarrow \pi \pi \pi \nu_{\tau})|_{exp} = (2\mathrm{.}11 \pm 0\mathrm{.}02) \, \times \, 
10^{-13}$ GeV \cite{Amsler:2008zzb}. From $F_V$ and $F_A$ in Eq.~(\ref{eq:set1}), and the second Weinberg sum rule we can also 
determine the value of $M_A = F_V M_V / F_A  \simeq 0\mathrm{.}94$ GeV, a result consistent with the one obtained in Ref.~\cite{Mateu:2007tr}.
 If, instead, we do not include the $\rho'$ contribution, the best agreement with experimental data is reached for the values of Set~2~:
\begin{eqnarray} \label{eq:set2}
F_V \, = \, 0\mathrm{.}206 \, \mathrm{GeV} \; & \; , \;  & \;  F_A \, = \, 0\mathrm{.}145 \, \mathrm{GeV} \; \; \; \; \; \; \;  , \; \; \; \;
\beta_{\rho'} \, = 0  \;,  \nonumber \\
M_V \, = \, 0\mathrm{.}775 \, \mathrm{GeV}  \; & \; , \;  & \;
M_{K^*} \, = \, 0\mathrm{.}8953 \, \mathrm{GeV}\;  \; \; , \; \; \; \;
M_{{\rm a}_1} \, = \, 1\mathrm{.}115 \, \mathrm{GeV}  \, ,
\end{eqnarray}
though the branching ratio is off by $15 \%$. A comparison between the results for the $\tau\to\pi\pi\pi\nu_\tau$ spectra obtained from Sets
 1, 2 and the data provided by $ALEPH$ is shown in Figure~\ref{fig:spectralfunction3pi}. Notice that we have corrected the results provided by Set~2 by a 
normalization factor of $1\mathrm{.}15$ in order to compare the shapes of the spectra. Though it is difficult to assign an error to our numerical values,
 by comparing Set~1 and Set~2 we consider that a $15\%$ should be on the safe side. Notice, however, that the error appears to be much smaller
 in the case of $M_{\mathrm{a}_1}$. We should remind, however, that the real part of the two-loop function contributing to the $a_1$ width was not included 
in our approach. This systematic error (that induces a shift in $M_{\mathrm{a}_1}$) prevents us finding the pole parameters associated to our model resonance results.\\
\hspace*{0.5cm}For Set 1 the width of the a$_1(1260)$ is $\Gamma_{\mathrm{a}_1}(M_{\mathrm{a}_1}^2) = 0\mathrm{.}483$ GeV, which, incidentally, is in agreement with the 
figure got in Ref.~\cite{GomezDumm:2003ku} from a fit to the data. The value of $\Gamma_{\mathrm{a}_1}(M_{\mathrm{a}_1}^2)$ quoted in the $PDG$ (2008)~\cite{Amsler:2008zzb}
goes from $250 \, \mathrm{MeV}$ up to $600 \, \mathrm{MeV}$.\\
Our preferred set of values in Eq.~(\ref{eq:set1}) satisfies reasonably well all the short distance constraints pointed out in 
Sect.~\ref{3pi_Shortdistance}, with a deviation from Weinberg sum rules of at most $10 \%$, perfectly compatible with deviations due to the single 
resonance approximation.\\
\subsection{Low-energy description}\label{3pi_LowE}
\begin{figure}[!h]
\begin{center}
\vspace*{1.4cm}
\includegraphics[scale=0.6,angle=-90]{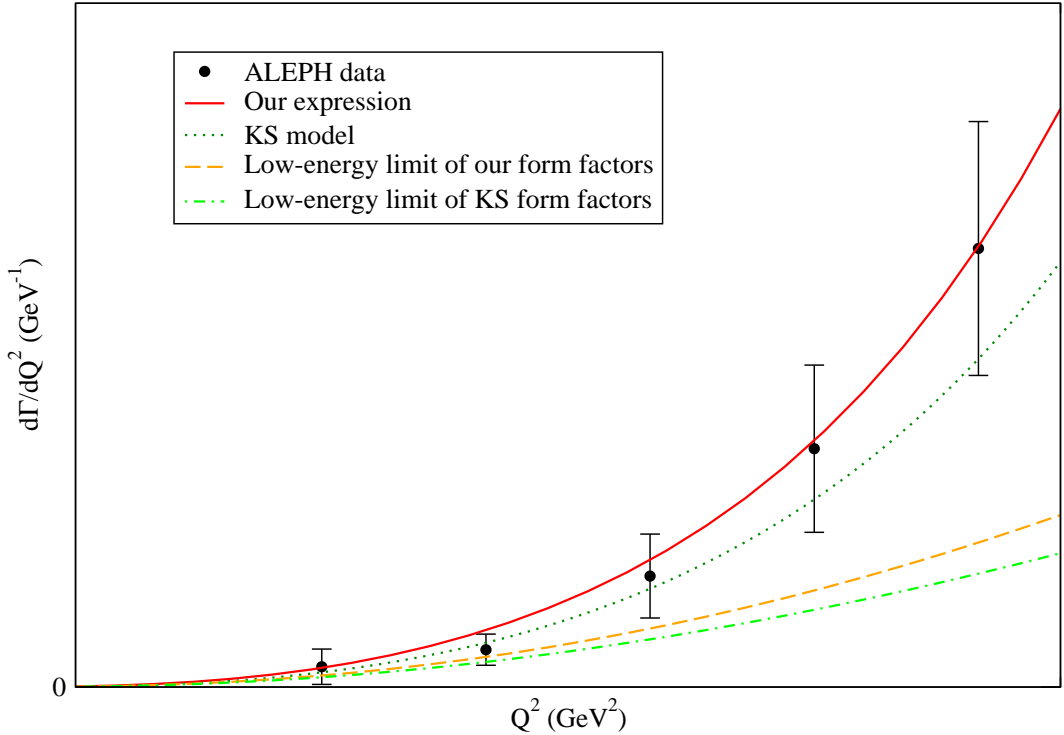}
\caption[]{\label{fig:lowQ23pi} \small{Comparison between the theoretical low-energy $M_{3\pi}^2$-spectra of the $\tau^- \rightarrow \pi^+ \pi^- \pi^-
\nu_{\tau}$  with $ALEPH$ data \cite{Barate:1998uf}. Our results (red solid line) correspond to Set~1, Eq.~(\ref{eq:set1}), and its 
corresponding low-energy limit (orange dashed line) to Eq.(\ref{chpt}). The green dotted line corresponds to the $KS$ results 
\cite{Kuhn:1990ad} and its low-energy limit (green dashed-dotted line) is given in Eq.~\ref{ks}. We observe that the wrong $KS$ description
 at $NLO$ in the chiral expansion is naturally carried on to the whole expression for the spectrum. Moreover, the excellent agreement of 
our prediction with data shows that our working hypothesis of neglecting the effect of $\mathcal{O}(p^4)$ chiral logs is well-based.}}
\end{center}
\end{figure}

\begin{figure}[!h]
\begin{center}
\vspace*{1.0cm}
\includegraphics[scale=0.6,angle=-90]{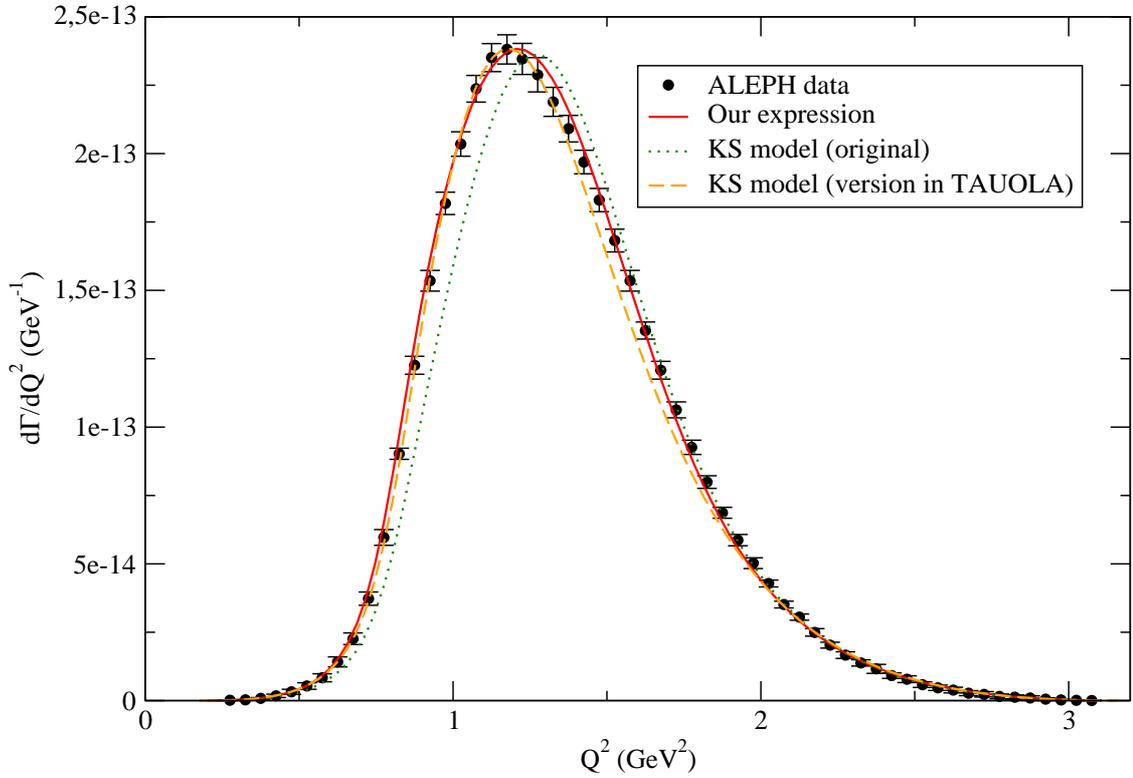}
\caption[]{\label{fig:3piWevsKS} \small{Comparison between the theoretical $M_{3\pi}^2$-spectra of the $\tau^- \rightarrow \pi^+ \pi^- \pi^- \nu_{\tau}$  
with $ALEPH$ data \cite{Barate:1998uf}. Our results correspond to Set~1, Eq.~(\ref{eq:set1}), and they are also compared to the $KS$ 
outcome, as they were given originally~\cite{Kuhn:1990ad} as indicated in Eq.~(\ref{ks}) (see also Fig. 1 of Ref.~\cite{Roig:2008xt}). We observe that the wrong $KS$ description at
 $NLO$ in the chiral expansion is naturally carried on as they are in $TAUOLA$ right now. In the original parameterization one sees that
 the wrong description of the $\mathcal{O}(p^4)$ $\CPT$ terms is carried naturally to the rest of the spectrum. This is corrected in the 
updated parameterization in TAUOLA at the price of including a noticeably large on-shell a$_1$ width ($0$.$6$ GeV) and an unnaturally large 
normalization factor of $1$.$38$.}}
\end{center}
\end{figure}

\hspace*{0.5cm}After that, we take a closer look to the low-$Q^2$ region of the spectrum. In fact, in our approach we have assumed that $\mathcal{O}(p^4)$ 
corrections arising from chiral logs are small, hence the dominant contributions to hadronic amplitudes arise from resonance exchange. In Figure~\ref{fig:lowQ23pi}, 
we can see that our expression fits the data remarkably well at low-$Q^2$ values without any need to improve it by adding the effect of the neglected chiral 
logs \footnote{Close to threshold (i.e.\ for $\sqrt{Q^2}$ well below $M_V$) one is able to explicitly calculate the contributions of $\mathcal{O}(p^4)$ 
chiral logs, therefore their impact can be numerically evaluated. In fact, Ref~\cite{GomezDumm:2003ku} considered this correction by using the
 results in Ref.~\cite{Colangelo:1996hs}.}. Our working hypothesis is thus confirmed. In this plot we also see that the form-factors proposed by the $KS$ model 
induced a systematic departure of the data points increasing with the energy. It is interesting to note that -as we asserted in Chapter \ref{Hadrondecays}- 
the wrong description at $NLO$ in the chiral expansion is naturally carried on to higher energies, once the full expression is included. Indeed, the 
low-energy limits of the $KS$ expressions for the form factors and ours \footnote{As described in Eqs.(\ref{chpt}) and (\ref{ks}).} already show that the $KS$ 
curve is systematically under ours getting farther as the energy increases.\\
\hspace*{0.5cm}In Figure~\ref{fig:3piWevsKS} we can see that the shift induced at low-energies in the original $KS$-model \footnote{The original $KS$ model
used the following values for the parameters: $M_{\mathrm{a}_1}\,=\,1\mathrm{.}251$ GeV and $\Gamma_{\mathrm{a}_1}\left( M_{\mathrm{a}_1}\right)\,=\, 0\mathrm{.}475$ GeV. With these parameters, the 
description of the experimental data available at that time \cite{Albrecht:1986kg} was very good. However, as the experimental errors were reduced ten 
years later by $CLEO-II$~\cite{Ackerstaff:1997dv}, $OPAL$~\cite{Browder:1999fr} and $ALEPH$~\cite{Barate:1998uf}, one noticed that there was a need of 
including another parameter to keep such a good description. In order to keep the expression for the off-shell width the solution adopted in $TAUOLA$ 
was to modify the on-shell a$_1$ width and include a normalization factor to correct the branching ratio. The later width reads $\Gamma_{\mathrm{a}_1}\left( M_{\mathrm{a}_1}
\right)\,=\, 0\mathrm{.}599$ GeV and the normalization factor enhances the decay rate by $\sim 1\mathrm{.}4$.} gets carried on naturally to higher energies. In the later
 $TAUOLA$ parameterization the agreement seemed to be better by introducing a large on-shell a$_1$ width ($0\mathrm{.}6$ GeV) that required to adjust the 
normalization by a factor of order $40\%$ ($1\mathrm{.}38$ in the curve) that appears to be quite unnatural. Even doing so, the description between $1$.$5$ 
and $1\mathrm{.}8$ GeV$^2$ was not good.\\
\subsection{$\frac{\mathrm{d}\Gamma}{\mathrm{d}s_{ij}}$ distributions}\label{3pi_dGdsij}
\begin{figure}[!t]
\begin{center}
\vspace*{1.2cm}
\includegraphics[scale=0.6,angle=-90]{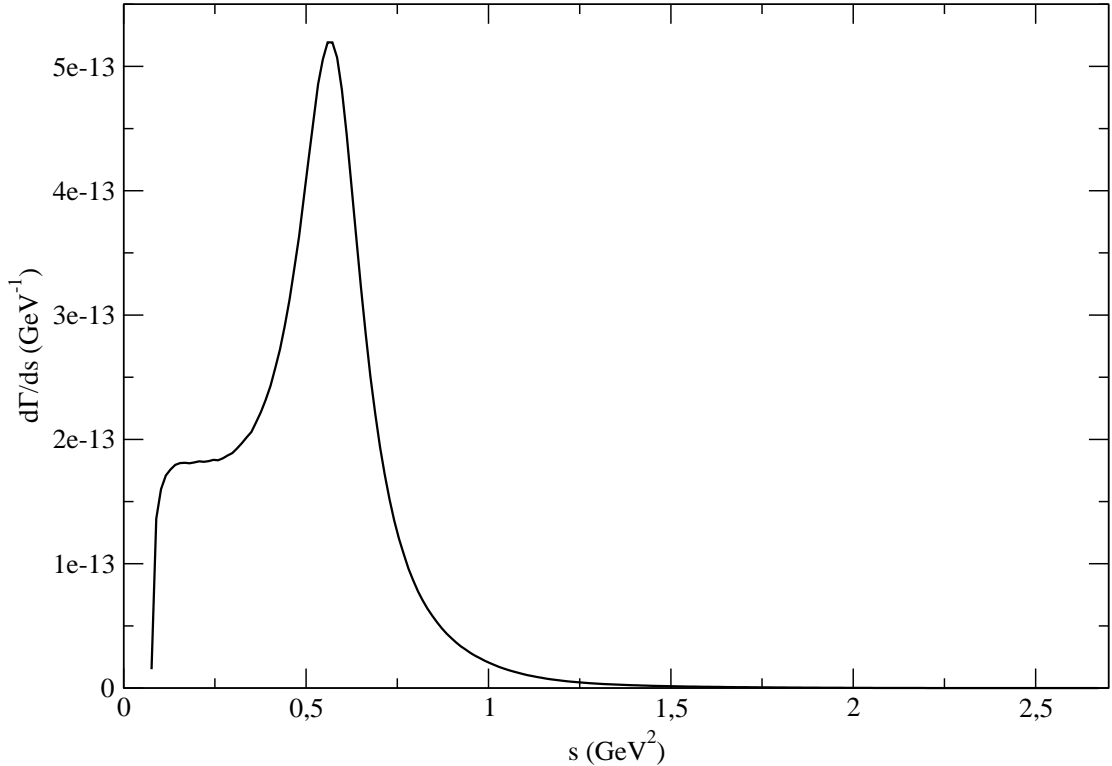}
\caption[]{\label{fig:dGds3pi} \small{Invariant mass distribution of the $\pi^+\pi^-$ pair in $\tau^-\to\pi^+\pi^-\pi^-\nu_\tau$ decays corresponding to our best fit values in eq. (\ref{eq:set1}).}}
\end{center}
\end{figure}

\hspace*{0.5cm}The differential distributions in the invariant masses of pairs of pions, $s_{ij}\,\equiv\,(p_i+p_j)^2\,=\,(Q-p_k)^2$ for $i\neq j\neq k$ and $i,j,k\,=\,1,\,2,\,3$
are also interesting. Neither Ref.~\cite{Barate:1998uf} nor any later publication made a dedicated study of these observables. We show in figure \ref{fig:dGds3pi} the $d\Gamma/ds$ distribution corresponding 
to our best fit values (\ref{eq:set1}). Not surprisingly, the $\rho(770)$ peak clearly shows up dominating the shape of the distribution. Other dynamical structures cannot be appreciated at first glance.

\begin{figure}[!t]
\begin{center}
\vspace*{1.2cm}
\includegraphics[scale=0.6]{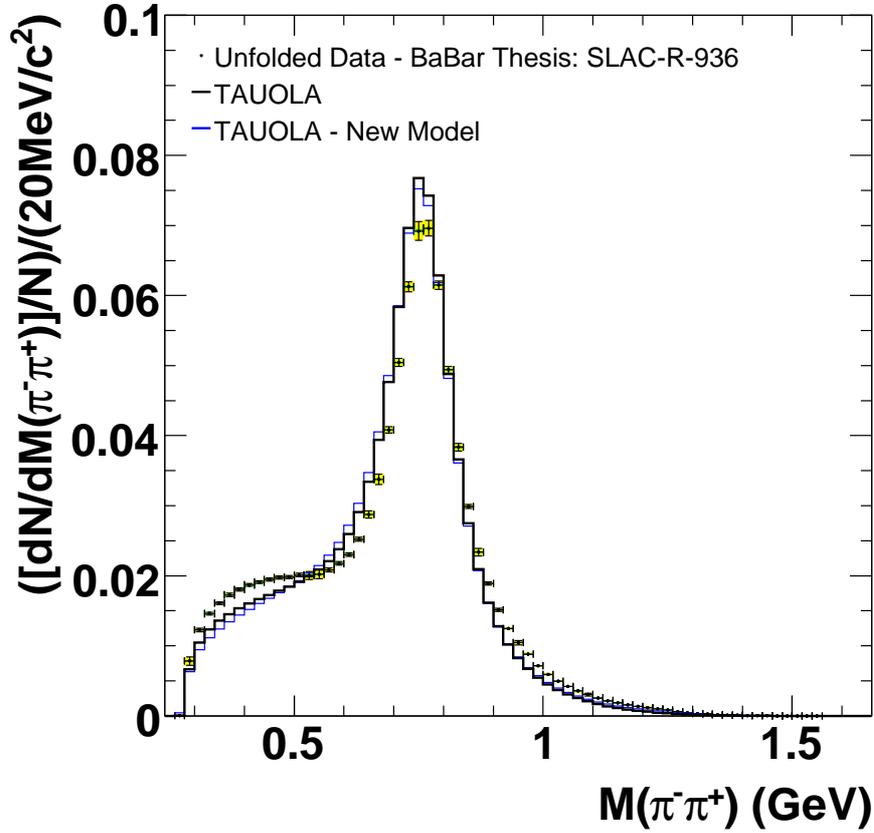}
\caption[]{\label{fig:dGds3piBB} \small{Invariant mass distribution of the $\pi^+\pi^-$ pair in $\tau^-\to\pi^+\pi^-\pi^-\nu_\tau$ decays. Ligther grey histogram corresponds to 
our prediction with the parameters of Set 1, Eq.~(\ref{eq:set1}). Darker grey is from default parametrization of TAUOLA CLEO. The unfolded BaBar data are taken from ref.~\cite{Nugent:2009zz}. 
Courtesy of Ian Nugent.}}
\end{center}
\end{figure}

Although when this Thesis was defended there were not data available to corroborate our predictions, now we have confronted them to BaBar data, as it appears in Ian M. Nugent's Thesis \cite{Nugent:2009zz}. This is 
shown in Fig.~\ref{fig:dGds3piBB}, taken from Ref.~\cite{Shekhovtsova:2012ra}. It is seen that neither parametrization is able to describe the data at low values of the invariant di-pion mass. This could 
be expected since the sigma -or $f_0(500)$- contribution, due to rescattering of the pion pair, is absent in both approaches. Its introduction is under study.\\

\begin{figure}[!h]
\begin{center}
\vspace*{1.0cm}
\includegraphics[scale=0.6,angle=-90]{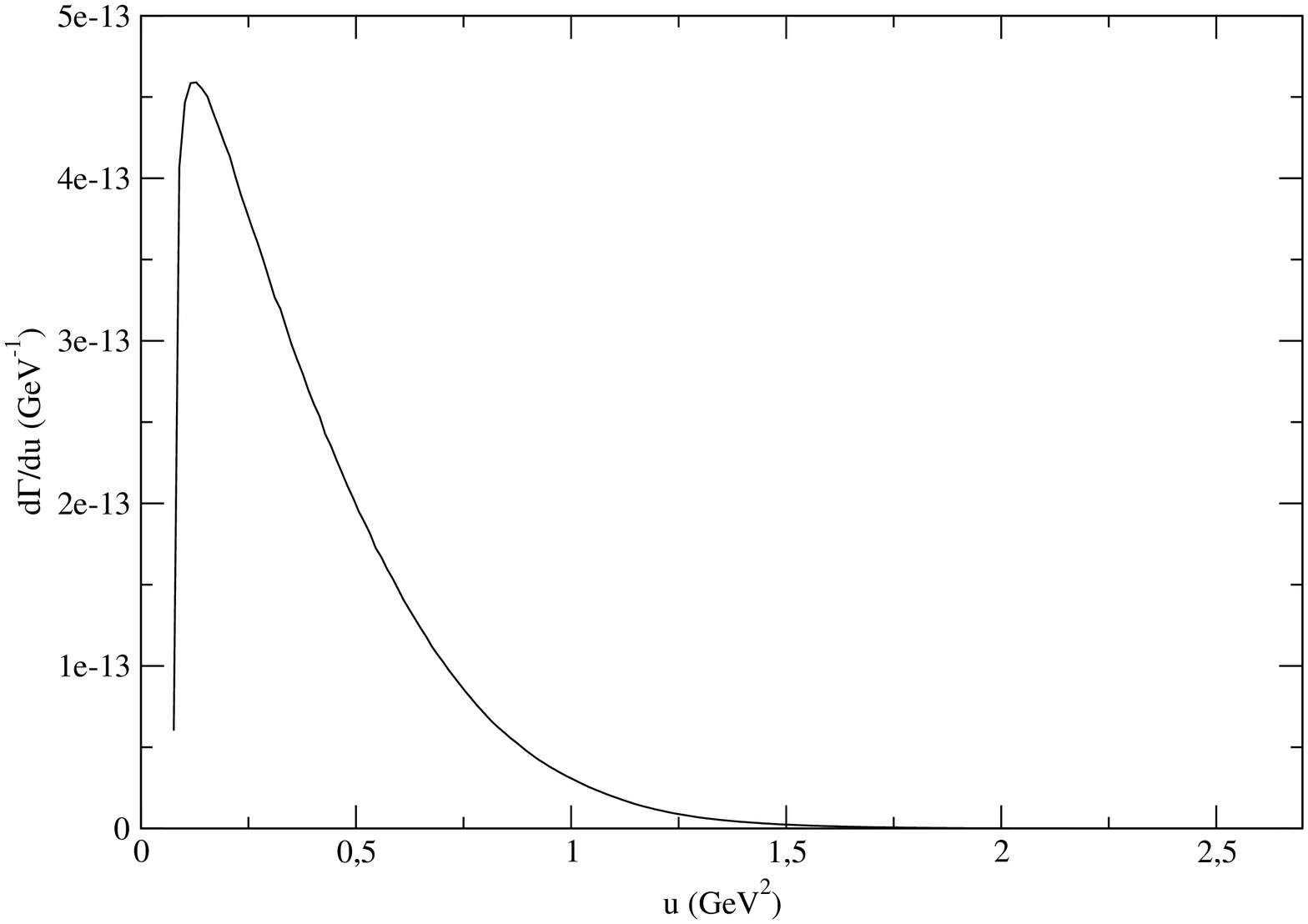}
\caption[]{\label{fig:dGdu3pi} \small{Our prediction for the $u$-spectrum of the $\tau^- \rightarrow \pi^+ \pi^- \pi^- \nu_{\tau}$ with the parameters of 
Set 1, Eq.~(\ref{eq:set1}).}}
\end{center}
\end{figure}

\hspace*{0.5cm}Similarly, in Fig
~\ref{fig:dGdu3pi} we plot our prediction for the distribution with respect to the 
 Mandelstam variable 
$u$ \footnote{Taking into account that the form factors are symmetric under the exchange $\left\lbrace1\leftrightarrow2,\,s\leftrightarrow t \right\rbrace$ due to the identity of two pions, the plot $d\Gamma/dt$ would 
be redundant. We have verified that the $s$- and $t$- plots are identical nonetheless.}. Although we are working in the isospin conserved limit where all three pions are identical, the plot in Figure~\ref{fig:dGdu3pi} is noticeably 
different because the two pions of equal electric charge cannot couple to a spin-one resonance which explains why no dynamical structure can be seen in this figure. We cannot forget that isospin symmetry breaking is not only induced 
by the difference of $u$ and $d$ quark masses compared to the value of the $s$ quark mass, but also by the different electric charges of the $u-$ and $d-$type quarks. 
When this results in a selection rule, the effects are sizeable as we have observed.\\
%
%
%
%
%
%
\subsection{Description of structure functions}\label{3pi_SF}
\begin{figure}[!h]
\vspace*{0.5cm}
\begin{center}
   \includegraphics[scale=0.6,angle=-90]{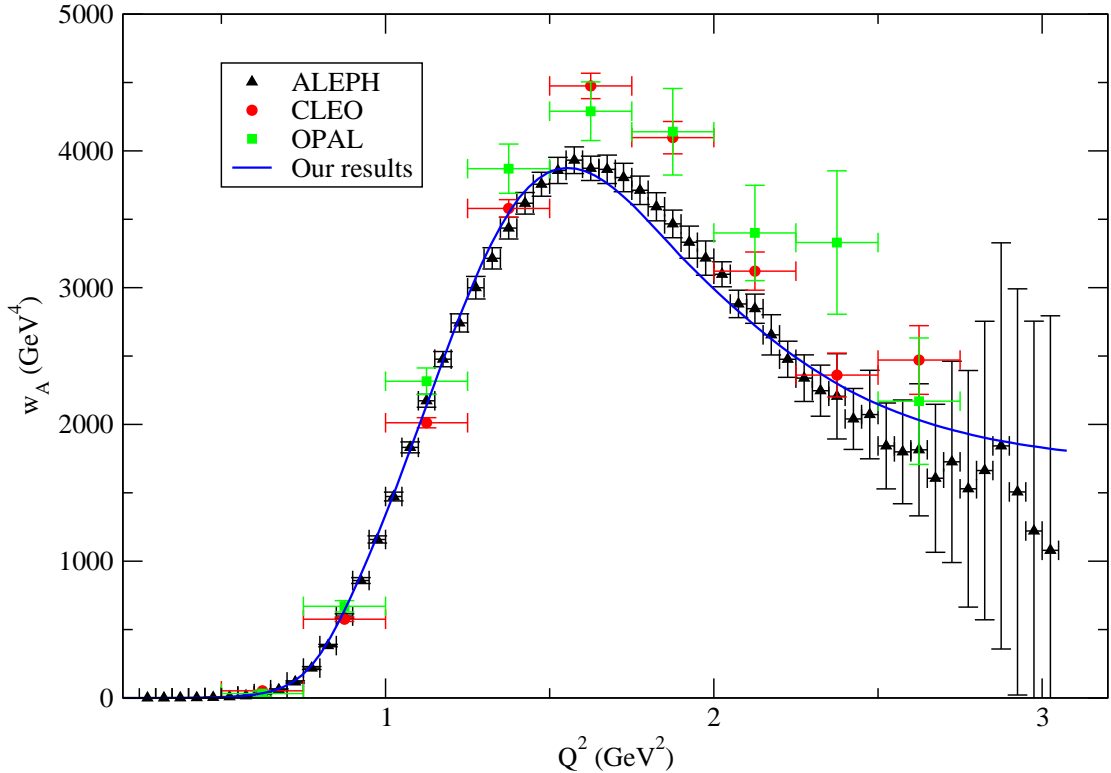}
 \caption{ \small{Comparison between the experimental data for $w_A$, from  $\tau^-\to \pi^-\pi^0\pi^0\nu_\t$, quoted by $CLEO-II$ and $OPAL$ \cite{Ackerstaff:1997dv,
 Browder:1999fr} and the values arising from $ALEPH$ measurements of $\tau^-\to \pi^-\pi^-\pi^+\nu_\t$ spectral functions \cite{Barate:1998uf}. The 
solid line is obtained using the values of Set 1, Eq.~(\ref{eq:set1}).}}
\label{fig:WA_ALL}
\end{center}
\end{figure}

\begin{figure}[!h]
\vspace*{0.5cm}
\begin{center}
   \includegraphics[scale=0.6,angle=-90]{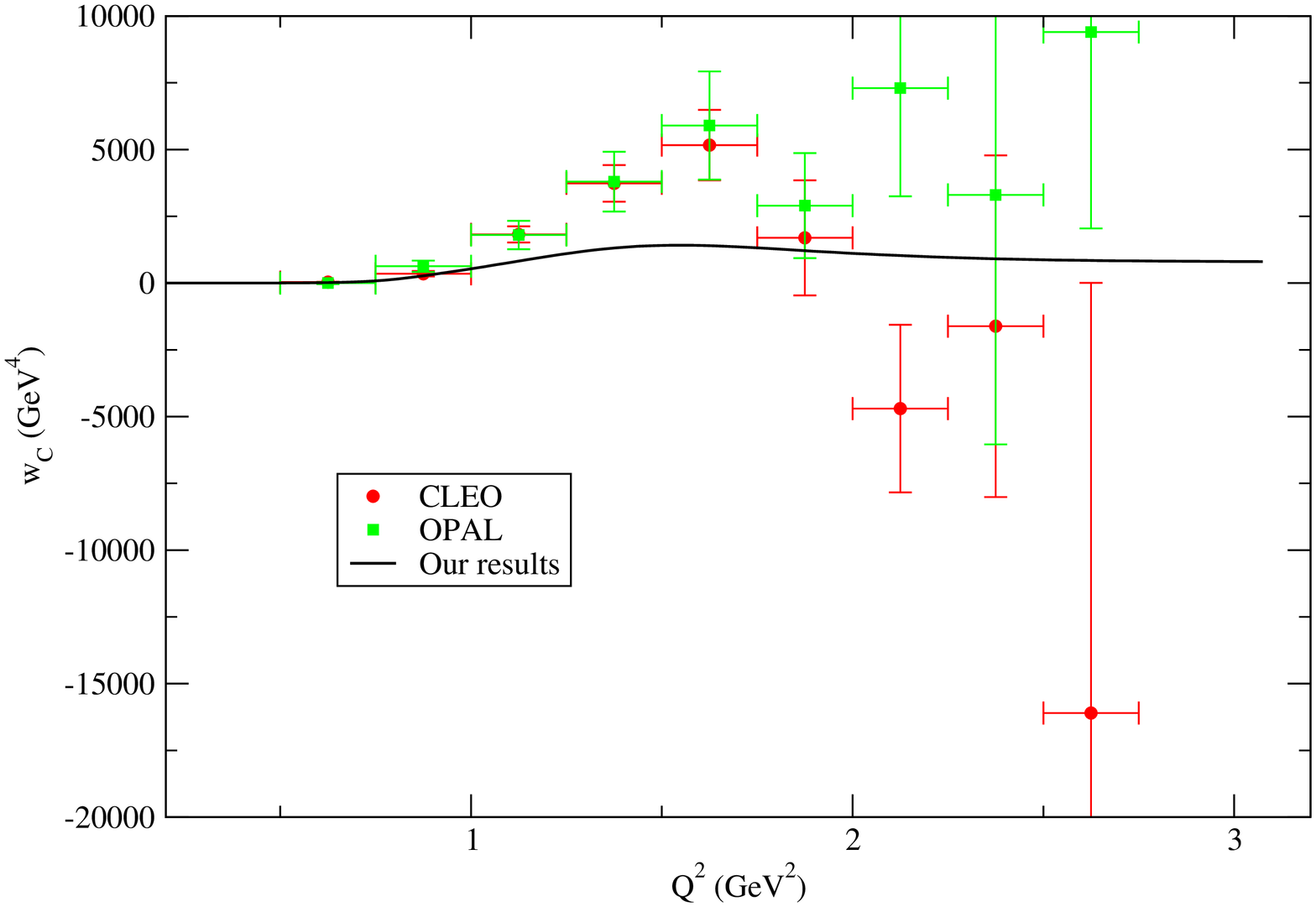}
 \caption{ \small{Comparison between the experimental data for $w_C$, from  $\tau^-\to \pi^-\pi^0\pi^0\nu_\t$, quoted by $CLEO-II$ and $OPAL$ \cite{Ackerstaff:1997dv,
 Browder:1999fr} and our results as obtained by using the values of Set 1, Eq.~(\ref{eq:set1}).}}
\label{fig:WC}
\end{center}
\end{figure}

\begin{figure}[!h]
\vspace*{0.5cm}
\begin{center}
   \includegraphics[scale=0.6,angle=-90]{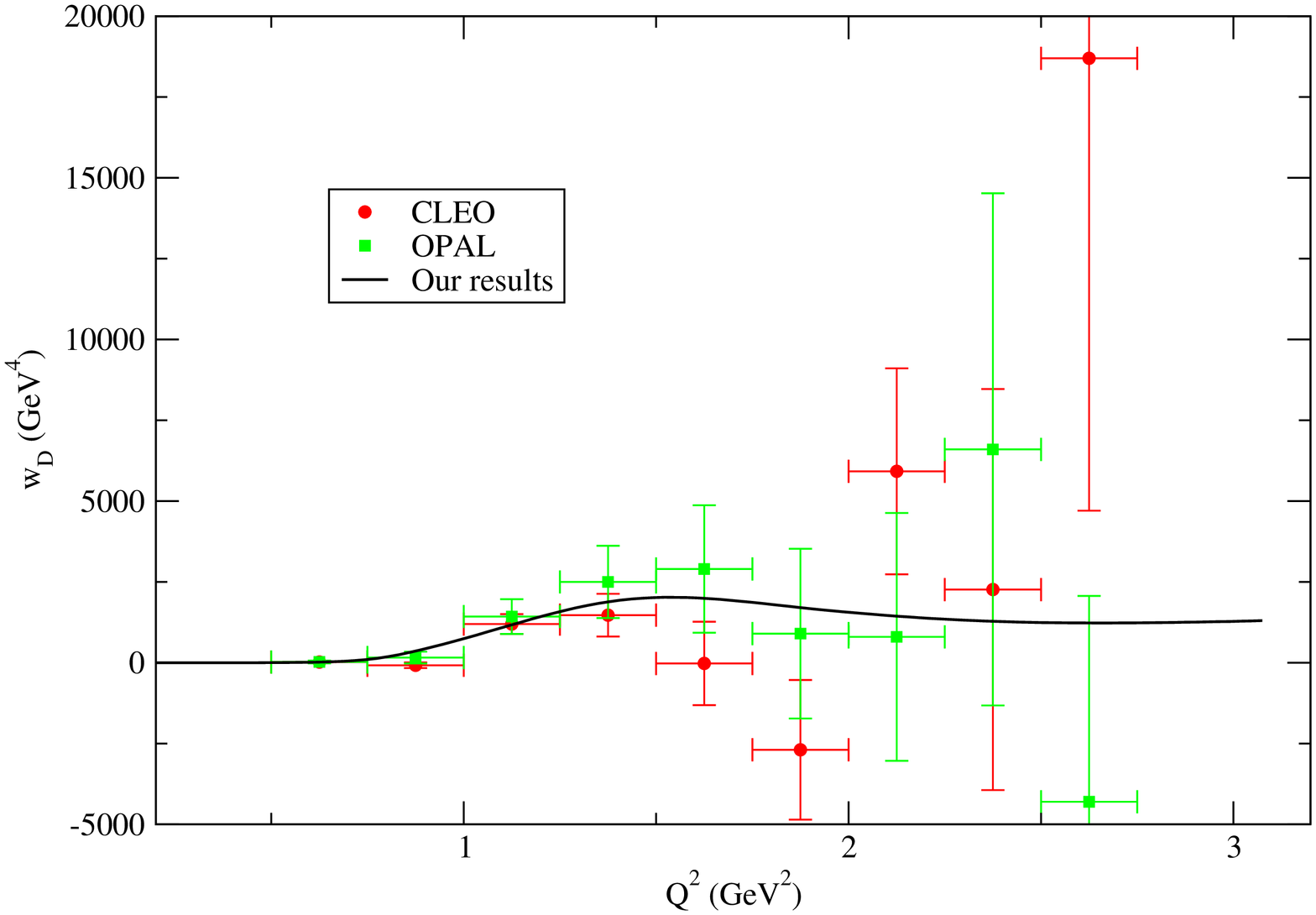}
 \caption{ \small{Comparison between the experimental data for $w_D$, from  $\tau^-\to \pi^-\pi^0\pi^0\nu_\t$, quoted by $CLEO-II$ and $OPAL$ \cite{Ackerstaff:1997dv,
 Browder:1999fr} and our results as obtained by using the values of Set 1, Eq.~(\ref{eq:set1}).}}
\label{fig:WD}
\end{center}
\end{figure}

\begin{figure}[!h]
\vspace*{0.5cm}
\begin{center}
   \includegraphics[scale=0.6,angle=-90]{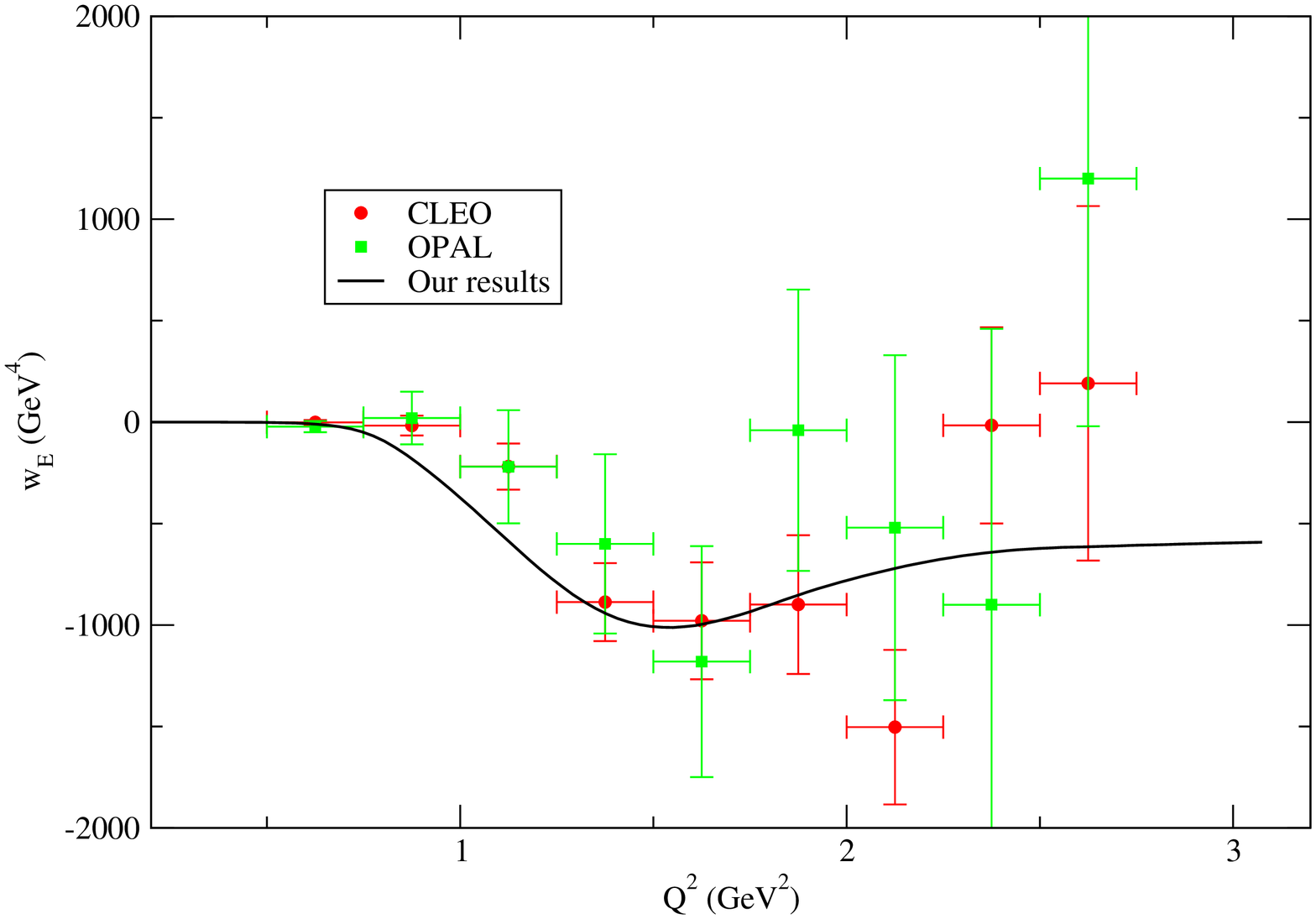}
 \caption{ \small{Comparison between the experimental data for $w_E$, from  $\tau^-\to \pi^-\pi^0\pi^0\nu_\t$, quoted by $CLEO-II$ and $OPAL$ \cite{Ackerstaff:1997dv,
 Browder:1999fr} and our results as obtained by using the values of Set 1, Eq.~(\ref{eq:set1}).}}
\label{fig:WE}
\end{center}
\end{figure}

\hspace*{0.5cm}Structure functions provide a full description of the hadronic tensor $T_{\mu} T_{\nu}^*$ in the hadron rest frame. There are $16$ real
valued structure functions in $\tau^- \rightarrow (P_1 P_2 P_3)^- \,\nu_{\tau}$ decays ($P_i$ is short for a pseudoscalar meson), most of which can be
determined by studying angular correlations of the hadronic system. Four of them carry information on the $J^P=1^+$ transitions only~: $w_A$, $w_C$, 
$w_D$ and $w_E$ (we refer to Ref.\ \cite{Kuhn:1992nz} and Eq.~\ref{W_i_3mesons} and to Appendix A for their precise definitions
 and discussion). Indeed, for the $\tau^-\to(\pi\pi\pi)^-\nu_\tau$ processes, other structure functions either vanish identically, or involve the 
pseudoscalar form factor $F_3^A$, which appears to be strongly suppressed above the very low--energy region due to its proportionality to the squared 
pion mass.\\
\hspace*{0.5cm}Unfortunately the $ALEPH$ collaboration data~\cite{Barate:1998uf} only allows to obtain $w_A$. However, both $CLEO-II$ \cite{Ackerstaff:1997dv} and 
$OPAL$ \cite{Browder:1999fr} studied all relevant structure functions. As a result, they have measured the four structure functions quoted above for the
 $\tau^-\to \pi^-\pi^0\pi^0\nu_\tau$ process, while concluding that other functions are consistent with zero within errors. Hence we can proceed to 
compare those experimental results with the description that our theoretical approach provides. In our expressions for the structure functions we input
 the values of the parameters of Set 1. This way we get the theoretical curves shown in Figs.~\ref{fig:WA_ALL}, \ref{fig:WC}, \ref{fig:WD} and 
\ref{fig:WE}. The latter are compared with the experimental data quoted by $CLEO$ and $OPAL$ \cite{Ackerstaff:1997dv, Browder:1999fr}. For $w_C$, $w_D$
 and $w_E$, it can be seen that we get a good agreement in the low $Q^2$ region, while for increasing energy the experimental errors become too large 
to state any conclusion (moreover, there seems to be a slight disagreement between both experiments at some points). It will be a task for the forthcoming
 experimental results from the $B$-factories to settle this issue.\\
\hspace*{0.5cm}On the other hand, in the case of the integrated structure function $w_A$, the quoted experimental errors are smaller, and the 
theoretical curve fits perfectly well the $ALEPH$ data -that is clearly the one with smaller error bars- and seems to lie somewhat below the $CLEO$ and 
$OPAL$ data for $Q^2 \lesssim 1$.$5$ GeV$^2$. However, it happens that $w_A$ contains essentially the same information about the hadronic amplitude as the 
spectral function $d\Gamma/dQ^2$, so it should not surprise us the excellent agreement with $ALEPH$ data, considering the curve obtained with Set 1 in 
Fig~\ref{fig:spectralfunction3pi}. This relation becomes clear by looking at Eq.~(\ref{fulldGammadQ2}) if the scalar structure function $W_{SA}$ is put to zero 
(remember that it should be suppressed by a factor $\mathcal{O}(m_\pi^2/Q^2)$). Taking into account that $w_A$ is given by Eq.~(\ref{integrated structure functions})
\begin{equation}
w_{A}(Q^2) = \int ds\, dt\; W_A(Q^2,s,t)\, \, \;,
\end{equation}
where $W_A$ is the structure function previously introduced in Eq.~(\ref{fulldGammadQ2}),  one simply has
\begin{equation}
\label{specwa}
\frac{d\Gamma}{dQ^2} \,  = \,  \frac{G_F^2\,|V_{ud}|^2}
{384\, (2\pi)^5\, M_\tau}\;
\left( \, \frac{M_\tau^2}{Q^2}-1 \, \right)^2 
\left(1 \, + \, 2 \,\frac{Q^2}{M_\tau^2}\right) \; \; w_A(Q^2)  \; \; .
\end{equation}
\hspace*{0.5cm}In this way one can compare the measurements of $w_A$ quoted by $CLEO-II$ and $OPAL$ with the data obtained by $ALEPH$ for the spectral
 function, conveniently translated into $w_A$. This is represented in Figure~\ref{fig:WA_ALL}, where it can be seen that some of the data from the different 
experiments do not agree with each other within errors. Notice that, due to phase space suppression, the factor of proportionality between $w_A(Q^2)$ 
and d$\Gamma /$d$Q^2$ in Eq.~(\ref{specwa}) goes to zero for $Q^2 \rightarrow M_{\tau}^2$, therefore the error bars in the $ALEPH$ points become 
enhanced toward the end of the spectrum. Notwithstanding, up to $Q^2\lesssim 2$.$5$ GeV$^2$, it is seen that $ALEPH$ errors are still smaller than those 
corresponding to the values quoted by $CLEO-II$ and $OPAL$. On this basis, we have chosen to take the data obtained by $ALEPH$ to select the parameters 
as indicated in Set 1 to better describe the hadronic amplitude. Finally, notice that a non vanishing contribution of $W_{SA}$ (which is a positive 
quantity) cannot help to solve the experimental discrepancies, as it would go in the wrong direction. Anyway we have estimated that it is orders of 
magnitude smaller than the axial-vector contribution.\\
\hspace*{0.5cm}In the analysis of data carried out by the $CLEO$ Collaboration \cite{Asner:1999kj} onto their $\t^-\to \pi^-\pi^0\pi^0\nu_\t$ results 
it was concluded that the data was showing large contributions from intermediate states involving the isoscalar mesons $f_0$(600), $f_0$(1370) and 
$f_2$(1270). Their analysis was done in a modelization of the axial--vector form factors that included Breit--Wigner functions in a K\"uhn and Santamar\'{\i}a
 inspired model. Our results in the Effective Theory framework show that, within the present experimental errors, there is no evidence of relevant 
contributions in $\t^-\to (\pi\pi\pi)^-\nu_\t$ decays beyond those of the $\rho$(770), $\rho$(1450) and a$_1$(1260) resonances.\\
\section{Conclusions}
\hspace*{0.5cm}The data available in $\tau\to\pi\pi\pi\nu_\tau$ decays provide an excellent benchmark to study the hadronization of the
axial-vector current and, consequently, the properties of the a$_1(1260)$ resonance. In this chapter we give a description of those decays 
within the framework of resonance chiral theory and the large-$N_C$ limit of $QCD$ that: 1) Satisfies all constraints of the asymptotic 
behaviour, ruled by $QCD$, of the relevant two and three point Green functions; 2) Provides an excellent description of the branching ratio 
and spectrum of the $\tau\to\pi\pi\pi\nu_\tau$ decays.\\
\hspace*{0.5cm}Though this work was started in Ref.~\cite{GomezDumm:2003ku}, later achievements showed that a deeper comprehension of the 
dynamics was needed in order to enforce the available $QCD$ constraints. To achieve a complete description we have defined a new off-shell 
width for the a$_1(1260)$ resonance in Eq.~(\ref{eq:Gamma_a1_tot}), which is one of the main results of this work. Moreover we have seen 
that the inclusion of the $\rho(1450)$ improves significantly the description of the observables. In passing we have also obtained the mass 
value $M_{{\rm a}_1} = 1$.$120 \, \mathrm{GeV}$ and the on-shell width $\Gamma_{\mathrm{a}_1}(M_{{\mathrm{a}}_1}^2) = 0$.$483 \,\mathrm{GeV}$.\\
\hspace*{0.5cm}With the description of the off-shell width obtained in this work we can now consider that the hadronization of the axial-vector
 current within our scheme is complete and it can be applied in other hadronic tau decays chnnels.\\
\hspace*{0.5cm}The a$_1$ resonance, its off-shell width and its coupling to $\rho$-$\pi$, play an important role \cite{Xiong:1992ui, Song:1993ae, Song:1994zs, 
Haglin:1994yv, Kim:1996nq, Gao:1997vm, Turbide:2003si} in the evaluation of the dilepton 
and photon production rates from a hadronic fireball assumed to be created in the relativistic heavy ion collisions. This would be important to be able to 
tell the electromagnetic radiation of the quark-gluon plasma from the hadronic sources, so that it could be regarded as an additional possible future application 
of our findings.\\
\chapter{$\tau^-\to (KK\pi)^- \nu_\tau$ decays}\label{KKpi}
\section{Introduction}\label{KKpi_Intro}
\hspace*{0.5cm}In this chapter we will discuss the hadronic form factors and related observables appearing in $\tau^-\to (KK\pi)^- \nu_\tau$
 decays. Once the main features of the hadronization of the axial-vector current in these kind of decays have been fixed in the previous chapter, 
we can deal with these channels where both vector and axial-vector current contribute at, in principle, comparable rates. In fact, one of the 
purposes of our work is to analyze what is the relative relevance of each of them.\\
\hspace*{0.5cm}We thus analyse the hadronization structure of both vector and axial-vector currents leading to $\tau\to KK\pi \,\nu_\tau$ 
decays.  At leading order in the $1/N_C$ expansion, and considering only the contribution of the lightest resonances, we work out, within the
 framework of the resonance chiral Lagrangian, the structure of the local vertices involved in those processes, that is richer than the one 
presented in previous chapters. The couplings in the resonance theory are constrained by imposing the asymptotic behaviour of vector and 
axial-vector spectral functions ruled by $QCD$. Noteworthy, the short-distance relations coming from $QCD$ constraints are compatible in all 
with those found in Chapter \ref{3pi} and to the ones we will find in Chapters \ref{eta} and \ref{Pgamma} as well, a feature that highlights the 
consistency of the whole description. In this way we predict the hadronic spectra and conclude that, contrarily to previous assertions, the vector 
contribution dominates by far over the axial-vector one in all $K K \pi$ charge channels.\\
\hspace*{0.5cm}Our study has a twofold significance. First, the study of branching fractions and spectra of those decays is a major goal of 
the asymmetric $B$ factories ($BaBar$, $BELLE$). These are supplying an enormous amount of quality data owing to their large statistics, and
 the same is planned for the near future at tau-charm factories such as $BES-III$. Second, the required hadronization procedures involve $QCD$ 
 in a non-perturbative energy region ($E \lesssim M_{\tau} \sim 1$.$8$ GeV) and, consequently, these processes are a clean benchmark, not spoiled 
by an initial hadronic state, where we can learn about the treatment of strong interactions when driven by resonances.\\
\hspace*{0.5cm}We recall that the analysis of these decays has to assume a model of hadronization, as discussed in Chapter \ref{Hadrondecays}.
 A very popular approach is due to the so-called K\"uhn-Santamar\'{\i}a model ($KS$) \cite{Pich:1987qq, Kuhn:1990ad} that, essentially, relies 
on the construction of form factors in terms of Breit-Wigner functions weighted by unknown parameters that are extracted from phenomenological 
analyses of data. This procedure, that has proven to be successful in the description of the $\pi \pi \pi$ final state, has been employed in 
the study of many two- and three-hadron tau decays \cite{Bruch:2004py, Finkemeier:1995sr, Decker:1992kj, Decker:1992rj, Decker:1993ay, GomezCadenas:1990uj}.
 The ambiguity related with the choice of Breit-Wigner functions \cite{Pich:1987qq, Kuhn:1990ad, Gounaris:1968mw} is currently being exploited 
to estimate the errors in the determination of the free parameters. The measurement of the $K K \pi$ spectrum by the $CLEO$ Collaboration 
\cite{Liu:2002mn} has shown that the parameterization described by the $KS$ model does not recall appropriately the experimental features
keeping, at the same time, a consistency with the underlying strong interaction theory \cite{Portoles:2004vr}. The solution provided by $CLEO$
based in the introduction of new parameters spoils the normalization of the Wess-Zumino anomaly, i.e. a specific prediction of $QCD$. Indeed, 
arbitrary parameterizations are of little help in the procedure of obtaining information about non-perturbative $QCD$. They may fit the data 
but do not provide us hints on the hadronization procedures. The key point in order to uncover the inner structure of hadronization is to 
guide the construction of the relevant form factors with the use of known properties of $QCD$.\\
\hspace*{0.5cm}The $TAUOLA$ library has been growing over the years \cite{Jadach:1990mv, Jadach:1990mz, Jezabek:1991qp, Jadach:1993hs, Golonka:2003xt,
Bondar:2002mw, Kuhn:2006nw, Davidson:2010rw, Golonka:2000iu} to be a complete library providing final state with full 
topology including neutrinos, resonances and lighter mesons and complete spin structure throughout the decay. In these works, the 
hadronization part of the matrix elements followed initially only assorted versions of the $KS$ model. At present, the $TAUOLA$ library has 
become a key tool that handles analyses of tau decay data and it has been opened to the introduction of matrix elements obtained with other 
models (like the results of this work, see Ref.~\cite{Shekhovtsova:2012ra}). Hence it has become an excellent tool where theoretical models confront experimental data. This or analogous libraries \cite{Gleisberg:2003xi,
 Thomas} are appropriate benchmarks where to apply the results of our research \cite{Roig:2008ev, Roig:2008xt, Roig:2008xta, Roig:2009ff}.\\
\hspace*{0.5cm}We will be assisted in the presented task by the recent analysis of $e^+ e^- \rightarrow K K \pi$ cross-section by $BABAR$ 
\cite{Aubert:2007ym} where a separation between isoscalar and isovector channels has been performed. Hence we will be able to connect both 
processes through $CVC$. The general framework for these kind of analyses is discussed in Appendix \ref{isosApp} and only the concrete 
application to this channel is included in this chapter. We have also used the process $\omega \to \pi^+\pi^-\pi^0$ to extract a given combination
 of Lagrangian couplings that will enter into the analysis. This computation constitutes Appendix \ref{omegaApp}. Although this chapter is 
based on Ref.~\cite{Dumm:2009kj}, the discussion of our predictions on the shape of the d$\Gamma/$d$s_{ij}$ and its comparison to the estimates 
of other models at the end of Sect. \ref{KKpi_Phenomenology} are included in this Thesis for the first time.\\
\section{Vector and axial-vector current form factors}\label{KKpi_FF}
\subsection{Form factors in $\tau^-\to \left(K\overline{K}\right)\pi^- \nu_\tau$ decays} \label{KKpi-_FF}
\begin{figure}
\begin{center}
\includegraphics[scale=0.8]{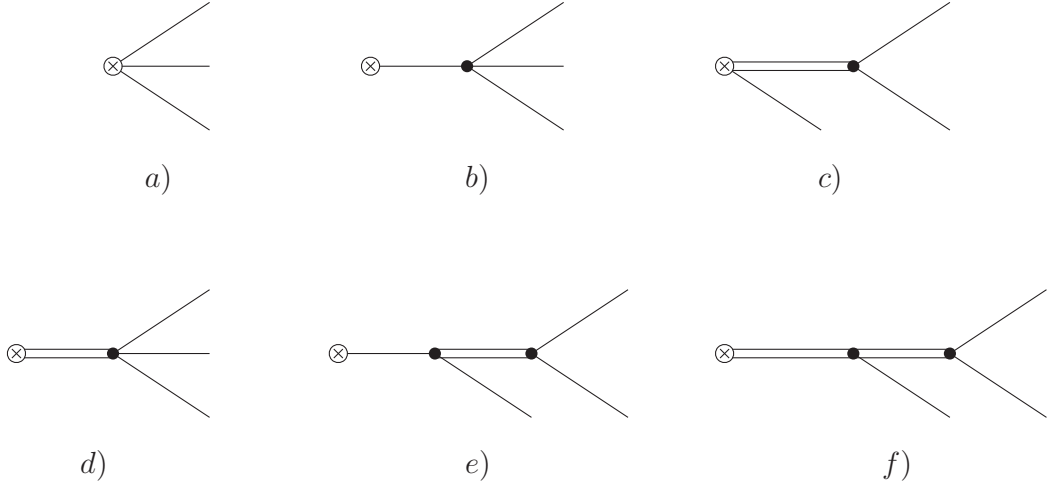}
\caption[]{\label{fig_diagsKKpi} \small{Topologies contributing to the final hadronic state in $\tau \rightarrow K K \pi \, \nu_{\tau}$ decays in the 
$N_C\rightarrow \infty$ limit. A crossed circle indicates the $QCD$ vector or axial-vector current insertion. A single line represents a 
pseudoscalar meson ($K$, $\pi$)  while a double line stands for a resonance intermediate state. Topologies $b)$ and $e) $ only contribute 
to the axial-vector driven form factors, while diagram $d)$ arises only (as explained in the text) from the vector current.}}
\end{center}
\end{figure}

\hspace*{0.5cm}Our effective Lagrangian will include the pieces given in Eqs. (\ref{p2-u}), (\ref{Z_WZW}) -$\CPT$ contributions-, (\ref{R}) 
\footnote{Again, the fact that the $SM$ coulings are of the type $V-A$ makes the spin-one resonances to rule hadronic tau decays. The 
contribution of scalar and pseudoscalar resonances to the relevant four-point Green function should be minor for $\tau \rightarrow K K \pi 
\nu_{\tau}$. Indeed the lightest scalar, namely $f_0(980)$, couples dominantly to two pions, and therefore its role in the $K \overline{K} 
\pi$ final state should be negligible. Heavier flavoured or unflavoured scalars and pseudoscalars are at least suppressed by their masses, 
being heavier than the axial-vector meson a$_1(1260)$ (like $K_0^*(1430)$ that couples to $K \pi$). The lightest pseudoscalar coupling to $K
\pi$ is the $K_0^*$ of $\kappa(800)$. As we assume the $N_C \rightarrow \infty$ limit, the nonet of scalars corresponding to the $\kappa(800)$
 is not considered. This multiplet is generated by rescattering of the ligthest pseudoscalars and then subleading in the $1/N_C$ expansion 
\cite{Cirigliano:2003yq}. In addition the couplings of unflavoured states to $K \overline{K}$ (scalars) and $K \overline{K} \pi$ (pseudoscalars) seem to be very small \cite{Amsler:2008zzb}. Thus in our 
description we include $J=1$ resonances only. Nevertheless, if the study of these processes requires a more accurate description, additional 
resonances could also be included in our scheme.}, (\ref{VJPops}), (\ref{LVPPP}) -operators with one resonance- (\ref{VVPops}) and 
(\ref{LVAPb}) -operators with two resonances-.\\
\hspace*{0.5cm}We write the decay amplitude for the considered processes as
\begin{equation}
 {\cal M} \, = \, - \, \frac{G_F}{\sqrt{2}} \, V_{ud} \, \overline{u}_{\nu_{\tau}} \,
\gamma^{\mu}  \left( 1 \, - \, \gamma_5 \right) \, u_{\tau} \, T_{\mu} \; ,
\end{equation}
where the model dependent part is the hadronic vector
\begin{equation}
 T_{\mu} \, = \, \langle K(p_1) \,  K(p_2) \,  \pi(p_3) \, | \, \left( V_{\mu} - A_{\mu} \right)
e^{i {\cal L}_{QCD}} \, | \, 0 \rangle \, ,
\end{equation}
that can be written in terms of four form factors $F_1$, $F_2$, $F_3$ and $F_4$, see Eq.~(\ref{generaldecomposition_3mesons}).\\
There are three different charge channels for the $KK\pi$ decays of the $\tau^-$ lepton, namely $K^+ (p_+) \, K^-(p_-) \, \pi^-(p_{\pi})$, 
$K^0(p_0)\, \overline{K}^0(\overline{p}_0) \, \pi^-(p_{\pi})$ and $K^-(p_-) \, K^0(p_0) \, \pi^0(p_{\pi})$. The definitions of 
Eq.~(\ref{generaldecomposition_3mesons}) correspond to the choice $p_3 = p_{\pi}$ in all cases, and~: $(p_1,p_2) =(p_+,p_-)$ for the 
$K^+ \, K^-$ case, $(p_1,p_2) = (\overline{p}_0,p_0)$ for $K^0 \, \overline{K}^0$  and $(p_1,p_2) = (p_-,p_0)$ for $K^- \, K^0$. In general, 
form factors $F_i$ are functions of the kinematical invariants~: $Q^2$, $s=(p_1+p_2)^2$ and $t=(p_1+p_3)^2$.\\
The general structure of the form factors, within our model, arises from the
diagrams displayed in Figure~\ref{fig_diagsKKpi}.
This provides the following decomposition~:
\begin{equation}
 F_i \, = \, F_i^{\chi} \, + \, F_i^{\mathrm{R}} \, + \, F_i^{\mathrm{RR}} \;, \; \;  \, i=1,\,2,\,3,\,4 \, ;
\end{equation}
where $F_i^{\chi}$ is given by the $\chi PT$ Lagrangian [topologies $a)$ and $b)$ in Figure~\ref{fig_diagsKKpi}], and the rest are the 
contributions of one [Figure~1$c)$, $d)$ and $e)$] or two resonances [Figure~1$f)$].\\
\hspace*{0.5cm}In the isospin limit, form factors for the $\tau^-\rightarrow K^+ K^- \pi^- \nu_{\tau}$ and $\tau^- \rightarrow K^0
\overline{K}^0 \pi^- \nu_{\tau}$ decays are identical. The explicit expressions for these are~:
\begin{eqnarray}
\label{eq:f11}
F_1^{\chi} &= & - \frac{\sqrt{2}}{3 \, F} \, , \nonumber \\
F_1^{\mathrm{R}}(s,t) & = & - \, \frac{\sqrt{2}}{6} \, \frac{F_V\,G_V}{F^3}  \, \left[
\, \frac{A^{\mathrm{R}}(Q^2,s,u,m_K^2,m_{\pi}^2,m_K^2)}{M_{\rho}^{2}-s}  \, +
\, \frac{B^{\mathrm{R}}(s,u,m_K^2,m_{\pi}^2)}{M_{K^{*}}^{2}-t} \,\right] ,  \, \nonumber \\[4mm]
F_1^{\mathcal{RR}}(s,t) & = & \frac{2}{3} \, \frac{F_A G_V}{F^3} \, \frac{Q^2}{M_{\mathrm{a}_1}^2-Q^2} \,
\, \left[ \, \frac{A^{\mathcal{RR}}(Q^2,s,u,m_K^2,m_{\pi}^2,m_K^2) }{M_{\rho}^2-s} \,
\right.  \\
& &  \qquad \qquad \qquad \qquad \qquad\left.
+ \, \frac{B^{\mathrm{RR}}(Q^2,s,u,t,m_K^2,m_{\pi}^2,m_K^2)}{M_{K^*}^2-t} \, \right]
\, \, , \nonumber
\end{eqnarray}
where the functions $A^{\mathrm{R}}$, $B^{\mathrm{R}}$, $A^{\mathrm{RR}}$ and $B^{\mathrm{RR}}$ are
\begin{eqnarray} \label{eq:AB}
A^{\mathrm{R}}(Q^2,x,y,m_1^2,m_2^2,m_3^2) & = &  3\, x \, + m_1^2 -m_3^2 +
\left( 1-\frac{2 G_V}{F_V} \right) \left[ 2\, Q^2-2\, x-y+m_3^2-m_2^2 \right] \, , \nonumber \\[3.5mm]
B^{\mathrm{R}}(x,y,m_1^2,m_2^2) & = & 2 \, \left( m_2^2-m_1^2 \right) \,
+ \, \left( 1-\frac{2 G_V}{F_V} \right) \left[ y - x + m_1^2-m_2^2\right] \, ,  
\end{eqnarray}
\begin{eqnarray}
A^{\mathrm{RR}}(Q^2,x,y,m_1^2,m_2^2,m_3^2) & = & \left( \lambda' + \lambda'' \right) \,
(-3\, x + m_3^2-m_1^2)\, \nonumber \\ &&  + \, \left( 2\, Q^2+x-y+m_1^2-m_2^2 \right)
F\left( \frac{x}{Q^2}\,,\,\frac{m_2^2}{Q^2} \right) \, , \nonumber \\ [3.5mm]
B^{\mathrm{RR}}(Q^2,x,y,z,m_1^2,m_2^2,m_3^2) & = & 2 \left(
\lambda'+\lambda'' \right) \left( m_1^2-m_2^2 \right)\nonumber\\
& &  + \left( y-x+m_2^2-m_1^2 \right)
F\left( \frac{z}{Q^2}\,,\,\frac{m_3^2}{Q^2} \right) \, . \nonumber
\end{eqnarray}
\hspace*{0.5cm}The dependence of the form factors with $t$ follows from the relation $u = Q^2 - s - t + 2 m_K^2 + m_\pi^2$. Moreover 
resonance masses correspond to the lowest states, $M_{\rho} = M_{\rho (770)}$, $M_{K^*} = M_{K^*(892)}$ and $M_{\mathrm{a}_1} = M_{\mathrm{a}_1(1260)}$. 
Resonance masses and widths within our approach are discussed in Appendix C
.\\
\hspace*{0.5cm}Analogously the $F_2$ form factor is given by~:
\begin{eqnarray}
\label{eq:f21}
 F_2^{\chi} &= & F_1^{\chi} \, ,  \\[3.5mm]
 F_2^{\mathrm{R}}(s,t) & = & - \, \frac{\sqrt{2}}{6} \, \frac{F_V\,G_V}{F^3}  \, \left[
 \, \frac{B^{\mathrm{R}}(t,u,m_K^2,m_K^2)}{M_{\rho}^{2}-s}  \,  +
 \, \frac{A^{\mathrm{R}}(Q^2,t,u,m_K^2,m_K^2,m_{\pi}^2)}{M_{K^{*}}^{2}-t} \,\right] ,  \, \nonumber \\[4mm]
F_2^{\mathrm{RR}}(s,t) & = & \frac{2}{3} \, \frac{F_A G_V}{F^3} \, \frac{Q^2}{M_{\mathrm{a}_1}^2-Q^2} \,
\, \left[ \, \frac{B^{\mathrm{RR}}(Q^2,t,u,s,m_K^2,m_K^2,m_{\pi}^2) }{M_{\rho}^2-s} \,
\right. \nonumber \\
& &  \qquad \qquad \qquad \qquad \qquad\left.
+ \, \frac{A^{\mathrm{RR}}(Q^2,t,u,m_K^2,m_K^2,m_{\pi}^2) }{M_{K^*}^2-t} \, \right] \, \, . \nonumber
\end{eqnarray}
The $F_3$ form factor arises from the chiral anomaly and the non-anomalous odd-intrinsic-parity amplitude. We obtain~:
\begin{eqnarray}
\label{eq:f31}
 F_3^{\chi} & = & - \frac{N_C \, \sqrt{2}}{12 \, \pi^2 \, F^3} \, , \nonumber \\[3.5mm]
 F_3^{\mathrm{R}}(s,t) & = & - \frac{4 \, G_V}{M_V \, F^3} \, \left[ \,
C^{\mathrm{R}}(Q^2,s,m_K^2,m_K^2,m_{\pi}^2) \,
\left( \sin^2 \theta_V \frac{1+ \sqrt{2}  \cot \theta_V}{M_{\omega}^2-s}
\right.\right. \nonumber \\
& & \qquad \qquad \; \; \left. + \, \cos^2 \theta_V
\frac{1- \sqrt{2} \tan \theta_V }{M_{\phi}^2-s} \right)
\, + \, \frac{C^{\mathrm{R}}(Q^2,t,m_K^2,m_\pi^2,m_K^2)}{M_{K^*}^2-t}
\nonumber \\
& & \qquad \qquad \; \; \left. \, - \, \frac{2 \, F_V}{G_V}
\, \frac{D^{\mathrm{R}}(Q^2,s,t)}{M_{\rho}^2-Q^2} \right] \, , \\[3.5mm]
F_3^{\mathrm{RR}}(s,t) & = & 4 \sqrt{2} \frac{F_V \, G_V}{F^3} \,  \frac{1}{M_{\rho}^2 - Q^2} \,
\left[ C^{\mathrm{RR}}(Q^2,s,m_{\pi}^2) \,
\left( \sin^2 \theta_V \frac{1+ \sqrt{2} \cot \theta_V}{M_{\omega}^2-s} \right. \right.
\nonumber \\
&& \qquad \qquad \qquad \qquad \qquad \; \left. \left. + \, \cos^2 \theta_V
\frac{1- \sqrt{2} \tan \theta_V }{M_{\phi}^2 -s} \right)
\, + \, \frac{C^{\mathrm{RR}}(Q^2,t,m_K^2)}{M_{K^*}^2-t} \right] \, ,
\nonumber
\end{eqnarray}
where $C^{\mathrm{R}}$, $D^{\mathrm{R}}$ and $C^{\mathrm{RR}}$ are defined as
\begin{eqnarray}\label{eq:cd}
 C^{\mathrm{R}}(Q^2,x,m_1^2,m_2^2,m_3^2) & = &
 (c_1-c_2+c_5) \, Q^2 - ( c_1-c_2-c_5+2 c_6) \, x \nonumber \\
 & & \, + (c_1+c_2 + 8 c_3 -c_5) \, m_3^2 + 8\, c_4\, (m_1^2-m_2^2) \, , \nonumber \\[3mm]
C^{\mathrm{RR}}(Q^2,x,m^2) & = & d_3 \, (Q^2+x)+ (d_1+8\, d_2-d_3) \, m^2 \, ,  \\ [3.5mm]
D^{\mathrm{R}}(Q^2,x,y) & = & (g_1+2 \, g_2-g_3)\, (x+y) -2 \,g_2 \, (Q^2+m_K^2)
\nonumber \\
&&  - (g_1-g_3)\, ( 3\,m_K^2+m_{\pi}^2 ) +2 \, g_4 \, (m_K^2+m_{\pi}^2) +2 \, g_5 \, m_K^2 \, , \nonumber
\end{eqnarray}
 and $\theta_V$ is the mixing angle between the octet and singlet vector states $\omega_8$ and $\omega_0$ that defines the mass eigenstates 
$\omega(782)$ and $\phi(1020)$~:
\begin{equation}
\left(  \begin{array}{c}
  \phi \\
  \omega
 \end{array} \right)\, = \,  \left(
\begin{array}{cc}
\cos \theta_V & - \sin \theta_V \\
\sin \theta_V &  \cos \theta_V
\end{array}  \right) \;  \left(
\begin{array}{c}
 \omega_8 \\
\omega_0
\end{array} \right)\, .
\end{equation}
For numerical evaluations we will assume ideal mixing,  i.e.\ $\theta_V = \tan^{-1}(1/\sqrt{2})$. In this case the contribution of the 
$\phi(1020)$ meson to $F_3$ vanishes \footnote{Although there are many phenomenological analysis supporting the picture of ideal mixing between the $\omega(782)$ 
and $\phi(1020)$ states \cite{Beringer:1900zz}, this result contradicts the observation of the Babar Collaboration \cite{Aubert:2007ym} of a sizable $\phi$ contribution in the isospin related decay $e^+e^-\to K^+K^-\pi^0$.}.\\
\hspace*{0.5cm}Finally, though we have not dwelled on specific contributions to the $F_4$ form factor, we quote for completeness the result 
obtained from our Lagrangian. Its structure is driven by the pion pole~:
\begin{eqnarray}
\label{eq:f41}
 F_4 & = & F_4^{\chi} + F_4^{\mathrm{R}} \, ,  \\[3mm]
 F_4^{\chi}(s,t) & = & \frac{1}{\sqrt{2} \, F} \frac{m_{\pi}^2}{m_{\pi}^2 - Q^2} \,
\left( 1+ \frac{m_K^2 - u}{Q^2} \right) \, , \nonumber \\[3mm]
 F_4^{\mathrm{R}}(s,t) & = & \frac{G_V^2}{\sqrt{2} \, F^3}
\frac{m_{\pi}^2}{Q^2 (m_{\pi}^2 - Q^2)} \, \left[ \frac{s(t-u)}{M_{\rho}^2 -s}
+ \frac{t(s-u) - (m_K^2 - m_\pi^2)(Q^2-m_K^2)}{M_{K^*}^2-t} \right] \, .\nonumber
\end{eqnarray}
\subsection{Form factors in $\tau^-  \rightarrow  K^- \,K^0 \, \pi^0 \, \nu_{\tau}$ decays} \label{KKpi0_FF}
\hspace*{0.5cm}The diagrams contributing to the $\tau^-  \rightarrow  K^- \,K^0 \, \pi^0 \, \nu_{\tau}$ decay amplitude are also those in 
Figure~\ref{fig_diagsKKpi}, hence once again we can write $F_i \, =\, F_i^{\chi} \, + \,F_i^{\mathrm{R}} \, + \, F_i^{\mathrm{RR}} \, + \, 
\dots$. However, the structure of the form factors for this process does not show the symmetry observed in $\tau \rightarrow K \overline{K} 
\pi \nu_{\tau}$. We find~:
\begin{eqnarray}
\label{eq:f12}
F_1^{\chi}  & = & -\frac{1}{F} \, , \nonumber \\[3mm]
F_1^{\mathrm{R}}(s,t) & = & - \frac{1}{6} \frac{F_V G_V}{F^3} \, \left[
\, \frac{B^{\mathrm{R}}(s,u,m_K^2,m_{\pi}^2)}{M_{K^{*}}^{2}-t} \, +
\, 2 \; \frac{A^{\mathrm{R}}(Q^2,s,u,m_K^2,m_\pi^2,m_K^2)}{M_\rho^{2}-s} \,
\right. \nonumber \\
& & \left. \qquad\qquad\qquad +
\, \frac{A^{\mathrm{R}}(Q^2,u,s,m_\pi^2,m_K^2,m_K^2)}{M_{K^{*}}^{2}-u} \, \right]
\, , \nonumber \\ [3.5mm]
F_1^{\mathrm{RR}}(s,t) & = &  \frac{\sqrt{2}}{3} \frac{F_A G_V}{F^3}
\frac{Q^2}{M_{\mathrm{a}_1}^2-Q^2} \, \left[
\, \frac{B^{\mathrm{RR}}(Q^2,s,u,t,m_K^2,m_{\pi}^2,m_K^2)}{M_{K^{*}}^{2}-t}
\right. \nonumber \\
& & \qquad\qquad\qquad\qquad\qquad
 + \, 2 \, \frac{A^{\mathrm{RR}}(Q^2,s,u,m_K^2,m_\pi^2,m_K^2)}{M_\rho^{2}-s} \,
\nonumber \\
& & \left.\qquad\qquad\qquad\qquad\qquad
 + \, \frac{A^{\mathrm{RR}}(Q^2,u,s,m_\pi^2,m_K^2,m_K^2)}{M_{K^{*}}^{2}-u} \, \right]
\, ,
\end{eqnarray}
\begin{eqnarray}
\label{eq:f22}
 F_2^{\chi} & = & 0 \, , \nonumber \\ [3mm]
F_2^{\mathrm{R}}(s,t) & = & - \frac{1}{6} \frac{F_V G_V}{F^3} \, \left[
\, \frac{A^{\mathrm{R}}(Q^2,t,u,m_K^2,m_K^2,m_{\pi}^2)}{M_{K^{*}}^{2}-t} \, +
\, 2 \; \frac{B^{\mathrm{R}}(t,u,m_K^2,m_K^2)}{M_\rho^{2}-s} \,
\right. \nonumber \\
& & \left. \qquad\qquad\qquad -
\, \frac{A^{\mathrm{R}}(Q^2,u,t,m_K^2,m_K^2,m_\pi^2)}{M_{K^{*}}^{2}-u} \, \right]
\, , \nonumber \\ [3.5mm]
F_2^{\mathrm{RR}}(s,t) & = &  \frac{\sqrt{2}}{3} \frac{F_A G_V}{F^3}
\frac{Q^2}{M_{\mathrm{a}_1}^2-Q^2} \, \left[
\, \frac{A^{\mathrm{RR}}(Q^2,t,u,m_K^2,m_K^2,m_{\pi}^2)}{M_{K^{*}}^{2}-t}
\right. \nonumber \\
& & \qquad\qquad\qquad\qquad\qquad
+ \, 2 \; \frac{B^{\mathrm{RR}}(Q^2,t,u,s,m_K^2,m_K^2,m_\pi^2)}{M_\rho^{2}-s} \,
\nonumber \\
& & \left. \qquad\qquad\qquad\qquad\qquad
- \, \frac{A^{\mathrm{RR}}(Q^2,u,t,m_K^2,m_K^2,m_\pi^2)}{M_{K^{*}}^{2}-u} \, \right] \, .
\end{eqnarray}
\hspace*{0.5cm}The form factor driven by the vector current is given by~:
\begin{eqnarray}
\label{eq:f32}
 F_3^{\chi} & = & 0  \, \nonumber \\ [3mm]
F_3^{\mathrm{R}}(s,t) & = & \frac{2 \sqrt{2} \, G_V}{M_V \, F^3}
\!\!\left[ \frac{C^{\mathrm{R}}(Q^2,t,m_K^2,m_\pi^2,m_K^2)}{M_{K^*}^2-t} -
\frac{C^{\mathrm{R}}(Q^2,u,m_K^2,m_\pi^2,m_K^2)}{M_{K^*}^2-u} \right. \nonumber\\
& &  \left. - \frac{2 F_V}{G_V}
\frac{E^{\mathrm{R}}(t,u)}{M_{\rho}^2-Q^2} \right] \, , \nonumber \\ [3mm]
F_3^{\mathrm{RR}}(s,t) & = & - 4 \frac{F_V G_V}{F^3} \frac{1}{M_{\rho}^2-Q^2}
\left[ \frac{C^{\mathrm{RR}}(Q^2,t,m_K^2)}{M_{K^*}^2-t} -
\frac{C^{\mathrm{RR}}(Q^2,u,m_K^2)}{M_{K^*}^2-u} \right] \, ,
\end{eqnarray}
with $E^{\mathrm{R}}$ is defined as
\begin{equation}
 E^{\mathrm{R}}(x,y)\,=\,(g_1+2 \, g_2-g_3)\, (x-y) \, .
\end{equation}
\hspace*{0.5cm}Finally for the pseudoscalar form factor we have~:
\begin{eqnarray}
\label{eq:f42}
 F_4^{\chi}(s,t) & = & \frac{1}{2 \, F} \frac{m_{\pi}^2 \, (t-u)}{Q^2 ( m_{\pi}^2 - Q^2 )} \, , \nonumber \\ [3mm]
F_4^{\mathrm{R}}(s,t) & = & \frac{1}{2} \frac{G_V^2}{F^3} \frac{m_{\pi}^2}{Q^2 (m_{\pi}^2 - Q^2 )}
\left[ \frac{t(s-u)-(m_K^2-m_{\pi}^2)(Q^2-m_K^2)}{M_{K^*}^2-t}
+ \frac{2 \, s (t-u)}{M_{\rho}^2-s} \right. \nonumber \\ [3mm]
& & \qquad \qquad \qquad \qquad \; \; \; \,\left. - \frac{u(s-t)-(m_{K}^2-m_{\pi}^2)(Q^2-m_K^2)}{M_{K^*}^2-u}
\right] \, .
\end{eqnarray}
\subsection{Features of the form factors}
\hspace*{0.5cm}Several remarks are needed in order to understand our previous results for the form factors related with the vector and 
axial-vector $QCD$ currents analysed above~:
\begin{itemize}
 \item[1/] Our evaluation corresponds to the tree level diagrams in Figure~\ref{fig_diagsKKpi} that arise from the $N_C \rightarrow \infty$ 
limit of $QCD$. Hence the masses of the resonances would be reduced to $M_V=M_{\rho}=M_{\omega}=M_{K^*}=M_{\phi}$ and $M_A=M_{\mathrm{a}_1}$ as they 
appear in the resonance Lagrangian (\ref{Rkin}), i.e. the masses of the nonet of vector and axial-vector resonances in the chiral and large
-$N_C$ limit. However it is easy to introduce $NLO$ corrections in the $1/N_C$ and chiral expansions on the masses by including the 
\textit{physical}  ones~: $M_{\rho}$, $M_{K^*}$, $M_{\omega}$, $M_{\phi}$ and $M_{\mathrm{a}_1}$ for the $\rho(770)$, $K^*(892)$, $\omega(782)$, 
$\phi(1020)$ and $\mathrm{a}_1(1260)$ states, respectively, as we have done in the expressions of the form factors. In this setting resonances also 
have zero width, which represents a drawback if we intend to analyse the phenomenology of the processes: Due to the high mass of the tau 
lepton, resonances do indeed resonate producing divergences if their widths are ignored. Hence we will include energy-dependent widths for the
$\rho(770)$, a$_1(1260)$ and $K^*(892)$ resonances, that are rather wide, and a constant width for the $\omega(782)$. This issue is 
discussed in the Appendix C
.\\
\hspace*{0.5cm}In summary, to account for the inclusion of $NLO$ corrections we perform the substitutions:
\begin{equation}
 \frac{1}{M_R^2-q^2} \; \; \longrightarrow \;\; \frac{1}{M_{phys}^2-q^2- \, i \, M_{phys} \, \Gamma_{phys}(q^2)} \; ,
\end{equation}
where $R=V,A$, and the subindex \textit{phys} on the right hand side stands for the corresponding \textit{physical} state depending on the 
relevant Feynman diagram.

\item[2/] If we compare our results with those of Ref.~\cite{Finkemeier:1995sr}, evaluated within the $KS$ model, we notice that the 
structure of our form factors is fairly different and much more intricate. This is due to the fact that the $KS$ model, i.e.\ a model 
resulting from combinations  of \textit{ad hoc} products of Breit-Wigner functions, does not meet higher order chiral constraints enforced 
in our approach.

\item[3/] As commented above the pseudoscalar form factors $F_4$ vanishes in the chiral limit. Indeed the results of Eqs.~(\ref{eq:f41}, 
\ref{eq:f42}) show that they are proportional to $m_{\pi}^2$, which is tiny compared with any other scale in the amplitudes. Hence the 
contribution of $F_4$ to the structure of the spectra is actually marginal.
\end{itemize}

\section{QCD constraints and determination of resonance couplings}\label{KKpi_QCDconstraints}
\hspace*{0.5cm}Our results for the form factors $F_i$ depend on several combinations of the coupling constants in our Lagrangian 
${\cal L}_{R \chi T}$, most of which are in principle unknown parameters. Now, if our theory offers an adequate effective description of
QCD at hadronic energies, the underlying theory of the strong interactions should give information on those constants. Unfortunately the
determination of the effective parameters from first principles is still an open problem in hadronic physics.\\
\hspace*{0.5cm}A fruitful procedure when working with resonance Lagrangians has been to assume that the resonance region, even when one does
 not include the full phenomenological spectrum, provides a bridge between the chiral and perturbative regimes \cite{Ecker:1989yg}. The 
chiral constraints supply information on the structure of the interaction but do not provide any hint on the coupling constants of the 
Lagrangian. Indeed, as in any effective theory \cite{Georgi:1991ch}, the couplings encode information from high energy dynamics. Our procedure
 amounts to match the high energy behaviour of Green functions (or related form factors) evaluated within the resonance theory with the
asymptotic results of perturbative $QCD$. This strategy has proven to be phenomenologically sound \cite{Ecker:1989yg, Cirigliano:2004ue, 
RuizFemenia:2003hm, Cirigliano:2005xn, Moussallam:1997xx, Knecht:2001xc,  Mateu:2007tr, Amoros:2001gf}, and it will be applied here in order 
to obtain information on the unknown couplings.\\
Two-point Green functions of vector and axial-vector currents $\Pi_{V,A}(q^2)$ were studied within perturbative $QCD$ in Ref.~
\cite{Floratos:1978jb}, where it was shown that both spectral functions go to a constant value at infinite transfer of momenta~:
\begin{equation}\label{FNdR}
 \Im m \, \Pi_{V,A}(q^2) \, \mapright{}{\; \; \; q^2 \rightarrow \infty \; \; \; } \, \, \frac{N_C}{12 \, \pi}
\, .
\end{equation}
\hspace*{0.5cm}By local duality interpretation the imaginary part of the quark loop can be understood as the sum of infinite positive 
contributions of intermediate hadronic states. Now, if the infinite sum is going to behave like a constant at $q^2 \rightarrow \infty$, it is 
heuristically sound to expect that each one of the infinite contributions vanishes in that limit. This deduction stems from the fact that 
vector and axial-vector form factors should behave smoothly at high $q^2$, a result previously put forward from parton dynamics in Refs.~
\cite{Brodsky:1973kr, Brodsky:1974vy, Lepage:1979zb, Lepage:1980fj}. Accordingly in the $N_C \rightarrow\infty$ limit this result applies to 
our form factors evaluated at tree level in our framework.\\
\hspace*{0.5cm}Other hints involving short-distance dynamics may also be considered. The analyses of three-point Green functions of $QCD$ 
currents have become a useful procedure to determine coupling constants  of the intermediate energy (resonance) framework \cite{Cirigliano:2004ue,RuizFemenia:2003hm,
Cirigliano:2005xn,Moussallam:1997xx,Knecht:2001xc}. The idea is to consider those functions (order parameters of the chiral symmetry breaking), evaluate them within 
the resonance framework and match this result with the leading term in the $OPE$ of the Green function.\\
\hspace*{0.5cm}In the following we collect the information provided by these hints on our coupling constants, attaching always to the 
$N_C \rightarrow \infty$ case \cite{Pich:2002xy} (approximated with only one nonet of vector and axial-vector resonances)~:
\begin{itemize}
 \item[i)] By demanding that the two-pion vector form factor vanishes at high $q^2$ one obtains the condition $F_V \, G_V = F^2$ \cite{Ecker:1989yg}.
\item[ii)] The first Weinberg sum rule \cite{Weinberg:1967kj} leads to $F_V^2 - F_A^2 = F^2$, and the second Weinberg sum rule gives $F_V^2 
\, M_V^2 \, = \, F_A^2 \, M_A^2$\cite{Ecker:1988te}.
\item[iii)] The analysis of the $VAP$ Green function \cite{Cirigliano:2004ue} gives for the combinations of couplings defined in 
Eq.~(\ref{lambdas0,',''}) the following results~:
\begin{eqnarray}
\label{eq:lambres}
 \lambda' & = & \frac{F^2}{2 \, \sqrt{2} \, F_A \, G_V} \; = \; 
\frac{M_A}{2 \, \sqrt{2} \, M_V} \,, \nonumber \\[3.5mm]
\lambda'' & = & \frac{2 \, G_V \, - F_V}{2 \, \sqrt{2} \, F_A} \; = \; \frac{M_A^2 - 2 M_V^2}{2 \, \sqrt{2} \, M_V \, M_A} \, , \nonumber \\[3.5mm]
4 \, \lambda_0 & = & \lambda' + \lambda'' \; ,
\end{eqnarray}
where, in the two first relations, the second equalities come from using relations i) and ii) above. Here $M_V$ and $M_A$ are the masses 
appearing in the resonance Lagrangian. Contrarily to what happens in the vector case where $M_V$ is well approximated by the $\rho(770)$ mass,
  in Ref.~\cite{Mateu:2007tr} it was obtained $M_A = 998 (49)$ MeV and in Ref.\cite{Pich:2010sm} $M_A = 920 (20)$ MeV: hence $M_A$ differs appreciably from the presently accepted value of 
$M_{\mathrm{a}_1}$. It is worth to notice that the two first relations in Eq.~(\ref{eq:lambres}) can also be obtained from the requirement that the 
$J=1$ axial spectral function in $\tau \rightarrow 3 \pi \nu_{\tau}$ vanishes for large momentum transfer \cite{GomezDumm:2003ku}.
\item[iv)] Both vector form factors contributing to the final states $K \overline{K} \pi^-$ and $K^- K^0 \pi^0$ in tau decays, when 
integrated over the available phase space, should also vanish at high $Q^2$. Let us consider $H_{\mu \nu}^3(s,t,Q^2) \equiv T_{\mu}^3 
T_{\nu}^{3 \, *}$, where $T_{\mu}^3$ can be inferred from Eq.~(\ref{generaldecomposition_3mesons}). Then we define $\Pi_V(Q^2)$ by~:
\begin{equation}
 \int \, \mathrm{d}\Pi_3 \, H_{\mu \nu}^3(s,t,Q^2) \, = \, \left( Q^2 g_{\mu \nu} \, - \,
Q_{\mu} Q_{\nu} \right) \, \Pi_V(Q^2) \, ,
\end{equation}
where \footnote{See Section \ref{Hadrondecays_Three_meson_decays_MI}.}
\begin{eqnarray}
 \int \mathrm{d}\Pi_3 \, & = & \, \int \frac{d^3 p_1}{2 E_1} \frac{d^3 p_2}{2 E_2} \frac{d^3 p_3}{2 E_3}
\delta^4\left( Q-p_1-p_2-p_3\right)  \delta \left( s-(Q-p_3)^2 \right) \delta \left( t-(Q-p_2)^2 \right) \,
\nonumber \\
\, & = &  \, \frac{\pi^2}{4 \,Q^2} \, \int ds \, dt \; .
\end{eqnarray}
Hence we find that
\begin{equation}
 \Pi_V(Q^2) \, = \, \frac{\pi^2}{12 \, Q^4} \, \int \, \mathrm{ds \, dt} \, g^{\mu \nu} \,
H_{\mu \nu}^3(s,t,Q^2) \, ,
\end{equation}
where the limits of integration can be obtained from Eq.~(\ref{integrationlimits}), should vanish at $Q^2 \rightarrow \infty$. This 
constraint determines several relations on the couplings that appear in the $F_3$ form factor, namely~:
\begin{eqnarray}
\label{eq:1c}
 c_1 \, - \, c_2 \, + \, c_5 \, & = & 0 \, , \\
\label{eq:2c}
c_1 \, - \, c_2 \, - \, c_5 \, + \, 2 c_6 \, & = & - \, \frac{ \,N_C}{96 \, \pi^2} \,
\frac{F_V \, M_V}{\sqrt{2} \, F^2} \, , \\
\label{eq:3c}
d_3 & = & - \frac{N_C}{192 \, \pi^2} \, \frac{M_V^2}{F^2} \, , \\
\label{eq:4c}
g_1 \, + \, 2 g_2 \, - g_3 & = & 0 \, , \\
\label{eq:5c}
g_2 & = & \frac{N_C}{192 \,\sqrt{2} \, \pi^2} \, \frac{M_V}{F_V} \, .
\end{eqnarray}
If these conditions are satisfied, $\Pi_V(Q^2)$ vanishes at high transfer of momenta for both $K \overline{K} \pi^-$ and $K^- K^0 \pi^0$
final states. We notice that the result in Eq.~(\ref{eq:1c}) is in agreement with the corresponding relation in Ref.~\cite{RuizFemenia:2003hm},
while Eqs.~(\ref{eq:2c}) and (\ref{eq:3c}) do not agree with the results in that work. In this regard we point out that the relations in
Ref.~\cite{RuizFemenia:2003hm}, though they satisfy the leading matching to the $OPE$ expansion of the $\langle VVP \rangle$ Green function 
with the inclusion of one multiplet of vector mesons, do not reproduce the right asymptotic behaviour of related form factors. Indeed it has 
been shown \cite{Knecht:2001xc,Mateu:2007tr} that two multiplets of vector resonances are needed to satisfy both constraints. Hence we will 
attach to our results above, which we consider more reliable~\footnote{One of the form factors derived from the $\langle VVP \rangle$ Green 
function is ${\cal F}_{\pi \gamma^* \gamma}(q^2)$, that does not vanish at high $q^2$ with the set of relations in Ref.~\cite{RuizFemenia:2003hm}.
 With our conditions in Eqs.~(\ref{eq:2c},\ref{eq:3c}) the asymptotic constraint on the form factor can be satisfied if the large-$N_C$ 
masses, $M_A$ and $M_V$, fulfill the relation $2 M_A^2 = 3 M_V^2$, that is again recovered in Chapter \ref{Pgamma}. It is interesting to 
notice the significant agreement with the numerical values for these masses mentioned above.}.
\item[v)] An analogous exercise to the one in iv) can be carried out for the axial-vector form factors $F_1$ and $F_2$. We have performed 
such an analysis and, using the relations in i) and ii) above, it gives us back the results provided in Eq.~(\ref{eq:lambres}) for $\lambda'$
 and $\lambda''$. Hence both procedures give a consistent set of relations.
\end{itemize}
After imposing the above constraints, let us analyse which coupling combinations appearing in our expressions for the form factors are still
unknown. We intend to write all the information on the couplings in terms of $F$, $M_V$ and $M_A$. From the relations involving $F_V$, $F_A$ 
and $G_V$ we obtain~:
\begin{eqnarray} \label{eq:fvfagv}
 \frac{F_V^2}{F^2} & = & \frac{M_A^2}{M_A^2-M_V^2} \, , \nonumber \\
\frac{F_A^2}{F^2} & = & \frac{M_V^2}{M_A^2-M_V^2} \, , \nonumber \\
\frac{G_V^2}{F^2} & = & 1 \, - \, \frac{M_V^2}{M_A^2} \, .
\end{eqnarray}
Moreover we know that $F_V$ and $G_V$ have the same sign, and we will assume that it is also the sign of $F_A$. Together with the relations 
in Eq.~(\ref{eq:lambres}) this determines completely the axial-vector form factors $F_{1,2}$. Now from Eqs.~(\ref{eq:1c}-\ref{eq:5c}) one 
can fix all the dominant pieces in the vector form factor $F_3$, i.e. those pieces that involve factors of the kinematical variables $s$, 
$t$ or $Q^2$. The unknown terms, that carry factors of $m_{\pi}^2$ or $m_K^2$, are expected to be less relevant. They are given by the 
combinations of couplings~:  $c_1 + c_2 + 8 \, c_3 - c_5$,  $d_1 + 8 \, d_2$,  $c_4$ , $g_4$ and  $g_5$. However small they may be, we will 
not neglect these contributions, and we will proceed as follows. Results in Ref.~\cite{RuizFemenia:2003hm} determine the first and the 
second coupling combinations. As commented above the constraints in that reference do not agree with those we have obtained by requiring 
that the vector form factor vanishes at high $Q^2$. However, they provide us an estimate to evaluate terms that, we recall, are suppressed by
pseudoscalar masses. In this way, from a phenomenological analysis of $\omega \rightarrow \pi^+ \pi^- \pi^0$ (see Appendix \ref{omegaApp})
 it is possible to determine the combination $2 \, g_4 + g_5$. Finally in order to evaluate $c_4$ and $g_4$ we will combine the recent analysis 
of $\sigma \left( e^+ e^- \rightarrow K K \pi \right)$ by $BaBar$ \cite{Aubert:2007ym} with the information from the $\tau \rightarrow K K 
\pi \nu_{\tau}$ width~\footnote{When this work was completed, Belle data on the $\tau \rightarrow K^+ K^- \pi^- \nu_{\tau}$\cite{:2010tc} decays was not available yet.}.\\
\subsection{Determination of $c_4$ and $g_4$}\label{determ_c4g4}
\begin{figure}[!t]
\begin{center}
\vspace*{0.2cm}
\includegraphics[scale=0.5,angle=-90]{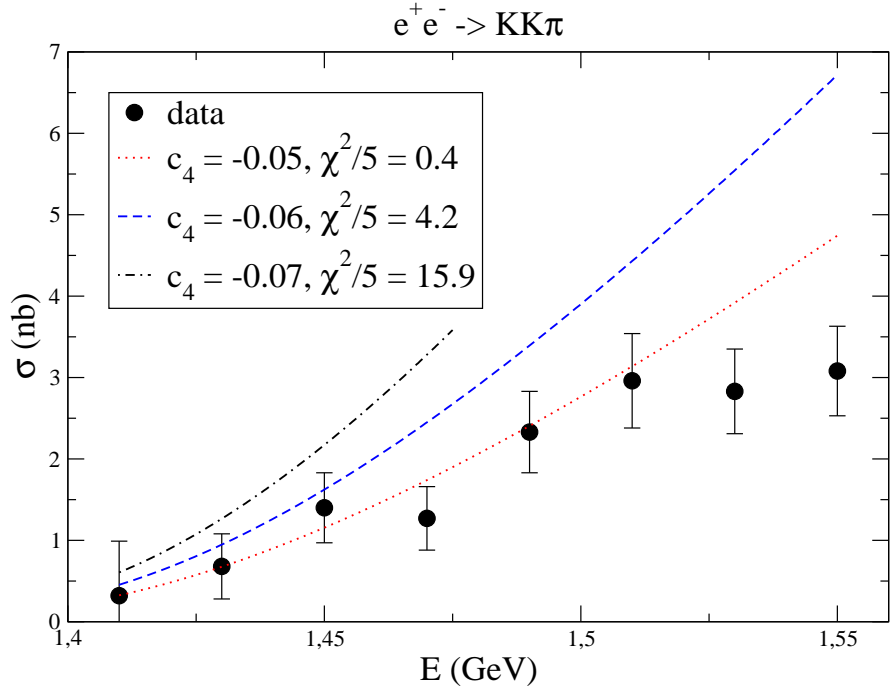}
\caption[]{\label{fig:e+e-KKpi} \small{Comparison of the experimental data \cite{Aubert:2007ym} with the theoretical prediction for the 
cross-section of the isovector component of $e^+ e^-  \rightarrow K^*(892) K \rightarrow K_S K^{\pm} \pi^{\mp}$ process, for different values
 of the $c_4$ coupling. The $\chi^2$ values are associated to the first 6 data points only.}}
\end{center}
\end{figure}

\hspace*{0.5cm}The separation of isoscalar and isovector components of the $e^+ e^- \rightarrow K K \pi$ amplitudes, carried out by $BaBar$ 
\cite{Aubert:2007ym}, provides us with an additional tool for the estimation of the coupling constant $c_4$ that appears in the hadronization
 of the vector current \cite{Roig:2008je, Roig:2009qw}. Indeed, using $SU(2)_I$ symmetry alone one can relate the isovector contribution to $\sigma 
\left( e^+ e^- \rightarrow K^- K^0 \pi^+ \right)$ with the vector contribution to $\Gamma \left( \tau^- \rightarrow K^0 K^- \pi^0 \nu_{\tau} \right)$
 through the relation~:
\begin{equation} \label{eq:CVC}
  \frac{\mathrm{d}}{\mathrm{d} \, Q^2} \, \Gamma \left( \tau^- \rightarrow K^0 K^- \pi^0 \nu_{\tau} \right)
\Bigg|_{F_3} \, = \, f(Q^2) \; \sigma_{I=1} \left( e^+ e^- \rightarrow K^- K^0 \pi^+ \right)
\; ,
\end{equation}
where $f(Q^2)$ and further relations are given in Appendix E
. Another relation similar to Eq.~(\ref{eq:CVC}) that has been widely used 
in the literature and the assumptions on which it relies are also discussed in this Appendix. In order not to lose the thread of our discourse, 
here we will only include the explanation of our methodology to determine $c_4$ and $g_4$.\\
\hspace*{0.5cm}Hence we could use the isovector contribution to the cross-section for the process $e^+ e^- \rightarrow K_S K^{\pm} \pi^{\mp}$
 determined by $BaBar$ and Eq.~(\ref{eq:CVC}) to fit the $c_4$ coupling that is the only still undetermined constant in that process. However
 we have to take into account that our description for the hadronization of the vector current in the tau decay channel does not, necessarily,
 provide an adequate description of the cross-section. Indeed the complete different kinematics of both observables suppresses the high-energy 
behaviour of the bounded tau decay spectrum, while this suppression does not occur in the cross-section. Accordingly, our description of the 
latter away from the energy threshold can be much poorer. As can be seen in Figure~\ref{fig:e+e-KKpi} there is a clear structure in the experimental 
points of the cross-section that is not provided by our description.\\
\hspace*{0.5cm}Taking into account the input parameters quoted in Eq.~(\ref{eq:set1}) we obtain~: $ c_4 \, = \, -0$.$047 \pm 0$.$002 $. The fit 
has been carried out for the first 6 bins (up to $E_{cm} \sim 1$.$52 \, \mathrm{GeV}$) using $MINUIT$ \cite{James:1975dr}. This result corresponds 
to $\chi^2 /dof = 0$.$3$ and the displayed error comes only from the fit.\\
\hspace*{0.5cm}We take into consideration now the measured branching ratios for the $K K \pi$ channels of Table~\ref{tab:KKpi_thvsexp} in 
order to extract information both from $c_4$ and $g_4$. We notice that it is not possible to reconcile a prediction of the branching ratios 
of $\tau \rightarrow K \overline{K} \pi \nu_{\tau}$ and $\tau \rightarrow K^- K^0 \pi^0 \nu_{\tau}$ in spite of the noticeable size of the 
errors shown in the Table~\ref{tab:KKpi_thvsexp}. Considering that the second process was measured long ago and that the $\tau^- \rightarrow
 K^+ K^- \pi^- \nu_{\tau}$ decay has been focused by both $CLEO-III$ and $BaBar$ we intend to fit the branching ratio of the latter. For the
 parameter values~:
\begin{eqnarray} \label{eq:c4g4I}
 c_4 & = & -0\mathrm{.}07 \pm 0\mathrm{.}01 \, , \nonumber \\
 g_4 & = & -0\mathrm{.}72 \pm 0\mathrm{.}20 \, ,
\end{eqnarray}
we find a good agreement with the measured widths $\Gamma(\tau^- \rightarrow K^+ K^- \pi^- \nu_{\tau})$ and $\Gamma(\tau \rightarrow K^- K^0
 \pi^0 \nu_{\tau})$ within errors (see Table~\ref{tab:KKpi_thvsexp}). Notice that the value of $|c_4|$ is larger than that obtained from the fit to the
 $e^+ e^- \rightarrow K_S K^{\pm} \pi^{\mp}$ data explained above. In Figure~\ref{fig:e+e-KKpi} we show the first 8 bins in the isovector 
component of $e^+ e^- \rightarrow K_S K^{\pm} \pi^{\mp}$ and the theoretical curves for different values of the $c_4$ coupling. As our 
preferred result we choose the larger value of $c_4$ in Eq.~(\ref{eq:c4g4I}), since it provides a better agreement with the present measurement
 of $\Gamma(\tau^- \rightarrow K^- K^0 \pi^0 \nu_{\tau})$. Actually, one can expect an incertitude in the splitting of isospin amplitudes
 in the $e^+ e^- \rightarrow K_S K^{\pm} \pi^{\mp}$ cross-section (as it is discussed in Appendix E). Taking into account this systematic
 error, it could be likely that the theoretical curve with $c_4 = -0$.$07$ falls within the error bars for the first data points.\\
\hspace*{0.5cm}Using $SU(2)_I$ symmetry, one can derive several relations between exclusive isovector hadronic modes produced in $e^+e^-$ 
collisions and those related with the vector current ($F_3$ form factor) in $\tau$ decays. One can read them in Appendix E
, where
 other relations for the three meson decays of interest are also derived.
\\
\section{Phenomenology of $\tau \rightarrow K K \pi \nu_{\tau}$~: Results and their analysis}\label{KKpi_Phenomenology}
\begin{figure}[!h]
\begin{center}
\vspace*{0.5cm}
\includegraphics[scale=0.6,angle=-90]{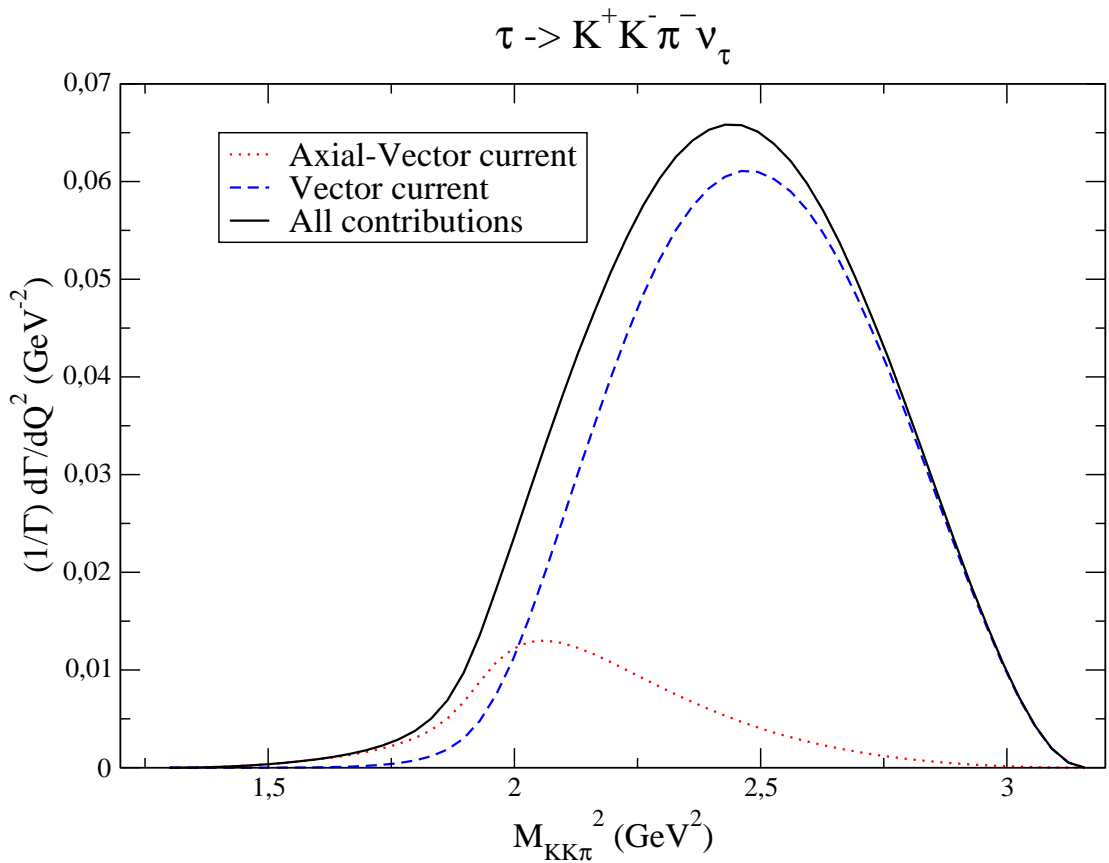}
\caption[]{\label{fig:K+K-pi-spectralfunction} \small{Normalized $M_{K K \pi}^2$-spectra for $\tau^- \rightarrow K^+ K^- \pi^- \nu_{\tau}$. Notice the dominance 
of the axial-vector current at very low values of $Q^2$.}}
\end{center}
\end{figure}

\begin{figure}[!h]
\begin{center}
\vspace*{0.5cm}
\includegraphics[scale=0.6,angle=-90]{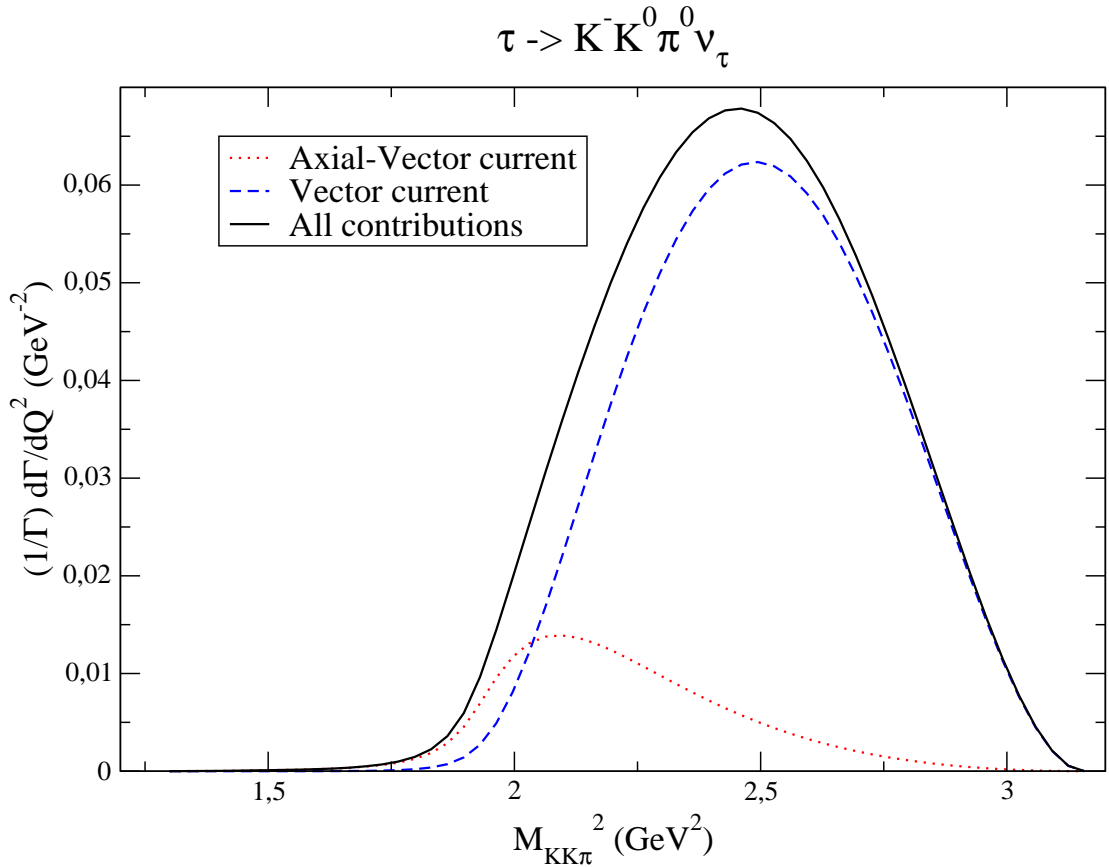}
\caption[]{\label{fig:K-K0pi0spectralfunction} \small{Normalized $M_{K K \pi}^2$-spectra for $\tau^- \rightarrow K^-K^0\pi^0\nu_{\tau}$. Notice the dominance of 
the axial-vector current at very low values of $Q^2$.}}
\end{center}
\end{figure}

\begin{figure}[!h]
\begin{center}
\vspace*{0.5cm}
\includegraphics[scale=0.6,angle=-90]{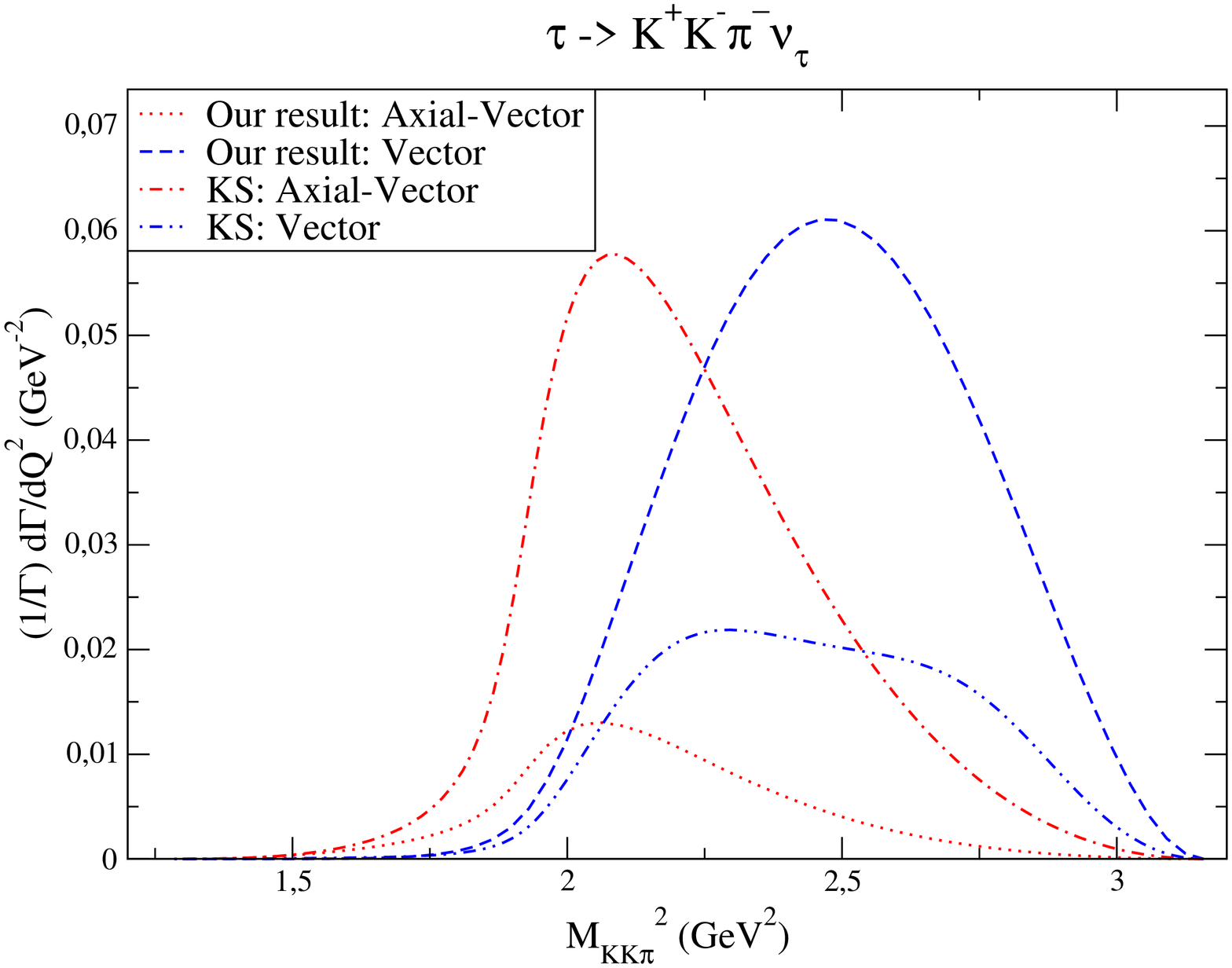}
\caption[]{\label{fig:comparisonwithKS1} \small{Comparison between the normalized $M_{K K \pi}^2$-spectra for the vector and axial-vector 
contributions to the $\tau^- \rightarrow K^+ K^- \pi^- \nu_{\tau}$ channel in the $KS$ model \cite{Finkemeier:1996hh} and in our approach.}}
\end{center}
\end{figure}

\begin{figure}[!h]
\begin{center}
\vspace*{0.5cm}
\includegraphics[scale=0.6,angle=-90]{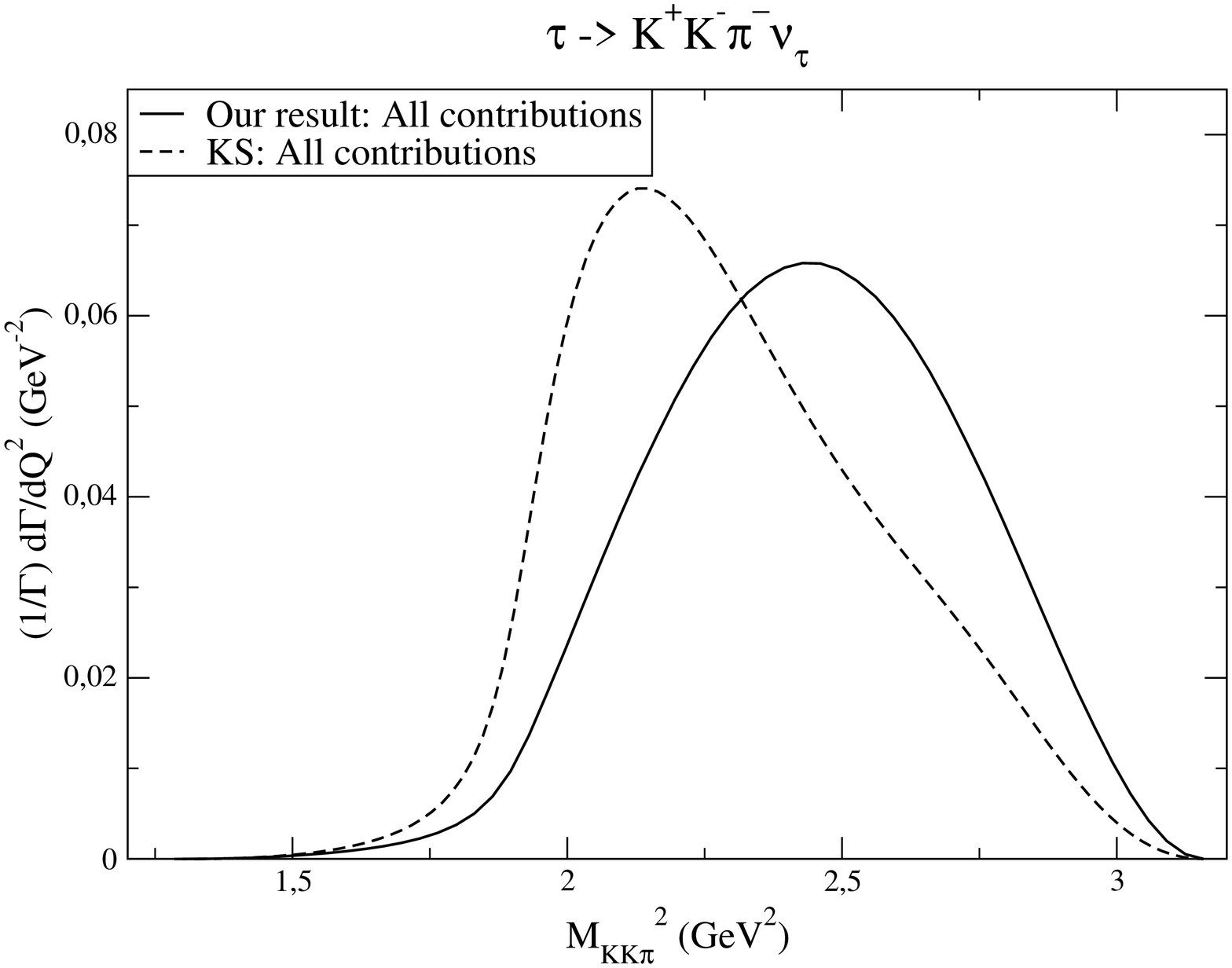}
\caption[]{\label{fig:comparisonwithKS2} \small{Comparison between the normalized $M_{K K \pi}^2$-spectra for $\tau^- \rightarrow K^+ K^- 
\pi^- \nu_{\tau}$ in the $KS$ model \cite{Finkemeier:1996hh} and in our approach.}}
\end{center}
\end{figure}

\begin{figure}[!h]
\begin{center}
\vspace*{0.5cm}
\includegraphics[scale=0.6,angle=-90]{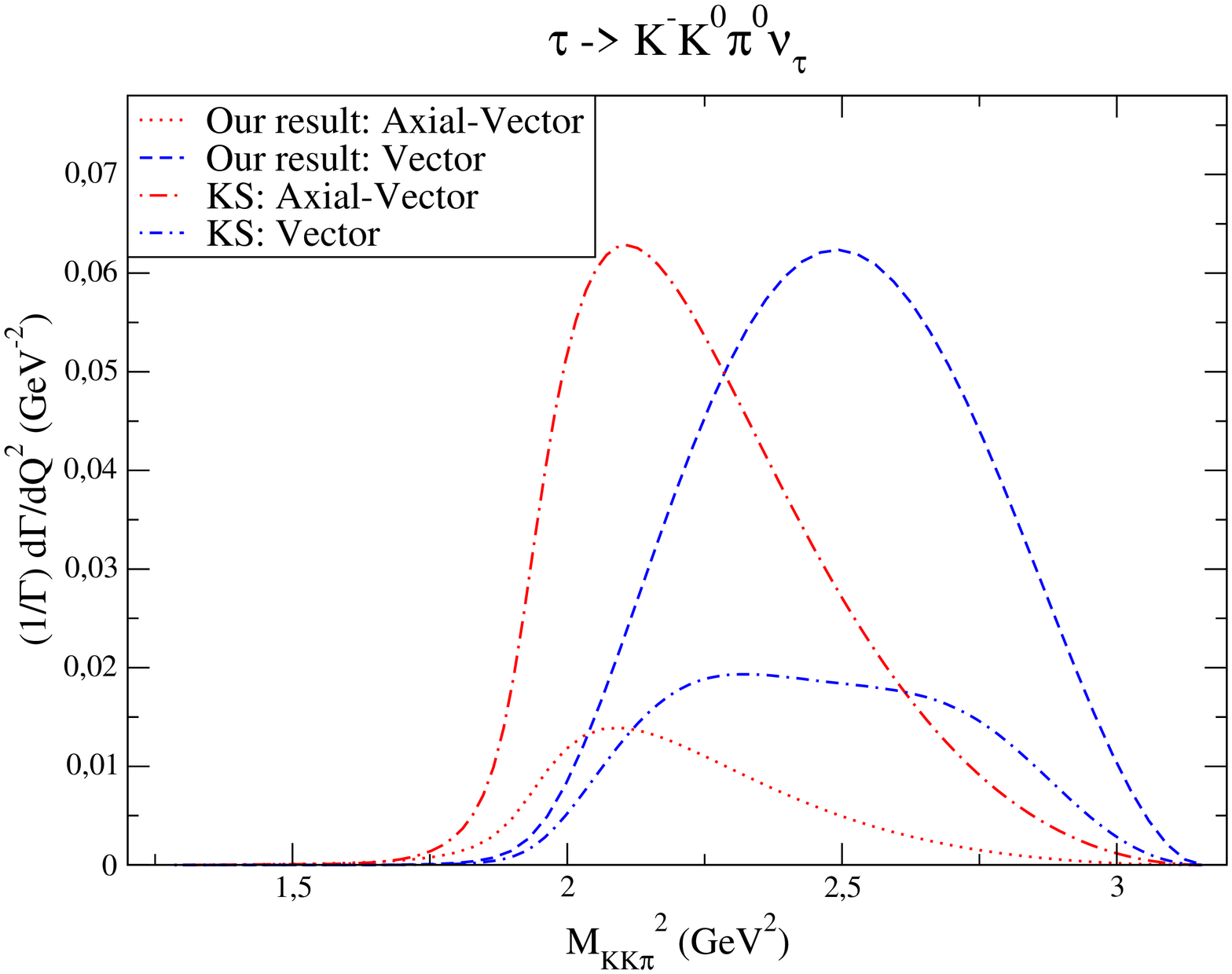}
\caption[]{\label{fig:comparisonwithKSextra} \small{Comparison between the normalized $M_{K K \pi}^2$-spectra for the vector and axial-vector 
contributions to the $\tau^- \rightarrow K^- K^0 \pi^0 \nu_{\tau}$ channel in the $KS$ model \cite{Finkemeier:1996hh} and in our approach.}}
\end{center}
\end{figure}

\begin{figure}[!h]
\begin{center}
\vspace*{0.4cm}
\includegraphics[scale=0.6,angle=-90]{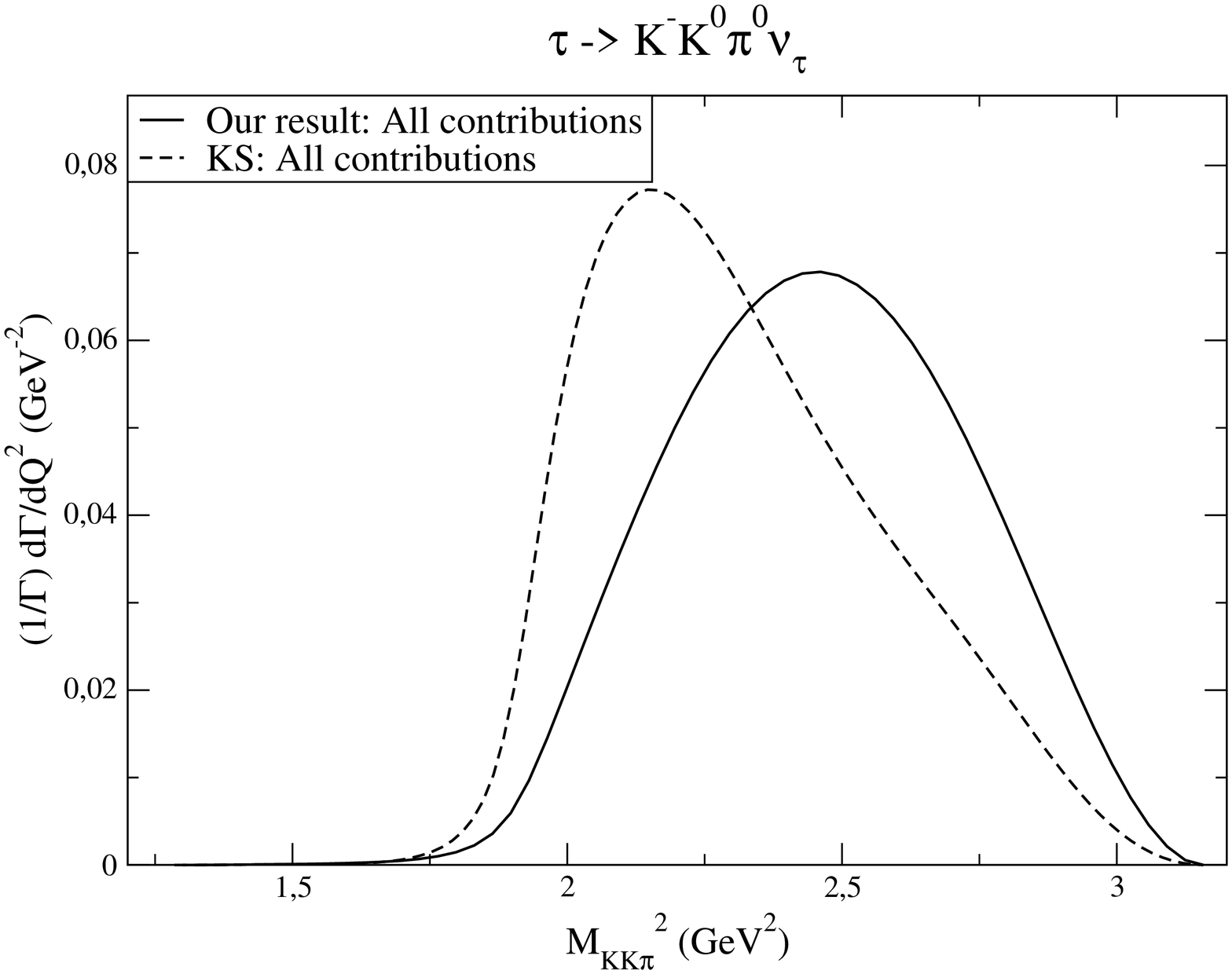}
\caption[]{\label{fig:comparisonwithKS3} \small{Comparison between the normalized $M_{K K \pi}^2$-spectra for $\tau^- \rightarrow K^- K^0 
\pi^0 \nu_{\tau}$ in the $KS$ model \cite{Finkemeier:1996hh} and in our approach.}}
\end{center}
\end{figure}

\begin{figure}[!h]
\begin{center}
\vspace*{1.0cm}
\includegraphics[scale=0.6]{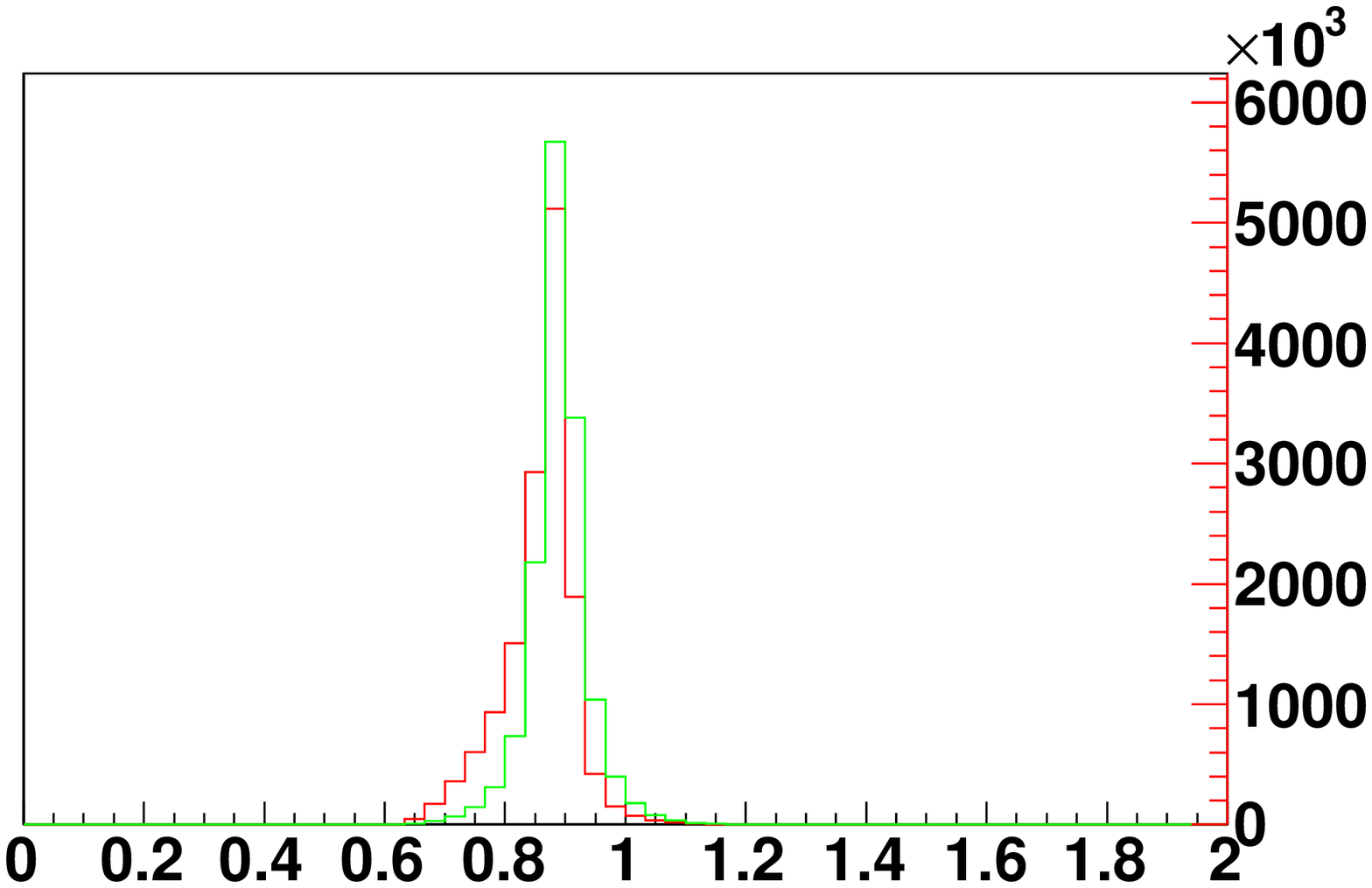}
\caption[]{\label{fig:K+pi-_KKpi} \small{Comparison of our predictions (green) and TAUOLA CLEO (red) for the $K^+\pi^-$ invariant mass (GeV) in the $\tau^- \rightarrow K^+ K^- \pi^- \nu_{\tau}$ decays. The normalization is arbitrary.}}
\end{center}
\end{figure}

\begin{figure}[!h]
\begin{center}
\vspace*{1.0cm}
\includegraphics[scale=0.6]{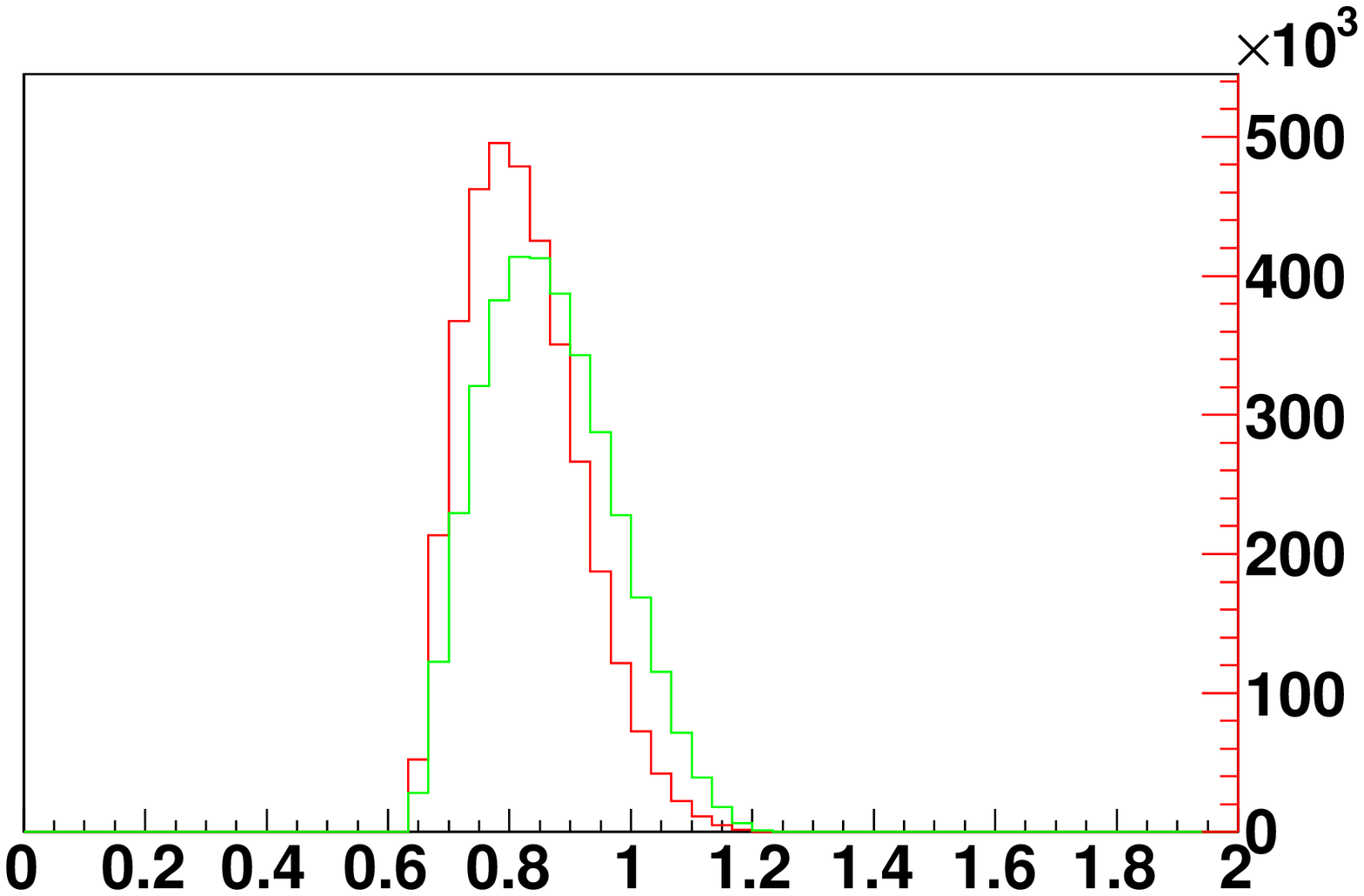}
\caption[]{\label{fig:KS0pi-_KKpi} \small{Comparison of our predictions (green) and TAUOLA CLEO (red) for the $K_S^0\pi^-$ invariant mass (GeV) in the $\tau^- \rightarrow K^0 \bar{K}^0 \pi^- \nu_{\tau}$ decays. The normalization is arbitrary.}}
\end{center}
\end{figure}

\begin{figure}[!h]
\begin{center}
\vspace*{1.0cm}
\includegraphics[scale=0.6]{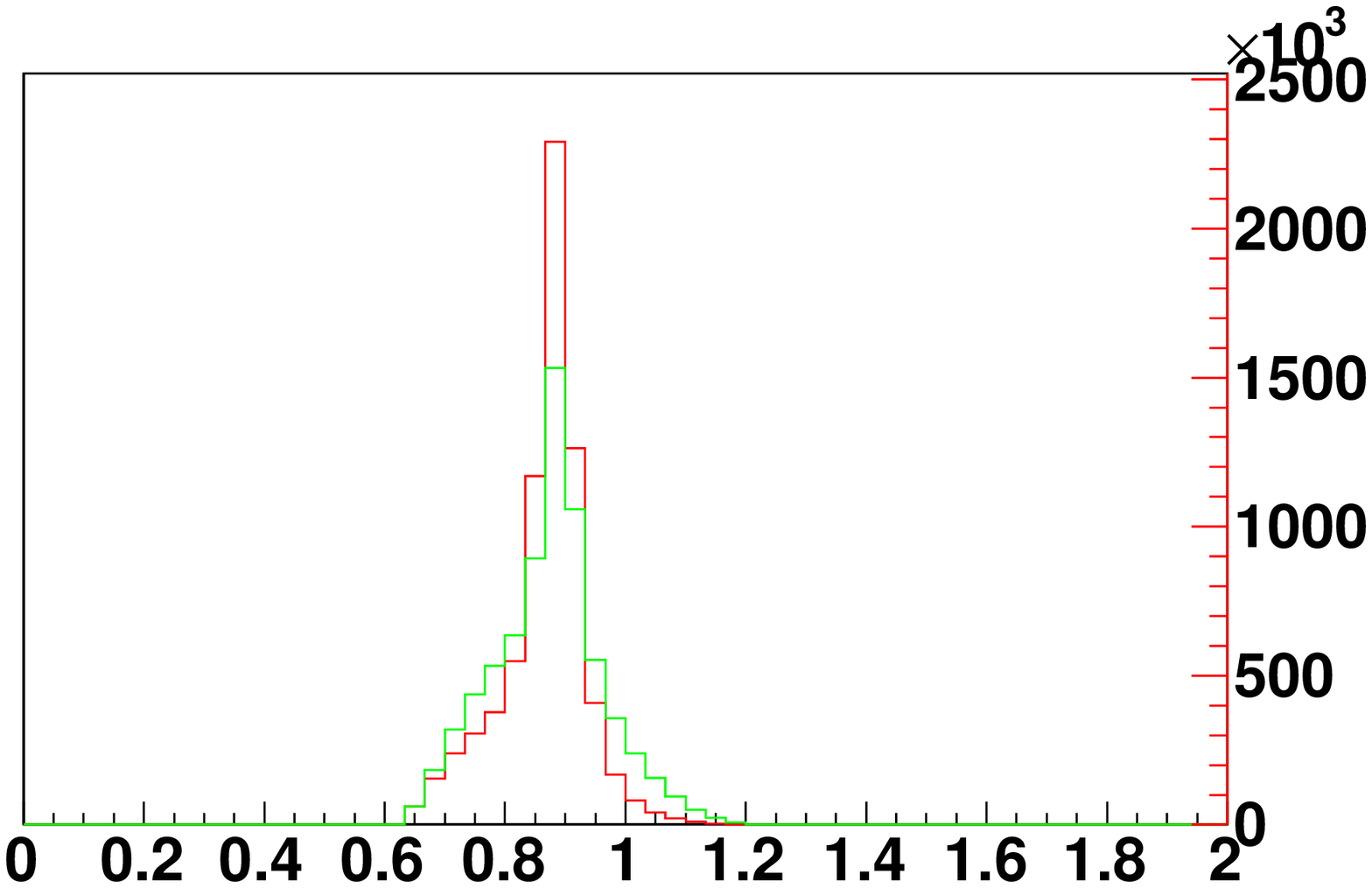}
\caption[]{\label{fig:K-pi^0_KKpi} \small{Comparison of our predictions (green) and TAUOLA CLEO (red) for the $K^-\pi^0$ invariant mass (GeV) in the $\tau^- \rightarrow K^- K^0 \pi^0 \nu_{\tau}$ decays. The normalization is arbitrary.}}
\end{center}
\end{figure}

\begin{figure}[!h]
\begin{center}
\vspace*{1.0cm}
\includegraphics[scale=0.5, angle=-90]{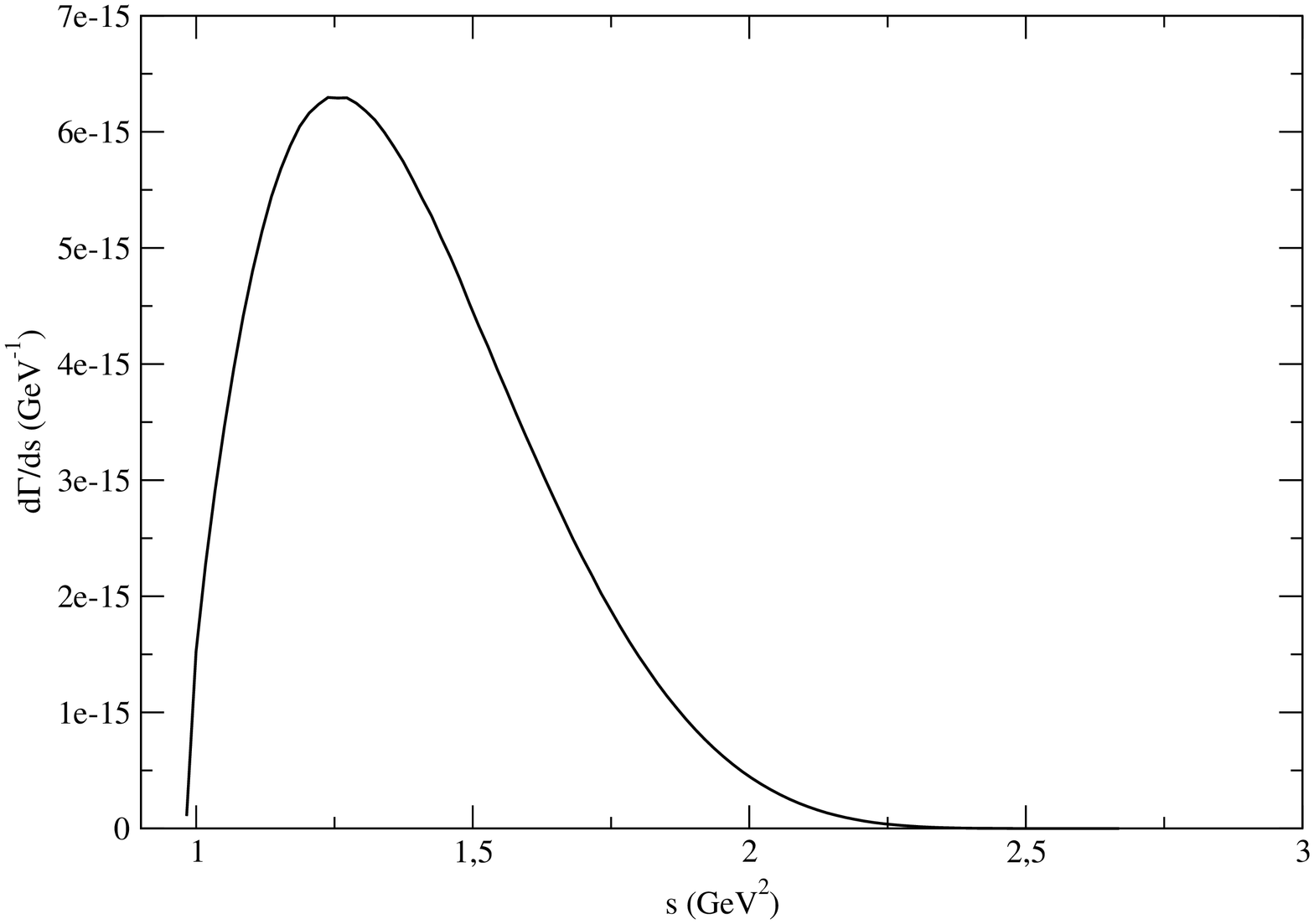}
\caption[]{\label{fig:dGdstkkp} \small{Our prediction for the $K^+K^-$ ($K^0\bar{K}^0$) invariant mass distribution in the $\tau^- \rightarrow K^+ K^- \pi^- \nu_{\tau}$ ($\tau^- \rightarrow K^0\bar{K}^0 \pi^- \nu_{\tau}$) 
decays is shown.}}
\end{center}
\end{figure}

\begin{figure}[!h]
\begin{center}
\vspace*{1.0cm}
\includegraphics[scale=0.45, angle=-90]{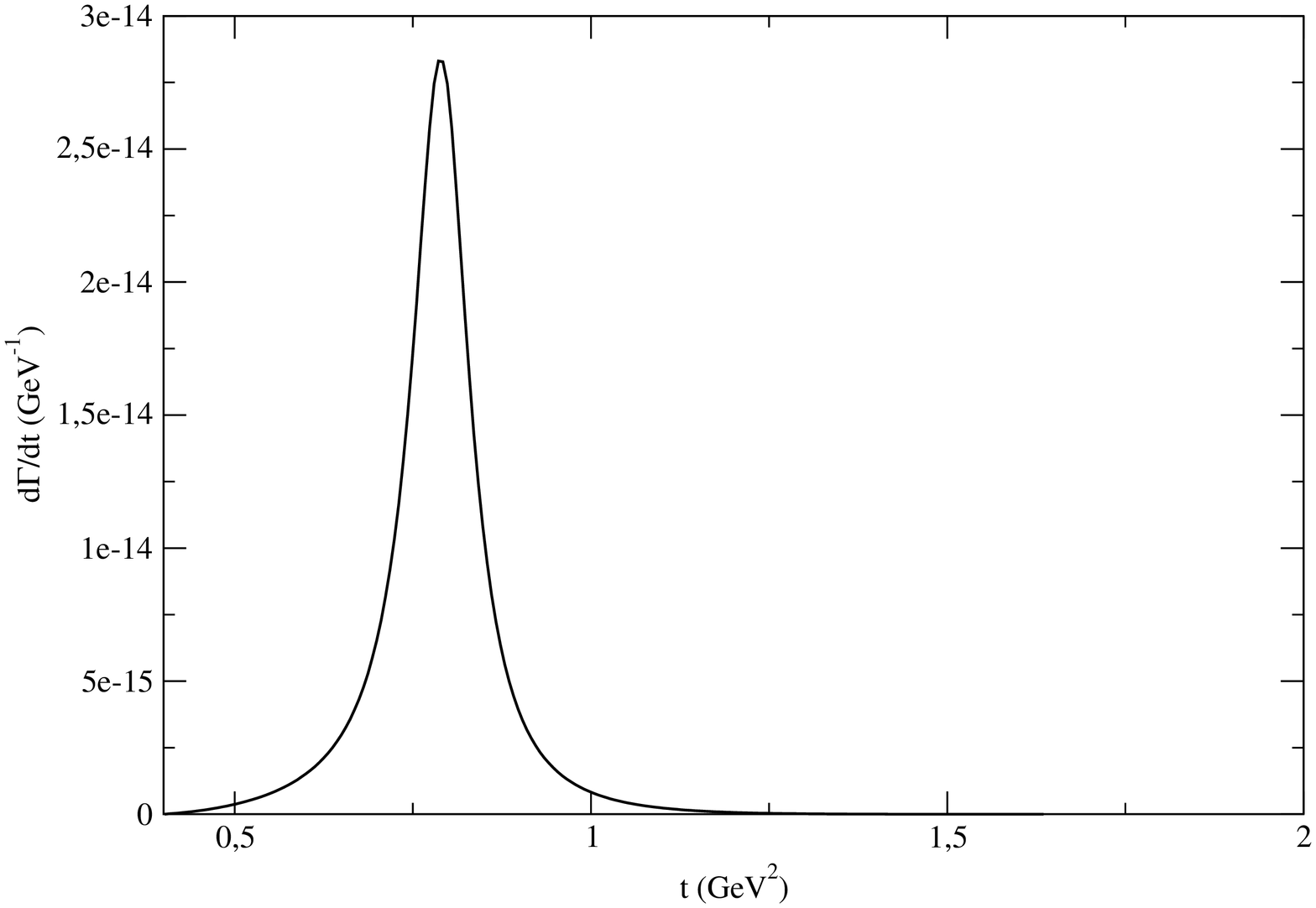}
\caption[]{\label{fig:dGdttkkp} \small{Our prediction for the $K^+\pi^-$ ($K^0\pi^-$) invariant mass distribution in the $\tau^- \rightarrow K^+ K^- \pi^- \nu_{\tau}$ ($\tau^- \rightarrow K^0\bar{K}^0 \pi^- \nu_{\tau}$) 
decays is shown.}}
\end{center}
\end{figure}

\begin{figure}[!h]
\begin{center}
\vspace*{1.0cm}
\includegraphics[scale=0.45, angle=-90]{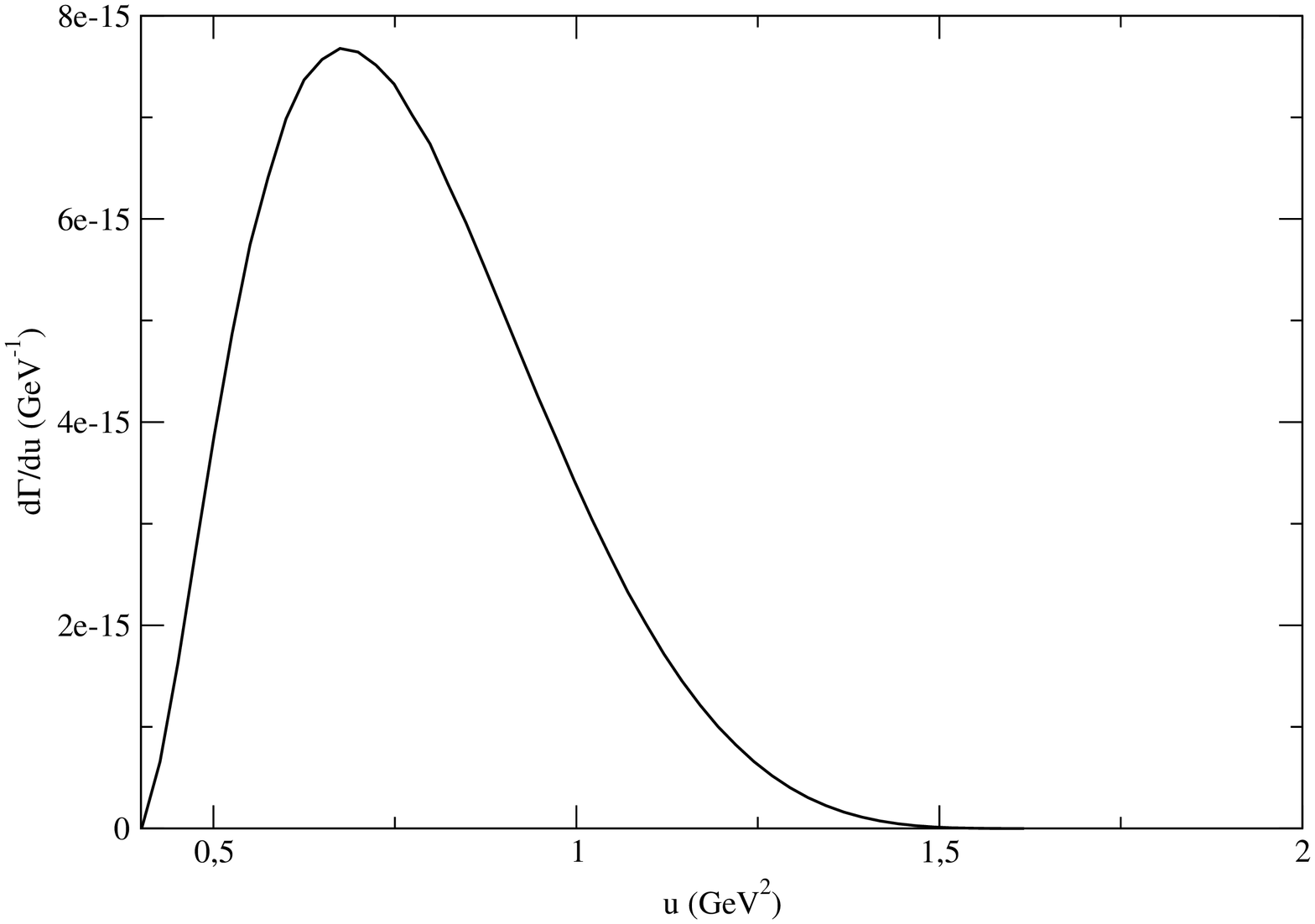}
\caption[]{\label{fig:dGdutkkp} \small{Our prediction for the $K^-\pi^-$ ($\bar{K}^0\pi^-$) invariant mass distribution in the $\tau^- \rightarrow K^+ K^- \pi^- \nu_{\tau}$ ($\tau^- \rightarrow K^0\bar{K}^0 \pi^- \nu_{\tau}$) 
decays is shown.}}
\end{center}
\end{figure}

\begin{figure}[!h]
\begin{center}
\vspace*{1.0cm}
\includegraphics[scale=0.45, angle=-90]{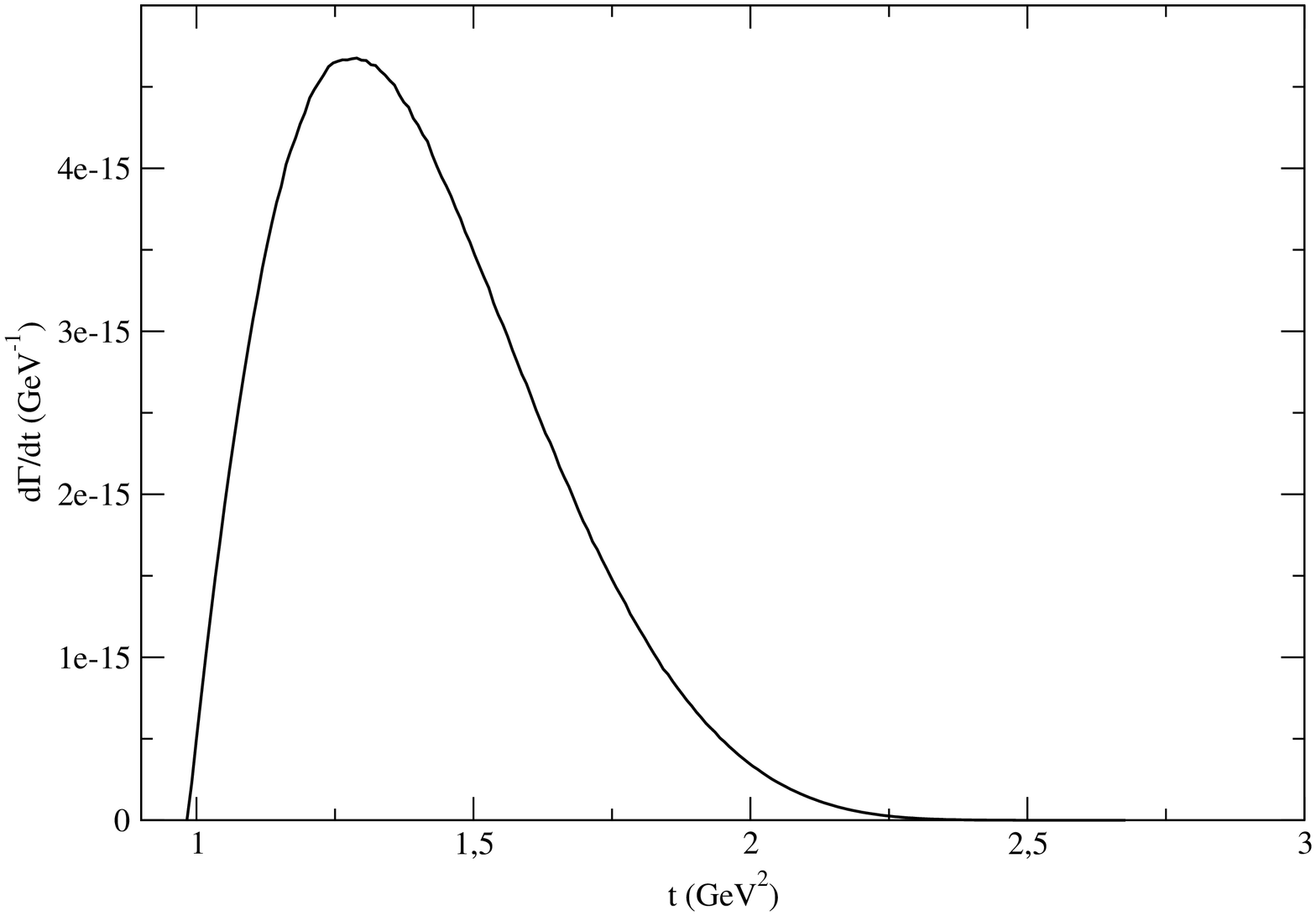}
\caption[]{\label{fig:dGdstk-k0p0} \small{Our prediction for the $K^-\pi^0$ invariant mass distribution in the $\tau^- \rightarrow K^- K^0 \pi^0 \nu_{\tau}$ decays is shown.}}
\end{center}
\end{figure}

\begin{figure}[!h]
\begin{center}
\vspace*{1.0cm}
\includegraphics[scale=0.45, angle=-90]{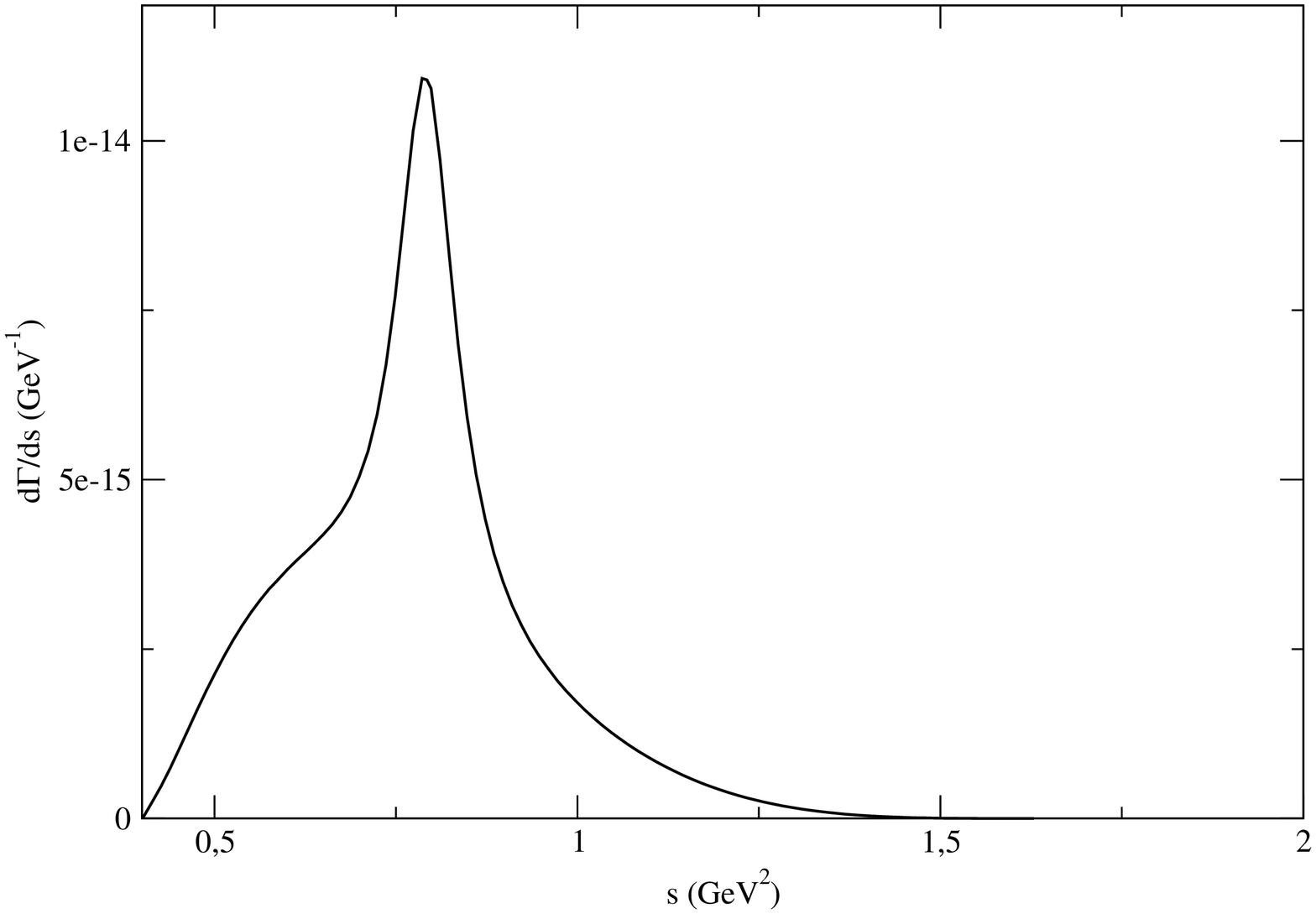}
\caption[]{\label{fig:dGdttk-k0p0} \small{Our prediction for the $K^-K^0$ invariant mass distribution in the $\tau^- \rightarrow K^- K^0 \pi^0 \nu_{\tau}$ decays is shown.}}
\end{center}
\end{figure}

\begin{figure}[!h]
\begin{center}
\vspace*{1.0cm}
\includegraphics[scale=0.45, angle=-90]{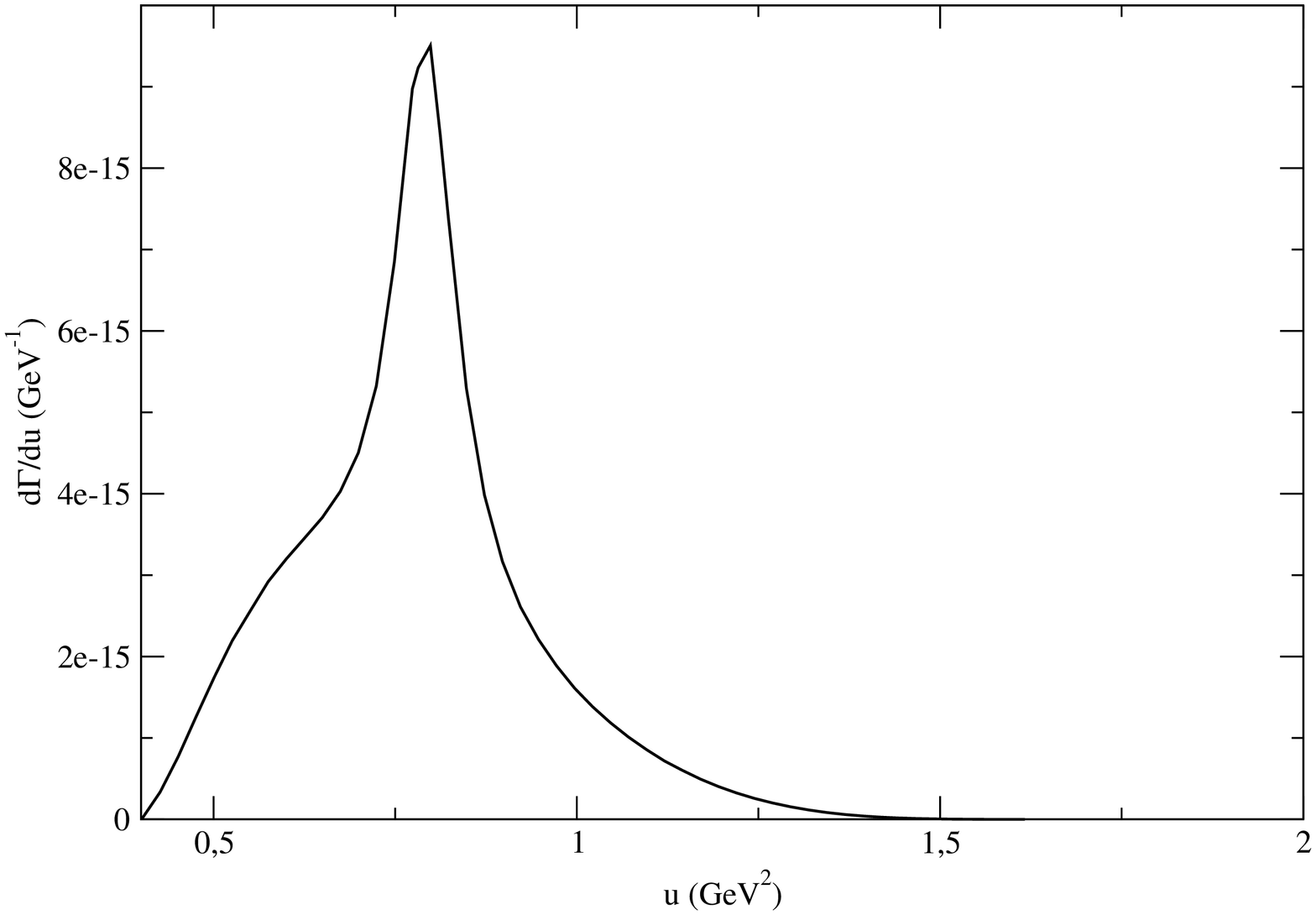}
\caption[]{\label{fig:dGdutk-k0p0} \small{Our prediction for the $K^0\pi^0$ invariant mass distribution in the $\tau^- \rightarrow K^- K^0 \pi^0 \nu_{\tau}$ decays is shown.}}
\end{center}
\end{figure}

\hspace*{0.5cm}Asymmetric $B$-factories span an ambitious $\tau$ programme that includes the determination of the hadronic structure of 
semileptonic $\tau$ decays such as the $K K \pi$ channel. As commented in the Introduction the latest study of $\tau^- \rightarrow K^+ K^- 
\pi^- \nu_{\tau}$ by the $CLEO-III$ Collaboration \cite{Liu:2002mn} showed a disagreement between the $KS$ model, included in $TAUOLA$, and 
the data. Experiments with higher statistics such as $BABAR$ and $Belle$ should clarify the theoretical setting.\\
\hspace*{0.5cm}For the numerics in this section we use the following values
\begin{eqnarray} 
F \, = \, 0\mathrm{.}0924 \, \mathrm{GeV} \; & \; , \;  & \;  F_V \, = \, 0\mathrm{.}180 \, \mathrm{GeV} \; \; \; \; \; \; \;  , \; \; \; \; 
F_A \, = \, 0\mathrm{.}149 \, \mathrm{GeV} \;,  \nonumber \\
M_V \, = \, 0\mathrm{.}775 \, \mathrm{GeV}  \; & \; , \;  & \; 
M_{K^*} \, = \, 0\mathrm{.}8953 \, \mathrm{GeV}\;  \; \; , \; \; \; \; 
M_{\mathrm{a}_1} \, = \, 1\mathrm{.}120 \, \mathrm{GeV}  \, .
\end{eqnarray}
Then we get $\lambda'$, $\lambda''$ and $\lambda_0$ from the first equalities in Eq.~(\ref{eq:lambres}).\\
\hspace*{0.5cm}At present no spectra for these channels is available and the determinations of the widths are collected in Table~\ref{tab:KKpi_thvsexp}
~\footnote{The Belle Collaboration has compared recently \cite{TalkTAU10} their spectra \cite{:2010tc} with our parametrization \cite{Dumm:2009kj}. Good agreement 
is seen at low-energies and a manifest deviation at $s\geq 2$ GeV$^2$ is observed. Their comparison shows our prediction for $c_4=-0.04$ and $g_4=-0.5$
, that corresponds to the values presented in Ref.~\cite{Roig:2008xt}. This points to a lower value of $c_4$, as obtained in the fit to $e^+e^-$ data 
(See Fig. \ref{fig:e+e-KKpi}) and also to a possible destructive interference of the higher-resonance states $\rho$' and $K^*$'. This issue will be investigated in detail.}.\\
\begin{table}
\begin{center}
\begin{tabular}{|c|c|c|c|}
\hline
&&& \\[-2.5mm]
Source & $\Gamma ( \tau^-\to K^+ K^- \pi^- \nu_\tau) \,$
& $ \Gamma ( \tau^-\to K^0 \overline{K}^0 \pi^- \nu_\tau) \,$
 & $\Gamma ( \tau^-\to K^- K^0 \pi^0 \nu_\tau) \,$  \\ [1.5mm]
\hline
&&&\\ [-2.5mm]
 $PDG$  \cite{Amsler:2008zzb} & $3$.$103\,(136)$& $3$.$465 \, (770) $& $3$.$262 \, (521)$ \\ [1.9mm]
$BaBar$ \cite{Aubert:2007mh} &$3$.$049 \, (85)$ & &  \\
[1.9mm]
$CLEO-III$ \cite{Liu:2002mn}&$3$.$511 \, (245)$ & &  \\ [1.9mm]
$Belle$ \cite{:2010tc} & $3$.$465 \, (136)$ && \\ [1.9mm]
Our prediction & $3$.$4^{+0\mathrm{.}5}_{-0\mathrm{.}2}$ & $3$.$4^{+0\mathrm{.}5}_{-0\mathrm{.}2}$ & $2$.$5^{+0\mathrm{.}3}_{-0\mathrm{.}2}$ \\ [2mm]
\hline
\end{tabular}
\caption{\small{Comparison of the measurements of partial widths (in units of $10^{-15} \, \mathrm{GeV}$) with our predictions for the set 
of values in Eq.~(\ref{eq:c4g4I}). For earlier references see \cite{Amsler:2008zzb}.}}
\label{tab:KKpi_thvsexp}
\end{center}
\end{table}
\hspace*{0.5cm}We also notice that there is a discrepancy between the $BaBar$ measurement of $\Gamma(\tau^- \rightarrow K^+ K^- \pi^- 
\nu_{\tau})$ and the results by $CLEO$ and $Belle$. Within $SU(2)$ isospin symmetry it is found that $\Gamma(\tau^- \rightarrow K^+ K^- \pi^-
 \nu_{\tau}) = \Gamma(\tau^- \rightarrow K^0 \overline{K}^0 \pi^- \nu_{\tau})$, which is well reflected by the values in Table~\ref{tab:KKpi_thvsexp} 
within errors. Moreover, as commented above, the $PDG$ data \cite{Amsler:2008zzb} indicate that $\Gamma(\tau^-\rightarrow K^- K^0 \pi^0 
\nu_{\tau})$ should be similar to $\Gamma(\tau^- \rightarrow K \overline{K} \pi \nu_{\tau})$. It would be important to obtain a more 
accurate determination of the $\tau^-\rightarrow K^- K^0 \pi^0 \nu_{\tau}$ width (the measurements quoted by the $PDG$ are rather old) in the
 near future.\\
\hspace*{0.5cm}In our analyses we include the lightest resonances in both the vector and axial-vector channels, namely $\rho(775)$, $K^*(892)$
 and a$_1(1260)$. It is clear that, as it happens in the $\tau \rightarrow \pi  \pi \pi \nu_{\tau}$ channel (see Chapter \ref{3pi}), a much 
lesser role, though noticeable, can be played by higher excitations on the vector channel. As experimentally only the branching ratios are 
available for the $K K \pi$ channel we think that the refinement of including higher mass resonances should be taken into account in a later 
stage, when the experimental situation improves.\\
\hspace*{0.5cm}In Figs.~\ref{fig:K+K-pi-spectralfunction} and \ref{fig:K-K0pi0spectralfunction} we show our predictions for the normalized $M_{K K \pi}^2-$spectrum of the $\tau^- 
\rightarrow K^+ K^- \pi^- \nu_{\tau}$ and $\tau^- \rightarrow K^-K^0\pi^0\nu_{\tau}$ decays, respectively. As discussed above we have taken 
$c_4 = -0$.$07\pm 0$.$01$ and $g_4 = -0$.$72 \pm 0$.$20$ (notice that the second process does not depend on $g_4$). We conclude that the vector 
current contribution ($\Gamma_V$) dominates over the axial-vector one ($\Gamma_A$) in both channels~:
\begin{eqnarray}
& &  \frac{\Gamma_A}{\Gamma_V} \, \Big|_{K \overline{K} \pi}  = \, 0\mathrm{.}16 \pm 0\mathrm{.}05\; , \quad \quad \quad \quad \quad
 \frac{\Gamma_A}{\Gamma_V} \, \Big|_{K^- K^0 \pi}  = \, 0\mathrm{.}18 \pm 0\mathrm{.}04\;  ,  \nonumber\\
& & \quad \quad \quad \quad \frac{\Gamma(\tau^- \rightarrow K^+ K^- \pi^- \nu_{\tau})}{\Gamma(\tau^- \rightarrow K^- K^0 \pi^0 \nu_{\tau})} \, = \, 1\mathrm{.}4  \pm 0\mathrm{.}3  \; ,
\end{eqnarray}
where the errors estimate the slight variation due to the range in $c_4$ and $g_4$. These ratios translate into a ratio of the vector current
 to all contributions of $f_v = 0$.$86 \pm 0$.$04$ for the $K \overline{K} \pi^-$ channel and $f_v = 0$.$85 \pm 0$.$03$ for $K^- K^0 \pi^0$ one, to be
 compared with the result in Ref.~\cite{Davier:2008sk}, namely $f_v(K\overline{K}\pi) = 0$.$20 \pm 0$.$03$. Our results for the relative 
contributions of vector and axial-vector currents deviate strongly from most of the previous estimates, as one can see in Table~\ref{tab:VvsA_thvsexp}. 
Only Ref.~\cite{GomezCadenas:1990uj} pointed already to vector current dominance in these channels, although enforcing just the leading 
chiral constraints and using experimental data at higher energies.\\
\hspace*{0.5cm}We conclude that for all $\tau \rightarrow K K \pi \nu_{\tau}$ channels the vector component dominates by far over the axial-vector 
one, though, as can be seen in the spectra in Figs.~\ref{fig:K+K-pi-spectralfunction}, \ref{fig:K-K0pi0spectralfunction}, the axial-vector current 
is dominating in the very-low $Q^2$ regime.\\
\begin{table}[!h]
\begin{center}
\begin{tabular}{|c|c|}
\hline
& \\[-2.5mm]
Source  & $\Gamma_V / \Gamma_A$  \\ [1.5mm]
\hline
& \\ [-2.5mm]
Our result &  $6  \pm 2$ \\ [1.9mm]
$KS$ model \cite{Finkemeier:1995sr} & $0$.$6 - 0$.$7$ \\ [1.9mm]
$KS$ model \cite{Finkemeier:1996hh} & $0$.$4 - 0$.$6$ \\ [1.9mm]
Breit-Wigner approach \cite{GomezCadenas:1990uj} & $\sim 9$ \\ [1.9mm]
$CVC$ \cite{Davier:2008sk} & $0$.$20 \pm 0$.$03$ \\ [1.9mm]
Data analysis \cite{Liu:2002mn} & $1$.$26 \pm 0$.$35$ \\[1.9mm]
\hline
\end{tabular}
\caption{\small{Comparison of the ratio of vector and axial-vector contribution for $\tau \rightarrow KK\pi\nu_{\tau}$ partial widths.
The last two lines correspond to the $\tau^- \rightarrow K^+ K^- \pi^- \nu_{\tau}$ process only. Results in Ref.~\cite{Finkemeier:1996hh} 
are an update of Ref.~\cite{Finkemeier:1995sr}. The result of Ref.~\cite{Davier:2008sk} is obtained by connecting the tau decay width with the $CVC$ 
related $e^+ e^- \rightarrow K_S K^{\pm} \pi^{\mp}$ (see Appendix \ref{isosApp}). The analysis in \cite{Liu:2002mn} was performed with a 
parameterization that spoiled the chiral normalization of the form factors. }} \label{tab:VvsA_thvsexp}
\end{center}
\end{table}
\hspace*{0.5cm}Next we contrast our spectrum for $\tau^- \rightarrow K^+ K^- \pi^- \nu_{\tau}$ with that one arising from the $KS$ model 
worked out in Refs.~\cite{Finkemeier:1995sr,Finkemeier:1996hh}. This comparison is by no means straight because in these references a second and even a third multiplet of
resonances are included in the analysis. As we consider that the spectrum is dominated by the first multiplet, in principle we could start by
 switching off heavier resonances. However we notice that, in the $KS$ model, the $\rho(1450)$ resonance plays a crucial role in the vector 
contribution to the spectrum. This feature depends strongly on the value of the $\rho(1450)$ width, which has been changed from Ref.~
\cite{Finkemeier:1995sr} to Ref.~\cite{Finkemeier:1996hh}~\footnote{Moreover within Ref.~\cite{Finkemeier:1995sr} the authors use two 
different set of values for the $\rho(1450)$ mass and width, one of them in the axial-vector current and the other in the vector one.
This appears to be somewhat misleading.}. In Figure~\ref{fig:comparisonwithKS1} we compare our results for the vector and axial-vector contributions with those
of the $KS$ model as specified in Ref.~\cite{Finkemeier:1996hh} (here we have switched off the seemingly unimportant $K^*(1410)$). As it can be seen there
are large differences in the structure of both approaches. Noticeably there is a large shift in the peak of the vector spectrum owing to the
inclusion of the $\rho(1450)$ and $\rho(1700)$ states in the $KS$ model together with its strong interference with the $\rho(770)$ resonance. 
In our scheme, including the lightest resonances only, the $\rho(1450)$ and $\rho(1700)$ information has to be encoded in the values of $c_4$
 and $g_4$ couplings (that we have extracted in Subsection \ref{determ_c4g4}) and such an interference is not feasible. It will be a 
task for the experimental data to settle this issue.\\
In Figure~\ref{fig:comparisonwithKS2} we compare the normalized full $M_{K K \pi}^2$ spectrum for the $\tau \rightarrow K \overline{K} \pi 
\nu_{\tau}$ channels in the $KS$ model \cite{Finkemeier:1996hh} and in our scheme. The most salient feature is the large effect of the vector 
contribution in our case compared with the leading role of the axial-vector part in the $KS$ model, as can be seen in Figure~\ref{fig:comparisonwithKS1}.
 This is the main reason for the differences between the shapes of $M_{K K \pi}^2$ spectra observed in Figure~\ref{fig:comparisonwithKS2}. We see 
in Figures \ref{fig:comparisonwithKSextra} and \ref{fig:comparisonwithKS3} that similar patterns are observed in the $K^-K^0\pi^0$ hadronic 
mode.\\
\hspace*{0.5cm}As we have taken advantage of in Chapter \ref{3pi}, the plot of the differential distribution of the decay rate versus the two-particle invariant masses 
is a very useful tool to learn about the dynamics of these processes. In Figure~\ref{fig:K+pi-_KKpi} (from Ref.~\cite{Shekhovtsova:2012ra}) we represent the $K^+\pi^-$ invariant mass distribution for the $\tau^- \rightarrow K^+ K^- 
\pi^- \nu_{\tau}$ decays, both for our prediction -there is no experimental data we can compare to- and the Finkemeier and Mirkes model. Figure~\ref{fig:K+pi-_KKpi} makes clear 
how different the dynamics contained in the $KS$ model and in our parameterization are.\\
\hspace*{0.5cm}Similarly, we present in Figs.~\ref{fig:KS0pi-_KKpi} and \ref{fig:K-pi^0_KKpi} (from Ref.~\cite{Shekhovtsova:2012ra}) the analogous plots for the $K_S^0\pi^-$ and $K^-\pi^0$ invariant mass distributions in the 
$\tau^- \rightarrow K^0 \bar{K}^0 \pi^- \nu_{\tau}$ and $\tau^- \rightarrow K^- K^0 \pi^0 \nu_{\tau}$ decays, respectively. Again, we observe that the physics contained in both approaches 
is pretty different. These are very interesting observables in which we expect data from the dedicated studies of $B$- and tau-charm-factories in the future \footnote{After this Thesis was defended, Belle data was compared to our 
prediction \cite{TalkTAU10}. Although nice agreement was seen at low values of $Q^2$, there was disagreement above the peak of the curve, where we predicted a higher decay rate. This difference is not surprising, since excited 
resonances are not included in our parametrization. Their incorporation is in progress.}.\\
Our predictions are depicted more clearly in Figs. \ref{fig:dGdstkkp}, \ref{fig:dGdttkkp}, \ref{fig:dGdutkkp} for the $\tau^- \rightarrow K^+ K^- \pi^- \nu_{\tau}$ decays and in 
Figs. \ref{fig:dGdstk-k0p0}, \ref{fig:dGdttk-k0p0}, \ref{fig:dGdutk-k0p0} for the $\tau^- \rightarrow K^- K^0 \pi^0 \nu_{\tau}$ decays. In Fig. \ref{fig:dGdstkkp} no significant 
$\phi(1020)$ contribution is seen, in agreement to the fact that we obtain a vanishing $\phi(1020)$ component for the ideal mixing case. On the contrary, in Fig. \ref{fig:dGdttkkp} 
the $K^{*0}(892)$ shows neatly in the $K^+\pi^-$ invariant mass spectrum. As expected, no dynamical structure is appreciated in the $K^-\pi^-$ invariant distribution because of 
the electrical charge involved. While the $K^{*}(892)$ resonance is clearly seen in Figs. \ref{fig:dGdstk-k0p0} and \ref{fig:dGdutk-k0p0} in the $K^-\pi^0$ and $K^0\pi^0$ 
invariant mass cuts, no similar structure can be appreciated in Fig. \ref{fig:dGdttk-k0p0} for the $K^-K^0$, which could be expected because no resonance couples strongly to 
a state with this quantum numbers.
\section{Conclusions}\label{KKpi_Conclusions}
\hspace*{0.5cm}Hadronic decays of the tau lepton are an all-important tool in the study of the hadronization process of $QCD$ currents, in a 
setting where resonances play the leading role. In particular the final states of three mesons are the simplest ones where one can test the 
interplay between different resonance states. At present there are three parameterizations implemented in the $TAUOLA$ library to describe 
the hadronization process in tau decays. Two are based on experimental data. The other alternative, namely the $KS$ model, though successfull
 in the account of the $\pi \pi \pi$ final state, has proven to be unsuitable \cite{Liu:2002mn} when applied to the decays into $K K \pi$ 
hadronic states. Our procedure, guided by large $N_C$, chiral symmetry and the asymptotic behaviour of the form factors driven by $QCD$, was 
already employed in the analysis of $\tau \rightarrow \pi \pi \pi \nu_{\tau}$ in Refs.~\cite{GomezDumm:2003ku} and \cite{Dumm:2009va}, which
only concern the axial-vector current. Here we have applied our methodology to the analysis of the $\tau \rightarrow K K \pi \nu_{\tau}$ 
channels where the vector current may also play a significant role.\\
\hspace*{0.5cm}We have constructed the relevant Lagrangian involving the lightest multiplets of vector and axial-vector resonances. Then we 
have proceeded to the evaluation of the vector and axial-vector currents in the large-$N_C$ limit of $QCD$, i.e.\ at tree level within our 
model. Though the widths of resonances are a next-to-leading effect in the $1/N_C$ counting, they have to be included into the scheme since 
the resonances do indeed resonate due to the high mass of the decaying tau lepton. We have been able to estimate the values of the relevant 
new parameters appearing in the Lagrangian with the exception of two, namely the couplings $c_4$ and $g_4$, which happen to be important in 
the description of $\tau \rightarrow K K \pi \nu_{\tau}$ decays. The range of values for these couplings has been determined from the 
measured widths $\Gamma(\tau^- \rightarrow K^+ K^- \pi^- \nu_{\tau})$ and $\Gamma(\tau^- \rightarrow K^- K^0 \pi^0 \nu_{\tau})$.\\
\hspace*{0.5cm}In this way we provide a prediction for the ---still unmeasured--- spectra of both processes. We conclude that the vector 
current contribution dominates over the axial-vector current, in fair disagreement with the corresponding conclusions from the $KS$ model 
\cite{Finkemeier:1996hh} with which we have also compared our full spectra. On the other hand, our result is also at variance with the 
analysis in Ref.~\cite{Davier:2008sk}. There are two all-important differences that come out from the comparison. First, while in the $KS$ 
model the axial-vector contribution dominates the partial width and spectra, in our results the vector current is the one that rules both 
spectrum and width. Second, the KS model points out a strong interference between the $\rho(770)$, the $\rho(1450)$ and the $\rho(1700)$
resonances that modifies strongly the peak and shape of the $M_{KK \pi}$ distribution depending crucially on the included spectra. Not 
having a second multiplet of vector resonances in our approach, we cannot provide this feature. It seems strange to us the overwhelming role 
of the $\rho(1450)$ and $\rho(1700)$ states but it is up to the experimental measurements to settle this issue.\\
\hspace*{0.5cm}Even if our model provides a good deal of tools for the phenomenological analyses of observables in tau lepton decays, it may
seem that our approach is not able to carry the large amount of input present in the $KS$ model, as the later includes easily many multiplets 
of resonances. In fact, this is not the case, since the Lagrangian can be systematically extended to include whatever spectra of particles 
are needed. If such an extension is carried out the determination of couplings could be cumbersome or just not feasible, but, on the same 
footing as the $KS$ model, our approach would provide a parameterization to be fitted by the experimental data. The present stage, however, 
has its advantages. By including only one multiplet of resonances we have a setting where the procedure of hadronization is controlled from 
the theory. This is very satisfactory if our intention is to use these processes to learn about $QCD$ and not only to fit the data to 
parameters whose relation with the underlying theory is unclear when not directly missing.\\

\chapter{$\tau^-\to \eta/\eta' \pi^- \pi^0 \nu_\tau$ and $\tau^-\to \eta/\eta' \eta \pi^- \nu_\tau$ decays}\label{eta}
\section{Introduction}\label{eta_Intro}
\hspace*{0.5cm}In this chapter we present the study of the three-meson $\tau$ decay modes containing $\eta$ mesons and pions. These are the decays $\tau^-\to\eta
\pi^-\pi^0\nu_\tau$ and $\tau^-\to\eta\eta\pi^-\nu_\tau$. They are really interesting: in the first one, if the exchange of spin-zero resonances is neglected, 
only the vector current participates allowing for a very precise study of the odd-intrinsic parity couplings \footnote{Instead, when scalar and pseudoscalar 
resonances are considered, their contributions are either forbidden by G-parity in the SU(2) symmetry limit or heavily suppressed by the dynamics. Isospin 
symmetry will be assumed in what follows.} and the second one is a rarer decay which, if discovered, would allow to study the contribution of scalar resonances, 
since the spin-one exchanges are suppressed by symmetries.\\
\hspace*{0.5cm}Although the computation of these modes is much simpler than that of the $2K\pi$ decay modes we can extract very precise information 
from them. As we advanced, the $\tau^-\to\eta \pi^-\pi^0\nu_\tau$ can only be produced, to a very good approximation, via vector current. This mode is measured with an error 
$\sim 13\%$ \cite{Amsler:2008zzb} \footnote{It is reduced to $\sim 7\%$ in the update on PDGlive, http://pdg.lbl.gov/2009/tables/rpp2009-sum-leptons.pdf.}, 
therefore it should be an ideal benchmark to learn about the hadronization of the vector current in presence of $QCD$ 
interactions \cite{Pich:1987qq} and, in particular, to test the determination of the couplings in the vector current resonance Lagrangian done in 
Chapter \ref{KKpi} and to confront it to the results in Chapter \ref{Pgamma}. However, the branching ratio for this mode in the PDG live disagrees with the 
value in the PDG 2008 within errors (the earlier value was $(1$.$81\pm0$.$24)\cdot10^{-3}$ while the new one is $(1$.$39\pm0$.$10)\cdot10^{-3}$), so one should be cautious 
about the strength of the conclusions we reach.  On the other side, the decay $\tau^-\to\eta\eta\pi^-\nu_\tau$ is privileged. There are no vector current 
contributions and the axial-vector current carries only pseudoscalar degrees of freedom in this case, being 
suppressed as $F_4\sim m_\pi^2/Q^2$, that is, as $\sim m_\pi^4/Q^4$ in the spectral function and branching ratios. This observation makes us to guess 
\footnote{We take into account the relative contribution of the pseudoscalar form factor in the $3\pi$ and $KK\pi$ $\t$ decay modes, where the relative
 suppression is identical.} a suppression at the level of four or five orders of magnitude with respect to the same observables in $\tau^-\to\eta \pi^-
\pi^0\nu_\tau$. This estimate would yield a branching ratio of $10^{-7}$ or smaller, four orders of magnitude less than the current lowest branching 
fraction obtained (See Table \ref{Listdecaystau}). Given this particular suppression, these decays may constitute a suitable place to characterize the effect of spin-zero resonance 
exchange. Once this is done the $SM$ contribution will be properly evaluated and, in future applications, one can even think of this decay mode as a very appealing channel to look for new physics.\\
\section{Form factors in $\tau^{-} \rightarrow \eta \pi^{-} \pi^{0} \nu_{\tau}$}\label{FFetaetapi}
\hspace*{0.5cm}We consider \cite{Roig:2010jp, Dumm:2012} the process $\tau^{-} \rightarrow \eta(p_1)\, \pi^{-}(p_2) \, \pi^{0}(p_{3})\, \nu_{\tau}$. The labeling of momenta 
corresponds to Eq. (\ref{Set_of_independent_Vectors_3meson}). The computation is made for $\eta = \eta_8$. The rest of definitions and normalizations 
are as usual.\\
\hspace*{0.5cm}Because of $G$-parity the axial-vector current form factors vanish 
\begin{equation}
T_{A_{1,2}\mu}^{\chi} \, = T_{A_{1,2}\mu}^{1R}\,=\,T_{A_{1,2}\mu}^{2R}=\, 0 \; .
\end{equation}
\hspace*{0.5cm}In order to see this \cite{Pich:1987qq}, one needs to consider the respective $G$-parities \footnote{G-parity is only exact in the limit
 of conserved isospin.} of pion and eta: $G_\eta\,=\,+1$, $G_\pi\,=\,-1$, and of the (axial-)vector currents $G_{A_\mu}\,=\,-1$ and $G_{V_\mu}\,=\,+1$. 
The vector form factors are not forbidden by any (approximate) symmetry and thus, for instance, one has the contribution of the $WZW$ term, Eq.~(\ref{Z_WZW}) and the resonance exchange contributions in the odd-intrinsic parity 
sector in Eqs.~(\ref{VJPops}), (\ref{VVPops}) and (\ref{LVPPP}) without any suppression factor vanishing in the isospin limit. Since any isospin-correction to 
the $G$-parity forbidden terms in the axial-vector current would contribute much less than all others in the vector current we neglect the former (as isospin breaking is neglected in any application considered in this Thesis).\\
\hspace*{0.5cm}For the vector form factor one needs to consider the diagrams analogous to Figures \ref{fig_diagsKKpi}.a), \ref{fig_diagsKKpi}.c), 
\ref{fig_diagsKKpi}.d) and \ref{fig_diagsKKpi}.f), where the solid single lines now correspond to $\pi$ and $\eta$ mesons. The vector form 
factors read
\begin{eqnarray} \label{T_etapipi_chi}
T_{V\mu}^{\chi} & = &  i \varepsilon_{\mu\nu\varrho\sigma} p_1^\nu p_2^\varrho p_3^\sigma \, 
\left[ \, \frac{N_C }{6\,\sqrt{6} \, \pi^2 \, F^{3}} \, \right] \; , 
\end{eqnarray}
\vspace*{0.2cm}
\begin{eqnarray} \label{T_etapipi_1R1}
T_{V\mu}^{1R(1)} & = &  i \varepsilon_{\mu\nu\varrho\sigma} p_1^\nu p_2^\varrho p_3^\sigma \, 
 \frac{8 \, G_V}{\sqrt{3}\,F^3 \, M_V} \,\frac{1}{M_\rho^2-u} \left[ (c_1-c_2+c_5)Q^2 \right.\times  \\
 & & \! \!\! \!\! \!\! \! \left.  - (c_1-c_2-c_5+2c_6) u +
(c_1+c_2+8c_3-c_5)m_\eta^2+8c_3\left(m_\pi^2-m_\eta^2\right) \right]  \, , \nonumber
\end{eqnarray}
\vspace*{0.2cm}
\begin{eqnarray} \label{T_etapipi_1R2}
T_{V\mu}^{1R(2)} & = & - i \varepsilon_{\mu\nu\varrho\sigma} p_1^\nu p_2^\varrho p_3^\sigma
\, \frac{16 F_V}{\sqrt{3}\,M_V \,F^3} \frac{1}{M_V^2 - Q^2} \,
\left[ \, \left(g_1\,+\, 2g_2\,-\,g_3\right)\,u\,\right. \\
& & \left. -\,g_2\, \left( Q^2\,+\,2 m_\pi^2\,-\,m_\eta^2\right)-\, \left(g_1\,-\,g_3\right)\,2\,m_\pi^2\,
+\,(2\,g_4\,+\,g_5) \,m_\pi^2 \right] \, . \nonumber
\end{eqnarray}
\vspace*{0.2cm}
\begin{eqnarray} \label{T_etapipi_2R}
T_{V\mu}^{2R} & = & - i \varepsilon_{\mu\nu\varrho\sigma} p_1^\nu p_2^\varrho p_3^\sigma \, 
\left\lbrace  \, \frac{8 \sqrt{2}}{\sqrt{3}} \, \frac{F_V G_V}{F^3} \, \frac{1}{M_V^2-Q^2} \, \times \right. \nonumber\\
& & \left.  \frac{1}{M_\rho^2-u} \, \left( d_3(Q^2+u)+(d_1+8d_2-d_3)m_\eta^2+8d_2(m_\pi^2
-m_\eta^2) \right) \right\rbrace \, .  
\end{eqnarray}
\hspace*{0.5cm}To obtain the $\eta_1$ contribution one simply has to multiply the above amplitudes by $\sqrt{2}$. Then, the matrix element for the decay 
into the physical hadronic states $\eta\pi^-\pi^0$ and $\eta^\prime\pi^-\pi^0$ can be obtained from the previous ones by applying the suitable mixing formalism. Here, we consider 
for simplicity the single-angle mixing scheme, so that $|\eta\ket=\mathrm{cos}\theta_P |\eta_8\ket-\mathrm{sin}\theta_P |\eta_1\ket$, 
$|\eta'\ket=\mathrm{sin}\theta_P |\eta_8\ket+\mathrm{cos}\theta_P |\eta_1\ket$~\footnote{A more realistic study would require the double-angle mixing framework \cite{DoubleAngleMixing} 
between the mass eigenstates $|\eta\ket$ and $|\eta'\ket$ and the flavour eigenstates $|\eta_1\ket$ and $|\eta_ 8\ket$, in a way consistent with the large-$N_C$ limit \cite{Kaiser:1998ds}.
 Ref.~\cite{Escribano:2010wt} includes a description consistent with $R\chi T$ \cite{Jamin:2001zq} using the latest values from KLOE \cite{Ambrosino:2009sc}. This approach is followed in our paper \cite{Dumm:2012}. 
For a given set of values of the Lagrangian couplings, the amplitude for the decay $\tau^-\to\eta(\eta^\prime)\pi^-\pi^0\nu_\tau$ increases (decreases) by $\sim22\%$($\sim9\%$) when considering the double-angle 
mixing scheme. The data favours this latter approach. See Ref. \cite{Dumm:2012} for more details.}.
In this way, we have
\begin{eqnarray}
 T_\eta = \mathrm{cos}\theta_P T_{\eta_8}+\mathrm{sin}\theta_P T_{\eta_1}=\left(\mathrm{cos}\theta_P +\mathrm{sin}\theta_P \sqrt{2}\right) T \sim 0\mathrm{.}600 T\nonumber\\
 T_{\eta'} = -\mathrm{sin}\theta_P T_{\eta_8}+\mathrm{cos}\theta_P T_{\eta_1}=\left(-\mathrm{sin}\theta_P+\mathrm{cos}\theta_P \sqrt{2}\right) T \sim 1\mathrm{.}625 T\,,
\end{eqnarray}
where $T$ stands for the amplitudes in Eqs.~(\ref{T_etapipi_chi}), (\ref{T_etapipi_1R1}), (\ref{T_etapipi_1R2}) and (\ref{T_etapipi_2R}), calculated 
for $\eta\,=\,\eta_8$ for a value of $\theta_P\sim-15^\circ$.\\
\section{Short-distance constraints on the couplings} \label{Shortdistance_etapipi}
\hspace*{0.5cm}Following the same procedure as in Sections \ref{3pi_Shortdistance} and \ref{KKpi_QCDconstraints} we have found the following 
constraints on the vector form factor:
\begin{eqnarray}
c_{125}\equiv c_1-c_2+c_5 &  = & 0 \; , \nonumber \\
c_{1256} \equiv c_1-c_2-c_5+2c_6& = &
- \, \frac{N_C}{96 \pi^2} \, \frac{M_V\,F_V}{\sqrt{2} \, F^2} \; , \nonumber \\
d_3 & = & - \, \frac{N_C}{192 \pi^2} \frac{M_V^2}{F^2} \, , \nonumber  \\
g_{123} \equiv g_1\,+\, 2g_2\,-\,g_3 & = & 0 \; , \nonumber \\
g_2\, & = & \, \frac{N_C}{192 \pi^2}\, \frac{M_V}{\sqrt{2} \,F_V} \; ,
\end{eqnarray}
that are consistent with the values found in the $3 \pi$ and $2K \pi$ tau decay channels previously analyzed and also to those to be found in the 
$P^-\gamma$ decays in the next Chapter.\\
\section{$\tau^{-} \rightarrow \eta \eta \pi^{-}\nu_\tau$} \label{etaetapi}
\hspace*{0.5cm}This mode is peculiar because in the chiral limit, it is not generated by the axial-vector current. This \cite{Pich:1987qq} 
can be understood by noticing that the axial-vector current coupling to three $pGs$ is built up from the structure generating the two-meson 
vector coupling that can not give either $\eta\eta$ (because it vanishes due to the antisymmetric structure in momenta) or $\eta\pi$ that would 
have $G$-parity $-1$, while that of the vector current is $+1$. This feature is preserved when passing from $\CPT$ to $R\chi T$ because it only 
depends on the couplings of spin-one currents to $pGs$ and selection rules. Moreover, $G$-parity also forbids all contributions to $F_4$ including 
the exchange of a vector or axial-vector resonance.\\
\hspace*{0.5cm}That is why we only get a contribution in the pseudoscalar form factor $F_4$, that is nothing more than the $\CPT$ result at 
$\mathcal{O}(p^2)$. That is
\begin{equation}\label{FFetaeta'pi}
 T_{A_4\,\mu}^{\CPT}\,=\,- i \frac{m_\pi^2}{3\sqrt{2} F (Q^2 - m_\pi^2)}\,Q_\mu\,. 
\end{equation}
\hspace*{0.5cm}This channel offers us the possibility to evaluate our assumption of neglecting the effect of the exchange of spin-zero resonances. 
Since the $\CPT$ result at $\cO(p^2)$ will give an irrelevantly small branching ratio, we can use this process to study in deep the relevance of scalar 
and pseudoscalar resonances in an appropriate environment where its r\^ole cannot be masked by any effect induced by vector or axial--vector resonances. 
Specifically, the intermediate states $f_0\pi^-$ and $\eta a_0^-$ contributions are not suppressed. If this decay channel was measured it would be 
possible to study their r\^ole. This, however, is not straightforward, since the scalar resonances in the Resonance Chiral Theory Lagrangian do not correspond 
to these ligther states ($f_0$, $a_0$), which have a predominant $q\bar{q}q\bar{q}$ component \cite{Cirigliano:2003yq}.\\
\section{Phenomenological analysis}
\begin{figure}[h!]
\begin{center}
\vspace*{0.9cm}
\includegraphics[scale=0.4,angle=-90]{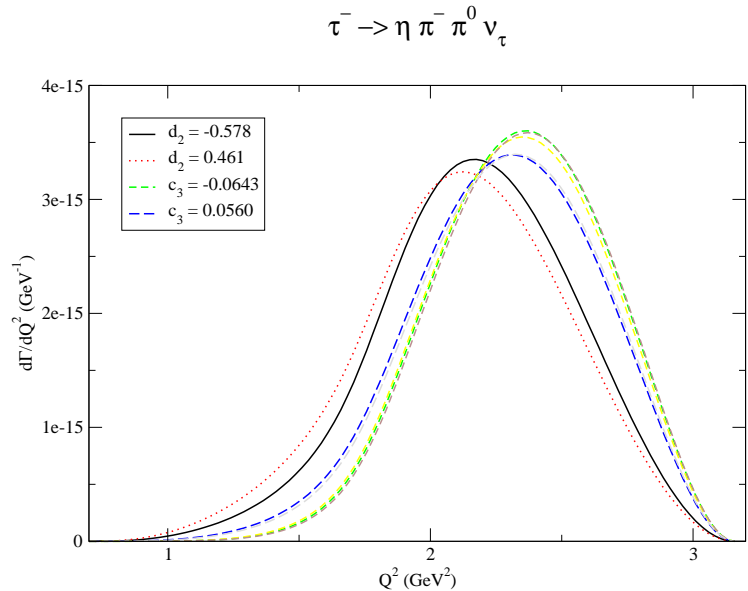}
\caption[]{\label{Fig:etapipi} \small{Spectral function for the decay $\tau^{-} \rightarrow \eta \pi^{-} \pi^{0} \nu_{\tau}$ using the values for the unknown 
couplings corresponding to the eight benchmark points as we define and explain in the text.}}
\end{center}
\end{figure}

\begin{figure}[h!]
\begin{center}
\vspace*{0.9cm}
\includegraphics[scale=0.4,angle=-90]{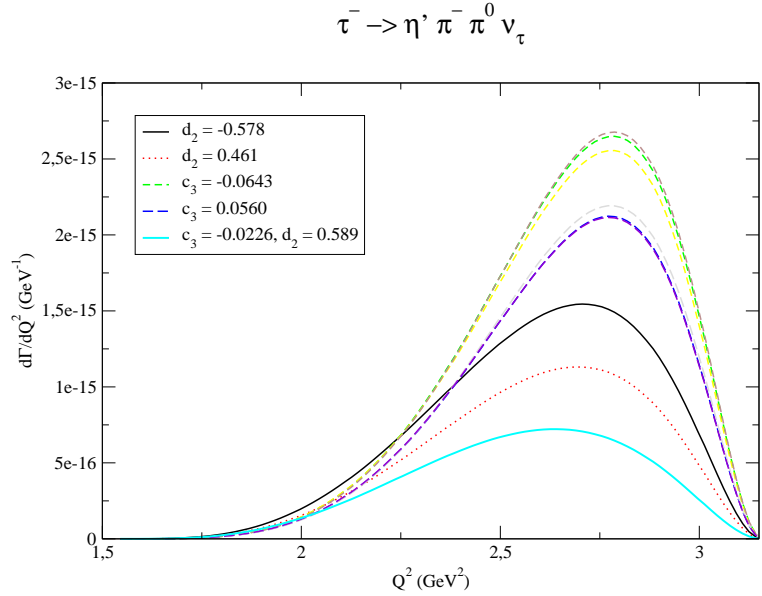}
\caption[]{\label{Fig:etaprimapipi} \small{Spectral function for the decay $\tau^{-} \rightarrow \eta' \pi^{-} \pi^0 \nu_{\tau}$ using the values for the unknown 
couplings corresponding to the eight benchmark points as we define and explain in the text.}}
\end{center}
\end{figure}

\hspace*{0.5cm}Unfortunately there is no available data for the spectra of any of the decays $\tau^{-} \rightarrow \eta' \pi^{-} \pi^{0} \nu_{\tau}$ and 
$\tau^{-} \rightarrow \eta^{(\prime)} \eta \pi^{-} \nu_\tau$. We will be thus guided in our study only by the figures given by the PDG live, 
that are: $\Gamma\left(\tau^{-} \rightarrow \eta \pi^{-} \pi^{0} \nu_{\tau}\right)\,=\,3\mathrm{.}15(23)\cdot10^{-15}$ GeV and 
$\Gamma\left(\tau^{-} \rightarrow \eta' \pi^{-} \pi^{0} \nu_{\tau}\right)\,\leq\,1\mathrm{.}81\cdot10^{-16}$ GeV. The first one is dominated by the recent 
measurement made by the $BELLE$ collaboration \cite{Inami:2008ar} (which provides us with the only available spectra), $3\mathrm{.}06(07)\cdot10^{-15}$ GeV with a high statistics $450$ million $\tau$-pair data sample. 
While this reference fixes an upper limit on the branching ratio for the mode $\tau^{-} \rightarrow \eta \eta \pi^{-} \nu_{\tau}$ consistent with the values given 
above, it does not provide any figure for the decay $\tau^{-} \rightarrow \eta' \pi^{-} \pi^{0} \nu_{\tau}$.\\
\hspace*{0.5cm}We will use the short-distance constraints obtained in Sect. \ref{Shortdistance_etapipi} and complement them with information got in Ref.~\cite{RuizFemenia:2003hm} 
as discussed in Chapters \ref{3pi} and \ref{KKpi}. We will employ the relevant values of the coupling constants fixed in Eq.(\ref{eq:set1}) and also the 
determination of $2\,g_4\,+\,g_5\,=\,-0\mathrm{.}60\pm0\mathrm{.}02$ in Chapter \ref{KKpi}. Notice that the determination of $c_4$ and $g_4$ in Sect. \ref{determ_c4g4} does 
not play any r\^ole here.\\
\hspace*{0.5cm}This way we are left with only two unknowns: the coupling constants $c_3$ and $d_2$, so our phenomenological analysis will we aimed to gain 
some information on them and on their relevance in the spectra of the considered decays.\\
\hspace*{0.5cm}First of all we notice that it is not possible to reach the PDG branching fraction for the $\eta\pi\pi$ mode with these couplings set to zero, 
since in this case we have $\Gamma\left(\tau^{-} \rightarrow \eta \pi^{-} \pi^{0} \nu_{\tau}\right)\,\sim\,6\mathrm{.}970\cdot10^{-16}$ GeV.\\
\hspace*{0.5cm}Then, a detailed study of the allowed region in parameter space for $c_3$ and $d_2$ yields that many possibilities are opened for 
$|c_3|\lesssim0$.$06$ and $|d_2|\lesssim0$.$5$, meaning that it is possible that one of them is zero while the other not and that it is possible that both do not 
vanish. In this last case, all possibilities of signs and relative signs are opened as we illustrate with the following eight benchmark points:
\begin{eqnarray} \label{benchmarkpointsc3d2}
 \left\lbrace c_3,\;d_2\right\rbrace  & = & \left\lbrace 0,\;-0\mathrm{.}578\right\rbrace ,\;\left\lbrace 0,\;0\mathrm{.}461\right\rbrace ,\;\left\lbrace -0\mathrm{.}0643,\;0\right\rbrace ,\;\left\lbrace 0\mathrm{.}0560,\;0\right\rbrace ,\\
& & \left\lbrace -0\mathrm{.}060,\;-0\mathrm{.}040\right\rbrace ,\;\left\lbrace -0\mathrm{.}067,\;0\mathrm{.}030\right\rbrace ,\;\left\lbrace 0\mathrm{.}060,\;-0\mathrm{.}038\right\rbrace ,\;\left\lbrace 0\mathrm{.}055,\;0\mathrm{.}011\right\rbrace\;.\nonumber
\end{eqnarray}
For all of them we reproduce the PDG live value within less than one sigma.\\
\hspace*{0.5cm}We have checked that for all allowed values of the parameters we obtain a value for the decay channel $\Gamma\left(\tau^{-} \rightarrow \eta' \pi^{-} \pi^0 \nu_{\tau}\right)$ that 
is above the PDG bound. We believe that the discovery of this mode will help to understand if that is a failure of our model or an issue in the detection of this mode. 
For this purpose, the analyses of the complete $BaBar$ and $Belle$ data samples will be useful. The values that we obtain for $\Gamma\left(\tau^{-} \rightarrow \eta' \pi^{-} \pi^0 \nu_{\tau}\right)$
 in units of $10^{-16}$ GeV for the eight benchmark points are: $10$.$92$, $8$.$035$, $16$.$36$, $13$.$32$, $15$.$90$, $16$.$45$, $13$.$65$ and $13$.$31$ (in the same order as given above).\\
\hspace*{0.5cm}In Figs. \ref{Fig:etapipi} and \ref{Fig:etaprimapipi} we can see that the coupling that has a bigger impact in the features of the spectrum is 
$c_3$ while $d_2$ is only relevant when the former is close to zero. This is the reason why in Figs. \ref{Fig:etapipi} and \ref{Fig:etaprimapipi} we are labeling 
only the curves with values of $c_3$ that are not close to each other and with $d_2$ only if $c_3\sim0$. We note a large correlation between these couplings. As pointed out in Ref.~\cite{Chen:2012vw}, 
there is anticorrelation between the dependence on these couplings in the radiative processes studied in the quoted reference and those we describe here, a feature that can be exploited to improve their determination. 
Analyzing a spectrum it should be possible to determine which of the 
four labeled curves is preferred. And even lacking of that, a measurement of the branching ratio for the $\eta'\pi^-\pi^0$ mode will serve for this purpose as well.\\

\begin{figure}[h!]
\begin{center}
\vspace*{0.9cm}
\includegraphics[scale=0.8]{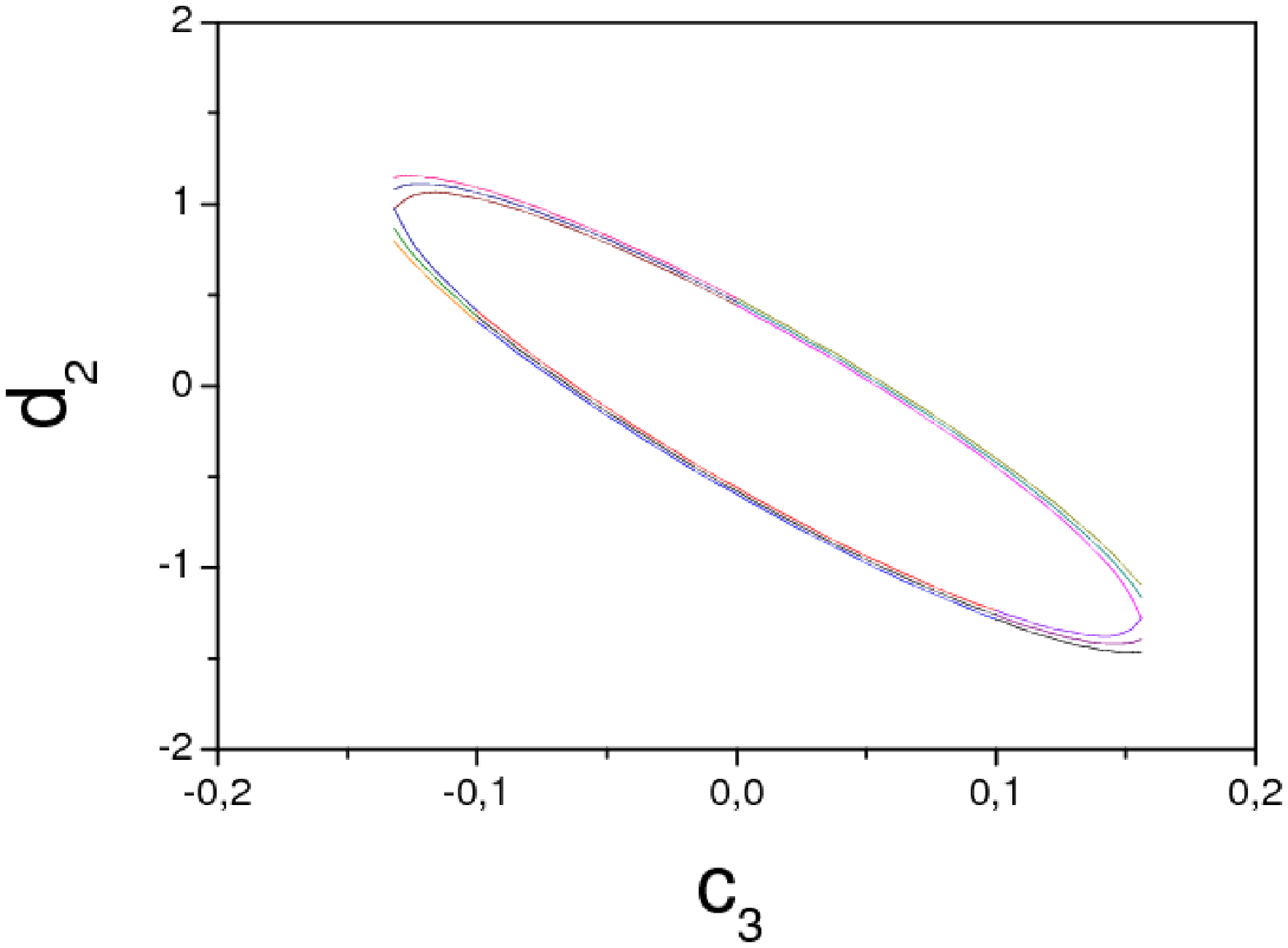}
\caption[]{\label{Fig:d2c3} \small{One-sigma contours for the branching ratio of the decay $\tau^{-} \rightarrow \eta \pi^{-} \pi^{0} \nu_{\tau}$ in the 
$c_3$-$d_2$ plane.}}
\end{center}
\end{figure}

\begin{figure}[h!]
\begin{center}
\vspace*{0.9cm}
\includegraphics[scale=0.75]{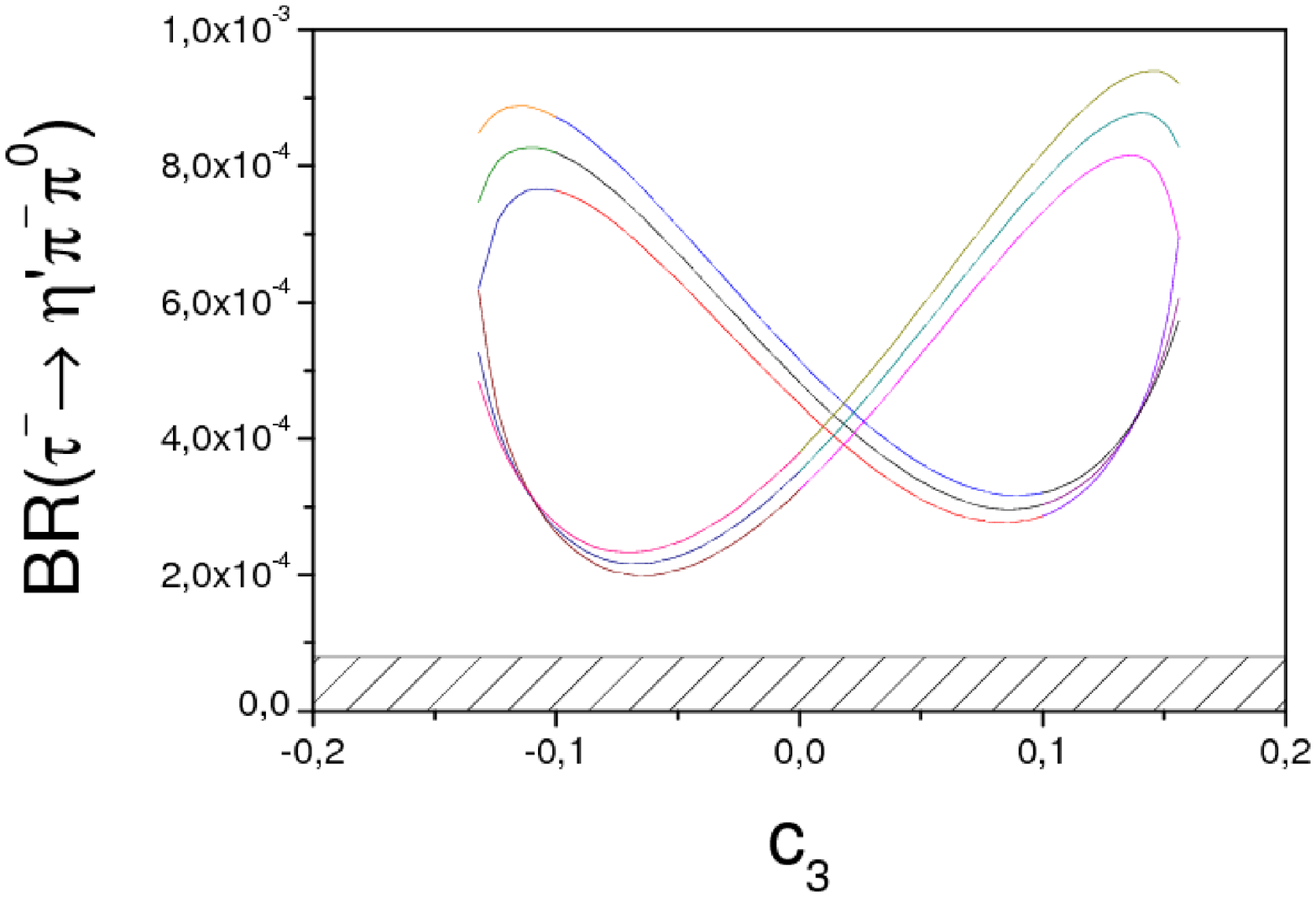}
\caption[]{\label{Fig:bretaprima} \small{The branching ratio for the mode $\tau^{-} \rightarrow \eta' \pi^{-} \pi^{0} \nu_{\tau}$ is plotted versus the value 
of $c_3$ for all values of $c_3$ (and $d_2$, whose value is not plotted) that yield a branching ratio for the decay $\tau^{-} \rightarrow \eta \pi^{-} \pi^{0} \nu_{\tau}$ consistent within 
one sigma with the pdg live bound. The horizontal line for a $br=0$.$8\cdot 10^{-4}$ represents the current pdg bound and the dashed area to the allowed region 
that excludes all our curves.}}
\end{center}
\end{figure}

\hspace*{0.5cm}In Figure \ref{Fig:d2c3} we show the one-sigma contour for the pdg live branching ratio for the mode $\tau^{-} \rightarrow \eta \pi^{-} \pi^0 \nu_{\tau}$ 
in the $d_2$-$c_3$ plane. In Figure \ref{Fig:bretaprima} we check that the branching ratio that we obtain for the mode $\tau^{-} \rightarrow \eta' \pi^{-} \pi^0 \nu_{\tau}$ 
is above the pdg bound for all allowed values of the parameters (labeled only by the parameter whose impact is bigger, $c_3$).\\

\begin{figure}[h!]
\begin{center}
\vspace*{0.9cm}
\includegraphics[scale=0.5,angle=-90]{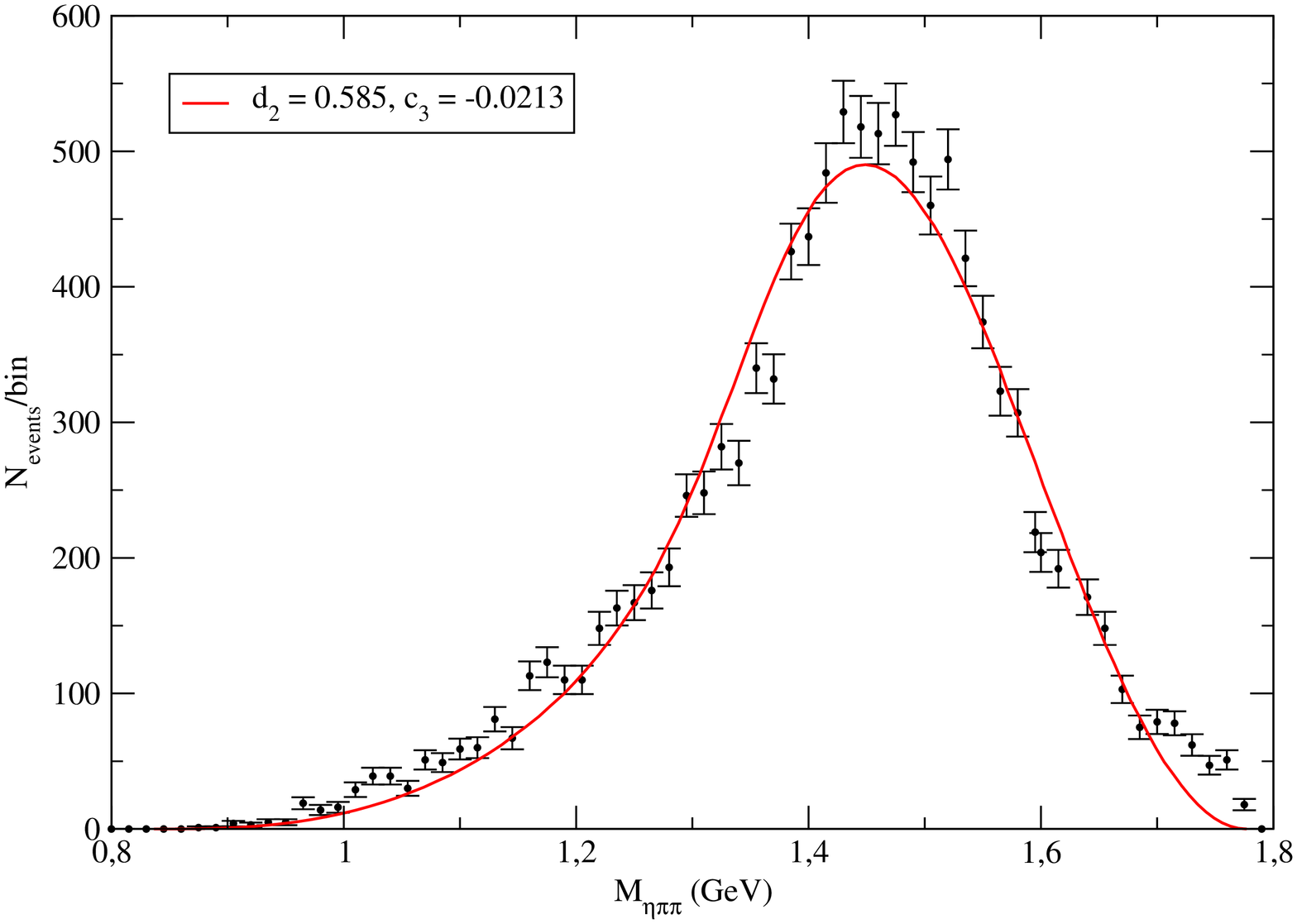}
\caption[]{\label{Fig:fitBelledata} \small{The two free parameters of our description of the $\tau^{-} \rightarrow \eta \pi^{-} \pi^{0} \nu_{\tau}$ decays 
are fitted to Belle data.}}
\end{center}
\end{figure}

\begin{figure}[h!]
\begin{center}
\vspace*{0.9cm}
\includegraphics[scale=0.45,angle=-90]{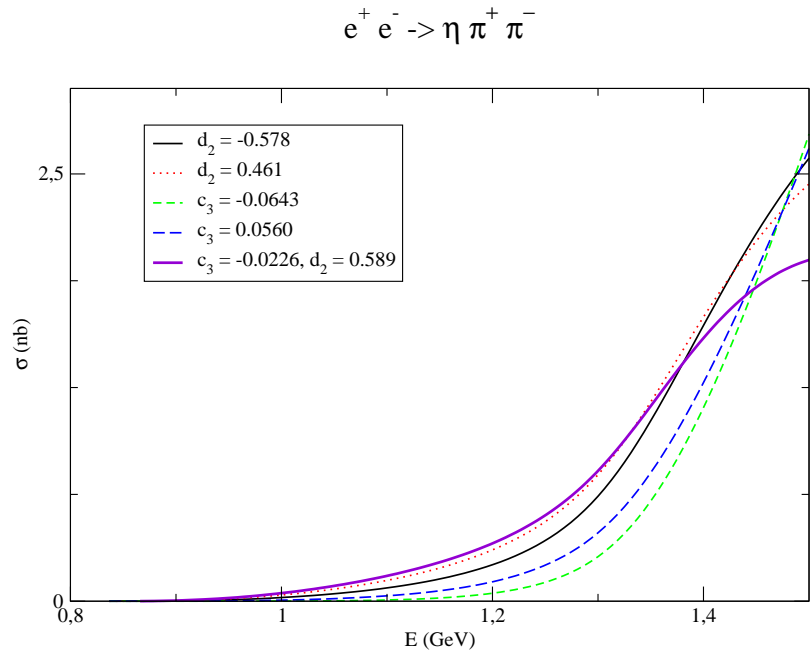}
\caption[]{\label{Fig:eeetapipi} \small{Having some cross-section data to compare with, one could tell which is the preferred scenario for $e^+ e^{-} \rightarrow 
\eta \pi^{+} \pi^{-}$. Noticeably, the curve corresponding to the fit parameters obtained in $\t^-\to\eta\pi^-\pi^0\nu_\t$ gives the smoothest behaviour in energy.}}
\end{center}
\end{figure}

\begin{figure}[h!]
 \begin{center}
  \vspace*{0.5cm}
\includegraphics[scale=0.45, angle=-90]{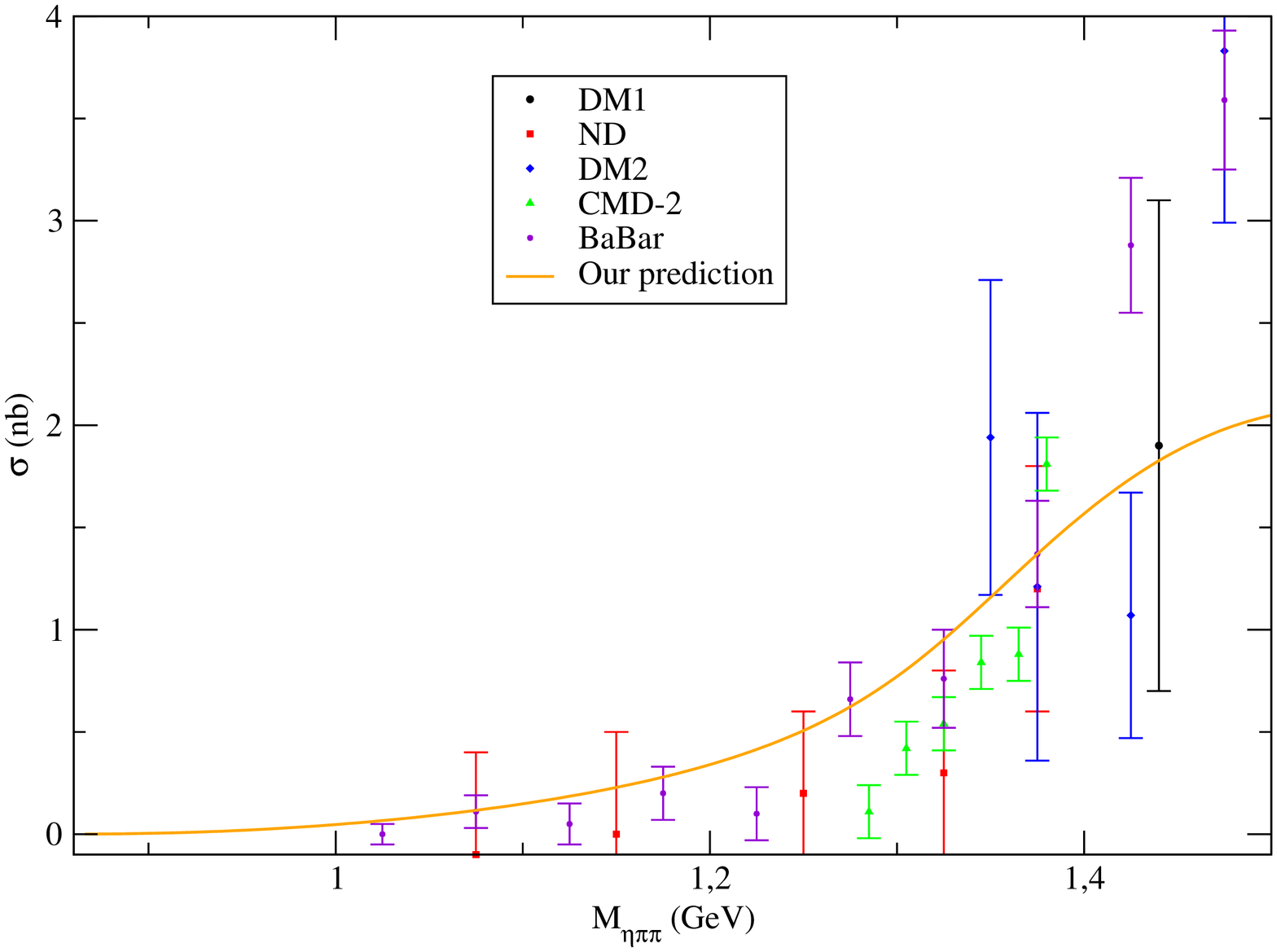}
\caption[]{\label{Fig:Comparisontoe+edata} \small{Prediction for the low-energy behaviour of $\sigma(e^+e^-\to \eta\pi^+\pi^-)$ based on the $\tau^{-} \rightarrow \eta \pi^{-} \pi^{0} \nu_{\tau}$ 
decays analysis compared to  DM1 \cite{Delcourt:1982sj}, ND \cite{Druzhinin:1986dk}, DM2 \cite{Antonelli:1988fw}, CMD-2 \cite{Akhmetshin:2000wv} and 
BaBar \cite{Aubert:2007ef} data.}}
 \end{center}
\end{figure}

\begin{figure}[!h]
\begin{center}
\vspace*{1.0cm}
\includegraphics[scale=0.45, angle=-90]{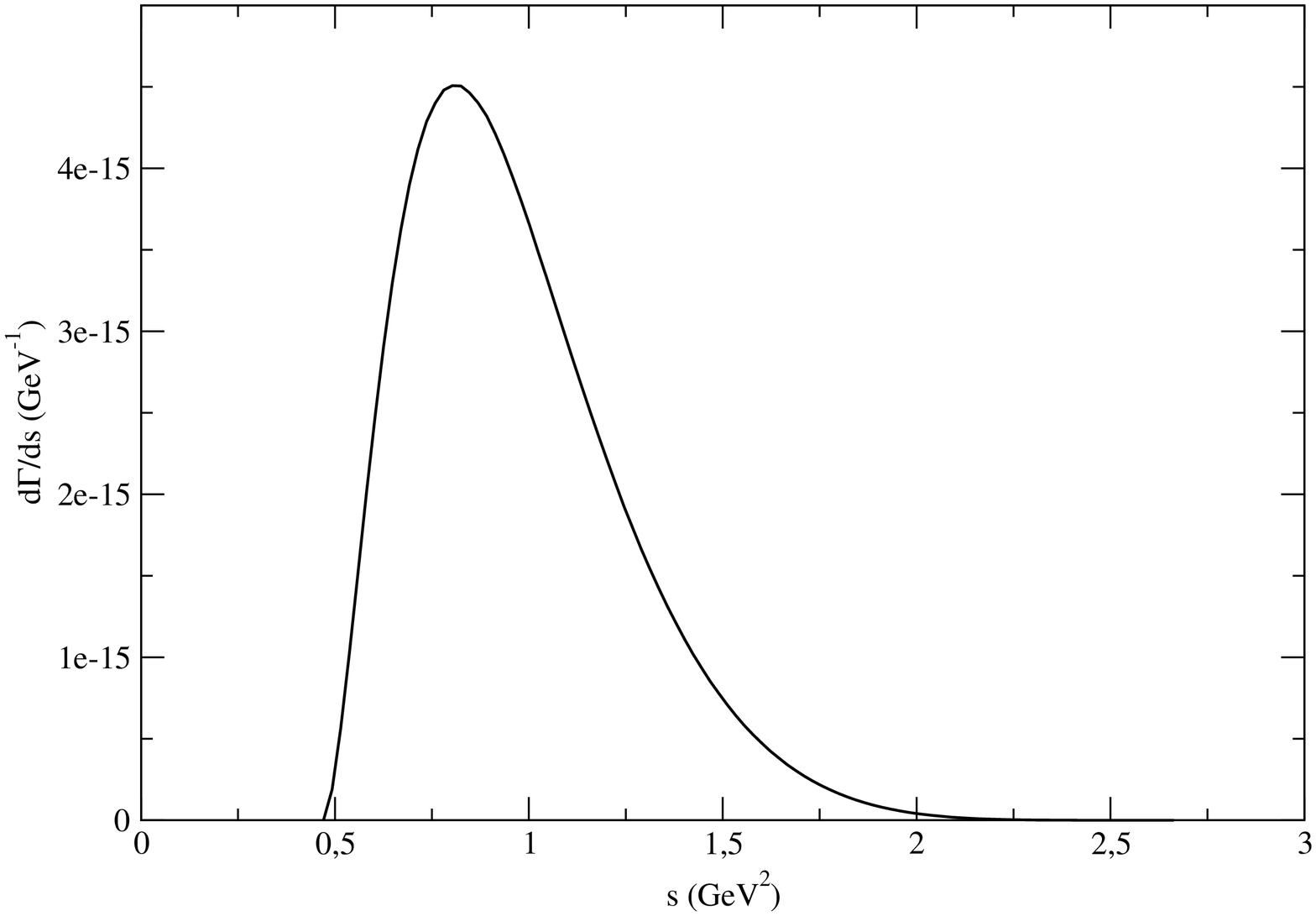}
\caption[]{\label{fig:dGdsetapp} \small{Our prediction for the $\eta\pi^-$ invariant mass distribution in the $\tau^- \rightarrow \eta \pi^- \pi^0 \nu_{\tau}$ decays is shown.}}
\end{center}
\end{figure}

\begin{figure}[!h]
\begin{center}
\vspace*{1.0cm}
\includegraphics[scale=0.45, angle=-90]{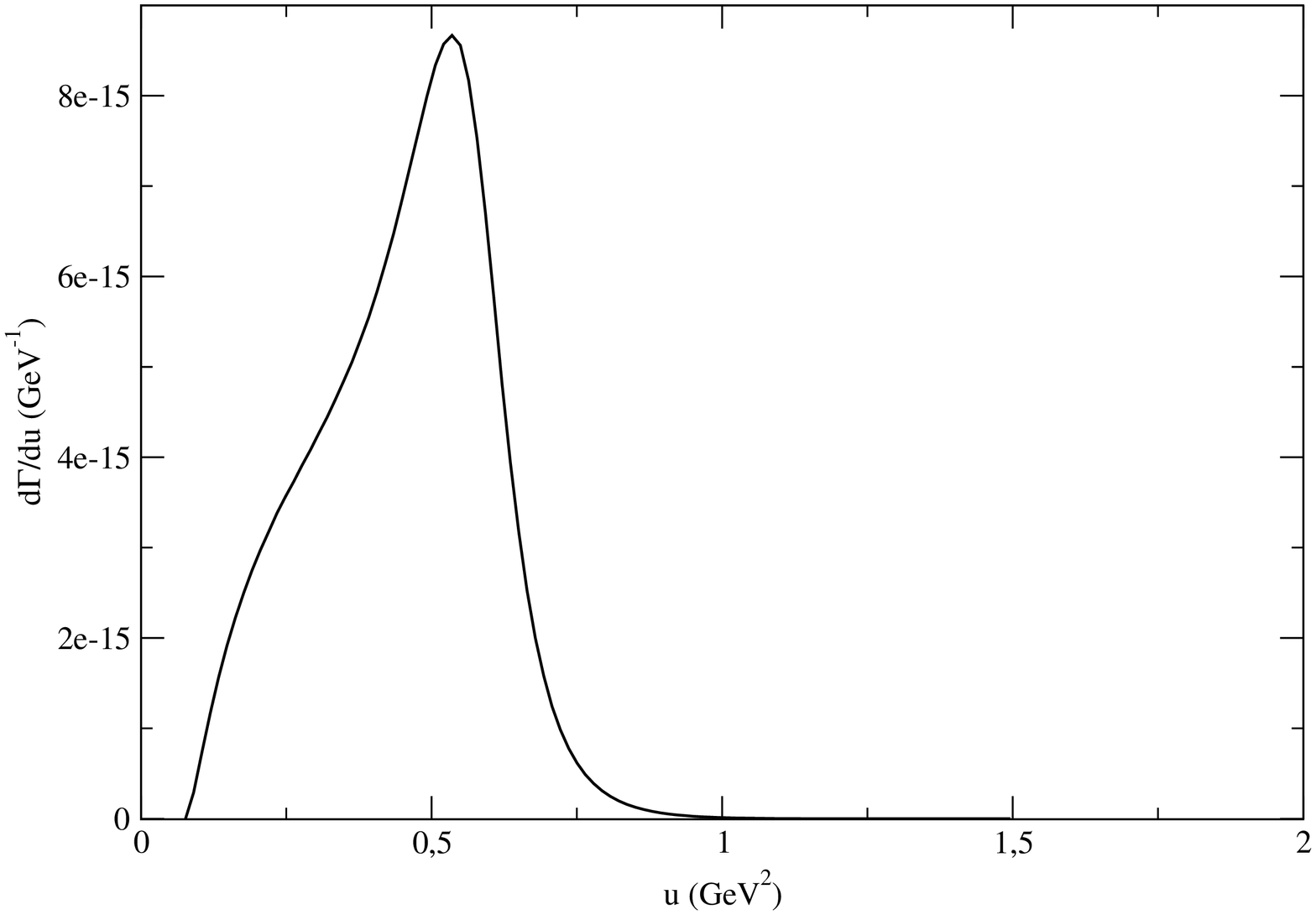}
\caption[]{\label{fig:dGduetapp} \small{Our prediction for the $\pi^-\pi^0$ invariant mass distribution in the $\tau^- \rightarrow \eta \pi^- \pi^0 \nu_{\tau}$ decays is shown.}}
\end{center}
\end{figure}

\begin{figure}[!h]
\begin{center}
\vspace*{1.0cm}
\includegraphics[scale=0.45, angle=-90]{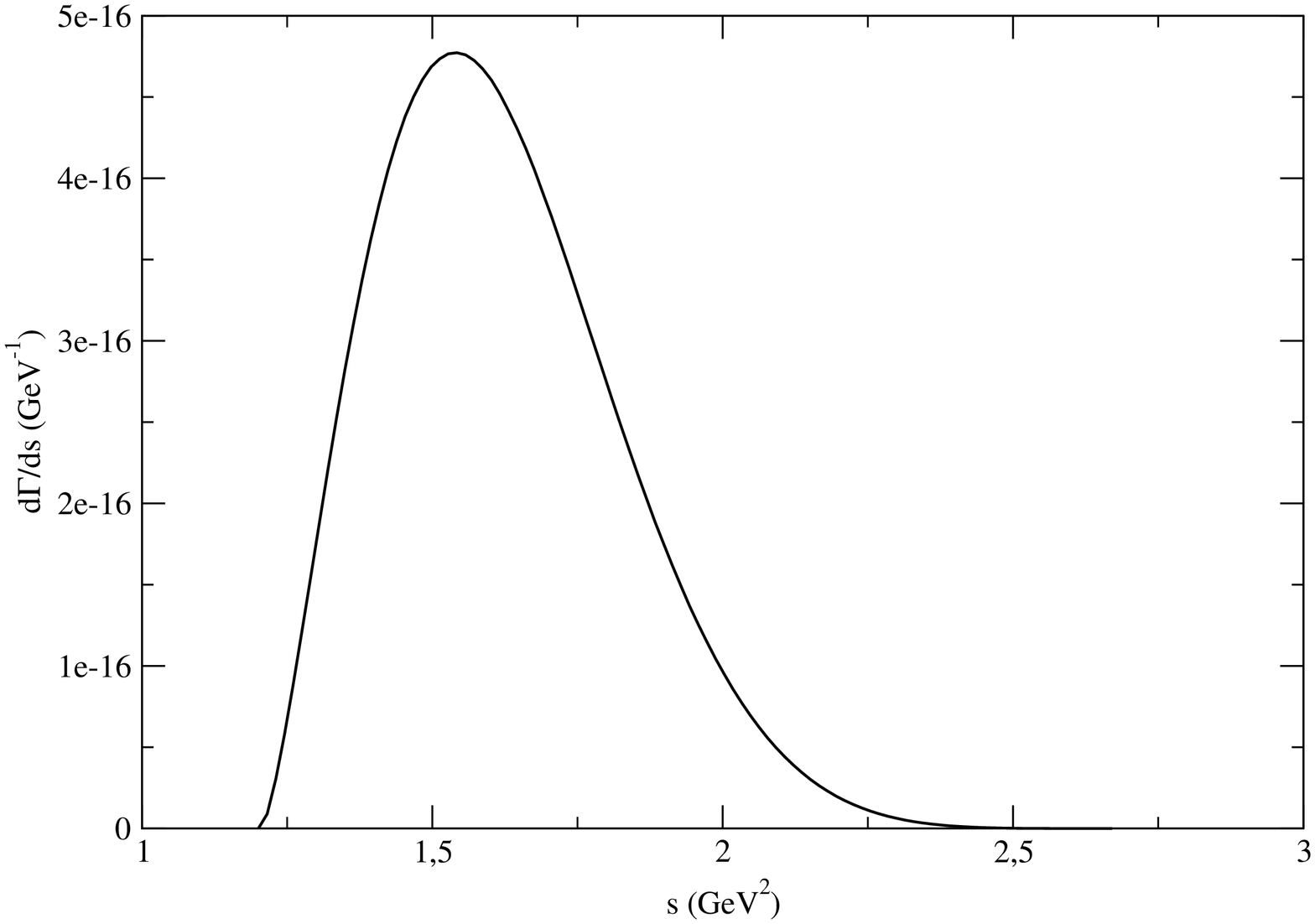}
\caption[]{\label{fig:dGdsetappp} \small{Our prediction for the $\eta^\prime\pi^-$ invariant mass distribution in the $\tau^- \rightarrow \eta^\prime \pi^- \pi^0 \nu_{\tau}$ decays is shown.}}
\end{center}
\end{figure}

\begin{figure}[!h]
\begin{center}
\vspace*{1.0cm}
\includegraphics[scale=0.45, angle=-90]{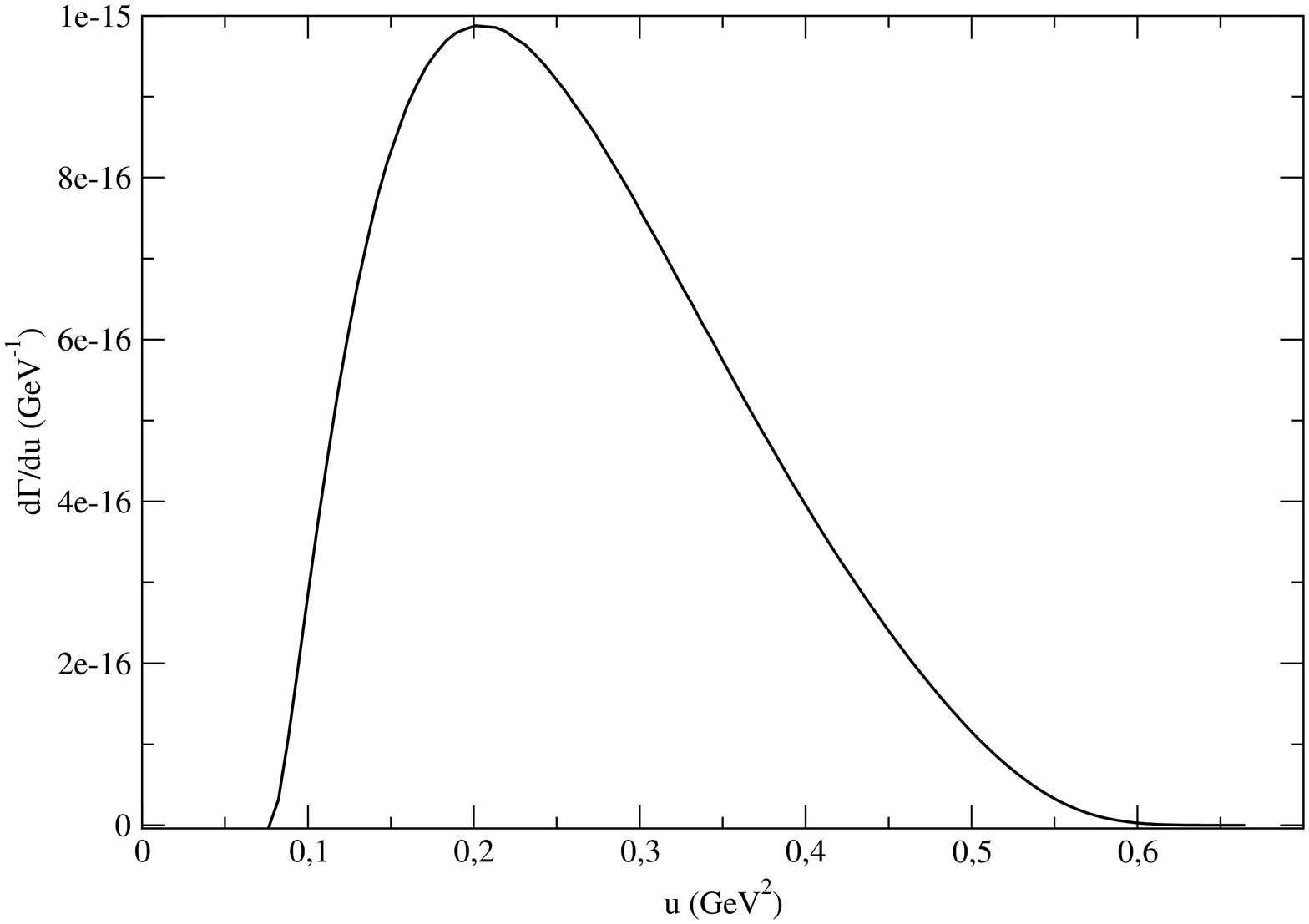}
\caption[]{\label{fig:dGduetappp} \small{Our prediction for the $\pi^-\pi^0$ invariant mass distribution in the $\tau^- \rightarrow \eta^\prime \pi^- \pi^0 \nu_{\tau}$ decays is shown.}}
\end{center}
\end{figure}

\hspace*{0.5cm}Next, we have analyzed Belle data on the $\tau^-\to\eta\pi^-\pi^0\nu_\tau$ decay spectra. In Figure \ref{Fig:fitBelledata} we can see the results 
of our fit, which yields the values $d_2 = 0$.$585\pm0.006$ and $c_3 = -0$.$0213\pm0$.$0026$ (the errors are only statistical) that is also the one giving one of the smallest decay widths for 
the $\eta'\pi^-\pi^0$ mode, that is nevertheless a factor of three larger than the PDG upper bound (last curve in Fig. \ref{Fig:etaprimapipi}) \footnote{Use of the double-angle mixing scheme \cite{Dumm:2012}
allows to obtain a branching ratio only $\sim30\%$ above the experimental upper bound while keeping a fit to $\tau^-\to\eta\pi^-\pi^0\nu_\tau$ data of similar quality. The values of the fitted couplings do not change much: 
$d_2 = 0$.$449$ and $c_3 = -0$.$0182$ are found, with slightly larger errors.}. Then, 
we conclude that the value of $d_2$ is much larger (in magnitude) than that of $c_3$ and the positive sign solution for 
$d_2$ is favoured by data. This could be confirmed by fitting the low-energy data on $\sigma(e^+e^-\to\eta\pi^+\pi^-)$ using the results in Appendix E 
\footnote{One can proceed conversely and use the data on $e^+e^-$ annihilation into hadrons to predict the corresponding semileptonic 
tau decays \cite{Eidelman:1990pb, Cherepanov:2009zz}.}
\footnote{Although the $\eta'$ meson decays to $\eta \pi^+ \pi^-$ about $45\%$ of the time, there is no significant contamination from the chain $\sigma(e^+e^-\to\eta'\to\eta\pi^+\pi^-)$
 since, because of $C$ parity it must occur at $NLO$ in the $\alpha$-expansion.}. In particular, the relation
\begin{equation}
 \frac{d\Gamma(\tau^-\to \eta\pi^-\pi^0\nu_\tau)}{\mathrm{d}Q^2}  =
  2\, f(Q^2) \sigma( e^+e^-\to \eta\pi^+\pi^-)\,,
\end{equation}
where $f(Q^2)$ is given by
\begin{equation} \label{fQ2}
 f(Q^2)=\frac{G_F^2 |V_{ud}|^2}{384(2\pi)^5
M_\tau}\left(\frac{M_\tau^2}{Q^2}-1\right)^2\left(1+2\frac{Q^2}{M_\tau^2}\right)
\left(\frac{\alpha^2}{48\pi}\right)^{-1}Q^6\, .
\end{equation}
This possibility is illustrated in Figure \ref{Fig:eeetapipi}, where we see that 
four representative benchmark points produce different predictions for this cross-section, probably enough to choose which scenario suits better if we had 
some experimental cross-section data to compare with. In addition, we consider also the curve obtained using the fit parameters for the spectrum of 
$\t^-\to\eta\pi^-\pi^0\nu_\t$. One observes that the latter curve has a clearly smoother behaviour in the highest-energy part of the figure, which is limited 
to $E\sim 1$.$5$ GeV since we cannot expect our parameterization to give a sensible description of the hadron $e^+e^-$ cross-section much beyond this energy \cite{Dumm:2009kj}.
In Fig.~\ref{Fig:Comparisontoe+edata} we compare our prediction to low-energy data from several experiments.\\
We finish this section by presenting the decay widths as functions of the two-particle invariant masses. Since this was not included in the paper \cite{Dumm:2012}, these figures correspond to the 
double angle mixing scheme and the best fit values found in that reference: $c_3 = -0.018$, $d_2 = 0.45$ \footnote{The differences with the preferred single angle mixing scheme results presented in this Thesis are tiny.}. 
In Fig. \ref{fig:dGdsetapp} we see the differential decay width as a function of the $\eta\pi^-$ invariant mass. In the isospin limit this coincides with the $\eta\pi^0$ invariant mass distribution. We also show, in Fig. 
\ref{fig:dGduetapp} the analogous plot as a function of the $\pi^-\pi^0$ invariant mass. While in the latter case the contribution of the $\rho(770)$ resonance is appreciated, no hints on underlying dynamical structures 
can be seen on the $\eta\pi^-$ and $\eta\pi^0$ distributions. Figs. \ref{fig:dGdsetappp} and \ref{fig:dGduetappp} are analogous for the $\tau^{-} \rightarrow \eta^\prime \pi^{-} \pi^{0} \nu_{\tau}$ decay. In this case, 
however, no dynamical structure can be seen: in the $s$ ($t$) distribution, because there is no resonance coupling significantly to the $\eta\pi$ system and, in the $u$ case because the high mass of the $\eta^\prime(958)$ 
makes that the maximum allowed value for $u$ is not enough to be sensitive to the $\rho(770)$ exchange.
\section{Conclusions}
\hspace*{0.5cm}We have worked out the decays $\tau^{-} \rightarrow \eta \pi^{-} \pi^0 \nu_{\tau}$, $\tau^{-} \rightarrow \eta' \pi^{-} \pi^0 \nu_{\tau}$ 
and $\tau^-\to\eta\eta\pi^-\nu_\tau$ within the framework of Resonance Chiral Theory guided by the large-$N_C$ expansion of $QCD$, the low-energy limit given by $\CPT$ and 
the appropriate asymptotic behaviour of the form factors that helps to fix most of the initially unknown couplings. Indeed only two remain free after 
completing this procedure and having used information acquired in the previous chapter.\\
\hspace*{0.5cm}We have seen that it is not possible to reproduce the decay width given by the PDG on the former mode with both couplings vanishing. Then, we have 
observed that it is quite easy to do that for natural values of these couplings in such a way that there is a a whole zone of allowed values in the parameter space 
for them. Using isospin symmetry, we provide a prediction for the low-energy behaviour of $\sigma( e^+e^-\to \eta\pi^+\pi^-)$. For any allowed value 
of the two unknowns in the previous study we can not, however, reconcile our prediction for $\Gamma\left(\tau^{-} \rightarrow \eta' \pi^{-} \pi^{0} \nu_{\tau}\right)$ 
with the PDG upper bound. We conclude that maybe there was not enough statistics yet for it to be detected and that this can happen soon analyzing 
the data from $BaBar$ and $Belle$. Finally, we find that until we characterize reliably the spin-zero contributions through resonance exchange to 
the process $\tau^-\to\eta\eta\pi^-\nu_\tau$ we cannot exploit the fact that the spin-one analogous contributions vanish, making thus this channel a 
very promising place to search for new physics once it is first detected. We plan to tackle this task elsewhere.\\

\chapter{$\tau^-\to P^- \gamma \nu_\tau$ decays ($P=\pi,\,K$)}\label{Pgamma}
\section{Introduction}\label{Pgamma_Intro}
\hspace*{0.5cm}In this chapter we will consider the structure dependent ($SD$) description of the $\tau^-\to P^- \gamma \nu_\tau$ 
decays ($P=\pi,\,K$) within the framework of $\RCT$ as discussed in earlier chapters. Until today these channels have not been observed, 
which is strange according to the most naïve expectations of their decay rates. To clarify this question is the main motivation of our study. 
We will also pursue a reliable determination of the one-meson tau decay widths including a radiated photon.\\
\hspace*{0.5cm}The structure independent part of the process has been discussed in Sect. \ref{Hadrondecays_One_meson_decays_MI}. We will compute 
the $SD$ part using the Lagrangians in Eqs. (\ref{p2-u}), (\ref{Z_WZW}), (\ref{R}) \footnote{We refer only to the part involving $A$ and $V$ 
resonances, as in any application in this Thesis. Given the vector character of the $SM$ couplings of the hadron matrix elements in $\tau$ decays, 
form factors for these processes are ruled by vector and axial-vector resonances. In the $\tau\to P^-\gamma\nu_\tau$ decays the relevant form
 factors are given by a two-point Green function. The study of these \cite{Ecker:1988te} showed that other quantum numbers play 
a negligible r\^ole.}, (\ref{LVAPb}), (\ref{VJPops}) and (\ref{VVPops}). This chapter is based on Ref.~\cite{GuoRoig}.\\
\hspace*{0.5cm}As we recall in Sect. \ref{Hadrondecays_One_meson_decays_BW}, the relative sign between the $IB$ and $SD$ part motivated an addendum 
to \cite{Decker:1993ut}. This confusion was motivated by the fact that they did not used a Lagrangian approach for the $SD$ part. In any Lagrangian approach 
this should not be an issue. In order to facilitate any independent check, we define the convention we follow as the one used by the $PDG$ \cite{Amsler:2008zzb} in order 
to relate the external fields $r_\mu, \ell_\mu$ with the physical photon field
\begin{equation}
r_\mu=\ell_\mu= -e\, Q \,A_\mu\,+...,
\end{equation}
where $e$ is the positron electric charge. Determining the relative sign between the model independent and dependent contributions is an added interest of our
 computation.\\
\section{Structure dependent form factors in $\tau^- \to \pi^- \gamma \nu_\tau$}
\begin{figure}[ht]
\begin{center}
\includegraphics[scale=0.7]{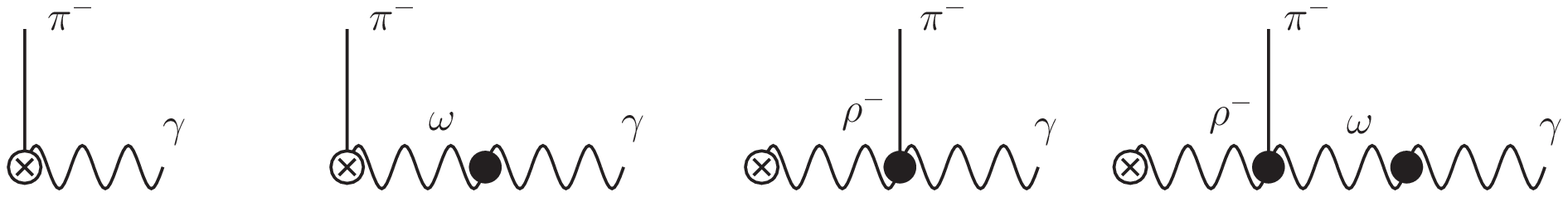}
\caption{Vector current contributions to $\tau^-\rightarrow  \pi^- \gamma \nu_\tau$. \label{fig.pi.v}}
\end{center}
\end{figure}

\hspace*{0.5cm}The Feynman diagrams, which are relevant to the vector current contributions to the $SD$ part of the 
$\tau^- \to \pi^- \gamma \nu_\tau$ processes are given in Figure \ref{fig.pi.v}. The analytical result is found to be
\begin{equation}
i \mathcal{M}_{\mathrm{SD}_{V}}= i G_F\, V_{ud}\, e \,\overline{u}_{\nu_\tau}(q)\,\gamma^\mu (1-\gamma_5)\,u_\tau(s)
\varepsilon_{\mu\nu\alpha\beta} \,\epsilon^{\nu}(k)\, k^\alpha p^\beta\, F_V^\pi(t)\,,
\end{equation}
where the vector form-factor $F_V^\pi(t)$ is
\begin{eqnarray}
F_V^\pi(t) &=& -\frac{N_C}{24\pi^2 F_\pi}+ \frac{2\sqrt2 F_V}{3 F_\pi M_V
}\bigg[ (c_2-c_1-c_5) t +
(c_5-c_1-c_2-8c_3) m_\pi^2 \bigg]\times\nonumber \\
& &  \left[ \frac{\mathrm{cos}^2\theta}{M_\phi^2}\left(1-\sqrt{2} \mathrm{tg}\theta \right)
+ \frac{\mathrm{sin}^2\theta}{M_\omega^2}\left(1+\sqrt{2} \mathrm{cotg}\theta \right)\right]
\nonumber \\
& & + \frac{2\sqrt2 F_V}{3 F_\pi M_V }\, D_\rho(t)\,  \bigg[ ( c_1-c_2-c_5+2c_6) t +
(c_5-c_1-c_2-8c_3) m_\pi^2 \bigg] \nonumber \\
& & + \frac{4 F_V^2}{3 F_\pi }\, D_\rho(t)\,  \bigg[ d_3 t +
(d_1+8d_2-d_3) m_\pi^2 \bigg]\times\nonumber \\
& & \left[ \frac{\mathrm{cos}^2\theta}{M_\phi^2}\left(1-\sqrt{2} \mathrm{tg}\theta \right)
+ \frac{\mathrm{sin}^2\theta}{M_\omega^2}\left(1+\sqrt{2} \mathrm{cotg}\theta \right)\right]\,.
\nonumber \\
\end{eqnarray}
Here we have defined $t=(k+p)^2=(s-q)^2$ and $D_R(t)$ as
\begin{equation}
D_R(t) = \frac{1}{M_R^2 - t - i M_R \Gamma_R(t)}\,.
\end{equation}
$\Gamma_R(t)$ stands for the decay width of the resonance $R$.\\
\hspace*{0.5cm}For the vector resonances $\omega$ and $\phi$, we will assume the ideal mixing case for them in any numerical application:
\begin{eqnarray}
\omega_1 = \mathrm{cos}\theta \;\omega - \mathrm{sin}\theta\;\phi \; \sim \sqrt{\frac{2}{3}} \omega - \sqrt{\frac{1}{3}} \phi \,, \nonumber \\
\omega_8 = \mathrm{sin}\theta \;\omega + \mathrm{cos}\theta\;\phi \; \sim \sqrt{\frac{2}{3}} \phi + \sqrt{\frac{1}{3}} \omega \,.
\end{eqnarray}
\\

\begin{figure}[ht]
\begin{center}
\includegraphics[scale=0.65]{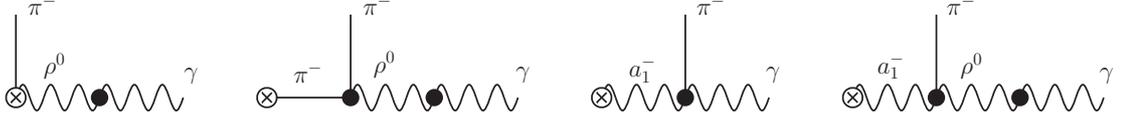}
\caption{Axial-vector current contributions to $\tau^-\rightarrow  \pi^- \gamma \nu_\tau$. \label{fig.pi.a}}
\end{center}
\end{figure}

\hspace*{0.5cm}The Feynman diagrams related to the axial-vector current contribution to the $SD$ part are given in Figure \ref{fig.pi.a}. 
The corresponding result is
\begin{equation}
i \mathcal{M}_{\mathrm{SD}_{A}} =  G_F\, V_{ud}\, e \,\overline{u}_{\nu_\tau}(q)\,\gamma^\mu (1-\gamma_5)\,u_\tau(s)
 \,\epsilon^{\nu}(k)\, \big[ (t-m_\pi^2)g_{\mu\nu}-2k_\mu p_\nu \big]\, F_A^\pi(t)\,,
\end{equation}
where the axial-vector form-factor $F_A^\pi(t)$ is
\begin{eqnarray} \label{fapit}
F_A^\pi(t) &=& \frac{F_V^2}{2F_\pi M_\rho^2}\left(1-\frac{2G_V}{F_V}\right) - \frac{F_A^2}{ 2 F_\pi} D_{\mathrm{a}_1}(t)
+ \frac{\sqrt2 F_A F_V}{ F_\pi M_\rho^2 }\, D_{\mathrm{a}_1}(t)\,  \bigg( - \lambda'' t +
\lambda_0 m_\pi^2 \bigg)\,,\nonumber\\
\end{eqnarray}
where we have used the notation from Eq. (\ref{lambdas0,',''}) for the relevant combinations of the couplings in $\mathcal{L}_2^{VAP}$, Eq. 
(\ref{LVAPa}).

\section{Structure dependent form factors in  $\tau^- \to K^- \gamma \nu_\tau$}
\begin{figure}[ht]
\begin{center}
\includegraphics[scale=0.7]{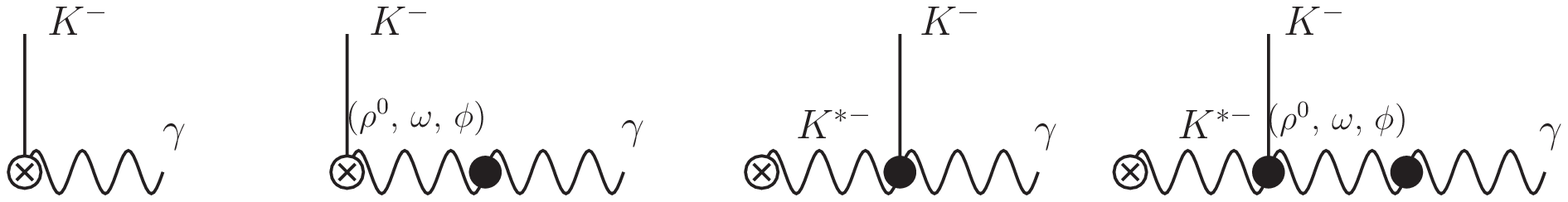}
\caption{Vector current contributions to $\tau^-\rightarrow  K^- \gamma \nu_\tau$. \label{fig.k.v}}
\end{center}
\end{figure}

\hspace*{0.5cm}Although one can read this from Eq.(\ref{Gral_IB}), let us emphasize that the model independent part 
$\mathcal{M}_{\mathrm{IB}_{\tau+K}}$ is the same as in the pion case by replacing the pion decay constant $F_\pi$ with the kaon decay constant
 $F_K$.  A brief explanation about this replacement is in order. The difference of  $F_\pi$ and $F_K$ is generated by the low energy constants
 and the chiral loops in $\CPT$~\cite{Gasser:1984gg}, while in the large $N_C$ limit of $\RCT$ this difference is due to
 the scalar resonances in an implicit way. Due to the scalar tadpole, one can always attach a scalar resonance to any of the $pG$ fields, which
 will cause the $pG$ wave function renormalization. A convenient way to count this effect is to make the scalar field redefinition before the
 explicit computation to eliminate the scalar tadpole effects. In the latter method, one can easily get the difference of $F_\pi$ and $F_K$.
 For details, see Ref.~\cite{SanzCillero:2004sk} and references therein. For the model dependent parts, the simple replacements are not 
applicable and one needs to work out the corresponding form factors explicitly.\\
\hspace*{0.5cm}The vector current contributions to the $SD$ part of the $\tau^- \to K^- \gamma \nu_\tau$ process are given in
Figure \ref{fig.k.v}. The analytical result is found to be
\begin{equation}
i \mathcal{M}_{\mathrm{SD}_{V}}= i G_F\, V_{us}\, e \,\overline{u}_{\nu_\tau}(q)\,\gamma^\mu (1-\gamma_5)\,u_\tau(s)
\varepsilon_{\mu\nu\alpha\beta} \,\epsilon^{\nu}(k)\, k^\alpha p^\beta\, F_V^K(t)\,,
\end{equation}
where the vector form-factor $F_V^K(t)$ is
\begin{eqnarray}
F_V^K(t) &=& -\frac{N_C}{24\pi^2 F_K}+ \frac{\sqrt2 F_V}{ F_K M_V }\bigg[ (c_2-c_1-c_5) t +
(c_5-c_1-c_2-8c_3) m_K^2 \bigg]\times\nonumber\\
& & \left[ \frac{1}{M_\rho^2}-\frac{\mathrm{sin}^2\theta}{3M_\omega^2}
\left( 1-2\sqrt{2} \mathrm{cotg}\theta\right) -\frac{\mathrm{cos}^2\theta}{3M_\phi^2} \left( 1+2\sqrt{2} \mathrm{tg}\theta\right)\right]
\nonumber \\ & &
+ \frac{2\sqrt2 F_V}{3 F_K M_V }\, D_{K^*}(t)\,  \bigg[ ( c_1-c_2-c_5+2c_6) t +
(c_5-c_1-c_2-8c_3) m_K^2 
\nonumber \\ & &
+24 c_4(m_K^2-m_\pi^2) \bigg] + \frac{2 F_V^2}{ F_K  }\, D_{K^*}(t)\,\bigg[ d_3 t +
(d_1+8d_2-d_3) m_K^2 \bigg]\times\nonumber\\
& &  \left[ \frac{1}{M_\rho^2}-\frac{\mathrm{sin}^2\theta}{3M_\omega^2}
\left( 1-2\sqrt{2} \mathrm{cotg}\theta\right) -\frac{\mathrm{cos}^2\theta}{3M_\phi^2} \left( 1+2\sqrt{2} \mathrm{tg}\theta\right)\right] \,.
\end{eqnarray}

\begin{figure}[ht]
\begin{center}
\includegraphics[scale=0.65]{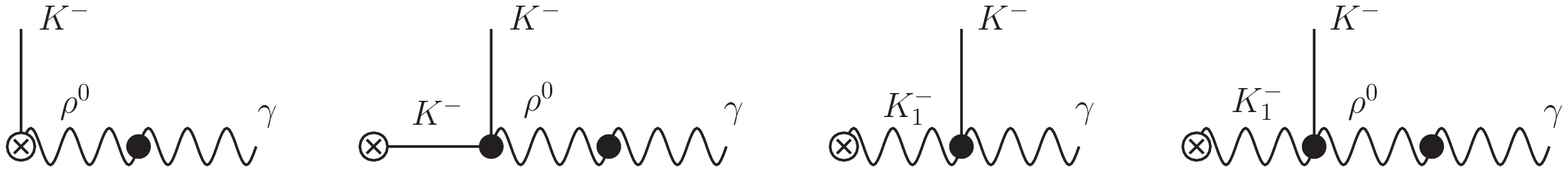}
\caption{Axial-vector current contributions to $\tau^-\rightarrow  K^- \gamma \nu_\tau$. \label{fig.k.a}}
\end{center}
\end{figure}

\hspace*{0.5cm}The axial-vector current contributions to $SD$ part are given in Figure \ref{fig.k.a}. The corresponding analytical result is
\begin{equation}
i \mathcal{M}_{\mathrm{SD}_{A}}=  G_F\, V_{us}\, e \,\overline{u}_{\nu_\tau}(q)\,\gamma^\mu (1-\gamma_5)\,u_\tau(s)
 \,\epsilon^{\nu}(k)\, \big[ (t-m_K^2)g_{\mu\nu}-2k_\mu p_\nu \big]\, F_A^K(t)\,,
\end{equation}
where the axial-vector form-factor $F_A^K(t)$ is
\begin{eqnarray} \label{fakt}
F_A^K(t) &=& \frac{F_V^2}{4F_K }\left( 1-\frac{2G_V}{F_V}\right)
\left( \frac{1}{M_\rho^2}+\frac{\mathrm{cos}^2\theta}{M_\phi^2}+\frac{\mathrm{sin}^2\theta}{M_\omega^2} \right)
- \frac{F_A^2}{ 2 F_K} \bigg[ \mathrm{cos}^2\theta_A D_{K_{1H}}(t) +\mathrm{sin}^2\theta_A D_{K_{1L}}(t) \bigg]
\nonumber \\ & &
+ \frac{ F_A F_V}{ \sqrt2 F_K }\,\bigg[ \mathrm{cos}^2\theta_A D_{K_{1H}}(t) +\mathrm{sin}^2\theta_A D_{K_{1L}}(t) \bigg]\,
\nonumber \\ &&\qquad\quad
\times\bigg( \frac{1}{M_\rho^2}+\frac{\mathrm{cos}^2\theta}{M_\phi^2}+\frac{\mathrm{sin}^2\theta}{M_\omega^2} \bigg)
\bigg( - \lambda'' t +\lambda_0 m_K^2 \bigg)\,.
\end{eqnarray}
\hspace*{0.5cm}We have used the notations of $K_{1H}$ and $K_{1L}$ for the physical states of $K_1(1400)$
and $K_1(1270)$ respectively and the mixing angle $\theta_A$ is defined in Eq.(\ref{thetaa}) as we explain in the following.\\
\hspace*{0.5cm}The $K_{1A}$ state appearing in Eq. (\ref{axial-res}) is related to the physical states $K_{1}(1270)$, $K_{1}(1400)$ through:
\begin{equation}\label{thetaa}
K_{1A} \,=\, \cos\theta_A\,\, K_{1}(1400) + \sin\theta_A \,\,K_{1}(1270) \,.
\end{equation}
About the nature of $K_1(1270)$ and $K_{1}(1400)$, it has been proposed in Ref.~\cite{Suzuki:1993yc} that they result from the mixing of 
the states $K_{1A}$ and $K_{1B}$, where $K_{1A}$ denotes the strange partner of the axial vector resonance a$_1$ with $J^{PC}=1^{++}$ and 
$K_{1B}$ is the corresponding strange partner of the axial vector resonance $b_1$ with $J^{PC}=1^{+-}$. However in this work, we will not 
include the nonet of axial vector resonances with $J^{PC}=1^{+-}$ \cite{Ecker:2007us}. As argued in Ref.~\cite{Suzuki:1993yc}, the contributions
 from these kind of resonances to tau decays are proportional to the $SU(3)$ symmetry breaking effects. Moreover, as one can see later, we 
will assume $SU(3)$ symmetry for both vector and axial-vector resonances in deriving the T-matrix always. For the $pGs$, physical masses will
 arise through the chiral symmetry breaking mechanism in the same way as it happens in $QCD$. For the (axial-)vector resonances, the experimental
 values will be taken into account in the kinematics.\\

\section{Constraints from $QCD$ asymptotic behavior}

\hspace*{0.5cm}In this part, we will exploit the asymptotic results of the form factors from perturbative $QCD$ to constrain the resonance couplings. When 
discussing the high energy constraints, we will work both in chiral and $SU(3)$ limits, which indicates we will not distinguish the form 
factors with pion and kaon, that are identical in this case.\\
\hspace*{0.5cm}For the vector form factor, the asymptotic result of perturbative $QCD$ has been derived in Ref.~\cite{Lepage:1979zb, Brodsky:1981rp}
\begin{equation}\label{lbhe}
F_V^P(t \to -\infty) = \frac{F}{t}\,,
\end{equation}
where $F$ is the pion decay constant in the chiral limit. From the above asymptotic behavior, we find three constraints on the 
resonance couplings
\begin{equation} \label{asymt1}
c_1-c_2+c_5 \,=\, 0 \,,
\end{equation}
\begin{equation} \label{asymt0}
c_2-c_1+c_5-2 c_6 \,=\, \frac{\sqrt{2} N_C M_V}{32 \pi^2 F_V} + \frac{\sqrt{2} F_V}{M_V} d_3 \,,
\end{equation}
\begin{equation} \label{asymtm1}
c_2-c_1+c_5-2 c_6 \,=\, \frac{3 \sqrt{2} F^2 }{4 F_V M_V} + \frac{\sqrt{2} F_V}{M_V} d_3 \,,
\end{equation}
where the constraints in Eqs.(\ref{asymt1}), (\ref{asymt0}) and (\ref{asymtm1}) are derived from $\mathcal{O}(t^1)$, $\mathcal{O}(t^0)$
 and $\mathcal{O}(t^{-1})$, respectively. Combining the above three constraints, we have
\begin{equation}\label{asymt0new}
c_5- c_6 \,=\, \frac{N_C M_V}{32\sqrt2 \pi^2 F_V} + \frac{F_V }{\sqrt2 M_V} d_3
\end{equation}
\begin{equation} \label{asymfmv}
F=\frac{M_V \sqrt{N_C}}{2\sqrt6 \pi}\,,
\end{equation}
where the constraint of Eq.(\ref{asymfmv}) has already been noticed in~\cite{Lepage:1979zb, Brodsky:1981rp, Decker:1993ut}.\\

It is worthy to point out different results for the asymptotic behavior of the vector form factor $F_{\pi\gamma}(t)$ have
also been noted in different frameworks, such as the ones given in Refs.~\cite{Manohar:1990hu, Gerard:1995bv}. In Refs.~\cite{Lepage:1979zb, Brodsky:1981rp},
the result was obtained in the parton picture and unavoidably the parton distribution
function for the pion has to be imposed to give the final predictions.  $OPE$ technique was exploited to
obtain its prediction in \cite{Manohar:1990hu}, which led to the conclusion that potentially large $QCD$ corrections could exist.
In \cite{Gerard:1995bv}, the form factor was discussed by using Bjorken-Johnson-Low theorem~\cite{BJL}. Variant methods have also been
used to analyze this form factor: the corrections from the transverse momentum of the parton were addressed in Ref.~\cite{Maboqiang:1996}; a
$QCD$ sum rule method was applied to derive the asymptotic behavior in Ref.~\cite{Radyushkin:1996}.
In the present discussion, we focus our attention on Refs.~\cite{Manohar:1990hu, Gerard:1995bv}, while
the study for the other results can be done analogously.
Although different results agree with the same leading power of the square momentum for large $t$, behaving as $1/t$,
they predict different coefficients, such as
\begin{equation} \label{otherhe1}
F_V^P(t \to -\infty)  =  \frac{2F}{3t}\,,
\end{equation}
from Ref.~\cite{Manohar:1990hu} and
\begin{equation}\label{otherhe2}
 F_V^P(t \to -\infty)  =   \frac{F}{3t}\,.
\end{equation}
from Ref.~\cite{Gerard:1995bv}.

By doing the same analyses as we have done by using the Lepage-Brodsky result in Eq.(\ref{lbhe}) to constrain
the resonance couplings, we can straightforwardly
get the constraints from the short-distance behaviors given in Eqs.(\ref{otherhe1})-(\ref{otherhe2}). Apparently the matching results given
in Eqs.(\ref{asymt1})-(\ref{asymt0}) will stay the same, since they are derived from $\mathcal{O}(t^1)$ and $\mathcal{O}(t^0)$.
Comparing with the result of Eq.(\ref{asymtm1}) from the matching of $\mathcal{O}(t^{-1})$ by using the coefficient in Eq.(\ref{lbhe}),
the corresponding results by using Eqs.(\ref{otherhe1}) and (\ref{otherhe2}) are respectively
\begin{equation} \label{asymtm1other}
c_2-c_1+c_5-2 c_6 \, =\, \frac{ F^2 }{ 2 \sqrt{2} F_V M_V} + \frac{\sqrt{2} F_V}{M_V} d_3 \,,   \\
c_2-c_1+c_5-2 c_6 \, =\, \frac{  F^2 }{4\sqrt{2} F_V M_V} + \frac{\sqrt{2} F_V}{M_V} d_3 \,,
\end{equation}
which lead to the following results, in order, by combining Eqs.(\ref{asymt1})-(\ref{asymt0})
\begin{eqnarray} \label{asymfmvother1}
F=\frac{M_V \sqrt{N_C}}{4\pi}\,,    \\
F=\frac{M_V \sqrt{N_C}}{2\sqrt2 \pi}\,.  \label{asymfmvother2}
\end{eqnarray}

The formulae displayed in Eq.(\ref{asymfmv}), Eq.(\ref{asymfmvother1}) and  Eq.(\ref{asymfmvother2}) provide a simple way to
discriminate between different asymptotic behaviors.
The chiral limit values for the pion decay constant $F$ and the mass of the lowest vector multiplet $M_V$ have been
thoughtfully studied at the leading order of $1/N_C$ in Ref.~\cite{guo:2009}, which predicts $F= 90.8$ MeV and $M_V= 764.3$ MeV.
The different results shown in Eq.(\ref{asymfmv}), Eq.(\ref{asymfmvother1}) and  Eq.(\ref{asymfmvother2})
 from  matching  different short-distance behaviors of Refs.~\cite{Lepage:1979zb,Manohar:1990hu,Gerard:1995bv}
deviate from the phenomenology study at the level of $5\%, 16\%, 64\%$ respectively \footnote{The results are mildly changed when estimating the values
of $F$ and $M_V$ by  $F_\pi$ and $M_\rho$ respectively, as expected.}, which implies that the short-distance behavior in
Eq.(\ref{lbhe}) is more reasonable than the ones in Eqs.(\ref{otherhe1})-(\ref{otherhe2}).
Hence for the matching result at the order of $1/t$, we will use the one in Eq.(\ref{asymfmv}) throughout the following discussion.
However we stress the inclusion of extra multiplets of vector resonances or the sub-leading corrections in $1/N_C$ may alter the
current conclusion.\\

\hspace*{0.5cm}The high energy constraints on the resonance couplings $c_i$ and $d_i$ 
have been studied in different processes. The $OPE$ analysis of the $VVP$ Green Function gives~\cite{RuizFemenia:2003hm}
\begin{equation}\label{asymvvpc5}
c_5- c_6 \,=\, \frac{N_C M_V}{64\sqrt2 \pi^2 F_V} \,,
\end{equation}
\begin{equation}\label{asymvvpd3}
d_3 \,=\, -\frac{N_C M_V^2}{64 \pi^2 F_V^2}+\frac{F^2}{8F_V^2}\,.
\end{equation}

\hspace*{0.5cm}The constraint from $\tau^- \to (V P)^- \nu_\tau$ study leads to
\begin{equation}\label{asymtauvp}
c_5- c_6 \,=\, -\frac{F_V }{\sqrt2 M_V} d_3\,,
\end{equation}
if one neglects the heavier vector resonance multiplet~\cite{Guo:2008sh}.\\
\hspace*{0.5cm}The results from the analysis of $\tau^- \to (K K \pi)^- \nu_\tau$ are~\cite{Roig:2007yp, Roig:2008je}
\begin{eqnarray} \label{asymtaukkpi}
c_5- c_6 \,=\, \frac{N_C M_V F_V}{192\sqrt2 \pi^2 F^2} \,, \nonumber \\
d_3 \,=\, -\frac{N_C M_V^2}{192 \pi^2 F^2}\,.
\end{eqnarray}
It is easy to check that the results of Eqs.(\ref{asymtauvp}) and (\ref{asymtaukkpi}) are consistent. Combining Eqs.(\ref{asymt0new})
and (\ref{asymtauvp}) leads to
\begin{eqnarray} \label{asym_taupg_tauvp}
c_5- c_6 &=& \frac{N_C M_V}{64\sqrt2 \pi^2 F_V} \,,\nonumber \\
d_3 &=& -\frac{N_C M_V^2}{64 \pi^2 F_V^2}\,,
\end{eqnarray}
where the constraint of $c_5 -c_6$ is consistent with the result from the $OPE$ analysis of the $VVP$ Green Function~\cite{RuizFemenia:2003hm},
 while the result of $d_3$ is not \footnote{However, the difference on the numerical value of both predictions is small, $\sim16\%$ and 
the impact of such a difference in any observable in the considered processes is extremely tiny.}.\\
\hspace*{0.5cm}By demanding the consistency of the constraints derived from the processes of $\tau^- \to P^- \gamma \nu_\tau$ and 
$\tau^- \to (V P)^- \nu_\tau$ given in Eq.(\ref{asym_taupg_tauvp}) and the results form $\tau^- \to (K K \pi)^- \nu_\tau$ given in 
Eq.(\ref{asymtaukkpi}), we get the following constraint~\footnote{It has also been obtained in Refs. \cite{Pich:2010sm,Nieves:2011gb}.}
\begin{equation}\label{ksrf3fv}
F_V=\sqrt3 F\,.
\end{equation}
If one combines the high energy constraint from the two pion vector form factor~\cite{Ecker:1989yg}
\begin{equation}
F_V G_V = F^2\,,
\end{equation}
and the result of Eq.(\ref{ksrf3fv}) we get here, the modified $KSRF$ will be derived
\begin{equation}
F=\sqrt3 G_V\,,
\end{equation}
which is also obtained in the partial wave dispersion relation analysis of $\pi \pi$ scattering~\cite{Guo:2007ff}\,.\\
\hspace*{0.5cm}Although the branching ratios for the modes $\t\to P \gamma\nu_\t$ we are discussing should be higher than for some modes 
that have been already detected, they have not been observed yet. Lacking of experimental data, we will make some theoretically and 
phenomenologically based assumptions in order to present our predictions for the spectra and branching ratios.\\
\hspace*{0.5cm}Taking into account the previous relations one would have $F_V^\pi(t)$ in terms of $c_1+c_2+8c_3-c_5$ and $d_1+8d_2-d_3$. 
For the first combination, $c_1+c_2+8c_3-c_5=c_1+4c_3$ ($c_1-c_2+c_5 = 0$ has been used), the prediction for $c_1+4c_3$ in \cite{RuizFemenia:2003hm} yields 
$c_1+c_2+8c_3-c_5=0$. In Ref.~\cite{RuizFemenia:2003hm} the other relevant combination of couplings is also restricted: $d_1+8d_2-d_3=\frac{F^2}{8F_V^2}$. In 
$F_V^\pi(t)$ $c_4$ appears, in addition. There is a phenomenological determination of this coupling in our work on the $KK\pi$ decay modes 
of the $\tau$~\cite{Roig:2008je}: $c_4=-0$.$07\pm0$.$01$.\\
\hspace*{0.5cm}Turning now to the axial-vector form factor, in both channels it still depends on four couplings: $F_A$, $M_A$, $\lambda''$ 
and $\lambda_0$. If one invokes the once subtracted dispersion for the axial vector form factor, as done in Ref.~\cite{Decker:1993ut}, one 
can not get any constraints on the resonance couplings from the axial vector form factors given in Eqs.(\ref{fapit}) and (\ref{fakt}). In 
fact by demanding the form factor to satisfy the unsubtracted dispersion relation, which guarantees a better high energy limit, we can get 
the following constraint
\begin{equation}
 \lambda''=\frac{2G_V-F_V}{2\sqrt{2}F_A}\,,
\end{equation}
which has been already noted in ~\cite{Cirigliano:2004ue}.\\
\hspace*{0.5cm}In order to constrain the free parameters as much as possible, we decide to exploit the constraints from the Weinberg sum 
rules ($WSR$) \cite{Weinberg:1967kj}: $F_V^2-F_A^2=F^2$ and $M_V^2F_V^2-M_A^2F_A^2=0$, yielding
\begin{equation}
 F_A^2=2F^2\;\;,\;M_A=\frac{6\pi F}{\sqrt{N_C}}\,.
\end{equation}
For the axial vector resonance coupling $\lambda_0$, we use the result from Ref.~\cite{Cirigliano:2004ue, Dumm:2009va}
\begin{equation}
\lambda_0=\frac{G_V}{4\sqrt{2}F_A}\,.
\end{equation}
\hspace*{0.5cm}To conclude this section, we summarize the previous discussion on the high energy constraints
\begin{eqnarray}
&&F_V=\sqrt3 F \,,\quad G_V=\frac{F}{\sqrt3}\,,\quad F_A=\sqrt2 F\,,\quad M_V=\frac{2\sqrt6 \pi F}{\sqrt{N_C}} \,,\quad
M_A=\frac{6\pi F}{\sqrt{N_C}}\,,
\nonumber \\ &&
\lambda_0=\frac{1}{8\sqrt3}\,,\quad \lambda''=-\frac{1}{4\sqrt3}\,,\quad
c_5-c_6=\frac{\sqrt{N_C}}{32\pi}\,,\quad d_3=-\frac{1}{8}\,.
\end{eqnarray}
In the above results, we have discarded the constraint in Eq.(\ref{asymvvpd3}), which is the only inconsistent result with the others.\\
\section{Phenomenological discussion}
\hspace*{0.5cm}Apart from the parameters we mentioned in the last section, there is still one free coupling $\theta_A$, which
describes the mixing of the strange axial vector resonances in Eq.(\ref{thetaa}). The value of $\theta_A$ has already been 
determined in literature~\cite{Suzuki:1993yc, Guo:2008sh, Cheng:2003bn}. We recapitulate the main results in the following.\\
\hspace*{0.5cm}In Ref.~\cite{Suzuki:1993yc}, it is determined $\theta_A\sim33^\circ$. In Ref.~\cite{Guo:2008sh}, $|\theta_A| \sim 58$.$1^\circ$
is determined through the considered decays $\tau^-\to (VP)^-\nu_\tau$. In Ref.~\cite{Cheng:2003bn}, 
the study of $\tau\to K_1 \nu_\tau$ gives $|\theta_A|=^{37^\circ}_{58^\circ}$ as the two possible solutions. The decay $D\to K_1 \pi$ allows to 
conclude that $\theta_A$ must be negative and it is pointed out that the observation of $D^0\to K_1^-\pi^+$ with a branching ratio $\sim5\cdot10^{-4}$ 
would imply $\theta_A\sim-58^\circ$, in agreement with \cite{Tayduganov:2011ui}, investigating the $K_1\to K \pi \pi$ decays. However, another analyses 
in Ref.~\cite{Cheng:2007mx} finds that the current measurement of $\bar{B}^0\to K_1^-(1400) \pi^+$ 
\cite{Amsler:2008zzb} favors a mixing angle of $-37^\circ$ over $-58^\circ$, a conclusion which is also supported by Ref.~\cite{Cheng:2011pb}. In this respect, the relation
\begin{equation}
 \Big|\Gamma\left(J/\Psi\to K_1^0(1400)\overline{K}^0\right)\Big|^2\,=\,\mathrm{tg}\theta_A^2\,\Big|\Gamma\left(J/\Psi\to K_1^0(1270)\overline{K}^0\right)\Big|^2
\end{equation}
would be very useful to get $\theta_A$, once these modes are detected.\\
\subsection{Results including only the WZW contribution in the SD part}
\begin{figure}[h!]
\begin{center}
\includegraphics[scale=0.5,angle=-90]{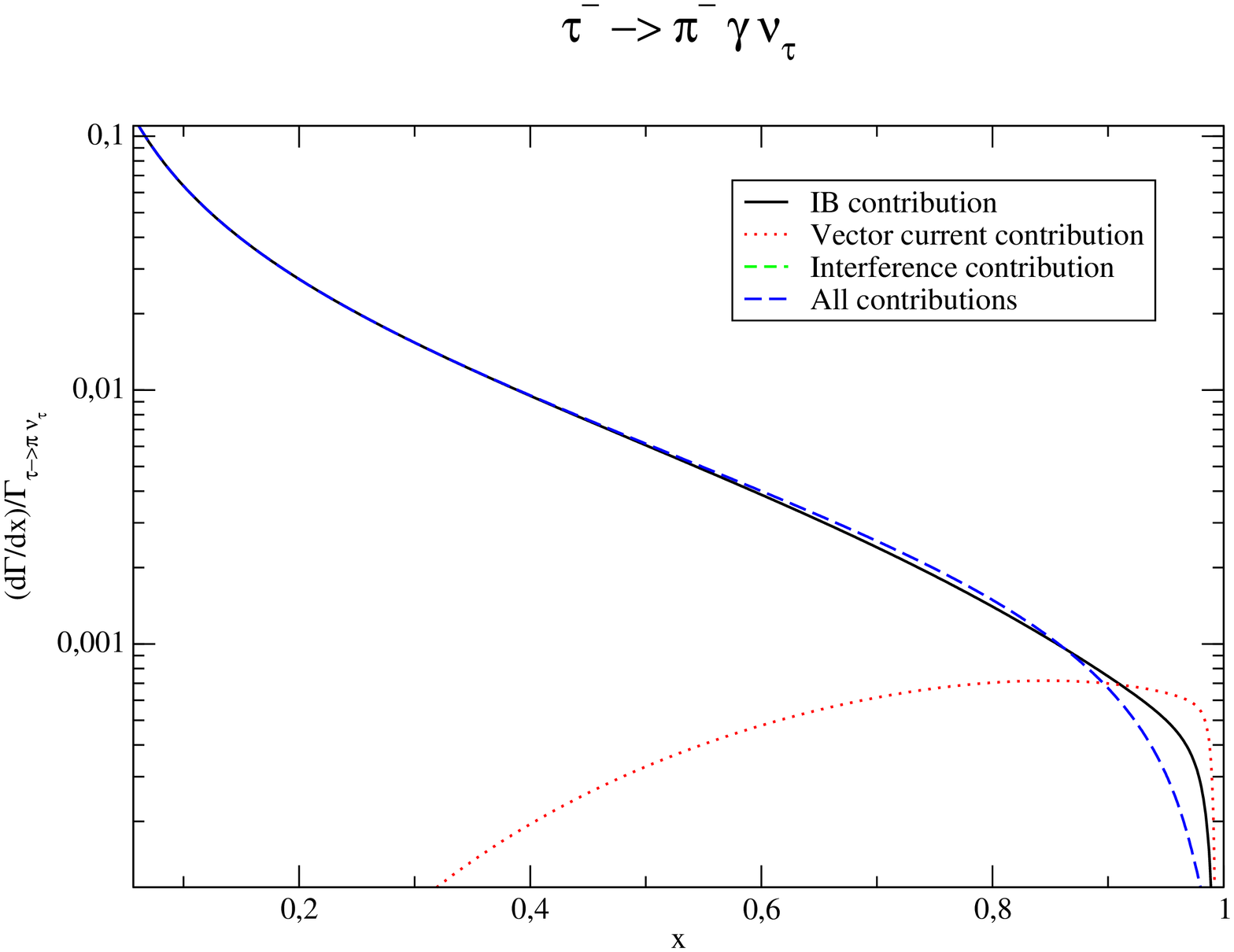}
\caption{Differential decay width of the process $\tau^-\rightarrow  \pi^- \gamma \nu_\tau$ including only the model independent contributions for a cut-off
 on the photon energy of $50$ MeV. In the vector form factor only the $WZW$ term is considered for this estimate.  The interference contribution is 
negative. It can be appreciated in Fig. \ref{MIpihigh}. \label{MIpilow}}
\end{center}
\end{figure}

\begin{figure}[h!]
\begin{center}
\includegraphics[scale=0.5,angle=-90]{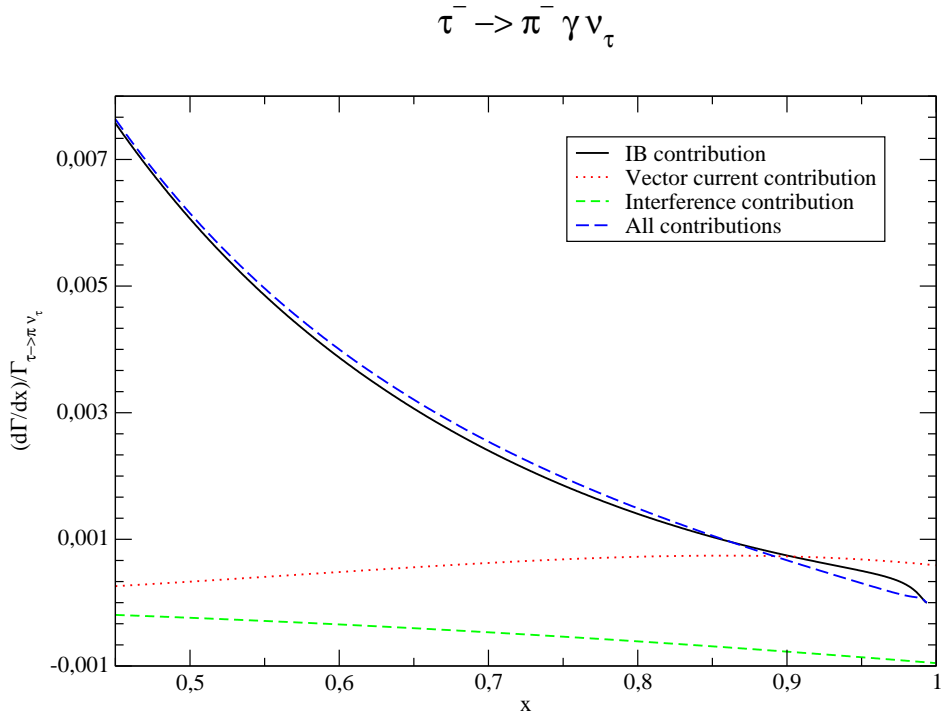}
\caption{Differential decay width of the process $\tau^-\rightarrow  \pi^- \gamma \nu_\tau$ including only the model independent contributions for a cut-off
 on the photon energy of $400$ MeV. In the vector form factor only the $WZW$ term is considered for this estimate. \label{MIpihigh}}
\end{center}
\end{figure}

\hspace*{0.5cm}As it was stated in Sects. \ref{Hadrondecays_One_meson_decays_BW} and \ref{Pgamma_Intro} it is strange that these modes have not been detected
 so far. The most na\"ive and completely model independent estimate would just include the $IB$ part and the $WZW$ contribution to the $VV$ part, as the latter 
is completely fixed by $QCD$. We know that doing this way we are losing the contribution of vector and axial-vector resonances, that should be important in 
the high-$x$ region. However, even doing so one is able to find that the radiative decay $\t^-\to\pi^-\gamma \nu_\t$ has a decay probability larger than 
the mode $\t^-\to K^+K^-K^-\nu_\t$ \footnote{see Table \ref{Listdecaystau}, $\Gamma\left(\t^-\to K^+K^-K^-\nu_\t\right)\,=\,3$.$579(66)\cdot 10^{-17}$ GeV.}. 
For a reasonably low cut on the photon energy this conclusion holds for the $\t^-\to K^-\gamma \nu_\t$ as well.\\
\hspace*{0.5cm}Before seeing this, we will discuss briefly the meaning of cutting on the photon energy. A cut on the photon energy was introduced in Sect. 
\ref{Hadrondecays_One_meson_decays_BW}. As it is well know \cite{Kinoshita:1962ur, Peskin:1995ev} the $IR$ divergences due to the vanishing photon mass cancel 
when considering at the same time the non-radiative and the radiative decays. In practice, this translates into mathematical language the physical notion that 
the detectors have a limited angular resolution that defines a threshold detection angle for photons. If one considers a photon emitted with a smaller angle 
it should be counted together with the non-radiative decay as it is effectively measured this way. The sum is of course an $IR$ safe observable. The splitting 
depends on the particular characteristics of the experimental setting. Obviously, the branching fraction for the radiative decay depends on this cut-off 
energy. We will consider here the case $E_{\gamma\,\mathrm{thr}}\,=\,50$ MeV, that corresponds to $x \, = \, 0$.$0565$. In order to illustrate the dependence 
on this variable, we will also show the extremely conservative case of  $E_{\gamma\,\mathrm{thr}}\,=\,400$ MeV ($x \, = \, 0$.$45$). In Figure~\ref{MIpilow} we 
see the radiative $\pi$ decay for a low value of $x$, while in Figure~\ref{MIpihigh} we plot it for the high-$x$ case. In the first case we obtain $\Gamma\left(
\t^-\to\pi^-\gamma \nu_\t\right)\,=\,3$.$182\cdot 10^{-15}$ GeV, and in the second one we are still above the bound marked by the $3K$ decay, $\Gamma\left(
\t^-\to\pi^-\gamma \nu_\t\right)\,=\,3$.$615\cdot 10^{-16}$ GeV. Proceeding analogously for the decays with a $K^-$, we find: $\Gamma\left(\t^-\to K^-\gamma 
\nu_\t\right)\,=\,6$.$002\cdot 10^{-17}$ GeV for $E_{\gamma\,\mathrm{thr}}\,=\,50$ MeV (Figure~\ref{MIKhigh}), and $\Gamma\left(\t^-\to K^-\gamma \nu_\t\right)
\,=\,4$.$589\cdot 10^{-18}$ GeV for $E_{\gamma\,\mathrm{thr}}\,=\,400$ MeV. For any reasonable cut on $E_{\gamma}$ these modes should have already been 
detected by the $B$-factories.\\
\\
\\
\begin{figure}[h!]
\begin{center}
\includegraphics[scale=0.5,angle=-90]{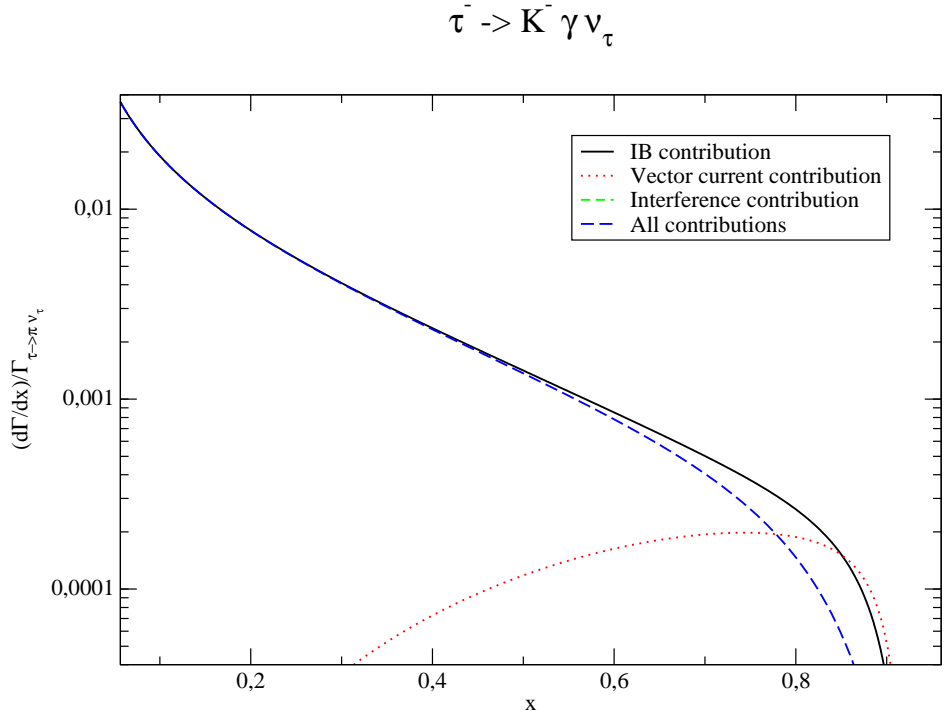}
\caption{Differential decay width of the process $\tau^-\rightarrow  K^- \gamma \nu_\tau$ including only the model independent contributions for a cut-off
 on the photon energy of $50$ MeV. In the vector form factor only the $WZW$ term (where the axial-vector contribution is absent) is considered for 
this estimate. The interference contribution is negative. It can be appreciated in Fig. \ref{MIKhigh}. \label{MIKlow}}
\end{center}
\end{figure}
\\
\begin{figure}[h!]
\begin{center}
\includegraphics[scale=0.5,angle=-90]{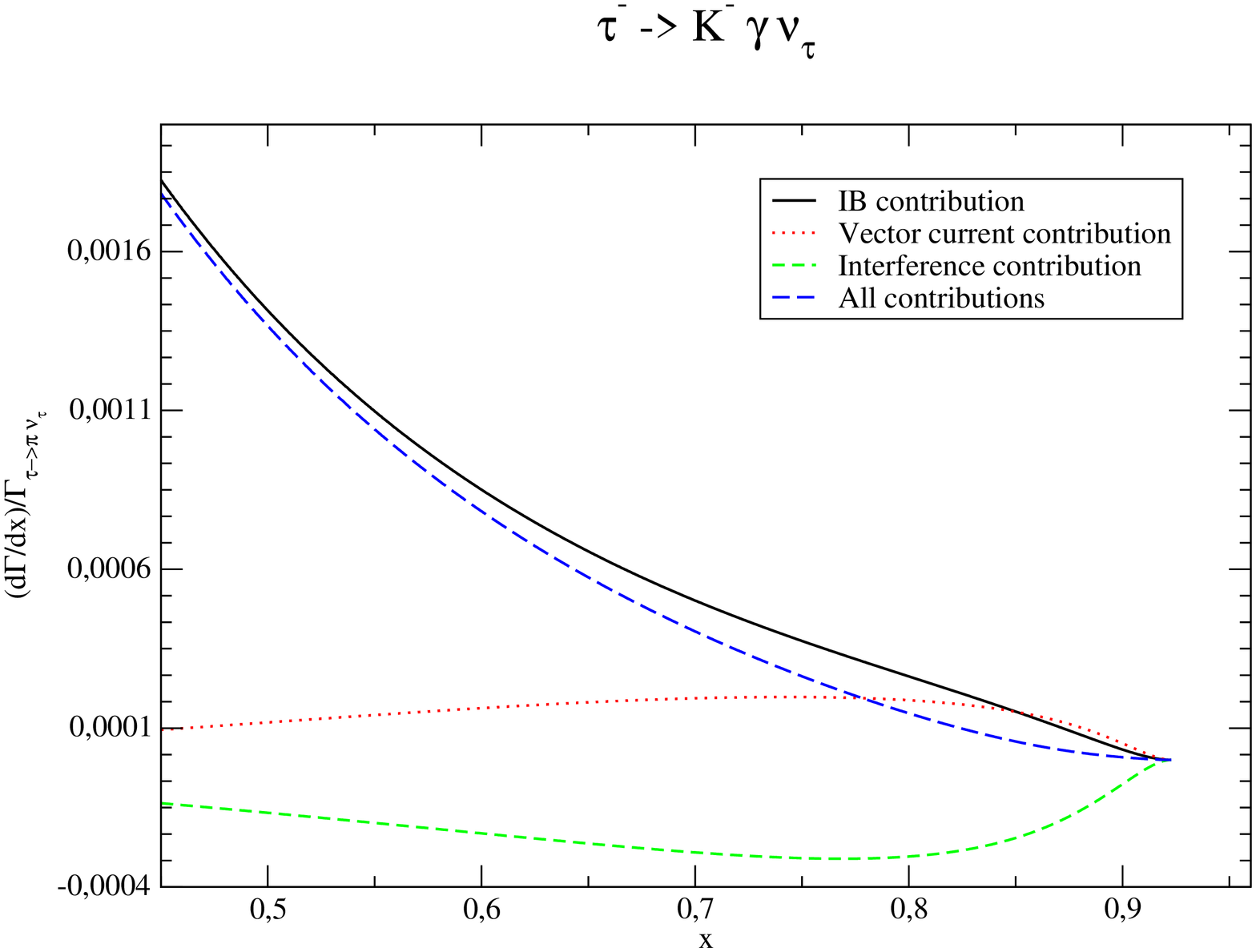}
\caption{Differential decay width of the process $\tau^-\rightarrow  K^- \gamma \nu_\tau$ including only the model independent contributions for a cut-off
 on the photon energy of $400$ MeV. In the vector form factor only the $WZW$ term (where the axial-vector contribution is absent) is considered for 
this estimate. \label{MIKhigh}}
\end{center}
\end{figure}
\hspace*{0.5cm}Already at this level of the phenomenological analysis, the question of the accuracy on the detection of soft photons at $B$-factories \cite{Becirevic:2009aq} arises. 
An error larger than expected (here and in some undetected particle interpreted as missing energy, in addition to a gaussian treatment
 of systematic errors) could enlarge the uncertainty claimed on the measurement of $B^-\to\tau^- \nu_\tau$ \cite{Amsler:2008zzb} when combining the 
$Belle$~\cite{Ikado:2006un, Hara:2010dk} and $BaBar$ measurements~\cite{Aubert:2007bx, Aubert:2007xj, Aubert:2009wt} taking it closer to the standard model expectations.\\
\subsection{Results including resonance contributions in the $\p$ channel}
\begin{figure}[h!]
\begin{center}
\includegraphics[scale=0.5,angle=-90]{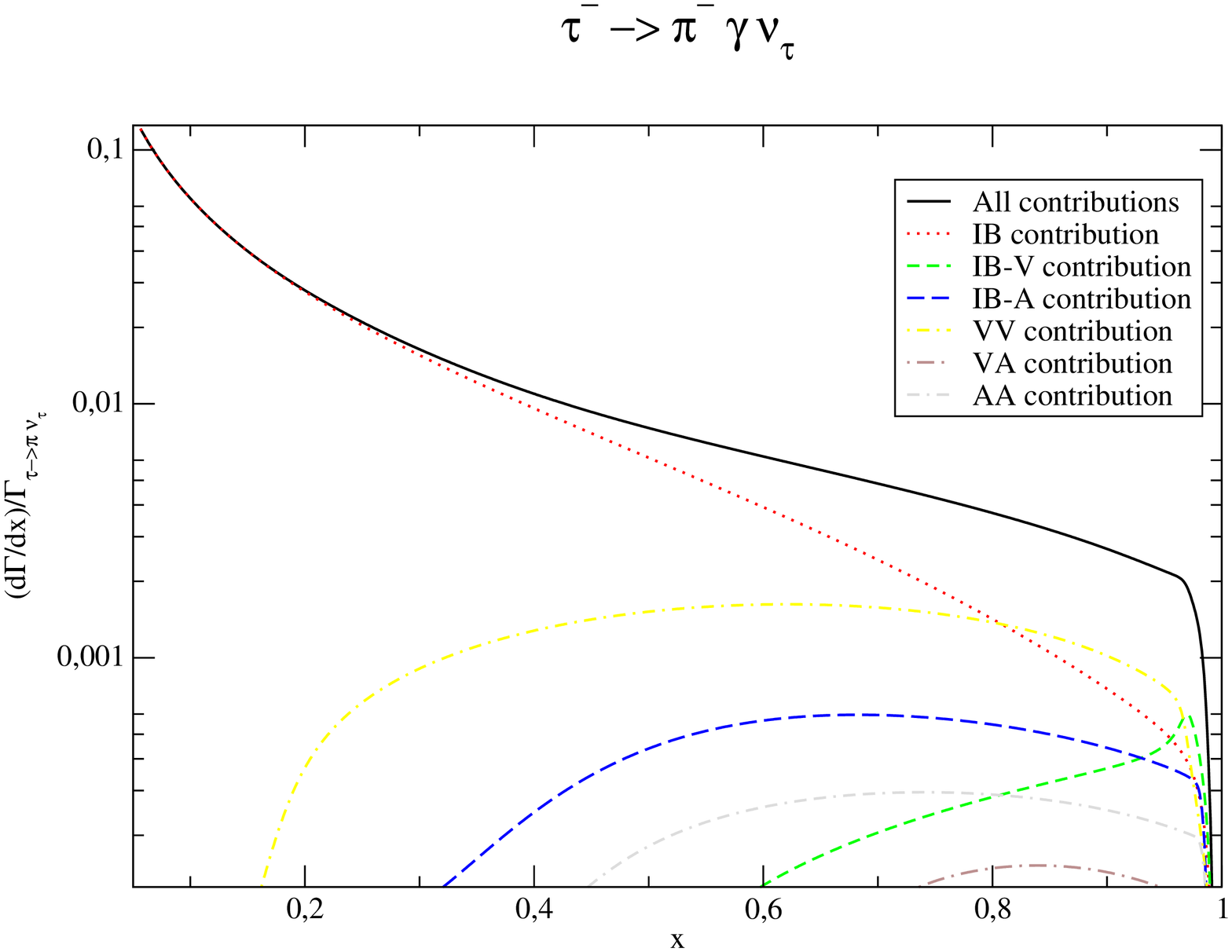}
\caption{Differential decay width of the process $\tau^-\rightarrow  \pi^- \gamma \nu_\tau$ including all contributions for a cut-off
 on the photon energy of $50$ MeV. \label{Allpigammalow}}
\end{center}
\end{figure}

\begin{figure}[h!]
\begin{center}
\includegraphics[scale=0.5,angle=-90]{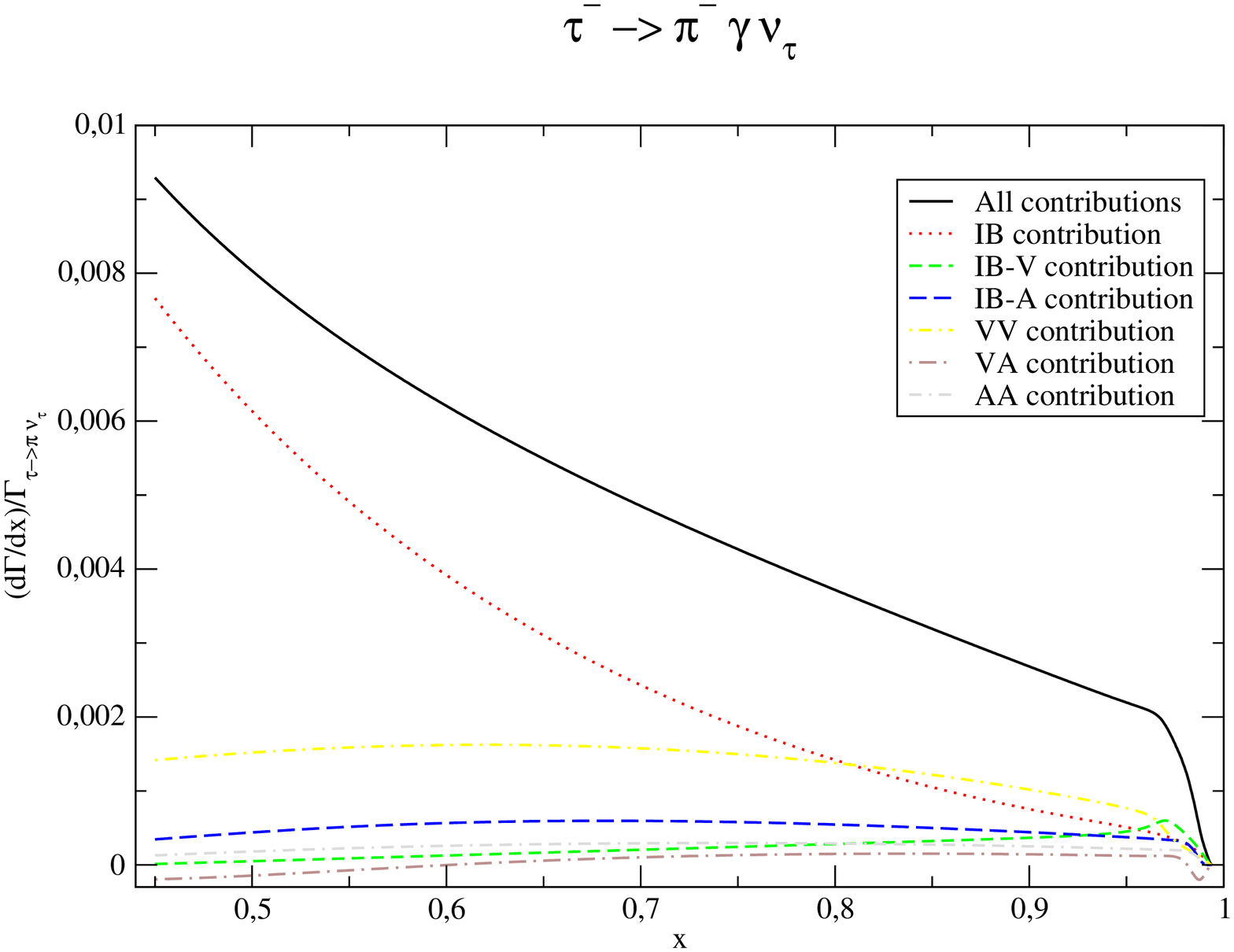}
\caption{Differential decay width of the process $\tau^-\rightarrow  \pi^- \gamma \nu_\tau$ including all contributions for a cut-off
 on the photon energy of $400$ MeV. \label{Allpigammahigh}}
\end{center}
\end{figure}

\begin{figure}[h!]
\begin{center}
\includegraphics[scale=0.5,angle=-90]{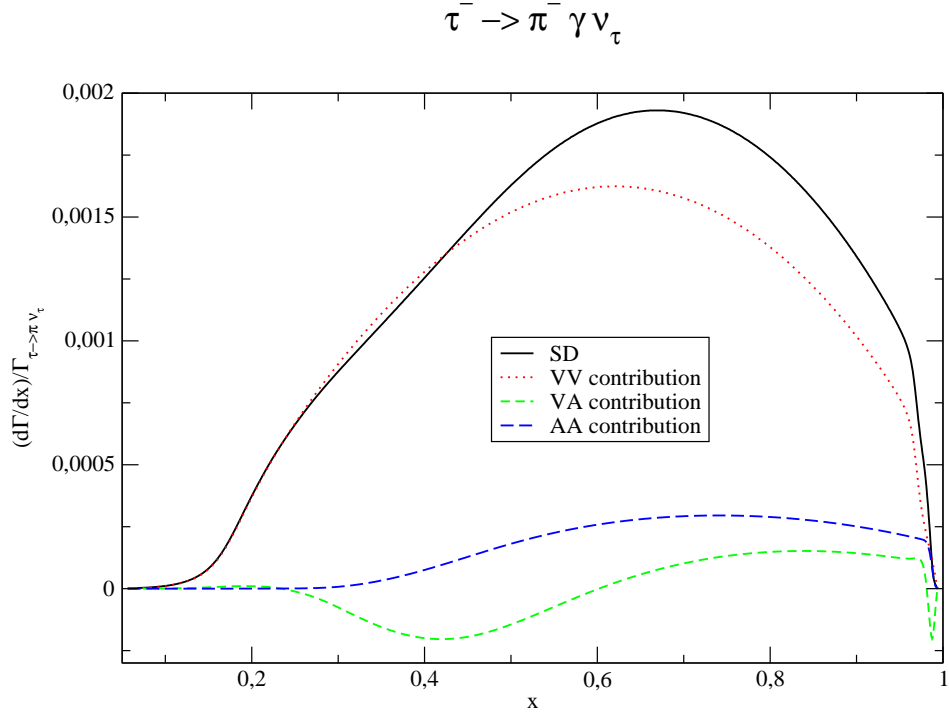}
\caption{Differential decay width of the process $\tau^-\rightarrow  \pi^- \gamma \nu_\tau$ including only the structure dependent contributions for a cut-off
 on the photon energy of $50$ MeV. \label{SDpigamma}}
\end{center}
\end{figure}

\begin{figure}[h!]
\begin{center}
\includegraphics[scale=0.5,angle=-90]{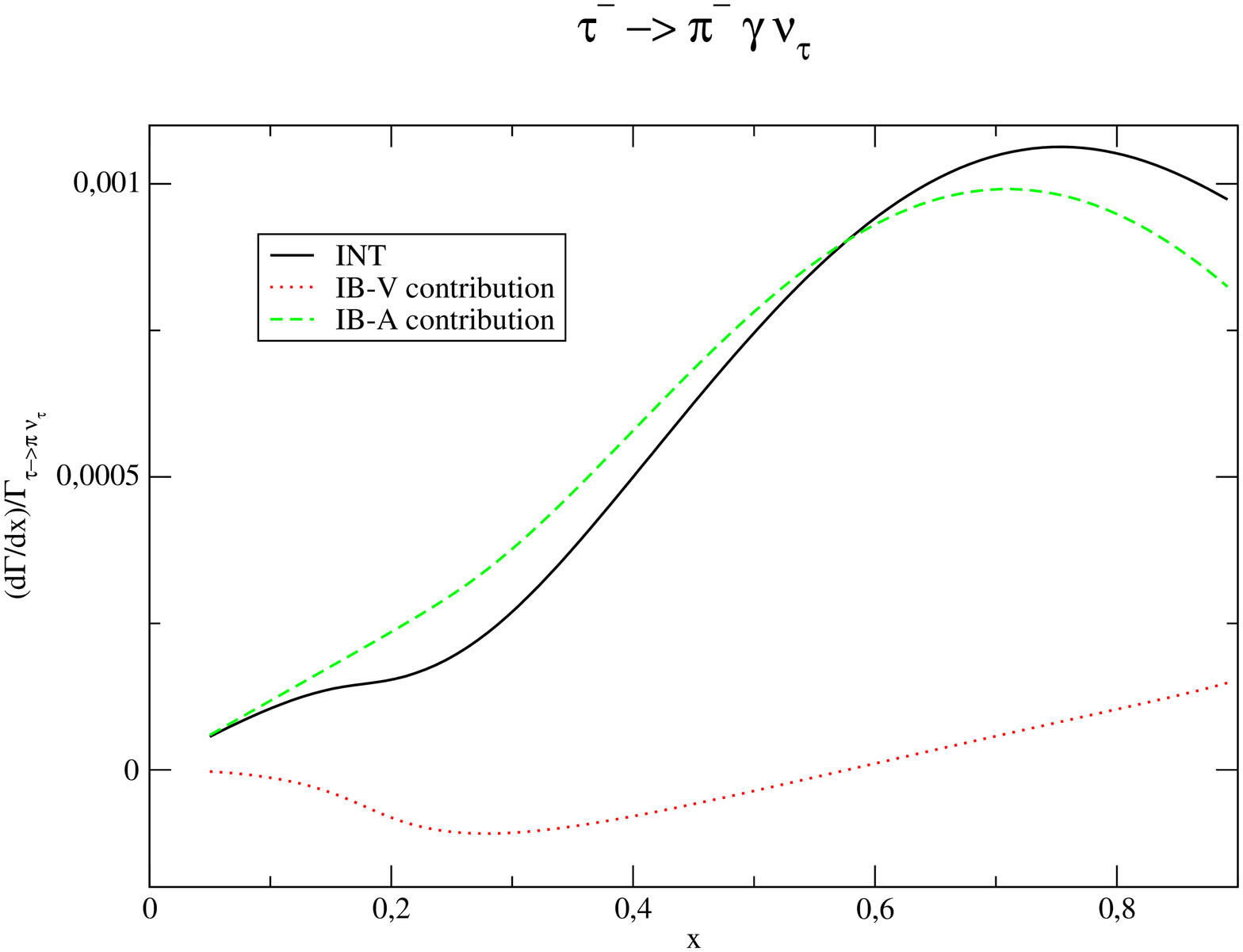}
\caption{Differential decay width of the process $\tau^-\rightarrow  \pi^- \gamma \nu_\tau$ including only the interference contributions for a cut-off
 on the photon energy of $50$ MeV. \label{INTpigamma}}
\end{center}
\end{figure}

\begin{figure}[h!]
\begin{center}
\vspace*{0.7cm}
\includegraphics[scale=0.5,angle=-90]{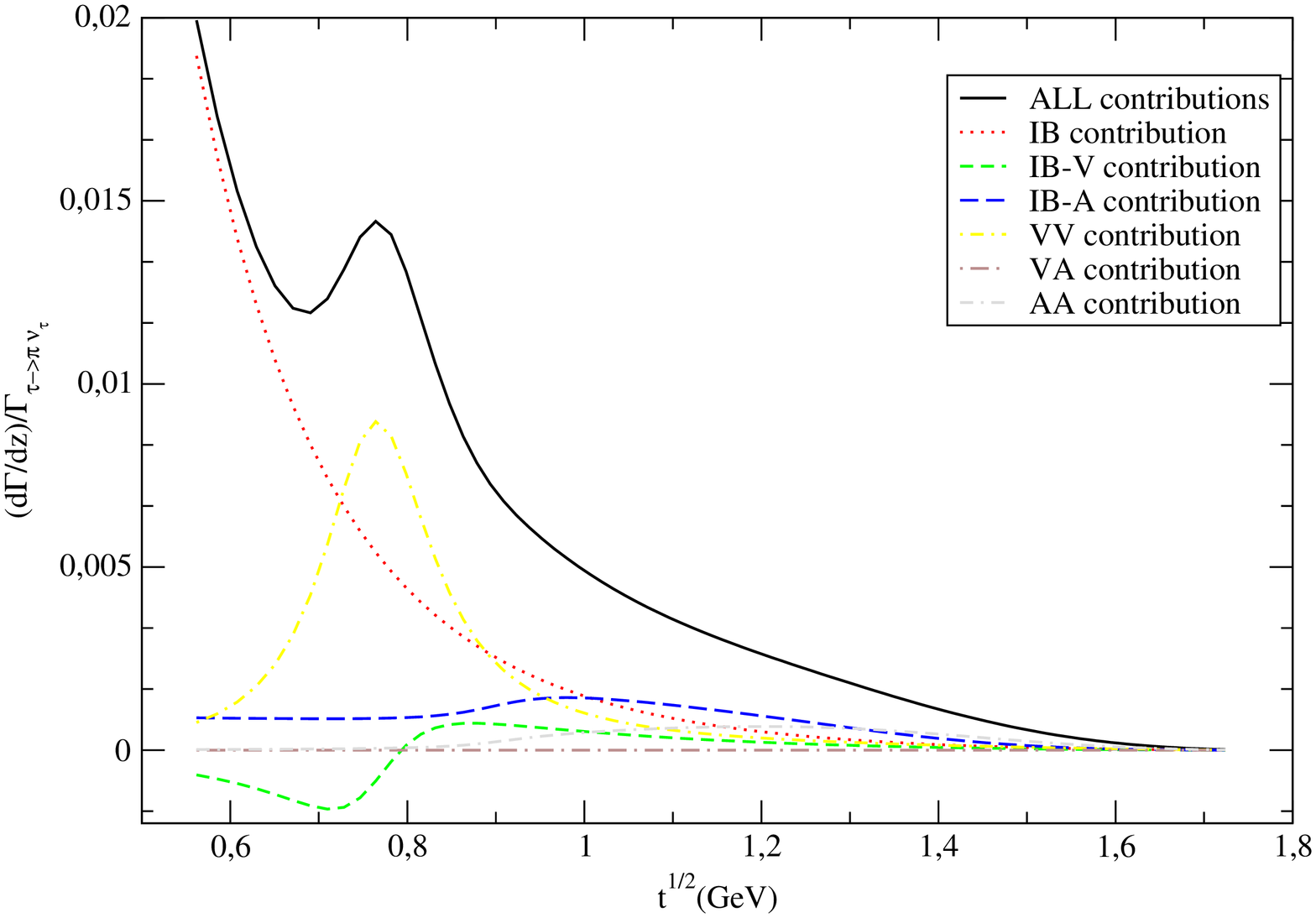}
\caption{Pion-photon invariant mass spectrum of the process $\tau^-\rightarrow  \pi^- \gamma \nu_\tau$ including all contributions. The $VA$ contribution 
vanishes identically as explained in the main text. \label{pigammatuno}}
\end{center}
\end{figure}

\begin{figure}[h!]
\begin{center}
\vspace*{0.7cm}
\includegraphics[scale=0.5,angle=-90]{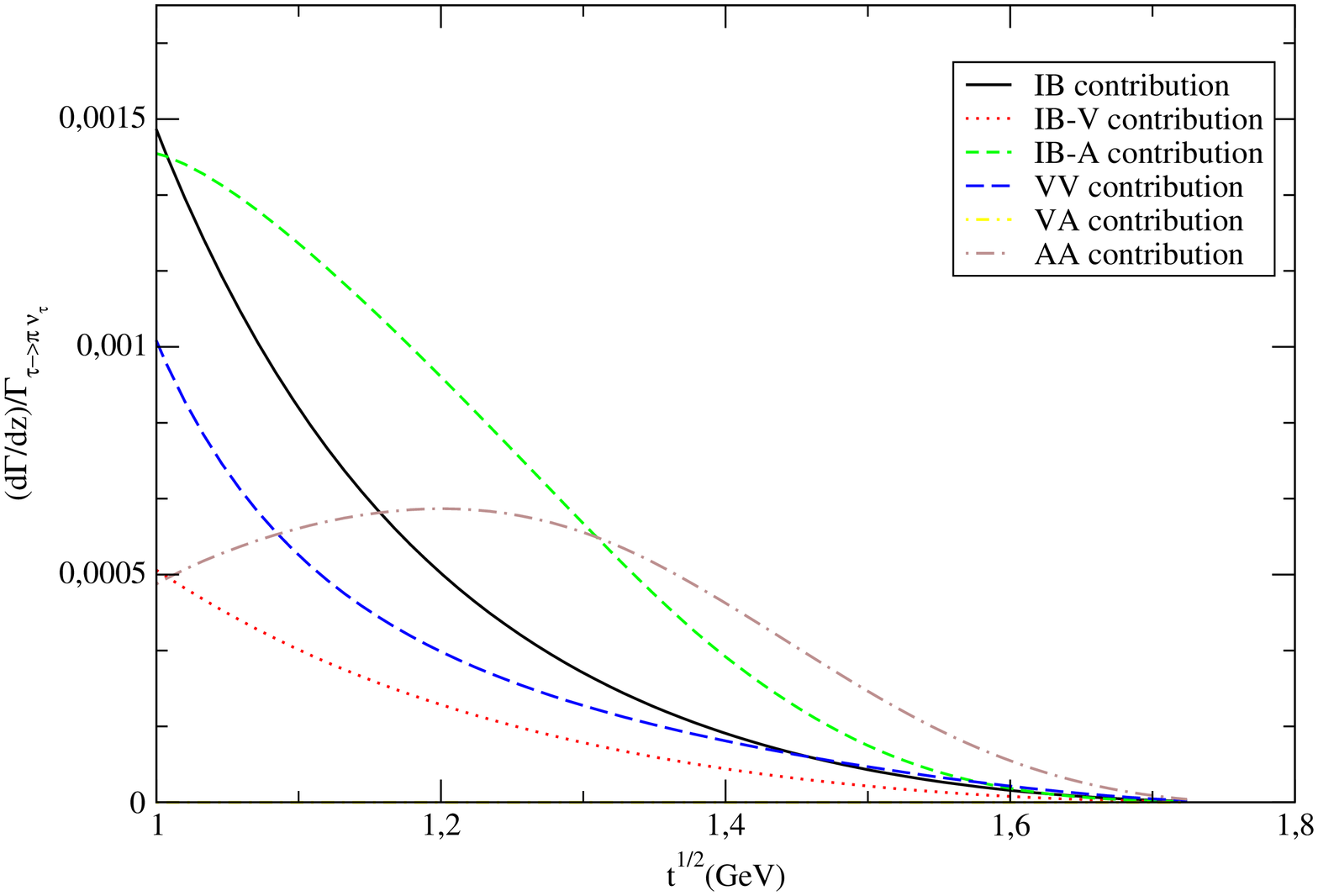}
\caption{Close-up of the pion-photon invariant mass spectrum  of the process $\tau^-\rightarrow  \pi^- \gamma \nu_\tau$ including all contributions for 
$\sqrt{t}\gtrsim1$ GeV. The $VA$ contribution 
vanishes identically as explained in the main text.\label{pigammatdos}}
\end{center}
\end{figure}

\begin{figure}[h!]
\begin{center}
\vspace*{0.7cm}
\includegraphics[scale=0.5,angle=-90]{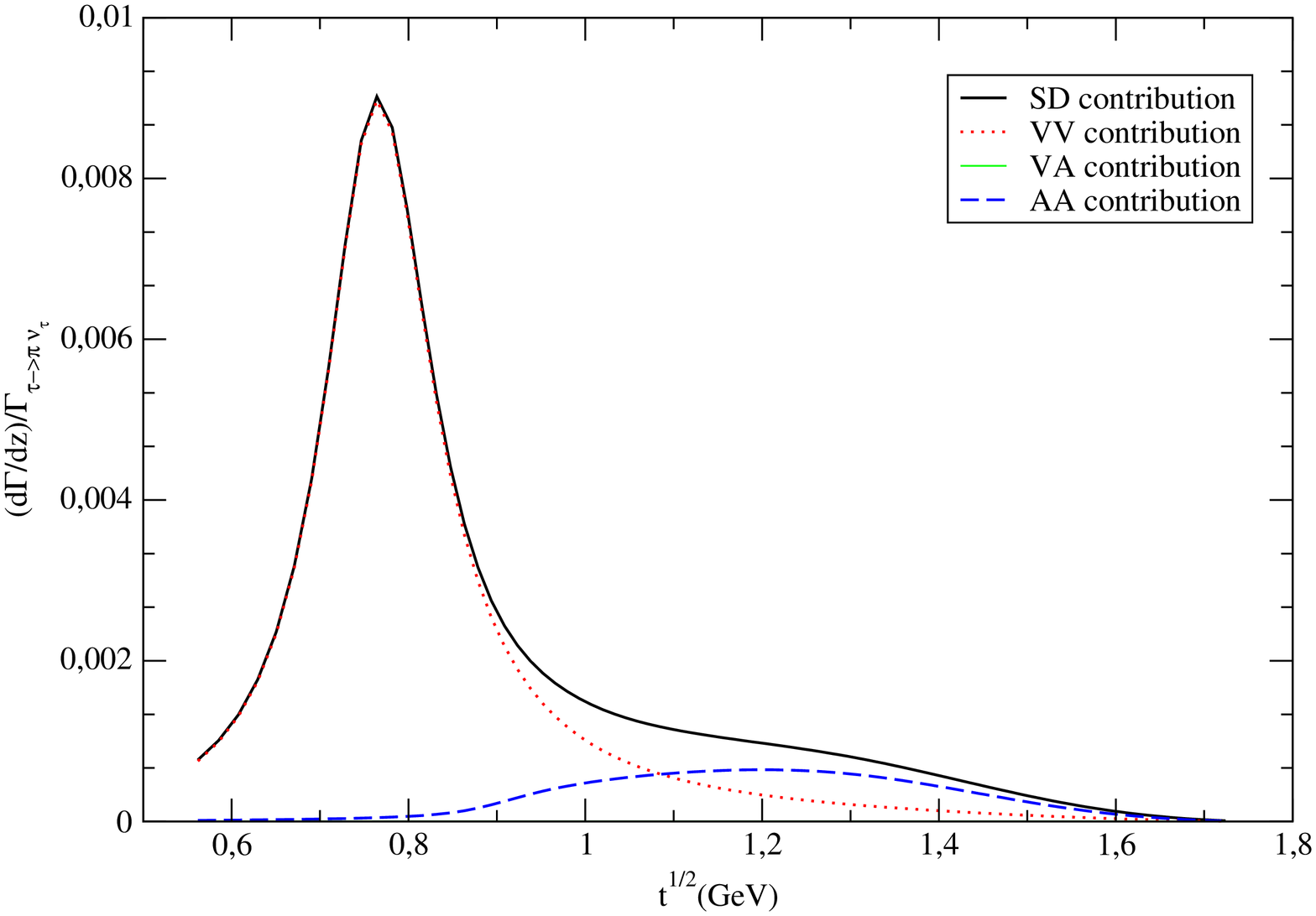}
\caption{Pion-photon invariant mass spectrum of the process $\tau^-\rightarrow  \pi^- \gamma \nu_\tau$ including only the $SD$ contributions. 
The $VA$ contribution 
vanishes identically as explained in the main text.\label{pigammattres}}
\end{center}
\end{figure}

\begin{figure}[h!]
\begin{center}
\vspace*{0.7cm}
\includegraphics[scale=0.5,angle=-90]{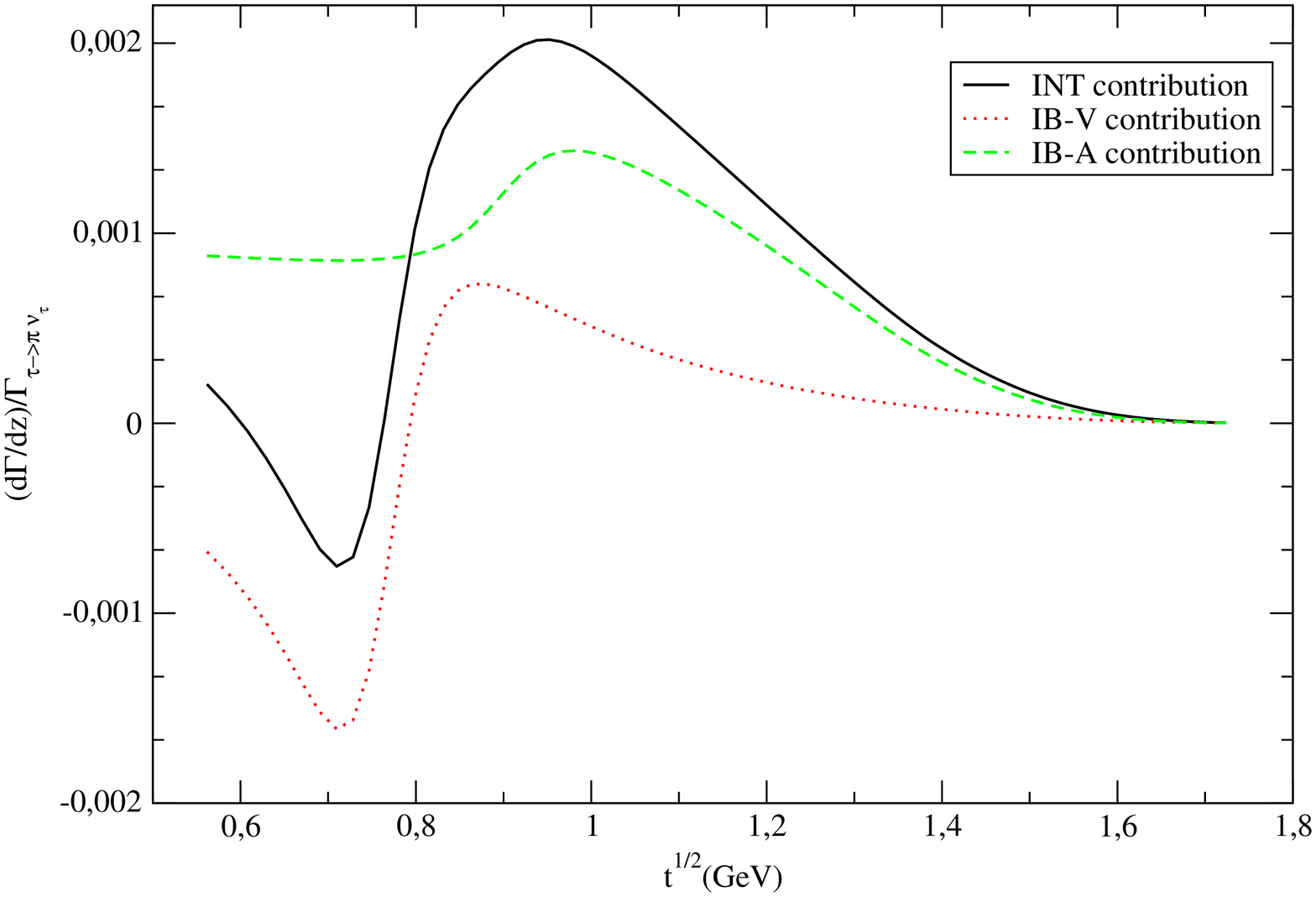}
\caption{Pion-photon invariant mass spectrum of the process $\tau^-\rightarrow  \pi^- \gamma \nu_\tau$ including only the interference contributions . \label{pigammatcuatro}}
\end{center}
\end{figure}

\hspace*{0.5cm}Next we include also the model-dependent contributions. Since in the Kaon channel there are uncertainties associated to the strange axial-vector 
off-shell width and to the mixing of the corresponding light and heavy states we will present first the pion channel where there is not any uncertainty of these 
type and everything is fixed in an analogous fashion to what discussed in the preceding chapters.\\
\hspace*{0.5cm}In Figs. \ref{Allpigammalow}-\ref{INTpigamma} the resulting photon spectrum in the process $\tau^-\rightarrow  \pi^- \gamma \nu_\tau$ is 
displayed. In Figure \ref{Allpigammalow}, all contributions are shown for a cutoff on the photon energy of $50$ MeV. For ''soft'' photons ($x_0\lesssim0$.$3$) the 
internal bremsstrahlung dominates completely. One should note that for very soft photons the multi-photon production rate becomes important, thus making that 
our $\cO(\alpha)$ results are not reliable too close to the $IR$ divergence $x\,=\,0$. We agree with the results in $DF$ papers, for the same value of $\alpha$ 
to the three significant figures shown in Ref. \cite{Decker:1993ut}.\\
\\
\hspace*{0.5cm}The spectrum is significantly enhanced by $SD$ contributions for hard photons ($x_0\gtrsim0$.$4$), as we can see in the close-up of Figure \ref{Allpigammalow}, 
in Figure \ref{Allpigammahigh}. In Figure \ref{SDpigamma} we show that the vector current contribution mediated by vector resonances dominates the $SD$ part, while 
in Figure \ref{INTpigamma} we plot the interference term between bremsstrahlung and $SD$ part. If we compare the predicted curves to those in Ref. \cite{Decker:1993ut} 
we see that the qualitative behaviour is similar: the $IB$ contribution dominates up to $x\sim0$.$75$. For larger photon energies, the $SD$ -that is predominantly 
due to the $VV$ contribution- overcomes it. We confirm the peak and shoulder structure shown at $x\sim1$ in the interference contribution, that is essentially 
due to $IB-V$ term, and also in the $VA$ term, that is in any case tiny.\\
\\
\\
\\
\hspace*{0.5cm}While the integration over the $IB$ needs an $IR$ cut-off, the $SD$ part does not. We have performed the integration over the complete phase 
space, yielding (all contributions to the partial decay width are given in units of the non-radiative decay, here and in what follows):
\begin{equation} \label{Gamma_SD_pi}
 \Gamma_{VV}=0\mathrm{.}99\cdot10^{-3}\;,\quad\Gamma_{VA}=1\mathrm{.}45\cdot10^{-9}\sim0\;,\quad\Gamma_{AA}=0\mathrm{.}15\cdot10^{-3}\,\Rightarrow \Gamma_{SD}=1\mathrm{.}14\cdot10^{-3}\,.
\end{equation}
 Our number for $\Gamma_{SD}$ lies between the results for the monopole and tripole parametrizations in Ref. \cite{Decker:1993ut}. However, they get a smaller(larger) 
$VV$($AA$) contribution than we do by $\sim20\%$($\sim200\%$). This last discrepancy is due to the off-shell a$_1$ width they use. In fact, if we use the constant width 
approximation we get a number very close to theirs for the $AA$ contribution. With our understanding of the a$_1$ width in the $\tau\to3\pi\nu_\tau$ observables, we 
can say that their (relatively) high $AA$ contribution is an artifact of the ad-hoc off-shell width used. Since the numerical difference in varied vector off-shell widths 
is not that high, the numbers for $VV$ are closer.\\
\hspace*{0.5cm}The numbers in Eq.(\ref{Gamma_SD_pi}) are translated into the following branching ratios
\begin{equation} \label{BR_SD_pi}
 \mathrm{BR}_{VV}\left(\tau\to\pi\gamma\nu_\tau\right)=1\mathrm{.}05\cdot10^{-4}\;,\quad\mathrm{BR}_{AA}\left(\tau\to\pi\gamma\nu_\tau\right)=0\mathrm{.}15\cdot10^{-4}\,.
\end{equation}
\hspace*{0.5cm}We can also compare the $VV$ value with the narrow width estimate: taking into account the lowest lying resonance $\rho$ we get
\begin{eqnarray} \label{est_VV_pi}
 \mathrm{BR}_{VV}\left(\tau\to\pi\gamma\nu_\tau\right)& \sim & BR(\tau\to\rho\nu_\tau)\times BR(\rho\to\pi\gamma)\sim BR(\tau\to\pi^-\pi^0\nu_\tau)BR(\rho\to\pi\gamma)\nonumber\\
& \sim & 25\mathrm{.}52\%\times4\mathrm{.}5\cdot10^{-4}=1\mathrm{.}15\cdot10^{-4}\,,
\end{eqnarray}
which is quite a good approximation.\\
In Table \ref{Tabledifferentcontributionspi} we give the display for two values of the photon energy cut-off how the different parts contribute to 
the total rate. For a low-energy cut-off the most of the rate comes from $IB$ while for a higher-energy one the $SD$ parts (and particularly the $VV$ contribution) 
gains importance. While the $VA$ contribution is always negligible, the $IB-V$, $IB-A$ and the $SD$ parts $VV$ and $AA$ have some relevance for a higher-energy 
cut-off.\\
\begin{table}[h!]  
\begin{center}
\renewcommand{\arraystretch}{1.2}
\begin{tabular}{|c|c|c|}
\hline
&$x_0=0$.$0565$&$x_0=0$.$45$\\
\hline
$IB$&$13$.$09\cdot10^{-3}$&$1$.$48\cdot10^{-3}$\\
$IB-V$&$0$.$02\cdot10^{-3}$&$0$.$04\cdot10^{-3}$\\
$IB-A$&$0$.$34\cdot10^{-3}$&$0$.$29\cdot10^{-3}$\\
$VV$&$0$.$99\cdot10^{-3}$&$0$.$73\cdot10^{-3}$\\
$VA$&$\sim 0$&$0$.$02\cdot10^{-3}$\\
$AA$&$0$.$15\cdot10^{-3}$&$0$.$14\cdot10^{-3}$\\
\hline
$ALL$&$14$.$59\cdot10^{-3}$&$2$.$70\cdot10^{-3}$\\
\hline
\end{tabular}
\caption{\small{Contribution of the different parts to the total rate, using two different cut-offs for the photon energy: $E_\gamma=50$ MeV ($x_0=0$.$0565$)
 and $E_\gamma=400$ MeV ($x_0=0$.$45$).}}
\label{Tabledifferentcontributionspi} 
\end{center}
\end{table}
\\
\hspace*{0.5cm}In Figs. \ref{pigammatuno}-\ref{pigammatcuatro} we show the pion-photon invariant spectrum. We find a much better separation between the $IB$ 
and $SD$ contributions as compared with the photon spectrum in the previous Figs. \ref{Allpigammalow} to \ref{INTpigamma}. Then, the pion-photon spectrum 
is better suited to study the $SD$ effects. In this case, the $VA$ is identically zero, since this interference vanishes in the invariant mass spectrum after 
integration over the other kinematic variable. Of course, in the $VV$ spectrum we see the shape of the $\rho$ contribution neatly, as one can see in Figure 
\ref{pigammattres} where, on the contrary, the a$_1$ exchange in $AA$ has a softer and broader effect. The $IB-SD$ radiation near the $a_1$ is 
dominated by $IB-A$, which gives the positive contribution to the decay rate. While near the energy region of the $\rho$ resonance, we find the $IB-SD$ 
contribution to be negative as driven by $IB-V$ there. In the whole spectrum only the $\rho$ resonance manifests as a peak and one can barely see the signal 
of the $a_1$, mainly due to its broad width and to the counter effect of interferences.\\
\\
\\
\\
\\
\subsection{Results including resonance contributions in the $K$ channel}
\hspace*{0.5cm}Next we turn to the $\tau^-\rightarrow  K^- \gamma \nu_\tau$ channel. In this case, there are several sources of uncertainty that make our 
prediction less controlled than in the $\tau^-\rightarrow  \pi^- \gamma \nu_\tau$ case. We comment them in turn.\\
\hspace*{0.5cm}Concerning the vector form factor contribution, there is no uncertainty associated to the vector resonances off-shell widths, that are implemented 
as done in previous applications and described in Appendix C. It turns out that the $SD$ part is extremely sensitive to $c_4$. We have observed that 
the $VV$ contribution is much larger (up to one order of magnitude, even for a low-energy cut-off) than the $IB$ one for $c_4\sim-0$.$07$, 
a feature that is unexpected. In this case, one would also see a prominent bump in the spectrum, contrary to the typical monotonous fall driven by the $IB$ term. 
For smaller values of $|c_4|$ (which are suggested by the comparison to $Belle$ data on $\tau\to K K\pi \nu_\tau$ decays) this bump reduces its magnitude 
and finally disappears. One should also not forget that the addition of a second multiplet of resonances may vary this conclusion.\\
\hspace*{0.5cm}The uncertainty in the axial-vector form factors is two-folded: on one side there is a broad band of allowed values for $\theta_A$, as discussed 
at the beginning of this section. On the other hand, since we have not performed the analyses of the decay $\tau\to K\pi\pi\nu_\tau$ modes yet, we do not have an 
off-shell width derived from a Lagrangian for the $K_{1A}$ resonances. In the $\tau\to3\pi\nu_\tau$ decays, $\Gamma_{\mathrm{a}_1}$ has the starring role. Since the 
$K_{1A}$ meson widths are much smaller ($90\pm20$ MeV and $174\pm13$ MeV, for the $K_{1}(1270)$ and $K_{1}(1400)$, respectively) and they are hardly close to 
the on-shell condition, the rigorous description of the width is not an unavoidable ingredient for a reasonable estimate.\\
\hspace*{0.5cm}With respect to the two uncertainties just commented, we have checked that the branching ratio contribution by $AA$ (that is subdominant) is 
$\sim20\%$ higher for the $|\theta_A|\sim37^\circ$ solution. In this case, the corresponding $AA$ differential distribution peaks at a slightly larger $x$, 
and the curve is lower in the $0$.$4\leftrightarrow 0$.$55$ region. In any case, different choices of $|\theta_A|$ can barely influence the final conclusions, as
 it is illustrated in Table \ref{Tabledifferentcontributionsk}.\\
 
\begin{table}[h!]
\begin{center}
\renewcommand{\arraystretch}{1.2}
\begin{tabular}{|c|c|c|c|c|}
\hline
&$x_0=0$.$0565$ & $x_0=0$.$0565$  &$x_0=0$.$45$  &$x_0=0$.$45$  \\
&$c_4=-0$.$07$ & $c_4=0$ & $c_4=-0$.$07$ & $c_4=0$ \\
& $|\theta_A|=58^\circ (37^\circ)$& $|\theta_A|=58^\circ (37^\circ)$& $|\theta_A|=58^\circ (37^\circ)$ & $|\theta_A|=58^\circ (37^\circ)$ \\
\hline
$IB$&$3$.$64\cdot10^{-3}$&$3$.$64\cdot10^{-3}$&  $0$.$31\cdot10^{-3}$  &$0$.$31\cdot10^{-3}$\\
$IB-V$&$0$.$69\cdot10^{-3}$&$0$.$10\cdot10^{-3}$  &$0$.$83\cdot10^{-3}$  &$0$.$12\cdot10^{-3}$\\
$IB-A$&$0$.$22(0$.$25)\cdot10^{-3}$&$0$.$22(0$.$25)\cdot10^{-3}$ &$0$.$15(0$.$18)\cdot10^{-3}$   & $0$.$15(0$.$18)\cdot10^{-3}$\\
$VV$&$58$.$55\cdot10^{-3}$  &$1$.$30\cdot10^{-3}$  &$29$.$04\cdot10^{-3}$  &$0$.$66\cdot10^{-3}$\\
$VA$&$ \sim 0 (\sim 0) $&$ \sim 0 (\sim 0) $&$0$.$09(0$.$09)\cdot10^{-3}$&$0$.$01(0$.$01)\cdot10^{-3}$\\
$AA$&$0$.$13(0$.$16)\cdot10^{-3}$&$0$.$13(0$.$16)\cdot10^{-3}$ &$0$.$12(0$.$15)\cdot10^{-3}$  &$0$.$12(0$.$15)\cdot10^{-3}$\\
\hline
$ALL$&$63$.$23(63$.$29)\cdot10^{-3}$&$5$.$39(5$.$45)\cdot10^{-3}$&$30$.$54(30$.$60)\cdot10^{-3}$&$1$.$37(1$.$43)\cdot10^{-3}$\\
\hline
\end{tabular}
\caption{\small{Contribution of the different parts to the total rate in the decay $\tau^-\to K^-\gamma\nu_\tau$ (in unit of $\Gamma_{\tau \to K \nu}$), 
using two different cut-offs for the photon energy: $E_\gamma=50$ MeV ($x_0=0$.$0565$) and $E_\gamma=400$ MeV ($x_0=0$.$45$) and also different values of the 
resonance couplings. The numbers inside the parentheses denote the corresponding results with $|\theta_A|= 37^\circ$, while the other numbers are obtained 
with $|\theta_A|= 58^\circ$.}}
\label{Tabledifferentcontributionsk}
\end{center}
\end{table}

\hspace*{0.5cm}For the $K_{1A}$ off-shell widths we will follow Ref.~\cite{Guo:2008sh} and use
\begin{equation}
 \Gamma_{K_{1A}}(t)\,=\,\Gamma_{K_{1A}}(M_{K_{1A}}^2)\,\frac{M_{K_{1A}}^2}{t}\frac{\sigma^3_{M_{K^*} m_\pi}(t)+\sigma^3_{M_\rho m_K}(t)}{\sigma^3_{M_{K^*} m_\pi}(M_{K_{1A}}^2)+\sigma^3_{M_\rho m_K}(M_{K_{1A}}^2)}\,,
\end{equation}
where
\begin{equation}
 \sigma_{P Q}(x)\,=\,\frac{1}{x}\,\sqrt{\left(x-(P+Q)^2\right)\left(x-(P-Q)^2\right)}\theta\left[ x-(P-Q)^2\right]\,.
\end{equation}
\hspace*{0.5cm}Considering all the sources of uncertainty commented, we will content ourselves with giving our predictions for the two limiting cases of 
$c_4=-0$.$07$ and $c_4=0$. We present the analogous plots to those we discussed in the $\tau^-\rightarrow  \pi^- \gamma \nu_\tau$ channel for both $c_4$ values.\\
\begin{figure}[h!]
\begin{center}
\vspace*{0.7cm}
\centerline{\includegraphics[scale=0.35,angle=-90]{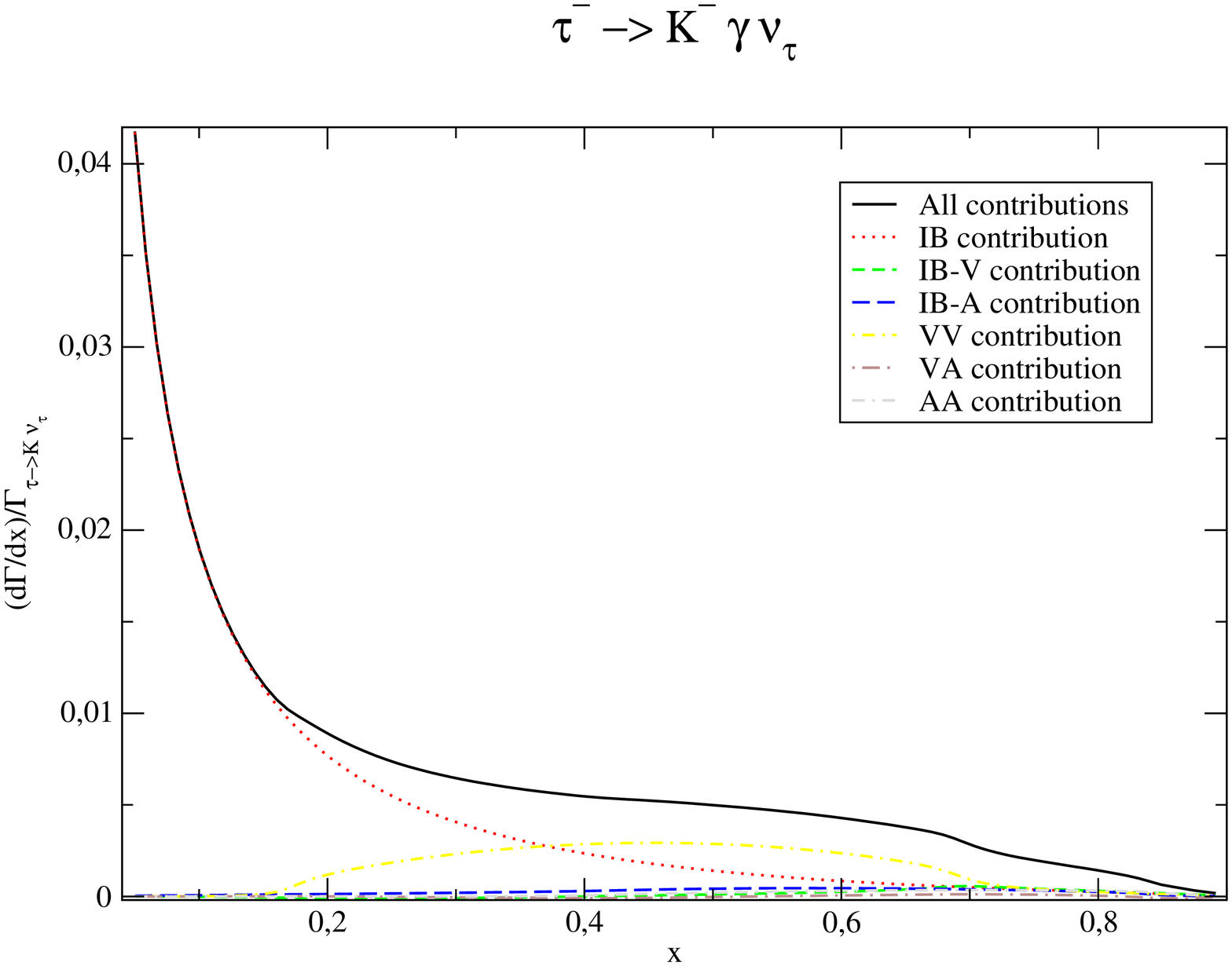}\quad \includegraphics[scale=0.35,angle=-90]{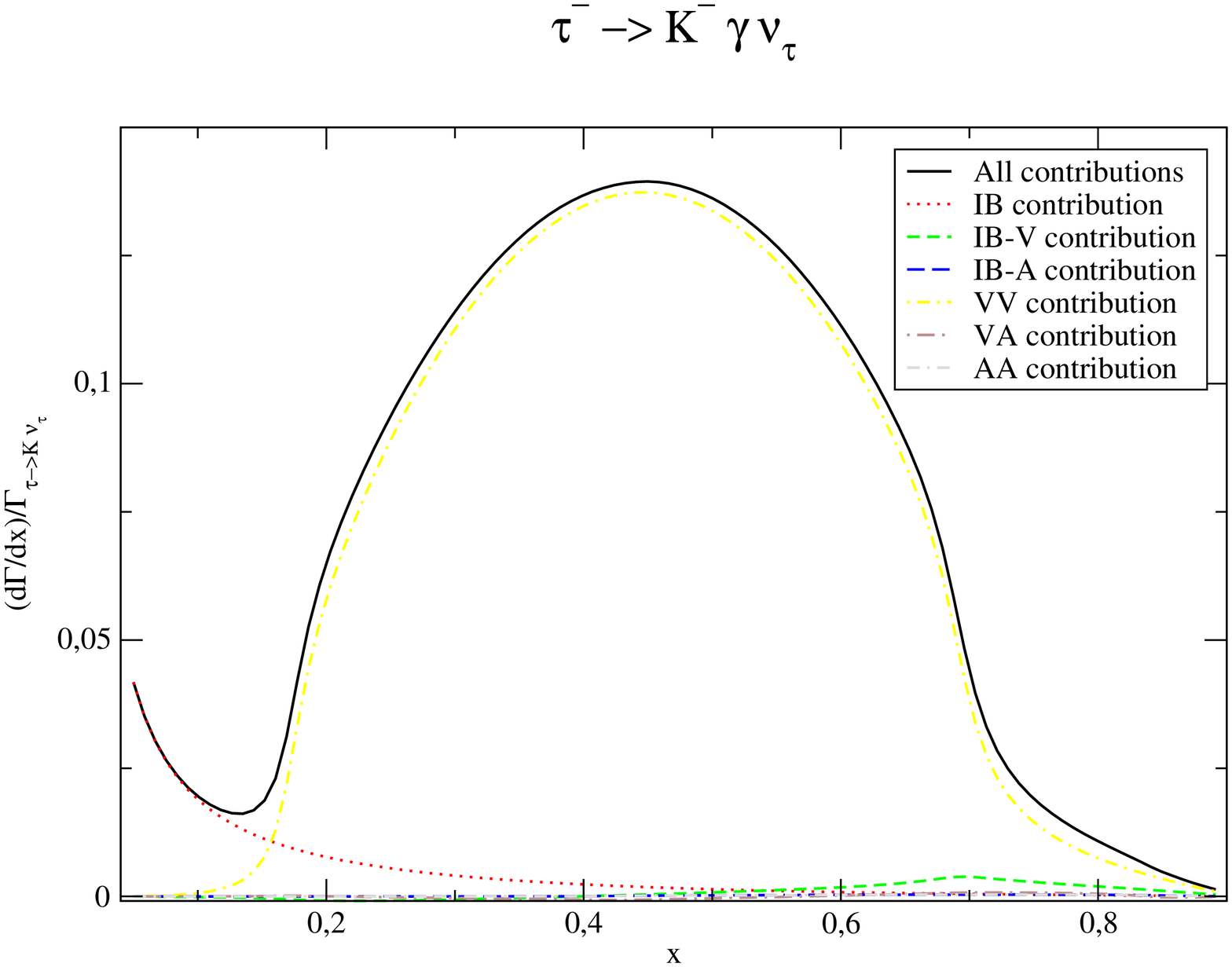}}
\caption{Differential decay width of the process $\tau^-\rightarrow  K^- \gamma \nu_\tau$ including all contributions for a cut-off
 on the photon energy of $50$ MeV and $c_4=0$ (left pane) and $c_4=-0$.$07$ (right pane). \label{AllKgammalow}}
\end{center}
\end{figure}
\\
\begin{figure}[h!]
\begin{center}
\vspace*{0.7cm}
\centerline{\includegraphics[scale=0.3,angle=-90]{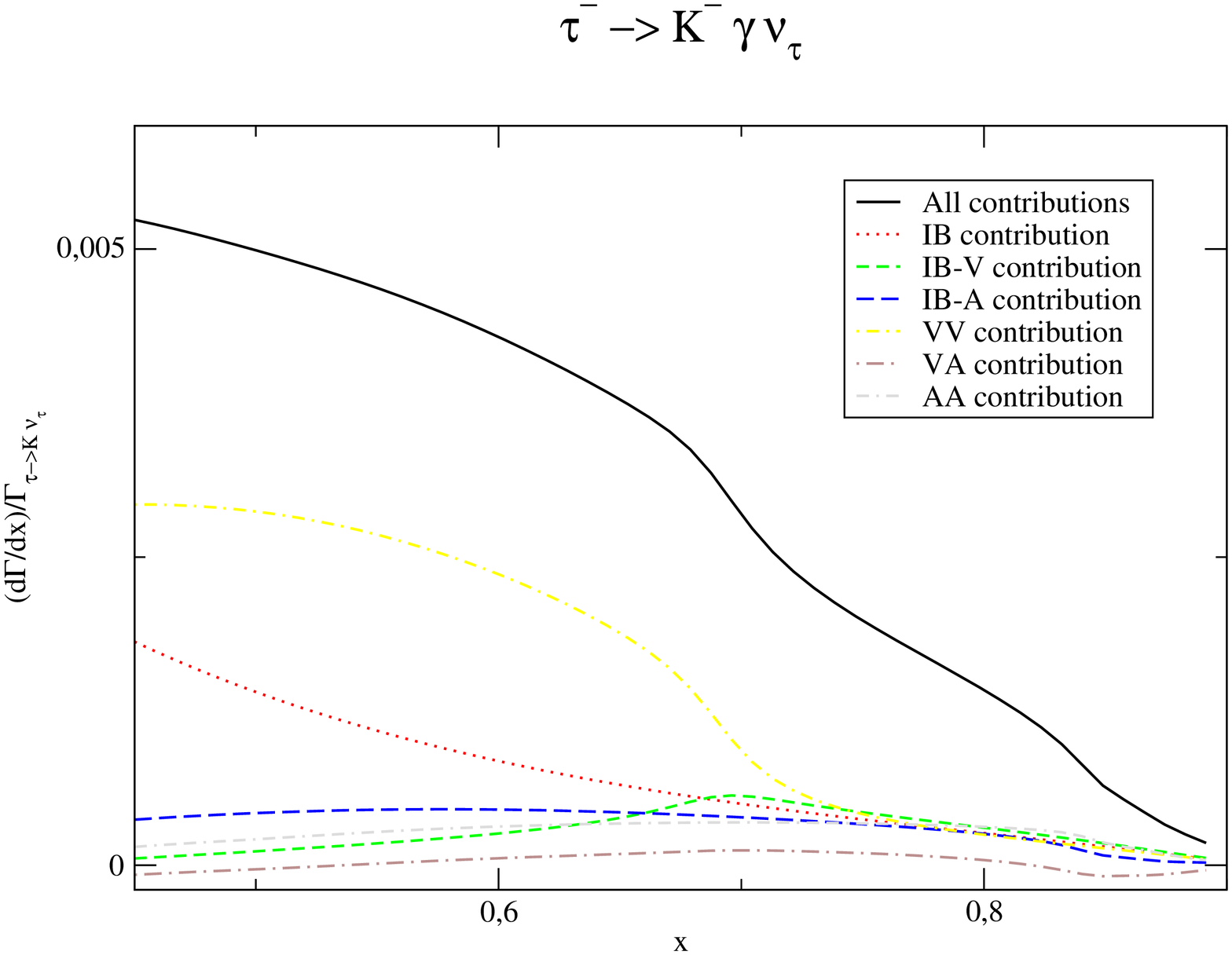}\quad \includegraphics[scale=0.3,angle=-90]{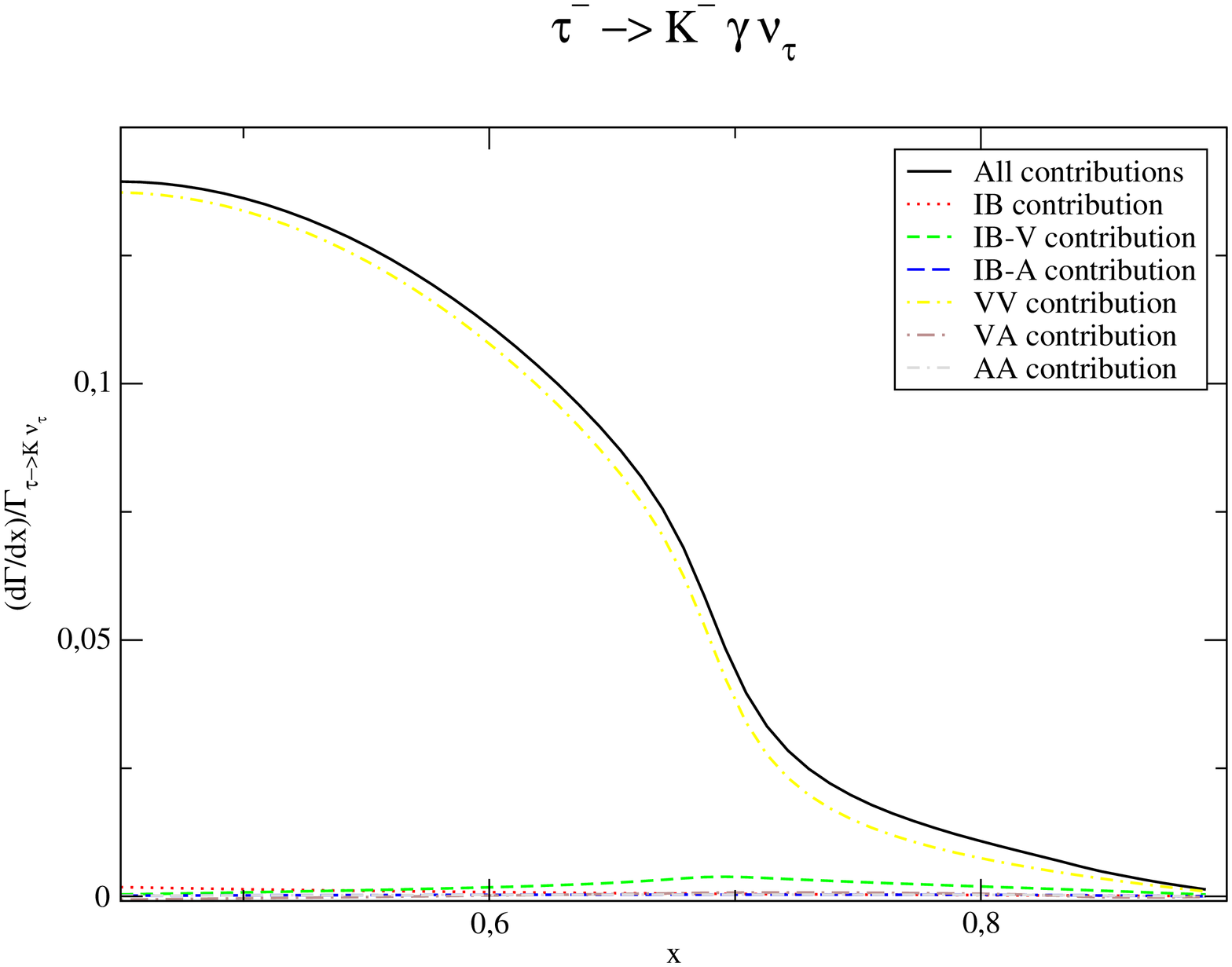}}
\caption{Differential decay width of the process $\tau^-\rightarrow  K^- \gamma \nu_\tau$ including all contributions for a cut-off
 on the photon energy of $400$ MeV and $c_4=0$ (left pane) and $c_4=-0$.$07$ (right pane). \label{AllKgammahigh}}
\end{center}
\end{figure}
\\
\begin{figure}[h!]
\begin{center}
\vspace*{0.7cm}
\centerline{\includegraphics[scale=0.35,angle=-90]{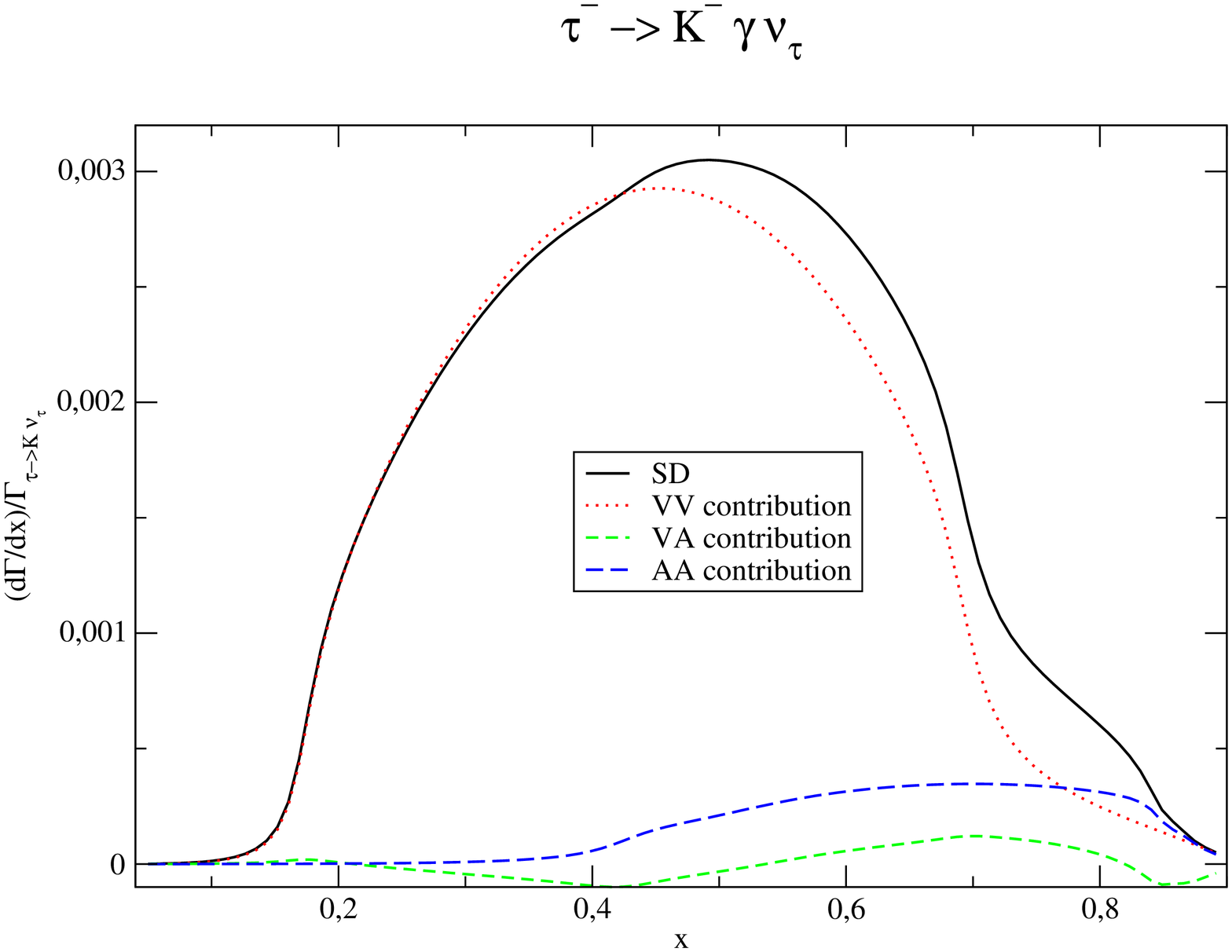}\quad \includegraphics[scale=0.35,angle=-90]{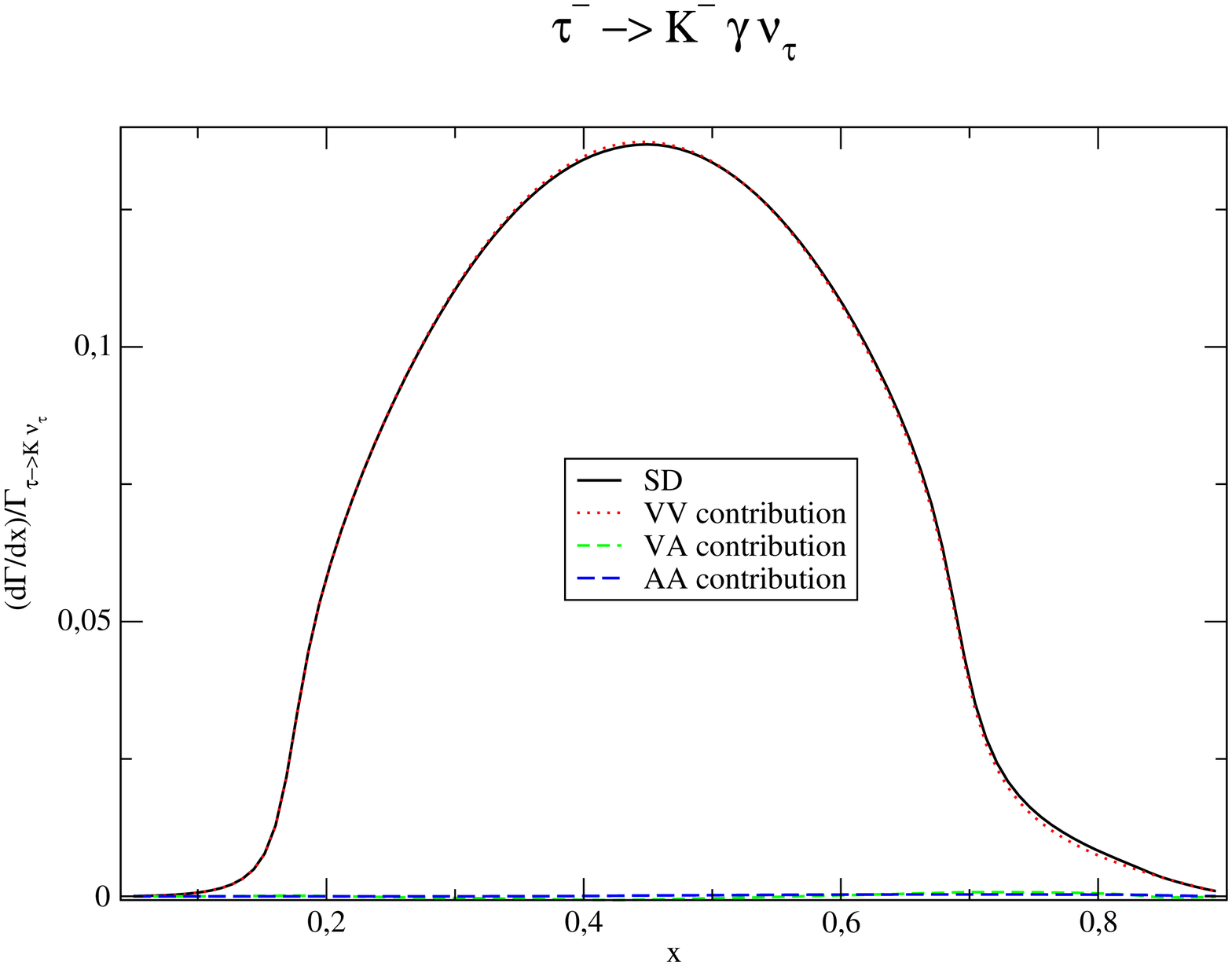}}
\caption{Differential decay width of the process $\tau^-\rightarrow  K^- \gamma \nu_\tau$ including only the structure dependent contributions for a cut-off
 on the photon energy of $50$ MeV and $c_4=0$ (left pane) and $c_4=-0$.$07$ (right pane). \label{SDKgamma}}
\end{center}
\end{figure}
\\
\begin{figure}[h!]
\begin{center}
\vspace*{0.7cm}
\centerline{\includegraphics[scale=0.35,angle=-90]{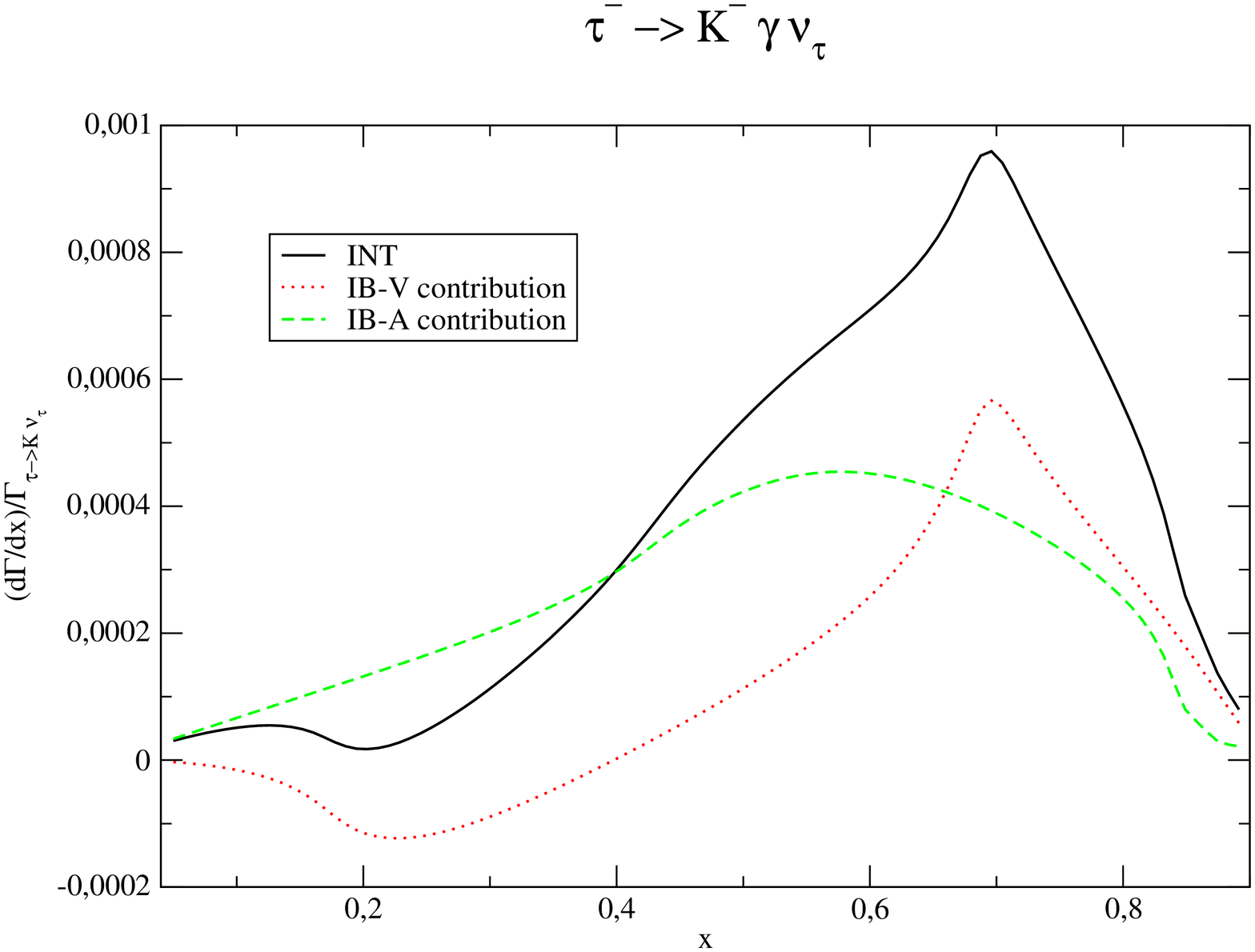}\quad \includegraphics[scale=0.35,angle=-90]{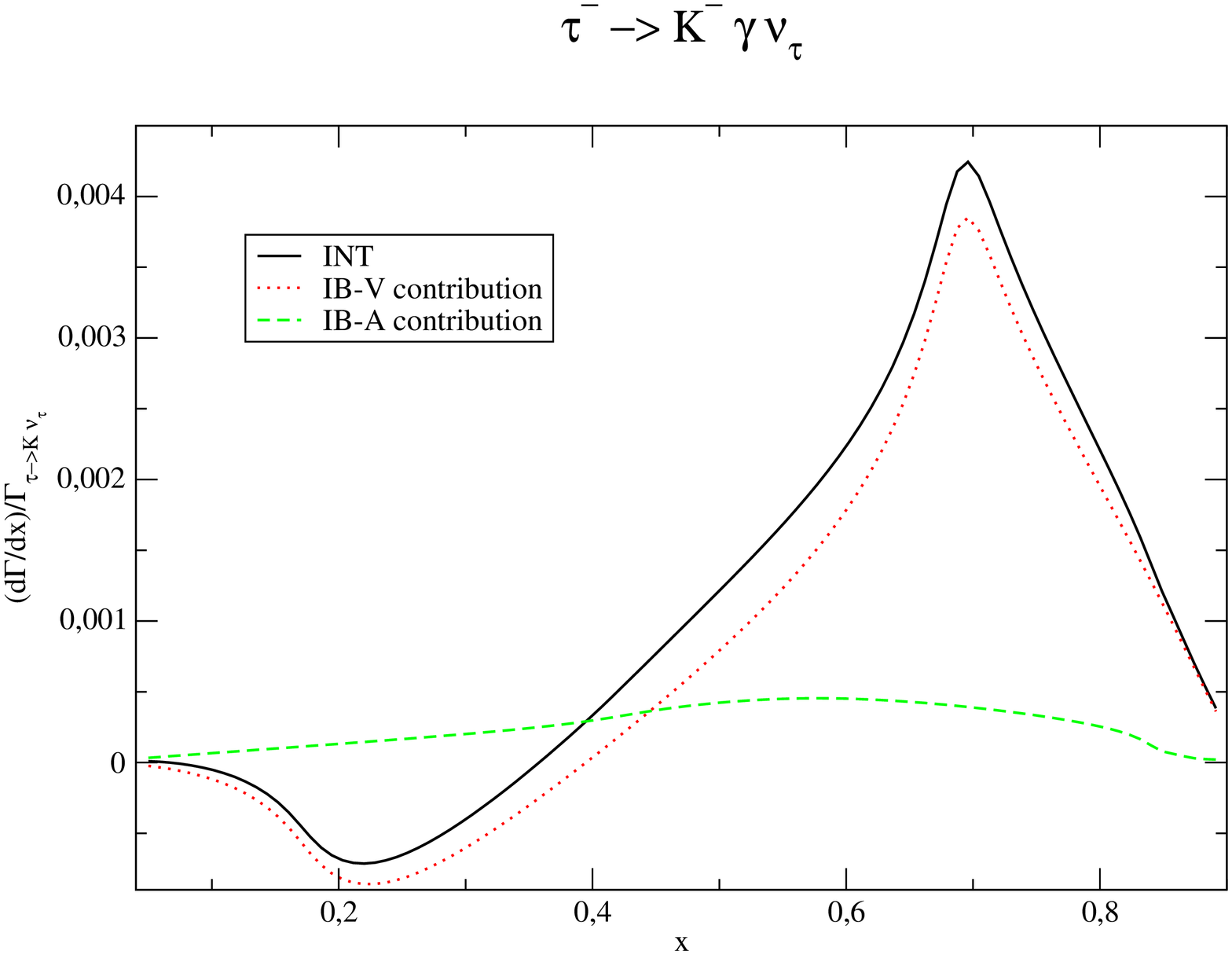}}
\caption{Differential decay width of the process $\tau^-\rightarrow  K^- \gamma \nu_\tau$ including only the interference contributions for a cut-off
 on the photon energy of $50$ MeV and $c_4=0$ (left pane) and $c_4=-0$.$07$ (right pane). \label{INTKgamma}}
\end{center}
\end{figure}
\\
\begin{figure}[h!]
\begin{center}
\vspace*{0.7cm}
\centerline{\includegraphics[scale=0.35,angle=-90]{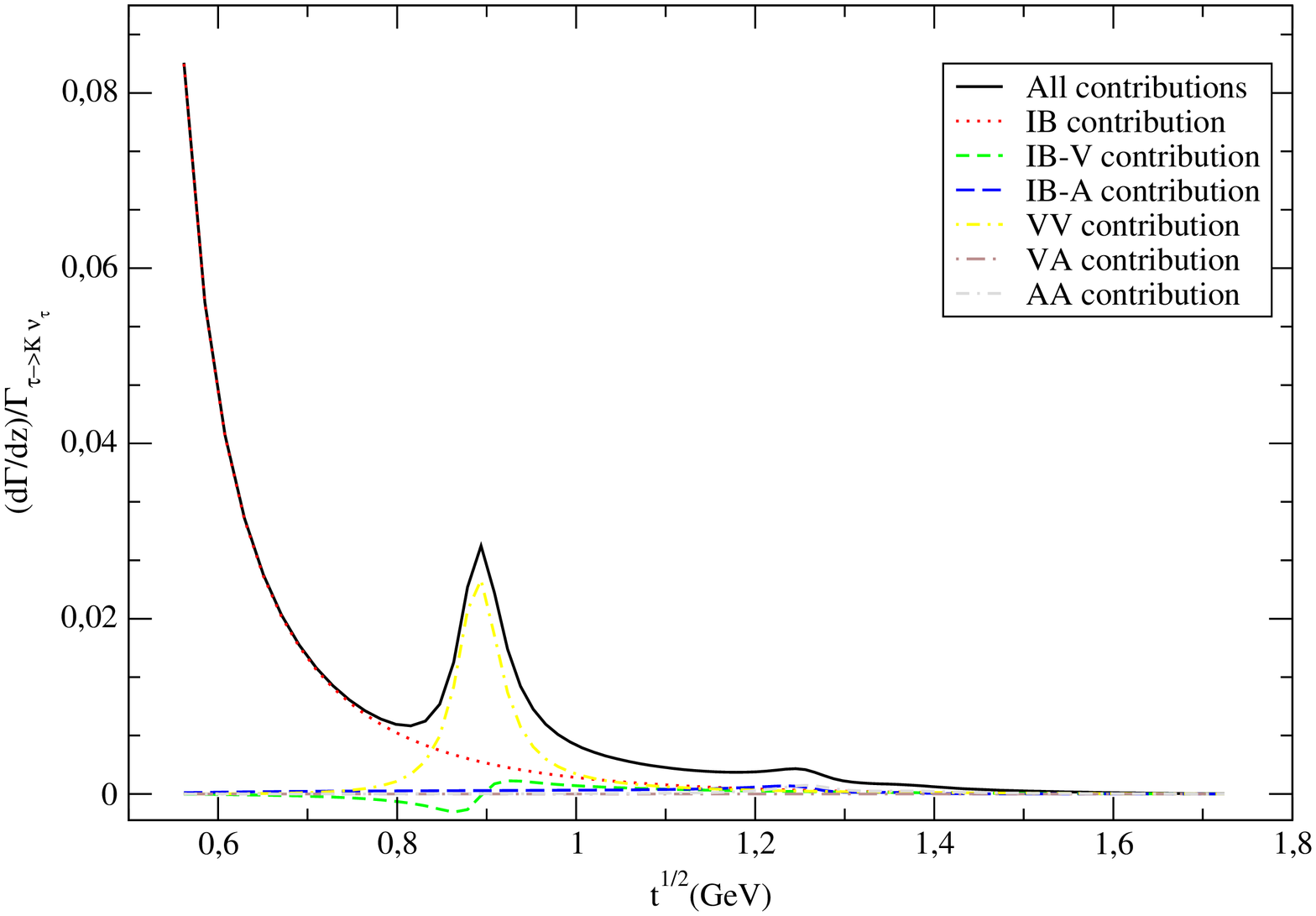}\quad \includegraphics[scale=0.35,angle=-90]{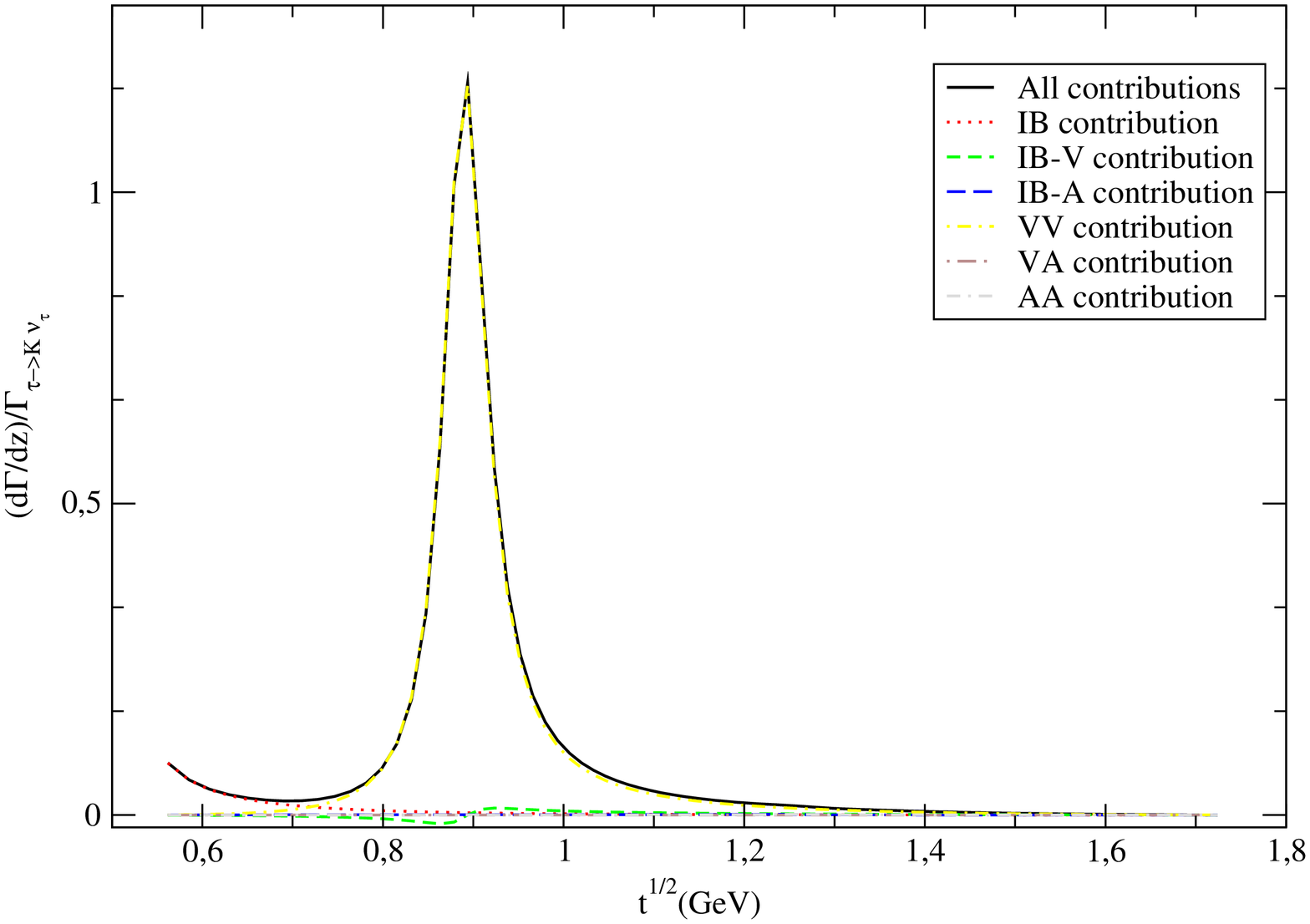}}
\caption{Kaon-photon invariant mass spectrum of the process $\tau^-\rightarrow  K^- \gamma \nu_\tau$ including all contributions for $c_4=0$ 
(left pane) and $c_4=-0$.$07$ (right pane). The $VA$ contribution vanishes identically as explained in the main text. \label{Kgammatuno}}
\end{center}
\end{figure}
\\
\begin{figure}[h!]
\begin{center}
\vspace*{0.7cm}
\centerline{\includegraphics[scale=0.35,angle=-90]{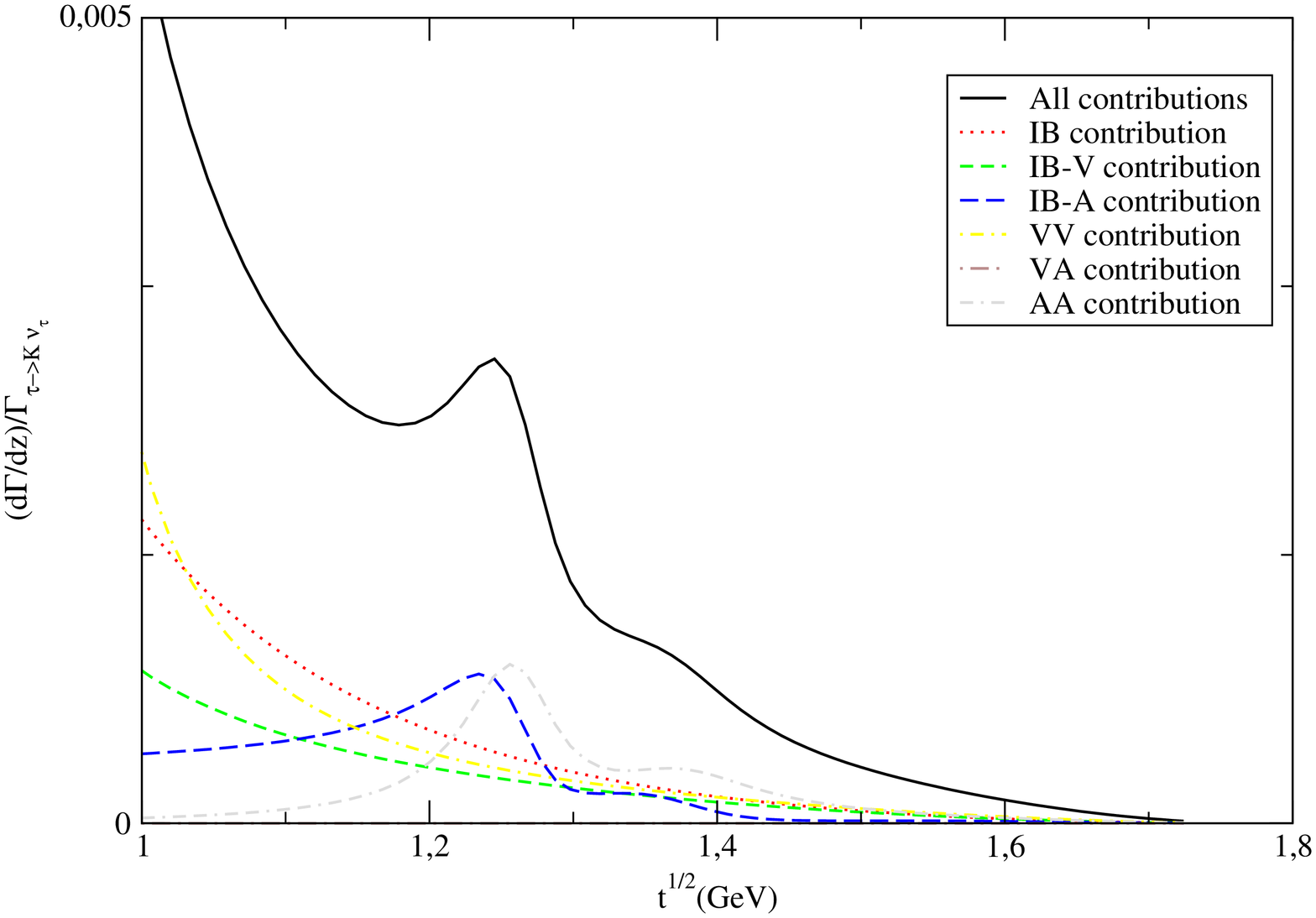}\quad \includegraphics[scale=0.35,angle=-90]{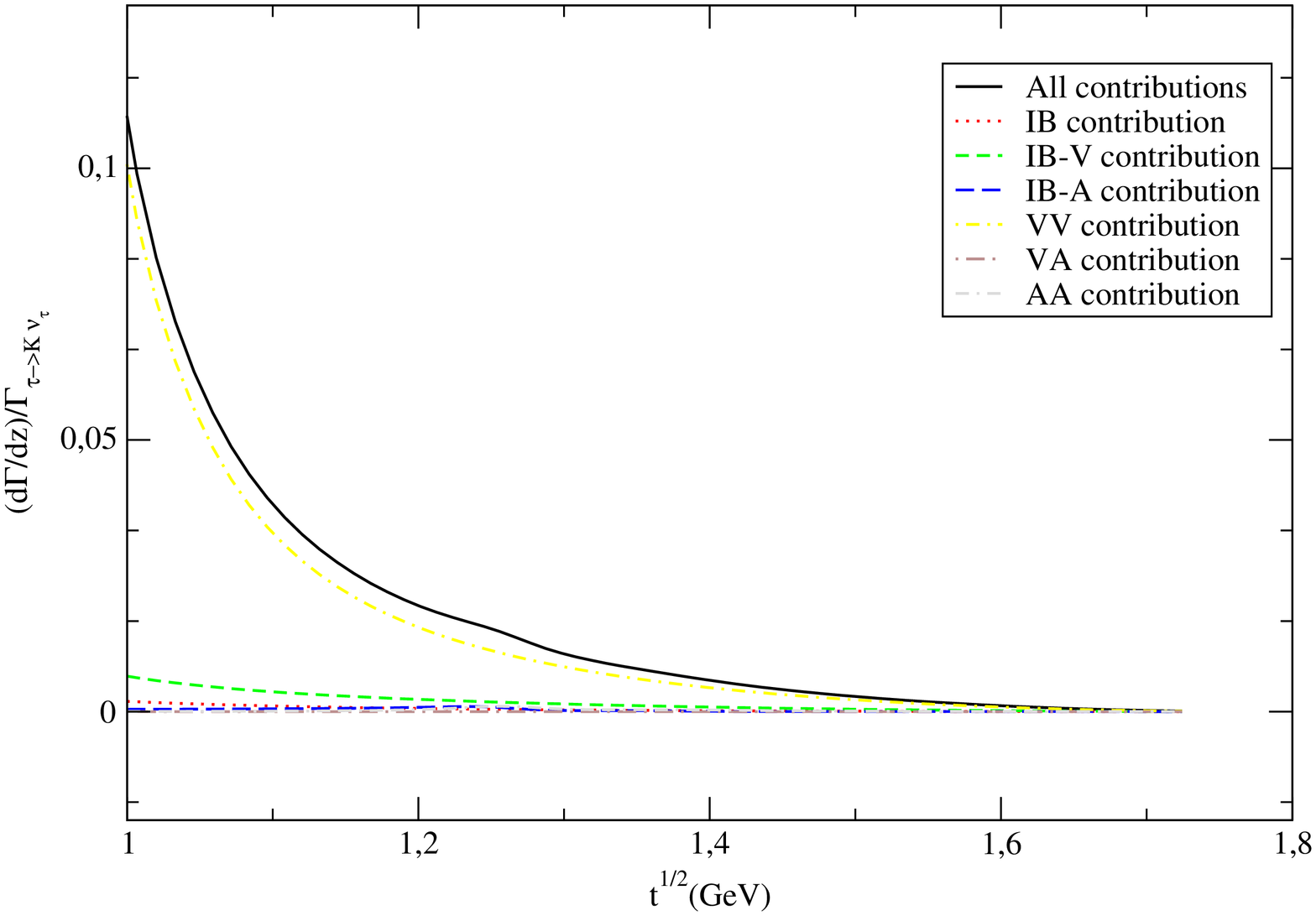}}
\caption{Close-up of the Kaon-photon invariant mass spectrum  of the process $\tau^-\rightarrow  K^- \gamma \nu_\tau$ including all contributions for 
$\sqrt{t}\gtrsim1$ GeV and  for $c_4=0$ (left pane) and $c_4=-0$.$07$ (right pane).  
The $VA$ contribution vanishes identically as explained in the main text.\label{Kgammatdos}}
\end{center}
\end{figure}
\\
\begin{figure}[h!]
\begin{center}
\vspace*{0.7cm}
\centerline{\includegraphics[scale=0.35,angle=-90]{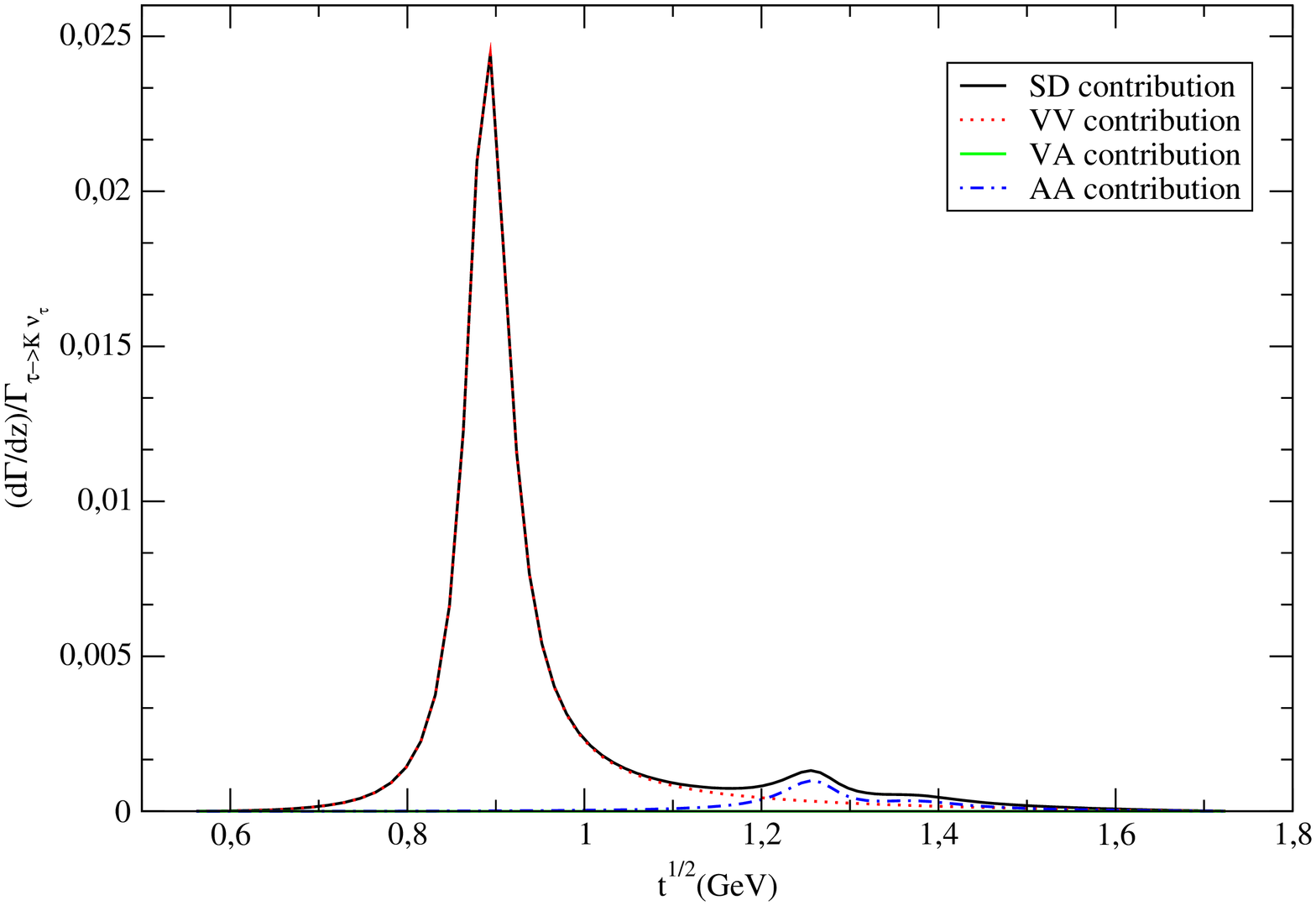}\quad \includegraphics[scale=0.35,angle=-90]{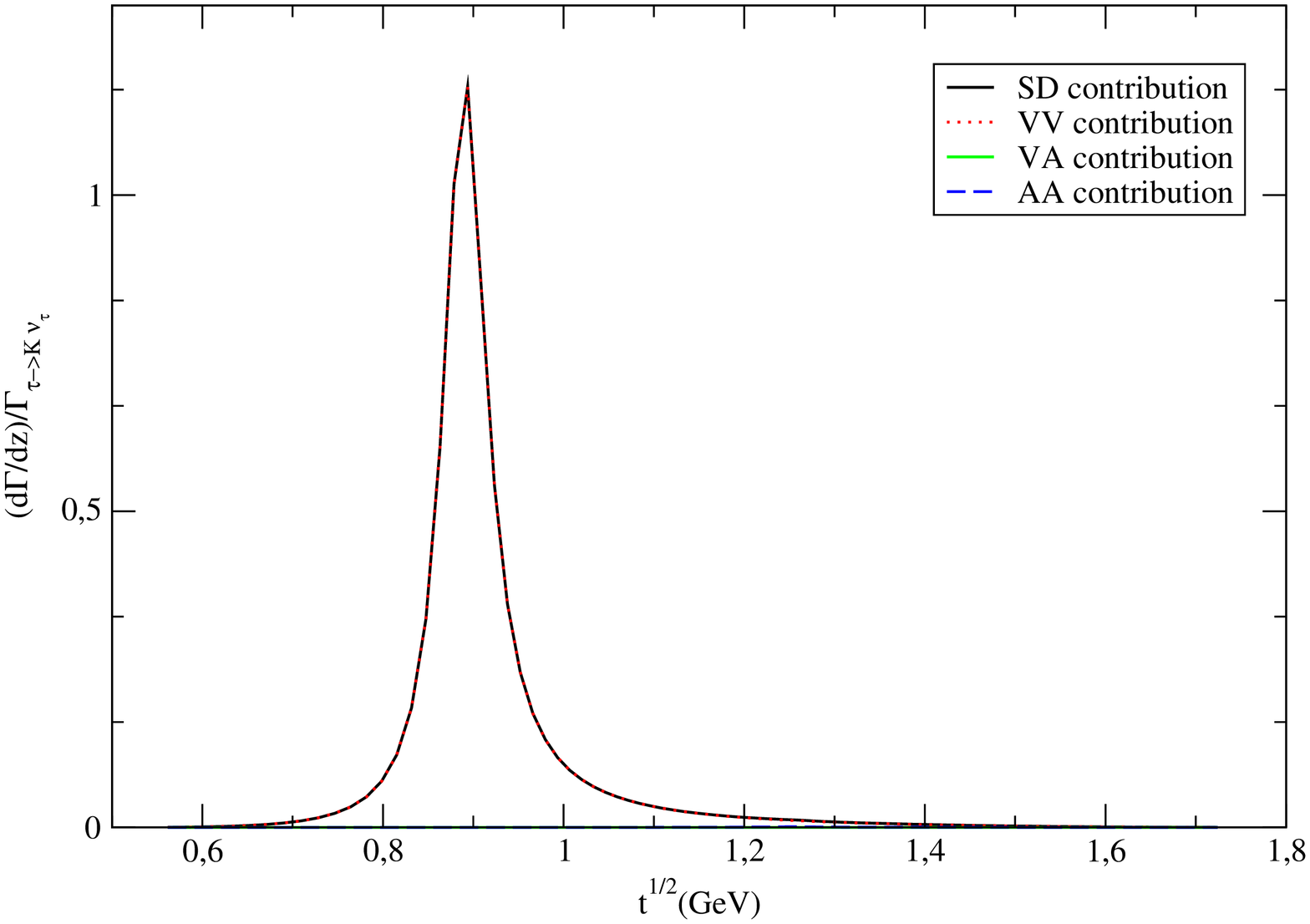}}
\caption{Kaon-photon invariant mass spectrum of the process $\tau^-\rightarrow  K^- \gamma \nu_\tau$ including only the $SD$ contributions for 
$c_4=0$ (left pane) and $c_4=-0$.$07$ (right pane).  The $VA$ contribution vanishes identically as explained in the main text.\label{Kgammattres}}
\end{center}
\end{figure}
\\
\begin{figure}[h!]
\begin{center}
\vspace*{0.7cm}
\centerline{\includegraphics[scale=0.35,angle=-90]{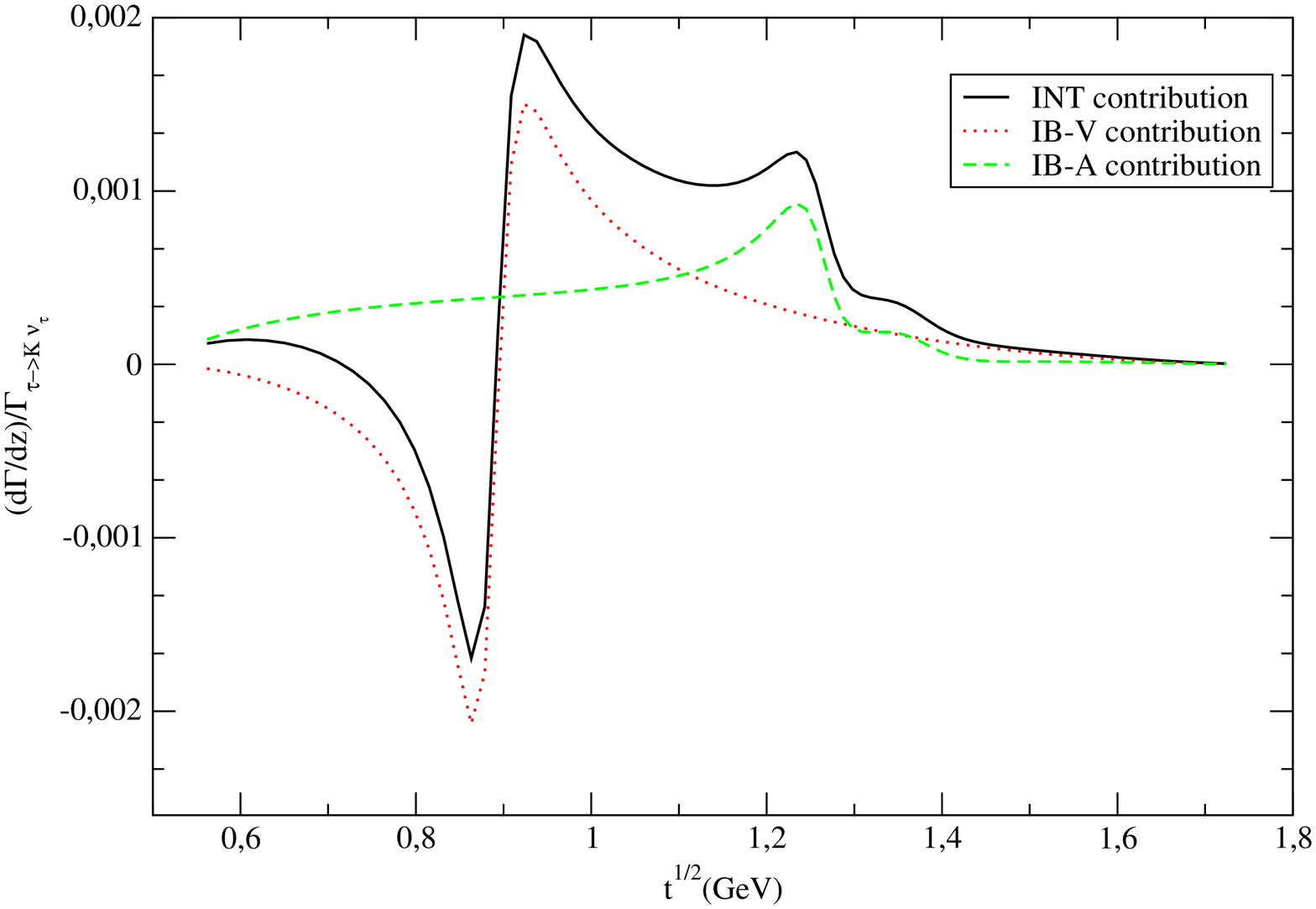}\quad \includegraphics[scale=0.35,angle=-90]{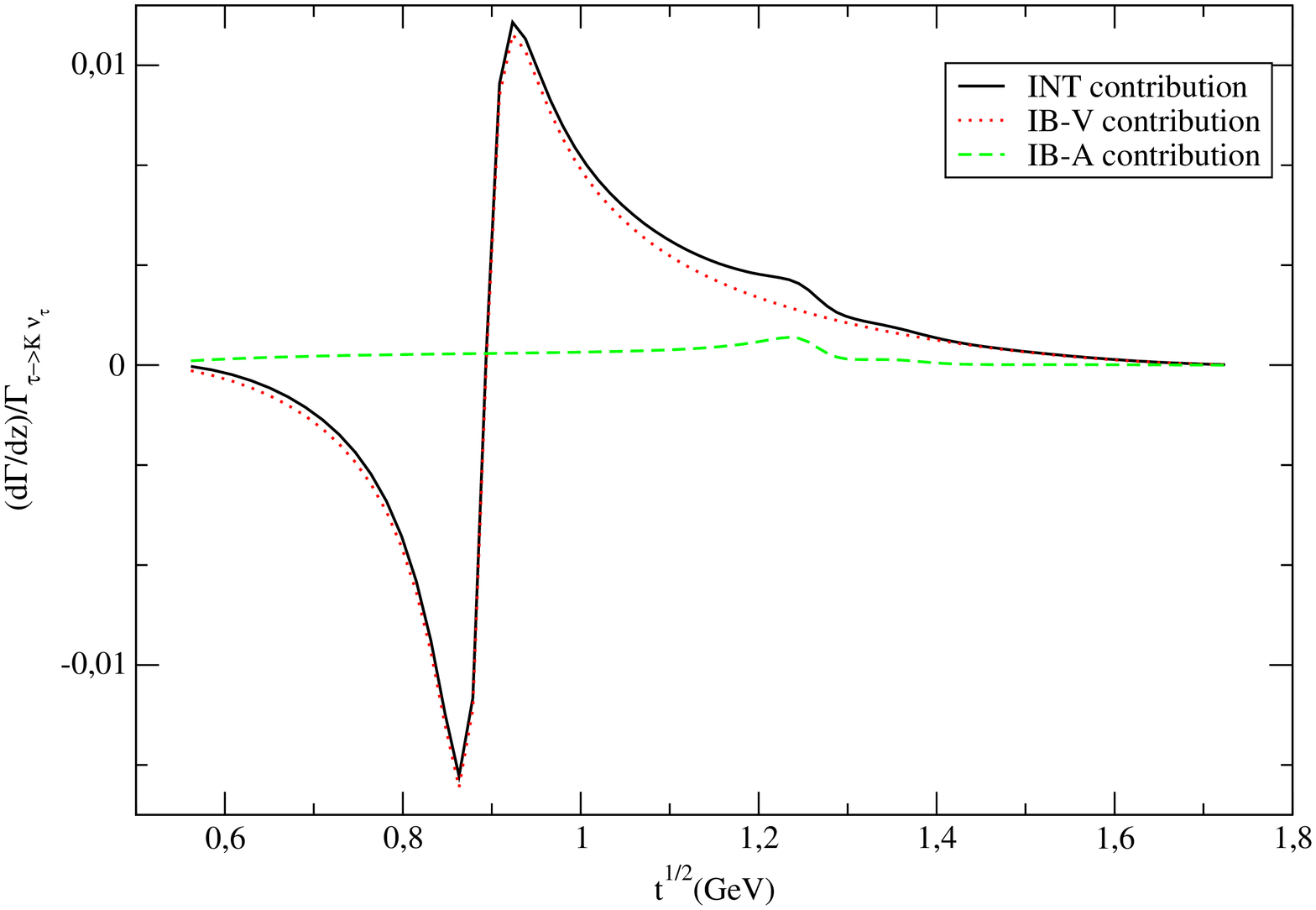}}
\caption{Kaon-photon invariant mass spectrum of the process $\tau^-\rightarrow  \pi^- \gamma \nu_\tau$ including only the interference contributions for $c_4=0$ (left pane) and $c_4=-0$.$07$ (right pane). \label{Kgammatcuatro}}
\end{center}
\end{figure}
\\
\section{Conclusions}
\hspace*{0.5cm}In this chapter we have studied \cite{GuoRoig} the radiative one-meson decays of the $\t$: $\tau^-\rightarrow  (\pi/K)^- \gamma \nu_\tau$. We 
have computed the relevant form factors for both channels and obtained the asymptotic conditions on the couplings imposed by the high-energy behaviour of 
these form factors, dictated by $QCD$. The relations that we have found here are compatible with those obtained in previous chapters in the other 
phenomenological applications considered in this Thesis.\\
\hspace*{0.5cm}One of our motivations to examine these processes is that they have not been detected yet, according to naive estimates 
or to Breit-Wigner parametrizations. We have checked the existing computations for the $IB$ part. Adding to it the $WZW$ contribution, that is the $LO$ contribution 
in $\CPT$ coming from the $QCD$ anomaly, we have estimated the model independent contribution to both decays, that could be taken as a lower bound. The values that we obtain 
for the $\pi$ channel are at least one order of magnitude above the already-observed $3K$ decay channel even for a high-energy cut-off on the photon energy. In 
the $K$ channel, the model independent contribution gives a $BR$ larger than that of the $3K$ decay channel, as well. Only imposing a large cut-off on $E_\gamma$ 
one could understand that the latter mode has not been detected so far. We expect, then, that future measurements at $B$-factories will bring us the discovery of these 
tau decay modes in the near future.\\
\hspace*{0.5cm}We do not have any free parameter in the $\tau^-\rightarrow  \pi^- \gamma \nu_\tau$ decay and that allowed us to make a complete study. Since the 
$IB$ contribution dominates, it will require some statistics to study the $SD$ effects. In this sense, the analysis of the $\pi-\gamma$ spectrum 
($t$-spectrum) is more promising than that of the pure photon spectrum ($x$-spectrum), as we have shown. We are eager to see if the discovery of 
this mode confirms our findings, since we believe that the uncertainties of our study are small for this channel.\\
\hspace*{0.5cm}Our most important result is that the inclusive one-pion decay of the tau gets a correction of $\delta_\pi=1$.$459\cdot10^{-2}$ for $E_\gamma^{min}=50$ MeV, in such a way that 
$\Gamma\left(\tau^-\to\pi^-(\gamma)\nu_\tau\right)=\Gamma\left(\tau^-\to\pi^-\nu_\tau\right)(1+\delta_\pi)$. Other values of $\delta_\pi(E_\gamma^{min})$ can be obtained from the authors upon request.\\
\hspace*{0.5cm}As expected, the higher mass of the Kaon makes easier the observation of $SD$ effects. However, there are several sources of uncertainty in 
the $\tau^-\rightarrow  K^- \gamma \nu_\tau$ decays that prevent us from having done any quantitative analysis. The most important one either rises some doubts 
about the value of $c_4$, a parameter describing the $SU(3)$ breaking effect, obtained in Ref.~\cite{Dumm:2009kj} or on the sufficiency of one multiplet of vector resonances to describe this decay. 
As we have shown, the value of this coupling affects drastically the strength of the $VV$ (and thus the whole $SD$) contribution. 
Besides, there is an uncertainty associated to the broad band of allowed values for $\theta_A$. However since the $AA$ contribution 
is anyway subleading, that one is negligible with respect to that on $c_4$. Even smaller is the error associated to the off-shell width behaviour of the axial-vector 
neutral resonance with strangeness, $K_{1H,L}$. Since we have not calculated the relevant three meson decay of the tau, we do not have this expression within $R\chi T$ 
yet. We took a simple parametrization including the on-shell cuts corresponding to the decay chains $K_{1H,L}\to (\rho K/K^* \pi)$. Since the effect of $c_4$ is so large, 
we expect that once it is discovered we will be able to bound this coupling.\\
\hspace*{0.5cm}As an application of this work, we are working out \cite{GuoRoig} the consequences of our study in lepton universality tests through the ratios $\Gamma\left( 
\tau^-\to\pi^-\nu_\tau \gamma\right)/\Gamma\left( \pi^-\to\mu^-\nu_\mu \gamma\right)$ and 
$\Gamma\left( \tau^-\to K^-\nu_\tau \gamma\right)/\Gamma\left( K^-\to\mu^-\nu_\mu \gamma\right)$ that were also considered by $DF$ \cite{Decker:1993py} 
and Marciano and Sirlin \cite{Marciano:1988vm, Sirlin:1981ie, Marciano:1993sh}. The ratio between the decays in the denominators within $\CPT$ have been studied by Cirigliano and 
Rosell \cite{Cirigliano:2007xi, Cirigliano:2007ga} and the radiative pion decay within $\RCT$ by Mateu and Portol\'es \cite{Mateu:2007tr} recently. The radiative pion/Kaon decay 
has been studied in $\CPT$ at $\cO(p^4)$ within $SU(2)$ \cite{Bijnens:1996wm} and $SU(3)$\cite{Geng:2003mt} symmetry groups. This result has been used in Ref.\cite{Unterdorfer:2008zz} 
to revise the radiative pion decay including electroweak radiative corrections \cite{DescotesGenon:2005pw} as well as $\alpha$-suppressed contributions to the $IB$ part \cite{Kuraev:2003gq}.\\

\chapter*{Conclusions}
\label{conclusions}

\addcontentsline{toc}{chapter}{Conclusions}
\appendix
\pagestyle{conclusions}
\hspace*{0.5cm}In this Thesis we have studied some decays of the $\t$ into hadrons. Besides their intrinsic interest we were motivated by the possibility 
of learning about $QCD$ hadronization in a clean environment provided by the $\t-\nu_\t-W$ coupling that keeps part of the process unpolluted from $QCD$. Moreover, since the resonances are not asymptotic states 
this kind of processes is an ideal tool to learn about their properties since its influence through exchange between the $L^\mu$ current and the final state mesons 
is sizable. Another target of our study was to provide the experimental community with an adequate theoretical tool to analyze these decays, in a time where 
there has been a lot of works from the $B$-factories $BaBar$ and $BELLE$ and the upgrade of the latter and the future results from $BES-III$ seem to point to 
an even more productive era. Since the description of hadron currents in the Monte Carlo generator $TAUOLA$ needed an improvement we wanted to work in this 
direction, as well. Finally, the low-energy $e^+e^-$ cross section in the Monte Carlo generator $PHOKHARA$ \cite{Rodrigo:2001kf} did not have all desirable low-energy 
constraints implemented for some modes \cite{Czyz:2000wh, Czyz:2005as, Czyz:2008kw, etasphokhara}.
 With the new efforts to measure with great precision this cross section exclusively in $VEPP$, $DA\Phi NE$ and the $B$-factories there was also a need to improve 
this low-energy interval of the form factors. We have worked in all these directions with the results that are summarized in the following.\\
\hspace*{0.5cm}Our task is rather non-trivial from the theoretical point of view since: first, the fundamental theory, $QCD$ is written in terms of hadrons, while we measure mesons. Second, a 
perturbative expansion in the coupling constant of the $QCD$ Lagrangian will not converge at the low and intermediate energies we are interested in, so that 
we need to find an alternative expansion parameter to work in an $EFT$ framework using the active fields in this range of energies as degrees of freedom and 
keeping the symmetries of the fundamental theory. Third, although it is clear how to build an $EFT$ for low-energy $QCD$ based on the approximate 
chiral symmetry of this subsector, it is not so when going to higher energies. Four, a promising parameter, as it is $1/N_C$, succeeds in explaining qualitatively 
the most salient features of meson phenomenology but it is difficult to apply it quantitatively since its predictions at lowest order are contradictory to the 
Weinberg's approach to $EFTs$: while $LO$ in the $1/N_C$ expansion predicts an infinite tower of infinitely narrow resonances, the $EFTs$ require just the relevant 
fields. Besides, only a few excited resonances are known for every set of quantum numbers so there is no model independent way to satisfy the $N_C\to\infty$ 
requirements, either. As a conclusion on this point we must admit that it will be necessary to model the $1/N_C$ expansion. One realizes that we have traded 
the problem of not having a suitable expansion parameter by having it, although lacking an unambiguous way to perform the expansion even cutting it at lowest 
order.\\
\hspace*{0.5cm}From the phenomenological point of view there are also some subtleties that require a caveat: while for some decay modes there is already a lot 
of very precise information that allows us to do a precision study ($\tau\to\pi\pi\nu_\tau$, $\tau\to K\pi\nu_\tau$, $\tau\to\pi\pi\pi\nu_\tau$), for other modes 
the situation is not that clear 
, like for instance the $\tau\to KK\pi\nu_\tau$ decays that would have helped to fix the vector current sector. Since that was not possible we turned to the channels $\tau\to\eta\pi\pi\nu_\tau$ 
where there is only vector current contribution in the isospin conserved limit that proved to be helpful. One should also bear in mind that the interplay 
between Monte Carlo generation and signal extraction is important and since the Monte Carlo relies on a given model for the signal to background splitting 
this brings in additional uncertainties, specially in the case where both currents can in principle contribute sizably to the decay as in $\tau\to KK\pi\nu_\tau$, 
or in rare decays where there can be an important background in some phase space corners from other modes. All this would suggest the following approach: The 
Monte Carlo generators having some variety of reasonable hadron currents and the fit to all relevant modes being made at a time. Since this is not possible yet, 
one should not take all conclusions from partial studies as definitive.\\
\hspace*{0.5cm}The considerations in the two previous paragraphs do not mean at all that there is no point in carrying these investigations on. The essential 
thing will be to recognize which conclusions are firm and which can be affected by any of the errors commented above. Our approach has as many $QCD$ features 
as we have been able to capture and they are more than in other approaches which justifies our labor and brings in its interest. We will emphasize its virtues 
next.\\
\hspace*{0.5cm}Our approach includes the right low-energy behaviour inherited from $\CPT$. This is essential because a mismatch there is carried on by the rest 
of the curves. It follows the ideas of the large-$N_C$ limit of $QCD$ and implements them in order to have a theory of mesons: including the lighter $pGbs$ and 
the light-flavoured resonances. Therefore it has the relevant degrees of freedom to describe the problem. The theory built upon symmetries does not have yet all 
$QCD$ features we can implement. To do so, we require a Brodsky-Lepage behaviour to the form factors. This warrants the right short-distance vanishing and 
determines some couplings which makes the theory more predictive. Our approach to the large-$N_C$ limit of $QCD$ is guided by simplicity on the spectrum (we include the 
least number of degrees of freedom that allows for a description of the data) and by the off-shell widths that are derived within $R\chi T$. In the remainder 
of the introduction we highlight the most relevant contributions we have made during our study.\\
\hspace*{0.5cm}First of all, we have fixed the axial-vector current sector making theoretical predictions and the description of observables compatible. There 
was an inconsistency between the relations obtained in the Green function $<VAP>$ and the description of the $\tau\to\pi\pi\pi\nu_\tau$ observables. 
We have found the way to understand both at a time and, remarkably, these relations do not only hold for all tau decays into three mesons and for the $\bra VAP \ket$ 
Green function but also for the radiative decays $\tau\to(\pi/K)\gamma\nu_\tau$ which we take as a confirmation of our picture and of the assumptions we have made 
concerning the modelization of the large-$N_C$ limit of $QCD$ in a meson theory.\\
\hspace*{0.5cm}A fundamental result coming from both the $\tau\to\pi\pi\pi\nu_\tau$ and $\tau\to KK\pi\nu_\tau$ studies is the off-shell width of the axial-vector 
meson $\Gamma_{\mathrm{a}_1}$. It incorporates all $3\pi$ and $KK\pi$ cuts and it neglects the $\eta \pi \pi$ cut that vanishes in the isospin limit because of 
$G$-parity and the $\eta \eta \pi$ cut whose upper br limit is tiny. $\Gamma_{\mathrm{a}_1}(Q^2)$ and the $\tau\to\pi\pi\pi\nu_\tau$ form factors have been implemented 
in $TAUOLA$. The agreement is better than the error associated to the statistical sampling and at the level of few per thousand.\\
\hspace*{0.5cm}We provide our prediction for additional observables both for $\tau\to\pi\pi\pi\nu_\tau$ and $\tau\to KK\pi\nu_\tau$ that could be confronted 
to forthcoming data. On the latter channel there is however some ambiguity because in addition to the short-distance relations we obtained we borrowed two of them 
from the study of the $<VVP>$ Green function and this is an assumption. However lacking of a prediction for the spectra (we could not even digitize the published plots 
since they corresponded to raw mass data without the efficiency corrections implemented) we decided to proceed this way since we have observed that the departure observed between 
short-distance relations affecting the vector current couplings in three-point Green functions and three-meson tau decays was small.\\
\hspace*{0.5cm}Within this approach we have given our prediction for the relevant observables in the $KK\pi$ channels and we showed that, contrary to some previous 
determinations, the vector current contribution can not be neglected in these decays. Since we have convincing reasons to believe that the description of the axial-vector 
current is pretty accurate, we think that this conclusion is firm. On the contrary, there can be some changes in the shape of the curves due to the assumption 
commented in the previous paragraph. This only data will tell. Our parametrization for the hadronic form factors in the $\tau\to KK\pi\nu_\tau$ has been implemented 
in $TAUOLA$ to a great level of precision. We have decided to allow for some freedom in the assumptions on the relations among couplings that we commented 
before.\\
\hspace*{0.5cm}The study of the decays $\tau\to\eta\pi\pi\nu_\tau$ has brought us information about two previously undetermined couplings. We have determined an 
allowed region where they can be and illustrated with some benchmark points the impact of them on the spectra. The experimental data has allowed us to favour one 
of the benchmark scenarios that we considered and refine our determination fitting $Belle$ data. If our description of the former process is correct we believe 
that the upper bound on the decay $\tau\to\eta'\pi\pi\nu_\tau$ may be wrong, and should be detected soon. Finally we have worked out isospin 
constraints and provided a prediction for the low-energy cross section $e^+e^-\to\eta\pi^+\pi^-$ that compares well to data in these regime. We have provided with our codes to the $PHOKHARA$ team. 
Symmetries make the $\tau\to\eta\eta\pi\nu_\tau$ decay a wonderful scenario to test the dominance of the spin-one resonances over those of spin $0$. All exchanges of vector 
and axial-vector resonances are forbidden and then one only has the contribution from $\CPT$ at $\cO(p^2)$. The discovery of this mode would allow to estimate 
in a clean environment the effect of scalar and pseudoscalar resonances in the future.\\
\hspace*{0.5cm}Finally, we have studied the radiative decays $\tau\to (\pi/K)\gamma \nu_\tau$. Since the axial-vector current is completely controlled we have very firm 
conclusions. First, we have confirmed the earlier estimations: this mode has much larger br than some that have indeed been measured. Second, we have obtained our 
predictions for the observables. We obtain a $VV$ contribution similar to earlier works but we obtain a much smaller $AA$ contribution as corresponds to the well-behaved $\Gamma_{\mathrm{a}_1}$ that we used.\\
\hspace*{0.5cm}If we assume that the $IB$ contribution should dominate up to the hard photon region, it seems that the channel $\tau\to K \gamma\nu_\tau$ suggests 
that the value of $c_4$ that was obtained from $\tau\to KK\pi\nu_\tau$ decays is too large. Since this coupling does not appear in the $\tau\to\eta^{(')}\pi\pi\nu_\tau$
decays one should wait to analyze experimental data on the $KK\pi$ mode to understand this issue.\\
\hspace*{0.5cm}Among the appendixes there is some technical material. 
 There are two appendixes concerning the relation of theoretical formulae to the experimental measurements. One can also find the complete set of isospin 
relations for three meson modes between $\sigma(e^+e^-)$ and $\tau$ decays and the computation of the process $\omega \to \pi^+\pi^-\pi^0$ that allowed us 
to fix another combination of couplings in the Lagrangian. For this purpose we needed to derive a new piece for the Resonance Chiral Theory Lagrangian. The 
other appendixes summarize theoretical information that helps to understand better the contents of this Thesis.\\
\hspace*{0.5cm}As final conclusions we would like to say that we are satisfied for many reasons: we have fixed the couplings of the axial-vector current 
sector of the $R\chi T$ Lagrangian and we have provided a precise description of the $\tau\to\pi\pi\pi\nu_\tau$ observables. This includes a sound description 
of the a$_1$ resonance width. We have improved the knowledge on the odd-intrinsic parity sector of the Resonance Lagrangian and applied it to some processes of 
interest. Future experimental data from $BaBar$, $Belle$ and $BES$ could help us to proceed further in this direction. We have also worked in the application 
of these findings to the Monte Carlo generators for low-energy Physics. In particular, for $TAUOLA$ in tau decays and $PHOKHARA$ in the $e^+e^-$ cross-section. 
Ideally this will result in a global fit to all relevant channels with an adequate splitting of signal and background accounting well for the pollution from 
other channels. We hope to be able to accomplish this program.\\
\appendix
\chapter*{Appendix A: Structure functions in tau decays}
\label{App_StructureFunctions}
\addcontentsline{toc}{chapter}{Appendix A: Structure functions in tau decays}
\appendix
\pagestyle{appendixa}
\newcounter{A}
\renewcommand{\theequation}{\Alph{A}.\arabic{equation}}
\setcounter{A}{1}
\setcounter{equation}{0}
\hspace*{0.5cm}Hadron and lepton tensors are Hermitian and can be expanded in terms of a set of $16$ independent elements:
\begin{equation}
\mathcal{L}_{\mu\nu}\mathcal{H}^{\mu\nu}\,=\,\mathcal{L}^{00}\mathcal{H}^{00}-\mathcal{L}^{i0}\mathcal{H}^{i0}-\mathcal{L}^{0j}\mathcal{H}^{0j}+\mathcal{L}^{ij}\mathcal{H}^{ij}\,.
\end{equation}
\hspace*{0.5cm}In order to isolate the different angular dependencies, it is convenient to introduce $16$ combinations of defined symmetry, the so-called 
lepton ($L_X$) and hadron ($W_X$) structure functions. This way,
\begin{equation}
\mathcal{L}_{\mu\nu}\mathcal{H}^{\mu\nu}\,=\,\sum_X L_X W_X\,=\,2(M_\tau^2-Q^2)\sum_X \overline{L}_X W_X\,,
\end{equation}
where $X$ stands for $A,\,B,\,C,\,D,\,E,\,F,\,G,\,H$ and $I$ -the structure functions that collect (axial-)vector contributions- and also for the ones 
including information on the pseudoscalar form factor: $SA,\,SB,\,SC,\,SD,\,SE,\,SF$ and $SG$. All of them are obtained through  \cite{Kuhn:1992nz}:
\begin{eqnarray} \label{structurefunctions general}
& & L_A \, = \, \frac{\mathcal{L}^{11}+\mathcal{L}^{22}}{2}\,,\quad W_A\,=\,\mathcal{H}^{11}+\mathcal{H}^{22}\,,\nonumber\\
& & L_B\,=\,\mathcal{L}^{33}\,,\quad W_B\,=\,\mathcal{H}^{33}\,,\nonumber\\
& & L_C\,=\,\frac{\mathcal{L}^{11}-\mathcal{L}^{22}}{2}\,,\quad W_C\,=\,\mathcal{H}^{11}-\mathcal{H}^{22}\,,\nonumber\\
& & L_D\,=\,\frac{\mathcal{L}^{12}+\mathcal{L}^{21}}{2}\,,\quad W_D\,=\,\mathcal{H}^{12}+\mathcal{H}^{21}\,,\nonumber\\
& & L_E\,=\,\frac{i}{2}(\mathcal{L}^{12}-\mathcal{L}^{21})\,,\quad W_E\,=\,-i(\mathcal{H}^{12}-\mathcal{H}^{21})\,,\nonumber\\
& & L_F\,=\,\frac{\mathcal{L}^{13}+\mathcal{L}^{31}}{2}\,,\quad W_F\,=\,\mathcal{H}^{13}+\mathcal{H}^{31}\nonumber\,,\\
& & L_G\,=\,\frac{i}{2}(\mathcal{L}^{13}-\mathcal{L}^{31})\,,\quad W_G\,=\,\,-i(\mathcal{H}^{13}-\mathcal{H}^{31})\,,\nonumber\\
& & L_H\,=\,\frac{\mathcal{L}^{23}+\mathcal{L}^{32}}{2}\,,\quad W_H\,=\,\mathcal{H}^{23}+\mathcal{H}^{32}\nonumber\,,\\
& & L_I\,=\,\frac{i}{2}(\mathcal{L}^{23}-\mathcal{L}^{32})\,,\quad W_I\,=\,\,-i(\mathcal{H}^{23}-\mathcal{H}^{32})\,;\nonumber\\
& & L_{SA}\,=\,\mathcal{L}^{00}\,,\quad W_{SA}\,=\,\mathcal{H}^{00}\,,\nonumber
\end{eqnarray}
\begin{eqnarray}
& & L_{SB}\,=\,-\frac{\mathcal{L}^{10}+\mathcal{L}^{01}}{2}\,,\quad W_{SB}\,=\,H^{01}+\mathcal{H}^{10}\,,\nonumber\\
& & L_{SC}\,=\,-\frac{i}{2}(\mathcal{L}^{01}-\mathcal{L}^{10})\,,\quad W_{SC}\,=\,-i(\mathcal{H}^{01}-\mathcal{H}^{10})\,,\nonumber\\
& & L_{SD}\,=\,-\frac{\mathcal{L}^{02}+\mathcal{L}^{20}}{2}\,,\quad W_{SD}\,=\,\mathcal{H}^{02}+\mathcal{H}^{20}\,,\nonumber\\
& & L_{SE}\,=\,-\frac{i}{2}(\mathcal{L}^{02}-\mathcal{L}^{20})\,quad \,W_{SE}\,=\,-i(\mathcal{H}^{02}-\mathcal{H}^{20})\,,\nonumber\\
& & L_{SF}\,=\,-\frac{\mathcal{L}^{03}+\mathcal{L}^{30}}{2}\,,\quad W_{SF}\,=\,\mathcal{H}^{03}+\mathcal{H}^{30}\,,\nonumber\\
& & L_{SG}\,=\,-\frac{i}{2}(\mathcal{L}^{03}-\mathcal{L}^{30})\quad ,\,W_{SG}\,=\,-i(\mathcal{H}^{03}-\mathcal{H}^{30})\,.
\end{eqnarray}
\hspace*{0.5cm}In the hadron rest frame, with axis $z$ and $x$ aligned with the normal to the hadron plane and $\widehat{q}_3$, respectively; one has the 
relations:
\begin{eqnarray}
& & q_3^{\mu}\,=\,(E_3,q_3^{x},0,0)\,\,,\,\,q_2^{\mu}\,=\,(E_2,q_2^{x},q_2^{y},0)\,\,,\,\,q_1^{\mu}\,=\,(E_1,q_1^{x},q_1^{y},0)\,\,\mathrm{with}\nonumber\\
& & E_i\,=\,\frac{Q^2-s_i+m_i^2}{2\sqrt{Q^2}}\,\,,\,\,q_3^x\,=\,\sqrt{E_3^2-m_3^2}\,,\nonumber\\
& & q_2^{x}\,=\,\frac{2E_2E_3-s_1+m_2^2+m_3^2}{2q_3^x}\,\,,\,\,q_2^{y}\,=\,-\sqrt{E_2^2-(q_2^{x})^2-m_2^2}\,,\nonumber\\
& & q_1^{x}\,=\,\frac{2E_1E_3-s_2+m_1^2+m_3^2}{2q_3^x}\,\,,\,\,q_1^{y}\,=\,\sqrt{E_1^2-(q_1^{x})^2-m_1^2}\,=\,-q_2^{y}\,.\nonumber\\
\end{eqnarray}
\hspace*{0.5cm}If we introduce the following variables:
\begin{equation}
x_1\,=\,V_1^x=q_1^x-q_3^x\,\,,\,\,x_2\,=\,V_2^x=q_2^x-q_3^x\,\,,\,\,x_3\,=\,V_1^y=q_1^y=-q_2^y\,\,,\,\,x_4\,=\,V_3^z\,=\,\sqrt{Q^2}x_3q_3^x\,,
\end{equation}
it is straightforward to see that both descriptions either in terms of form factors \footnote{As defined in Eq.~(\ref{generaldecomposition_3mesons}).} or 
structure functions are completely equivalent:
\begin{eqnarray} \label{W_i_3mesons}
& & W_A \, = \, (x_1^2+x_3^2)^2|F_1^A|^2+(x_2^2+x_3^2)^2|F_2^A|^2+2(x_1x_2-x_3^2)\Re e(F_1^AF_2^{A*})\,,\nonumber\\
& & W_B\,=\,x_4^2|F_4^V|^2\,,\nonumber\\
& & W_C\,=\,(x_1^2-x_3^2)^2|F_1^A|^2+(x_2^2-x_3^2)^2|F_2^A|^2+2(x_1x_2+x_3^2)\Re e(F_1^AF_2^{A*})\,,\nonumber\\
& & W_D\,=\,2\left[x_1x_3|F_1^A|^2-x_2x_3|F_2^A|^2+x_3(x_2-x_1)\Re e(F_1^AF_2^{A*})\right]\,,\nonumber\\
& & W_E\,=\,-2x_3(x_1+x_2)\Im m(F_1^AF_2^{A*})\,,\nonumber\\
& & W_F\,=\,2x_4\left[x_1\Im m(F_1^AF_4^{V*})+x_2\Im m(F_2^AF_4^{V*})\right]\,,\nonumber\\
& & W_G\,=\,-2x_4\left[x_1\Re e(F_1^AF_4^{V*})+x_2\Re e(F_2^AF_4^{V*})\right]\,,\nonumber
\end{eqnarray}
\begin{eqnarray}
& & W_H\,=\,2x_3x_4\left[\Im m(F_1^AF_4^{4*})-\Im m(F_2^AF_4^{V*})\right]\,,\nonumber\\
& & W_I\,=\,-2x_3x_4\left[\Re e(F_1^AF_4^{V*})-\Re e(F_2^AF_4^{V*})\right]\,;\nonumber\\
& & W_{SA}\,=\,Q^2 |F_3^A|^2\,,\nonumber\\
& & W_{SB}\,=\,2\sqrt{Q^2}\left[x_1 \Re e(F_1^AF_3^{A*})+x_2\Re e(F_2^AF_3^{A*})\right]\,,\nonumber\\
& & W_{SC}\,=\,-2\sqrt{Q^2}\left[ x_1 \Im m(F_1^AF_3^{A*})+x_2\Im m(F_2^AF_3^{A*})\right]\,,\nonumber\\
& & W_{SD}\,=\,2\sqrt{Q^2}x_3\left[ \Re e(F_1^AF_3^{A*})-\Re e(F_2^AF_3^{A*})\right]\,,\nonumber\\
& & W_{SE}\,=\,-2\sqrt{Q^2}x_3\left[ \Im m(F_1^AF_3^{A*})-\Im m(F_2^AF_3^{A*})\right]\,,\nonumber\\
& & W_{SF}\,=\,-2\sqrt{Q^2}x_4\left[ \Im m(F_4^VF_3^{A*})\right]\,,\nonumber\\
& & W_{SG}\,=\,-2\sqrt{Q^2}x_4\left[ \Re e(F_4^VF_3^{A*})\right]\,.
\end{eqnarray}
The corresponding formulae for the two-meson decays of the tau can be found in Eq. (\ref{W_i in two meson decay}) and for the decays into one $pG$ and a 
spin-one resonance in Ref.~\cite{Decker:1992jy}.\\
\chapter*{Appendix B: $\frac{\mathrm{d}\Gamma}{\mathrm{d}s_{ij}}$ in three-meson tau decays}
\label{DGammadSij}
\addcontentsline{toc}{chapter}{Appendix B: $\frac{\mathrm{d}\Gamma}{\mathrm{d}s_{ij}}$ in three-meson tau decays}
\appendix
\pagestyle{appendixb}
\newcounter{B}
\renewcommand{\theequation}{\Alph{B}.\arabic{equation}}
\setcounter{B}{2}
\setcounter{equation}{0}
\hspace*{0.5cm}In this appendix we provide the formulae used to obtain the differential decay width as a function of the invariant masses of the different meson pairs, $\frac{\mathrm{d}\Gamma}{\mathrm{d}s_{ij}}$.\\
\hspace*{0.5cm}In a general four-body decay $P(p,M)\to P_1(p_1,m_1)P_2(p_2,m_2)P_3(p_3,m_3)P_4(p_4,m_4)$ the differential decay width reads \cite{Alain}
\begin{equation}
 \mathrm{d}\Gamma = \frac{X\beta_{12}\beta_{34}}{4(4\pi)^6M^3} |\overline{\mathcal{M}}|^2 \mathrm{d}s_{12} \mathrm{d}s_{34} \mathrm{d\,(cos}\theta_1) \mathrm{d\,(cos}\theta_3) \mathrm{d}\phi_3\,,
\end{equation}
with $s_{ij}=(p_i+p_j)^2$, $\beta_{ij}=\lambda^{1/2}(s_{ij},m_i^2,m_j^2)/s_{ij}$, $X=\lambda^{1/2}(M^2,s_{12},s_{34})/2$. The integrations limits are $_0^{2\pi}$ for $\phi_3$, $_{-1}^{+1}$ for cos($\theta_1$) and 
cos($\theta_3$), $s_{34}^{min}=(m_3+m_4)^2$, $s_{34}^{max}=(M-\sqrt{s_{12}})^2$, $s_{12}^{min}=(m_3+m_4)^2$ and $s_{12}^{max}=(M-m_3-m_4)^2$.

Using the structure functions introduced in appendix A, the matrix element summed over polarizations and averaged over final-state polarizations reads
\begin{eqnarray}
 |\overline{\mathcal{M}}|^2 & = & G_F^2 |V_{ij}^{CKM}|^2 M_\tau^2 \left(\frac{M_\tau^2}{Q^2}-1\right)\left\lbrace W_{SA}(Q^2,s,t)+\right.\nonumber\\
& & \left. \frac{1}{3}\left(1+\frac{2Q^2}{M_\tau^2}\right)\left[W_{A}(Q^2,s,t)+W_{B}(Q^2,s,t)\right]\right\rbrace\,.
\end{eqnarray}
It is straightforward to express $Q^2$, $s$ and $t$ in terms of $s_{12}$, $s_{34}$, $\theta_1$, $\theta_3$ and $\phi_3$.
If we are interested in the $\mathrm{d}\Gamma/\mathrm{d}t$ or $\mathrm{d}\Gamma/\mathrm{d}u$ distributions we have to exchange $s\leftrightarrow t$ and $s\leftrightarrow u$ in all expressions, respectively. The integration limits 
for the two outermost integrations have to be changed accordingly.

It is worth to notice that the proposed expression is efficient and fast when computing the integration, even with rather ellaborated structures for the form factors and realistic off-shell widths for the resonances.\\

\chapter*{Appendix C: Off-shell width of Vector resonances} \label{widthApp}
\addcontentsline{toc}{chapter}{Appendix C: Off-shell width of Vector resonances}
\appendix
\pagestyle{appendixc}
\newcounter{C}
\renewcommand{\thesection}{\Alph{C}.\arabic{section}}
\renewcommand{\thesubsection}{\Alph{C}.\arabic{subsection}}
\renewcommand{\theequation}{\Alph{C}.\arabic{equation}}
\renewcommand{\thefigure}{\Alph{C}.\arabic{figure}}
\setcounter{C}{3}
\setcounter{section}{0}
\setcounter{subsection}{0}
\setcounter{equation}{0}
\setcounter{figure}{0}
\section{Introduction} \label{widthApp_Intro}
\hspace*{0.5cm}Resonance widths have a big importance in any process whose energy is able to reach its on-mass shell, specially if they are
 rather wide. Any sensible modelization of the process must take this into account. Masses and widths of particles depend on the conventions 
one employs and on the chosen formalism. We will explain in this appendix the approach we use and show that it is consistent with $R\chi T$
 and general field theory arguments.\\
\hspace*{0.5cm}Since our work only includes spin-one resonances, we will not consider the case of scalar and pseudoscalar resonances. We will 
start by the easier case of vector resonances that involves, at lowest order, two-particle intemediate states. Then, we will consider the 
case of axial-vector resonances, where the three particle cuts give the first contribution.
\section{Definition of a hadronic off-shell width for vector resonances}  \label{widthApp_DefWidth}
\hspace*{0.5cm}In the antisymmetric tensor formulation \footnote{For further details, see Appendix \ref{antisymApp}.} the bare 
propagator of vector mesons is given by
\begin{equation} 
\bra 0|T\left\{V_{\mu\nu}(x),\,V_{\rho\sigma}(y)\right\}|0\ket \,=\,
\int
\frac{\mathrm{d}^4k}{(2\pi)^4} e^{-ik(x-y)} 
\left\{\frac{2i}{M^2-q^2}\,\Omega^L_{\mu\nu,\rho\sigma} \,+\, \frac{2i}{M^2}\,
\Omega_{\mu\nu,\rho\sigma}^T\right\} \,,
\end{equation}
with $\Omega^{L(T)}_{\mu\nu,\rho\sigma}$ the projectors over longitudinal (transverse) polarizations.\\
\hspace*{0.5cm}There is no doubt that physical observables are insensitive to the field representation. But here we are concerned about the
 off-shell behaviour of resonances so, in principle, the issue of independence on field redefinitions should be studied for the proposed 
width.\\ 
\hspace*{0.5cm}Ref.~\cite{GomezDumm:2000fz} proposes to define the spin-$1$ meson width as the imaginary part of the pole generated by resuming
 those diagrams, with an absorptive contribution in the $s$ channel, that contribute to the two-point function of the corresponding vector 
current. That is, the pole of
\begin{equation}
\Pi^{jk}_{\mu\nu}\,=\,i\,\int\mathrm{d}^4x e^{iqx}\bra0|T\left[ V_\mu^j(x)V_\nu^k(0)\right] |0\ket\,,
\end{equation}
with
\begin{equation}
V_\mu^j\,=\,\frac{\delta S_{R\chi T}}{\delta v^\mu_j}\,,
\end{equation}
where $S_{R\chi T}$ is the action that generates the Lagrangian of R$\chi$T.\\
\hspace*{0.5cm}The widths obtained in this way are shown to satisfy the requirements of analyticity, unitarity and chiral symmetry 
prescribed by $QCD$.\\
\subsection{$\rho$ off-shell width} \label{widthApp_rho}
\begin{figure}[h]
\centering
\includegraphics[scale=1.2]{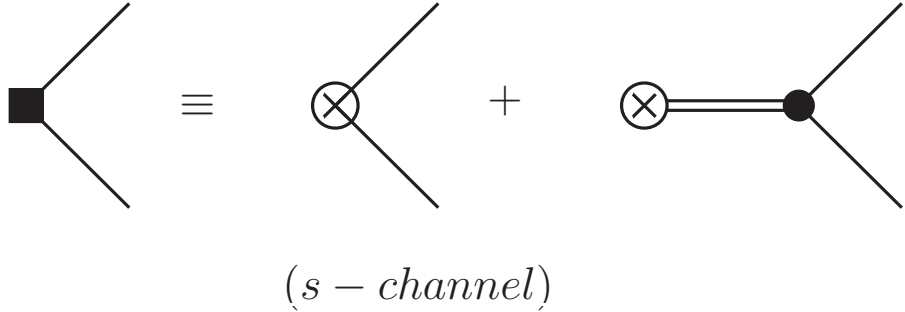}
\caption{Effective vertices contributing to vector transitions in the $s$ channel that are relevant for the vector form factor of the pion
. The crossed circle stands for an external vector current insertion. A double 
line indicates the vector meson and a single one the $pG$. Local vertices on the RHS are provided, at $LO$, by $\mathcal{L}_\chi$ at 
$\cO(p^2)$.}
\label{effverticesforvectortransitions}
\end{figure}

\begin{figure}[h]
\centering
\includegraphics[scale=1.0]{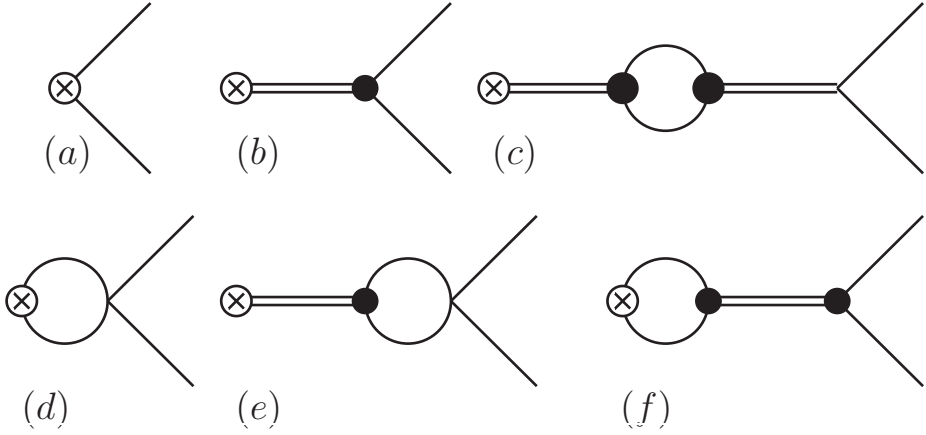}
\caption{Diagrams contributing to the vector form factor of the pion up to one loop within $\RCT$ that have an absorptive part in the $s$
 channel.}
\label{diagsVFFpiRCT}
\end{figure}

\begin{figure}[h]
\centering
\includegraphics[scale=1.2]{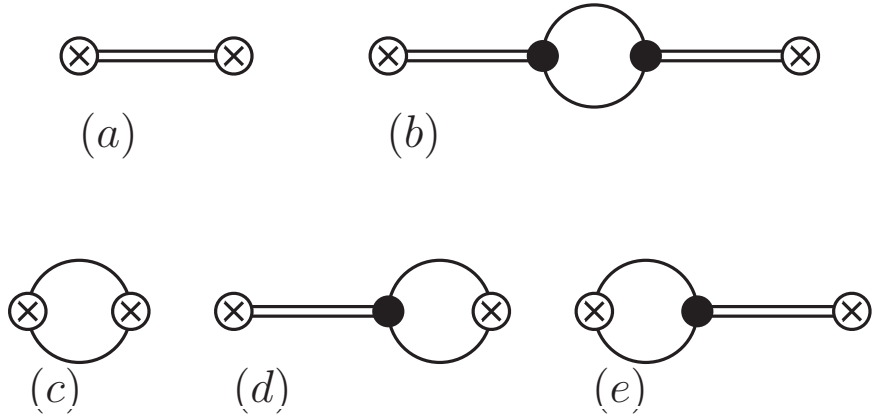}
\caption{Diagrams contributing to the vector-vector correlator $\Pi^{33}_{\mu\nu}$ up to one loop within $\RCT$.}
\label{diagsVVcorrelRCT}
\end{figure}

\hspace*{0.5cm}In order to construct the dressed propagator of the $\rho^0$ (770) meson, we should consider -for a definite intermediate 
state- all the contributions carrying the appropriate quantum numbers. In this case, the first cut corresponds to a two-$pGs$ absorptive 
contribution that happens to saturate its width. We will neglect the contribution of higher multiplicity states that is suppressed by 
phase space and ordinary chiral counting. 
The procedure will not reduce to the computation of self-energy diagrams. The counting in the $EFT$ will rule what effective vertices are
 to be used to obtain the relevant contributions to the off-shell width.\\
\hspace*{0.5cm}The effective vertices that will contribute to $\pi\,\pi$ scattering and to the pion vector form factor, are those corresponding
 to an external vector current coupled to two $pGs$, and to a vector transition in the $s$ channel contributing to the four $pGs$-vertex.
 The construction of the effective vertices goes as sketched in Figure \ref{effverticesforvectortransitions}
 where, at the lowest chiral order, the local vertices on the RHS of the equivalence are provided by the $\cO(p^2)$ $\CPT$ Lagrangian. The
 diagrams contributing to the physical observables will be constructed taking into account all possible combinations of these two 
effective vertices.\\
\hspace*{0.5cm}In Ref.~\cite{GomezDumm:2000fz}, it was proposed to construct a Dyson-Schwinger-like equation through a perturbative loop 
expansion. At tree level, one has to take into account the amplitude provided by Figs.\ref{diagsVFFpiRCT}(a) and \ref{diagsVFFpiRCT}(b)
, that is, the effective vertex in Figure \ref{effverticesforvectortransitions}
. For the one-loop corrections, we are only interested in those contributions with absorptive parts in the $s$ channel, generated by
 inserting a $pG$-loop using the two effectives vertices in Figure \ref{effverticesforvectortransitions} which leads to the four 
contributions in Figs. \ref{diagsVFFpiRCT}(c), \ref{diagsVFFpiRCT}(d), \ref{diagsVFFpiRCT}(e) and \ref{diagsVFFpiRCT}(f). In this way
 the computation is complete up to two loops. The resulting infinite series happens to be geometric and its resummation gives
\begin{equation} \label{resummedVFF}
F_V(q^2)\,=\,\frac{M_V^2}{M_V^2\left[1\,+\,2\frac{q^2}{F^2} \Re e \overline{B_{22}}\right]\,-\,q^2\,-\,i\,M_V\,\Gamma_\rho(q^2) }\,,
\end{equation}
where $M_V$ is the common mass for all the multiplet of vector mesons in the chiral limit,
\begin{equation}
\overline{B_{22}}\, \equiv B_{22} \, (q^2,\,m_\pi^2,\,m_\pi^2)\,+\,\frac{1}{2}\, B_{22} (q^2,\,m_K^2,\,m_K^2)\,,
\end{equation}
and $B_{22} (q^2,\,m_K^2,\,m_K^2)$ is defined through
\begin{equation}
\int \frac{\mathrm{d}^D\ell}{i(2\pi)^D}\,\frac{\ell_\mu \ell_\nu}{\left[ \ell^2\,-\,m_K^2\right] \left[ (\ell-q)^2\,-\,m_K^2\right] } \, \equiv \,q_\mu q_\nu\, B_{21}\,+\,q^2 g_{\mu\nu}\,B_{22}\,,
\end{equation}
as
\begin{eqnarray} \label{B22}
B_{22}\, (q^2,\,m_\pi^2,\,m_\pi^2)& = & \frac{1}{192\pi^2}\,\left[ \left(1\,-\,6\frac{m_\pi^2}{q^2}\right)\left[ \lambda_{\infty}\,+\,\mathrm{ln}\left(\frac{m_\pi^2}{\mu^2}\right)\right]\right.\nonumber\\
& & \left.+\,8\frac{m_\pi^2}{q^2}\,-\,\frac{5}{3}\,+\,\sigma_\pi^3 \mathrm{ln}\left(\frac{\sigma_\pi\,+\,1}{\sigma_\pi\,-\,1}\right) \right]\,,
\end{eqnarray}
where $\sigma_P\,=\,\sqrt{1-\frac{4m_P^2}{q^2}}$ and $\lambda_\infty\,=\,\left[ \frac{2}{D-4}\right] \mu^{D-4}\,-\,\left[ \Gamma'(1)\,+\,\mathrm{ln}(4\pi)\,+\,1\right]$.\\
\hspace*{0.5cm}The $q^2$-dependent width of the $\rho^0$ (770) meson is given by
\begin{eqnarray} \label{rhowidth}
\Gamma_\rho(q^2) & = & -2M_V\frac{q^2}{F^2}\,\Im m \,\overline{B_{22}}\nonumber\\
& = & \frac{M_V\,q^2}{96\,\pi\,F^2}\left[\sigma_\pi^3\,\theta(q^2\,-\,4m_\pi^2)\,+\,\frac{1}{2}\, \sigma_K^3\,\theta(q^2\,-\,4m_K^2)\right]\,,
\end{eqnarray}
in complete agreement with the expression in Ref.~\cite{Guerrero:1997ku}.\\
\\
\\
\hspace*{0.5cm}The real part of the pole of $F_V(q^2)$ in Eq. (\ref{resummedVFF}) needs still to be regulated through wave function and 
mass renormalization of the vector field. The local part of $\Re e B_{22}$ can be fixed by matching Eq. (\ref{resummedVFF}) with the 
$\cO(p^4)$ $\CPT$ result.\\
\hspace*{0.5cm}An analogous procedure can be applied to the study of the vector component of $\pi\,\pi$ scattering. We will be concerned
 about the $s$-channel amplitude of $\pi^+\,\pi^-\,\to\,\pi^+\pi^-$, that is dominated by $\rho$ exchange, so that one can construct a 
Dyson-Schwinger equation as in the case of the pion form factor. Consequently, analogous diagrams to those in Figure \ref{diagsVFFpiRCT} are
 considered, replacing external vector current insertions by two pion legs, according to all possible contributions in Figure \ref{effverticesforvectortransitions}
. Projecting the $p$ wave, it is found a geometric series, which can be resummed to give
\begin{eqnarray}
\mathcal{A}(\pi^+\,\pi^-\,\to\,\pi^+\pi^-)|_{J=1}& = & \frac{-i}{2F^2}\,(u-t)\\
& & \times \frac{M_V^2}{M_V^2\left[1\,+\,2\frac{q^2}{F^2} \Re e \overline{B_{22}}\right]\,-\,q^2\,-\,i\,M_V\,\Gamma_\rho(q^2)}\,,\nonumber
\end{eqnarray}
where $u$ and $t$ are the usual Mandelstam variables ($q^2\,=\,s$). Remarkably, the pole of the amplitude coincides with the one obtained
 for the vector form factor of pion and, therefore, gives the same width for the $\rho^0$ meson.\\
\hspace*{0.5cm}When one applies the definition proposed at the beginning of the section for spin-one meson widths to the case of the $\rho^0$
 (770), its quantum numbers correspond to $j\,=\,k\,=3$ for the flavour index. Lorentz covariance and current conservation allow to define
 the two-point function of the considered vector current in terms of an invariant function of $q^2$ through
\begin{eqnarray}
\Pi^{33}_{\mu\nu}\,=\,(q^2\,g_{\mu\nu}\,-\,q_\mu\,q_\nu)\,\Pi^\rho(q^2)\,,\nonumber\\
\\
\Pi^\rho(q^2)\,=\,\Pi^\rho_{(0)}\,+\,\Pi^\rho_{(1)}\,+\,\Pi^\rho_{(2)}\,,\,\dots\nonumber\,,
\end{eqnarray}
where $\Pi^\rho_{(0)}$ corresponds to the tree level contribution of Figure \ref{diagsVVcorrelRCT} (a), $\Pi^\rho_{(1)}$ to the one-loop 
amplitudes and so forth. Up to one loop, and considering again the two-particle absorptive contributions only, all the diagrams generated
 by the effective vertices in Figure \ref{effverticesforvectortransitions} are shown in Figure \ref{diagsVVcorrelRCT}. One finds, in the 
isospin limit,
\begin{eqnarray}
\Pi^\rho_{(0)} & = & \frac{F_V^2}{M_V^2\,-\,q^2}\,,\\
\Pi^\rho_{(1)} & = & \Pi^\rho_{(0)}\left[ -\frac{M_V^2}{F_V^2}\,\frac{M_V^2}{M_V^2\,-\,q^2}\,4\overline{B_{22}}\right]\,. 
\end{eqnarray}
\\
\begin{figure}[h]
\centering
\includegraphics[scale=1.5]{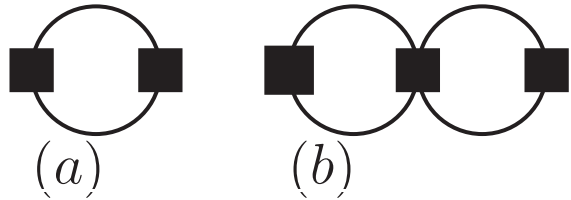}
\caption{One- and two-loop diagrams leading to $\Pi^\rho_{(1)}$ (a) and $\Pi^\rho_{(2)}$ (b). The effective squared vertices are given in Figure \ref{effverticesforvectortransitions}.}
\label{1L2LdiagsforPi1Pi2}
\end{figure}
\\
\hspace*{0.5cm}At this point, one realizes that the resummation procedure cannot consist only of self-energy diagrams. $QCD$ predicts 
the two-point spectral function of vector currents to go to a constant value as $q^2\,\to\,\infty$ \cite{Floratos:1978jb}. The loop diagram
 in Figure \ref{diagsVVcorrelRCT} (b) behaves itself as a constant value in this limit, which is against of expectations because it corresponds
 to only one of the infinite number of possible intermediate states. In order to satisfy the $QCD$-ruled behaviour, one would foresee that
 all the individual (positive) contributions from the intermediate states should vanish in the infinite $q^2$ limit. Indeed, this is 
achieved when one adds the diagrams depicted in Figs.\ref{diagsVVcorrelRCT}(c), \ref{diagsVVcorrelRCT}(d) and \ref{diagsVVcorrelRCT}(e).
 The requirement of vanishing at $q^2\,\to\,\infty$ for the $\Pi^\rho_i$ is also fulfilled for $i\geq2$ provided one considers, at a 
given order, all possible diagrams with absorptive contributions in the $s$-channel, and not just self-energies.\\
\hspace*{0.5cm}Iterating in all possible ways the one-loop diagrams in Figure \ref{diagsVVcorrelRCT}, one obtains all possible contributions 
to the two-loop computation, as sketched in Figure \ref{1L2LdiagsforPi1Pi2}. The result of the calculation was found to be
\begin{equation}
\Pi^\rho_{(2)} \, = \, \Pi^\rho_{(1)}\left[ -\frac{q^2}{F_V^2}\,\frac{M_V^2}{M_V^2\,-\,q^2}\,4\overline{B_{22}}\right]\,. 
\end{equation}
\hspace*{0.5cm}In Ref. \cite{GomezDumm:2000fz}, it was checked explicitly up to three loops that the invariant two-point function 
$\Pi^\rho(q^2)$, generated by resuming effective loop diagrams with an absorptive amplitude in the $s$ channel \footnote{Note that the 
procedure employed implies that the only significant result of $\Pi^\rho(q^2)$ is its imaginary part.} is perturbatively given by
\begin{eqnarray}
\Pi^\rho(q^2) & = & \Pi^\rho_{(0)}\,+\,\, \Pi^\rho_{(1)}\sum_{n=0}^{\infty} \left[ -\frac{q^2}{F_V^2}\,\frac{M_V^2}{M_V^2\,-\,q^2}\,4\overline{B_{22}}\right]^n\nonumber\\
& & = \Pi^\rho_{(0)}\,\left[ 1\,+\,\omega \sum_{n=0}^{\infty}\left( \frac{q^2}{M_V^2}\,\omega\right)^n \right]\,, 
\end{eqnarray}
where
\begin{equation} \label{omega}
\omega\,=\,-\frac{M_V^2}{F_V^2}\,\frac{M_V^2}{M_V^2\,-\,q^2}\,4\overline{B_{22}}\,.
\end{equation}
Using that $F_V^2\,=\,2F^2$, and substituting (\ref{omega}) for performing the resummation, one finally gets
\begin{eqnarray} \label{VVcorrel}
\Pi^\rho(q^2) & = & \frac{2F^2}{M_V^2\left[1\,+\,2\frac{q^2}{F^2} \Re e \overline{B_{22}}\right]\,-\,q^2\,-\,i\,M_V\,\Gamma_\rho(q^2)}\nonumber\\
& & \times \left[1\,-\,2\frac{M_V^2}{F^2} \overline{B_{22}}\right]\,,
\end{eqnarray}
where the off-shell $\rho^0$ width $\Gamma_\rho(q^2)$ is given by (\ref{resummedVFF}). The consistency of the resummation procedure shows 
up neatly because the residue in $\Pi^\rho(q^2)$ satisfies the required unitarity condition
\begin{eqnarray}
\Im m\Pi^\rho(q^2) & = & \frac{1}{48\pi}\,\left[ \sigma_\pi^3\,\theta(q^2\,-\,4m_\pi^2)\,+\,\frac{1}{2}\,\sigma_K^3\,\theta(q^2\,-\,4m_K^2)\right]\nonumber\\
& & \times |F_V(q^2)|^2\,, 
\end{eqnarray}
with $F_V(q^2)$ given by (\ref{resummedVFF}).\\
\hspace*{0.5cm}The last comment to be made concerns the independence of the definition of the spin-one meson width on the chosen 
representation for the fields. To see this, it is enough to realize that the effective vertices in Figure \ref{effverticesforvectortransitions}
 are universal. Different theoretical descriptions of the spin-one mesons lead to resonance-exchange contributions that differ by local 
terms. Since the physical amplitudes are required to satisfy the $QCD$-ruled behaviour at short distances, this difference is necessarily 
counterbalanced by explicit local terms \cite{Ecker:1989yg}. Including these local terms in the local vertices of Figure \ref{effverticesforvectortransitions}
, the resulting effective vertices (which are the building blocks of the described resummation) are formulation independent and thus, the
 whole procedure is.\\
\subsection{$K^*$ off-shell width} \label{widthApp_K*}
\hspace*{0.5cm}Being the definition of spin-one resonance width completely general, what is left now is simply to employ it for any 
resonance we are interested in. In particular, for the case of the $K^*$ resonance, we have
\begin{eqnarray} \label{K*width}
\Gamma_{K^*}(q^2) & = & \frac{M_{K^*}\,q^2}{128\pi F^2}\,\left[ \lambda^{3/2}\left(1,\,\frac{m_K^2}{q^2},\,\frac{m_\pi^2}{q^2}\right)\,\theta \left(q^2\,-\,(m_K\,+\,m_\pi)^2\right)\right.\nonumber\\
& &\left. +\,\lambda^{3/2}\left(1,\,\frac{m_K^2}{q^2},\,\frac{m_\eta^2}{q^2}\right)\,\theta \left(q^2\,-\,(m_K\,+\,m_\eta)^2\right)\right]\,,
\end{eqnarray}
in agreement with \cite{Jamin:2006tk}.\\
\subsection{$\omega$-$\phi$ off-shell width}\label{widthApp_omegaphi}
\hspace*{0.5cm}The full widths PDG \cite{Amsler:2008zzb} reports for the vector resonances we are interested in are: $\Gamma_\rho\,=\,149$.$4\pm1$.$0$
 MeV, $\Gamma_\omega\,=\,8$.$49\pm0$.$8$ MeV, $\Gamma_\Phi\,=4$.$26\pm0.04\,$ MeV,  and $\Gamma_{K^*}\,\sim\,50$.$5\pm1$.$0$ MeV. Based on this, we have decided
 to neglect the off-shell width of the isospin zero resonances $\omega$ (782) and $\Phi(1020)$, because it is a tiny effect compared both
 to that of the $\rho$ (770) and $K^*$ (892) widths and to the uncertainties we still have in the determination of the coupling constants
 or the error introduced by other approximations. We have used the values reported as the constant $\omega$ (782) and $\Phi$ (1020) widths
 in our study.\\
\chapter*{Appendix D: $\omega\rightarrow \pi^+\pi^-\pi^0$ within $R\chi T$}
\label{omegaApp}
\addcontentsline{toc}{chapter}{Appendix D: $\omega\rightarrow \pi^+\pi^-\pi^0$ within $R\chi T$}
\appendix
\pagestyle{appendixd}
\newcounter{D}
\renewcommand{\theequation}{\Alph{D}.\arabic{equation}}
\renewcommand{\thefigure}{\Alph{D}.\arabic{figure}}
\setcounter{D}{4}
\setcounter{equation}{0}
\setcounter{figure}{0}
\hspace*{0.5cm}The decay of the $\omega$(782) into three pions, $\omega \rightarrow \pi^{+}(k_1)\, \pi^{-}(k_2) \, \pi^{0}(k_3)$, has been
 a useful source of information on the odd-intrinsic parity couplings. Within the framework of $\RCT$ it was first studied in Ref.~\cite{RuizFemenia:2003hm},
 where the contribution of the $VVP$ vertices was already found and the need to account for $VPPP$ vertices was put forward \footnote{This process has been 
studied recently in the vector formalism \cite{Gudino:2011ri}.}.\\
\hspace*{0.5cm}We will denote the polarization vector of the $\omega$ as $\varepsilon_\omega^\sigma$ and use the kinematic invariants 
$s_{ij}\,=\,(k_i\,+\,k_j)^2$. In our work we have included for the first time the contribution of the decay via a direct vertex.\\
\hspace*{0.5cm}The amplitude associated to the diagram of Figure~\ref{fig1_omega3pi} -that should be the leading one according to vector meson dominance-
including cyclic permutations among $k_1,\,k_2$ and $k_3$, reads
\begin{figure}[h!]
\begin{center}
\hspace*{-0.5cm}
\includegraphics[scale=0.90]{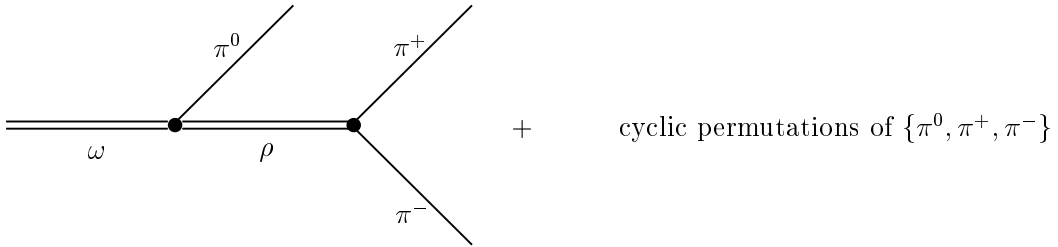}
\caption[]{\label{fig1_omega3pi} The $\omega\to\pi^+\pi^-\pi^0$ decay amplitude
via an intermediate $\rho$ exchange.}
\end{center}
\end{figure}

\begin{eqnarray}
i\,{\cal M}_{\omega\to3\pi}^{\mathrm{VMD}}=i\,\epsilon_{\alpha\beta\rho\sigma}
\,k_1^{\alpha}k_2^{\beta}k_3^{\rho}\epsilon^{\sigma}_{\omega}
\frac{8\,G_V}{M_{\omega}F^3}\,\Bigg[&&\!\!\!\!\!\!\!\!
\frac{m^2_{\pi}(d_1+8d_2-d_3)
+(M_{\omega}^2+s_{12}) \, d_3}{M_V^2-s_{12}}
\nonumber\\[3mm]
&&\!\!\!\!\!\!\!\!+\,\{s_{12}\to s_{13}\}+\{s_{12}\to s_{23}\}\,
\Bigg]
\,.
\label{eq:w-3pi}
\end{eqnarray}
Now, the contribution via a direct vertex $VPPP$ yields
\begin{eqnarray}
 i\,\mathcal{M}_{\omega \rightarrow \pi^+ \pi^- \pi^0}^{VPPP} \, = \, i\,\varepsilon_{\alpha\beta\rho\sigma}k_1^\alpha k_2^\beta k_3^\rho \varepsilon_\omega^\sigma\, & &  \left\lbrace \frac{8G_V}{M_\omega F^3}  \frac{8 \sqrt{2}}{F^3 M_\omega M_V}\left[ (g_1\,-\,g_2\,-\,g_3)(M_\omega^2\,-\,3\,m_\pi^2)\,\right.\right.\nonumber\\
& & \left.\left. +\,3\,m_\pi^2(2\,g_4\,+\,g_5)\right] \right\rbrace\,.
\end{eqnarray}
In the above expression, we have assumed ideal mixing between the states $ \mid \omega_8\rangle $ and $ \mid \omega_1 \rangle$:
\begin{equation}
 \mid \omega \rangle\,=\,\sqrt{\frac{2}{3}}  \mid \omega_1\rangle+\sqrt{\frac{1}{3}}  \mid \omega_8\rangle \,,
\end{equation}
and
\begin{equation}
 \mid \phi \rangle\,=\,-\sqrt{\frac{1}{3}}  \mid \omega_1 \rangle  +\sqrt{\frac{2}{3}}  \mid \omega_8\rangle\,.
\end{equation}
The relation between the amplitudes of the singlet and octet states is the following one:
\begin{equation}
\mathcal{M}_{\omega_1 \rightarrow \pi^+ \pi^- \pi^0} \, =\,\sqrt{2}\,\mathcal{M}_{\omega_8 \rightarrow \pi^+ \pi^- \pi^0}\,.
\end{equation}
\hspace*{0.5cm}The decay width is then obtained as
\begin{eqnarray}
\label{eq:Gamma_w3pi} 
\Gamma(\omega\to \pi^+ \pi^- \pi^0) & = & 
\frac{G_V^2}{4 \, \pi^3 \, M_{\omega}^5 \, F^6}\,
\, \int_{4m_{\pi}^2}^{(M_\omega-m_{\pi})^2}ds_{13}
\int_{s_{23}^{min}}^{s_{23}^{max}}ds_{23}\,{\cal P}(s_{13},s_{23}) \, \times
\\[3mm]
&& \times\Bigg[  \, \frac{m^2_{\pi}(d_1+8d_2-d_3)
+(M_{\omega}^2+s_{12}) \, d_3}{M_V^2-s_{12}}
+\,\{s_{12}\to s_{13}\}+\{s_{12}\to s_{23}\}
\, \nonumber\\
&& + \frac{8 \sqrt{2}}{F^3 M_\omega M_V}\left[ (g_1\,-\,g_2\,-\,g_3)(M_\omega^2\,-\,3\,m_\pi^2)\, +\,3\,m_\pi^2(2\,g_4\,+\,g_5) \right]
\Bigg]^2\,, \nonumber 
\end{eqnarray}
where the function ${\cal P}$ is the polarization average of the tensor structure of ${\cal M}_{\omega\to3\pi}$, 
\begin{equation}
{\cal P}(s_{13},s_{23})=\frac{1}{12}\left\{ -m_{\pi}^2(m_{\pi}^2-
M_{\omega}^2)^2-
s_{13}s_{23}^2+(3m_{\pi}^2+M_{\omega}^2-s_{13})s_{13}s_{23}\right\}
\,.
\end{equation} 
With $G_V = F / \sqrt{2}$ and the relations obtained by the short-distance matching. In the analyses of the $VVP$ Green function all couplings 
appearing in Eq. (\ref{eq:w-3pi}) were predicted
\begin{eqnarray}
 d_1 + 8 \, d_2&=& -\frac{N_C}{64\pi^2}\frac{M_V^2}{F_V^2} \, + 
\, \frac{F^2}{4F_V^2}  \; \; \; ,
\nonumber\\[3mm]
d_3&=&-\frac{N_C}{64\pi^2}\frac{M_V^2}{F_V^2} \, + \, \frac{F^2}{8F_V^2}\,.
\label{eq:cond}
\end{eqnarray}
\hspace*{0.5cm}Taking just this piece of the amplitude into account, one obtains a decay width that is only one fifth \cite{RuizFemenia:2003hm} of the experimental value.\\
\hspace*{0.5cm}Additionally, the following relations were obtained studying the decays $\tau^-\to (KK\pi)^- \nu_\tau$ \cite{Dumm:2009kj, Roig:2007yp}:
\begin{eqnarray} \label{morereliable}
d_3 & = & - \, \frac{N_C}{192 \pi^2} \frac{M_V^2}{F_V \,G_V} \, , \nonumber  \\
g_1\,+\,2\,g_2\,-\,g_3 & = & 0 \; , \nonumber \\
g_2\, & = & \, \frac{N_C}{192 \pi^2}\, \frac{M_V}{\sqrt{2} \,F_V} \; .
\end{eqnarray}
As explained in Chapter \ref{KKpi} we find more reliable the determination of the coupling $d_3$ in Eq.(\ref{morereliable}), that we will follow. 
Taking all these pieces of information into account we are able to match the experimental value reported by the $PDG$ \cite{Amsler:2008zzb} 
$\Gamma(\omega\to \pi^+ \pi^- \pi^0)|_{\mathrm{exp}}= (7$.$57\, \pm \, 0$.$06) \,\mathrm{MeV}$ with $2g_4+g_5=-0$.$60\pm0$.$02$ that has been used 
in the hadronic tau decays studied in this Thesis.\\
\chapter*{Appendix E: Isospin relations between $\tau^-$ and $e^+e^-$ decay channels}
\label{isosApp}
\addcontentsline{toc}{chapter}{Appendix E: Isospin relations between $\tau^-$ and $e^+e^-$ decay channels}
\appendix
\pagestyle{appendixe}
\newcounter{E}
\renewcommand{\thesection}{\Alph{E}.\arabic{section}}
\renewcommand{\thesubsection}{\Alph{E}.\arabic{subsection}}
\renewcommand{\theequation}{\Alph{E}.\arabic{equation}}
\setcounter{E}{5}
\setcounter{section}{0}
\setcounter{subsection}{0}
\setcounter{equation}{0}
\section{Introduction} \label{isosApp_Intro}
\hspace*{0.5cm}In this appendix we provide the derivation of several relations between $\tau^-$ and $e^+e^-$ decay channels that are related 
by an isospin rotation. Since $SU(2)$ is a very accurate symmetry whose violations are smaller than the typical errors of the experimental 
measurements and our theoretical assumptions, the conclusions we draw should hold. At the low energies we are interested in, one can safely 
neglect the $Z$ contributions to the hadron $e^+e^-$ cross-section. In this limit, the process will only be due to vector current via photon 
exchange. Therefore, the relations that we obtain will relate $\sigma_{e^+e^-\to \rm{hadrons}}$ to the vector current contribution in the corresponding
 tau decay. Depending on the channel, the importance of the latter will vary from being the only one to be forbidden by symmetry arguments, 
like $G$-parity. Thus this study will be interesting for some of the channels and irrelevant for others~\footnote{As two immediate examples 
of this we quote the three pion tau decay channel where there is no vector current contribution because of $G$-parity and the $3 \eta$ state that 
cannot be a decay product of the $\tau$.}.\\
\hspace*{0.5cm}In this introduction we will first give the conventions we follow for the relations between one-particle charge and isospin 
states and the ladder operators. Then we will recall the general formula for the tau decay width into a given final state of three mesons 
and the tau neutrino and give the derivation of the analogous expression for the $e^+e^-$ cross-section into a three-hadron system. We will 
finish this section with the projection of the weak and electromagnetic currents into its isospin components. Charge conjugated relations 
are understood and most of the times not written.\\
\hspace*{0.5cm}The triplet of pions is related to the isospin states, $|I,I_3\ket$, in the following way:
\begin{equation}
 \pi^+ \sim \overline{d}u \sim -|1,+1\ket\,,\quad
 \pi^0 \sim \frac{\overline{u}u-\overline{d}d}{\sqrt{2}} \sim |1,0\ket\,,\quad
 \pi^- \sim \overline{u}d \sim |1,-1\ket\,.
\end{equation}
\hspace*{0.5cm}It is important to understand that the four kaon states group into two doublets according to its strangeness:
\begin{equation}
 K^+ \sim \overline{s}u \sim \Big| \frac{1}{2},+\frac{1}{2}\Big>\,,\quad
 K^0 \sim \overline{s}d \sim \Big| \frac{1}{2},-\frac{1}{2}\Big>\,,
\end{equation}
and
\begin{equation}
  \overline{K}^0 \sim \overline{d}s \sim -\Big| \frac{1}{2},+\frac{1}{2}\Big>\,,\quad
  K^- \sim \overline{u}s \sim \Big| \frac{1}{2},-\frac{1}{2}\Big>\,.
\end{equation}
\hspace*{0.5cm}Defining the isospin operators $T_{\pm}=\frac{T_1\pm
iT_2}{\sqrt{2}}$, we get the following relations:
\begin{equation} \label{rels_iso_ops_1}
 \left[T_+,T_-\right] \,=\,T_3,\qquad
 \left[T_+,u\right]\,=\,-\frac{d}{\sqrt{2}},\qquad
 \left[T_+,d^\dagger\right]\,=\,\frac{u^\dagger}{\sqrt{2}},
\end{equation}
\begin{equation}\label{rels_iso_ops_2}
T_+|I,+I\ket = T_-|I,-I\ket=0\ ,
\end{equation}
\begin{equation}\label{rels_iso_ops_3}
 T^+\overline{d}u|0\ket=\left[ T_+,\overline{d}u\right]|0\ket
  =\frac{\overline{u}u-\overline{d}d}{\sqrt{2}}|0\ket\,\,,
\end{equation}
that will allow to relate the neutral and charged current weak decays.\\
\\
\hspace*{0.5cm}Before analysing the most interesting channels, we will need Eq.~(\ref{fulldGammadQ2vanishingnumass}) for the tau decay width
 into a given three meson and a tau-neutrino final state and the analogous formula for the $e^+e^-$ cross-section into a three hadron final 
state. The latter is obtained in the following.\\
\hspace*{0.5cm}We consider the decay $e^+(\ell_1,s_1)e^-(\ell_2,s_2)\to h_1(p_1) h_2(p_2) h_3(p_3)$. The amplitude of the process is splitted 
into its lepton and hadron tensors as in Chapter \ref{Hadrondecays}.One has
\begin{equation}
 \mathcal{L}_{\mu\nu}=\sum_{s_1,s_2}\overline{v}(\ell_2,s_2)\gamma_\mu u(\ell_1,s_1) \overline{u}(\ell_1,s_1)\gamma_\nu v(\ell_2,s_2)\,.
\end{equation}
Its transverse projection reads ($Q\,=\,p_1+p_2+p_3\,=\,\ell_1+\ell_2$)
\begin{equation}
\left( g^{\mu\nu}-\frac{Q^\mu Q^\nu}{Q^2}\right) \mathcal{L}_{\mu\nu} = -4 \,(Q^2\,+\,2\,m_e^2)\,.
\end{equation}
On the other hand, the hadron tensor is decomposed as
\begin{equation}
 \mathcal{H}^{\mu\nu}=S\left( Q^2,\,s,\,t\right)\frac{Q^\mu Q^\nu}{Q^2}+\left( g^{\mu\nu}-\frac{Q^\mu Q^\nu}{Q^2}\right)V\left( Q^2,\,s,\,t\right)\,.
\end{equation}
Since the process is mediated by vector current only the scalar component vanishes ($S=0$) and we have
\begin{equation}
 V\left( Q^2,\,s,\,t\right)=\frac{\mathcal{H}^{\mu\nu}}{3}\left( g^{\mu\nu}-\frac{Q^\mu Q^\nu}{Q^2}\right)\,.
\end{equation}
Using the expression for $\int\mathrm{d}\Pi_3$ in Eq.~(\ref{d3Pi}) one immediately has
\begin{eqnarray}
\sigma_{e^+e^-\to h_1h_2h_3}(Q^2) & = & \frac{1}{2^2}\left( \frac{e^2}{Q^2}\right)^2 \,\frac{1}{2\lambda^{1/2}(Q^2,m_e^2,m_e^2)}\,\frac{1}{32\pi^3}
\,\frac{1}{4Q^2}\int\,\mathrm{d}s\mathrm{d}t  \times \nonumber\\
& & \frac{4}{3}\,(Q^2\,+\,2\,m_e^2)\,|F_3|^2\,\left(-V_{3\mu} V_3^{\mu *}\right) \nonumber\\
& \sim & \frac{\alpha^2}{48\pi}\,\frac{1}{Q^6}\,\int\,\mathrm{d}s\mathrm{d}t\,|F_3|^2\,\left(-V_{3\mu} V_3^{\mu *}\right)\,.
\end{eqnarray}
\hspace*{0.5cm}So that the $e^+e^-$ cross-section into three hadrons is given by
\begin{equation} \label{e+e-xsect3hads}
\sigma_{e^+e^-\to h_1h_2h_3}(Q^2)=\frac{e^4}{768\,\pi^3}\frac{1}{Q^6}\int
\mathrm{d}s \mathrm{d}t |F_3|^2 \left(-V_{3\mu} V_3^{\mu *}\right)\,.
\end{equation}
This way one ends up with the desired relation
\begin{equation}
 \frac{\mathrm{d}\Gamma\left( \tau^-\to (3 h)^-\nu_\tau\right) }{\mathrm{d}Q^2}\,=\,f(Q^2)\,\sigma_{e^+e^-\to (3 h)^0}(Q^2)|_{I=1}\,,
\end{equation}
where $(3h)^-$ and $(3h)^0$ are related via an isospin rotation and
\begin{equation} \label{fQ2}
 f(Q^2)=\frac{G_F^2 |V_{ij}^{CKM}|^2}{384(2\pi)^5
M_\tau}\left(\frac{M_\tau^2}{Q^2}-1\right)^2\left(1+2\frac{Q^2}{M_\tau^2}\right)
\left(\frac{\alpha^2}{96\pi}\right)^{-1}Q^6\, .
\end{equation}
\\
\hspace*{0.5cm}The $W$ bosons can couple to $\overline{u}s$, $\overline{s}u$ (that are two components of different multiplets with $I=1/2$), 
and to $\overline{d}u$, $\overline{u}d$. Both have $I=1$ and differ by a relative global sign.\\
\hspace*{0.5cm}The $Z$ boson can couple to the $I=0,1$ combinations that are $\frac{\overline{u}u\mp \overline{d}d}{\sqrt{2}}$
 where the upper signs correspond to the neutral component of $I=1$.\\
\hspace*{0.5cm}Now let us consider the electromagnetic current. One can decompose it into its $I=0$ and $I=1$ pieces:
\begin{equation}
 \Gamma^\mu=\frac{1}{3}\left( 2\overline{u}\gamma^\mu
u-\overline{d}\gamma^\mu d-\overline{s}\gamma^\mu s\right) =
\Gamma^\mu_{(0)}+\Gamma^\mu_{(1)}\,,
\end{equation}
where
\begin{equation} \label{Gammas_em}
 \Gamma^\mu_{(0)}=\frac{1}{6}\left(\overline{u}\gamma^\mu
u+\overline{d}\gamma^\mu d-2\overline{s}\gamma^\mu
s\right)\,,\quad\Gamma^\mu_{(1)}=\frac{1}{2}\left(\overline{u}\gamma^\mu
u-\overline{d}\gamma^\mu d\right)\,.
\end{equation}
\hspace*{0.5cm}As it is well-known, there are no tree level $FCNC$. Moreover, the electromagnetic current conserves strangeness. This implies 
that the strangeness changing channels $K\pi\pi$, $K\pi\eta$, $K\eta\eta$, $KKK$ can only be reached via $W^\pm$-mediated loops. Therefore, they are
 very much suppressed -even more at the low-energies we are interested in- and it makes no sense to analyze them in this context~\footnote{A 
complementary reasoning in terms of isospin can also be made: Neither the $Z$ nor the $\gamma$ couple to $I=1/2$. This prevents a study of 
this type for the $|\eta\eta K\ket\sim| K\ket$ state -with $I=1/2$- because there is only one accesible state in $\tau^-$ decays, in addition,
 so that no relation can be established. Similarly, the three kaon state has half-integer $I$, so it can only be produced in $\tau$ decays.
 There is only a trivial isospin relation $A_{+--}=A_{0\overline{0}-}$ in this case. The notation employed uses as subscripts the electric charges of the 
particles involved for a given mode. In this case, for instance, this would correspond to $A_{K^+K^-K^-}=A_{K^0\bar{K}^0K^-}$.}. This short-hand writing will 
be used in the remainder of the chapter.\\
\section{$KK\pi$ channels} \label{isosApp_KKpi}
\hspace*{0.5cm}One must realize that in all charge channels both kaons belong to different isospin multiplets because they have opposite 
strangeness. At the practical level this implies that there is an additional label implying that the ordering when writing the states is 
irrelevant. We give an example to illustrate this: $|\pi^+ \pi^-\ket$ and $|\pi^- \pi^+\ket$ are different isospin states corresponding to 
$| 1,+1 \ket\otimes |1,-1 \ket$ and $|1,-1 \ket \otimes | 1,+1 \ket$, respectively. However, $K^- K^0$ is $|-1, \frac{1}{2}, 
-\frac{1}{2}\ket\otimes |1,\frac{1}{2},-\frac{1}{2} \ket$, where the first label is the strangeness of the state ($-1$ for $s$) making 
manifest that they belong to different subspaces.\\
\hspace*{0.5cm}We will consider first the product of the two kaons states. This can give either the isoscalar channel '$\omega$' or the 
isovector channel '$\rho$'. Then we will consider the product of the produced states with the remaining $\pi$. Some signs may vary by 
considering first the product of one of the kaons to the pion and then that of the resulting states with the kaon left. However, the 
relations we will find are independent of the procedure we follow.\\
\hspace*{0.5cm}The pair of kaons can couple as
\begin{eqnarray} \label{kaon_otimes_kaon}
\frac{1}{\sqrt{2}}\,\left(K^+K^-+K^0\overline{K}^0\right) & \quad & I=0\,,\nonumber\\
-K^+\overline{K}^0 & \quad & I=1,\,I_3=+1\,,\nonumber\\
\frac{1}{\sqrt{2}}\,\left(K^+K^--K^0\overline{K}^0\right) & \quad & I=1,\,I_3=0\,,\nonumber\\
K^0 K^- & \quad & I=1,\,I_3=-1\,.
\end{eqnarray}
\hspace*{0.5cm}Now we consider the direct product of the '$\rho$' and '$\omega$' states with the appropriate pion. We will have $I=1$ in 
'$\omega$' channel and $I=0,\,1,\,2$ in '$\rho$' channel. One has
\begin{eqnarray}
\frac{1}{\sqrt{2}}\left(K^+ K^- \pi^+ + K^0 \overline{K}^0\pi^+\right) & \quad & I=1,\,I_3=+1\,,\nonumber\\
\frac{1}{\sqrt{2}}\left(K^+ K^- \pi^0 + K^0 \overline{K}^0\pi^0\right) & \quad & I=1,\,I_3=0\,,\nonumber\\
\frac{1}{\sqrt{2}}\left(K^+ K^- \pi^- + K^0 \overline{K}^0\pi^-\right) & \quad & I=1,\,I_3=-1\,,
\end{eqnarray}
in $\omega$ channel, while the states produced in $\rho$ channel are
\begin{eqnarray}
\frac{1}{\sqrt{3}}\left[K^0 K^- \pi^+ - \frac{1}{\sqrt{2}} \left(K^+ K^- \pi^0 - K^0 \overline{K}^0\pi^0\right) - K^+ \overline{K}^0 \pi^-\right] & \quad & I=0,\,I_3=0\,,\nonumber\\
\frac{1}{\sqrt{2}}\left[-K^+ \overline{K}^0 \pi^0 + \frac{1}{\sqrt{2}} \left(K^+ K^- \pi^+ - K^0 \overline{K}^0\pi^+\right) \right] & \quad & I=1,\,I_3=+1\,,\nonumber\\
\frac{1}{\sqrt{2}}\left[-K^+ \overline{K}^0 \pi^- - K^0 K^-\pi^+\right] & \quad & I=1,\,I_3=0\,,\nonumber\\
\frac{1}{\sqrt{2}}\left[ \frac{1}{\sqrt{2}} \left(K^+ K^- \pi^- - K^0 \overline{K}^0\pi^-\right)-K^0 K^-\pi^0 \right] & \quad & I=1,\,I_3=-1\,,\nonumber
\end{eqnarray}
\vspace*{2.0cm}
\begin{eqnarray}
-K^+\overline{K}^0\pi^+ & \quad & I=2,\,I_3=+2\,,\nonumber\\
\frac{1}{2}\left(-\sqrt{2}K^+\overline{K}^0\pi^0 + K^+K^-\pi^+ - K^0\overline{K}^0\pi^+\right) & \quad & I=2,\,I_3=+1\,,\nonumber\\
\frac{1}{\sqrt{6}}\left(-K^+\overline{K}^0\pi^- + \sqrt{2} K^+K^-\pi^0 - \sqrt{2} K^0\overline{K}^0\pi^0 + K^0 K^- \pi^+\right) & \quad & I=2,\,I_3=0\,,\nonumber\\
\frac{1}{2}\left(K^+K^-\pi^- - K^0 \overline{K}^0\pi^- + \sqrt{2} K^0 K^- \pi^0\right) & \quad & I=2,\,I_3=-1\,,\nonumber\\
K^0 K^- \pi^- & \quad & I=2,\,I_3=-2\,.\nonumber\\
\end{eqnarray}
\hspace*{0.5cm}Since the operator $\overline{d}\,\Gamma^\mu\, u$ has $I=1$ we have \footnote{$\Gamma^\mu=\gamma^\mu,\gamma^\mu \gamma_5$. Since 
the spinor structure is unrelated to isospin, one has separate relations holding for both for the vector and axial-vector currents. This 
will be understood in what follows.}
\begin{equation}
_{(2,-1)}\bra KK\pi|\overline{d}\Gamma_\mu u|0\ket\,=\,0\,,
\end{equation}
for the charged weak current. Thus
\begin{equation} \label{charged_weak_KKpi_1}
\bra K^+K^-\pi^- -K^0\overline{K}^0\pi^- +\sqrt{2}K^0 K^- \pi^0|\overline{d}\,\Gamma_\mu \,u|0\ket\,=\,0\,.
\end{equation}
If we denote the correponding hadron amplitudes as $A_\mu^{+--}$, $A_\mu^{00-}$ and $A_\mu^{0-0}$, Eq.~(\ref{charged_weak_KKpi_1}) implies
\begin{equation} \label{charged_weak_KKpi_2}
A_\mu^{+--} - A_\mu^{00-}\,=\,-\sqrt{2} A_\mu^{0-0}\,.
\end{equation}
One can proceed analogously for the neutral current weak operator $\frac{\overline{u}\Gamma_\mu u-\overline{d}\Gamma_\mu d}{\sqrt{2}}$. 
Since it carries isospin $I=1$, we have the relations
\begin{equation}
_{(2,0)}\bra KK\pi|\overline{u}u - \overline{d}d|0\ket\,=\,_{(0,0)}\bra KK\pi|\overline{u}u - \overline{d}d|0\ket\,=\,0\,.
\end{equation}
Therefore, writing the isospin states in terms of charge states one has
\begin{eqnarray}
A_\mu^{0-+} - A_\mu^{+0-}\,=\,-\frac{1}{\sqrt{2}} \left(A_\mu^{000} - A_\mu^{+-0}\right)\,=\,0\,,\nonumber\\
A_\mu^{0-+} - A_\mu^{+0-}\,=\,-\sqrt{2} \left(A_\mu^{+-0} - A_\mu^{000}\right)\,=\,0\,,
\end{eqnarray}
where the respective amplitudes were denoted using the same convention for the indices as before.\\
\hspace*{0.5cm}Finally, the relations (\ref{rels_iso_ops_1}), (\ref{rels_iso_ops_2}), (\ref{rels_iso_ops_3}) allow to write
\begin{equation}
_{(1,-1)}\bra KK\pi|\overline{d}\Gamma_\mu u|0\ket=-_{(1,0)}\Big<
KK\pi\Big|\frac{\overline{u}\Gamma_\mu u-\overline{d}\Gamma_\mu d}{\sqrt{2}}\Big|0\Big>\,.
\end{equation}
This yields relations for the hadron amplitudes in '$\rho$' and '$\omega$' channels, between the charged and neutral current weak processes:
\begin{eqnarray}
A_\mu^{0-0} \, = \, A_\mu^{+0-}\,=\,A_\mu^{0-+}\,=\,\frac{1}{\sqrt{2}} \left(A_\mu^{00-} - A_\mu^{+--}\right)\,,\nonumber\\
A_\mu^{+--} + A_\mu^{00-}\,=\,2 A_\mu^{+-0} =\,2 A_\mu^{000}\,.
\end{eqnarray}
\hspace*{0.5cm}Now we consider also the electromagnetic processes with $I=1$. We will factor out the Lorentz structure in the hadron matrix elements:
\begin{eqnarray} \label{factoring_Lorentzstructure_Vectorcurrent}
& & A_\mu^{-}\,\equiv\,\bra (KK\pi)^-|\overline{d}\gamma_\mu u|0\ket\,=\,A^-\epsilon_ {\mu\nu\rho\sigma}p_1^\nu p_2^\rho p_3^\sigma\,,\nonumber\\
& & A_\mu^{0}\,\equiv\,\Big< (KK\pi)^0\Big|\frac{\overline{u}\gamma_\mu u-\overline{d}\gamma_\mu d}{\sqrt{2}}\Big|0\Big>\,=\,A^0\epsilon_ {\mu\nu\rho\sigma}p_1^\nu p_2^\rho p_3^\sigma\,.
\end{eqnarray}
\hspace*{0.5cm}In general we will have
\begin{equation}
\bra (KK\pi)^-|=\frac{a_1}{\sqrt{2}}\bra K^+K^-\pi^- + K^0 \overline{K}^0 \pi^-|+\frac{a_2}{\sqrt{2}}\bra K^+K^-\pi^- - K^0 \overline{K}^0 \pi^-|+a_3\bra K^0K^-\pi^0|\,,
\end{equation}
in such a way that
\begin{equation}
A^-\,=\,a_1 A_1^-+a_2 A_2^-+a_3 A_3^-\,=\,\frac{a_1+a_2}{\sqrt{2}} A^{+--}\,+\,\frac{a_1-a_2}{\sqrt{2}} A^{00-}\,+\,a_3 A^{0-0}\,.
\end{equation}
Since
\begin{equation}
\frac{A^{+--}-A^{00-}}{\sqrt{2}}\,=\,-A^{0-0}\Rightarrow A_2^-=-A^{0-0}\,,
\end{equation}
we can easily solve for the $A_i^-$:
\begin{equation}
A_1^-\,=\,\frac{A^{+--}+A^{00-}}{\sqrt{2}}\,,\quad A_2^-\,=\,\frac{A^{+--}-A^{00-}}{\sqrt{2}}\,,\quad A_3^-\,=\,A^{0-0}\,.
\end{equation}
Following Eq.~(\ref{factoring_Lorentzstructure_Vectorcurrent}) for naming the amplitudes of the different charge channels we have finally
\begin{equation} \label{relation_all_amplitudes_KKpi}
A^{+0-}\,=\,A^{0-+}\,=\,A^{0-0}\,,\quad A^{+-0}\,=\,A^{000}=\frac{1}{2}\left(A^{+--}+A^{00-}\right)\,=\,\frac{1}{\sqrt{2}}\,A_1^-\,.
\end{equation}
Summing over all $\tau^-$ charge channels one has
\begin{equation}
|A^{+--}|^2+|A^{00-}|^2+|A^{0-0}|^2\,=\,|A_1^-|^2+|A_2^-|^2+|A_3^-|^2\,=\,2|A_3^-|^2+|A_1^-|^2\,,
\end{equation}
whereas doing it over the four neutral channels reached in $e^+e^-$ annihilations with $I=1$ we have
\begin{equation}
|A^{+0-}|^2+|A^{0-+}|^2+|A^{+-0}|^2+|A^{000}|^2\,=\,2|A^{0-0}|^2+2\left(\frac{1}{2}\right)^2|A^{+--}+A^{00-}|^2\,=\,2|A_3^-|^2+|A_1^-|^2\,.
\end{equation}
Using the above relations one can obtain the isovector component of the process $e^+e^-\to K^0 K^- \pi^+$ using the form factors computed for 
$\Gamma\left(\tau^-\to K^0K^-\pi^0\nu_\tau\right)$.
\begin{equation}
\frac{\mathrm{d}\Gamma\left(\tau^-\to K^0 K^- \pi^0 \nu_\tau\right)}{\mathrm{d}Q^2}|_{\mathrm{Vector}}\,=\,f(Q^2)\,\sigma|_{I=1}\left(e^+e^-\to K^0 K^-\pi^+\right)\,,
\end{equation}
where $f(Q^2)$ was defined in Eq.~(\ref{fQ2}). One can also establish a similar relation including linear combinations of decay channels. Namely
\begin{eqnarray}
& & \frac{\mathrm{d}\Gamma\left(\tau^-\to K^0 K^- \pi^0 \nu_\tau\right)}{\mathrm{d}Q^2}|_{\mathrm{Vector}}\,+\,\frac{\mathrm{d}\Gamma\left(\tau^-\to K^0 K^- \pi^0 \nu_\tau\right)}{\mathrm{d}Q^2}|_{\mathrm{Vector}}\,=\nonumber\\
& & f(Q^2)\,\left[\sigma|_{I=1}\left(e^+e^-\to K^0 K^-\pi^+\right)\,+2\sigma|_{I=1}\left(e^+e^-\to K^0 K^-\pi^+\right)\right]\,.
\end{eqnarray}
Summing all charge channels one finds
\begin{equation} \label{masterformula_e+e-tau_KKpi}
\sum_ {i=1}^3 \frac{\mathrm{d}\Gamma\left(\tau^-\to (K K \pi)^- \nu_\tau\right)}{\mathrm{d}Q^2}|_{\mathrm{Vector}}\,=\,f(Q^2)\,\sum_{i=1}^4\sigma|_{I=1}\left(e^+e^-\to (K K\pi)^0\right)\,.
\end{equation}
As a byproduct we have obtained the relations
\begin{equation}
F_i\left(\tau^-\to K^+ K^- \pi^- \nu_\tau\right)-F_i\left(\tau^-\to K^0 \overline{K}^0 \pi^- \nu_\tau\right)\,=\,-\sqrt{2} F_i\left(\tau^-\to K^0 K^- \pi^0 \nu_\tau\right)\,.
\end{equation}
We have checked that our form factors in Chapter \ref{KKpi} satisfy this constraint.\\
\hspace*{0.5cm}We emphasize that isospin symmetry alone is not able to relate the isovector component of $\sigma\left(e^+ e^- \to K_S K^\pm \pi^\mp\right)$
 with the sum of all $KK\pi$ $I=1$ contributions of the $e^-e^+$ cross section. For that, the experimental collaborations use to employ the relation\\
\begin{equation} \label{factor of 3}
\sigma \left(e^+e^-\to K K\pi\right)\,=\,3\, \sigma \left(e^+e^-\to K_S K^\pm\pi^\mp\right)\,.
\end{equation}
\hspace*{0.5cm}However, even using all available isospin relations it is not possible to express $\sigma \left(e^+e^-\to K K\pi\right)$ in terms of the corresponding 
cross section for a single charge channel.
\hspace*{0.5cm}However, Eq.~(\ref{factor of 3}) can be justified following the arguments that we explain at the end of this section.\\
\hspace*{0.5cm}For this we will need to take into account not only the $I=1$ component of the electromagnetic current -as before- but also the isoscalar part. 
The application of the corresponding current, $\Gamma^\mu_{(0)}$ in Eq. (\ref{Gammas_em}), on the $|2,0\ket$ and $|1,0\ket$ states allows to obtain nontrivial 
relations between the corresponding isoscalar ($^{(0)}$) electromagnetic amplitudes:
\begin{equation}
 A_{+-0}^{(0)}\,=\,-A_{0\bar{0}0}^{(0)}\,,\quad A_{+\bar{0}-}^{(0)}\,=\,-A_{0-+}^{(0)}\,,\quad \sqrt{2} (A_{+-0}^{(0)}-A_{0\bar{0}0}^{(0)})\,=\,A_{+\bar{0}-}^{(0)}-A_{0-+}^{(0)}\,,
\end{equation}
which yields
\begin{equation}
 \sqrt{2}A_{+-0}^{(0)}\,=\,A_{+\bar{0}-}^{(0)}\,=\,-A_{0-+}^{(0)}\,=\,-\sqrt{2}A_{0\bar{0}0}^{(0)}\,.
\end{equation}
 Adding this information to the relations found previously one is still unable to reproduce Eq.~(\ref{factor of 3}).\\
However, now we proceed in a different way. We do not consider the $KK\pi$ state as a $1/2\times1/2\times1$ isospin state in our reasoning. Since the 
$K^*$ contribution dominates over that of the $\rho$, $\omega$ and $\phi$ in the hadronic matrix elements of interest, we can consider the composition 
$K \times \pi$ and then keep only its $I=1/2$ component, corresponding to the $K^*$. In addition, the processes with charged and neutral pions can be distinguished 
at detection, which makes that the following amplitudes should be considered independently \cite{Aubert:2007ym} (we will be writing the $K\pi$ pair making 
the $K^*$ as the last two particles until the end of this section): $K^+K^-\pi^0$, $K^0\bar{K}^0\pi^0$, $K^0K^-\pi^+$ and $K^+\bar{K}^0\pi^-$. In addition, 
we will considered the $C$-parity conjugated decays. Proceeding this way one finds the following amplitudes (we call $B_0$ and $B_1$ the participating 
isoscalar and isovector amplitudes):
\begin{eqnarray} \label{amplitudesKKpi_B}
 A\left(K^+K^-\pi^0\right)\,=\,-\frac{B_0+B_1}{\sqrt{6}}\,,\quad A\left(K^0\bar{K}^0\pi^0\right)\,=\,\frac{B_0-B_1}{\sqrt{6}}\,,\;\nonumber\\
 A\left(K^0K^-\pi^+\right)\,=\,\frac{B_1-B_0}{\sqrt{3}}\,,\quad A\left(K^+\bar{K}^0\pi^-\right)\,=\,-\frac{B_0+B_1}{\sqrt{3}}\,,
\end{eqnarray}
while for the $C$-conjugated amplitudes one finds (we introduce the amplitudes $C_0$ and $C_1$):
\begin{eqnarray} \label{amplitudesKKpi_C}
 A\left(K^-K^+\pi^0\right)\,=\,\frac{C_1-C_0}{\sqrt{6}}\,,\quad A\left(\bar{K}^0K^0\pi^0\right)\,=\,\frac{C_0+C_1}{\sqrt{6}}\,,\;\nonumber\\
 A\left(\bar{K}^0K^+\pi^-\right)\,=\,-\frac{C_1+C_0}{\sqrt{3}}\,,\quad A\left(K^-K^0\pi^+\right)\,=\,\frac{C_1-C_0}{\sqrt{3}}\,,
\end{eqnarray}
Summing up the first, second, third and four relations in Eqs. (\ref{amplitudesKKpi_B}) and (\ref{amplitudesKKpi_C}) in pairs one obtains the relations
\begin{eqnarray}\label{crosssections_Davier}
& & \sigma\left(e^+e^-\to K^+\bar{K}^0\pi^- + e^+e^-\to K^-K^+\pi^0\right)\,=\,\frac{1}{6}\,|A_0-A_1|^2\,,\nonumber\\
& & \sigma\left(e^+e^-\to K^0\bar{K}^0\pi^0 + e^+e^-\to \bar{K}^0K^0\pi^0\right)\,=\,\frac{1}{6}\,|A_0+A_1|^2\,,\nonumber\\
& & \sigma\left(e^+e^-\to K^0K^-\pi^+ + e^+e^-\to \bar{K}^0K^+\pi^-\right)\,=\,\frac{1}{3}\,|A_0+A_1|^2\,,\nonumber\\
& & \sigma\left(e^+e^-\to \bar{K}^0K^+\pi^- + e^+e^-\to K^-K^0\pi^+\right)\,=\,\frac{1}{3}\,|A_0-A_1|^2\,,
\end{eqnarray}
where $A_0 \equiv B_0 + C_0$ and $A_1 \equiv – B_1 + C_1$ have been introduced. Summing up all Eqs. in (\ref{crosssections_Davier}) gives
\begin{equation}
 \sigma\left(e^+e^-\to K K \pi\right) = |A_0|^2 + |A_1|^2\,,
\end{equation}
and adding the last two Eqs. in (\ref{crosssections_Davier}) yields
\begin{equation}
 \sigma\left(e^+e^-\to KK \pi^\pm\right) = \frac{2}{3} \left(|A_0|^2 + |A_1|^2\right)
\end{equation}
and using that $K_S = \left(K^0 - \bar{K}^0\right)/\sqrt{2}$ one gets finally
\begin{equation}
 3 \,\sigma\left(e^+e^-\to K_S K^\pm \pi^\mp\right) = |A_0|^2 + |A_1|^2 = \sigma\left(e^+e^-\to K K \pi\right)\,,
\end{equation}
which is Eq. (\ref{factor of 3}). Since $BaBar$ \cite{Aubert:2007ym} manages to split the $I=0$ and $I=1$ components of $\sigma\left(e^+e^-\to K_S K^\pm \pi^\mp\right)$, 
and thus to measure $|A_0|^2/3$ and $|A_1|^2/3$, it is straightforward to obtain $\sigma|_{I=1}\left(e^+e^-\to K K \pi\right)$.\\
What are the approximations employed in order to get this relation? In addition to the well supported $SU(2)$ symmetry and $K^*$ dominance, there is a 
source of error given by the definition employed for the $K^*$. To give an example, and as commented at the beginning of this section, the states $K^+ K^- \pi^0$ and 
$K^- K^+ \pi^0$ are the same in a $KK\pi$ analysis while this is not the case in a $KK^*$ study. One could argue that since the $K^*$ is quite narrow, this 
approximation is justified.\\
\section{$\eta\pi\pi$ channels} \label{isosApp_etapipi}
\hspace*{0.5cm}Since both the $\eta_8$ and the $\eta_1$ are
$SU(2)$-singlets, we can compute the isospin relations between $\eta_{1,8}\pi\pi$
channels just by taking into account the isospin of the $\pi\pi$ states. We will use $\eta$ to denote either state irrespectively. The first study 
of isospin relations for this and related modes was carried out in Ref.~\cite{Gilman:1987my}. Our results are, to our knowledge, new.\\
\hspace*{0.5cm}The different states $\eta_{1,8}\pi\pi$ that can be produced in
$\tau$ decays and $e^+e^-$ collisions are the following:
\begin{eqnarray} \label{charge states function of isospin states}
 |\pi^+\pi^-\ket & = & |1,+1\ket \otimes |1,-1\ket= \frac{1}{\sqrt{6}}|2,0\ket+\frac{1}{\sqrt{2}}|1,0\ket+\frac{1}{\sqrt{3}}|0,0\ket\,,\nonumber\\
 |\pi^-\pi^+\ket & = & |1,-1\ket \otimes  |1,+1\ket=\, \frac{1}{\sqrt{6}}|2,0\ket-\frac{1}{\sqrt{2}}|1,0\ket+\frac{1}{\sqrt{3}}|0,0\ket,\nonumber\\
 |\pi^0\pi^0\ket & = & |1,0\ket \otimes |1,0\ket=\,\sqrt{\frac{2}{3}}|2,0\ket-\frac{1}{\sqrt{3}}|0,0\ket,\nonumber\\
 |\pi^-\pi^0\ket & = & |1,-1\ket \otimes |1,0\ket=\,\frac{1}{\sqrt{2}}\left( |2,-1\ket-|1,-1\ket\right) ,\nonumber\\
 |\pi^0\pi^-\ket & = & |1,0\ket \otimes |1,-1\ket=\,\frac{1}{\sqrt{2}}\left( |2,-1\ket+|1,-1\ket\right),\nonumber\\
 |\pi^+\pi^0\ket & = & |1,+1\ket \otimes |1,0\ket=\,\frac{1}{\sqrt{2}}\left( |2,+1\ket+|1,+1\ket\right),\nonumber\\
 |\pi^0\pi^+\ket & = & |1,0\ket \otimes |1,+1\ket=\,\frac{1}{\sqrt{2}}\left( |2,+1\ket-|1,+1\ket\right).
\end{eqnarray}
\hspace*{0.5cm}Solving for the $|I,I_3\ket$ states yields:
\begin{eqnarray} \label{isospin states function charge states}
 |2,0\ket & = & \frac{1}{\sqrt{6}}\left( |\pi^+\pi^-\ket+|\pi^-\pi^+\ket+2|\pi^0\pi^0\ket\right) \,,\nonumber\\
 |1,0\ket & = & \frac{1}{\sqrt{2}}\left( |\pi^+\pi^-\ket-|\pi^-\pi^+\ket\right) \,,\nonumber\\
 |0,0\ket & = & \frac{1}{\sqrt{3}}\left( |\pi^+\pi^-\ket+|\pi^-\pi^+\ket-|\pi^0\pi^0\ket\right) \,,\nonumber\\
 |2,-1\ket & = & \frac{1}{\sqrt{2}}\left( |\pi^-\pi^0\ket+|\pi^0\pi^-\ket\right) \,,\nonumber\\
 |1,-1\ket & = & \frac{1}{\sqrt{2}}\left( |\pi^0\pi^-\ket-|\pi^-\pi^0\ket\right) \,,\nonumber\\
 |2,+1\ket & = & \frac{1}{\sqrt{2}}\left( |\pi^+\pi^0\ket+|\pi^0\pi^+\ket\right) \,,\nonumber\\
 |1,+1\ket & = & \frac{1}{\sqrt{2}}\left( |\pi^+\pi^0\ket-|\pi^0\pi^+\ket\right) \,.
\end{eqnarray}
Using Eqs.~(\ref{rels_iso_ops_1}), (\ref{rels_iso_ops_2}), (\ref{rels_iso_ops_3}) one has:
\begin{equation}
_{(I,-1)}\bra \eta\pi\pi|\overline{d}\,\Gamma_\mu\, u|0\ket=-_{(I,0)}\Big<
\eta\pi\pi\Big|\frac{\overline{u}\,\Gamma_\mu\, u\,-\,\overline{d}\,\Gamma_\mu \,d}{\sqrt{2}}\Big|0\Big>\,.
\end{equation}
\hspace*{0.5cm} Now if we denote by $T_{-0},T_{0-},T_{+-},T_{-+},T_{00}$ the
amplitudes $\bra \eta\pi\pi|\overline{d}\Gamma_\mu u|0\ket$ and $\Big< \eta\pi\pi\Big|\frac{\overline{u}\Gamma_\mu u-\overline{d}\Gamma_\mu d}{\sqrt{2}}\Big|0\Big>$ for 
charge $\eta\pi\pi$ states, we obtain the following relations
\begin{eqnarray}
& & \frac{1}{\sqrt{2}}\left(T_{-0}+T_{0-}\right) = -\frac{1}{\sqrt{6}}\left(
T_{+-}+T_{-+}-2T_{00}\right) = 0\,,
\nonumber\\
& & \frac{1}{\sqrt{2}}\left( T_{0-}-T_{-0}\right) =
-\frac{1}{\sqrt{2}}\left( T_{-+}-T_{+-}\right)\,,
\nonumber\\
& &  \sqrt{3}\left( T_{+-}+T_{-+}-T_{00}\right) =  0\,,
\end{eqnarray}
which lead to
\begin{equation}
 T_{00}=0\;\;,T_{+-}=-T_{-+}=T_{0-}=-T_{-0}\,.
\end{equation}
\hspace*{0.5cm}Now let us consider the electromagnetic current. One can
decompose it into $I=0$ and $I=1$ pieces:
\begin{equation}
 \Gamma^\mu=\frac{1}{3}\left( 2\overline{u}\gamma^\mu
u-\overline{d}\gamma^\mu d-\overline{s}\gamma^\mu s\right) =
\Gamma^\mu_{(0)}+\Gamma^\mu_{(1)}\,,
\end{equation}
where
\begin{equation}
 \Gamma^\mu_{(0)}=\frac{1}{6}\left(\overline{u}\gamma^\mu
u+\overline{d}\gamma^\mu d-2\overline{s}\gamma^\mu
s\right)\,,\,\,\Gamma^\mu_{(1)}=\frac{1}{2}\left(\overline{u}\gamma^\mu
u-\overline{d}\gamma^\mu d\right)\,.
\end{equation}
\hspace*{0.5cm}In general we will have
\begin{equation}
 \bra (\eta\pi\pi)^0| = A_{+-}\bra \eta\pi^+\pi^-| + A_{-+}\bra \eta\pi^-\pi^+| + A_{00}\bra \eta\pi^0\pi^0|\,.
\end{equation}
Using the decomposition in Eq.~(\ref{charge states function of isospin states}) one can relate the amplitudes $\bra
\eta\pi\pi|\Gamma^\mu|0\ket$ for charge and isospin $|\eta\pi\pi\ket$ states
as:
\begin{equation}
A_{+-}=\frac{1}{\sqrt{2}}A_1+\frac{1}{\sqrt3}A_0\,,\quad
A_{-+}=-\frac{1}{\sqrt{2}}A_1+\frac{1}{\sqrt3}A_0\,,\quad
A_{00}=-\frac{1}{\sqrt{3}}A_0\,.
\end{equation}
Moreover the vanishing of the amplitude $A_2$ implies
\begin{equation}
 2A_{00}+A_{+-}+A_{-+}=0\,.
\end{equation}
\hspace*{0.5cm}In this way one obtains the following relations:
\begin{eqnarray}
A_{+-}+A_{00}=\frac{1}{\sqrt{2}}A_1\,,\;
A_{-+}+A_{00}=-\frac{1}{\sqrt{2}}A_1\,,\;
A_1=\frac{A_{+-}-A_{-+}}{\sqrt{2}}\,,
\nonumber\\
A_0=\frac{A_{+-}+A_{-+}-A_{00}}{\sqrt{3}}=-\sqrt{3}A_{00}=\frac{\sqrt{3}}{2}\left(
A_{+-}+A_{-+}\right) \,,
\end{eqnarray}
which lead to
\begin{equation}
|A_{+-}+A_{-+}|^2+|A_{+-}-A_{-+}|^2= 2\left(
|A_{+-}|^2+|A_{-+}|^2\right)=4|A_{00}|^2+2|A_1|^2\,,
\end{equation}
\begin{equation}
 |A_1|^2=|A_{+-}|^2+|A_{-+}|^2-2|A_{00}|^2\,.
\end{equation}
\hspace*{0.5cm}The corresponding cross sections are related by \footnote{Because of $C$-parity, $\sigma\left(e^+e^-\to \eta\pi^0\pi^0\right)=0$ 
to $\mathcal{O}(\alpha)$, since $C_\gamma=-$ and $C_{\pi^0,\eta,\eta'}=+$. Although it is non-vanishing at higher orders, it can be safely neglected 
in all the low-energy applications we are considering.}
\begin{eqnarray} \label{I=1 component}
\!\!\!\!\!\!\!\sigma\left( e^+e^-\to \eta\pi\pi\right)|_{I=1} & =&
\sigma\left( e^+e^-\to \eta\pi^+\pi^-\right)+\sigma\left( e^+e^-\to \eta\pi^-\pi^+\right) \nonumber \\
 & & -\ 2\times2\;\sigma\left( e^+e^-\to
\eta\pi^0\pi^0\right)\,,\nonumber \\
& = & 2\,\sigma\left( e^+e^-\to \eta\pi^+\pi^-\right) - 4\,\sigma\left(e^+e^-\to \eta\pi^0\pi^0\right) \sim\nonumber \\
& \sim & 2\,\sigma\left( e^+e^-\to \eta\pi^+\pi^-\right) \,,
\end{eqnarray}
where the additional factor of $2$ in the above relation comes from the
identity of particles in the final state, introducing a factor of $1/2$ in
the angular integration. For the isoscalar part we have
\begin{equation}\label{I=0 component}
 \sigma\left( e^+e^-\to \eta\pi\pi\right)|_{I=0} = 6\, \sigma\left(
e^+e^-\to \eta\pi^0\pi^0\right)\sim0\,.
\end{equation}
\hspace*{0.5cm}Finally, one has
\begin{equation}
 \frac{1}{\sqrt{2}}\left(T^{0-}-T^{-0}\right)=\sqrt{2} T^{0-}=-\langle
1,0|\frac{\bar{u}u-\bar{d}d}{\sqrt{2}} |0\rangle =-\sqrt{2}A_1\,.
\end{equation}
\hspace*{0.5cm}Taking into account Eq.~(\ref{e+e-xsect3hads}) the cross-sections for the different modes read 
($|A_{+-}|^2\,=\,\frac{|A_{1}|^2}{2}\,=\,\frac{|A_{-0}|^2}{2}$)
\begin{eqnarray}
 \sigma\left( e^+e^-\to \eta\pi^+\pi^-\right) & = &
\frac{\alpha^2}{96\pi}\frac{1}{Q^6} \int \mathrm{d}s\; \mathrm{d}t\;
|T_{-0}|^2 \left( V_{3\mu} V^{3\mu *}\right) \,,
 \nonumber\\
 \sigma\left( e^+e^-\to \eta\pi^0\pi^0\right) & =  &
\frac{\alpha^2}{48\pi}\frac{1}{Q^6} \int \mathrm{d}s\; \mathrm{d}t\;
\frac{1}{2}|T_{00}|^2 \left( V_{3\mu} V^{3\mu *}\right)\sim0\,,
\end{eqnarray}
where the additional factor of $1/2$ in the second line comes from having
identical particles in the final state. \\
\hspace*{0.5cm}Using the former isospin relations one finally obtains
\begin{eqnarray}
 \frac{d\Gamma(\tau\to \eta\pi^-\pi^0\nu_\tau)}{\mathrm{d}Q^2} & = &
 f(Q^2) \, \sigma(e^+e^-\to \eta\pi\pi)|_{I=1} \nonumber \\
 & = & 2 f(Q^2) \left[\sigma( e^+e^-\to \eta\pi^+\pi^-) - 2\,\sigma
 (e^+e^-\to \eta\pi^0\pi^0)\right] \nonumber \\
 & \sim & 2 f(Q^2) \,\sigma( e^+e^-\to \eta\pi^+\pi^-)
\end{eqnarray}
where $f(Q^2)$ is given in Eq.~(\ref{fQ2}).\\
\section{Other channels} \label{isosApp_Others}
\subsection{$\eta\eta\pi$ channels} \label{isosApp_etaetapi}
\hspace*{0.5cm}Since the $\eta_{1,8}$ are $SU(2)$ singlets it is as if it was just $\eta\eta\pi\sim\pi$, where $\eta$ will be referring 
either to the singlet or the octet state here and in following sections.\\
\hspace*{0.5cm}Here we are concerned with the processes $e^+e^-\to\eta\eta\pi^0\sim|1,0\ket$ and $\tau^-\to\eta\eta\pi^-\nu_\tau\sim|1,-1\ket$.
 Considering that
\begin{equation}
_{(1,-1)}\bra\eta\eta\pi|\overline{d}\,\Gamma_\mu\, u|0\ket\,=\,-_{(1,0)}\Big<\eta\eta\pi\Big|\frac{\overline{u}\,\Gamma_\mu\, u-\overline{d}\,\Gamma_\mu\, d}{\sqrt{2}}\Big|0\Big>\,,
\end{equation}
the respective amplitudes ($T_0$ and $T_-$) are the same up to a sign. Using also that there is only isovector component in the considered 
$e^+e^-$ cross section we have
\begin{equation}
\frac{\mathrm{d}\Gamma\left(\tau\to \eta\eta\pi^-\nu_\tau\right)}{\mathrm{d}Q^2} \, = \,
 f(Q^2) \, \sigma\left(e^+e^-\to \eta\eta\pi^0\right)\,.
\end{equation}
\subsection{$\eta K K$ channels} \label{isosApp_etaKK}
\hspace*{0.5cm}Again, the $\eta$ can be ignored as far as isospin is concerned, so that we have $e^+e^-\to K^+ K^- \eta\sim| K^+K^-\ket$, 
$e^+e^-\to K^0 \overline{K}^0 \eta\sim| K^0\overline{K}^0\ket$ and $\tau \to \eta K^- K^0\sim| K^- K^0\ket$. Note that both kaons belong 
to different isospin multiplets, so the order is not important. Using the results in Eq.~(\ref{kaon_otimes_kaon}) we see that
\begin{equation}
| 0,0\ket\,=\,\Big|\frac{K^+K^- + K^0\overline{K}^0}{\sqrt{2}}\Big>\,,\quad |1,0 \ket\,=\,\Big|\frac{K^+K^- - K^0\overline{K}^0}{\sqrt{2}}\Big>\,,\quad |1,-1 \ket\,=\, |K^-K^0\ket\,.
\end{equation}
Using that
\begin{equation}
_{(I,-1)}\bra K K \pi|\overline{d}\,\Gamma_\mu\, u|0\ket\,=\,-_{(I,0)}\Big< K K \pi\Big|\frac{\overline{u}\,\Gamma_\mu\, u-\overline{d}\,\Gamma_\mu\, d}{\sqrt{2}}\Big|0\Big>\,,
\end{equation}
one gets
\begin{equation}
 T_{-0}\,=\,\frac{1}{\sqrt{2}}\left(T_{0\overline{0}}\,-\,T_{+-}\right)\,,
\end{equation}
for the weak amplitudes. Since the quark operators carry $I=1$, the amplitude associated to the production of $| 0,0\ket$ vanishes, so that
$T_{+-}+T_{0\overline{0}}\,=\,0$ and we have the following relations between the weak amplitudes
\begin{equation}
T_{-0}\,=\,-\sqrt{2}T_{+-}\,=\,\sqrt{2}T_{0\overline{0}}\,,
\end{equation}
that one can use to obtain the low-energy description of $e^+e^-\to \eta KK$ using the vector form factor computed in the $CVC$-related $\tau$ 
decay, $\tau\to\eta K^- K^0$, with amplitude $T_{-0}$.\\
\hspace*{0.5cm}Now we consider the electromagnetic current. The isospin amplitudes are
\begin{equation}
 A_1\,=\,\frac{A_{+-}\,-\,A_{0\overline{0}}}{\sqrt{2}}\,,\quad A_0\,=\,\frac{A_{+-}\,+\,A_{0\overline{0}}}{\sqrt{2}}\,.
\end{equation}
\hspace*{0.5cm}The isoscalar piece vanishes and we are only left with the isovector one. We have
 thus
\begin{equation}
\sigma\left(e^+e^-\to KK\eta\right)\,=\,2\sigma\left(e^+e^-\to K^+K^-\eta\right)\,.
\end{equation}
\\
\chapter*{Appendix F: Antisymmetric tensor formalism for meson resonances}
\label{antisymApp}
\addcontentsline{toc}{chapter}{Appendix F: Antisymmetric tensor formalism for meson resonances}
\appendix
\pagestyle{appendixf}
\newcounter{F}
\renewcommand{\theequation}{\Alph{F}.\arabic{equation}}
\setcounter{F}{6}
\setcounter{equation}{0}
\hspace*{0.5cm}The antisymmetric tensor formalism for spin-one fields was already developed in the sixties \cite{Kyriakopoulos:1969zm, Takahashi:1970ev}, 
although its generalization needed to wait until Gasser and Leutwyler proposed it to introduce the starring $\rho$ resonance in the chiral Lagrangian \cite{Gasser:1983yg}
 and a few years later, Ecker \textit{et al.} \cite{Ecker:1988te} took adventage of it for including the resonances in $\RCPT$. These benefits will be explained
 throughout this Appendix together with the main features of the formalism.\\
\hspace*{0.5cm}A crucial understanding was that -provided consistency with $QCD$ asymptotic behaviour- the physics given by the $EFT$ does not depend on the
 chosen formalism \cite{Ecker:1989yg}, which authorizes us to choose the tensor formalism for convenience.\\
\hspace*{0.5cm}In Ref.~\cite{Capri:1984dn,Capri:1986vb} it was proved that for massive antisymmetric tensor fields there are (up to multiplicative factors 
and a total divergence) only two possible Lagrangians of second order in derivatives, if one assumes the existence of a Klein-Gordon divisor. They correspond
 to having either the Lorentz condition or else the Bianchi identity satisfied by the fields. In the case of spin-$1$ particles one has the following two 
options ($W_{\mu\nu}=-W_{\nu\mu}$),
\begin{enumerate}
 \item The subsidiary condition is the Bianchi identity, i.e. $\epsilon^{\mu\lambda\rho\sigma}\partial_\lambda W_{\rho\sigma}=0$, and $W_{ik}$ are frozen,
 so the three dynamical degrees of freedom are $W_{i0}$, where $i=1,2,3$.
 \item The subsidiary condition is the Lorentz condition, $\partial^\rho W_{\rho\nu}=0$ and $W_{i0}$ are frozen, so the degrees of freedom are 
$W_{ij}$.
\end{enumerate}
\hspace*{0.5cm}For historical reasons the first option was chosen, as we will see in the following. We consider a Lagrangian quadratic in the antisymmetric 
tensor field $W_{\mu\nu}$,
\begin{equation} \label{massiveLag}
\mathcal{L}\,=\,a\,\partial^\mu W_{\mu\nu} \partial_\rho W^{\rho\nu} \,+\, b\,\partial^\rho W_{\mu\nu}
\partial_\rho W^{\mu\nu}\,+\,c\,W_{\mu\nu}W^{\mu\nu} \, ,
\end{equation}
where $a$, $b$ and $c$ are arbitrary constants. The field $W^{\mu\nu}$ contains six degrees of freedom. To describe massive spin-one particles we must reduce
 them to three corresponding to the physical polarizations these particles have. This can be done with a clever choice of $a$ and $b$. Indeed, consider the 
$EOM$
\begin{equation}
a\, (\partial^\mu\partial_\sigma W^{\sigma\nu}\,-\,\partial^\nu\partial_\sigma
W^{\sigma\mu})\,+\,2b\,\partial^\sigma\partial_\sigma W^{\mu\nu} \,-\,2c\,W^{\mu\nu}\,=\,0 \,  ,
\end{equation}
that can be splitted up into the time-spatial and spatial-spatial components:
\begin{eqnarray}
(a+2b)\ddot{W}^{0i}\,+\,a\partial_l \dot{W}^{li}\,-\,a\partial^i\partial_l W^{l0}\,-\,2(b\square\,+\,c)W^{0i}\,=\,0\,,\nonumber\\
2b\ddot{W}^{ik}\,+\,a\left[ \partial^i(\dot{W}^{0k}\,+\,\partial_l W^{lk})\right]\,-\,2(b\square\,+\,c)W^{ik}\,=\,0\,,
\end{eqnarray}
where the dots denote time derivatives and $\square$ stands for the Dalembertian operator. For $a\,+\,2b\,=\,0$, the three fields $W^{0i}$ do not propagate
 ($b\,=\,0$ freezes the spatial-spatial components, on the contrary). The $W^{\mu\nu}$ propagator, defined to be the inverse of the differential operator in 
(\ref{massiveLag}) contains poles in $k^2=-c/b$ and $k^2\,=\,-2c/(a+2b)$, which disappear for $b\,=\,0$, or $a\,+\,2b\,=\,0$, respectively. To maintain only 
one pole and reduce the number of degrees of freedom to three, we must choose among these two options. In \cite{Ecker:1988te}, it was preferred to fix $b\,=\,0$,
 and to choose $a$ and $c$ for the pole to correspond to the particle mass, that is, $a\,=\,-1/2$, and $c\,=\,M^2/4$. Then, the Lagrangian (\ref{massiveLag}) 
becomes
\begin{equation} \label{massiveLagdef}
\mathcal{L}\,=\,-\frac{1}{2}\partial^\mu W_{\mu\nu} \partial_\rho W^{\rho\nu} \,
+\frac{1}{4}M^2W_{\mu\nu}W^{\mu\nu} \, ,
\end{equation}
from which the free-case $EOM$ is
\begin{equation} \label{EOMfree}
\partial^\mu\partial_\sigma W^{\sigma \nu}\,-\,\partial^\nu\partial_\sigma
W^{\sigma\mu}\,+\,M^2W^{\mu\nu}\,=\,0 \, ,
\end{equation}
where only three degrees of freedom corresponding to a spin-one particle resonance of mass $M$ are described. Notice that the definition
\begin{equation}
W_\mu\,=\,\frac{1}{M}\,\partial^\nu W_{\nu\mu}\,,
\end{equation}
allows to recover from (\ref{EOMfree}) the familiar Proca equation
\begin{equation}
\partial_\rho(\partial^\rho W^\mu-\partial^\mu W^\rho)\,+\,M^2 W^\mu\,=\,0\,.
\end{equation}
\hspace*{0.5cm}From the Lagrangian (\ref{massiveLagdef}), one can derive the explicit expression for the resonance propagator
\begin{equation} \label{propagator}
\bra 0|T\left\{W_{\mu\nu}(x),\,W_{\rho\sigma}(y)\right\}|0\ket \,=\,
\int
\frac{\mathrm{d}^4k}{(2\pi)^4} e^{-ik(x-y)} 
\left\{\frac{2i}{M^2-q^2}\,\Omega^L_{\mu\nu,\rho\sigma} \,+\, \frac{2i}{M^2}\,
\Omega_{\mu\nu,\rho\sigma}^T\right\} \,,
\end{equation}
where the antisymmetric tensors
\begin{eqnarray}\label{omega1}
\Omega^L_{\mu\nu,\rho\sigma}(q)&=&\frac{1}{2q^2}\left( g_{\mu\rho}q_\nu q_\sigma\,-\,g_{\rho\nu}q_\mu
q_\sigma\,-\,(\rho\leftrightarrow\sigma)\right)\,,\nonumber\\
\Omega_{\mu\nu,\rho\sigma}^T(q)&=&-\frac{1}{2q^2}\left( g_{\mu\rho}q_\nu q_\sigma\,-\,g_{\rho\nu}q_\mu
q_\sigma\,-\,q^2g_{\mu\rho}g_{\nu\sigma}\,-\,(\rho\leftrightarrow\sigma)\right)\,,
\end{eqnarray}
have been defined. Upper-indices mean longitudinal or transversal polarizations. In order to identify the preceding operators with projectors over these 
polarizations, one needs to consider as a generalized identity in this space the tensor $\mathcal{I}_{\mu\nu,\rho\sigma}$,
\begin{eqnarray}\label{omega2}
\mathcal{I}_{\mu\nu,\rho\sigma}&=&\frac{1}{2}\left(g_{\mu\rho}g_{\nu\sigma}\,
-\,g_{\mu\sigma}g_{\nu\rho}\right)\,,
\end{eqnarray}
because any antisymmetric tensor, $\cal{A}_{\mu\nu}\,=\,-\cal{A}_{\nu\mu}$, fulfills
\begin{equation}
\mathcal{A}\cdot\mathcal{I}\,=\,\mathcal{I}\cdot\mathcal{A}\,=\,\mathcal{A} \,,
\end{equation}
and therefore the $\Omega^{L(T)}$ indeed verify projector properties
\begin{eqnarray}
\Omega^T\,+\,\Omega^L\,=\,\mathcal{I}\,,\,\,\,\,\Omega^T\cdot\Omega^L\,=\,\Omega^L\cdot\Omega^T\,=\,\,0\,,\nonumber \\
\Omega^T\cdot\Omega^T\,=\,\Omega^T\,,\,\,\,\,\Omega^L\cdot\Omega^L=\,\Omega^L\,.
\end{eqnarray}
\hspace*{0.5cm} The propagator (\ref{propagator}) corresponds to the normalization
\begin{equation}
\bra 0 | \, W_{\mu\nu} \,|W,p\ket \,=\,  
\frac{i}{M}[p_\mu\epsilon_\nu(p)\,-\,p_\nu\epsilon_\mu(p)] \, .
\end{equation}
\hspace*{0.5cm} Once we have seen the general properties of the antisymmetric tensor formalism and how it works, let us move to the second important issue: 
What is the advantage of using it instead of the more familiar Proca formalism? Working with the antisymmetric tensor formalism there is no need to consider
 $\mathcal{L}_4$ from $\CPT$ to give the $EFT$ the asymptotic behaviour ruled by $QCD$.\\
\hspace*{0.5cm} As an example of that, I will consider the same taken in Ref.~\cite{Ecker:1989yg}: the vector form factor of the pion.\\
\hspace*{0.5cm} Tree-level computation with (\ref{R}) -with antisymmetric tensor formalism, then- gives:
\begin{equation} \label{AtfpiFF}
\mathcal{F}(q^2)\,=\,1\,+\frac{F_VG_V}{F^2}\frac{q^2}{M_V^2-q^2} \, .
\end{equation}
\hspace*{0.5cm} Let us consider now the corresponding Lagrangian written in the Proca formalism that describes meson resonances \cite{Meissner:1987ge, Donoghue:1988ed} 
\footnote{A mixed formalim has been considered in Ref.~\cite{Kampf:2006yf}.},
\begin{equation}
\mathcal{L}^{\mathrm{Proca}}\,=\,\mathcal{L}^{\mathrm{Proca}}_{\mathrm{kin}}\,+
\,\mathcal{L}^{\mathrm{Proca}}_2
\,,
\end{equation}
where it has been defined
\begin{eqnarray} \label{ProcaR} 
\mathcal{L}^{\mathrm{Proca}}_{\mathrm{kin}}&=&-\frac{1}{4}\bra 
\hat{V}_{\mu\nu}\hat{V}^{\mu\nu}\,-\,2M_V^2\hat{V}_\mu\hat{V}^\mu \ket \,, \nonumber \\
\Delta{L}^{\mathrm{Proca}}_2&=&-\,\frac{1}{2\sqrt{2}}\left(
f_V\bra\hat{V}^{\mu\nu}f^+_{\mu\nu}\ket\,+\,ig_V\bra\hat{V}_{\mu\nu}[u^\mu,u^\nu]\ket \right) \, , \nonumber\\
\hat{V}_{\mu\nu}&=&\nabla_\mu \hat{V}_\nu \,-\, \nabla_\nu \hat{V}_\mu \, ,
\end{eqnarray}
and the hat identifies Proca formalism. For simplicity, only that part of the Lagrangian contributing to the considered form factor has been written. The 
result (\ref{ProcaR}) gives for the vector form factor of the pion is
\begin{equation} \label{PpiFF}
\widetilde{\mathcal{F}}^{\mathrm{Proca}}(q^2)\,=\,1\,+\,\frac{f_Vg_V}{F^2}\frac{(q^2)^2}{M_V^2-q^2} \, .
\end{equation}
\hspace*{0.5cm}$QCD$ short-distance behaviour ($q^2\rightarrow\infty$) dictates that the pion form factor must vanish in this limit \footnote{Formally, this comes
 from the analysis of the spectral function $\Im$m $\Pi_V(q^2)$ of the $I\,=\,1$ vector current two-point function. In the framework of $QCD$, one finds 
\cite{Floratos:1978jb} $\Im m\, \Pi_V(q^2)\,\to\,$constant as $q^2\to\infty$, from which it follows that $F(q^2)$ obeys a dispersion relation with at most 
one subtraction. In the narrow-width approximation for the exchanged $\rho$, all this drives directly to 
$\mathcal{F}(q^2)\,=\,1\,+\frac{\mathrm{const.}\times q^2}{M_V^2-q^2}$, as (\ref{AtfpiFF}).}. For (\ref{AtfpiFF}) this relates three $LECs$ as in Eq. 
(\ref{rest1}), $F_VG_V=F^2$, but for (\ref{PpiFF}) this behaviour is not possible unless we add to (\ref{ProcaR}) a local term. This one must have the structure
 of the term whose coefficient is $L_9$ in Eq. (\ref{L-Op4-u}). One then needs at the same time that $L_9^{\mathrm{Proca}}=\frac{1}{2}f_Vg_V$ and 
$f_Vg_V=F^2/M_V^2$ happen
\begin{eqnarray}
\mathcal{F}^{\mathrm{Proca}}(q^2)&=&1\,+\,\frac{f_Vg_V}{F^2}\frac{(q^2)^2}{M_V^2-q^2}\,
+\,\frac{2}{F^2}L_9^{\mathrm{Proca}}q^2 \,, \nonumber \\
L_9^{\mathrm{Proca}}&=&\frac{1}{2}f_Vg_V \, ,
\end{eqnarray}
where the tilde over the form factor has been removed because it has been corrected by the needed local terms discussed previously to guarantee the right 
asymptotic behaviour.\\
\hspace*{0.5cm} It can be shown \cite{Ecker:1989yg} that this finding in the case of the pion vector form factor is a general fact: Working with the tensor 
formalism there is no need to include the terms of $\mathcal{L}_4$ from $\CPT$,
\begin{equation}
L_i\,=\,0 \qquad i=1,2,3,9,10 \, \,,
\end{equation}
whereas for the Proca case we must include them fulfilling
\begin{eqnarray}
L_1^{\mathrm{Proca}}\,=\,\frac{1}{8}g_V^2 \,,\qquad
L_2^{\mathrm{Proca}}\,=\,\frac{1}{4}g_V^2 \,,\qquad
L_3^{\mathrm{Proca}}\,=\,-\frac{3}{4}g_V^2 \,,\nonumber\\
L_9^{\mathrm{Proca}}\,=\,\frac{1}{2}f_Vg_V \,,\qquad
L_{10}^{\mathrm{Proca}}\,=\,-\frac{1}{4}f_V^2 \,,
\end{eqnarray}
for vector resonances. Something similar \cite{Ecker:1989yg} happens for the axial-vectors, the other resonances that dominate phenomenology whenever they 
can be involved. Then, it is clear we can choose the formalism for describing resonances, and justify that it is more convenient to take the antisymmetric
 tensor formalism.\\
\hspace*{0.5cm} For completeness, I mention that there is another way of treating resonances: the so-called hidden-gauge formalism \cite{Bando:1984ej, Fujiwara:1984mp, Bando:1985rf, Bando:1987br}.
 This method is based on the freedom that exists to choose the representative of the coset $G/H$ of the chiral group $G$ over the vector subgroup.
 In the Hidden Local Symmetry model, vector mesons are regarded as authentic gauge bosons of a hidden symmetry of the Lagrangian that relates the different
 possible choices of the coset representative. However, it is not clear at all that vector mesons stand out from axial-vectors (in fact, VMD involves both),
 nor the gauge nature of resonances is not an artifact. At the end of the day, (pseudo)scalar resonances do exist and there is no natural procedure to 
include them in this model. The loop corrections in these theories \cite{Harada:1992bu, Harada:2003jx} give an ultraviolet behaviour that is much simpler
 than the one found in $\RCT$ \cite{Rosell:2007kc}.\\

\chapter*{Appendix G: Successes of the large-$\N$ limit of $QCD$}
\label{SuccessesApp}
\addcontentsline{toc}{chapter}{Appendix G: Successes of the large-$\N$ limit of $QCD$}
\appendix
\pagestyle{appendixg}
\newcounter{G}
\renewcommand{\thesection}{\Alph{G}.\arabic{section}}
\renewcommand{\thesubsection}{\Alph{G}.\arabic{subsection}}
\renewcommand{\theequation}{\Alph{G}.\arabic{equation}}
\renewcommand{\thetable}{\Alph{G}.\arabic{table}}
\setcounter{G}{7}
\setcounter{section}{0}
\setcounter{subsection}{0}
\setcounter{equation}{0}
\setcounter{table}{0}
\section{Introduction} \label{Sucess_LargeN_Intro}
\hspace*{0.5cm}In this appendix we will review the most important phenomenological successes of the large-$\N$ limit of $QCD$. First we will consider the results obtained in the 
limit $\N\to\infty$ limit of $QCD$ to understand some characteristic features of meson phenomenology. After that, we will review the r\^ole of the $1/\N$ expansion 
in the effective theories of $QCD$ for low and intermediate energies: $\CPT$ and $\RCT$. They come to complement the most relevant success of the large-$\N$ limit 
of $QCD$ for us, namely that it provides us with a framework able to describe exclusive hadron decays of the $\t$ as we have seen in this Thesis.\\
\section{Phenomenological successes of the large-$N_C$ expansion} \label{Sucess_LargeN_Phenomenological}
\begin{itemize}
\item There is a supression in hadron physics of the sea quark pairs, $\overline{q}\,q$. Therefore, mesons are pure $\overline{q}\,q$ states,
 thus exotic states such as $\overline{q}\,q\,\overline{q}\,q$ are eliminated in practice. In short, this is due to the fact that there are much more gluon
 than quark states. In the large-$\N$ limit sea quarks are negligible. Apart from that, in this limit, mesons do not interact, so any candidate to an exotic
 state must be, in fact, a set of ordinary states. Being exotic requires interaction for the state to be seen as a composite one; but mesons do not interact
 in the $\N\to\infty$ limit.
\item Confinement restricts hadron states to be singlets of colour. From the group theoretical point of view, it is clear \cite{Georgi:1982jb} that a quark
-antiquark state can be decomposed into a direct sum in the following way (all representations are in colour space):
\begin{equation} 
\underline{3}\otimes\underline{\overline{3}}\,=\,\underline{1}\oplus\underline{8}\,.
\end{equation}
\hspace*{0.5cm}Of course, the octet $\underline{8}$ cannot live as meson non-singlet of colour state. Still, it can combine with a partner to become 
$\underline{1}$:
\begin{equation}
\underline{8}\otimes\underline{8}\,=\,\underline{27}\oplus\underline{10}\oplus\underline{\overline{10}}\oplus\underline{8}\oplus\underline{8}\oplus\underline{1}\,.
\end{equation}
\hspace*{0.5cm}Zweig rule \cite{Lipkin:1981pq} states that this possibility is strongly suppressed, being greatly exceeded by one-gluon exchange among a meson 
and a sea quark-antiquark pair. For instance, this together with the conservation of all internal quantum numbers explains why the $J/\Psi$ has such a narrow 
decay width: six strong vertices are required for its decay. Zweig rule has the consequence that mesons are better described, in the large-$\N$ limit, as 
flavour $U(3)$-nonets, rather than as singlets plus octets, because this splitting involves annihilation diagrams among them that are suppressed in this limit
 \footnote{For $\N\rightarrow\infty$, the axial anomaly disappears and $U(n_f)_L\otimes U(n_f)_R$ is restored. Moreover, under very general assumptions, it 
can be shown that in the large-$\N$ limit, $U(n_f)_L\otimes U(n_f)_R$ is broken down to $U(n_f)_V$ \cite{Coleman:1980mx}.}. To conclude, gluon states decoupling
 is a result of all that: because they cannot be produced at $LO$ in $1/N_C$ as a product of reactions starting from hadrons or electroweak currents. Therefore,
 these states are not seen experimentally.
\item Meson decays are mostly of two-body type, because many final-state particle processes are less probable than those resonant decays into two intermediate
 particles. This is a natural consequence in the $1/N_C$-expansion. For a particle decaying into three mesons both processes are globally $\cO(1/\N)$; the 
point is that decaying directly is $\cO(1/\N)$, while the first vertex in the two-step process is $\cO(1/\sqrt{\N})$, and thus the two-body intermediate 
decay dominates over the direct one.
\item At first sight, it seemed a strange feature that the number of resonances becomes infinite in this limit. 
The big number of resonances that has been discovered and their relatively thin width can be taken as another phenomenological support of the
 large-$\N$ arguments.
\item Last but not least, the success of phenomenology describing strong interaction in the intermediate-energy region in terms of tree-level Feynman diagrams
 with hadrons as degrees of freedom, that is, the success of $EFTs$ -and particularly of $\RCPT$-, based on large$-\N$ arguments; is a recognition of the 1/$\N$
-expansion. $NLO$ corrections correspond to loop diagrams involving hadrons and provide the resonances with a finite width.\\
\end{itemize}
\section{1/$\N$ expansion for $\CPT$} \label{Sucess_LargeN_ChPT}
\hspace*{0.5cm}The 1/$\N$ expansion provides a well-defined counting scheme for $EFTs$ of $QCD$, exactly what we needed to develop an $EFT$ of the strong interaction in the 
intermediate-energy region involving light quarks. Although we do not need $1/N_C$ as an expansion parameter at very-low energies, it is justified to 
apply it to $\CPT$, to check the consistency of the expansion with a theory known to be successful. After that, we will have a strong test for any new $EFT$
 extending to higher energies. As we will see, for $\RCPT$ one gets reasonable results when subjected to this exam.\\
\hspace*{0.5cm}The main features of the effective theory relevant for the meson sector of $QCD$ in the large-$N_C$ limit were discovered long ago 
\cite{Witten:1980sp, Di Vecchia:1980ve, Rosenzweig:1979ay, Kawarabayashi:1980dp, Di Vecchia:1980sq}.
 The systematic analysis in the framework of $\CPT$ was taken up in Ref. \cite{Gasser:1984gg}, where the Green functions of $QCD$ were studied by means of a simultaneous 
expansion in powers of momenta, quark masses and $1/N_C$ (with $1/N_C\sim p^2\sim m_q$).\\
\hspace*{0.5cm}The dominant terms should be $\cO(\N)$, as they are the corresponding correlation functions among quark bilinears. A quark loop means a 
trace in Dirac, colour and flavour space. The last one supresses these kind of terms with respect to those without quark loops. One quark loop is needed to 
provide the quantum numbers of a meson, but each additional quark loop will be suppressed by a 1/$\N$ factor.\\
\hspace*{0.5cm}In the large-$N_C$ limit of $QCD$ the axial anomaly disappears, so that the spectrum does not correspond to $SU(3)$-multiplets anymore (octet
 and singlet, each one on its own), but to $U(3)$-multiplets, that is, nonets.\\
\hspace*{0.5cm}Therefore, at $LO$ in 1/$\N$ the axial anomaly vanishes and the eta singlet becomes the ninth $pG$:
\begin{eqnarray} \label{PhiGoldstonesLO1/N} 
\Phi(x)& = & \frac{1}{\sqrt{2}} \sum_{a=0}^8  \lambda_a \Phi^a\\
& = & \left( \begin{array} {ccc}
\frac{1}{\sqrt{2}}\pi^{0}+\frac{1}{\sqrt{6}}\eta_8+\frac{1}{\sqrt{3}}\eta_1 & \pi^{+} & K^{+} \\
\pi^{-} & -\frac{1}{\sqrt{2}}\pi^{0}+\frac{1}{\sqrt{6}}\eta_8+\frac{1}{\sqrt{3}}\eta_1 & K^{0} \\
K^{-} & \overline{K}^{0} & -\frac{2}{\sqrt{6}}\eta_8+\frac{1}{\sqrt{3}}\eta_1 \end{array} \right) \,,\nonumber
\end{eqnarray}
where the set of Gell-Mann matrices that are the generators of $SU(3)$ in the fundamental representation, has been enlarged by including the extra-generator
 of $U(3)$ proportional to the identity matrix: $\lambda_0\,=\,\sqrt{\frac{2}{3}}I_3$ \footnote{One can work however the large-$N_C$ limit either for a 
$SU(3)\otimes SU(3)$ theory \cite{Gasser:1984gg} of for a $U(3)\otimes U(3)$ theory \cite{HerreraSiklody:1996pm}. 
The first approach is taken in what follows. The matching between both was studied in Refs.~\cite{Kaiser:2000gs, HerreraSiklody:1998cr}.}.\\
\hspace*{0.5cm}At $LO$ in the chiral expansion, $\mathcal{L}_2$ has only two $LECs$: $F$ and $B$. The first one was defined in (\ref{F}), and it is the 
analogue of $f_n$ in the above discussion, so it is $F$ $\sim\cO(\sqrt{\N})$. $B$ was defined in (\ref{F,B}). Both the LHS and the RHS \footnote{The matrix 
element is of order $\cO\left(\sqrt{\N}^2\right)$, exactly the same as $F^2$.} are $\cO(\N)$, so $B\sim\cO(1)$. With these dependencies, we can check that 
scattering amplitudes behave as explained before. For instance, for $\pi\,\pi$ scattering, we have:
\begin{equation}
T\,=\,\frac{s-m_\pi^2}{F^2}\,\sim\,\frac{1}{\N}\,,
\end{equation}
$s\,=\,(p_{\pi,1}+p_{\pi,2})^2$, and has the right dependence. We conclude also that $\mathcal{L}_2$ has a global dependence of $\cO(\N)$ due to the common 
factor $\frac{F^2}{4}$. Each Goldstone field we add comes from the exponential divided by a factor $F$, giving thus the expected suppresion of $\cO(\sqrt{\N})$
 for every additional $pG$. $m$-meson interaction vertices go as $F^{2-m}$, so they are $\cO(\N^{1-n/2})$. Because of the global $\N$ factor in $\mathcal{L}_2$
 and the independence of the exponential on $N_C$, the expansion in $1/\N$ is equivalent to a semiclassical expansion for an $EFT$ whose degrees of freedom 
are hadrons \footnote{At first glance, (\ref{L_QCD_redefinedfields}) seems to imply -because of the global $N_C$ factor- that $QCD$ also reduces to a semiclassical
 theory in terms of quark and gluon fields in the large-$N_C$ limit, but this is not true, because the number of them increases as $N_C$ and $\sim N_C^2$, respectively. This is not the case for $\CPT$. As commented, the exponential including the $pGs$ does not introduce additional factors of $N_C$ as we 
include more and more of them.}. Quantum corrections computed with this Effective Lagrangian are suppressed by $1/\N$ for each loop.\\
\hspace*{0.5cm}The ten phenomenologically relevant $LECs$ at this chiral order are not expected to be of the same order in this expansion, because there are
 terms with only one trace in flavour space, and others with two; as it has been explained, each additional loop receives a supression of $1/\N$. Therefore,
 $L_3$, $L_5$, $L_8$, $L_9$ and $L_{10}$ would be $\,\cO(N_C)$, whereas $L_1$, $L_2$, $L_4$, $L_6$ and $L_7$ would be $\,\cO(1)$ -see Eq. (\ref{L-Op4-u})-.
 There is, however, a relation holding for traces of $3\times3$ matrices that modifies this na\"ive reasoning warning us that although $L_1$ and $L_2$ are,
 separately, $\,\cO(1)$; when considering the relation mentioned before, they get modified by $\delta L_i$:
\begin{equation}
2\delta L_1\,\sim\,\delta L_2\,\sim\, -\frac{1}{2} \delta L_3 \,\sim\, \cO(N_C)\,.
\end{equation}
\hspace*{0.5cm}Deciding not to consider the new term, both $L_1$ and $L_2$ become $\cO(N_C)$, but their combination $2L_1-L_2$ persists to be $\,\cO(1)$.
 Table \ref{L_i_LargeN} displays how the experimental values obtained for the $\cO(p^4)$ $LECs$ do agree with large-$N_C$ predictions \cite{Ecker:1994gg, Pich:2002xy} 
\footnote{A new global fit of the $L^r_i$ at next-to-next-to-leading order in Chiral Perturbation Theory has been performed recently \cite{Bijnens:2011tb}. One big change, 
with respect to the older predictions presented here in Table \ref{L_i_LargeN}, is that the fits do not have the expected behaviour in the limit of large $N_c$ as well as before.}.
\begin{table} \label{L_i_LargeN}
\begin{center}
\begin{tabular}{|c|c|c|c|c|}
\hline
i & $L_i^r(M_\rho)$ & $\cO(N_C)$ & source & $L_i^{N_C\rightarrow \infty}$ \\
\hline
$2L_1-L_2$ & $-0$.$6\pm0$.$6$  & $\cO(1)$ & $K_{e4}$, $\pi \pi \rightarrow \pi \pi$ &$0$.$0$ \\
$L_2$ & $1$.$4\pm0$.$3$ & $\cO(N_C)$ &$K_{e4}$, $\pi \pi \rightarrow \pi \pi$ &$1$.$8$ \\
$L_3$ & $-3$.$5\pm1$.$1$ & $\cO(N_C)$ &$K_{e4}$, $\pi \pi \rightarrow \pi \pi$ &$-4$.$3$\\
$L_4$ & $-0$.$3\pm0$.$5$ & $\cO(1)$ & Zweig's rule  &$0$.$0$ \\
$L_5$ & $1$.$4\pm0$.$5$ & $\cO(N_C)$ & $F_K\,:F_\pi$ &$2$.$1$ \\
$L_6$ & $-0$.$2\pm0$.$3$ & $\cO(1)$ & Zweig's rule &$0$.$0$ \\
$L_7$ & $-0$.$4\pm0$.$2$ & $\cO(1)$ & GMO, $L_5$, $L_8$ &$-0$.$3$\\
$L_8$ & $0$.$9\pm0$.$3$ & $\cO(N_C)$ &$M_\phi$, $L_5$ &$0$.$8$\\
$L_9$ & $6$.$9\pm0$.$7$ & $\cO(N_C)$ & $\bra r^2 \ket^\pi_V$ &$7$.$1$ \\
$L_{10}$ & $-5$.$5\pm0$.$7$ & $\cO(N_C)$ &$\pi \rightarrow e \nu \gamma$ &$-5$.$4$ \\
\hline
\end{tabular}
\caption{\small{Experimental values of the coupling constants $L_i^r(M_\rho)$ from the Lagrangian $\mathcal{L}_4$ in units of $10^{-3}$~\cite{Pich:2002xy}. The 
fourth column shows the experimental source employed. The fifth column shows the predictions that are obtained in the large-$N_C$ limit using the one-resonance approximation.}}
\end{center}
\end{table}
\hspace*{0.5cm}Summing up, all the $LECs$ appearing at $LO$ and $NLO$ in the chiral Lagrangians obey the following 1/$\N$ counting (the case of $L_7$ will 
be commented later on):
\begin{eqnarray}
B_0,\,\mathcal{M},\,m_{\pi,K,\eta},\,2L_1-L_2,\,L_4,\,L_6\, \sim \cO(1) \, , \nonumber \\
L_1,\,L_2,\,L_3,\,L_5,\,L_8,\,L_9,\,L_{10} \sim \cO(N_C) \, , \nonumber \\
F  \sim \cO\left(\sqrt{N_C}\right) \, .
\end{eqnarray}
\hspace*{0.5cm}The $LO$ chiral Lagrangian in the odd-intrinsic parity sector does not introduce any new $LEC$, but it has a global factor of $\N$ generated
 by the triangular quark loop over which the different number of colours run.\\
\section{1/$\N$ expansion for $\RCPT$} \label{Sucess_LargeN_RChT}
\hspace*{0.5cm}There are three kinds of checks we can perform for $\RCPT$. On the one hand, we can restore to phenomenology to fit the couplings entering its 
Lagrangian and predict another observables with the obtained values. This way has been exploited throughout the Thesis with optimistic results. On the other
 hand, as we have derived the $1/N_C$ expansion for $QCD$, we can apply it to $\RCPT$ in much the same way we did it for $\CPT$ and verify that large-$N_C$ 
estimates are not at variance with phenomenology. Finally, one can also explicitly check the convergence of the expansion by comparing the leading and 
next-to-leading orders in $1/N_C$, whenever the latter are available.\\
\hspace*{0.5cm}First of all, we will see the $1/\N$ expansion for $\RCPT$ and the relations that are derived among $\RCPT$ couplings in (\ref{Rkin}), (\ref{R}).
 The theory built upon the symmetries of $QCD^{n_f=3}$ that reproduces its low-energy behaviour is still not complete. A capital step is the matching 
procedure, as we have explained in Sec. \ref{EFT_Example}. We must enforce the theory to yield the asymptotic behaviour of the underlying theory, as it has 
been done repeatedly throughout this Thesis.\\
\hspace*{0.5cm}We aim to characterize the couplings appearing in $\mathcal{L}_R$. One can work with (\ref{Rkin}), (\ref{R}) written in a way that splits the 
singlet and the octet terms \cite{Ecker:1988te}. The obtained result is in complete agreement with convergence of octet plus singlet into nonet.\\
\hspace*{0.5cm}First of all, and according to the fact that meson decay constants are $\cO\left(\sqrt{\N}\right)$, and decay processes are given at $LO$ by 
tree level amplitudes, it is clear that couplings creating a resonance from the vacuum will be $\cO\left(\sqrt{\N}\right)$:  $F_V$, $F_A$, $c_m$, $\tilde{c}_m$,
 $d_m$ and $\tilde{d}_m$.\\
\hspace*{0.5cm}The other kind of processes we need to consider for completing the study are decays of resonances into $pGs$. Again, the decay of one vector
 or scalar resonance into two $pGs$ is $\cO\left(\sqrt{\N}\right)$, so $G_V$, $c_d$ and $\tilde{c}_d$ will be also of this order.\\
\hspace*{0.5cm}It was shown that masses have smooth limits in the large-$\N$ limit: they are $\cO(1)$.\\
\hspace*{0.5cm}In summary, the couplings entering $\RCPT$ Lagrangian are:
\begin{equation}
F_V,\,G_V,\,F_A,\,c_d,\,\tilde{c}_d,\,c_m,\,\tilde{c}_m,\,d_m,\tilde{d}_m \,\,\, \sim \, \cO\left(\sqrt{N_C}\right) \, ,
\qquad M_i \, \sim \, \cO(1) \, .
\end{equation}
\hspace*{0.5cm}At $LO$ in $1/\N$ Zweig rule becomes exact, the axial anomaly disappears and $U(3)_L\otimes U(3)_R$ is restored. For hadron spectroscopy this
 implies that particles fill nonet representations of $U(3)$ instead of octet plus singlet of $SU(3)$. In the large-$N_C$ limit one has the relations
\begin{equation} \label{nonet}
M_{S_1}\,=\,M_S\,, \quad |\tilde{c}_d|\,=\,\frac{|c_d|}{\sqrt{3}} \,, \quad 
|\tilde{c}_m|\,=\,\frac{|c_m|}{\sqrt{3}} \,, \quad 
M_{P_1}\,=\,M_P\,, \quad |\tilde{d}_m|\,=\,\frac{|d_m|}{\sqrt{3}} \,.
\end{equation}
\\
\hspace*{0.5cm}Now, I turn to examine how (axial-)vector contributions to $\cO(p^4)$ saturate the $LECs$ $L_i$ when integrated out, reproducing the notion 
of Vector meson dominance, proposed long ago \cite{Sakurai}. In fact, independent large-$\N$ analyses of $\CPT$ and $\RCPT$ yielded that most of the $L_i$ were 
$\cO(\N)$ and those in $\mathcal{L}_R$ were $\cO(\sqrt{\N})$. Because resonance exchange is then giving an $\cO(\N)$ contribution coming from the two vertices,
 it is plausible that in the large-$\N$ limit $\RCPT$ $LECs$ saturate $\CPT$ couplings.\\
\hspace*{0.5cm}Considering the $sra$, the obtained contributions are all $\cO(\N)$
\begin{eqnarray}
L_1 & = & \frac{G_V^2}{8M_V^2}\,-\,\frac{c_d^2}{6M_S^2}\,+\,\frac{\tilde{c}_d^2}{2M_{S_1}^2}\,,\,\, 
L_2 \, = \, \frac{G_V^2}{4M_V^2}\,,\,\,
L_3 \, = \, -\frac{3G_V^2}{4M_V^2}\,+\,\frac{c_d^2}{2M_S^2}\, , \nonumber \\
L_4 & = & -\frac{c_dc_m}{3M_S^2}\,+\,\frac{\tilde{c}_d\tilde{c}_m}{M_{S_1}^2}\,,\,\,
L_5 \, = \, \frac{c_dc_m}{M_S^2}\,,\,\,
L_6 \, = \, -\frac{c_m^2}{6M_S^2}\,+\,\frac{\tilde{c}_m^2}{2M_{S_1}^2}\,, \nonumber \\
L_7 & = & \frac{d_m^2}{6M_P^2}-\frac{\tilde{d}_m^2}{2M_{P_1}^2} \,,\,\,
L_8 \, = \, \frac{c_m^2}{2M_S^2}\,-\,\frac{d_m^2}{2M_P^2} \,,\,\,
L_9 \, = \, \frac{F_VG_V}{2M_V^2}\,, \nonumber \\
L_{10} & = & -\frac{F_V^2}{4M_V^2}\,+\,\frac{F_A^2}{4M_A^2}\,,\,\,
H_1 \, = \, -\frac{F_V^2}{8M_V^2}-\frac{F_A^2}{8M_A^2}\,,\,\,
H_2 \, = \, \frac{c_m}{M_S^2}\,+\,\frac{d_m^2}{M_P^2} \,. 
\end{eqnarray}
\hspace*{0.5cm}Taking into account the large-$\N$ relations (\ref{nonet}); $L_4$, $L_6$ and $L_7$ contributions vanish, while for $L_1$ only that coming 
from vectors survives. The suppresion of these $LECs$ and the saturation of all $L_i$ by (axial-)vector contributions are shown in Table \ref{VMDTable}.\\
\\
\begin{table}[h!]  
\begin{center}
\renewcommand{\arraystretch}{1.2}
\begin{tabular}{|c|c|c|c|c|c|c|c|c|}
\hline
i & $L_i^r\,(M_\rho)$ & $V$ & $A$ & $S$ & $\eta_1$ & Total & Total$^b$ & Total$^c$\\
\hline
$1$  & $0$.$4\pm0$.$3$  & $0$.$6$ & $0$.$0$ & $0$.$0$ & $0$.$0$ & $0$.$6$ & $0$.$9$ & $0$.$9$\\
$2$  & $1$.$4\pm0$.$3$  & $1$.$2$ & $0$.$0$ & $0$.$0$ & $0$.$0$ & $1$.$2$ & $1$.$8$ & $1$.$8$\\
$3$  & $-3$.$5\pm1$.$1$ & $-3$.$6$ & $0$.$0$ & $0$.$6$ & $0$.$0$ & $-3$.$0$ & $-4$.$9$ & $\lbrace -3$.$2,-4$.$3,-5$.$0\rbrace$ \\
$4$  & $-0$.$3\pm0$.$5$ & $0$.$0$ & $0$.$0$ & $0$.$0$ & $0$.$0$ & $0$.$0$ & $0$.$0$ & $0$.$0$\\ 
$5$  & $1$.$4\pm0$.$5$  & $0$.$0 $& $0$.$0$ & $1$.$4^a$ & $0$.$0$ & $1$.$4$ & $1$.$4$ & $2$.$2$\\
$6$  & $-0$.$2\pm0$.$3$ & $0$.$0$ & $0$.$0$ & $0$.$0$ & $0$.$0$ & $0$.$0$ & $0$.$0$ & $0$.$0$\\
$7$  & $-0$.$4\pm0$.$2$ & $0$.$0$ & $0$.$0$ & $0$.$0$ & $-0$.$3$ & $-0$.$3$ & $-0$.$3$ & $\lbrace -0$.$2,-0$.$3,-0$.$3\rbrace$\\
$8$  & $0$.$9\pm0$.$3$  & $0$.$0$ & $0$.$0$ & $0$.$9^b$ & $0$.$0$ & $0$.$9$ & $0$.$9$ & $\lbrace 0$.$6,0$.$8,1$.$5\rbrace$\\
$9$  & $6$.$9\pm0$.$7$  & $6$.$9^a$ & $0$.$0$ & $0$.$0$ & $0$.$0$ & $6$.$9$ & $7$.$3$ & $7$.$2$\\
$10$ & $-5$.$5\pm0$.$7$ & $-10$.$0$ & $4$.$0$ & $0$.$0$ & $0$.$0$ & $-6$.$0$ & $-5$.$5$ & $-5$.$4$\\
\hline
\end{tabular}
\caption{\small{Comparison between phenomenological values of the coupling constants $L_i^r(M_\rho)$ in units of $10^{-3}$ and the contributions given by 
resonance exchange~\cite{Pich:1998xt}. For scalar resonances it is considered a nonet and the contribution of pseudoscalar resonances is neglected with respect
 to the $\eta_1$ contribution. $^a$ stands for inputs and 
$^b,^c$ means that short-distance $QCD$ corrections have been taken into account. The last column corresponds to the reanalysis of Ref.\cite{Kaiser:2005eu}, where 
three different values for the parameter $d_m$ are considered. Essentially this is possible because there are less restrictions from high-energy $QCD$ behaviour 
in the spin-zero sector than in that with spin-one \cite{Pich:2002xy, Golterman:1999au}.}}
\label{VMDTable}
\end{center}
\end{table}
\\
\hspace*{0.5cm}Due to the $U(1)$ anomaly, even in the chiral limit, the $\eta_1$ has -apart from the common contribution coming from the trace anomaly \cite{Adler:1976zt}- the
 anomalous extra-term which motivates that commonly it is also integrated out from the standard $\CPT$ Lagrangian. Using the same notation for this coupling both in $\CPT$
 and in its large-$\N$ limit we can say that this provokes a change in $L_7$ due to pseudoscalar $\eta_1$-exchange. With the notation introduced before, we 
can simply write
\begin{equation}
L_7\,=\,-\frac{\tilde{d}_{\eta_1}^2}{2M^2_{\eta_1}}\,, \qquad 
\tilde{d}_{\eta_1}\,=\,-\frac{F}{\sqrt{24}} \, :
\end{equation}
being the extra contribution to $M_{\eta_1}\,\sim\,\cO(1/\N)$, $L_7$ -that is $\cO(1)$ in $1/\N$- grows to reach the value $\cO(\N^2)$ 
for what we
have written as $L_7$ in the previous equation. Still, the $1/\N$-counting is not so clear at this point \cite{Peris:1994dh}: I have explained how out of the chiral limit $M_{\eta_1}$ receives three comparable
 contributions: from explicit chiral symmetry breaking, from the singlet-axial anomaly and from the trace anomaly. To integrate the $\eta_1$ out amounts to 
admit that the axial anomaly contribution is much greater than the other two and this does not seem to be the case. For further discussions on $\eta_1$/$\eta_8$
 and their mixings $\eta$/$\eta'$, see \cite{DoubleAngleMixing,Escribano:2010wt,HerreraSiklody:1997kd, HerreraSiklody:1999ss}.
\\
\hspace*{0.5cm}In Table \ref{VMDTable} the experimental value of these couplings and the contributions got from resonance exchange \cite{Pich:1998xt} are 
presented. We see that there is good agreement between them and that Vector meson dominance emerges as a natural result of the analyses. There is no reason
 to include additional multiplets of resonances looking only at $\CPT$ at $\cO(p^4)$. The comparison has been made at a renormalization scale $\mu\,=\,M_\rho$ 
-for the $\CPT$ loops-, but similar results are found for any value belonging to the region of interest: $0$.$5$ GeV $\leq\,\mu\,\leq$ $1$ GeV.\\
\hspace*{0.5cm}Finally, we can see how these conclusions change when considering the evaluations of the $L_i$ at $NLO$ in $1/N_C$ within $\RCT$. The most 
important acquaintance we gain is that now one keeps full control of the renormalization scale dependence of these $LECs$. References \cite{Rosell:2006dt, Portoles:2006nr, Pich:2008jm, Rosell:2009yb, SanzCillero:2009ap, Pich:2010sm}
 constitute the study of this question within $\RCT$. By imposing $QCD$ short-distance constraints, the chiral couplings can be written in terms of the resonance
 masses and couplings and do not depend explicitly on the coefficients of the chiral operators in the Goldstone boson sector of $\RCT$. This is the counterpart
 formulation of the resonance saturation statement in the context of the resonance lagrangian. As an illustration, the values of the couplings $L_8$ and $L_{10}$ 
at $NLO$ in $1/N_C$ evaluated at $\mu=M_\rho$ are given: $L_8^r(M_\rho)=0$.$6\pm0$.$4$, $L_{10}^r(M_\rho)=-4$.$4\pm0$.$9$ (always in units of $10^{-3}$). We see that the
 corrections of the $NLO$ term amount to a reasonable $\left(20 \leftrightarrow30\right)\%$.\\
\chapter*{Appendix H: Comparing theory to data}\label{App_theorydata}
\addcontentsline{toc}{chapter}{Appendix H: Comparing theory to data}
\appendix
\pagestyle{appendixh}
\newcounter{H}
\renewcommand{\theequation}{\Alph{H}.\arabic{equation}}
\setcounter{H}{8}
\setcounter{equation}{0}
\hspace*{0.5cm}We include a brief note on how we have normalized our theoretical spectrum in order to compare it with experimental measurements.
 Let $P$ be the process $\t^-\to (\pi\pi\pi)^-\nu_\t$ and $x$ (an energy) the variable in which the spectrum is given. The experiment provides us
with the total number of events of process $P$, $N_P$,  and its spectrum in $x$, i.e., $\frac{Events}{bin}$ versus $x$, where $bin\,=\,\Delta x$.\\
\hspace*{0.5cm}Our theoretical computation yields $\frac{\mathrm{d}\Gamma_P}{\mathrm{d}x}$, whose integral over the whole $x$-spectrum gives
 the partial width of process $P$:
\begin{equation}
\int^{x_{min}}_{x_{max}}\mathrm{d}x\frac{\mathrm{d}\Gamma_P}{\mathrm{d}x}\,=\,\Gamma_P\longleftrightarrow \int^{x_{min}}_{x_{max}}\mathrm{d}x\frac{1}{\Gamma_P}\frac{\mathrm{d}\Gamma_P}{\mathrm{d}x}\,=\,1\,,
\end{equation}
and this allows us to compare with the experiment provided
\begin{equation}
\int^{x_{min}}_{x_{max}}\mathrm{d}x\frac{Events}{\Delta x}\,=\,N_P\,
\end{equation}
and therefore
\begin{equation}
N_P\int^{x_{min}}_{x_{max}}\mathrm{d}x\frac{1}{\Gamma_P}\frac{\mathrm{d}\Gamma_P}{\mathrm{d}x}\,=\,\int^{x_{min}}_{x_{max}}\mathrm{d}x\frac{Events}{\Delta x}\,,
\end{equation}
or in differential form:
\begin{equation} \label{zero}
\underbrace{N_P}_{exp}\left[ \frac{1}{\Gamma_P}\frac{\mathrm{d}\Gamma_P}{\mathrm{d}x}\right]_{th} \,=\,\left[ \frac{Events}{\Delta x}\right]_{exp} \,,
\end{equation}
where $exp$ and $th$ are a reminder of the source of each term: either the experiment, or the theoretical computation.\\
\hspace*{0.5cm}Usually, $x$ is a dimensionful variable, so we can write  $\Delta x\,=\,n\left[ x\right]$, where the $x$ in square brackets 
stands for the dimensions of $\Delta x$. Thus, the LHS of (\ref{zero}) has dimension $x^{-1}$ and, finally:
\begin{equation} \label{one}
\underbrace{N_P\,n}_{exp} \left[ \frac{1}{\Gamma_P}\frac{\mathrm{d}\Gamma_P}{\mathrm{d}x}\right]_{th} \,=\,\left[ \frac{Events}{\Delta x}\right]_{exp} \,,
\end{equation}
and $n$ is chosen according to the experimental information as we will see next.\\
\hspace*{0.5cm}Our computation yields $\frac{\mathrm{d}\Gamma_{\t^-\to \pi^+\pi^-\pi^-\nu_\t}}{\mathrm{d}Q^2}$, so what we have for comparing
 with the experiment is
\begin{equation}
\frac{1}{\Gamma_{\t^-\to (\pi\pi\pi)^-\nu_\t}}\frac{\mathrm{d}\Gamma_{\t^-\to (\pi\pi\pi)^-\nu_\t}}{\mathrm{d}Q^2}\,=\,\sum_{i\,=\,SA,\,A,}\frac{\mathrm{d}\Gamma_i}{\mathrm{d}Q^2}\,.
\end{equation}
One can reason similarly for other three meson modes and for other observables, like $d\Gamma/ds_{ij}$.\\
\backmatter
\chapter*{Bibliography}
\label{references}

\addcontentsline{toc}{chapter}{Bibliography}
\appendix
\pagestyle{references}

\chapter{Acknowledgements}
\pagestyle{acknowledgements}
\hspace*{0.5cm}Llega el momento m\'as deseado de la escritura de la Tesis, en el que no quedan ni siquiera peque\~{n}os retoques y uno se decide a escribir 
por fin los agradecimientos y dejar caer el punto y final.\\
\hspace*{0.5cm}Vull agrair-li a Jorge Portol\'es, el director d'aquesta Tesi, el haver pogut treballar amb ell aquestos anys, en un tema que des del 
comen\c{c}ament m'ha semblat molt interessant (percepci\'o que s'ha anat afian\c{c}ant amb el temps). Agraix de la seua direcci\'o que estimulara la meua 
independ\`encia i que en les discussions mai no em donara la ra\'o fins no estar totalment segur si la tenia (aspecte on hi ha hagut sana reciprocitat, 
val a dir-ho). Sense la seua revisi\'o aquesta Tesi contindria infinitament m\'es errates de les que hagen pogut quedar despr\'es del seu esfor\c{c}at 
recorregut sobre un document que --cal recon\`eixer-lo-- ha eixit un p\`el llarg. Li done tamb\'e les gr\`acies per haver-me donat llibertat per emprendre 
altres projectes durant aquesta Tesi i molt especialment, que sempre haja estat disposat a donar un bon consell i a ajudar-me.\\
\hspace*{0.5cm}A Toni Pich, el meu tutor, li dec la entrada al grup d'investigaci\'o PARSIFAL, primer que res. Les excepcionals classes, seues i de Jorge, de 
Part\'{\i}cules Elementals van ser un motiu de pes per entrar a aquest grup, eren un aut\`entic desperta-vocacions, des del meu punt de vista. Li agra\"isc 
tamb\'e que m'haja guiat en algunes decissions dif\'icils, sobre les estades finals de la Tesi i que sempre haja tingut un moment, malgrat tots els seus 
compromisos, per les cosses importants. Tant a Toni com a Jorge els hi dec el haver-me donat suport a l'hora d'encetar projectes i col·laboracions internacionals 
que han de resultar molt profitosos.\\
\hspace*{0.5cm}A Daniel G\'omez Dumm le agradezco su colaboraci\'on a lo largo de estos a\~{n}os. Fue un placer. Especial menci\'on requiere su revisi\'on del 
ap\'endice sobre relaciones de isosp\'in y gran parte del material del cap\'itulo octavo y por supuesto todo su esfuerzo para entender qu\'e aproximaciones 
hac\'ian los experimentales (y cu\'ales nosotros) al aplicar CVC en los canales $KK\pi$.\\
\hspace*{0.5cm}I thank Zhi-Hui Guo for the collaboration giving rise to the material included in chapter nine of this Thesis. I do not expect that many people 
understand this, but I remember how great it was to achieve very accurate agreement in the numerics while sitting on the floor in a corridor outside the 
last Flavianet meeting (of course during a coffee break).\\
\hspace*{0.5cm}I wish to thank Nora Brambilla and Antonio Vairo, whom I have visited in two different long periods. One while in Milano and the other one 
in M\"unchen. We also enjoyed time of collaboration in Valencia, where they came for some time as well. Our study of the lineshape in $J/\Psi\to\gamma \eta_c$ 
and related decays revealed as very interesting and offered me the oportunity to learn a lot about radiative transitions of Quarkonia.\\
\hspace*{0.5cm}I would like to thank S\'ebastien Descotes-Genon for his time and interest in our study of duality violations in tau decays and $e^+e^-$ annihilation 
into hadrons while I was in Orsay, from which I have learnt a lot.\\
\hspace*{0.5cm}I have been enjoying the collaboration with Olga Shekhovtsova, Tomasz Przedzi\'nski and Zbigniew Was for three years. I am happy with the results that we are obtaining 
and I hope that the project can extend for long since we will need it in order to complete a global description of the most important tau decay channels in TAUOLA. 
For this, the task of our experimental colleagues, Ian Nugent, Denis Epifanov and Vladimir Cherepanov, among others, is crucial.\\
\hspace*{0.5cm}It is a pleasure to thank Simon Eidelman for useful discussions and comments that made me learn a lot and for his help in accessing data and 
understanding its suitable interpretation (which was always the most difficult part).\\
\hspace*{0.5cm}Two years and a half ago I became member of the Working Group on Monte Carlo Generators and Radiative Corrections for Low-Energy Physics and 
since the first meeting I have felt like in a family (one of those in which peaceful harmony reins, I mean). I thank all more faithful participants for the 
interesting and participative discussions that were held and especially the WG convenors Henryk Czyz, Guido Montagna and Graziano Venanzoni for their 
always kind invitation to attend every meeting (despite the icelandic volcano) and Germ\'an Rodrigo, who helped my adaptation at the beginning.\\
\hspace*{0.5cm}I want to thank some people in PARSIFAL: Vicent Mateu, con el que compart\'i despacho tres a\~{n}os y con el que extra\~{n}amente nunca llegu\'e a
 hablar en Valenciano pues a las pocas palabras salt\'abamos inevitablemente al castellano. Con \'el sostuve much\'isimas discusiones muy interesantes en ese 
tiempo. Tambi\'en agradezco las discusiones con otros ex-miembros del grupo: Pedro Ruiz Femen\'{\i}a, Juanjo Sanz Cillero y Natxo Rosell. Con todos ellos he 
tenido la ocasi\'on de seguir debatiendo diversas cuestiones de Teor\'ias Quiral, de Resonancias y de QCD no relativistas a lo largo de los a\~{n}os. I 
remember with gratitude all the efforts Fr\'ed\'eric Jugeau made in order to help me prepare my first presentation to the group.\\
\hspace*{0.5cm}I also acknowledge the following researchers by useful advices and illuminating conversations at different conferences: Bogdan Malaescu, Ximo 
Prades -God rest his soul-, Matthias Jamin, Gabriel L\'opez Castro, Kim Maltman, Hisaki Hayashii, Alberto Lusiani, George Lafferty, Michel Davier, 
Kenji Inami, Rafel Escribano, Jaroslav Trnka, Hans K\"uhn, Maurice Benayoun, Bachir Moussallam, Gerard Ecker, Heiri Leutwyler, Eduardo de Rafael, Santi 
Peris, Joan Soto, Antonio Pineda, Gino Isidori, Jimmy MacNaughton, Kiyoshi Hayasaka, Ryan Mitchell, Yu Jia and Miguel \'Angel Sanch\'{\i}s.\\
\hspace*{0.5cm}Hay mucha otra gente a la que quiero agradecer su presencia a mi lado a lo largo de este periplo, creo que a todos les debo alguna ayuda 
inestimable o conversaci\'on valiosa (probablemente m\'as de una en cada caso): Miguel Nebot, Patricia Aguar, Josep Canet, Emma Torr\'o, Mercedes Mi\~{n}ano, 
Loli Jord\'an, Neus L\'opez, Joan Catal\`a, Carlos Escobar, Paco Salesa, Mar\'ia Moreno, Avelino Vicente, Alberto Aparici, Pere Masjuan, Oscar Cat\`a, 
Francisco Flores, Miguel \'Angel Escobedo, Max Stahlhofen, Xavier Garcia i Tormo, Emilie Passemar, Fabio Bernardoni, Alberto Filipuzzi, Paula Tuz\'on, 
David Greynat, David Palao, Alfredo Vega, Pancho Molina, Joel Jones, Jacopo Ghiglieri, Guillaume Toucas, Pablo Arnalte, Jos\'e Cuenca, Pepe R\'odenas, David Ruano, 
Juan Carlos Algaba, Javier Ors, Brian Mart\'inez, Sergio Fern\'andez, Carlos Navarrete, Julia Garayoa, Iv\'an Agull\'o, Jacobo D\'iaz, Enrique Fern\'andez, 
Ana C\'amara, Elena Fern\'andez, Antonio Lorenzo, Gin\'es P\'erez, Lorena Marset, Guillermo R\'ios, Rub\'en Garc\'ia, Jos\'e Miguel No y Dani Pe\~{n}a 
(a ellos tres no voy a incluirlos despu\'es en el apartado dedicado al Colegio de Espa\~{n}a, para evitar double counting), Diogo Boito, Alberto Ramos, 
Jorge Mond\'ejar, Xian-Wei Kang, Yoko Usuki, Anna Vinokurova, Houssine El-Mezoir y Thomas Laubrich.\\
\hspace*{0.5cm}Agradezco tambi\'en a Carlos Mart\'inez que resucitara mi port\'atil en un par ocasiones para evitar p\'erdida de documentos y que consiguiera 
hacer funcionar el ca\~{n}\'on del proyector tras un fallo inesperado previo a la presentaci\'on de mi Tesina. Tambi\'en le doy las gracias a Lauren por revivir 
mi port\'atil la \'ultima vez, con una quinta parte de la Tesis dentro y sin copias de seguridad hechas (como mandan los c\'anones).\\
\hspace*{0.5cm}Agradezco mi buena formaci\'on a los Profesores del Colegio Salesiano San Juan Bosco de la Avenida de la Plata y a los del Instituto Extensi\'on 
del San Vicente Ferrer (actual Blasco Ib\'a\~{n}ez). Muy especialmente a los buenos profesores que he tenido en la carrera (sin repetir nombres ya mencionados) que 
supieron recuperar mi inter\'es despu\'es de un inicio muy rutinario y aburrido: Jos\'e Bernab\'eu, Benito Gimeno, Jos\'e Adolfo de Azc\'arraga, 
Jos\'e Navarro Salas (con quien disfrut\'e una beca de introducci\'on a la investigaci\'on), Nuria Rius y Arcadi Santamar\'ia. A G\"unter Werth y Alexandros 
Drakoudis, con quienes tuve mi primera experiencia de iniciaci\'on a la investigaci\'on, en Mainz, a nuestros \textit{monis} entonces, a toda la gente querida 
que se qued\'o por all\'i, por Mil\'an, por Munich y por Orsay.\\
\hspace*{0.5cm}Por supuesto una parte importante de mis agradecimientos van para quienes no tienen nada que ver con la F\'isica (m\'as all\'a de lo que les 
haya salpicado indirectamente por conocerme). Lo primero agradecerle a mi madre (y a mi abuela, que en paz descanse) la buena educaci\'on que me dieron. Nada 
marca tan decisivamente a una persona y siempre agradecer\'e el que se me educara para serlo integralmente, tanto en conocimientos y habilidades como emocionalmente 
y en ser humano.\\
\hspace*{0.5cm}A Maru le agradezco que sea exactamente como es, hermosa \textit{flod} que perfectamente acopla con su arbolito, amorosos ellos.\\
\hspace*{0.5cm}Y, por supuesto, a todos mis amigos de la falla y de Salesianos, los que me han visto evolucionar y crecer, ser el Boig de las primeras salidas, 
el Tanque de las pachangas o Tarik el emigrante: Nacho, Manolo, Morgan, Dobi, Goti, Rafa, Boro, Mar\'{\i}a, Mar\'{\i}a (el lector atento 
podr\'a observar que la diferente graf\'ia denota dos personas distintas), Zalo, Gonzalo (nada que ver con la puntualizaci\'on anterior), Taribo, Pablo, 
Carmen, Elena, Virgi, Mar\'{\i}a, Inma, Josep, Bea, Falo, Vicent, Juanma, \'Angela, Capi, Luque, Paco, Toni, Primo, Dani, Cubatas (¿c\'omo poner \'Angel?).\\
\hspace*{0.5cm}La experiencia del Colegio de Espa\~{n}a ha sido maravillosamente enriquecedora (al margen de haber conocido a Maru en \'el, acontecimiento que 
se sale de la escala). Agradezco a su Director, Javier de Lucas, su implicaci\'on en la jornada divulgativa que all\'i coordin\'e y a Stephanie Mignot, por nuestra 
colabraci\'on en la organizaci\'on de charlas a cargo de los residentes durante todo el a\~{n}o. Tambi\'en agradezco el trato recibido de Ram\'on Sol\'e y 
V\'ictor Matamoro. Por supuesto agradezco a Juanjo Ru\'e, Jadra Mosa, Jovi Siles, Mar\'{\i}a Jos\'e L\'opez, Mercedes S\'anchez, Juliana Saliba, Paloma Palau, 
Paloma Guti\'errez y Santi Rello (se ha mantenido la pol\'itica de eliminar el doble agradecimiento) nuestra experiencia en el Comit\'e de Residentes del 
Colegio. Aunque hubo mucho trabajo, lo hicimos bien y lo pasamos mejor. Tambi\'en quiero tener un recuerdo especial para Gerardo, Ana, Sergio, Idoia-Idoia 
(una sola persona en este caso), Evita, Joaqu\'in, Marta, Laura, Geles, Estrella, los Patitos, Roberto, Ricard, el Cordob\'es, Roc\'io, Marta, Neus, Betlem, 
Mourad, Elipe, Chus, Javi, ...
\hspace*{0.5cm}A todos, gracias.

\end{document}